\pdfoutput=1
\documentclass[a4paper, 12pt]{report}
\usepackage[top=1.4in, bottom=1.4in, left=1.0in, right=1.0in]{geometry}
\usepackage{fancyhdr}
\usepackage{afterpage}
\usepackage{slashed}
\usepackage{textpos}
\usepackage[nottoc]{tocbibind}
\newcommand{\changefont}{%
    \fontsize{11}{13}\selectfont
}
\usepackage{ifthen}
\newboolean{pdflatex}
\setboolean{pdflatex}{true} 

\newboolean{uprightparticles}
\setboolean{uprightparticles}{false} 

\newboolean{inbibliography}
\setboolean{inbibliography}{false} 

\pdfoutput=1

\usepackage{microtype}
\usepackage{lineno}  
\usepackage{xspace} 
\usepackage{caption} 

\usepackage{graphicx}  
\usepackage{color}
\usepackage{colortbl}
\definecolor{bordeau}{rgb}{0.3515625,0,0.234375}
\graphicspath{{./figs/ppbar/}{./figs/mass/}{./figs/exp/}{./figs/synops/}{./figs/preprint/}{./figs/phiphi/}{./figs/lhcb/}{./Cover_PHENIICS/images/}{./Cover_PHENIICS/}{./figs/theory/}{./figs/thFits/}} 

\usepackage{amsmath} 
\usepackage{amssymb}
\usepackage{amsfonts}
\usepackage{upgreek} 

\newcommand*\patchAmsMathEnvironmentForLineno[1]{%
\expandafter\let\csname old#1\expandafter\endcsname\csname #1\endcsname
\expandafter\let\csname oldend#1\expandafter\endcsname\csname
end#1\endcsname
 \renewenvironment{#1}%
   {\linenomath\csname old#1\endcsname}%
   {\csname oldend#1\endcsname\endlinenomath}%
}
\newcommand*\patchBothAmsMathEnvironmentsForLineno[1]{%
  \patchAmsMathEnvironmentForLineno{#1}%
  \patchAmsMathEnvironmentForLineno{#1*}%
}
\AtBeginDocument{%
\patchBothAmsMathEnvironmentsForLineno{equation}%
\patchBothAmsMathEnvironmentsForLineno{align}%
\patchBothAmsMathEnvironmentsForLineno{flalign}%
\patchBothAmsMathEnvironmentsForLineno{alignat}%
\patchBothAmsMathEnvironmentsForLineno{gather}%
\patchBothAmsMathEnvironmentsForLineno{multline}%
\patchBothAmsMathEnvironmentsForLineno{eqnarray}%
}

\usepackage{hyperref}

\usepackage{xspace} 
\usepackage{upgreek}

\def\lhcb {\mbox{LHCb}\xspace}
\def\atlas  {\mbox{ATLAS}\xspace}
\def\cms    {\mbox{CMS}\xspace}
\def\alice  {\mbox{ALICE}\xspace}
\def\babar  {\mbox{BaBar}\xspace}
\def\belle  {\mbox{Belle}\xspace}
\def\cleo   {\mbox{CLEO}\xspace}
\def\cdf    {\mbox{CDF}\xspace}

\def\delphi {\mbox{DELPHI}\xspace}

\def\lthree {\mbox{L3}\xspace}

\def\cern {\mbox{CERN}\xspace}
\def\lhc    {\mbox{LHC}\xspace}
\def\lep    {\mbox{LEP}\xspace}
\def\tevatron {Tevatron\xspace}

\def\velo   {VELO\xspace}
\def\rich   {RICH\xspace}
\def\richone {RICH1\xspace}
\def\richtwo {RICH2\xspace}

\def\MagUp {\mbox{\em Mag\kern -0.05em Up}\xspace}

\def\hltone {HLT1\xspace}
\def\hlttwo {HLT2\xspace}

\ifthenelse{\boolean{uprightparticles}}%
{

 \def\Peta        {\ensuremath{\upeta}\xspace}

 \def\Pmu         {\ensuremath{\upmu}\xspace}

 \def\Ppi         {\ensuremath{\uppi}\xspace}

 \def\Pphi        {\ensuremath{\upphi}\xspace}                 
                  
 \def\Pchi        {\ensuremath{\upchi}\xspace}                 
 \def\Ppsi        {\ensuremath{\uppsi}\xspace}

 \def\PDelta      {\ensuremath{\Delta}\xspace}                 
 \def\PXi      {\ensuremath{\Xi}\xspace}                 
 \def\PLambda      {\ensuremath{\Lambda}\xspace}                 
 \def\PSigma      {\ensuremath{\Sigma}\xspace}                 
 \def\POmega      {\ensuremath{\Omega}\xspace}                 
 \def\PUpsilon      {\ensuremath{\Upsilon}\xspace}                 
 
 \def\PB      {\ensuremath{\mathrm{B}}\xspace}                 
                  
 \def\PD      {\ensuremath{\mathrm{D}}\xspace}

 \def\PJ      {\ensuremath{\mathrm{J}}\xspace}                 
 \def\PK      {\ensuremath{\mathrm{K}}\xspace}

 \def\PQ      {\ensuremath{\mathrm{Q}}\xspace}

 \def\PW      {\ensuremath{\mathrm{W}}\xspace}

 \def\PZ      {\ensuremath{\mathrm{Z}}\xspace}                 
                  
 \def\Pb      {\ensuremath{\mathrm{b}}\xspace}                 
 \def\Pc      {\ensuremath{\mathrm{c}}\xspace}                 
 \def\Pd      {\ensuremath{\mathrm{d}}\xspace}                 
 \def\Pe      {\ensuremath{\mathrm{e}}\xspace}

 \def\Pi      {\ensuremath{\mathrm{i}}\xspace}

 \def\Pp      {\ensuremath{\mathrm{p}}\xspace}                 
 \def\Pq      {\ensuremath{\mathrm{q}}\xspace}                 
                  
 \def\Ps      {\ensuremath{\mathrm{s}}\xspace}                 
 \def\Pt      {\ensuremath{\mathrm{t}}\xspace}                 
 \def\Pu      {\ensuremath{\mathrm{u}}\xspace}

}
{

 \def\Peta        {\ensuremath{\eta}\xspace}

 \def\Pmu         {\ensuremath{\mu}\xspace}

 \def\Ppi         {\ensuremath{\pi}\xspace}

 \def\Pphi        {\ensuremath{\phi}\xspace}                 
                  
 \def\Pchi        {\ensuremath{\chi}\xspace}                 
 \def\Ppsi        {\ensuremath{\psi}\xspace}                 
                  
 \mathchardef\PDelta="7101
 \mathchardef\PXi="7104
 \mathchardef\PLambda="7103
 \mathchardef\PSigma="7106
 \mathchardef\POmega="710A
 \mathchardef\PUpsilon="7107
                  
 \def\PB      {\ensuremath{B}\xspace}                 
                  
 \def\PD      {\ensuremath{D}\xspace}

 \def\PJ      {\ensuremath{J}\xspace}                 
 \def\PK      {\ensuremath{K}\xspace}

 \def\PQ      {\ensuremath{Q}\xspace}

 \def\PW      {\ensuremath{W}\xspace}

 \def\PZ      {\ensuremath{Z}\xspace}                 
                  
 \def\Pb      {\ensuremath{b}\xspace}                 
 \def\Pc      {\ensuremath{c}\xspace}                 
 \def\Pd      {\ensuremath{d}\xspace}                 
 \def\Pe      {\ensuremath{e}\xspace}

 \def\Pi      {\ensuremath{i}\xspace}

 \def\Pp      {\ensuremath{p}\xspace}                 
 \def\Pq      {\ensuremath{q}\xspace}                 
                  
 \def\Ps      {\ensuremath{s}\xspace}                 
 \def\Pt      {\ensuremath{t}\xspace}                 
 \def\Pu      {\ensuremath{u}\xspace}

}

\makeatletter
\ifcase \@ptsize \relax
  \newcommand{\miniscule}{\@setfontsize\miniscule{4}{5}}
\or
  \newcommand{\miniscule}{\@setfontsize\miniscule{5}{6}}
\or
  \newcommand{\miniscule}{\@setfontsize\miniscule{5}{6}}
\fi
\makeatother

\DeclareRobustCommand{\optbar}[1]{\shortstack{{\miniscule (\rule[.5ex]{1.25em}{.18mm})}
  \\ [-.7ex] $#1$}}

\def\epem       {{\ensuremath{\Pe^+\Pe^-}}\xspace}

\def\mup        {{\ensuremath{\Pmu^+}}\xspace}

\def\W      {{\ensuremath{\PW}}\xspace}
\def\Wp     {{\ensuremath{\PW^+}}\xspace}

\def\Z      {{\ensuremath{\PZ}}\xspace}

\def\quark     {{\ensuremath{\Pq}}\xspace}
\def\quarkbar  {{\ensuremath{\overline \quark}}\xspace}
\def\qqbar     {{\ensuremath{\quark\quarkbar}}\xspace}
\def\QQbar     {{\ensuremath{\PQ\bar{\PQ}}}\xspace}
\def\uquark    {{\ensuremath{\Pu}}\xspace}

\def\dquark    {{\ensuremath{\Pd}}\xspace}

\def\DDbar     {{\ensuremath{\PD\overline\PD}}\xspace}
\def\BBbar     {{\ensuremath{\PB\overline\PB}}\xspace}
\def\squark    {{\ensuremath{\Ps}}\xspace}
\def\squarkbar {{\ensuremath{\overline \squark}}\xspace}
\def\ssbar     {{\ensuremath{\squark\squarkbar}}\xspace}
\def\cquark    {{\ensuremath{\Pc}}\xspace}
\def\cquarkbar {{\ensuremath{\overline \cquark}}\xspace}
\def\ccbar     {{\ensuremath{\cquark\cquarkbar}}\xspace}
\def\bquark    {{\ensuremath{\Pb}}\xspace}
\def\bquarkbar {{\ensuremath{\overline \bquark}}\xspace}
\def\bbbar     {{\ensuremath{\bquark\bquarkbar}}\xspace}
\def\tquark    {{\ensuremath{\Pt}}\xspace}
\def\tquarkbar {{\ensuremath{\overline \tquark}}\xspace}
\def\ttbar     {{\ensuremath{\tquark\tquarkbar}}\xspace}

\def\pion   {{\ensuremath{\Ppi}}\xspace}
\def\piz    {{\ensuremath{\pion^0}}\xspace}

\def\pip    {{\ensuremath{\pion^+}}\xspace}
\def\pim    {{\ensuremath{\pion^-}}\xspace}

\def\kaon    {{\ensuremath{\PK}}\xspace}
  \def\Kbar    {{\kern 0.2em\overline{\kern -0.2em \PK}{}}\xspace}

\def\KorKbar    {\kern 0.18em\optbar{\kern -0.18em K}{}\xspace}
\def\Kz      {{\ensuremath{\kaon^0}}\xspace}

\def\Kp      {{\ensuremath{\kaon^+}}\xspace}
\def\Km      {{\ensuremath{\kaon^-}}\xspace}

\def\KS      {{\ensuremath{\kaon^0_{\mathrm{ \scriptscriptstyle S}}}}\xspace}

\def\Kstarz  {{\ensuremath{\kaon^{*0}}}\xspace}

\def\Kstar   {{\ensuremath{\kaon^*}}\xspace}

\def\Kstarp  {{\ensuremath{\kaon^{*+}}}\xspace}

\newcommand{\phiz}{\ensuremath{\Pphi}\xspace}

  \def\Dbar    {{\kern 0.2em\overline{\kern -0.2em \PD}{}}\xspace}
\def\D       {{\ensuremath{\PD}}\xspace}

\def\DorDbar    {\kern 0.18em\optbar{\kern -0.18em D}{}\xspace}
\def\Dz      {{\ensuremath{\D^0}}\xspace}

\def\Dp      {{\ensuremath{\D^+}}\xspace}

\def\Ds      {{\ensuremath{\D^+_\squark}}\xspace}

\def\B       {{\ensuremath{\PB}}\xspace}
\def\Bbar    {{\ensuremath{\kern 0.18em\overline{\kern -0.18em \PB}{}}}\xspace}

\def\BorBbar    {\kern 0.18em\optbar{\kern -0.18em B}{}\xspace}
\def\Bz      {{\ensuremath{\B^0}}\xspace}
\def\Bzb     {{\ensuremath{\Bbar{}^0}}\xspace}
\def\Bu      {{\ensuremath{\B^+}}\xspace}
\def\Bub     {{\ensuremath{\B^-}}\xspace}
\def\Bp      {{\ensuremath{\Bu}}\xspace}
\def\Bm      {{\ensuremath{\Bub}}\xspace}
\def\Bpm     {{\ensuremath{\B^\pm}}\xspace}

\def\Bd      {{\ensuremath{\B^0}}\xspace}
\def\Bs      {{\ensuremath{\B^0_\squark}}\xspace}
\def\Bsb     {{\ensuremath{\Bbar{}^0_\squark}}\xspace}

\def\Bc      {{\ensuremath{\B_\cquark^+}}\xspace}
\def\Bcp     {{\ensuremath{\B_\cquark^+}}\xspace}
\def\Bcm     {{\ensuremath{\B_\cquark^-}}\xspace}

\def\jpsi     {{\ensuremath{{\PJ\mskip -3mu/\mskip -2mu\Ppsi\mskip 2mu}}}\xspace}
\def\psitwos  {{\ensuremath{\Ppsi{(2S)}}}\xspace}

\def\etac     {{\ensuremath{\Peta_\cquark}}\xspace}
\def\etactwos {{\ensuremath{\Peta_\cquark{(2S)}}}\xspace}
\def\chiczero {{\ensuremath{\Pchi_{\cquark 0}}}\xspace}
\def\chicone  {{\ensuremath{\Pchi_{\cquark 1}}}\xspace}
\def\chictwo  {{\ensuremath{\Pchi_{\cquark 2}}}\xspace}

  \def\Y#1S{\ensuremath{\PUpsilon{(#1S)}}\xspace}
\def\OneS  {{\Y1S}}
\def\TwoS  {{\Y2S}}

\def\FourS {{\Y4S}}
\def\FiveS {{\Y5S}}

\def\chic  {{\ensuremath{\Pchi_{c}}}\xspace}

\def\proton      {{\ensuremath{\Pp}}\xspace}
\def\antiproton  {{\ensuremath{\overline \proton}}\xspace}

\def\Xires       {{\ensuremath{\PXi}}\xspace}

\def\Lz          {{\ensuremath{\PLambda}}\xspace}
\def\Lbar        {{\ensuremath{\kern 0.1em\overline{\kern -0.1em\PLambda}}}\xspace}
\def\LorLbar    {\kern 0.18em\optbar{\kern -0.18em \PLambda}{}\xspace}

\def\Lb      {{\ensuremath{\Lz^0_\bquark}}\xspace}

\def\Xib     {{\ensuremath{\Xires_\bquark}}\xspace}

\def\BF         {{\ensuremath{\mathcal{B}}}\xspace}

\def\BR         {\BF}
\newcommand{\decay}[2]{\ensuremath{#1\!\to #2}\xspace}         
\def\ra                 {\ensuremath{\rightarrow}\xspace}
\def\to                 {\ensuremath{\rightarrow}\xspace}

\def\eps   {{\ensuremath{\varepsilon}}\xspace}

\def\CP                {{\ensuremath{C\!P}}\xspace}

\newcommand{\phis}{{\ensuremath{\phi_{\squark}}}\xspace}

\def\AT#1     {\ensuremath{A_{\mathrm{T}}^{#1}}\xspace}           

\def\C#1      {\ensuremath{\mathcal{C}_{#1}}\xspace}                       
\def\Cp#1     {\ensuremath{\mathcal{C}_{#1}^{'}}\xspace}                    
\def\Ceff#1   {\ensuremath{\mathcal{C}_{#1}^{\mathrm{(eff)}}}\xspace}        
\def\Cpeff#1  {\ensuremath{\mathcal{C}_{#1}^{'\mathrm{(eff)}}}\xspace}       
\def\Ope#1    {\ensuremath{\mathcal{O}_{#1}}\xspace}                       
\def\Opep#1   {\ensuremath{\mathcal{O}_{#1}^{'}}\xspace}                    

\newcommand{\tev}{\ifthenelse{\boolean{inbibliography}}{\ensuremath{~T\kern -0.05em eV}}{\ensuremath{\mathrm{\,Te\kern -0.1em V}}}\xspace}
\newcommand{\gev}{\ensuremath{\mathrm{\,Ge\kern -0.1em V}}\xspace}
\newcommand{\mev}{\ensuremath{\mathrm{\,Me\kern -0.1em V}}\xspace}
\newcommand{\kev}{\ensuremath{\mathrm{\,ke\kern -0.1em V}}\xspace}
\newcommand{\ev}{\ensuremath{\mathrm{\,e\kern -0.1em V}}\xspace}
\newcommand{\gevc}{\ensuremath{{\mathrm{\,Ge\kern -0.1em V\!/}c}}\xspace}
\newcommand{\mevc}{\ensuremath{{\mathrm{\,Me\kern -0.1em V\!/}c}}\xspace}
\newcommand{\gevcc}{\ensuremath{{\mathrm{\,Ge\kern -0.1em V\!/}c^2}}\xspace}
\newcommand{\gevgevcccc}{\ensuremath{{\mathrm{\,Ge\kern -0.1em V^2\!/}c^4}}\xspace}
\newcommand{\mevcc}{\ensuremath{{\mathrm{\,Me\kern -0.1em V\!/}c^2}}\xspace}

\def\km   {\ensuremath{\mathrm{ \,km}}\xspace}
\def\m    {\ensuremath{\mathrm{ \,m}}\xspace}

\def\cm   {\ensuremath{\mathrm{ \,cm}}\xspace}

\def\mm   {\ensuremath{\mathrm{ \,mm}}\xspace}

\def\mum  {\ensuremath{{\,\upmu\mathrm{m}}}\xspace}

\def\mub{\ensuremath{{\mathrm{ \,\upmu b}}}\xspace}
\def\nb {\ensuremath{\mathrm{ \,nb}}\xspace}

\def\pb {\ensuremath{\mathrm{ \,pb}}\xspace}

\def\invfb   {\ensuremath{\mbox{\,fb}^{-1}}\xspace}

\def\ns   {\ensuremath{{\mathrm{ \,ns}}}\xspace}
\def\ps   {\ensuremath{{\mathrm{ \,ps}}}\xspace}
\def\fs   {\ensuremath{\mathrm{ \,fs}}\xspace}

\newcommand{\chisq}{\ensuremath{\chi^2}\xspace}
\newcommand{\chisqndf}{\ensuremath{\chi^2/\mathrm{ndf}}\xspace}
\newcommand{\chisqip}{\ensuremath{\chi^2_{\text{IP}}}\xspace}

\def\gsim{{~\raise.15em\hbox{$>$}\kern-.85em
          \lower.35em\hbox{$\sim$}~}\xspace}
\def\lsim{{~\raise.15em\hbox{$<$}\kern-.85em
          \lower.35em\hbox{$\sim$}~}\xspace}

\def\PDF {PDF\xspace}

\def\sPlot{\mbox{\em sPlot}\xspace}

\def\sqs   {\ensuremath{\protect\sqrt{s}}\xspace}

\def\ptot       {\mbox{$p$}\xspace}
\def\pt         {\mbox{$p_{\mathrm{ T}}$}\xspace}
\def\kt         {\mbox{$k_{\mathrm{ T}}$}\xspace}

\def\mrad{\ensuremath{\mathrm{ \,mrad}}\xspace}

\newcommand{\lum} {\ensuremath{\mathcal{L}}\xspace}
\newcommand{\Lagr} {\ensuremath{\mathcal{L}}\xspace}

\def\evtgen     {\mbox{\textsc{EvtGen}}\xspace}

\def\geant      {\mbox{\textsc{Geant4}}\xspace}

\def\photos     {\mbox{\textsc{Photos}}\xspace}

\def\pythia     {\mbox{\textsc{Pythia}}\xspace}

\def\tell1  {TELL1\xspace}
\def\ukl1   {UKL1\xspace}

\usepackage{cite} 
\usepackage{mciteplus}

\usepackage{longtable} 
\usepackage{subfigure}
\usepackage{booktabs}
\usepackage{framed,color}
\usepackage{epigraph}
\usepackage{footnote}
\usepackage{rotating}
\usepackage{footnotehyper}
\usepackage[utf8x]{inputenc}
\definecolor{shadecolor}{rgb}{1,1,1}

\title{Study of charmonium production using decays to hadronic final states with the LHCb experiment}
\author{Andrii Usachov}

\usepackage{textcomp}
\usepackage{tikz}
\usetikzlibrary{calc}
\usepackage[french,british]{babel}
\usepackage{ragged2e}
\usepackage{float}
\usepackage{setspace}
\usepackage{pdfpages}

\begin{document}
\newcommand{\tzfit}{\textit{$t_z$-fit technique}\xspace}
\newcommand{\tzcut}{\textit{separation technique}\xspace}
\newcommand{\ppbar}{\proton\antiproton}
\newcommand{\phiphi}{$\phi\phi$\xspace}
\newcommand{\JpsiToPpbar}{\decay{\jpsi}{\ppbar}}
\newcommand{\JpsiToPpbarPiz}{\decay{\jpsi}{\ppbar\piz}}
\newcommand{\EtacToPpbar}{\decay{\etac}{\ppbar}}
\newcommand{\bToEtacX}{\decay{\bquark}{\etac(1S) X}}
\newcommand{\bToJpsiX}{\decay{\bquark}{\jpsi X}}
\newcommand{\etacPromptRelativeXsec}{1.88 \pm 
                                     0.16_{stat} \pm 
                                     0.14_{syst} \pm 
                                     0.21_{norm}}
\newcommand{\etacPromptAbsoluteXsec}{(1.41 \pm
                                     0.12_{stat}\pm
                                     0.10_{syst}\pm
                                     0.16_{norm}) \mub} 
\newcommand{\jpsiPromptAbsoluteXsec}{(0.749 \pm
                                      0.005 \pm
                                      0.028 \pm
                                      0.037) \mub} 
\newcommand{\etacSecondaryRelativeBR}{ 0.48 \pm 0.03_{stat} \pm 0.03_{syst} \pm 0.05_{norm}} 
\newcommand{\etacSecondaryAbsoluteBR}{(5.51 \pm 0.32_{stat} \pm 0.29_{syst} \pm 0.77_{norm})\times 10^{-3}} 

\newcommand{\etacMassDiffTzFitSim}{(111.2\pm1.1) \mev}
\newcommand{\jpsiMassTzFitSim}{(3096.6\pm0.1) \mev}

\newcommand{\etacMassDiffTzCutInt}{(112.7\pm0.8) \mev}
\newcommand{\jpsiMassTzCutInt}{(3096.6\pm0.1) \mev}

\newcommand{\resoSim}{7.78\pm0.12}

\newcommand{\epsRatioJpsipppizJpsipp}{0.062\pm0.002}

\newcommand{\etacMassDiff}{(112.99 \pm
                            0.67_{stat} \pm
                            0.11_{syst}) \mev}

\newcommand{\etacMassDiffInBins}{(113.22\pm0.67) \mev}

\newcommand{\etacMassDiffPDG}{(113.5\pm0.5) \mev}
\newcommand{\jpsiMassPDG}{(3096.900\pm0.006) \mev}
\newcommand{\brEtacpp}{(1.50\pm0.16)\times10^{-3} }
\newcommand{\brJpsipp}{(2.120\pm0.029)\times10^{-3} }
\newcommand{\brJpsipppiz}{(1.19\pm0.08)\times10^{-3} }
\newcommand{\brRatioJpsipppizJpsipp}{0.56\pm0.04 }

\pagenumbering{roman}
\newpage
\pagenumbering{arabic}

\includepdf[pages=-]{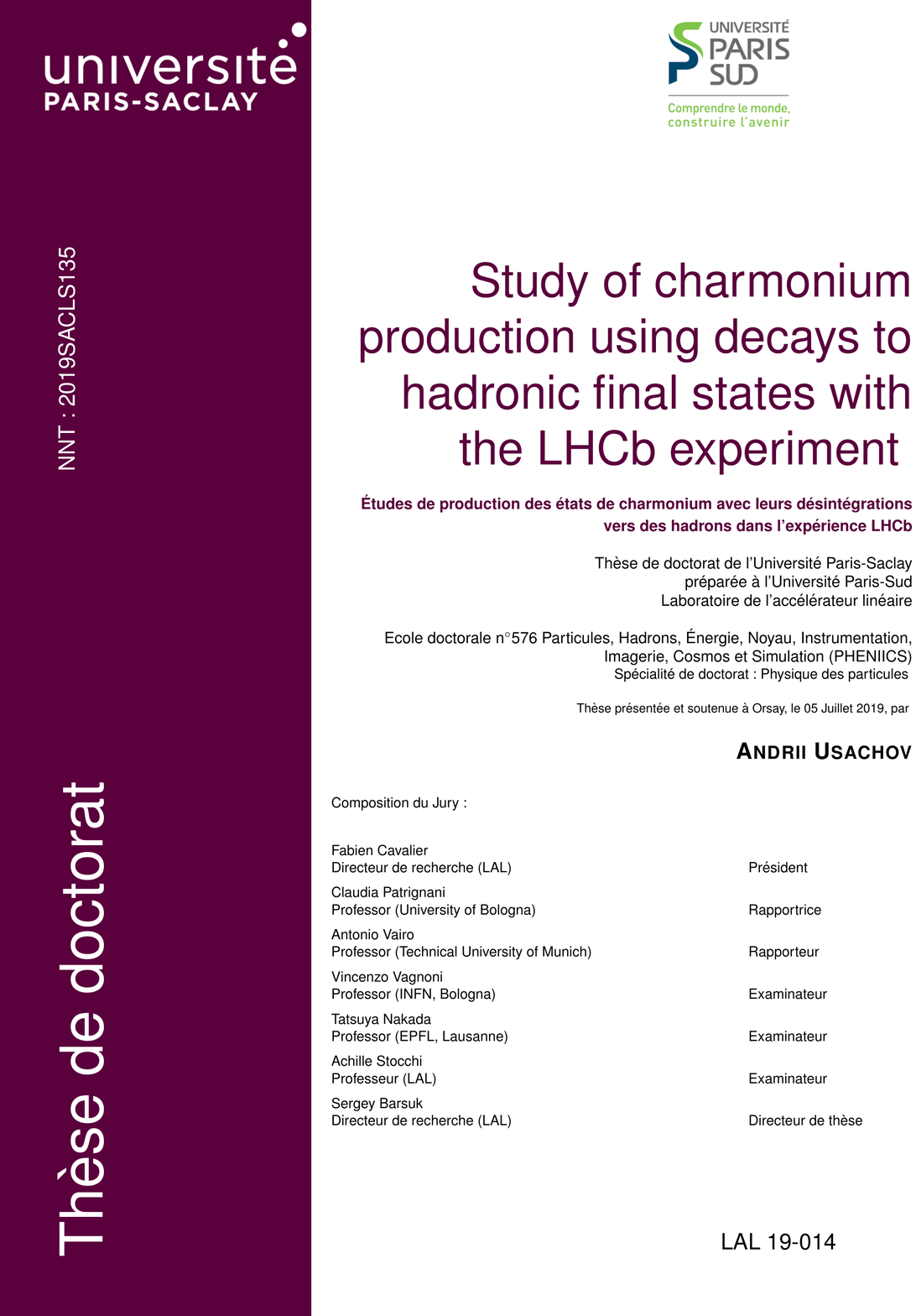}
\clearpage
\setcounter{page}{1}

\begin{singlespace}
\tableofcontents
\listoffigures
\listoftables
\end{singlespace}
\chapter{Introduction}

Studies of charmonium production provide stringent tests of Non-Relativistic QCD-based models. 
The so far theory tests come mostly from the measurements of experimentally clean $1^{-\,-}$ charmonium states, the \jpsi and \psitwos, decaying to a pair of muons. In addition, $\chicone$ and $\chictwo$ states are reconstructed via their radiative transitions to \jpsi, which however requires a reconstruction of low-energy photons. 
Reconstructing charmonium decays to hadrons allows to perform production studies of all known charmonium states. Using charmonia decays to \ppbar and $\phi\phi$ final states I study production of $\etac(1S)$, $\etactwos$ and $\chic_J$ states with the \lhcb experiment. For that I proposed a technique to select pure multi-$\phi$ final states free from kaon combinatorial background. This analysis report first measurement of \chiczero and \etactwos production in \bquark-hadron inclusive decays and the most precise \chicone and \chictwo production in the mixture of all \bquark-hadrons. In addition, the evidence of the decay $\decay{\etactwos}{\phi\phi}$ is reported for the first time.
Two different techniques have been employed to measure the $\etac(1S)$ production using the $\etac(1S)\to\ppbar$ decay. In addition, using this decay, the most precise single determination of the $\etac(1S)$ mass is also performed in the thesis. The first measurement of the $\etac(1S)$ prompt production in proton-proton collision at $\sqs=13\tev$ is reported together with the most precise determination of the branching fraction $\decay{\bquark}{\etac(1S) X}$. Also, reported measurement of the $\etac(1S)$ mass is the most precise measurement from a single experiment to date.

In order to compare the obtained result to theory predictions, I proposed to use a simultaneous fit of the measured production for charmonium states with linked long-distance matrix elements using prompt charmonium production and production in \bquark-hadron inclusive decays.
This allows to strongly restrict the allowed phase-space of the matrix elements describing charmonium production. This also demonstrates a limit of theory application and calls for further model development. 

In addition, a search for other charmonium(-like) states production in \bquark-hadron inclusive decays is performed relative to the production of charmonium states with similar quantum numbers. I measure the branching fraction of the $\decay{\etac(1S)}{\phi\phi}$ decay to resolve a tension in other existing measurements. 

Finally, \Bs mesons are reconstructed via decays to two or three $\phi$ mesons. This allows to perform an independent measurement of $\BR(\Bs\to\phi\phi)$ and the first evidence of the decay $\Bs\to\phi\phi\phi$. A resonance structure of the $\Bs\to\phi\phi\phi$ decay as well as $\phi$ meson polarization is studied with limited available decay sample.

This thesis is organised as follows.
Chapter~\ref{ch:theory} introduces selected available phenomenological approaches to describe charmonium production and confronts their predictions to the production observables measured at different facilities. Chapter~\ref{ch:decays} summarises charmonium decays channels to hadrons, which can potentially be used to reconstruct charmonium at \lhcb. Chapter~\ref{ch:lhcb} describes the \lhcb detector and shows how hadronic final states can be reconstructed and triggered.
Chapter~\ref{ch:ppbar} describes the analysis of $\etac(1S)$ production at \sqs=13~\tev using decays to \ppbar. Chapter~\ref{ch:phiphi} addresses the analysis of \chic and \etactwos production measurement in \bquark-hadron inclusive decays using decays to $\phi\phi$. Chapter~\ref{ch:pheno} compares obtained experimental results with theory predictions. A simultaneous fit of $S$-wave charmonium states production is also reported. 
Chapter~\ref{ch:mass} documents measurements of charmonium resonance parameters using both \ppbar and $\phi\phi$ decay channels. Chapter~\ref{ch:bs} describes a study of \Bs meson decays to $\phi$ mesons.
Finally, Chapter~\ref{ch:prospects} summarises the study and discusses future prospects of charmonium production measurements at \lhcb. Other studies requiring reconstruction of charmonium states using their decays to hadrons are also addressed.

\begin{singlespace}
\chapter{Charmonium production}
\label{ch:theory}
\end{singlespace}
This chapter describes the state of art of charmonium production.
The charmonium production is a branch of heavy flavour production studies, which is essential for understanding of the dynamics of strong interactions.
From a theory point of view, the production of charmonium or bottomonium is a problem involving several energy scales and to be solved by QCD. The interplay between different scales makes this problem more complex and requires accurate calculations of the entire production process by using different approaches to describe effects happening at different scales.

The experimental studies are being performed since more than 40 years and include many measured observables. As will be shown in this chapter, the theory aims at simultaneous description of most of the measured experimental observables.
Despite a significant progress from both theory and experimental sides, a comprehensive description of observables remains a challenge. It will be shown, that the \lhcb measurements of the \etac\footnote{The $\etac(1S)$ meson is denoted as \etac throughout the thesis. In some places, to be more explicit the $\etac(1S)$ denotion is used for clarity.} and \chic production play an outstanding role in formulating the charmonium production puzzle. Besides, further measurements requested by theory can be performed at \lhcb.

After the introducing quarkonium in Section~\ref{sec:introQQprod}, different theoretical approaches to describe quarkonium production are addressed in Section~\ref{sec:qqbarProd}. The current status of charmonium production puzzle is given in Section~\ref{sec:measurements} by confronting available measurements of charmonium production observables at many facilities to theory predictions.
\clearpage
\section{Introduction}
The Quantum Chromodynamics (QCD) is a sector of Standard Model (SM) aiming to describe strong interactions. The QCD originates from the first theories addressing a structure of hadrons such as Gell-Mann's quark model~\cite{GellMann:1964nj} and  parton model~\cite{Feynman:1969ej,Bjorken:1969ja}.
A development of the theory describing interactions of the hadron constituents was triggered by first experimental results probing an internal structure of a proton~\cite{Bloom:1969kc}. At the same time a color charge of strong interactions has been introduced~\cite{Greenberg:1964pe} considering hadrons as colourless objects.

The discovery of the first charmonium state \jpsi in 1974 - so-called November revolution - happened only 10 years after the initial Gell-Mann's paper. The \jpsi meson was discovered by the experiments at BNL~\cite{Aubert:1974js} and SLAC~\cite{Augustin:1974xw}.
This was a great success of the Gell-Mann's quark model~\cite{GellMann:1964nj} and the first observation of \cquark-quark. The existence of fourth quark was predicted~\cite{Bjorken:1964gz} to explain a suppresion of flavour-changing neutral currents, and in particular the $\decay{K_L}{\mup\mu^{-}}$ decay. 
The suppression has been explained only one year before the \jpsi discovery by Glashow, Iliopoulos and Maiani by so-called GIM mechanism~\cite{Glashow:1970gm}.
Systematic studies of charmonium properties started shortly after its discovery.

The strong dynamics is modulated by a strong coupling constant $\alpha_s$.
The behaviour of $\alpha_s$ depending on the energy scale is such, that at high energies (short distances) the $\alpha_s$ is small, which causes the asymptotical freedom regime. It also means that at large energies, a strong dynamics can be described perturbatively using an expansion on $\alpha_s$. 
Contrary to the electomagnetic coupling constant, the $\alpha_s$ becomes large at small energies and confinement regime takes place. The confinement explains the existence of the color field only inside hadron matter in mesons and baryons. However, the confinement has never been obtained analythically. At low energies, the expansion on $\alpha_s$ has not much sense and perturbative methods don't work anymore.
The non-perturbative dynamics can be derived from the first principles using for example lattice calculations~\cite{Callaway:1982eb}. However, the predicting power of lattice calculations remains limited. 

The quarkonium - charmonium and bottomonium - production is a complex process involving several well-separated energy scales. A number of phenomenological approaches aims at its description by introducing factorization, expansion on scales etc.
The interplay between the scales and the treatment of the initial state plays a crucial role in the QCD phenomenology.

The first measurements at Tevatron~\cite{Abe:1997jz} demonstrated that existing theoretical framework within Color Singlet (CS) model underestimates the measured \jpsi production cross-section by an order of magnitude, which was explained by a large Color Octet (CO) contribution.
Moreover, further measurements of the \jpsi polarisation in hadron-hadron collisions showed that \jpsi meson is produced almost unpolarized contrary to the CO theory prediction. This is known as the \jpsi polarization puzzle.

Existing theoretical frameworks give links between production observables of different quarkonim states.
This work follows the first measurement of the \etac production at \lhcb~\cite{LHCb-PAPER-2014-029} in 2014, which has to be described by theory simultaneously with the \jpsi production and polarization. Contrary to expectations, Color Octet contributions largely overestimate the measured \etac production cross-section.
This is an example how current phenomenological approaches are challenged by a limited number of measured observables. 
Finally, a perspective approach aims at simultaneous description of charmonium production in different collision processes.

A significant experimental progress can be achived by performing measurements of new charmonium production observables at \lhc following an example of the \etac production measurement. The \lhcb experiment is probably the only \lhc experiment, which is capable to provide a set of new measurements using signatures of charmonium decays to hadrons. However, further investigations are still needed.

\clearpage
\section{Quarkonium}
\label{sec:introQQprod}
The quarkonium is a bound state consisting of a heavy \cquark (charmonium) or \bquark (bottomonium) quark-antiquark pair. Quarkonium is as much important object for Quantum Chromodynamics (QCD) as positronium or hydrogen atom for Quantum Electrodynamics (QED).
The quark $Q$ is considered to be heavy if its mass $m_Q$ is much larger than the QCD scale $\Lambda_{QCD}\approx200~\mev$. Only \cquark, \bquark and \tquark quarks satisfy this requirement. Note that here only a qualitative relation is discussed, while for strict description one needs to define a quark mass value and always estimate the corresponding uncertainty.

Quarkonium is a non-relativistic object such that the values of $v^2$ is about 0.3 (0.1) for charmonium (bottomonium) states, where $v$ is the heavy quark velocity in the charmonium rest frame. Hence, the mass of ground state quarkonium is comparable with the $2m_Q$. 
Note, that only two flavours form quarkonium. The \tquark quark is the heaviest particle in the Standard Model (SM), so in principle it could form a non-relativistic \ttbar bound state (toppomonium) with $v^2\sim 0.01$. However, the lifetime of the \tquark quark is about $5 \times 10^{-25}\,s$, which is one order of magnitude smaller than the time scale of the strong interaction. It means, that the \tquark quark decays before its hadronisation to bound state. The \squark quark is much lighter than \cquark quark and hence the \ssbar mesons are rather relativistic. Another issue is that the lightest known "pure" \ssbar state, $\phi(1020)$, is not a bound state but a resonance since its mass is above the \Kp\Km threshold differently from charmonium or bottomonium states below \DDbar and \BBbar mass threshold, respectively.

\begin{figure}[h]
\begin{center}
\protect\includegraphics[width=0.7\textwidth]{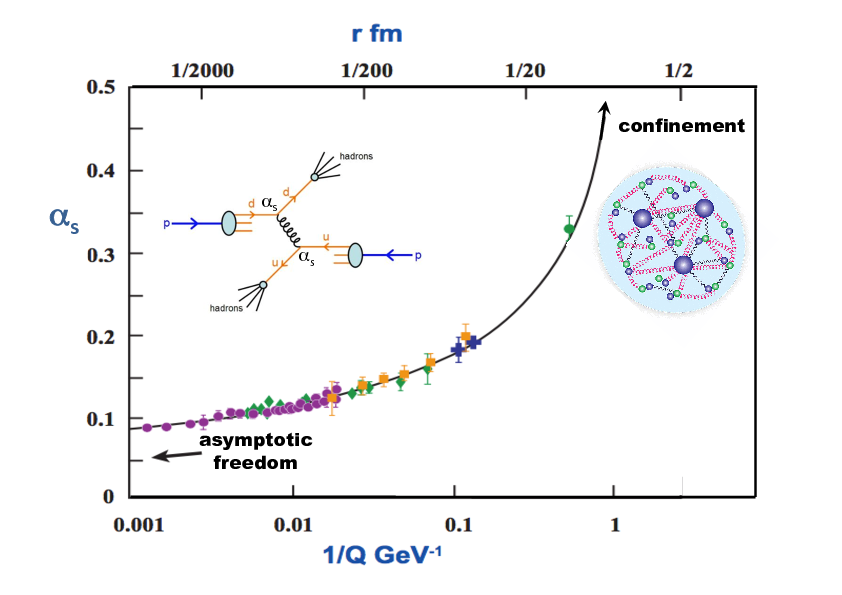}
\caption{The dependence of $\alpha_s$ on distance $r$.}
\label{fig:alphas}       
\end{center}
\end{figure}

A first approach to describe quarkonium was done with potential model describing non-relativistic quark-antiquark interaction. A generic central potential can be written as
\begin{equation}
   V(r) = -\frac{4}{3}\frac{\alpha_s}{r} +b\,r,
\label{eq:V}
\end{equation}
where $\alpha_s$ is the strong coupling constant, $r$ is a radial distance between quark and antiquark and $b$ is a parameter.
The first term represents a Coulomb potential with a quark color factor $4/3$. The asymptotics of the first term represents an asymptotic freedom of quark at small distance. The dependence of $\alpha_s$ on the scale has to be taken into account to describe the running constant as
\begin{equation}
   \alpha_s(r) = \frac{2\pi}{9\,ln\frac{1}{r\Lambda_{QCD}}}. 
\end{equation} 
An illustrative dependence of the $\alpha_s$ on $r$ is shown on Fig.~\ref{fig:alphas}.
The last expression is done in one-loop approximation, while for taking into account higher order corrections a specific renormalization scheme should be used (see for example Refs.~\cite{Peter:1997me,Schroder:1998vy}).
The second term in Eq.~\ref{eq:V} represents a long-distance interaction and is related to confinement. The dependence of second term on $r$ can be different; the only trend is that it should grow with inrease of $r$. The choice of linear dependence is coming from the description of interaction as string-like at long distances. The most popular non-relativistic potential of this shape is Cornell potential~\cite{Eichten:1975ag,Eichten:1978tg,Eichten:1979ms,Eichten:2004uh}.

Similarly to QED, the potential can be improved in order to take into account relativistic spin interactions as discused in Ref.~\cite{Voloshin:2007dx} as
\begin{equation}
   V_1(r) = V_{LS}(\vec{L}\vec{S})+V_T(r)[S(S+1)-3\frac{(\vec{S}\cdot\vec{r})(\vec{S}\cdot\vec{r})}{r^2}] + V_{SS}(r)[S(S+1)-3/2],
\end{equation}
where $V_{LS}$, $V_T$ and $V_{SS}$ described spin-orbit, tensor and spin-spin interactions, respectively; $S$, $L$ are spin and orbital momentum quantum numbers. The $V_{SS}$ is responsible for mass splitting between singlet and triplet quarkonium states, for example between \etac and \jpsi. After $V_{LS}$, $V_T$ and $V_{SS}$ terms are defined, the solution of Shr\"{o}dinger equation will  produce a quarkonium spectrum.
A general review of charmonium potential models is given in Ref.~\cite{Voloshin:2007dx}. 

Another model is Buchm\"{u}ller-Tye model~\cite{Buchmuller:1980su} developed in 1980. Results of this model are often used as an estimate charmonium wave function at origin. Interestingly, that original paper~\cite{Buchmuller:1980su} predicts triplet-singlet mass of ground state charmonium splitting (\jpsi-\etac mass difference) to be about 100~\mev, which is not far from current world average value of 113.3~\mev.

In potential models the potential should reflect also non-perturbative effects and hence needs to be tuned in order to describe quarkonium spectrum.

\begin{figure}[h]
\begin{center}
\protect\includegraphics[width=0.6\textwidth]{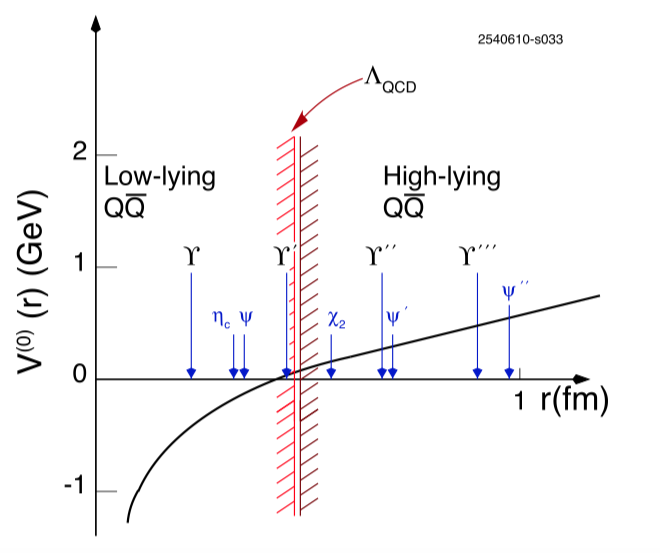}
\put(-100,210){\colorbox{shadecolor}{\makebox(100,20){\textcolor{white}{a}}}}
\caption{Static \QQbar potential as a function of quarkonium radius $r$~\cite{Brambilla:2010cs}.}
\label{fig:rVsV}       
\end{center}
\end{figure}

It is important to emphasize that for lowest level $S$-wave quarkonia (\OneS, \jpsi, \etac) and \Bc, the binding energy is relatively large such that $m_Q v^2 \gtrsim \Lambda_{QCD}$, which is not the case for excited quarkonium states (Fig.~\ref{fig:rVsV}).  This allows to apply perturbative theory for $S$-wave quarkonia since non-perturbative corrections are small.
Moreover, it means that precision physics is possible for these states to extract important model parameters, such as masses of \bquark or \cquark quarks, strong coupling constant, hyperfine splittings, natural width, leptonic decay widths, etc.
For excited quarkonium states the computations are more sophisticated since non-perturbative effects are large and an input from lattice calculations is needed. However, both spin-dependent and spin-independent potentials can been computed on lattice.

Several approaches have been used in order to obtain spin-dependent and spin-independent potentials from QCD without relying on perturbation theory.  
The spin-dependent and spin-independent \QQbar potentials up to $O(1/m_Q^2)$ were obtained for example in Refs.~\cite{Barchielli:1986zs,Brown:1979ya,Szczepaniak:1996tk,Wilson:1974sk,Gromes:1984ma} and then investigated using lattice~\cite{Bali:2000gf}.
It has been pointed out that some potentials are missed~\cite{Lucha:1991vn,Brambilla:2000gk}. Also, the infra-red divergences in the perturbative computations of $P$-wave quarkonium decays were impossible to accommodate in the framework of potential models.

In general framework, well-distinguished scales of quarkonium physics such as $m_Q$, relative momentum of heavy quarks $m_Q v$ and binding energy $m_Q v^2$ are treated with a help of Nonrelativistic Effective Field Theories. Indeed, for quarkonium the following hierarchy of scales takes place $m_Q \gg m_Q v \gg m_Q v^2$. 
The EFTs take an advantage from scales separation by integrating out higher energy scales in order to describe observables at lower energy regions. Non-relativistic EFTs are originated from QCD by systematically integrating out the high energy scales and replacing QCD by suitable expansions. The EFT should be equivivalent to QCD if all orders of the scale expansion are considered. As will be shown later, the potential picture of quarkonium dynamics can be obtained as a particular case of Nonrelativistic EFTs.

The following EFTs have been developed for quarkonium physics:
\begin{itemize}
\item Non-Relativistic QCD (NRQCD)~\cite{Caswell:1985ui,Bodwin:1994jh}, factorizing contibutions from the scale $m_Q$ (see Section~\ref{sec:nrqcd}); 
\item potential NRQCD (pNRQCD)~\cite{Pineda:1997bj,Brambilla:1999xf}, dedicated to describe quarkonium states close to threshold. The pNRQCD arises from QCD by integrating out all energy scales above $mv^2$ such as $m$ and $mv$.
\end{itemize}
The pNRQCD provides a description, which is close to Shr\"{o}dinger description.
The Lagrangian of pNRQCD can be written as a sum of static potential lagrangian, corrections to potential and interactions with other low-energy degrees of freedom.

Specific EFTs have been also developed to describe charmonium-like states above \DDbar threshold, where additional degrees of freedom (open heavy flavour, molecule, hybrid, etc.) can play an important role. Examples for $X(3872)$ state can be found in Refs.~\cite{Fleming:2008yn,Fleming:2007rp,Braaten:2007dw}.



\begin{figure}[h]
\begin{center}
\protect\includegraphics[width=1.0\textwidth]{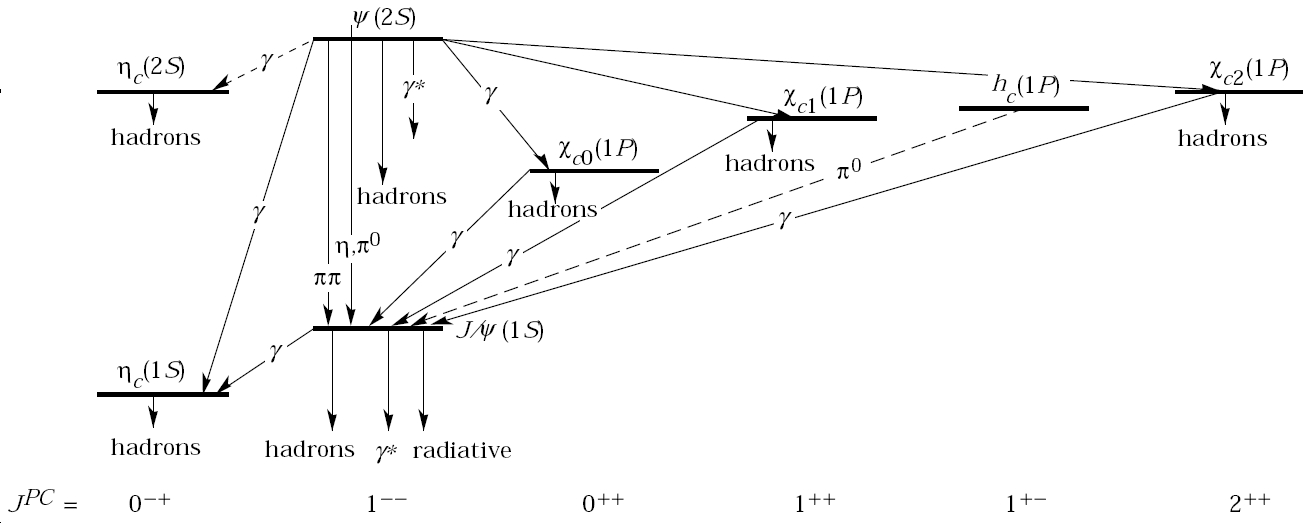}
\caption{Quantum numbers and decays of charmonium states below $D\bar{D}$ threshold.}
\label{fig:charmonium}       
\end{center}
\end{figure}

In this work I will focus on most of charmonium states below \DDbar threshold. Namely, $S$-wave charmonium states \etac and \jpsi together with their radial excitations \etactwos and \psitwos and $P$-wave states \chiczero, \chicone, \chictwo and to some extend $h_c$ will be discussed.
A scheme of charmonium family under \DDbar threshold together with charmonium states quantum numbers and dominant transitions are shown on Fig.~\ref{fig:charmonium}. The notation of charmonium states follows traditional form of atomic physics $^{2S+1}L_J$, where $J$ is a total angular momentum.
Currently, all charmonium states with a mass below the $D\bar{D}$ threshold have been observed and have their quantum numbers $J^{PC}$ well established.    

Generally, quarkonium provides many important observables for understanding nature of strong interactions. Both perturbative and non-perturbative effect are involved. 
Quarkonium spectroscopy and decays provide important information for QCD due to natural annihilation diagrams allowing separation between initial and final states in the first approximation. In the following sections, charmonium production in different processes is addressed. 

\newpage
\section{Theoretical formalism of quarkonium production}
\label{sec:qqbarProd}
Generally, quarkonium can be produced in many processes, and each of them provides an important observable for various theoretical formalisms based on QCD. In this work, I focus on inclusive production of single quarkonium in hard processes: parton scattering and decays. 

Among the complementary observables not mentioned in this work, one can highlight multiple quarkonium production in hard processes, jet-associated production, production in ion collisions, various exclusive production processes.
The multiple quarkonium production in hard processes provides important complementary observables to the theoretical framework addressed in this section, 
which come however through the description of multiple parton scattering.
Quarkonium production in medium (e.g. ion collisions) measures nuclear modification factors to study properties of cold nuclear matter or quark-gluon plasma (QGP) by comparing the quarkonium production in $pp$, $pA$ and $AA$ collisions, where $A$ denotes an ion with the mass number $A$. The central exclusive quarkonium production in soft processes are essential for soft QCD and should be described by entirely different theoretical approaches. It has an advantage of most direct theory interpretation by for example Regge-based theory. The only color singlet state can be created in the central exclusive production process.

The inclusive production of quarkonium states implies at least three well-distinguished intrinsic momentum scales: the mass of the heavy quark, $m_Q$; the relative momentum of heavy quarks of order $m_Q v$ and the binding energy $m_Q v^2$. For quarkonium produced in the scattering process, the scattering scale $p_{hard}$ also enters the description of quarkonium production. Below, I consider the case of charmonium, while similar considerations apply also for bottomonium. However, the bottomonium production description can differ from charmonium one. Since two of the mentioned scales involve relative quark velocity $v$, one naively expects that higher-order corrections on these scales are smaller for bottomonium. 
Therefore, the color octet mechanism described in this section is likely to be relevant to a lesser extend for bottomonium than for charmonium. Nevertheless, with more available data on different bottomonium states production, theory will utilize powerful comparisons of charmonium and bottomonium production under expansions in $v$.

Quarkonium can be inclusively produced in following hard processes:
\begin{itemize}
\item Transitions from higher mass quarkonium states (feed-down),
\item \bquark-hadron decays\footnote{relevant for charmonium only\label{ft:onlyCC}},
\item Bottomonium decays\textsuperscript{\ref{ft:onlyCC}},
\item \epem collisions,
\item $ep$ collisions,
\item Initial parton scattering in hadron-hadron collisions (hadroproduction),
\item $Z$, $W$, Higgs or \tquark-quark decays.
\end{itemize}
Experimentally, the measured production cross-section includes the feed-down from higher mass quarkonium states, which can be produced in the same production process. In the case of hadron-hadron collision, the total production process (sum of hadroproduction and the feed-down contributions) constitute prompt production. The feed-down subtracted production cross-section is often referred as direct production.

The feed-down contributes to the most of production cases and can be estimated using theoretical and experimental input. 
The amount of the feed-down contribution depends on the production cross-section of higher mass charmonium states and on the branching fractions of the feed-down transitions. 
The feed-down can be experimentally subtracted if the production of states, which are its dominant sources, is measured in the same kinematical regime and input branching fractions are known.
The total feed-down contribution can be quite sizeable. For example, about 30\% of promptly produced \jpsi at Tevatron or \lhc are coming from \chic and \psitwos transitions to \jpsi. In this case, the feed-down contributions should be taken into account carefully because theoretical uncertainty on \jpsi hadroproduction cross-section is comparable. At the same time, the experimental observables are rather well measured.
While measurement of \chic and \psitwos production cross-sections can be accessed experimentally and then used in the \jpsi production description, the experimental determination of the feed-down contribution to \etac production cross-section is more complicated. The dominant expected source of the feed-down to \etac state is the $h_c$ charmonium state, which decays to $\etac\gamma$ with a branching fraction of about 50\%. The production of the $h_c$ state has been never measured at hadron machines. Hence, the feed-down from $h_c$ state can be addressed only theoretically and using experimental upper limits if any.
On the other hand, the heaviest charmonium states below the \DDbar threshold are feed-down free, since the resonances above the threshold would rather decay strongly to \DDbar with the branching fraction close to 100\%.
Consequently, all effects of the feed-down contributions to any \jpsi production observable can be studied by measuring the same observable for radially excited \psitwos state, which is feed-down free.

The $Z$, $W$, Higgs or \tquark-decays can provide an important test of quarkonium production but are extremely complicated to measure at available facilities. Both production cross-section and branching fraction of $Z$, $W$, Higgs or \tquark-decays to quarkonium are very small.
Hence, these cases of quarkonium production suffer from the lack of data. 
Also, not many experimental measurements are available for bottomonium decays to charmonium. 

Apart from the production cross-section, another important observable is the quarkonium polarisation, which should be described simultaneously within the same theoretical framework.

The hard scale parameter is estimated differently for different production processes: in the case of quarkonium production in hadron-hadron collisions, the scale is usually defined as the order of charmonium transverse momentum, \pt, while for the \epem collisions or production in decays the quarkonium momentum in the \epem rest frame, $p^*$, is used. 

Naively one might expect that the good separation between the scales would lead to splitting quarkonium production process into two independent stages of \QQbar pair creation and its hadronisation to the quarkonium state. 
The latter is known as the factorisation assumption, where the amplitude of the entire production process can be written as the sum of products of the short-distance and long-distance matrix elements.
The \QQbar pair creation is a short-distance process happening at the $p_{hard}$ scale and can be calculated perturbatively using an expansion in $\alpha_s$.
The hadronisation is a long-distance process, and its dynamics is characterised by the scales $m_Q v$, $m_Q v^2$ and $\Lambda_{QCD}$. The long-distance matrix elements (LDME) describing hadronisation cannot be calculated perturbatively and are expanded in terms of $m_Q v$ and $m_Q v^2$. The LDME values are obtained from phenomenology or lattice calculations. The independence of stages of the production leads to the universality assumption, that the values LDMEs are the same for production processes, whose scale $p_{hard}$ is large enough.

\subsection{Factorization and PDFs}
The short distance process of quarkonium production is firstly described at the level of parton interactions.
The quarkonium hadroproduction at \lhc energies is happening predominantly via gluon-gluon fusion. Precise theoretical description also need to take into account other partonic processes (e.g. take into account quark-quark process) since their contribution is not negligible.

In order to obtain the hadroproduction cross-section, the partonic cross-sections should be convoluted with corresponding non-perturbative probability density functions (PDFs) of partons (e.g. gluons and quarks in the case of hadron-hadron collisions). The PDFs of gluons and quarks of different flavours are extracted from global fits to many measured production observables.

The treatment of the partons of initial state can be different. 
Most of the theoretical calculation of quarkonium production is performed within collinear factorisation~\cite{Collins:1989gx, Brock:1993sz}. In the collinear factorisation the transverse momentum of initial partons is neglected. Hence, the PDFs do not depend on parton transverse momentum.
Within collinear factorisation, full Next-to-Leading-Order in $\alpha_s$ calculations of quarkonium production are available as will be shown in Section~\ref{sec:measurements}.

The \kt-factorisation~\cite{Gribov:1984tu,Catani:1990eg,Lipatov:1995pn} is another approach to perform factorisation to describe quarkonium production.  
The \kt-factorisation approach takes into account a dependence of partonic PDFs on their transverse momentum \kt and longitudinal momentum fraction $x$ carried by parton. The gluon dynamics is described by Balitsky-Fadin-Kuraev-Lipatov (BFKL) evolution equation~\cite{Kuraev:1976ge,Balitsky:1978ic} resumming logarithmic contributions by introducing reggeized gluons. The \kt-factorisation works in high energy regime, i.e. $\sqs\rightarrow \infty$ or small-$x$ limit.
In the collinear factorisation approach, the initial state parton can receive some transverse momentum at NLO by emitting additional parton.
The same term appears in the \kt-factorisation at LO so that there is an interplay between \kt-factorisation and collinear factorisation. Physically, one can interpret that higher order terms appear due to taking into account initial state radiation. 
The latter leads to more accurate LO calculations with \kt-factorisation than LO predictions made with collinear factorisation. 
On the other hand, \kt-dependence is poorly constrained since it requires a special transverse momentum dependent PDFs (TMDs), while collinear factorization uses integrated PDFs. Another issue of \kt-factorisation is that only the LO calculations are available so far.

Similarly, TMD factorisation, firstly introduced in Ref.~\cite{Collins:1981uw}
and discussed in Refs.~\cite{Aybat:2011zv} works at lower energy limit and resumes many parton emissions from initial state. 
The complications of TMD factorisation is that some cancellations appear only after the integration over transverse momentum. The TMD factorisation works better at low-\pt range compared to collinear factorisation. The NLO computation within TMD factorization for quarkonium production is not available, however it is done for Higgs production~\cite{Ji:2005nu}.

Below, different theoretical models describing inclusive quarkonium production in hard processes are summarised. The key difference between models is the approach to the long-distance hadronisation description.

\subsection{Color evaporation model}
\label{sec:CEM}
The colour evaporation model (CEM) is an easy and historically one of the first phenomenological models of quarkonium production~\cite{Fritzsch:1977ay,Halzen:1977rs,Gluck:1977zm}.

The CEM assumes that produced \QQbar pair hadronizes into quarkonium if the initial mass of \QQbar is below the threshold of the two open-flavour mesons creation (e.g. \DDbar threshold in the case of charmonium). Hence, in the CEM the total inclusive production cross-section of charmonium state $H$ in $A+B$ collision is expressed as
\begin{equation}
\sigma_{A+B\to H+X} = F_H\int^{4M^2}_{4m^2_Q}dm^2_{\QQbar}\frac{d\sigma_{A+B\to H+X}}{dm^2_{\QQbar}}, 
\label{eq:CEM_sigma}
\end{equation}
where $m_{\QQbar}$ is the mass of \QQbar pair, the $M$ is the mass of the lightest open flavour meson containing quark $Q$, the $d\sigma_{A+B\to H+X}/dm^2_{\QQbar}$ is the differential production cross-section and the $F_H$ is the probability of hadronization of \QQbar to a given quarkonium $H$.

The $F_H$ is a non-perturbative constant, which does not depend on the momentum nor on the process. The $F_H$ is the only parameter of the model and can be determined using a measurement of total $H$ production cross-section. Once $F_H$ is determined, the prediction of differential production cross-section in any process and kinematical conditions can be obtained in a straightforward way. The Eq.~\ref{eq:CEM_sigma} implies internal sum over spin and colour states of \QQbar. The model assumes that the colour of \QQbar system is neutralised by the surrounding field (colour evaporation). In other words, CEM assumes that the requirement of quarkonium colourlessness
does not imply any constraint on colour states of \QQbar and all of them contribute to overall production.
Note, that the sum of $F_H$ for different charmonium states is less than unity because the \QQbar system can receive some energy from a surrounding medium during the hadronisation stage.
Naively, taking into account the probability to produce a colourless object among $3\times3$ possible color combinations of \QQbar one can expect that 
\begin{equation}
\sigma_{A+B\to H+X} = \frac{1}{9}\int^{4M^2}_{4m^2_Q}dm^2_{\QQbar}\frac{d\sigma_{A+B\to \ccbar+X}}{dm^2_{\QQbar}}, 
\label{eq:CEM_color}
\end{equation}
while the remaining part of the $\ccbar$ pair production cross-section should be accounted by the open flavour production.
In the case of hadron-hadron collisions, the Eq.~\ref{eq:CEM_sigma} should be written for parton-parton interactions, $i+j\to H+X$, and then convoluted with PDFs.

The CEM at LO predicts the \pt-differential production cross-section of charmonium production in hadron-hadron collisions to be proportional to a $\delta$-function or, in other words, charmonium is produced with $\pt = 0$. The production processes of $i+j \to k\QQbar$ can produce charmonium with non-zero \pt, where $k$ is another quark or gluon. Hence, in order to describe \pt-differential production cross-section, first complete NLO calculations for CEM have been performed for hadron-hadron collisions~\cite{Schuler:1996ku,Mangano:1992kq}.

The CEM has been extensively tested and compared with other theoretical models and measurements. For a corresponding review see Refs.~\cite{Amundson:1996qr,Bodwin:2005hm}. A straightforward prediction of CEM is that the production ratio of any pair of quarkonium states is the same for different production processes. The apparent violation of this prediction is observed in the comparison of the feed-down contribution from \chic states to \jpsi prompt production to \jpsi production in \bquark-hadron decays as will be discussed in next section. 
In addition, CEM predicted a qualitative description of the \jpsi, \psitwos and \chic \pt-differential production cross-sections obtained for example at \cdf~\cite{Amundson:1996qr}. However, the quality of fits to data is poor. CEM predicts the production rate of $\chic_J$ states to be proportional to $2J+1$, which is strongly violated as shown in Chapter~\ref{ch:pheno}. 

The independence of production on the spin of the \QQbar pair leads to the prediction of non-polarisation of quarkonium, which contradicts to the observed non-zero polarisation of \jpsi meson in many processes (see Section~\ref{sec:measurements}). The recently developed Improved Color Evaporation Model (ICEM) aims at describing both quarkonium production and polarisation more appropriately without increasing the number of parameters of the model~\cite{Ma:2016exq}. Particularly, the ICEM distinguishes soft emitted gluons by the \QQbar system from exchanged gluons. Therefore, the interaction of \QQbar pair with surrounding strong field is described in more details.

The ICEM gives a reasonable basic description of the relative charmonium production contrary to naive CEM. In addition to that, ICEM predicts non-zero polarisation of \jpsi mesons~\cite{Vogt:2019zmr}. Further improvement of ICEM is ongoing by exploiting \kt-factorisation for calculations.

One can conclude that CEM is an easy illustrative model, which depends on a single parameter for each charmonium state, with however limited predicting power. Significant improvements came from introducing ICEM. 

\clearpage
\subsection{Non-Relativistic QCD (NRQCD)}
\label{sec:nrqcd}
The non-relativistic QCD (NRQCD)~\cite{Bodwin:1994jh} is so far the most successful theoretical framework predicting inclusive quarkonium production. Contrary to CEM, NRQCD recognises separate contributions from different spin and colour states of \QQbar. The latter is achieved by taking into account a description of the hadronisation via expansion in $m_Q v$ and $m_Q v^2$.

A generic expansion for quarkonium production in NRQCD can be written as
\begin{equation}
d\sigma_{A+B\to H+X}=\sum_{n}d\sigma_{A+B\to \QQbar[n]+X}\langle O^H(n)\rangle,
\end{equation}
where $n$ denotes the color and spin state of \QQbar and the $\langle O^H(n)\rangle$ is the LDME describing the evolution of $\QQbar[n]$ to a quarkonium state $H$. 

Historically NRQCD is an extension of Color Singlet (CS) Model (CSM) - the first model describing quarkonium production~\cite{Einhorn:1975ua,Ellis:1976fj,Carlson:1976cd,Chang:1979nn}. The CSM assumes that only colourless \QQbar state contributes to the quarkonium production. CSM also requires \QQbar to have the same spin state as the resulting $H$. 

The most profound internal theoretical evidence of the incompleteness of the CS model comes from the presence of infrared divergences in the production cross
sections and decay rates of P-wave quarkonium.
The presence of infrared divergences implies a failure of the simple factorisation assumption, upon which the CS model is based.

The NRQCD approach provides a natural solution by introducing a Color Octet (CO) mechanism in addition to nominal CS. Within CO mechanism, the colour and spin states of \QQbar and quarkonium can be different and are adjusted during the hadronisation stage.
Heavy quark pairs that are produced at short distances in a CO state can evolve into physical charmonium via emitting soft gluons when the quark pair has already expanded to the charmonium size.
According to the power counting rules described below, all CO matrix elements for the production (or decay) of S-wave quarkonia are suppressed by powers of the velocity compared to the CS contribution. Hence, the CS model is naturally included in NRQCD and represents the first term of expansion on $v$.
However, CO processes can become significant, and even dominant, if the short-distance cross section for producing $\QQbar$ in a CO state is enhanced.

Contrary to CEM and CSM, NRQCD is a rigorous EFT, which aims at describing of quarkonium at scales smaller than $m_Q$. The Lagrangian NRQCD can be derived from the QCD one by using an expansion in $1/m_Q$. The NRQCD Lagrangian up to $O(1/m_Q^2)$ can be written as.
\begin{equation}
\begin{aligned}
\Lagr_{NRQCD} &= \Lagr_g+\Lagr_l+\Lagr_{\psi}+\Lagr_{\chi}+\Lagr_{\psi\chi},
\end{aligned}
\end{equation}
where the interaction term $\Lagr_{\psi\chi}$ is expressed in terms of singlet and octet operators $O_1(^1S_0)$, $O_1(^3S_1)$, $O_8(^1S_0)$, $O_8(^3S_1)$ as 
\begin{equation}
\begin{aligned}
\Lagr_{\psi\chi} &= \frac{f_1(^1S_0)}{m_Q^2} O_1(^1S_0) +
\frac{f_1(^3S_1)}{m_Q^2} O_1(^3S_1)+ \frac{f_8(^1S_0)}{m_Q^2} O_8(^1S_0) + \frac{f_8(^3S_1)}{m_Q^2} O_8(^3S_1),
\end{aligned}
\end{equation} 
\begin{equation}
\begin{aligned}
 O_1(^1S_0) &= \psi^+ \chi \chi^+ \psi,\\
 O_1(^3S_1) &= \psi^+ \boldsymbol{\sigma} \chi \chi^+ \boldsymbol{\sigma} \psi,\\
 O_8(^1S_0) &= \psi^+ T^a \chi \chi^+ T^a \psi,\\
 O_8(^3S_1) &= \psi^+ T^a \boldsymbol{\sigma} \chi \chi^+ T^a \boldsymbol{\sigma} \psi,\\
\end{aligned}
\end{equation}
where $\psi$ is a spinor that annihilates the quark and $\chi$ is a spinor that creates the antiquark, $T^a$ is a basis generator of the fundamental representation of the $SU(3)$ group.
As was already mentioned, NRQCD introduces CO LDMEs. The color-octet contributions can not be incorporated using potential models.

\begin{figure}[h]
\centering
\protect\includegraphics[width=1.\textwidth]{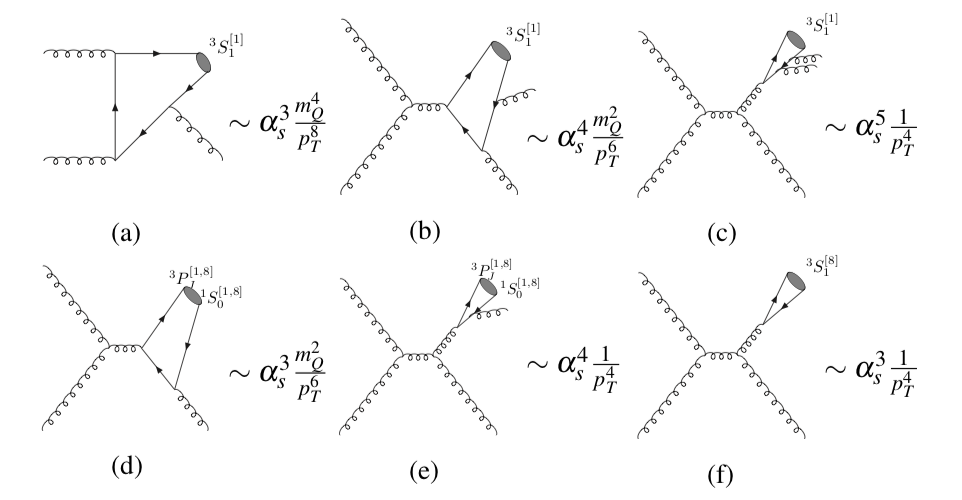}
\caption[The CS and CO diagrams contributing to charmonium production at leading orders.]{The CS (a-e) and CO (d-f) diagrams contributing to charmonium production at leading orders~\cite{Chao:2012upa}.}
\label{fig:DiagrNRQCD}
\end{figure}

The NRQCD factorization predicts a scale dependence of short-distance matrix elements. For example, the contributions corresponding to different S-wave LDMEs relevant for \jpsi production have the following asymptotic behaviour
\begin{alignat}{2}
d\sigma^{\jpsi}_{^3S_1^{[8]}}/d p_{hard} \sim 1/p_{hard}^4\\
d\sigma^{\jpsi}_{^1S_0^{[8]}}/d p_{hard} \sim 1/p_{hard}^6\\
d\sigma^{\jpsi}_{^3P_J^{[8]}}/d p_{hard} \sim 1/p_{hard}^6 \\ 
d\sigma^{\jpsi}_{^3S_1^{[1]}}/d p_{hard} \sim 1/p_{hard}^6.
\end{alignat}
That is why in a  high momentum region, the production is sensitive to $O_{8}^{\jpsi}(^3S_1)$, while at lower momenta, two matrix elements $O_{8}^{\jpsi}(^1S_0)$ and $O_{1}^{\jpsi}(^3S_1)$ have similar asymptotic behavior. 
For illustration, typical diagrams for charmonium hadroproduction via both CO and CS mechanisms are given on Fig.~\ref{fig:DiagrNRQCD} together with their asymptotical behaviour.

It has to be stated that NRQCD factorization hypothesis has not been rigorously proven for quarkonium production yet (contrary to quarkonium annihilation). Hence, the universality assumption for LDMEs has not been strictly proven neither. Its possible violation can be related to effects, which are neglected under a definition of LDME. 
The complication of NRQCD is that at least two remaining scales $mv$ and $mv^2$ should be taken into account in a single expansion. Therefore, NRQCD does not have unique power counting rules.

Since LDMEs cannot be calculated perturbatively and are usually taken as parameters extracted from the fits to data, the NRQCD has an infinite number of parameters.
However, the importance of the various LDMEs of NRQCD can be assessed with power counting rules \cite{Fleming:2000ib} using an expansion of $v$. 
They can be derived by considering the Fock state decomposition of a quarkonium state
$|H\rangle$ in Coulomb gauge,
\begin{equation}
 |H\rangle=\psi^H_{\QQbar}|\QQbar\rangle + \psi^H_{\QQbar g}|\QQbar g\rangle + ... .
\end{equation}
The dominant component $|\QQbar\rangle$ comprises a heavy quark pair in a colour-singlet state and with angular momentum quantum numbers $^{2S+1}L_J$ that are consistent with the
quantum numbers of the physical quarkonium. The higher Fock states, such as $|\QQbar g\rangle$,
contain dynamical gluons or light $\qqbar$ pairs. Thus heavy quark pair can be in either a
CS or a CO state with spin $S=0,1$ and angular momentum $L=0, 1, 2,$ etc. All higher Fock states have probabilities suppressed by powers of $v$ compared
to that of $|\QQbar\rangle$. The $|\QQbar g\rangle$ states with the highest probability of $\mathcal{O}(v)$ are those that can be reached from the dominant $|\QQbar\rangle$ state through a chromoelectric interaction.
Higher Fock states $|\QQbar g\rangle$ which can be reached from the dominant $|\QQbar\rangle$ state through the chromomagnetic interaction have probabilities of $\mathcal{O}(v^2)$.
Both chromoelectric and chromomagnetic transitions change the colour state of the $\QQbar$ pair from CS to CO, and from CO to either CS or CO. 

A difference between NRQCD and CEM in velocity suppression factors up to order of $v^4$ has been demonstated in Ref.~\cite{Bodwin:2005hm} and is shown in Table~\ref{tab:powCEM_NRQCD}.
\begin{table}[h]
\resizebox{\textwidth}{!}{
\centering
\begin{tabular}{c|cccc|cccccccc}
\hline\hline
\multicolumn{13}{c}{NRQCD}\\ \hline
 &$^1S_0^{1}$ & $^3S_1^{1}$ & $^1S_0^{8}$ & $^3S_1^{8}$ & $^1P_1^{1}$ & $^3P_0^{1}$ & $^3P_1^{1}$ & $^3P_2^{1}$ & $^1P_1^{8}$ & $^3P_0^{8}$ & $^3P_1^{8}$ & $^3P_2^{8}$ \\ \hline
$\eta_{c,b}$ & 1 & & $v^4$ & $v^3$ & & & & & $v^4$ & & & \\
\jpsi, $\Upsilon$ & & 1 & $v^3$ & $v^4$ & & & & & & $v^4$ & $v^4$ & $v^4$ \\ \hline
\hline
\multicolumn{13}{c}{CEM}\\ \hline
$H$ & 1 & 1 & 1 & 1 & $v^2$ & $v^2$& $v^2$& $v^2$& $v^2$ & $v^2$& $v^2$& $v^2$\\ \hline\hline
\end{tabular}
}
\caption[Velocity suppression factors for LDMEs in S-wave \QQbar in the NRQCD and in the CEM.]{Velocity suppression factors for LDMEs in S-wave \QQbar in the NRQCD and in the CEM. The $^{2S+1}L_J^{1}$ indicates the CS and the $^{2S+1}L_J^{8}$ indicated the CO states, respectively~\cite{Bodwin:2005hm}.}
\label{tab:powCEM_NRQCD}
\end{table}


Experimental differential production cross-section can be fitted to the theory model to extract the information about relative contributions of CS and CO mechanisms. By taking into account dominant CS and CO contributions, NRQCD provides a basic description of available experimental information in a significant \pt range.

Moreover, the heavy quark spin symmetry (HQSS) provides relations between LDMEs of different charmonium states. Therefore, it also creates intrinsic links between the production observables of different quarkonium states. Namely, states with the same orbital angular momentum and radial quantum numbers are linked.
Investigation of complementary charmonium states with different $J$ quantum numbers is consequently
 a powerful tool to further constrain available theoretical descriptions.

One of the first candidates is the simultaneous study of the lowest charmonium states \etac and \jpsi since both are experimentally accessible and a link between the $\etac(1S)$ and \jpsi matrix elements can be established.
Spin symmetry gives the following relation between the $\etac(1S)$ and the $\jpsi$ color-singlet matrix elements:
\begin{alignat}{4}
O_{1}^{\jpsi}(^3S_1) \approx 3\times O_{1}^{\eta_c}(^1S_0).
\end{alignat}
Relation between color-octet matrix elements are shown below.
\begin{alignat}{4}
O_{8}^{\jpsi}(^3S_1) \approx 3\times O_{8}^{\etac}(^1S_0),\\
O_{8}^{\jpsi}(^1S_0) \approx O_{8}^{\etac}(^3S_1),\\
O_{8}^{\jpsi}(^3P_J) \approx \frac{2J+1}{3}\times O_{8}^{\etac}(^1P_1).
\end{alignat}
The links between LDMEs are not exact, and are satisfied up to $o(v^2)$ precision.

Note, that LDMEs of the CS mechanism are related to the quarkonium wave function and can be extracted either from the potential model (for example Ref.~\cite{Buchmuller:1980su}, from lattice calculations or from the measurements of branching fractions of quarkonium decays. The CS LDMEs are considered as well known with 10-20\% precision.

In summary, development of NRQCD yielded a framework that reasonably describes hadroproduction of the measured quarkonium states in a wide range of transverse momentum (\pt) and rapidity.
However, a comprehensive simultaneous description of the production and polarisation of the \jpsi state at \tevatron and \lhc energies in an entire \pt range remains a challenge.
Similarly, NRQCD describes the quarkonium production in other processes.

The NRQCD is often used with collinear factorization. The NLO calculations are necessary for NRQCD as will be seen below.
The comparisons of NRQCD predictions to experimental measurements are shown in Section~\ref{sec:measurements}. Once main experimental results obtained in this thesis are presented, Chapter~\ref{ch:pheno} will outline a systematic discussion of their description by NRQCD.

\newpage
\section{Theory vs experiment: state of art}
\label{sec:measurements}
In this section, I compare earlier measured observables of charmonium production with theoretical predictions. For the predictions, we will focus on the NRQCD as the most successful framework to date.
A more complete review on quarkonium production can be found in Refs.~\cite{Andronic:2015wma,Brambilla:2014jmp,Brambilla:2010cs}.
The most up to date review including the last \lhc results and the latest progress in quarkonium production phenomenology can be found in Ref.~\cite{Lansberg:2019adr}.

The charmonium production in hard processes provides several important observables, which allow selective comparison to theoretical predictions: 
\begin{itemize}
\item total production cross-section,
\item the shape of differential production cross-section in \pt and $p^*$ and rapidity,
\item polarisation of vector or tensor charmonium states.
\end{itemize}
The production cross-section is the first powerful observable to understand the production mechanism, since mesurements of the differential cross-section is naturally more complicated to perform. As will be shown below, often, the cross-section is measured to be much larger than the CSM prediction, which indicates a need for CO contribution or taking into account higher order calculations. Qualitatively, the CO mechanism includes possiblity of charmonium creation from a single gluon, which leads to transverse polarization of produced vector charmonium. The shape of \pt- or $p^*$-differential cross-sections provides an additional constraint on CS and different CO contributions, which have different asymptotical behaviour.
The polarisation of charmonium is also a powerful observable to distinguish CS and CO contributions since the predictions of CS and CO lead to the opposite expected polarisations. 
Experimentally, the polarisation is conveniently accessed by measuring
the angular distribution of charmonium decays, which is customarily parametrized
using the polarization observables $\lambda_\theta$, $\lambda_\phi$, and $\lambda_{\theta\phi}$ as
\begin{equation}
W(\theta,\phi)\sim 1+\lambda_\theta cos^2(\theta) + \lambda_\phi sin^2(\theta)cos(2\phi) + \lambda_{\theta\phi}sin(2\theta)cos(\phi).
\end{equation}
Here, at the example of the $\decay{\jpsi}{\mup\mu^{-}}$ probe, $\theta$ and $\phi$ are respectively the polar and azimuthal angles of muons momenta in the \jpsi rest frame. The values $\lambda_\theta=0,+1,-1$ correspond to unpolarized, fully transversely polarized, and fully longitudinally polarized \jpsi mesons, respectively. This defines polarisation observables in helicity frame. Alternatively, the polarisation can be measured at different frames: Collins-Soppers~\cite{Collins:1977iv}, target frame~\cite{Beneke:1998re}, which is the case for polarisation analyses at HERA, as will be shown below.

Different production observables of different production processes can be used in simultaneous studies. Namely, the joint fits aim at using the same set of LDMEs to describe all observables in different production processes for linked charmonium states.
Within the NRQCD description, the four independent LDMEs are used to describe the production of S-wave charmonium \etac and \jpsi. Only two LDMEs are used to describe P-wave charmonium (\chiczero, \chicone, \chictwo and $h_c$).
The relevant LDMEs together with HQSS relations are summarised in Table~\ref{tab:specLDMEs}. Anologous relations apply for radially excited states (\etactwos and \psitwos).
\begin{table}[ht]
\begin{center} 
\begin{tabular}{l|c|c} 
                              & HQSS relations & independent LDMEs \\ \hline
S-wave &      
 $\langle O_1^{\etac}(^1S_0)\rangle = \frac{1}{3} \langle O_1^{\jpsi}(^3S_1)\rangle$
& $\langle O_1^{\jpsi}(^3S_1)\rangle$ \\
(\etac and \jpsi)  &

 $\langle O_8^{\etac}(^1S_0)\rangle = \frac{1}{3} \langle O_8^{\jpsi}(^3S_1)\rangle$
& $\langle O_8^{\jpsi}(^3S_1)\rangle$ \\
&

 $\langle O_8^{\etac}(^3S_1)\rangle = \langle O_8^{\jpsi}(^1S_0)\rangle$
& $\langle O_8^{\jpsi}(^1S_0)\rangle$ \\
&

 $\langle O_8^{\etac}(^1P_1)\rangle = 3 \langle O_8^{\jpsi}(^3P_0)\rangle$
& $\langle O_8^{\jpsi}(^3P_0)\rangle$ \\ \hline

P-wave &      

 $\langle O_1^{\chic_J}(^3P_J)\rangle = (2J+1)\langle O_1^{\chiczero}(^3P_0)\rangle$
& $\langle O_1^{\chiczero}(^3P_0)$ \\
($\chic_J$ and $h_c$) &

 $\langle O_8^{\chic_J}(^3S_1)\rangle = (2J+1)\langle O_8^{\chiczero}(^3S_1)\rangle$
& $\langle O_8^{\chiczero}(^3S_1)\rangle$ \\
&

 $\langle O_1^{h_c}(^1P_1\rangle = 3\langle O_1^{\chiczero}(^3P_0)\rangle$
& \\
&

 $\langle O_8^{h_c}(^1S_0)\rangle = 3\langle O_8^{\chiczero}(^3S_1)\rangle$
& \\
\end{tabular} 
\end{center} 
 \caption{The LDMEs relevant for joint description of charmonium states.} 
\label{tab:specLDMEs}
\end{table} 

\newpage
\subsection{Production in \bquark-hadron inclusive decays}
The \bquark-hadron decays provide a good opportunity to study charmonium since the branching fractions of inclusive \bquark-hadron decays to charmonium are relatively large (order of 1\% for S-wave charmonium) and large \bquark-hadron samples have been accumulated.
The decays of \bquark hadrons are studied at \epem (B- or Z- factories) or hadronic machines. Among the available approaches to exploit inclusive $\decay{\bquark}{(\ccbar) X}$ transitions, the most precise studies can be done for decays integrating over all available \bquark-hadrons since the resonstruction of exclusive decays has smaller efficiency.
The inclusive branching fractions of \bquark-hadron mixtures to charmonium have been measured at LEP~\cite{Abreu:1994rk,Adriani:1993ta,Buskulic:1992wp}. 
The CLEO collaboration measured the branching fraction of the $B \to \jpsi X$ decay for the first time~\cite{Alam:1986ic}.
Later B-factories succeeded to measure the branching fractions of the light B-mesons (\Bp, \Bz and sometimes \Bs) mixture to charmonium using clean event samples. The resulting measurements by \babar and CLEO2 have outstanding precision of about 1\%~\cite{Aubert:2002hc,Anderson:2002md} and report, in addition, feed-down subtracted direct branching fraction.

However, in general, the available experimental results on inclusive charmonium production from $b$-hadron decays are limited and
are shown in Table~\ref{BRinclusive}. 
\begin{table}[h]
\centering
\begin{tabular}{r|c|c}
        & $\Bub / \Bzb$ mixture   & $\Bub / \Bzb / \Bsb / b$-baryon mixture \\
\hline
\etac(1S)       & $< 0.9  @ 90 \% CL$   & $0.488 \pm 0.097$ \\
\jpsi(1S)       & $1.094 \pm 0.032$     & $1.16 \pm 0.10$ \\
\chiczero(1P)   & --                    & -- \\
\chicone(1P)    & $0.355 \pm 0.027$     & $1.4 \pm 0.4$   \\
$h_c$(1P)       & --                    & -- \\
\chictwo(1P)    & $0.100 \pm 0.017$     & --  \\
\etac(2S)       & --                    & -- \\
\psitwos        & $0.307 \pm 0.021$     & $0.286 \pm 0.028$ \\
\end{tabular}
\caption[Branching fractions of the inclusive \bquark-hadron decays into charmonium states.]{Branching fractions (in \%) of the inclusive \bquark-hadron decays into charmonium states~\cite{PDG2018}, excluding results reported in this work (Chapter~\ref{ch:phiphi}). 
The mixture of light \Bu and \Bzb mesons is shown for the measurements of the \epem experiments 
operating at centre-of-mass energy around \FourS resonance, 
while mixtures of \Bub, \Bzb, \Bsb and \bquark-baryons are considered 
for measurements from experiments at LEP, Tevatron and \lhc.}
\label{BRinclusive}
\end{table}
According to the experimental conditions, these measurements involve different mixtures of \bquark-hadron species. 
At the time, where the majority of \bquark-physics results were coming from the experiments operating around \FourS\ resonance energy, 
the \bquark-samples comprised light \Bub and \Bzb mesons. 
The results from the \cleo and \belle experiments operating around \FiveS\ resonance energy, can involve in addition \Bsb mesons. 
At \lep experiments, operating around \Z resonance region, 
and the \tevatron and \lhc, \tev scale machines, all \bquark-hadron species are produced, including weakly decaying 
\Bub, \Bzb, \Bsb, \Bcm mesons and \bquark-baryons. 

The world average values for charmonium branching fractions in the inclusive decays of mixture of light \B-mesons 
are dominated by 
\cleo~\cite{Anderson:2002md,Chen:2000ri}, \belle~\cite{Abe:2002wp, Bhardwaj:2015rju} and \babar~\cite{Aubert:2002hc} results. 
While the measurement of \jpsi, \psitwos and \chicone branching fractions are consistent across different experiments, yielding 
the average of better than 10\% precision, the \cleo
result~\cite{Chen:2000ri} on the \chictwo branching fraction 
is significantly smaller with respect to those by \belle~\cite{Abe:2002wp} and \babar~\cite{Aubert:2002hc}, and PDG gives a $3\sigma$ precise average value~\cite{PDG2016}. 

An upper limit on the inclusive \etac meson production in \bquark-hadron (\Bub and \Bzb mesons) decays was established by \cleo experiment, 
$\BF ( \Bub , \Bzb \to \etac (1S) X ) < 9 \times 10^{-3}$ at the $90 \%$~confidence level~\cite{Balest:1994jf}.
Recently, \lhcb measurement reached a precision allowing first measurement of this decay (involving all \bquark-species) of 
$\BF ( \bquark \to \etac (1S) X ) = ( 4.88 \pm 0.64 \pm 0.29 \pm 0.67 ) \times 10^{-3}$, 
where the third uncertainty is associated to the 
$\bquark \to \jpsi X$ and $\etac (1S) \to \proton \antiproton$ branching fractions~\cite{LHCb-PAPER-2014-029}. 
The world average values for the branching fraction of the \jpsi and \psitwos inclusive production in \bquark-hadron decays, 
where all \bquark-species are involved, are known at a 10\% level, with the results dominated by the measurements at \lep~\cite{Abreu:1994rk,Adriani:1993ta,Buskulic:1992wp}. 
The ratio of \psitwos and \jpsi yields have been measured at the \lhc, by the \lhcb and \cms experiments, 
to a 5\% level~\cite{LHCb-PAPER-2011-045,Chatrchyan:2011kc}.
The only PDG input for the \chic family, is the \chicone inclusive production in \bquark-hadron decays, which is a 3.5 $\sigma$ average between \delphi and \lthree measurements~\cite{Abreu:1994rk,Adriani:1993ta}. This reflects a difficulty to reconstruct low-energy photons in high multiplicity events, 
and in particular in a hadron machine environment. 
However, many measurements of charmonium production at \lhc reviewed in the next section don't report the branching fractions and hence didn't enter the PDG list.

The branching fractions of $B \to \psi X$ measured at B-factories are significantly larger than the predictions of NRQCD at LO~\cite{Ko:1995iv}.
The full NLO analysis of the inclusive B-decays to charmonium has been performed for S-wave charmonium states~\cite{Beneke:1998ks,Ma:2000bz}. Ref.~\cite{Beneke:1998ks} provides in addition a description of the P-wave charmonium states. These predictions will be used in Chapter~\ref{ch:pheno}. In both cases, authors extracted linear combinations of LDMEs. The obtained values of CO LDMEs are smaller than the ones obtained from the fit to \tevatron and \lhc hadroproduction data.

The shape of the $p^*$-distribution of charmonium in \bquark-decays is sensitive to the production mechanism including potential contributions from intermediate states. The $p^*$-distribution has been studied in Ref.~\cite{Beneke:1999gq}.

The polarisation of \jpsi produced in B-decays has been studied in Refs.~\cite{Fleming:1996pt,Ko:1999zx} predicting the values of $\lambda_{\theta}$ parameter. The result is consistent with the measurement performed by CLEO collaboration~\cite{Fleming:1996pt}. 
Generally, NRQCD is able to describe observables of S-wave charmonium produced in \bquark-decays.

Ref.~\cite{Bodwin:1992qr} provides a prediction of the \chic states production in B-decays. Note that \chiczero and \chictwo states cannot be produced at LO in CSM~\cite{Kuhn:1979zb,Kuhn:1983ar}, while the CO LDME contribution is proportional to $2J+1$. As will be shown later, the description of the \chic states production in \bquark-hadron decays is challenging, which was expected by the authors of Ref.~\cite{Beneke:1998ks}. Charmonium production in \bquark decays has been extensively studied during the era of the first generation of B-factories in 90s. Each campaign of new precision measurements of charmonium production observables may potentially indicate a demand to revisit their theoretical description. Particularly the relative \chic production rate in \bquark-hadron decays is not accomodated by available predictions.
A detailed comparison of the measurements of $S$- and $P$-wave charmonium production in \bquark-decays, performed in this thesis, is given in Chapter~\ref{ch:pheno}.

\newpage
\subsection{Hadroproduction}
\subsubsection{\jpsi and \psitwos prompt production and polarisation}
Study of prompt charmonium production and especially measurement of \jpsi total and differential production cross-section is an essential part of the physics program at \tevatron and \lhc. The first measurement of prompt \jpsi production has been performed by \cdf experiment at \sqs=1.8~\tev~\cite{Abe:1997jz,Abe:1997yz}. The comparison with CS model shown that the measured cross-section is an order of magnitude larger than the prediction made at LO, which triggered the development of the CO concept followed by the introduction of the NRQCD approach. All these theory considerations equally apply for the \psitwos state.

\begin{figure}[h]
\begin{center}
\protect\includegraphics[width=0.6\textwidth]{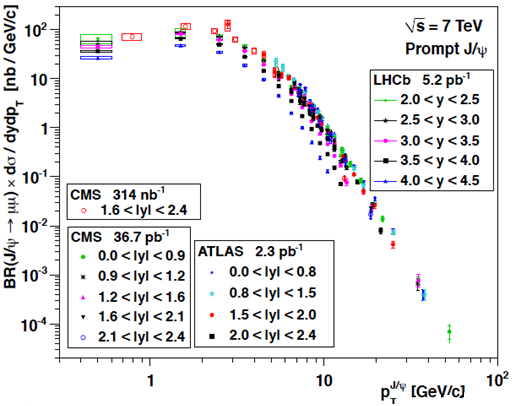}
\caption[The \pt-differential cross-section of prompt \jpsi production at the \lhc experiments at $\sqrt{s}$=7 TeV.]{The \pt-differential cross-section of prompt \jpsi production at the \lhc experiments~\cite{Aaij:2011jh,Aad:2011sp,Chatrchyan:2011kc} at $\sqrt{s}$=7 TeV.}
\label{fig:LHC_jpsi_PT}       
\end{center}
\end{figure}

At the \lhc, the differential cross-section measurements have been performed by the \lhcb, \atlas and \cms experiments at \sqs=2.76, 7, 8 and 13 \tev yielding well-consistent results, complementary to each other~\cite{LHCb-PAPER-2012-039,LHCb-PAPER-2011-003,LHCb-PAPER-2013-016,LHCb-PAPER-2015-037,Aad:2011sp,Aad:2015duc,Chatrchyan:2011kc,Sirunyan:2017qdw,Abelev:2012kr,Aamodt:2011gj,LHCb-PAPER-2013-008,LHCb-PAPER-2013-067,Chatrchyan:2013cla,Abelev:2011md,Abulencia:2007us}. The results of prompt \jpsi production cross-section in bins of transverse momentum for \lhc experiments~\cite{Aaij:2011jh,Aad:2011sp,Chatrchyan:2011kc} are shown on Fig.~\ref{fig:LHC_jpsi_PT}. Measurements from these experiments cover different regions in \pt and partially overlap, which allows a partial direct comparison.

\begin{figure}[t]
\begin{center}
\protect\includegraphics[width=1.\textwidth]{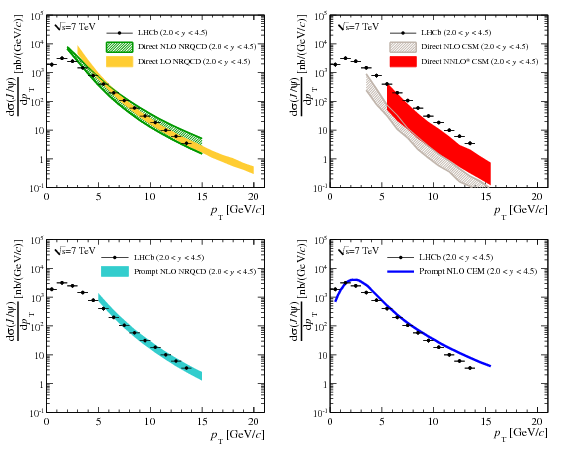}
\caption[The \pt-differential cross-section of prompt \jpsi production at \lhcb at \sqs=7 TeV compared to theory.]{The \pt-differential cross-section of prompt \jpsi production at \lhcb at \sqs=7 TeV compared to theory: direct NLO and LO NRQCD (top left), NLO and NNLO CS model (top right), prompt NLO NRQCD (bottom left), CEM (bottom right)~\cite{LHCb-PAPER-2011-003}.}
\label{fig:LHCb_vs_th}
\end{center}
\end{figure}

The comparison of measured \pt-differential production cross-sections with different theoretical models is is shown taking as an example the \lhcb measurement of \jpsi prompt production at \sqs=7~\tev~\cite{LHCb-PAPER-2011-003} on Fig.~\ref{fig:LHCb_vs_th}. 
The measurement is compared to direct NRQCD predictions at LO~\cite{Artoisenet:2010zz} and NLO~\cite{Butenschoen:2010rq}; 
CS model prediction at NLO and NNLO*~\cite{Lansberg:2008gk}, where NNLO* denotes NLO calculations with taking into account additional NNLO contributions; 
NRQCD prediction at LO taking also into account feed-down contributions~\cite{Ma:2010yw};
CEM prediction~\cite{Frawley:2008kk}.
The comparison shows that CSM cannot describe \jpsi production at NLO and NNLO* underestimating production cross-section. The CEM model provides a description of the experimentally available \pt-range only at a qualitative level. 

NRQCD provides the best description at NLO. 
One can note that NRQCD is applicable above $\pt>6~\gev$. Possible interpretations of a poor description at low-\pt region by NRQCD at NLO are that the production process is not hard enough to satisfy the factorisation assumption (collinear factorisation was used) or that the convergence of the $v$ expansion is not perfect because charmonium is still too light. The latter statement should be tested since NRQCD does not directly predict the \pt-region of its applicability. 

Recently, a \kt-factorisation prediction, which used NRQCD, showed a good description of \jpsi and \psitwos \pt-differential prompt production measurements at \lhcb as discussed in Ref.~\cite{Baranov:2015laa,Baranov:2016clx}. Due to a different factorization approach, a good description is achieved for entire experimentally measured \pt-range.



\begin{figure}[h]
\begin{center}
\protect\includegraphics[width=0.6\textwidth]{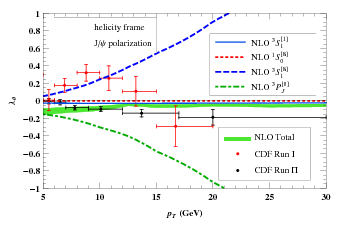}
\caption[The $\lambda_{\theta}$ polarisation parameter measured by \cdf and compared to NRQCD prediction.]{The $\lambda_{\theta}$ polarisation parameter measured by \cdf and compared to NRQCD prediction~\cite{Chao:2012iv}.}
\label{fig:jpsipolCDF}
\end{center}
\end{figure}

The first measurement of \jpsi prompt polarisation has been performed at \tevatron by \cdf collaboration~\cite{Affolder:2000nn,Abulencia:2007us} and then updated with a larger data sample.
This represents so-called the \cdf polarisation puzzle since the measurement performed using \cdf Run I data is not compatible with the \cdf Run II measurement. The reason of this incompatibility is, however, not well understood. As an illustation, a comparison of \cdf measurements to the NRQCD prediction~\cite{Chao:2012iv} is given on Fig.~\ref{fig:jpsipolCDF}. The \cdf Run II measurement is consistent to the results at LHC described below.

\begin{figure}
    \subfigure[\jpsi polarisation measurements at \lhc~\cite{Aaij:2013nlm,Abelev:2011md,Chatrchyan:2013cla}.]{
        \centering
        \protect\includegraphics[width=0.42\textwidth]{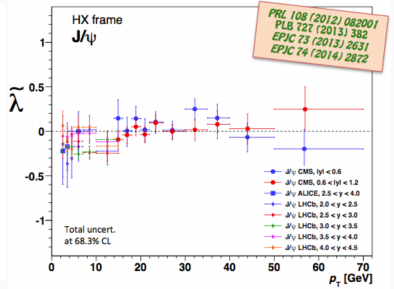}
        \label{fig:LHC_polarisation}
        }
    \quad
    \subfigure[LHCb measurement compare to theory~\cite{Aaij:2013nlm}.]{
        \centering
        \protect\includegraphics[width=0.48\textwidth]{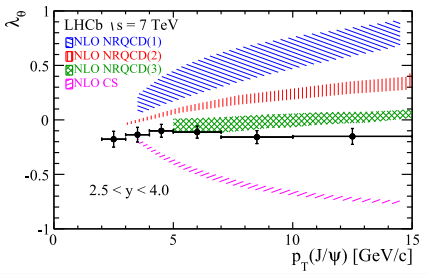}
        \label{fig:JpsiLHCbPol}
        }
    \caption{\jpsi polarisation ($\lambda_{\theta}$) measurements at \lhc as a function of \pt.}
    \label{fig:JpsiPol}
\end{figure}

Polarization of \jpsi was investigated by the \lhcb~\cite{Aaij:2013nlm}, \alice~\cite{Abelev:2011md} and \cms~\cite{Chatrchyan:2013cla} experiments. All results show small polarization and are consistent with each other (Fig. ~\ref{fig:LHC_polarisation}).
Comparison of polarisation measurement at \lhcb~\cite{Aaij:2013nlm} to theory predictions by the models developed to describe charmonia production~\cite{Butenschoen:2011yh,Chao:2012iv,Gong:2012ug} is shown on Fig.~\ref{fig:JpsiPol}. 

The NRQCD factorization framework predicts a strong polarization for CS mechanism. In addition, fits to the \jpsi production cross-section shown that the CO is the dominant process for \jpsi hadroproduction at large \pt, having  less sharp \pt spectrum than the CS contribution. In the CO, a \QQbar pair can be produced from a single gluon and hence a transverse polarisation of prompt \jpsi mesons is expected at large \pt. 

The tension between the NRQCD and the polarisation measurements can be reduced (as shown e.g. in Ref.~\cite{Gong:2012ug}) by taking into account that a large fraction (about 30\%) of prompt \jpsi mesons is coming from the feed-down of \chic and \psitwos states. Its polarization is substantially different from the polarisation of the directly produced \jpsi mesons. 

\begin{figure}[h]
\centering
  \subfigure[]{ 
    \centering
    \protect\protect\includegraphics[width=0.465\textwidth]{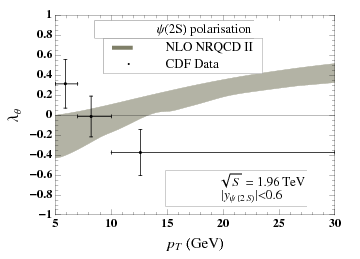}
     \label{fig:polPsiCDF}
  }
  \subfigure[]{
    \centering
    \protect\protect\includegraphics[width=0.465\textwidth]{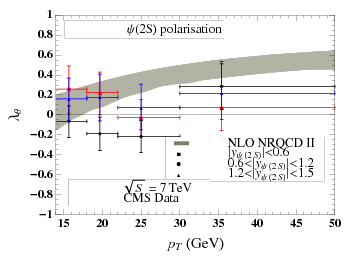}
    \label{fig:polPsiCMS}
  }
  \subfigure[]{
    \centering
    \protect\protect\includegraphics[width=0.465\textwidth]{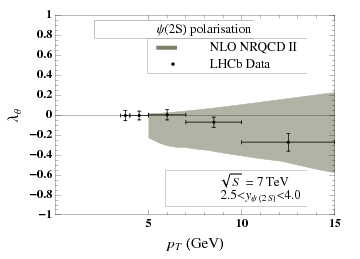}
    \label{fig:polPsiLHCb}
  }
\caption
[Theory description of measured \psitwos polarization ($\lambda_{\theta}$) at CDF, CMS and LHCb.]
{Theory description~\cite{Shao:2014yta} of measured \psitwos polarization ($\lambda_{\theta}$) at CDF~\subref{fig:polPsiCDF}, CMS~\subref{fig:polPsiCMS} and LHCb~\subref{fig:polPsiLHCb}.} 
\label{fig:psiPol}
\end{figure}

The puzzle of the \jpsi polarisation and the impact of the feed-down contributions receive more information from the studies of \psitwos polarization since no feed-down sources are expected in this case. At the same time, the production rate of \psitwos is much smaller than the one of \jpsi, and hence the measurement precision is reduced. The polarisation of the \psitwos has been measured by \cdf experiment~\cite{Abulencia:2007us} and the comparisons with theoretical predictions are not conclusive due to large experimantal uncertainties. 
After that, the \psitwos polarisation has been measured at \lhc by \cms~\cite{Chatrchyan:2013cla} and \lhcb~\cite{Aaij:2014qea}.
In Ref.~\cite{Shao:2014yta} authors report a good description of \psitwos polarization measured at \cdf, \cms and \lhc. The comparison of measurements with theoretical description is given on Fig.~\ref{fig:psiPol}. Despite reasonable description, a more precise measurement of \psitwos polarization (especially at large \pt) is needed due large experimental uncertainties of existing results.

A polarization of \jpsi was known to be problematic for description by \kt-factorizarion approach due to large predicted polarization contrary to measurements.
Refs.~\cite{Baranov:2015laa,Baranov:2016clx} provide a reasonable description of mentioned \jpsi and \psitwos polarization measurements by using \kt-factorization approarch together with NRQCD.  A good description of \jpsi polarization is achieved due to the cancellation of contributions from $^3P_1^{[8]}$ and $^3P_2^{[8]}$ CO terms in the process $^3S_1^{[8]} \to ^3P_J^{[8]} \to \jpsi$.

\subsubsection{\etac and \etactwos prompt production}
The \etac state is much less studied due to complications of its reconstruction at \lhc. At charm factories a sample of \etac mesons is reduced due to small branching fraction of \psitwos and \jpsi decays to \etac. A large data sample and selective trigger is needed in order to observe a signal from prompt \etac mesons. The only measurement of prompt and b-decays production cross-section has been performed so far.

The \lhcb collaboration measured for the first time a cross-section 
of the \etac meson prompt production in proton-proton collisions 
at $\sqs = 7$ and $8 \tev$~\cite{LHCb-PAPER-2014-029}. 
Due to challenging background conditions and limited trigger bandwidth, some bins of the measurement of \pt-differential production cross-section have uncertainties larger than the uncertainties in NRQCD predictions. The experimental uncertainties are dominated by statistical ones. A more precise measurement of the \etac production with large data set at higher \sqs with increased production cross-section is required to validate the observed effect and study its energy dependence. A larger data sample allows to improve a technique of the measurement by explicit modelling its decay-time distribution.
For \jpsi production studies~\cite{LHCb-PAPER-2015-037}, a measurement of the ratio of production cross-sections at different \sqs has largely reduced theory and experimental uncertainties. Contrary to that, a similar ratio for \etac production cross-section would be less precise than a single measurement, since experimental uncertanties are strongly dominated by statistical ones. In principle, a large enough data sample would allow to extend presently studied \pt-range.

The measurement has been compared to four NRQCD predictions~\cite{Butenschoen:2011yh,Chao:2012iv,Gong:2012ug,Bodwin:2014gia}. The predictions are obtained by projecting the \jpsi production cross-section using HQSS relations. More details are discussed in Chapter~\ref{ch:pheno}.
The comparison shows that the CS contribution already saturates the observed cross-section and CO contribution projected from the \jpsi production studies would largely overshoot the measured \pt-differential production cross-section. 

The tension between the theory and experimental result is clear for all available predictions.
\begin{figure}[b]
\centering{
        \subfigure[~\cite{Butenschoen:2011yh}]{ 
          \protect\protect\includegraphics[width=0.22\textwidth]{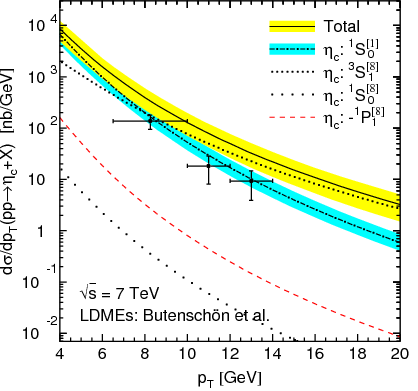}
          \label{fig:etac1S_Butten_7tev}}
        \subfigure[~\cite{Chao:2012iv}]{ 
          \protect\protect\includegraphics[width=0.22\textwidth]{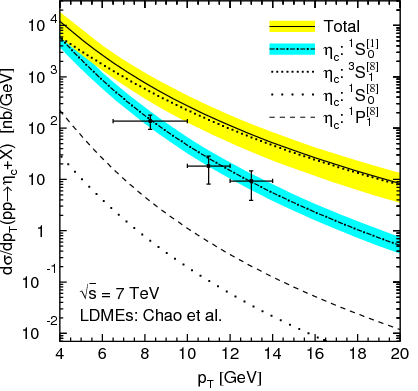}
          \label{fig:etac1S_Chao_7tev}}
        \subfigure[~\cite{Gong:2012ug}]{ 
          \protect\protect\includegraphics[width=0.22\textwidth]{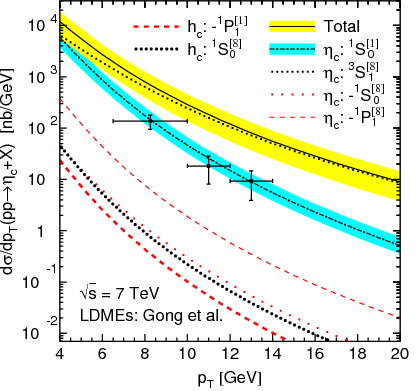}
          \label{fig:etac1S_Gong_7tev}}
        \subfigure[\cite{Bodwin:2014gia}]{ 
          \protect\protect\includegraphics[width=0.22\textwidth]{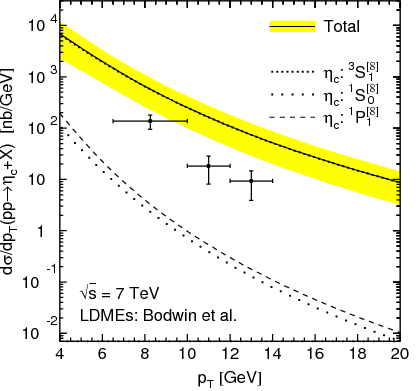}
          \label{fig:etac1S_Bodwin_7tev}}
        \subfigure[~\cite{Butenschoen:2011yh}]{ 
          \protect\protect\includegraphics[width=0.22\textwidth]{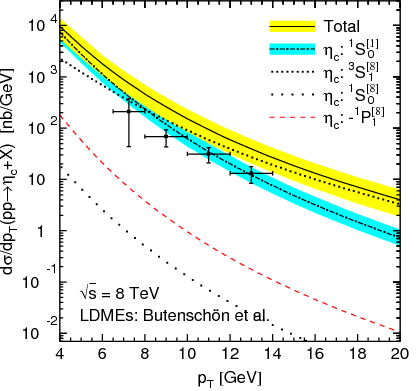}
          \label{fig:etac1S_Butten_8tev}}
        \subfigure[~\cite{Chao:2012iv}]{ 
          \protect\protect\includegraphics[width=0.22\textwidth]{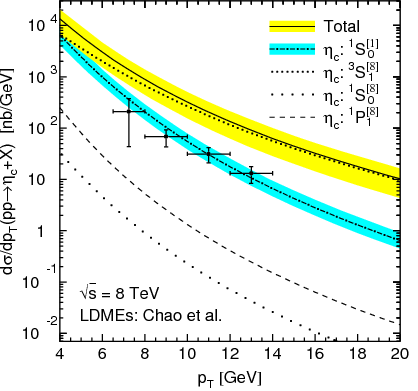}
          \label{fig:etac1S_Chao_8tev}}
        \subfigure[~\cite{Gong:2012ug}]{ 
          \protect\protect\includegraphics[width=0.22\textwidth]{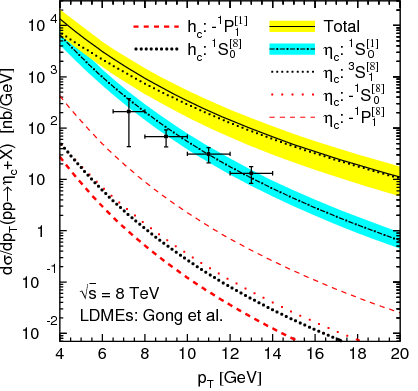}
          \label{fig:etac1S_Gong_8tev}}
        \subfigure[~\cite{Bodwin:2014gia}]{ 
          \protect\protect\includegraphics[width=0.22\textwidth]{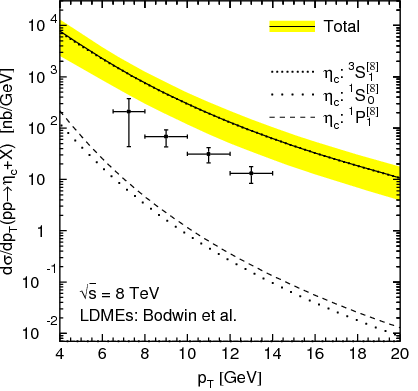}
          \label{fig:etac1S_Bodwin_8tev}}
 }
\label{fig:etacPmtComp}
\caption
[Comparison of theory predictions to the \etac prompt production measurements at \sqs=7 and 8 TeV by \lhcb.] 
{Comparison of predictions from Refs.~\cite{Butenschoen:2011yh,Chao:2012iv,Gong:2012ug,Bodwin:2014gia} to the \etac prompt production measurements at \sqs=7 (a-d) and 8 (e-h) TeV by \lhcb. The yellow (blue) band represents the CO (CS)contribution. Figure is taken from Ref.~\cite{Butenschoen:2014dra}.} 
\end{figure}
The \lhcb measurement demonstrated a lack of comprehensive theoretical models, which are able 
to simultaneously describe a production cross-section of the \jpsi and \etac 
states and a polarisation of the \jpsi meson.
The links between the LDMEs corresponding to the \jpsi and \etac production within a heavy-quark spin symmetry assumption as discussed in Section~\ref{sec:qqbarProd} allow to make a prediction of \etac prompt production using LDMEs determined from the fit to \jpsi production measurement. 

\begin{figure}[htb]
\centering{
        \subfigure[]{ 
          \protect\protect\includegraphics[width=0.4\textwidth]{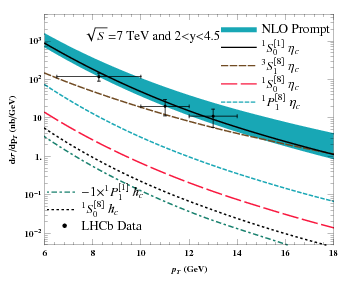}
          \label{fig:etac1S_HS_7tev_new}}
        \subfigure[]{ 
          \protect\protect\includegraphics[width=0.4\textwidth]{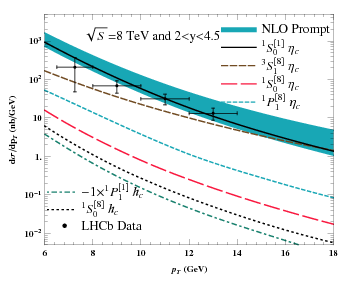}
          \label{fig:etac1S_HS_8tev_new}}
 }
\caption
[The \etac production measurement compared to the prediction.]
{The \etac production measurement compared to the prediction from Ref.~\cite{Han:2014jya}.} 
\label{fig:etacHSnew}
\end{figure}

\begin{figure}[htb]
\centering{
        \subfigure[]{ 
          \protect\protect\includegraphics[width=0.45\textwidth]{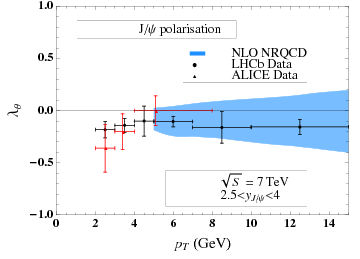}
          \label{fig:PolNoConstr}}
        \subfigure[]{ 
          \protect\protect\includegraphics[width=0.45\textwidth]{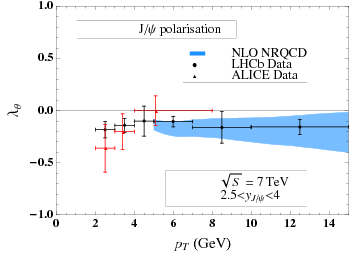}
          \label{fig:PolConstr}}
 }
\caption
[The NRQCD prediction of \jpsi polarisation compared to the \lhcb and \alice measurements without and with the contraint from the \etac production measurement.]
{The NRQCD prediction of \jpsi polarisation compared to the \lhcb~\cite{Aaij:2013nlm} and \alice~\cite{Abelev:2011md} measurements without~\subref{fig:PolNoConstr} and with~\subref{fig:PolConstr} the contraint from the \etac production measurement as discussed in the text. Figures are taken from Refs.~\cite{Shao:2014yta,Han:2014jya}.}
\label{fig:polEtac}
\end{figure}

The \lhcb measurement triggered new efforts to describe S-wave charmonium production.
A revisiting of the theoretical framework followed~\cite{Feng:2015cba,Sun:2015pia,Likhoded:2015qyl,Gao:2016ihc,Faccioli:2017hym,Butenschoen:2017iks, Baranov:2016clx}.
Recent progress by theorists~\cite{Han:2014jya} yielded a good description of \etac production in a limited \pt-range by taking into account both CS and CO contributions (Fig.~\ref{fig:etacHSnew}). Currently, this is the only available successful description of the \etac production by NRQCD; and values od LDMEs were constrained by the \etac prompt production measurement. A good description of the data points is achieved by two CO contributions cancelling each other, which creates a hierarchy problem. This calls for further development of theory models describing S-wave charmonium production. Note, that a blue band on Fig.~\ref{fig:etacHSnew} doesn't represent a theory uncertainty but represents an uncertainty of 100\% due to reasons described below. Authors didn't use the \etac prompt production measurement in the simulataneous fit. Instead of that, they neglected the dominant CS contribution in the \etac production to obtain a very conservative constraint (upper limit) on relevant CO LDMEs. 
The obtained CO LDMEs were projected to the prompt \jpsi polarisation measurements by \lhcb~\cite{Aaij:2013nlm} and \alice~\cite{Abelev:2011md}. Figure~\ref{fig:polEtac} compares the NRQCD prediction with and without a constraint obtained using \etac prompt production measurement.

\begin{figure}[h]
\centering
\subfigure[]{
\protect\protect\includegraphics[width=0.4\textwidth]{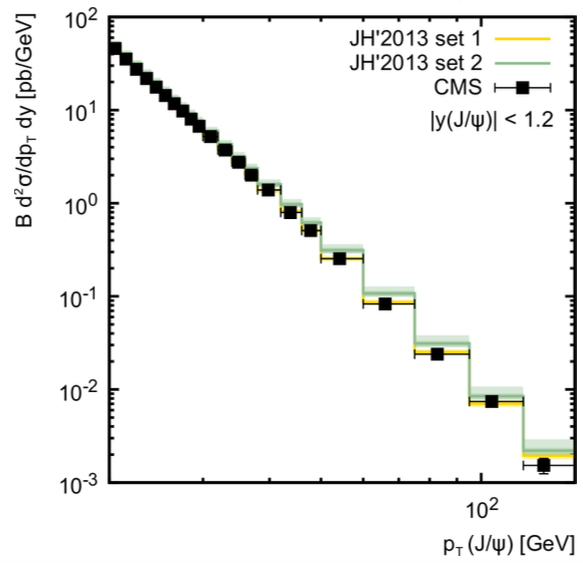}
\label{fig:jpsiKT}
}
\quad
\quad
\subfigure[]{
\protect\protect\includegraphics[width=0.4\textwidth]{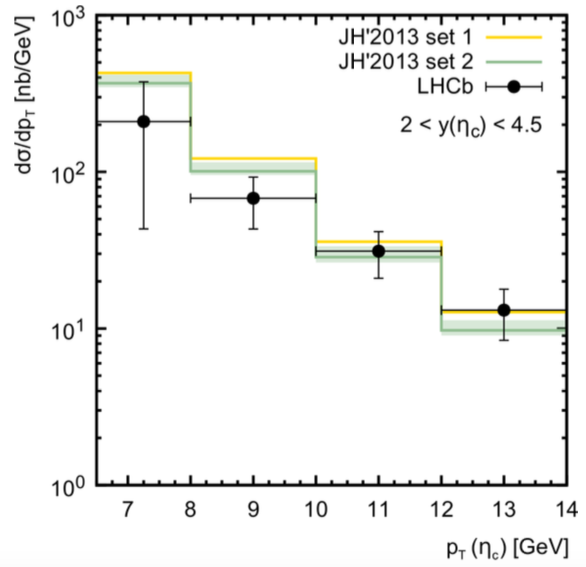}
\label{fig:etacKT}
}
\caption
[A simultaneous fit of the \jpsi and the \etac prompt production measured at \cms and \lhcb compared to the \kt-factorization prediction incorporating NRQCD.]
{A simultaneous fit of the \jpsi~\subref{fig:jpsiKT} and the \etac~\subref{fig:etacKT} prompt production measured at \cms~\cite{Khachatryan:2015rra} and \lhcb~\cite{Aaij:2014bga} compared to the \kt-factorization prediction incorporating NRQCD~\cite{Baranov:2019joi}.}
\label{fig:KT}
\end{figure}

\begin{figure}[h]
\centering
\subfigure[]{
\protect\protect\includegraphics[width=0.4\textwidth]{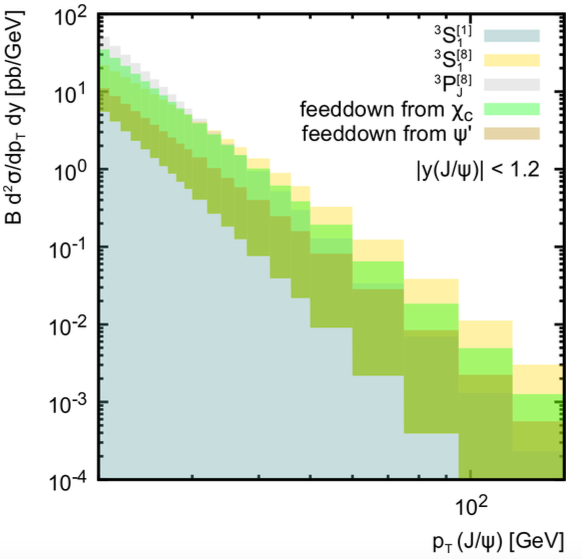}
\label{fig:jpsiKT_frac}
}
\quad
\quad
\subfigure[]{
\protect\protect\includegraphics[width=0.4\textwidth]{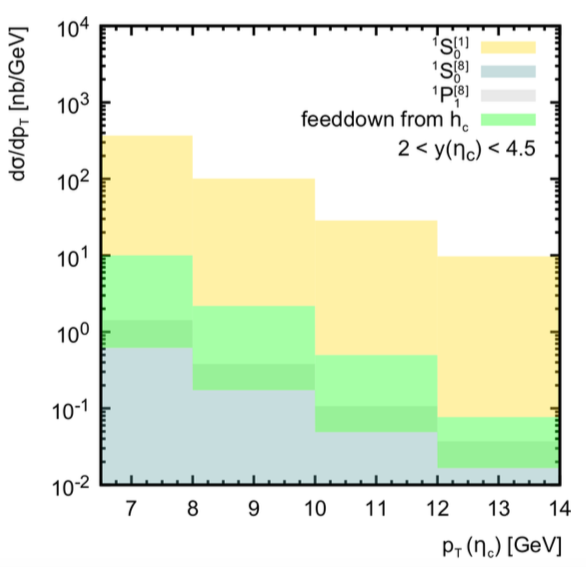}
\label{fig:etacKT_frac}
}
\caption
[Contributions to \jpsi and \etac prompt production within \kt-factorization prediction incorporating NRQCD.]
{Contributions to \jpsi~\subref{fig:jpsiKT_frac} and \etac~\subref{fig:etacKT_frac} prompt production within \kt-factorization prediction incorporating NRQCD~\cite{Baranov:2019joi}.}
\label{fig:KT_frac}
\end{figure}

Recently, authors of Ref.~\cite{Baranov:2019joi} provided a description of the \etac prompt production with \kt-factorization incorporating NRQCD.
This work provides a simultaneous fit of \jpsi~\cite{Khachatryan:2015rra} and \etac~\cite{Aaij:2014bga} prompt production measurements at \lhc contrary to the NRQCD prediction made within collinear factorisation addressed above. In addition, no kinematical requirements on the \pt or rapidity range were used. The fit takes into account feed-down contributions from \chic and \psitwos to \jpsi and from $h_c$ to \etac.
The results of simultaneous fit on Fig.~\ref{fig:KT} for \jpsi prompt production are compared to \cms measurement~\cite{Khachatryan:2015rra} and for \etac prompt production compared to \lhcb measurement.
The contributions from different CS and CO states and feed-down sources are shown on Fig.~\ref{fig:KT_frac} to \jpsi prompt production and for \etac prompt production.
This result showed a good simultaneous description of both \jpsi and \etac prompt production measurements at \lhc in a considered \pt range. Note, that on the plots a contrinution from $^1S_0^{[8]}$ to the \etac production is not present due to cancellations in CO processes similar to those mentioned for \jpsi polarisation description within \kt-factorization. Within this fit a theoretical description is strongly constrained. This is reflected by small theoretical uncertaintites displayed, which include uncertaintites due to scale and LDME values only. 

\clearpage
No difference in the relation between $2S$ states and $1S$ states production is expected.
Since similar links between LDMEs apply also between \psitwos and \etactwos, Lansberg and Shao suggested to measure prompt \etactwos at \lhcb using $\decay{\etactwos}{\ppbar}$ decay.
The measurement of the \etactwos prompt production would be a further stringent test of the NRQCD model developed for the \etac and \jpsi production. 
The advantage of \etactwos and \psitwos states is that both are expected to be feed-down free.
The reconstruction of the \etactwos state is however more complicated than that of \etac. Not many \etactwos decays have been observed, and only a few measurements of branching fractions are available. The discussion on applicable decay channels to reconstruct \etactwos at \sqs=13~\tev at \lhcb is given in Chapter~\ref{ch:decays}.

\begin{figure}[h]
\centering{
\protect\protect\includegraphics[width=0.48\textwidth]{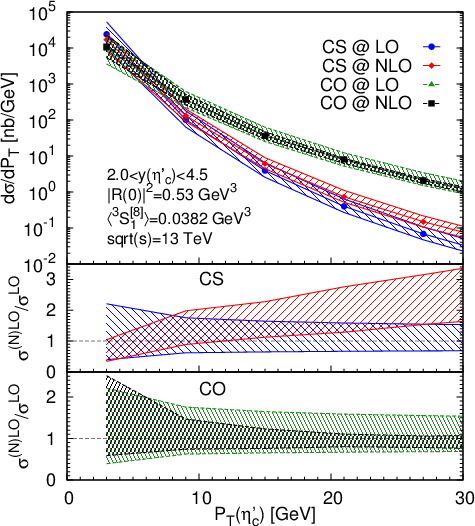}
}
\caption
[The NRQCD prediction of the \pt-differential \etactwos production cross-section for the \lhcb fiducial region at \sqs=13 TeV.] 
{The NRQCD prediction of the \pt-differential \etactwos production cross-section for the \lhcb fiducial region at \sqs=13 TeV~\cite{Lansberg:2017ozx}.} 
\label{fig:etac2_JPL}
\end{figure}

A predictions for the \etactwos production can be done in a similar way as for \etac and is available in Ref.~\cite{Lansberg:2017ozx} together with projections to the \lhcb fiducial region. The NRQCD prediction for the \pt-differential prompt production cross-section at \sqs=13 TeV of \etactwos at LO and NLO is shown on Fig.~\ref{fig:etac2_JPL}.

\begin{figure}[h]
\centering{
        \subfigure[]{ 
          \protect\protect\includegraphics[width=0.4\textwidth]{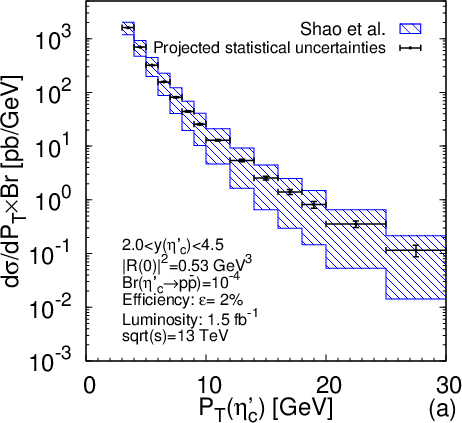}}
        \quad
        \quad
        \subfigure[]{ 
          \protect\protect\includegraphics[width=0.4\textwidth]{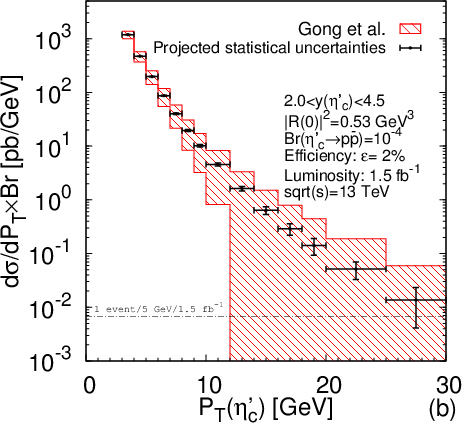}}
        \quad
        \quad
        \subfigure[]{ 
          \protect\protect\includegraphics[width=0.4\textwidth]{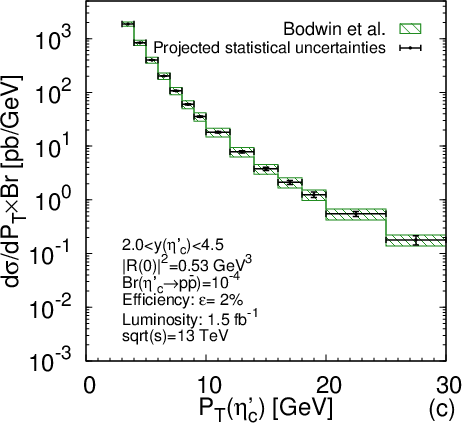}}
          }
\caption
[The NRQCD predictions of the \pt-differential \etactwos production cross-section for the \lhcb fiducial region at \sqs=13 TeV.]
{The NRQCD predictions of the \pt-differential \etactwos production cross-section for the \lhcb fiducial region at \sqs=13 TeV from Refs.~\cite{Shao:2014yta} (a)
~\cite{Gong:2012ug} (b)
~\cite{Bodwin:2015iua} (c).
Figure is taken from Ref~\cite{Lansberg:2017ozx}.}
\label{fig:etac2sProj}
\end{figure}

The predictions for the \etac production using LDMEs from three theoretical groups~\cite{Shao:2014yta,Gong:2012ug,Bodwin:2015iua} are shown on Fig.~\ref{fig:etac2sProj}. The prediction from Ref.~\cite{Gong:2012ug} has the largest uncertainty due to allowed negative values of LDMEs contrary to two other predictions. The prediction from Ref.~\cite{Bodwin:2015iua} has the smallest uncertainty. A measurement of the \etactwos prompt production is an important test of mentioned predictions.

\clearpage
\subsubsection{\chic prompt production}
The prompt production of \chic mesons is conventionally studied using $\decay{\chic_{1,2}}{\jpsi\to\mup\mu^{-} \gamma}$ decays. In addition, a reconstruction of \jpsi decay to pair of muons, requires a reconstruction of a photon with an energy of a few handred~\mev. The \chicone (\chictwo) states have relatively large  branching fractions of radiative transition to \jpsi of about 30\%(20\%).
The \chiczero is more complicated to reconstruct due to a smaller branching fraction (1.4\%) and lower photon energy. The photon energy is reconstructed using calorimeter or tracking detectors for photon conversions to a \epem pair takes place. Calorimeter resolution for low-energy photons is often compromised, as in the case of LHCb calorimeter optimized to resolve photons from radiative \bquark-decays. Since masses \chicone and \chictwo states are separated by 40~\mev, limited detector resolution can lead to overlapping peaks and hence \chicone and \chictwo signals will be complicated to separate. 
Below, the available results on the \chicone and \chictwo prompt production are discussed, while there is no measurement of \chiczero \pt-differential production cross-section.

The prompt production of \chicone and \chictwo states has been measured by \cdf~\cite{Abe:1997yz} at \sqs=1.8 \tev; \atlas~\cite{ATLAS:2014ala}, \cms~\cite{Chatrchyan:2012ub} and \lhcb~\cite{Aaij:2013dja} at \sqs=7~\tev. In the same paper, the \lhcb collaboration also reported a value of integral \chiczero relative production with a significane of about 4 $\sigma$.

\begin{figure}[h]
\centering{
\protect\protect\includegraphics[width=0.6\textwidth]{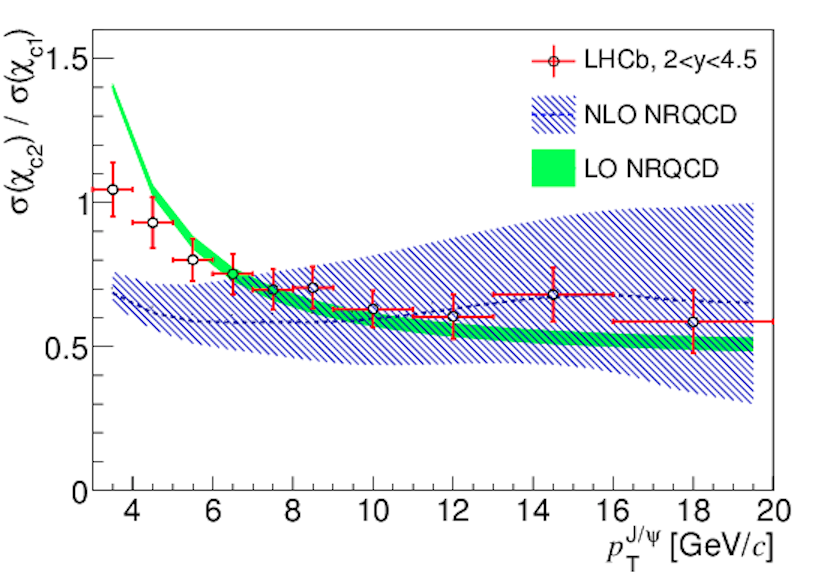}
}
\caption
[The \chictwo-to-\chicone prompt production ratio measured at \lhcb compared to the NRQCD prediction at NLO and LO.] 
{The \chictwo-to-\chicone prompt production ratio measured at \lhcb~\cite{Aaij:2013dja} compared to the NRQCD prediction at NLO~\cite{Ma:2010vd} and LO~\cite{Likhoded:2013aya}. Figure is taken from Ref.~\cite{Aaij:2013dja}.} 
\label{fig:chicRat}
\end{figure}

The NRQCD prediction at NLO~\cite{Ma:2010vd} well describes data point measured by \atlas. It can also be compared to \kt-factorisation prediction~\cite{Baranov:2010zz}, showing that \kt-factorisation overshoot data points. Later calculations have been updated with incorporating NRQCD~\cite{Baranov:2015yea}, which showed a good description of measurements.
Another powerful observable is the relative \chictwo-to-\chicone production ratio, which has to be also addressed. 
The comparison of the \lhcb measurement of the ratio to the NLO NRQCD prediction~\cite{Ma:2010vd} is shown on Fig.~\ref{fig:chicRat}. The NRQCD at NLO describes well the ratio for $\pt>6~\gev$ only. 

Note, that all measurements of \chic prompt production is done under the assumption that \chic states are produced unpolarized. In addition, the NRQCD fit two production observables with two LDMEs. Additional observables would over-constrain the $P$-wave charmonium production description. Unfortunately, the \chiczero and $h_c$ hadroproduction and $\chic_J$ and $h_c$ prompt polarisation have not been measured so far.
A measurement of \chic production down to small \pt  can be done by exploiting recently discovered $\decay{\jpsi}{\chic\mup\mu^{-}}$ decays to study the low-\pt region, where the NRQCD doesn't provide reliable description of data points. 

\subsection{Photoproduction in $ep$ collisions}
\begin{figure}[h]
\centering
\protect\includegraphics[width=1.\textwidth]{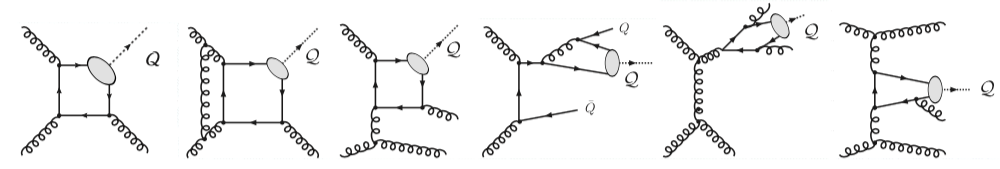}
\caption
[Diagrams representing \jpsi hadroproduction within CS mechanism.]
{Diagrams representing \jpsi hadroproduction within CS mechanism. Figure is taken from Ref.~\cite{Lansberg:2019adr}.}
\label{fig:DiagrHadro}
\centering
\protect\includegraphics[width=1.\textwidth]{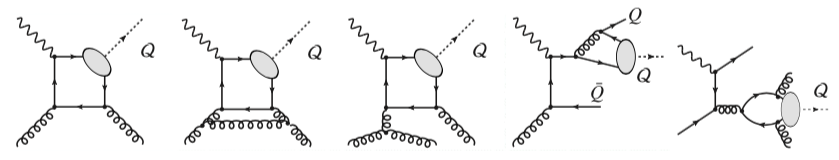}
\caption
[Diagrams representing \jpsi photoproduction within CS mechanism.]
{Diagrams representing \jpsi photoproduction within CS mechanism. Figure is taken from Ref.~\cite{Lansberg:2019adr}.}
\label{fig:DiagrPhoto}
\end{figure}

The photoproduction of charmonium can also be studied using $ep$ collisions. 
The production process at $ep$ collisions is characterised by a dynamical variable $z$ (elasticity) that is defined as a fraction of the virtual photon momentum carried by the final state charmonium.
Depending on the value of $z$, the production can happen at different regimes.
The direct photoproduction regime takes place for small photon virtuality $q^2$ and $z\gsim0.3$. Another important observable is an invariant mass of $\gamma p$ system, $W_{\gamma p}$ or $W$, which reflects the energy of incoming photon. Hence the differential production cross-section in $W$ is also measured.
In this case, the electron scattering angle is small, and the photon can be treated as quasi-real. 
Naively, one can expect that the description of the photoproduction is easier than the hadroproduction. Indeed, the diagrams representing photo- and hadro-production are similar. One needs to replace a gluon in the initial state by a photon in hadroproduction diagram to obtain a diagram for photoproduction. For example, diagrams relevant for \jpsi within CS mechanism are shown for hadroproduction on Fig.~\ref{fig:DiagrHadro} and for photoproduction on Fig.~\ref{fig:DiagrPhoto}.

However, in addition to direct photoproduction, the photon can interact with \cquark quark via hadronic component (e.g. resolved process at $z\lsim0.3$). The resolved process is more complicated to describe theoretically. In addition to that, the exclusive and diffractive production cross-section is comparable for the photoproduction. Hence, all the processes mentioned above should be considered to achieve a comprehensive theory description.

From the experimental side, due to small values of total cross-section, large collected luminosities are needed to study charmonium photoproduction.
The \jpsi and \psitwos inelastic production cross-sections in $ep$ collisions have been measured at HERA by ZEUS and H1 collaborations~\cite{Adloff:2002ex,Aaron:2010gz,Chekanov:2002at,Chekanov:2009ad}. Later, $z$ and \pt-differential cross-sections have been reported~\cite{Abramowicz:2012dh}

The first NLO calculation of the charmonium photoproduction has been performed in Refs.~\cite{Kramer:1995nb,Kramer:2001hh} and shows a reasonable description of the measured cross-sections. 
Later, it has been shown that the values of factorisation and renormalisation scales used in Ref.~\cite{Kramer:1995nb} are probably too low and after correcting the scale, the NLO prediction underestimates the H1 and ZEUS~\cite{Butenschoen:2009zy,Artoisenet:2009xh,Chang:2009uj,Butenschon:2009zz,Butenschon:2010iy,Butenschoen:2012qh} results.
The \jpsi photoproduction at HERA has also been studied using \kt-factorisation approach at LO~\cite{Baranov:2002cf,Kniehl:2006sk,Martin:2009ii} with CSM. The CSM can explain measured \pt and $z$ distributions within large uncertainties arising from parton PDFs.
The first complete NLO analysis with the CO contribution considered is reported in Ref.~\cite{Butenschoen:2009zy}. The obtained predictions are compatible with the H1 measurement of both \pt- and $z-$differential cross-section. 
This result has been used in the simultaneous fit of the \jpsi hadroproduction and photoproduction~\cite{Butenschoen:2011yh}. 
The NLO NRQCD fit is performed to prompt production cross-section measurements at RHIC~\cite{Adare:2009js}, Tevatron~\cite{Acosta:2004yw,Abe:1997jz} and LHC~\cite{Khachatryan:2010yr,ATLAS:2010sca,Abelev:2014qha,Aaij:2011jh}, and photoproduction at HERA \cite{Adloff:2002ex,Aaron:2010gz,Chekanov:2002at}. The \chisqndf of the fit is $\chisqndf = 857/194 = 4.42$. For most of cases, theoretical uncertaintites are larger that experimental ones. The fit resonably describes hadroproduction measurements with a slight tension with \cms measurement at large \pt. The worst description takes place for $z$-differential photoproduction cross-section measurements due to complications mentioned above.
The obtained values of LDMEs will be compared to the result of simultaneous fit to hadroproduction and production in \bquark-hadron decays in Chapter~\ref{ch:pheno}. The results in Ref.~\cite{Butenschoen:2011yh} are also compared to \jpsi production measurements in $\gamma\gamma$ collisions addressed in the next section.
\begin{figure}[h!]
\centering
\protect\protect\includegraphics[width=0.95\textwidth]{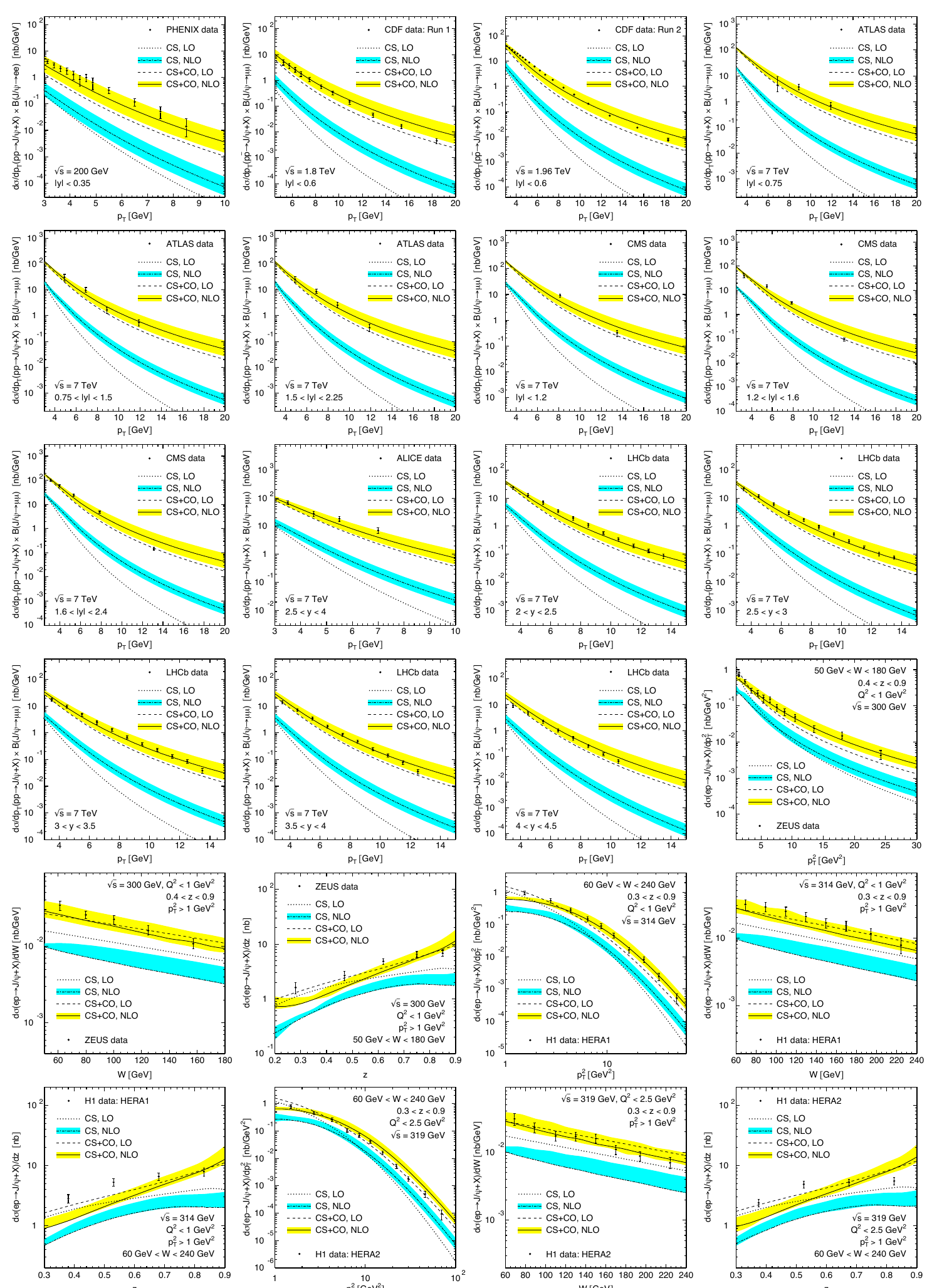}
\caption
[The NRQCD NLO fit to hadroproduction at RHIC, Tevatron and \lhc and protoproduction measurements at HERA.]
{The NRQCD NLO fit to hadroproduction at RHIC, Tevatron and \lhc and protoproduction measurements at HERA ~\cite{Butenschoen:2011yh}.} 
\label{fig:globalFit}
\end{figure}

The polarisation of the \jpsi has also been measured at HERA. 
The H1 collaboration measured both $\lambda$ and $\nu$ parameters in helicity and Collins-Soppers frames and required $0.3<z<0.9$ to suppress diffractive contributions~\cite{Aaron:2010gz}. The ZEUS collaboration performed a measurement in target spin-quantisation frame.
The measurements show small longitudinal polarisation decreasing with \pt.
The description of the polarisation is reasonable for both CSM~\cite{Artoisenet:2009xh,Chang:2009uj,Aaron:2010gz,Chekanov:2009ad} and with considering entire NRQCD at NLO~\cite{Butenschoen:2011ks}. Taking into account uncertainties, it is not clear whether the CO contribution is needed to describe the observed polarisation.


\clearpage
\subsection{Inclusive production in $\gamma\gamma$ collisions}
Another important observable of charmonium production comes from $\gamma\gamma$ collisions. 
Similarly to the photoproduction case, in addition to direct production, the resolved photons can contribute to the production mechanism.
Moreover, one can distinguish the processes with a single resolved photon ($i\gamma \to (\ccbar) i$) and with two resolved photons (double-resolved) $ij \to (\ccbar) k$, where $i$, $j$ and $k$ denote either the light quark or a gluon.
A single resolved process is similar to a photoproduction, while the double-resolved process is similar to the hadroproduction.

The integral inclusive production of \jpsi in $\gamma\gamma$ collisions has been measured at \lep by \delphi experiment~\cite{Abdallah:2003du}. The limited data sample size did not allow to perform precise measurement of differential production cross-section.
The measurement has been compared to the CSM predictions at LO~\cite{Ma:1997bi,Japaridze:1998ss,Godbole:2001pj,Klasen:2001mi,Klasen:2001cu} showing that theory underestimates the measured cross-section.
The first complete NLO prediction was found to be similar to the LO one~\cite{Butenschoen:2011yh}. 
In Ref.~\cite{Chen:2016hju}, the contribution from $\gamma\gamma\to\jpsi\ccbar X$ has been addressed at NLO level, and it has been shown that this process might dominate the CS production.

All above suggests that the CO contribution can be dominant, but the existing prediction~\cite{Butenschoen:2011yh} is several times smaller than the experimental result. This, however, is not conclusive given large uncertainties from both theory and experimental sides. This calls for a new precise measurement of the \jpsi production in $\gamma\gamma$ collisions.

\subsection{Inclusive production in \epem collisions}
The prompt inclusive production cross-section of the \jpsi meson in \epem collisions, $\epem \to \jpsi X$, has been measured most precisely by  \babar~\cite{Aubert:2001pd} and \belle~\cite{Abe:2001za} experiments. However, results from other \epem experiments are also available.
To describe this process, theory predictions should also take into accoung contributions from specific $\epem \to \jpsi \ccbar$, $\epem \to \jpsi gg$ and $\epem \to \jpsi \qqbar + gg$ processes, where $q$ denotes \uquark, \dquark or \squark quark~\cite{Yuan:1996ep,Cho:1996cg,Baek:1998yf,Schuler:1998az,Kiselev:1994pu,Liu:2003zr}.
The LO calculations using CSM predict the cross-section, which is 3-5 times smaller than the measured values. In addition, the measurement of the $\epem \to \jpsi \ccbar$ cross-section by \belle~\cite{Abe:2002rb} is 5 times larger than the LO NRQCD prediction with both CS and CO~\cite{Cho:1996cg,Baek:1998yf,Schuler:1998az,Kiselev:1994pu,Liu:2003zr,Liu:2003jj} mechanisms considered. 

Later, the \belle collaboration measured the cross-section of $\epem \to \jpsi X$, $\epem \to \jpsi \ccbar$ and $\epem \to \jpsi X_{non-\ccbar}$~\cite{Pakhlov:2009nj}. The value of the $\epem \to \jpsi X$ cross-section was found to be almost twice smaller than the first measurement. Nevertheless, the updated measurement of the $\epem \to \jpsi X$ cross-section is still larger than  theory predictions. 

Authors of Ref.~\cite{Zhang:2006ay} shown that NLO calculations lead to significant enhancement by about factor 2 in the production cross-section compared to the LO result when using the same set of input parameters. 
Further investigations have been done by taking into account the QED contributions from  the $\epem\to 2\gamma^* \to \jpsi \ccbar X$ and $\epem\to \chic \ccbar X$ processes~\cite{Zhang:2006ay} and the feed-down contribution from the \psitwos state, which produces a contribution of about 35\% to the $\epem \to \jpsi \ccbar X$ cross-section. After taking into account all contributions mentioned above, the discrepancy between theory and measurements is largely reduced.
Similarly, corrections to the $\epem \to \jpsi gg$ process at NLO have been calculated~\cite{Ma:2008gq,Gong:2009kp}. Finally the result became consistent with the latest \belle measurement of the $\epem \to \jpsi X_{non-\ccbar}$ cross-section as shown below.
\begin{equation}
\begin{aligned}
\sigma^{Belle}(\epem\to \jpsi X_{non-\ccbar}) &= 0.43 \pm 0.09 \pm 0.09 \pb, \\
\sigma^{NLO}(\epem\to \jpsi gg) &= 0.29-0.41 \pb.
\end{aligned}
\end{equation}
In addition, the measured $p^*$ distributions of the $\epem \to \jpsi \ccbar$ and  $\epem \to \jpsi X_{non-\ccbar}$ processes~\cite{Pakhlov:2009nj} are reasonably compatible with the NLO prediction of $\epem \to \jpsi gg$~\cite{Ma:2008gq,Gong:2009kp}.

Later, it has been shown that the relativistic corrections produce an enhancement in the $\sigma(\epem \to \jpsi gg)$ cross-section~\cite{He:2009uf,Jia:2009np}. If one takes into account these corrections together with the NLO computation, the CS contribution will saturate the measured $\epem \to \jpsi X_{non-\ccbar}$ cross-section. It has been understood that a poor description of the measurements originated from the values of LDMEs, which were extracted from the fit to \jpsi hadroproduction at \tevatron~\cite{Zhang:2009ym} at LO. 

To conclude, the studies of \jpsi production in \epem collisions also challenge NRQCD. It took more than ten years to achieve a reasonable description of the production observables. The important consequence of the studies is that the LO calculations cannot describe \jpsi production observables and at least the NLO level is needed.

In the next chapter, I will discuss the charmonium decays to hadronic final states, which can be used to study charmonium production in the \lhcb experiment.

\begin{singlespace}
\chapter{Charmonium decay channels}
\label{ch:decays}
\end{singlespace}
This chapter summarises charmonium decay channels, which are used or can potentially be used to reconstruct different charmonium states and measure their production and properties. The listed decay channels are promising for studies at the \lhcb experiment. Therefore they are most useful for further discussion. 
Not many charmonium decays to hadrons have been reconstructed in the hadron machine environment. 
Therefore, it is often difficult to predict the corresponding physics reach before proceeding to the actual data analysis. 
In should be stated, that in order to study promptly produced charmonium with a specific hadronic decay channel, a dedicated online trigger line should be developed.

The decays including neutral particles in the final state are used at B-factories but are more challenging at LHCb, have not been used in the studies performed within the thesis and hence are omitted. 

The charmonium decays receive much attention from theory since measurements of their branching fractions often challenge theoretical predictions. This topic is also explicitly excluded from the discussion.
\clearpage
The most precise charmonium studies employ decays into clean dimuon final state, 
which is possible for $J^{PC} = 1^{-\,-}$ charmonia. 
In addition to the $1^{-\,-}$ states, the $\chic$ family can be accessed via radiative transitions to \jpsi, 
$\chi_{c}\to \jpsi(\to\mu^{+}\mu^{-})\gamma$. 
However, low-energy photon reconstruction is required.
Other states from the charmonium family cannot be explored using decays to a pair of muons. Therefore other final states should be investigated~\cite{Barsuk:2012ic}.

In this chapter possible decay channels to study the $\etac$, $\chiczero$, $h_c$ and $\etac(2S)$ mesons, which can't be accessed using their decays to $\mu^+\mu^-$ or $(\jpsi\to\mup\mum) \gamma$ are discussed. The known branching fractions~\cite{PDG2017} of promising decays discussed below are summarised in Table~\ref{tab:Brs}. Many of these  branching fractions can be measured more precisely at Belle, Belle II, BES III, or future high-luminosity tau-charm experiments.  


The charmonia decays to $\ppbar$ have been proposed to measure charmonium production at LHC~\cite{Barsuk:2012ic}.
The first measurement of the $\etac$ production at the LHCb experiment has been performed using the $\etac \to \ppbar$ decay~\cite{Aaij:2014bga}. This demonstrated that the $\ppbar$ final state is 
powerful to reconstruct the $\etac$ meson, even though the measurement is performed only for transverse momentum larger than 6.5~\gev due to available trigger bandwidth. This decay is also used to study exotic candidates decaying to $(\etac\to\ppbar) \pi^-$~\cite{Aaij:2018bla}. The branching fraction of the $\etac \to \ppbar$ channel is known to about 10\% precision~\cite{PDG2017}. The studies of $\etac$ would benefit from more precise measurement of $\BR(\etac \to \ppbar)$ or $\BR(\etac \to \ppbar)/\BR(\jpsi \to \ppbar)$. Branching fractions of $\chic_J \to \ppbar$ and $\psitwos \to \ppbar$ decays have been measured to about 3-5\% precision. Recently, LHCb observed the $\etac(2S) \to \ppbar$ decay channel using a data sample of exclusive $B^+ \to \ppbar K^+$ decays~\cite{Aaij:2016kxn}. Together with the measurement of $\BR(B^+ \to \etac(2S) K^+)$ by Belle~\cite{Kato:2017gfv}, the branching fraction of the $\etac(2S) \to \ppbar$ is indirectly determined to be about $0.7\times10^{-4}$. Therefore, the decay $\etac(2S) \to \ppbar$ is promising for the $\etac(2S)$ hadroproduction studies at \lhc.

Another promising final state to study prompt production of charmonium is $\phi\phi$. The $1^{-}$ charmonium states are forbidden to decay to $\phi\phi$. The LHCb measured the $\chi_{c0,1,2}$ and $\etac(2S)$ production in inclusive b-hadron decays using the $\phi\phi$ final state with the first evidence of the $\etac(2S) \to \phi\phi$ decay~\cite{Aaij:2017tzn}. In Section~\ref{sec:brEtac2PhiPhi}, a tension between the PDG fit value of $\BR(\etac \to \phi\phi)$ and the PDG average value~\cite{PDG2017} by about two times was pointed out and the ratio of branching fractions $\BR(\etac \to \phi\phi)/\BR(\etac \to \ppbar)$ was measured. Further measurements are needed to establish a robust value of the $\BR(\etac \to \phi\phi)$. Following the evidence of $\etac(2S) \to \phi\phi$, this channel is also promising to study a hadroproduction of the $\etac(2S)$. Similarly, the $\phi K^+ K^-$ and the $\phi \pi^+ \pi^-$ final states can potentially be used including final states with intermediate resoncances such as $\phi f$, where $f$ decays to $\pi^+ \pi^-$ or $\Kp\Km$.

The branching fractions of charmonium decays to long-lived baryons such as $\Lambda\bar{\Lambda}$ and $\Xi^+\Xi^-$ are measured for most charmonium states. Reconstruction of these decay channels is challenging for LHCb due to a flight distance of these baryons, so that they escape the Vertex Locator (VELO), which causes a reduced reconstruction and trigger efficiency.

Decays involving short-lived baryons are reconstructed by LHCb with better efficiency. The decays $\chi_{c0,2} \to \Lambda(1520)\bar{\Lambda}(1520)$ have been observed by the BES III collaboration~\cite{Ablikim:2011uf} while the $\jpsi \to \Lambda(1520)\bar{\Lambda}(1520)$ decay is not observed so far. This channel becomes another candidate to measure hadroproduction of charmonium states~\cite{Jacques}.

The least studied charmonium state below the \DDbar threshold is the $h_c$ meson with only a few of $h_c$ decays observed so far. The $h_c$ meson is expected to decay to $\ppbar$, however, the upper limit on $\BR(h_c \to \ppbar)$ reported by the BES III collaboration~\cite{Ablikim:2013hdv} is more than an one order of magnitude smaller than the theoretical prediction~\cite{Barsuk:2012ic}.  Alternatively, the $h_c$ can be reconstructed using its radiative transition $h_c \to \etac \gamma$ with the branching fraction about 50\%, which requires a reconstruction of the photon in addition to the $\etac$ state. Recently, LHCb observed new clean decays $\chi_{c1,2} \to \jpsi \mu^+ \mu^-$, and measured precisely the $\chictwo$ mass and natural width~\cite{Aaij:2017vck}. Following this observation, the $h_c \to \etac \mu^+ \mu^-$ decay can also be searched. Recently, BES III has observed the $h_c \to \ppbar \pi^+\pi^-$ decay and measured its branching fraction to be $(2.89\pm0.32\pm0.55)\times 10^{−3}$~\cite{Ablikim:2018ewr}, which makes it promising for studies at LHCb.
   
The reconstruction of various charmonium states is important for systematic studies of charmonium production and properties. Many measurements in \bquark-physics (searches of hadron exotics, \bquark-anomalies, etc.) study decays with clean signatures from \jpsi decays to leptons in the final state. Many studies would benefit from the analogous measurements exploiting other charmonium states in the final state.
As a conclusion, a number of hadronic final states are promising to simultaneously reconstruct charmonium states. Incorporating charmonium states other than $1^{-\,-}$ implies a systematic measurements of hadronic branching fractions of their decays.
Some charmonium states are poorly studied and not many decays have been observed so far, which makes expectations of their signal significances more complicated.

In this thesis, I study charmonia using their decays to hadrons with the \lhcb detector. In the next chapter I will describe the \lhcb experiment, and more specifically the detector features essential to reconstruct charmonia via hadronic decays. 

\begin{savenotes}
\begin{table}[h!]
\newsavebox{\tablebox}
\begin{center}
\begin{lrbox}{\tablebox}
\rotatebox{90}{
\begin{tabular}{|c||c|c|c|c|c|c|c|c|c|}
        \hline\          
& \multicolumn{8}{ c |}{$\BR\times 10^3$}\\
&$\ppbar$
&$\phi\phi$ 
&$\phi K^+K^-$ 
&$\phi\pi^+\pi^-$
&$\Lambda\overline{\Lambda}$ 
&$\Xi^+{\Xi}^-$ 
&$\Lambda(1520)\overline{\Lambda}(1520)$
&$\etac\gamma$
&$\ppbar\pi^+\pi^-$
\\
\hline \hline 
$\etac$ 
&$1.52\pm0.16$ 			
&$1.79\pm0.20$ 			
&$2.9\pm1.4$ 			
&unknown 				
&$1.09\pm0.24$ 			
&$9.0\pm2.6$ 			
& - 					
& - 					
&$5.3\pm1.8$			
\\				

$\jpsi$ 
&$2.12\pm0.03$ 			
&forbidden 				
&$0.83\pm0.12$ 	
&$0.87\pm0.09$ 			
&$1.89\pm0.08$			
&$0.97\pm0.08$			
&unknown 				
&$17\pm4$ 				
&$6.0\pm0.5$ 			
\\

$\chiczero$ 
&$0.22\pm0.01$			
&$0.80\pm0.07$			
&$0.97\pm0.25$			
&unknown				
&$0.33\pm0.02$			
&$0.48\pm0.07$			
&$0.31\pm0.12$ 			
&forbidden				
&$2.1\pm0.7$			
\\

$h_c$ 
&$<0.15$				
&forbidden          	
&unknown				
&unknown 				
&unknown				
&unknown				
&unknown 				
&$510\pm60$ 			
&unknown	  			
\\

$\chicone$ 
&$0.076\pm0.003$		
&$0.42\pm0.05$			
&$0.41\pm0.15$			
&unknown				
&$0.11\pm0.01$		
&$0.08\pm0.02$		
&$<0.09$  				
&forbidden				
&$0.50\pm0.19$  		
\\

$\chictwo$ 
&$0.073\pm0.003$		
&$1.06\pm0.09$			
&$1.42\pm0.29$			
&unknown				
&$0.18\pm0.02$		
&$0.14\pm0.03$		
&$0.46\pm0.15$  		
&forbidden				
&$1.32\pm0.34$			
\\

$\etac(2S)$				
&$0.07$\footnote{Indirect determination} 
&unknown          		
&unknown				
&unknown 				
&unknown				
&unknown				
&unknown 				
&forbidden 				
&unknown    			
\\

$\psitwos$ 
&$0.29\pm0.01$		
&forbidden				
&$0.07\pm0.02$		
&$0.12\pm0.03$		
&$0.38\pm0.01$		
&$0.29\pm0.01$		
&unknown 				
&$3.4\pm0.5$			
&$0.60\pm0.04$			
\\
\hline
\end{tabular}
}
\end{lrbox}
\resizebox{0.26\textwidth}{!}{\usebox{\tablebox}}
\end{center} 
\caption
[The branching fractions of charmonium decays to hadrons and radiative decays to $\etac \gamma$.]
{The branching fractions $\times 10^3$ of charmonium decays to hadrons and radiative decays to $\etac \gamma$.}
\label{tab:Brs}
\end{table} 
\end{savenotes}

\begin{singlespace}
\chapter{\lhcb detector}
\label{ch:lhcb}
\end{singlespace}
The analyses of charmonium production using decays to hadrons, described in Chapters~\ref{ch:ppbar} and~\ref{ch:phiphi} have been performed using data collected by the \lhcb experiment. The \lhcb experiment is well suited and is the most performant among the experiments at hadron machines to reconstruct hadronic decays of charmonium states. Even though such studies have not been considered as a part of the core program of the experiment, a flexible trigger of the \lhcb experiment provides an opportunity to measure \etac, $\chic_J$ and \etactwos production observables using \ppbar and $\phi\phi$ final state signatures. 

After introducing the Large Hadron Collider (\lhc) in the CERN accelerator complex in Section~\ref{sec:lhc}, the \lhcb detector is described in Section~\ref{sec:lhcb} with the accent on the detector features most relevant for charmonium reconstruction via decays to hadrons. The vertex and track reconstruction at \lhcb are discussed in Section~\ref{sec:TrAndVtx}. The particle identification within \lhcb experiment is described in Section~\ref{sec:pid}. Finally, the \lhcb trigger
together with dedicated selections for prompt charmonium reconstruction via decays to hadrons is addressed in Section~\ref{sec:trigger}.
\clearpage
\section{Large Hadron Collider}
\label{sec:lhc}
The Large Hadron Collider (\lhc)~\cite{Evans:2008zzb} is a synchrotron with a circumference of 27~\km located near Geneva Swiss-French border located about 100\m underground at CERN exploiting proton-proton, proton-lead and lead-lead collisions.
During the core proton-proton collisions program of the \lhc Run I and II in 2011-2018 years, the counter-propagating proton beams were accelerated to an energy of 7, 8 and 13~\tev.

Before the injection into the \lhc proton beams pass several steps of acceleration. The protons obtained from hydrogen atoms are firstly accelerated by the linear accelerator LINAC2 up to an energy of 50~\mev, then the BOOSTER accelerates protons to an energy of 1.4~\gev. After that, protons are injected into the Proton Synchrotron (PS) and accelerated to an energy of 26~\gev, which followed by the injection into the Super Proton Synchrotron (SPS), yielding the proton beams with an energy of 450~\gev. 

The beams from SPS are then injected into \lhc and accelerated to the final energy using 16 Radio-Frequency (RF) cavities located along the \lhc ring. The conduction of the beams along the ring is performed by 12300 superconducting dipole magnets providing a magnetic field of 8.3 T. The coils of magnets are cooled by a liquid helium cryogenic system to a temperature of 1.9 K. The focusing of the beams is ensured using about 400 quadrupole magnets.

\begin{figure}[ht]
\centering
\protect\protect\includegraphics[width=0.8\textwidth]{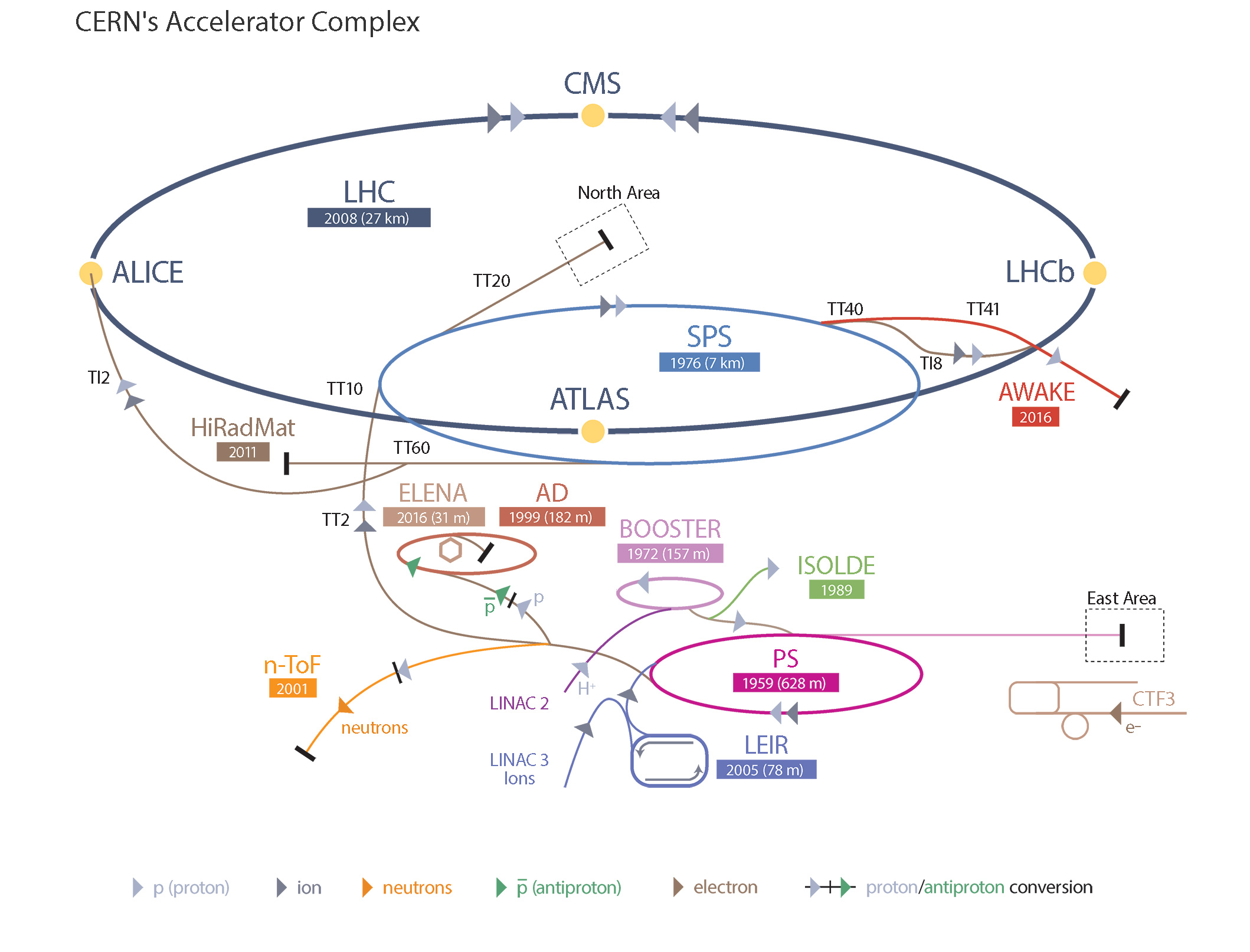}
\caption
[The \lhc and \cern accelerating complex.]
{The \lhc and \cern accelerating complex~\cite{Christiane:1260465}.} 
\label{fig:lhc}
\end{figure}

For the nominal proton-proton program, \lhc provides proton beams of $1.3\times10^{11}$ protons per bunch with a collision rate of $40 MHz$ and an instanteneous luminosity up to $10^{34} cm^{-2} s^{-1}$. The four main \lhc experiments are placed around four collision points. The scheme of the \cern accelerating complex is shown on Fig.~\ref{fig:lhc}.

The \lhc experiments are:
\begin{itemize}
\item \atlas (A Toroidal LHC ApparatuS)~\cite{Aad:2008zzm},
\item \cms (Compact Muon Solenoid)~\cite{Chatrchyan:2008aa},
\item \alice (A Large Ion Collider Experiment)~\cite{Aamodt:2008zz},
\item \lhcb (Large Hadron Collider beauty) (see Section~\ref{sec:lhcb}),
\item LHCf (Large Hadron Collider forward)~\cite{Adriani:2008zz},
\item TOTEM (TOTal Elastic and diffractive cross-section Measurement)~\cite{Anelli:2008zza},
\item MoEDAL (Monopole and Exotics Detector At the LHC)~\cite{Pinfold:2009oia}.
\end{itemize}
The \atlas and \cms are so-called $4\pi$ experiments performing  direct studies of Standard Model (SM) particles and searches for New Physics (NP). Their core physics programs overlap and aim at studying the Higgs boson and \tquark quark properties and decays and direct searches for new supersymmetric (SUSY) particles, additional Higgs bosons, etc.
Another important part of the \atlas and \cms programs is dedicated for heavy flavour physics in \bquark and \cquark quark sectors.

The \alice is an experiment designed to exploit in lead-lead, proton-lead and lead-proton collisions. The main goal of the \alice experiment is to look for signatures of the deconfined state of hadronic matter Quark Gluon Plasma (QGP). The studies are performed by measuring, for example, the heavy flavour production suppression and comparing it for different kinds of collisions. Besides, the studies of other in-matter production effects, such as cold nuclear matter effect, are performed.

The TOTEM experiment is designed for measuring elastic, diffractive and dissociative proton scattering cross-sections. Their measurements are essential for soft Quantum Chromodynamics (QCD).
The LHCf experiment is designed for measurements aiming to simulate the cosmic rays in the laboratory conditions.
The MoEDAL is performing searches for Dirac magnetic monopole.

\section{\lhcb experiment}
\label{sec:lhcb}
The \lhcb experiment~\cite{Alves:2008zz,Aaij:2014jba} is designed for studies of heavy flavour - \bquark and \cquark quark - sectors in the forward region.
The core physics program of the \lhcb experiment is dedicated to precision measurements or searches of:
\begin{itemize}
\item CP-violation in \bquark and \cquark quark mixing and decays including measurement of the unitarity triangle parameters,
\item Search for indirect contributions of NP to (rare) processes, including contributions to $B$-meson decays involving a lepton pair in the final state and tests of lepton universality,
\item \bquark- and \cquark-hadrons spectroscopy (\Bs, \Bc mesons, \bquark and \cquark-baryons, quarkonium etc.) and searches for hadron exotics (tetraquarks, pentaquarks, etc.), 
\item QCD effects in \bquark-decays to open charm particles or charmonium,
\item Heavy flavour production and soft QCD processes,
\item Electroweak physics,
\item Heavy ion physics.
\end{itemize}

The \lhcb experiment is a forward single-arm spectrometer 
covering the pseudorapidity range of $2 < \eta < 5$.
The angular acceptance of the \lhcb is 10-300 \mrad on $x$-axis and 10-250 \mrad on $y$-axis.
Typically, \lhcb is capable to detect particles in the \pt range of $0.25~\gev < \pt < 20~\gev$.

\begin{figure}[t]
\centering
\protect\protect\includegraphics[width=0.4\textwidth]{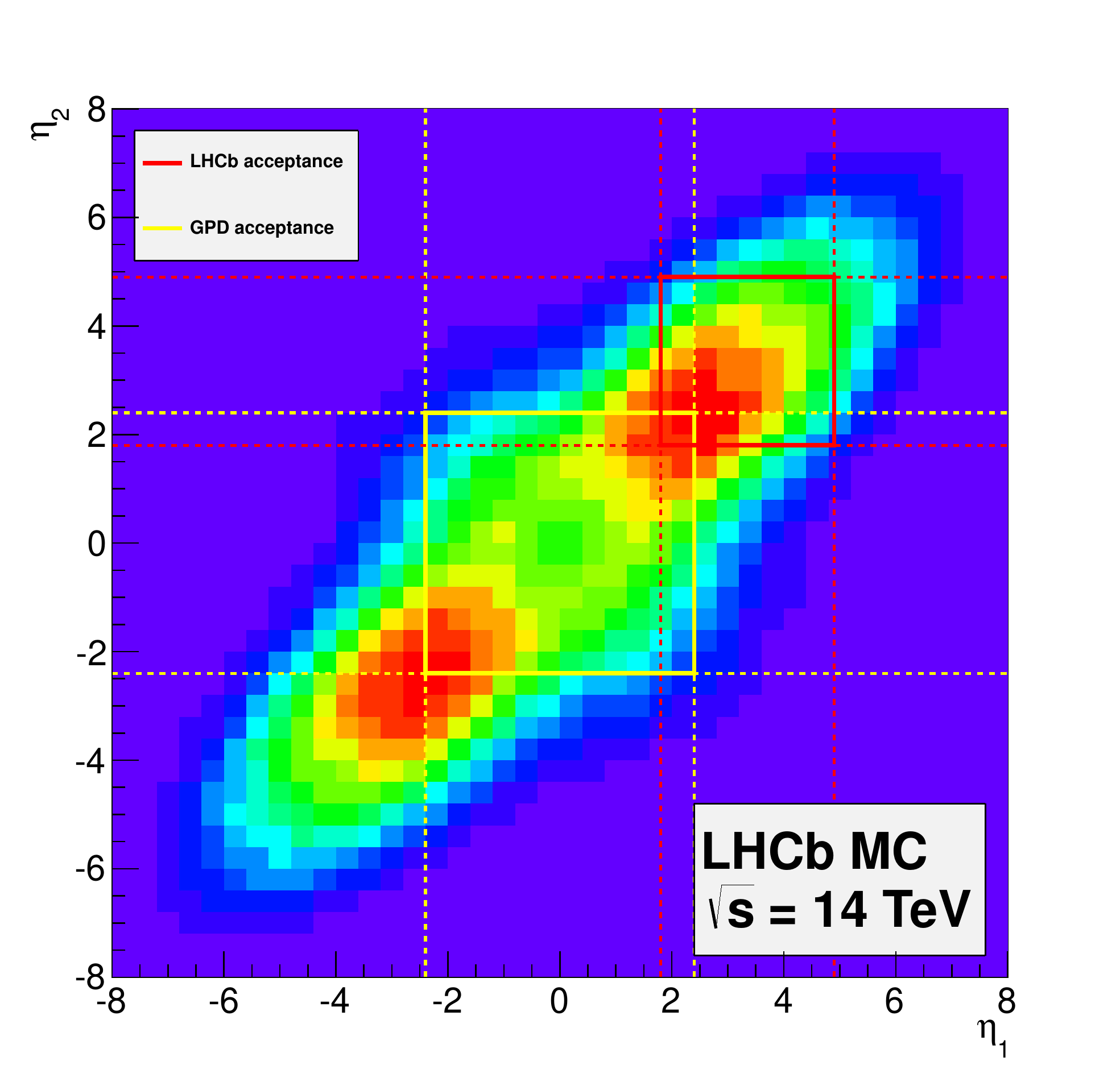}
\caption
[The \bbbar production as a function of quarks pseudorapidity.]
{The \bbbar production as a function of quarks pseudorapidity. The red rectangle shows the \lhcb acceptance. The yellow rectangle shows the \atlas and \cms acceptance~\cite{bbangles}.} 
\label{fig:bbarProd}
\end{figure}

In hadron-hadron collisions, a \bbbar (or similarly \ccbar) quark pair production is dominated by $gg\to\bbbar$, $\qqbar\to\bbbar$, $gg\to\bbbar g$ and $\qqbar\to\bbbar g$ processes, where $g$ denotes a gluon and $q$ denotes a light quark. At the \lhc energies, the processes mentioned above lead to the \bbbar production predominantly in the forward region. The \bbbar production cross-section as a function of the \bquark and \bquarkbar quarks pseudorapidity is shown on Fig.~\ref{fig:bbarProd} and is compared with \lhcb, \atlas and \cms acceptances. By covering only about of 4\% of the full solid angle, the \lhcb detector receives fraction of total \bbbar production cross-section comparable with the one of a $4\pi$ experiment, such as \atlas or \cms. This feature together with precise vertex reconstruction, powerful particle identification and selective trigger (see Sections~\ref{sec:TrAndVtx}, ~\ref{sec:pid} and~\ref{sec:trigger}) makes \lhcb exclusive or at least more profitable to study most of heavy flavour physics observables compared to other \lhc experiments. In addition, \lhcb covers the \pt and rapidity ranges complementary to those of \atlas and \cms. In total, about $5\times10^{11}$ \bbbar pairs and about $3\times10^{12}$ \ccbar pairs are created within \lhcb acceptance per \invfb of integrated luminosity at \sqs=14~\tev.

In the heavy flavour physics program, \lhcb is competing with the so-called B-factories (\bquark-sector) and charm factories (\cquark-sector). The B-factories,  such as \belle~\cite{Abashian:2000cg} and \babar~\cite{Aubert:2001tu}, are the asymmetric \epem experiments operating at the centre-of-mass energy of $\FourS$ resonance mass, which decays strongly to the \Bp\Bm or \Bz\Bzb with a branching fraction close to 100\%. The advantage of B-factories is that the underlying experimental environment is clean and most of the events contain a pair of light B-mesons. Due to its detector design choice, the \lhcb experiment has reduced reconstruction performance of neutral particles contrary to B-factories. Hence, the reconstruction of B-meson decay modes involving neutral particles in the final state is much better accessed by B-factories. At the same time, the production cross-section of $\FourS$ resonance in \epem collisions is three orders of magnitude smaller compared to the \bbbar production cross-section in hadron-hadron collisions at \tev energies. Also, \lhcb is capable to study \bquark-hadrons other than lightest B-mesons, i.e. \Bs, \Lb, \Bc, etc.
Charm factories (for example BES~\cite{Ablikim:2009aa}) operate at \epem collision energies in the charmonium mass region to produce \jpsi, \psitwos charmonium states. The charm factories provide many of the most precise measurements of  charmonium decays. 
The most precise measurements of resonance parameters of many charmonium states are performed at $\bar{\proton} \proton$-collision experiments (E760 and E835~\cite{Patrignani:2004vt}) and also \epem experiments such as KEDR~\cite{Anashin:2013twa}.

During the data taking, \lhcb reduces the nominal \lhc instantaneous luminosity (luminosity levelling) by two orders of magnitude. This leads to both reduced total yield of the \bbbar production and smaller detector occupance. The latter is crucial for the online trigger, which has limited bandwidth. Also, the luminosity reduction improves the track and vertex reconstruction performance and timing characteristics. Another benefit from the luminosity levelling is that the detector components ageing is reduced. 
The luminosity levelling is made in the way that the instantaneous luminosity is constant during the \lhc fill and is adjusted by adding an offset between the beams at the collision point. The typical target value of the number of interactions per beam crossing (pile-up) at \lhcb is $\mu=1.5$. Since the \lhc instantaneous luminosity during the fill is decreasing, the offset is adjusted by using information from online luminosity monitoring. More information about the luminosity levelling with offset beam is given in Ref.~\cite{Follin:2014nva}. 

\begin{figure}[h]
\centering
\protect\protect\includegraphics[width=0.8\textwidth]{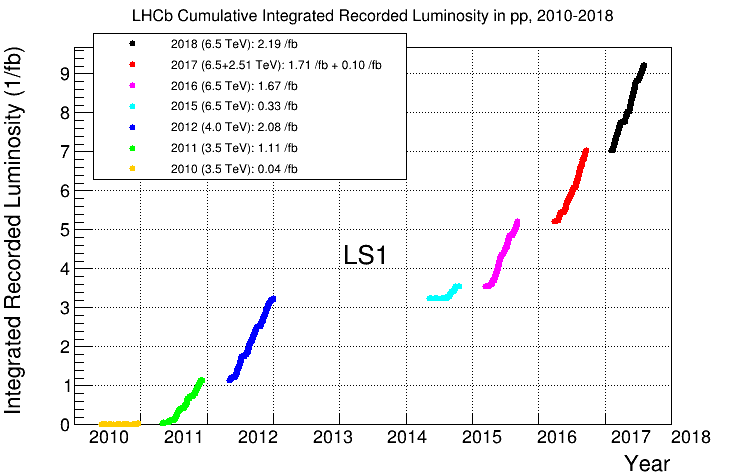}
\caption{Integrated luminosity collected by \lhcb during \lhc Runs I and II.} 
\label{fig:Lumi}
\end{figure}

During the \lhc Run I and Run II, \lhcb recorded integrated luminosity of:
\begin{itemize}
\item 0.04~\invfb at \sqs=7~\tev in 2010,
\item 1.11~\invfb at \sqs=7~\tev in 2011,
\item 2.08~\invfb at \sqs=8~\tev in 2012,
\item 0.33~\invfb at \sqs=13~\tev in 2015,
\item 1.67~\invfb at \sqs=13~\tev in 2016,
\item 1.71~\invfb at \sqs=13~\tev in 2017,
\item 2.19~\invfb at \sqs=13~\tev in 2018
\end{itemize} 
in proton-proton collisions (Fig.~\ref{fig:Lumi}).

\begin{figure}[t]
\centering
\protect\protect\includegraphics[width=1.0\textwidth]{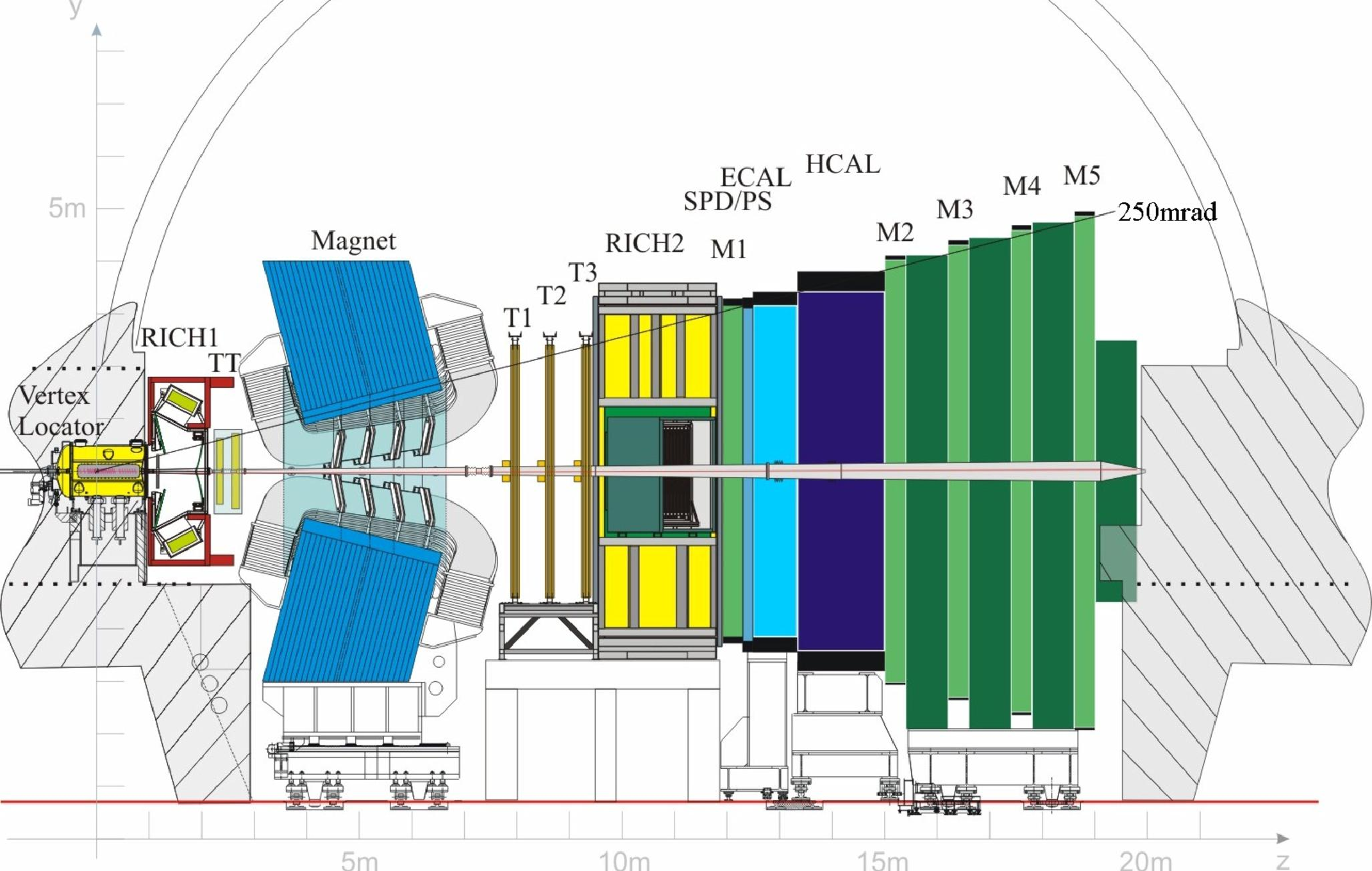}
\caption
[The LHCb detector.]
{The LHCb detector~\cite{Alves:2008zz}.} 
\label{fig:lhcb_sketch}
\end{figure}

The sketch of the \lhcb detector is shown on Fig.~\ref{fig:lhcb_sketch}.
The detector comprises a high-precision tracking system
consisting of a silicon-strip vertex detector surrounding the $pp$
interaction region~\cite{LHCb-DP-2014-001}, a large-area silicon-strip detector located upstream of a dipole magnet, and three stations of silicon-strip detectors and straw
drift tubes~\cite{LHCb-DP-2013-003} placed downstream of the magnet.
Different types of charged hadrons are distinguished using information
from two ring-imaging Cherenkov detectors~\cite{LHCb-DP-2012-003}. 
Photons, electrons and hadrons are identified by a calorimeter system consisting of
scintillating-pad and preshower detectors, an electromagnetic
calorimeter and a hadronic calorimeter. Muons are identified by a
system composed of alternating layers of iron and multiwire
proportional chambers~\cite{LHCb-DP-2012-002}.

To reduce the LHC bunch-crossing frequency of 40 MHz to storable event rates 
LHCb employs a two-level trigger system, 
including a hardware (L0) trigger and a software (HLT) trigger implemented in a processor farm. 
A general idea of the most of online (L0 and HLT1) trigger requirements is to select particles, which have large \pt and/or are well displaced from any collision vertex since weakly decaying \bquark and \cquark-hadrons fly a significant distance before the decay.
The L0 reduces the rate to about 1 MHz, and 
the L0 triggered events are passed to the online stage of the software trigger (HLT1), which 
partially reconstructs events, confirming (or not) the L0 decision.
The second level of software trigger (HLT2) processes fully reconstructed events and stores relevant information about selected decay candidates.

\clearpage
\section{Vertex and track reconstruction}
\label{sec:TrAndVtx}
\subsection{Vertex reconstruction}
\label{sec:vertexing}
The goal of the vertex reconstruction is to recognise vertices, distinguish primary (PV) from \bquark (or \cquark)-decays ones and measure the flight distance of the decaying \bquark-hadron and assign tracks to vertices.

\begin{figure}[b]
\centering
\protect\protect\protect\includegraphics[width=0.85\linewidth]{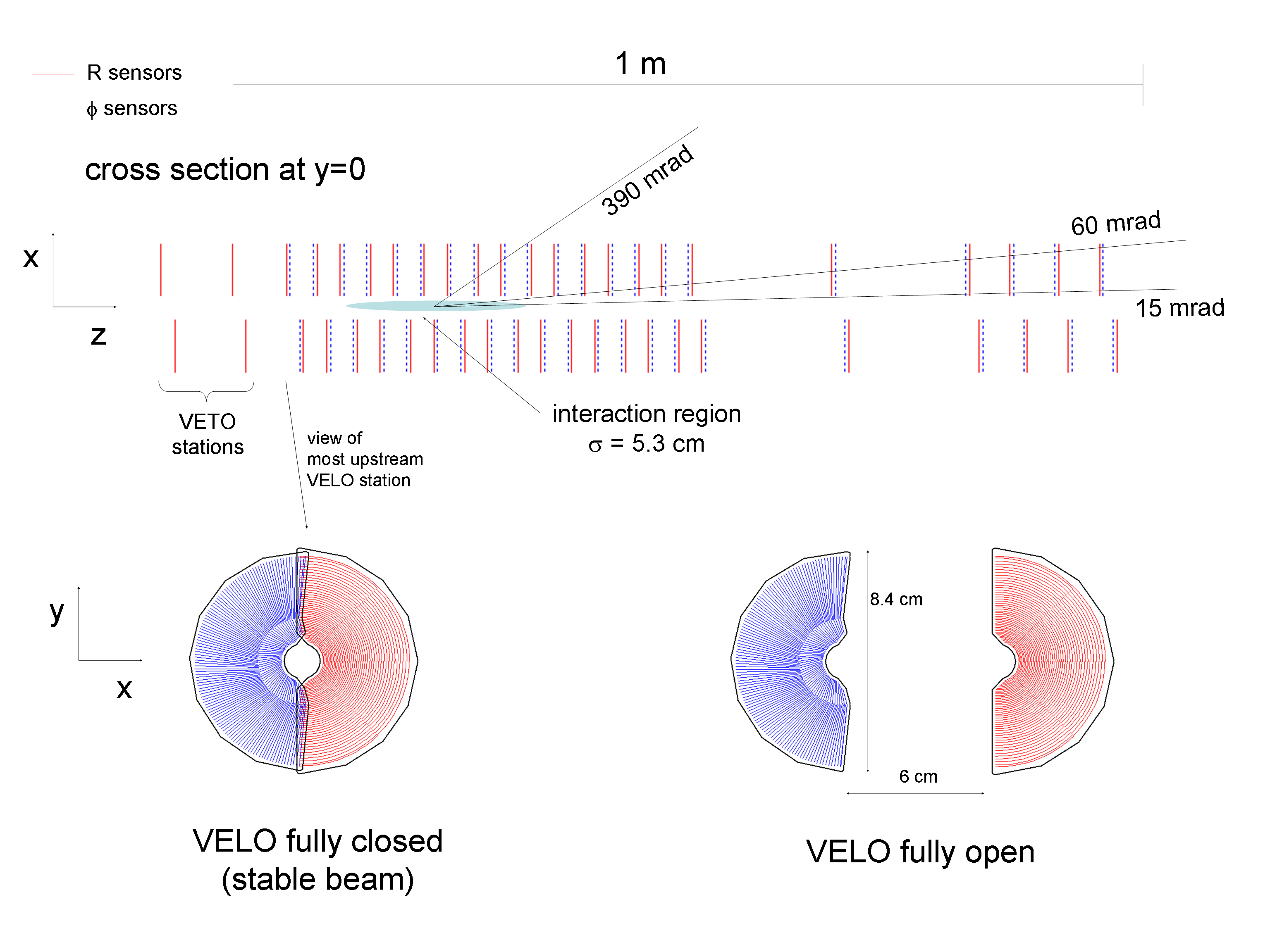}
\caption
[Sketch of the VELO detector.]
{Sketch of the VELO detector~\cite{veloPlots}.} 
\label{fig:velo2}
\centering
\protect\protect\protect\includegraphics[width=0.5\linewidth]{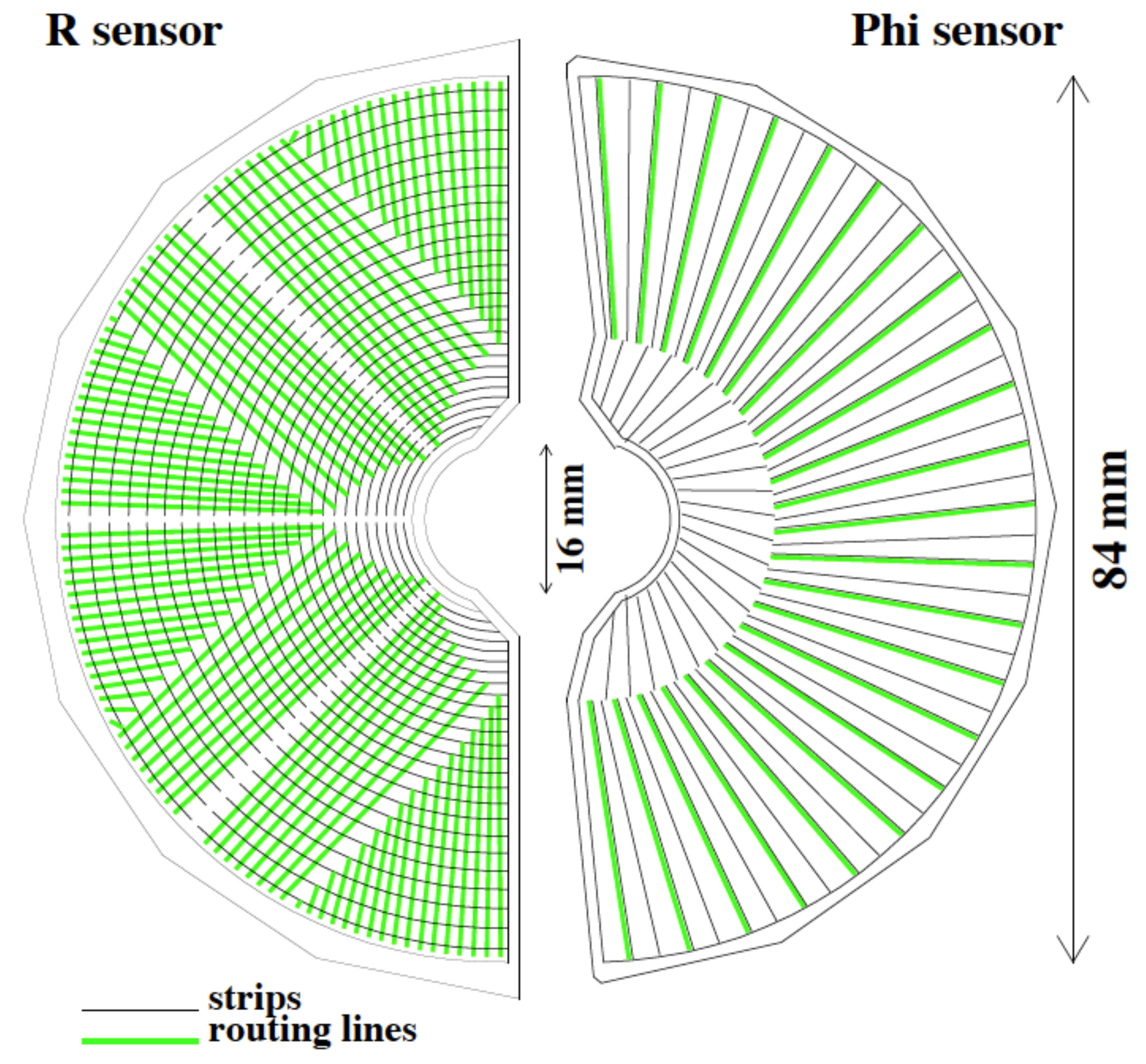}
\caption
[The $r$ and $\phi$ sensors of the \velo detector.]
{The $r$ (left) and $\phi$ (right) sensors of the \velo detector~\cite{veloPlots}.} 
\label{fig:velosensors}
\end{figure}

The vertex reconstruction in \lhcb is performed thanks to the Vertex Locator (VELO)~\cite{LHCb:2001aa}. 
VELO is a silicon strip detector, which measures trajectories of charged particles close to the interaction point. The sketch of VELO is shown on Fig.~\ref{fig:velo2}. VELO consists of 42 semicircular silicon modules along the beam forming two halves of the detector, which can approach or move away from each other. Each module consists of $r$ and $\Phi$ sensors measuring track hit polar coordinates $r$ and $\phi$, respectively. This configuration allows making faster track reconstruction than the geometry of the rectangular strips.
The strips of $r$-sensor are concentric rings with a variable pitch that increases linearly from 38~\mum at the inner edge to 102~\mum at the outer edge. 
The strips of $\phi$-sensor strips are divided into two regions at $r$ = 17.25\mm in order to reduce the occupancy and to avoid large strip pitches at the outer edge of the sensors. 
The strips have a pitch of 38\mum in the inner region (increasing to 78\mum at the outer edge), while the strips in the outer region have a pitch of 39\mum (increasing to 97\mum at the outer edge). The sketch of both sensors is shown on Fig.~\ref{fig:velosensors}.

The inner radius (i.e. distance to beam axis) of VELO module is about 8\mm. At this distance, severe radiation can cause destruction of the modules.
A mechanical moving system is designed to open or close VELO modules when needed. During the phase of stable beams of data taking the VELO modules are closed, while during other phases when beams are circulating in the \lhc unsqueezed, VELO is kept in the safe opened state. The monitoring system ensures that VELO can be closed by using the online information about the number of reconstructed vertices. The time needed for VELO to close from a completely opened state is about 3 minutes 
via iterative procedure of refining vertex position.

The inner faces of the vessels (RF-foils) separate the VELO vacuum from the \lhc vacuum. The RF-foils are designed to minimise the material traversed by particles before crossing VELO sensors. Furthermore, the geometry of the RF-foils is such that it makes the two halves of the VELO overlap when it is moved to the closed position. 

\begin{figure}[b]
\centering{
        \subfigure[The PV $x$-coordinate resolution.]{ 
          \protect\protect\protect\includegraphics[width=0.35\textwidth]{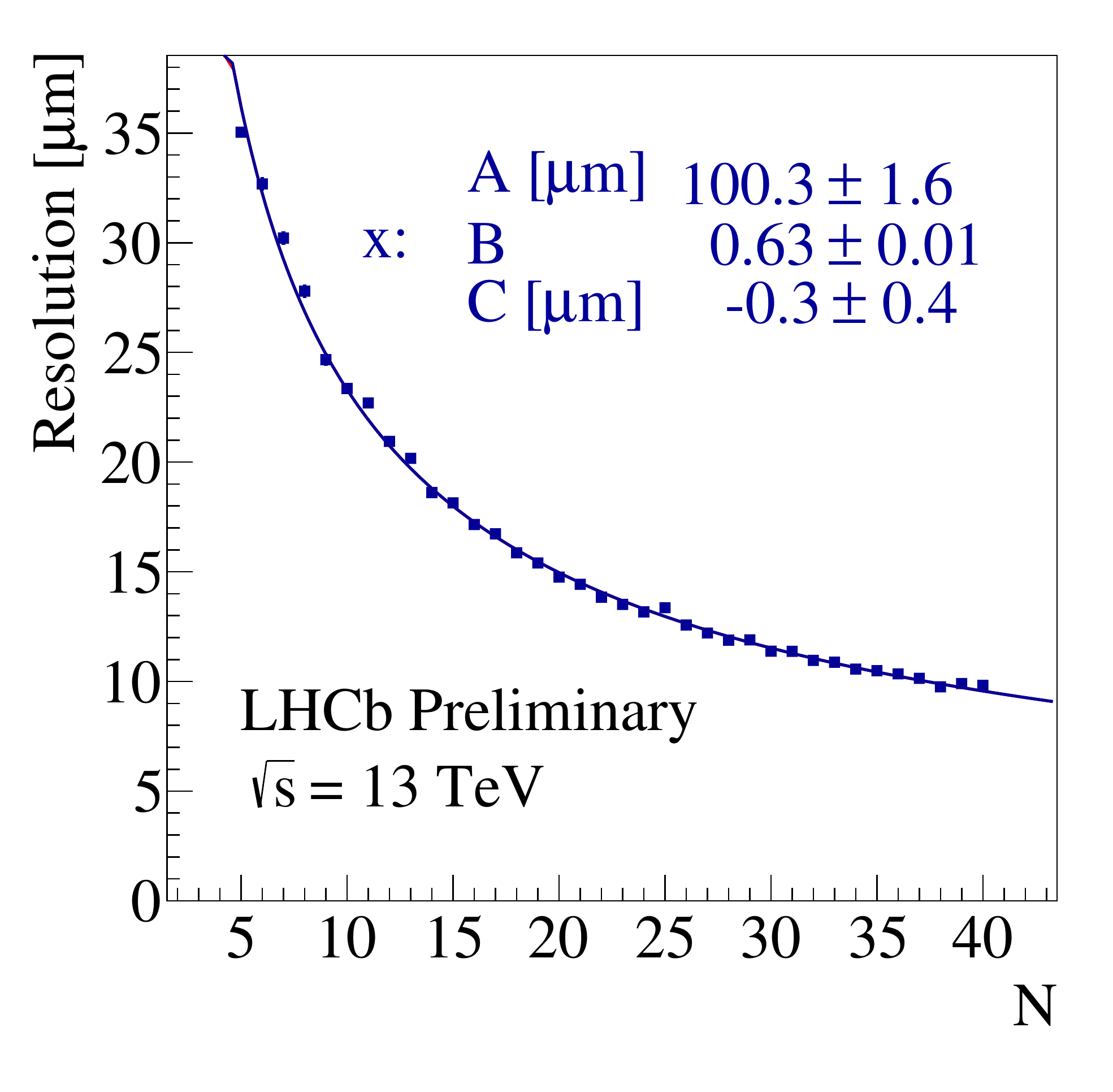}
          \label{fig:sigmaRelPrompt}}
\quad
        \subfigure[The PV $z$-coordinate resolution.]{
          \protect\protect\protect\includegraphics[width=0.35\textwidth]{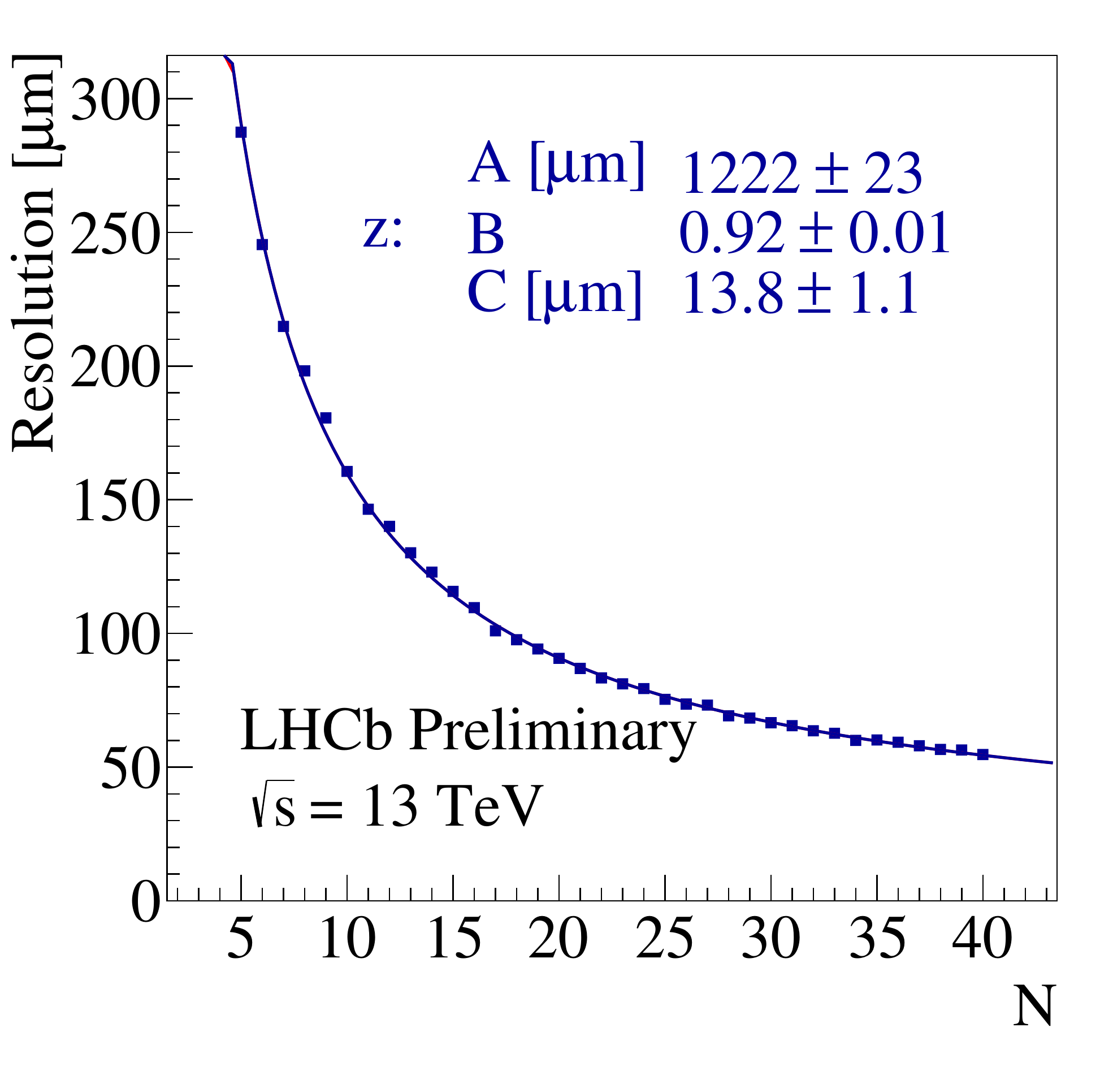}
          \label{fig:sigmaRelFromB}}
\quad
        \subfigure[The IP resolution.]{
          \protect\protect\protect\includegraphics[width=0.4\linewidth]{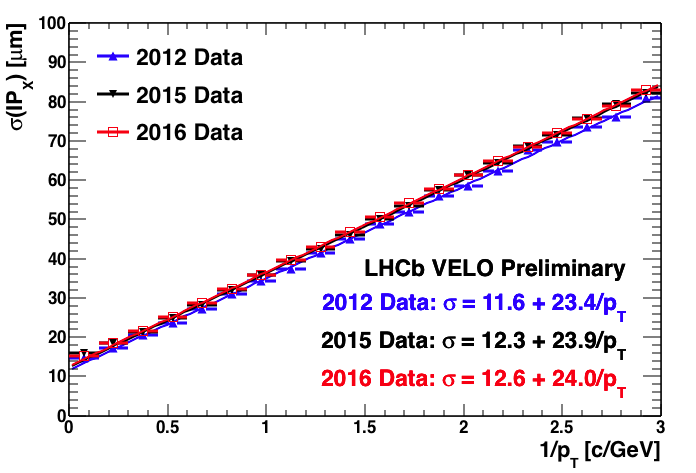}
          \label{fig:IPXreso}}
 }
\caption
[The resolution of PV position and IP provided by \velo.]
{The resolution of PV position and IP provided by \velo~\cite{veloPlots}.} 
\label{fig:veloReso}
\end{figure}

\newpage
VELO dominates the measurements of the PV position and the track's impact parameter (IP) with respect to PV, which is crucial for trigger and further event selections.
The IP is measured with a resolution of $(15+29/\pt)\mum$,
where \pt is the component of the momentum transverse to the beam, in\,\gev.
The IP resolution determination driven by VELO is the best among all \lhc experiments.
The PV position and resolution along and across the beam as a function of the vertex multiplicity and the IP resolution are shown on Fig.~\ref{fig:veloReso}. 

Since \velo opens and closes many times, its alignment is important during operations to match online and offline reconstruction. During the Run I, the alignment and calibration have been performed offline. The online reconstruction in Run I was more simple compared to that of Run II. Therefore, the data passing trigger was reprocessed every year to take into account and correct possible effects of alignment and calibration.
During \lhcb Run II the alignment and calibration were performed online. The data recorded at the beginning of the \lhc fill was used to update alignment and calibration constants if needed. The alignment of all detectors (\velo, trackers, \rich mirros) takes in total about 20-30 minutes, while the \velo alignment takes only few minutes. 

\begin{figure}[t]
\centering
\protect\protect\protect\includegraphics[width=0.7\linewidth]{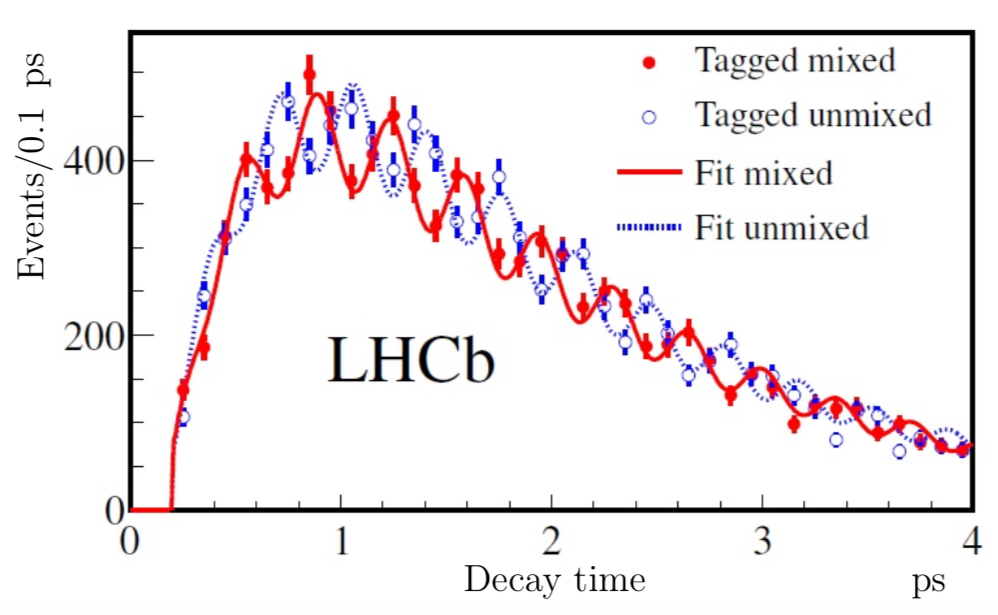}
\caption
[Decay time distribution for \Bs candidates tagged as mixed or unmixed.]
{Decay time distribution for \Bs candidates tagged as mixed (different flavour at decay and production; red, continuous line) or unmixed (same flavour at decay and production; blue, dotted line). The data and the fit projections are plotted in a signal window around the reconstructed \Bs mass~\cite{Aaij:2013mpa}.} 
\label{fig:BsOsc}
\end{figure}

\clearpage
Vertex reconstruction precision using the LHCb tracking system is illustrated by resolving rapid \Bs-$\bar{\Bs}$ oscillations~\cite{Aaij:2013mpa}. Fig.~\ref{fig:BsOsc} shows decay time distribution for \Bs candidates, reconstructed via $\decay{\Bs}{\Ds \pip}$, with the \Ds decaying via $\decay{\Ds}{\Kp \Km \pim}$, $\decay{\Ds}{K^{*0} \Km}$,  $\decay{\Ds}{\Km \pip \pim}$ and $\decay{\Ds}{\pim \pip \pim}$, tagged as mixed or unmixed.



\clearpage
\subsection{Track reconstruction}
The \lhcb tracking system is designed for the reconstruction of the tracks of stable charged particles and measurement of their charge and momenta.
The momenta of particles are determined by measuring the curvature of tracks bent in the magnetic field of the \lhcb dipole magnet.

The \lhcb experiment uses a warm dipole magnet~\cite{LHCB:2000ac} with a total weight of about 1,600 tons.
The magnetic field is created by two identical trapezoidal coils located symmetrically in the magnets yoke.
The magnet provides an integrated magnetic field of about $4{\mathrm{\,T}}$ along the $y$-axis mainly. The non-uniformities of the field amount to about 1\% and are important for track reconstruction. Therefore, the map of the magnetic field is used for track reconstruction.
The opposite magnet polarities "Up" and "Down" are alternated during the data taking in order to reduce systematic uncertainties related to detector asymmetries, which could lead to potential asymmetries in reconstruction of particles with opposite charge. Approximately the same amount of integrated luminosity is collected with two different polarities.

The tracking system of the \lhcb experiment comprises four tracker stations: Tracker Turicensis (TT) placed upstream the \lhcb magnet and three stations T1, T2 and T3 placed downstream the magnet. 
Two different technologies are used in the inner (Inner Tracker) and outer (Outer Tracker) regions of the T1-T3 stations in order to withstand the different particle flux. 
The TT and IT together costitute the \lhcb Silicon Tracker (ST), since the same technology is used for both detector systems.

\begin{figure}[h]
\centering
\protect\protect\protect\includegraphics[width=0.7\linewidth]{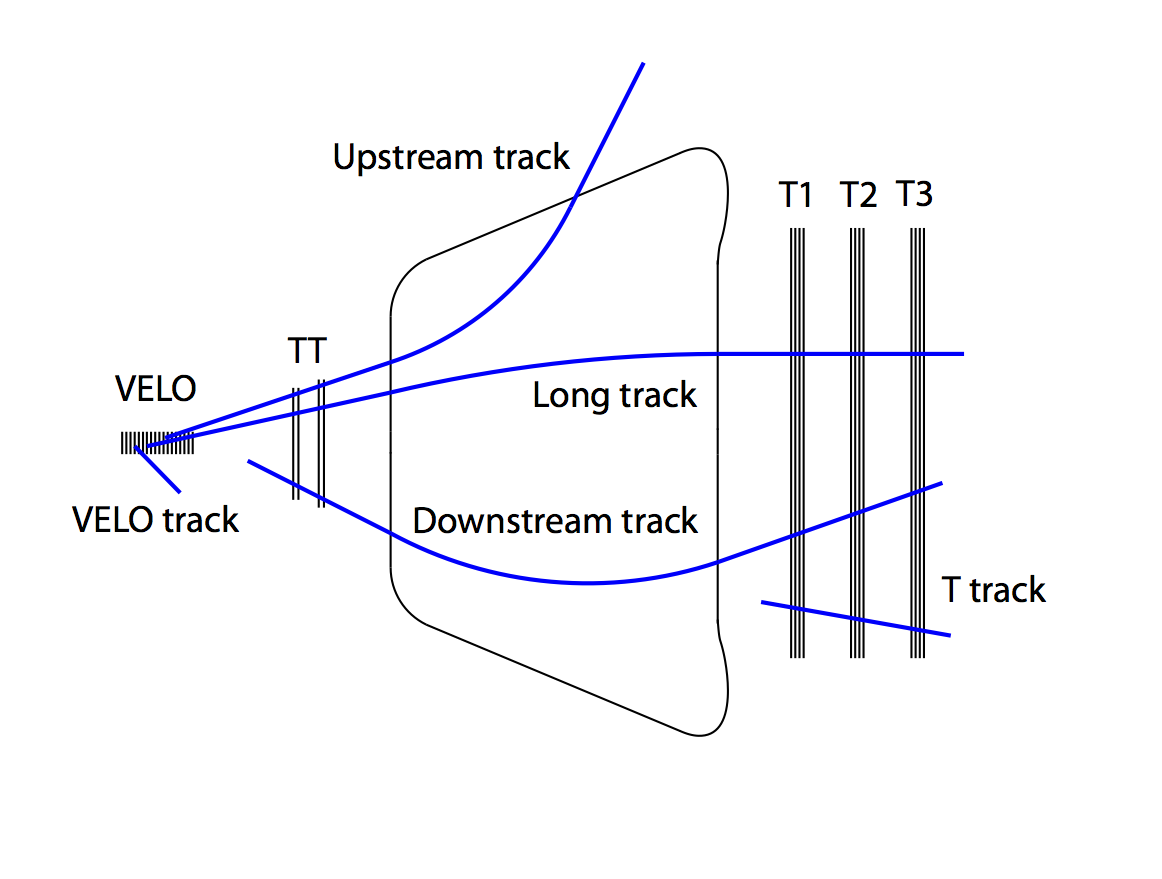}
\caption
[Track categories at \lhcb.]
{Track categories at \lhcb.} 
\label{fig:tracks}
\end{figure}

\clearpage
The different kinds of tracks at \lhcb are categorised as:
\begin{itemize}
\item Long tracks consisting of hits in VELO, TT and T1-3 detectors. The reconstruction of long tracks has the best performance. This kind of tracks are used in data analyses described in Chapters~\ref{ch:ppbar} and~\ref{ch:phiphi}.
\item Upstream tracks, which are reconstructed from hits in VELO and TT. These tracks belong to particles with low momentum such that they escape the T1-3 acceptance due to the magnetic field.
\item Downstream tracks consisting of hits in TT and T1-3 but not in VELO. These tracks can belong to long-lived particles such as \KS, \Lz, \Xib, etc., whose lifetime is enough to leave the VELO detector.
\item T track composed from hits in T1-3 stations only. These tracks can belong to secondary particles created due to the interaction of the primary particle with a material of detector. 
\item VELO tracks consisting of hits in VELO detector only.
\end{itemize} 
The different categories of tracks in \lhcb are illustrated on Fig.~\ref{fig:tracks}.

\subsubsection{Tracker Turicensis (TT)}
The TT is located upstream of the magnet and improves the precision of VELO tracks. Another important goal of TT is to reconstruct vertices formed by downstream tracks, e.g. corresponding to decays of long-lived strange hadrons.

\begin{figure}[b]
\centering
\protect\protect\protect\includegraphics[width=0.65\linewidth]{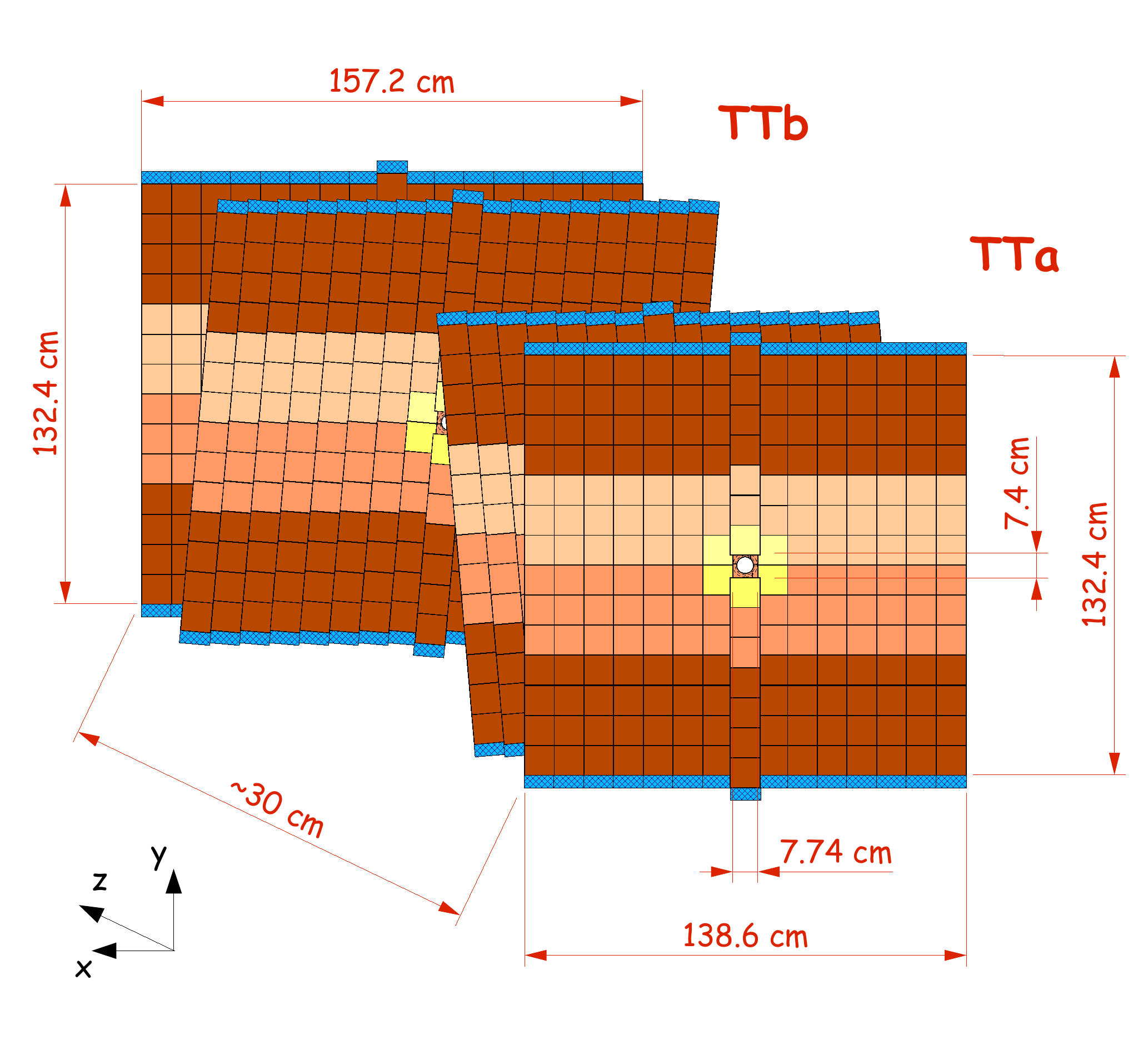}
\caption
[Sketch of Tracker Turicensis.]
{Sketch of Tracker Turicensis~\cite{STPlots}.} 
\label{fig:TT}
\end{figure}

The TT is a silicon microstrip detector with a pitch between sensors of 183~\mum. 
The TT consists of four rectangular detector layers places in so-called $x-u-v-x$ configuration. The $x$-layers are located vertically, while the $u(v)$ stereo layers are rotated by $-5^{\circ}(+5^{\circ})$ relative to the vertical position. Such arrangement improves a spatial resolution of the detector.
The TT comprises two substations $x-u$ (TTa) and $v-x$ (TTb) separated by a distance of 27~\cm along $z$-axis. The total active area of TT is $8~\m^2$. The sketch of TT is shown on Fig.~\ref{fig:TT}. The TT is designed to cover the entire acceptance of the magnet. 
Each layer is made of $9.44\cm\times9.64\cm$ rectangular sensors 0.5\mm thick with 512 strips in total. The sensors are organised into half-modules containing 7 sensors each, which are then grouped into read-out sectors. The central sectors are smaller due to higher detector occupancy. The important feature of TT is that the front-end electronics and the cooling system are located outside of the \lhcb acceptance. The TT provides a spatial resolution of 50\mum. 

\subsubsection{Inner Tracker (IT) and Outer Trackers (OT)}
\begin{figure}[h]
\centering
\protect\protect\protect\includegraphics[width=0.5\linewidth]{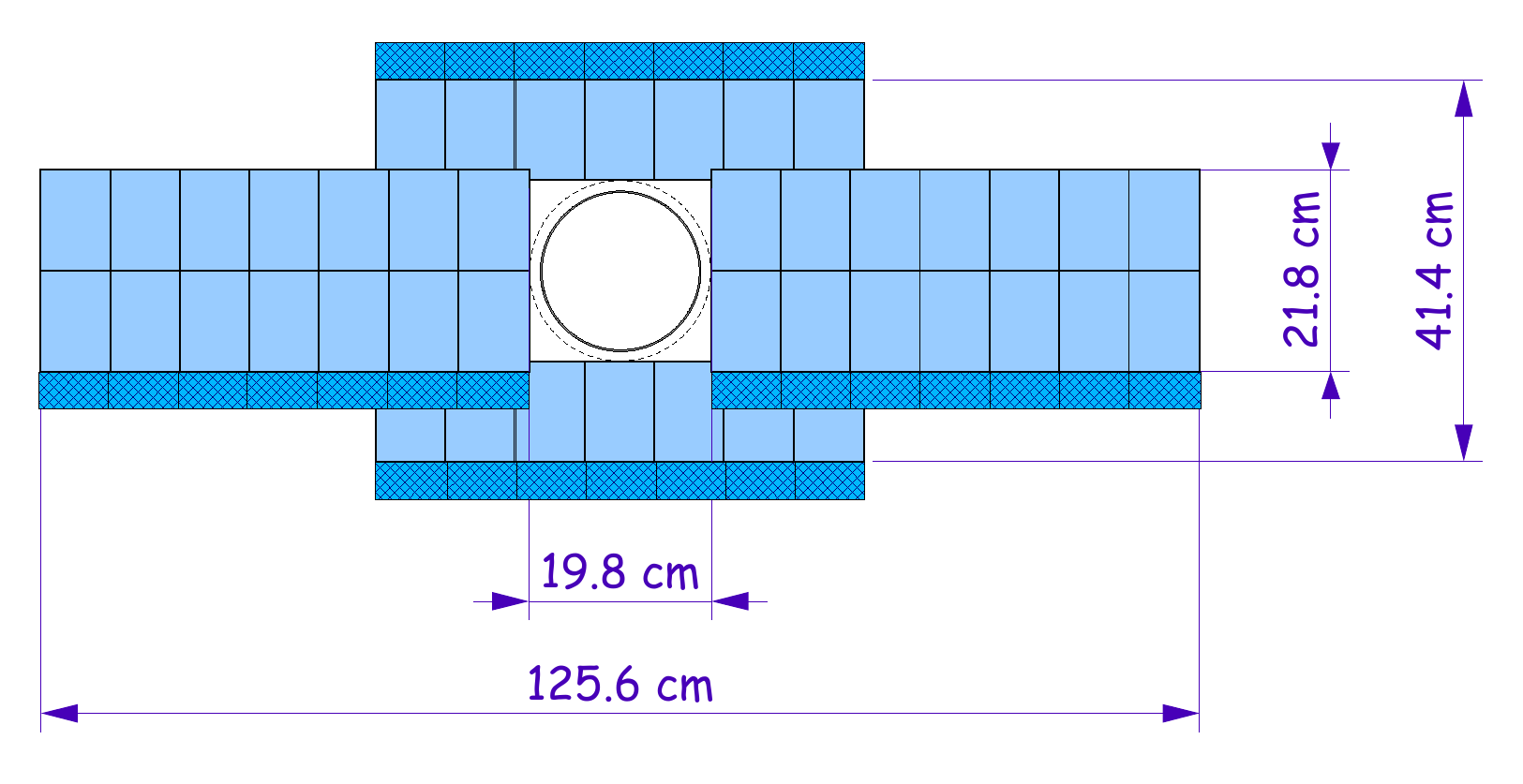}
\caption
[Sketch of Inner Tracker station.]
{Sketch of Inner Tracker station~\cite{STPlots}.} 
\label{fig:IT}
\end{figure}

The tracking stations T1-3 measure the tracks deflected by the magnet and hence are important for the determination of the particles' momenta.
Three stations of IT are arranged similarly to TT.
The IT covers about 2\% of total acceptance area of the tracking station, which corresponds to about 20\% of particle flux.
The IT comprises four detector boxes consisting of four layers similarly to TT as shown on Fig.~\ref{fig:IT}. 

The IT modules consist of two or three sensors, which are shifted along the beam direction with respect to OT detector modules from the same tracking station.
The IT provides a spatial resolution of about 50\mum depending on the detector occupancy. The IT is described in more details in Ref.~\cite{Barbosa-Marinho:582793}.

The OT detector is a gaseous detector based on straw tubes and covering the total area of $597\cm\times485\cm$. 
The straw tubes have a length of 2.4\m and the inner diameter of 4.9\mm filled with a mixture of $Ar$ and $CO_2$ gases with a small fraction of $O_2$. Such gas mixture provides a drift time of 50\ns and a tolerable ageing.
The vertical positioning of straw tubes avoids the sagging of the anode, which is located at the centre of the straw tube. 
The tubes are fixed to carbon-fibre panels forming gas-tight boxes enclosing detector modules. 

\begin{figure}[h]
\centering{
  \subfigure[Module cross section.]{ 
    \protect\protect\protect\includegraphics[width=0.6\textwidth]{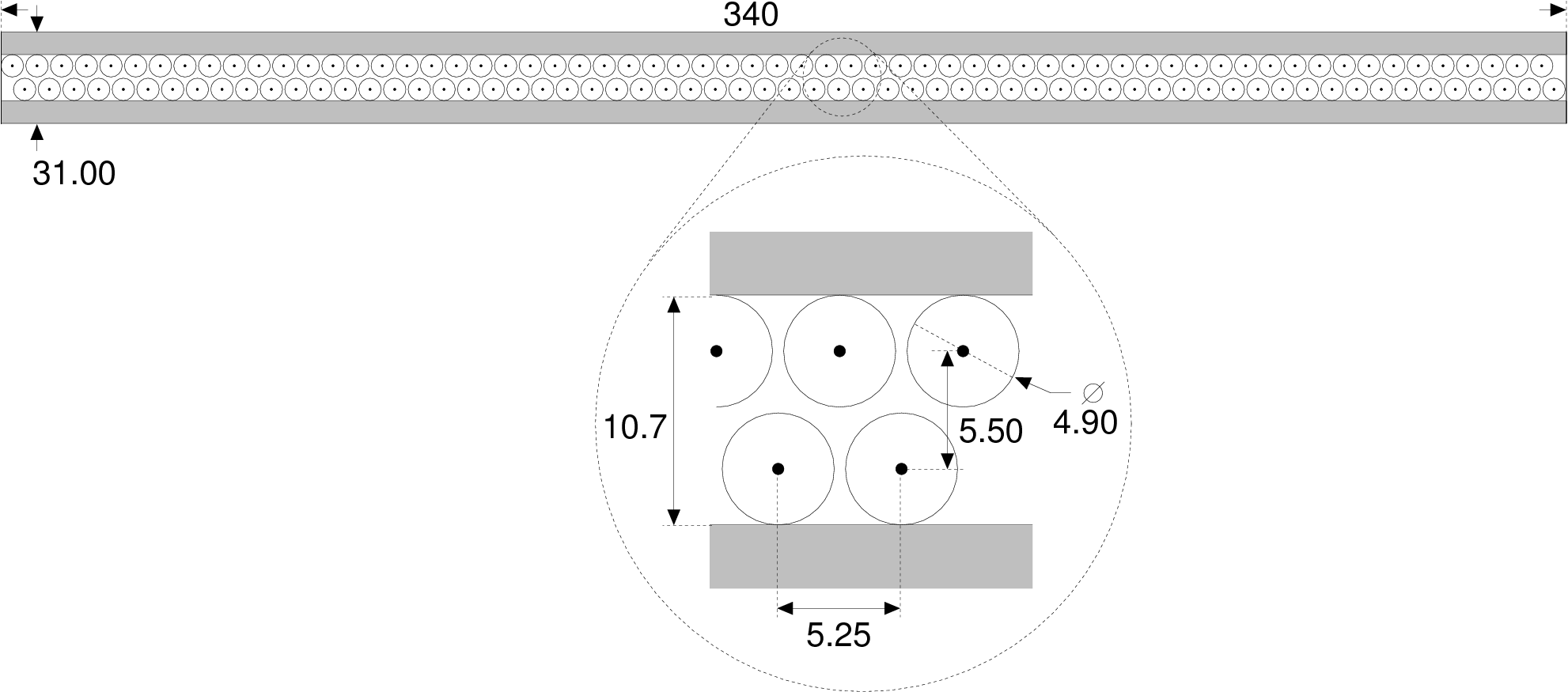}
    \label{fig:OTa}}
  \quad
  \subfigure[Arrangement of OT drift tube modules in layers and stations]{
    \protect\protect\protect\includegraphics[width=0.6\textwidth]{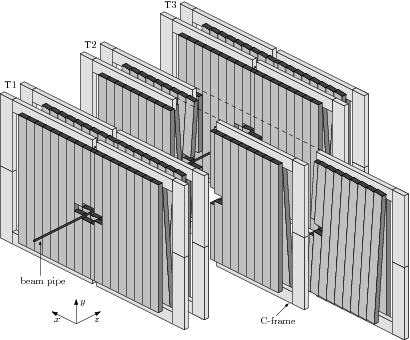}
    \label{fig:OTb}}
}
\caption
[Outer Tracker detector.]
{Outer Tracker detector~\cite{Arink:2013twa}.} 
\label{fig:OT}
\end{figure}

The OT layer is composed of 14 long and eight short modules, containing two staggered layers of straw tubes each.
The OT layers are located vertically with the same $x-v-u-x$ configuration as for TT and IT and form the OT station. 
The schematical view of OT and its module is shown on Fig.~\ref{fig:OT}.

The OT measures the time of arrival of the signal with respect to the \lhcb clock, which provides a measurement of the drift length and improves position resolution to about 200\mum.
The performance of the OT is given in Ref.~\cite{Arink:2013twa}.

\clearpage
\section{Particle identification}
\label{sec:pid}
The particle identification (PID) is a complex task of distinguishing among different kind of limited number of (quasi) stable particles measuring their energy or momentum and studying their interaction with detector material.

The Electromagnetic Calorimeter (ECAL) performs PID of electrons, photons and $\pi^0$ in the decay $\pi^0\to \gamma \gamma$. Also, ECAL measures photon energy and corrects the energy of electron by identifying emitted bremsstrahlung photons.
At trigger level, charged hadrons are identified by fast Hadronic calorimeter (HCAL), while precise hadron ID of charged pions, kaons, protons and deuterons is performed by two Ring Imaging Cherenkov (\rich) detectors. The Muon detector is designed for identifying muons.


\subsection{Calorimeters}
Calorimeter system comprises SPD, PS, ECAL and HCAL and is organised in a preudo-projective geometry. In all detectors, the light from scintillating tiles is transmitted to photon detectors by optical fibres. The four detectors play a key role in the \lhcb trigger. In addition, they provide particle ID and energy measurement for neutral particles.
\subsubsection{SPD and PS}
\begin{figure}[ht]
\centering
\protect\includegraphics[width=0.4\textwidth]{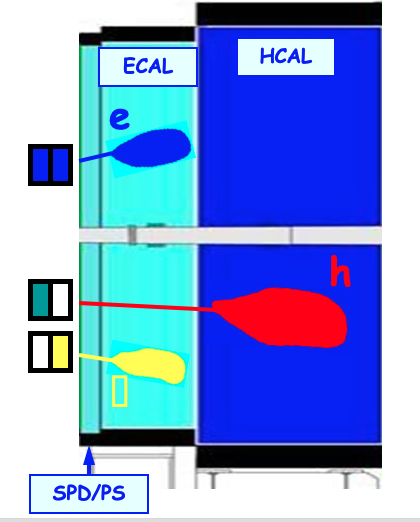}
\caption
[Principle of PID with the \lhcb calorimeter system comprising PS, SPD, ECAL and HCAL.]
{Principle of PID with the \lhcb calorimeter system comprising PS, SPD, ECAL and HCAL. Hadronic (red) and electromagnetic (yellow and blue) showers are illustrated.} 
\label{fig:caloPrinc}
\end{figure}

Apart from the ECAL and HCAL, the calorimeter system of \lhcb~\cite{Amato:494264} comprises the Pre-Shower detector (PS) and Scintillating Pad Detector (SPD). The main goal of the PS is to initiate the shower in front of the ECAL by the electromagnetic particles. The task of SPD is to distinguosh charged particles from uncharged ones. The PS distinguishes electrons from photons.
An illustration of the principle of PID with the \lhcb calorimeters is shown on Fig.~\ref{fig:caloPrinc}.

Both SPD and PS are planar scintillating pad detectors separated by a distance of 56\mm. A 15\mm thick layer of lead is inserted between the two detectors. The amount of lead corresponds to 2.5 electromagnetic interaction length ($X_0$) and a small fraction of hadron interaction length ($0.1~\lambda_I$). Within this configuration, hadrons cross SPD and PS without losing sizeable fractions of their energies, while electrons and photons create electromagnetic showers in PS. Contrary to photons, electrons leave signal in the SPD detector. It is also important to say that the number of hits in the SPD detector is used as a proxy for event multiplicity. 
The total detection area of SPD and PS is $6.2\m \times 6.6\m$.
Similarly to other detectors, the granularity of PS and SPD decreases from inner to the outer region. The size of the cell is about $40~\mm \times 40~\mm$ in inner section, $60~\mm \times 60~\mm$ in the middle section and $120~\mm \times 120~\mm$ in the outer section. The light in scintillator planes is conducted by wavelength shifting fibres connected to Multi-Anode PhotoMultiplier Tubes (MaPMTs). 

\subsubsection{Electromagnetic Calorimeter (ECAL) and Hadronic Calorimeter (HCAL)}
ECAL measures the energy of electromagnetic particles by absorbing their showers.
The ECAL is a "shashlyk"-type calorimeter made from the alternate detector and absorbed layers. The detector layers are made of polystyrene scintillator planes 4~\mm thick. For an absorber, lead layers with a thickness of 2~\mm are used. 
The total depth of ECAL (42~\cm) corresponds to about $25\,X_0$ to ensure that electron and photon showeers are entirely absorbed. On the other hand, the depth of ECAL corresponds to about $1.1\,\lambda_I$, which means that ECAL is effectively a pre-shower detector for HCAL.
The total detection area of ECAL is $7.8\m \times 6.3\m$. The granularity of ECAL is also split into three different regions with effectively the same cell sizes as for SPD and PS.

The ECAL provides a relative energy resolution of
\begin{equation}
\frac{\sigma_E}{E}=\frac{10\%}{\sqrt{E}}+1\%,
\end{equation}
where the energy $E$ is expressed in \gev.

\begin{figure}[t]
\centering{
  \subfigure[SPD, PS and ECAL.]{ 
    \protect\protect\protect\includegraphics[width=0.5\textwidth]{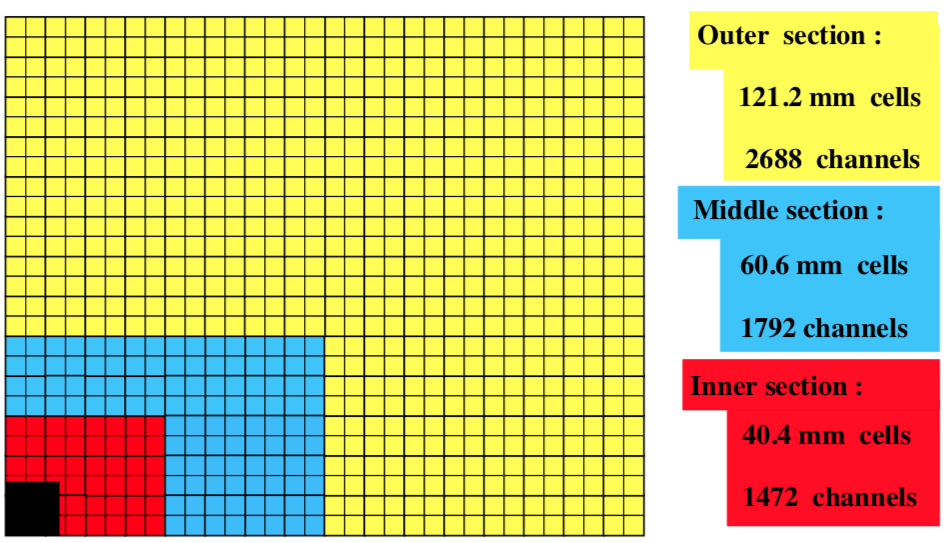}
    \label{fig:ecal}}
  \quad
  \subfigure[HCAL.]{
    \protect\protect\protect\includegraphics[width=0.5\textwidth]{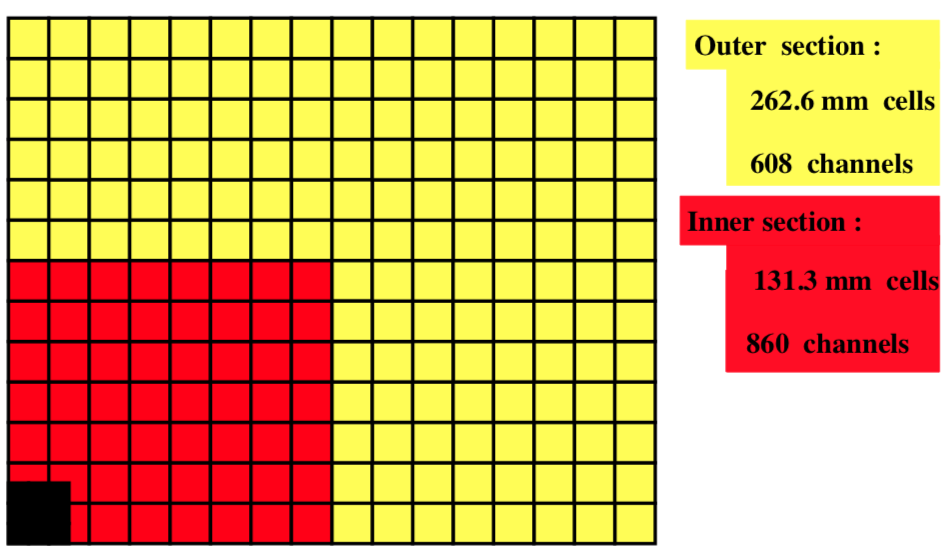}
    \label{fig:hcal}}
}
\caption
[Segmentation of \lhcb calorimeters.]
{Segmentation of \lhcb calorimeters~\cite{Amato:494264}.} 
\label{fig:calo}
\end{figure}

The HCAL is a sampling detector made of scintillating (3~\mm thick) and iron absorber (16~\mm thick) tiles, which are glued to master plates. Contrary to ECAL geometry, the tiles are oriented along the beam axis. 
The depth of HCAL is defined by 1.65 m of absorber, which corresponds to about $5.6\,\lambda_I$.
The total detection area of HCAL is $8.4\m \times 6.8\m$.
The size of the HCAL cell is $121~\mm \times 121~\mm$ in the inner region and $263~\mm \times 263~\mm$ in the outer region.
The segmentation of \lhcb calorimeters is summarised on Fig.~\ref{fig:calo}.

The worse energy resolution of HCAL compared to the one of ECAL is caused by fluctuations of hadron showers and is shown below.
\begin{equation}
\frac{\sigma_E}{E}=\frac{69\%}{\sqrt{E}}+9\%,
\end{equation}
where the energy $E$ is expressed in \gev.
The HCAL is not used for offline measurement of charged hadron energy since it can be measured more precisely using the information from both tracking system and \rich detectors. On the other hand, HCAL can be used for measurement of neutron energy. However, HCAL provides very fast information about the presence of sizeable transverse energy deposit, which corresponds to the hadron with large \pt. This information is used in the first hardware level of the trigger.

\subsection{\rich detectors}
Charged hadron identification plays a crucial role in the reconstruction of charmonium using its hadronic decays. It is thanks to charge hadron ID, that \lhcb experiment can reconstruct decays, which are not accessible by \atlas and \cms experiments.

\begin{figure}[ht]
\centering
\protect\protect\includegraphics[width=0.6\textwidth]{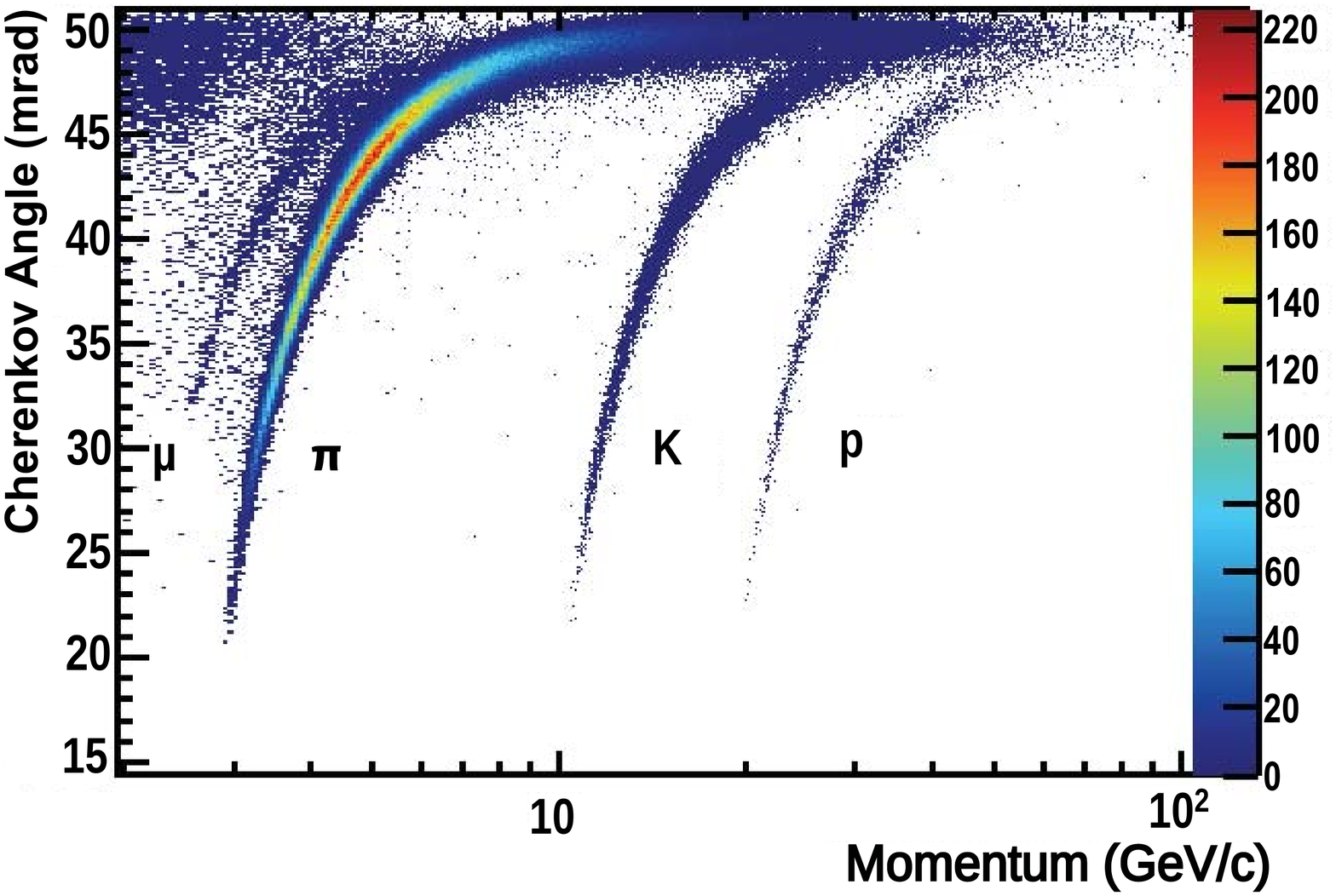}
\caption
[Cherekov angle in $C_4F_{10}$ gas as a function of particle momentum for different kinds of charged hadrons.]
{Cherekov angle in $C_4F_{10}$ gas as a function of particle momentum for different kinds of charged hadrons~\cite{Adinolfi:2012qfa}.} 
\label{fig:rich_angle}
\end{figure}

In \lhcb, charged pions, kaons and protons are distinguished with two Ring Imaging CHerenkov detectors (\rich).  
When a charged particle tranverses a dielectric medium with speed $\beta$ that is larger than a speed of light in this medium ($1/n$, where $n$ is a refractive index of the medium), a cone of the Cherenkov light is emitted.
The angle of the light emission is a function of the particle velocity:
\begin{equation}
cos(\theta_C)=\frac{1}{n\beta}.
\end{equation}
The choice of the radiator is crucial for detector performance. The radiator defines a hadron momentum range, where \rich detector possesses a separating power.
The most illustrative separation between hadrons is achieved near the threshold of Cherenkov light emition.
As an example, the dependence of the Cherenkov angle on track momentum is shown on Fig.~\ref{fig:rich_angle} for $C_4F_{10}$ radiator.
\clearpage

The \rich detector makes use of this radiation by projecting the cone of Cherenkov light onto planar photodetectors by use of a spherical mirrors.
The radius of the obtained ring is a function of the particle velocity. 
Having a measurement of the track momentum from the tracking detector and the measured radius of the associated ring in the \rich detector, one can calculate the mass of the particle. 
The only charged particles produced that can reach the \rich detectors are $e^{-}$, $\mu^{-}$, $\pim$, $\Km$, $\proton$ and $d$ (and nuclei), thus one categorises the rings according to all possible PID hypothesis. 
\lhcb uses two \rich detectors (\richone and \richtwo) to distinguish three kinds of charged hadrons: pions, kaons and protons in a wide range of momentum. 

The \richone is located upstream the magnet before the TT and aims at providing the PID of particles with lower momentum from 1 to 60~\gev, which also includes particles leaving upstream tracks. For that, the silica aerogel and $C_4F_{10}$ gas were used as radiators. For the \lhcb Run II, the aerogel was removed from the \richone detector, which increased the effective lower limit of momentum for PID provided by \richone. The amount of material of \richone corresponds to only about $0.08\,X_0$.

The \richtwo is located downstream of the magnet and covers higher momentum range from 15 to 100~\gev. In \richtwo, the $CF_4$ gas with a small fraction of $CO_2$ is used as a radiator.
The amount of material of \richtwo corresponds to only about $0.015\,X_0$.

\begin{figure}[ht]
\centering
\protect\protect\includegraphics[width=0.8\textwidth]{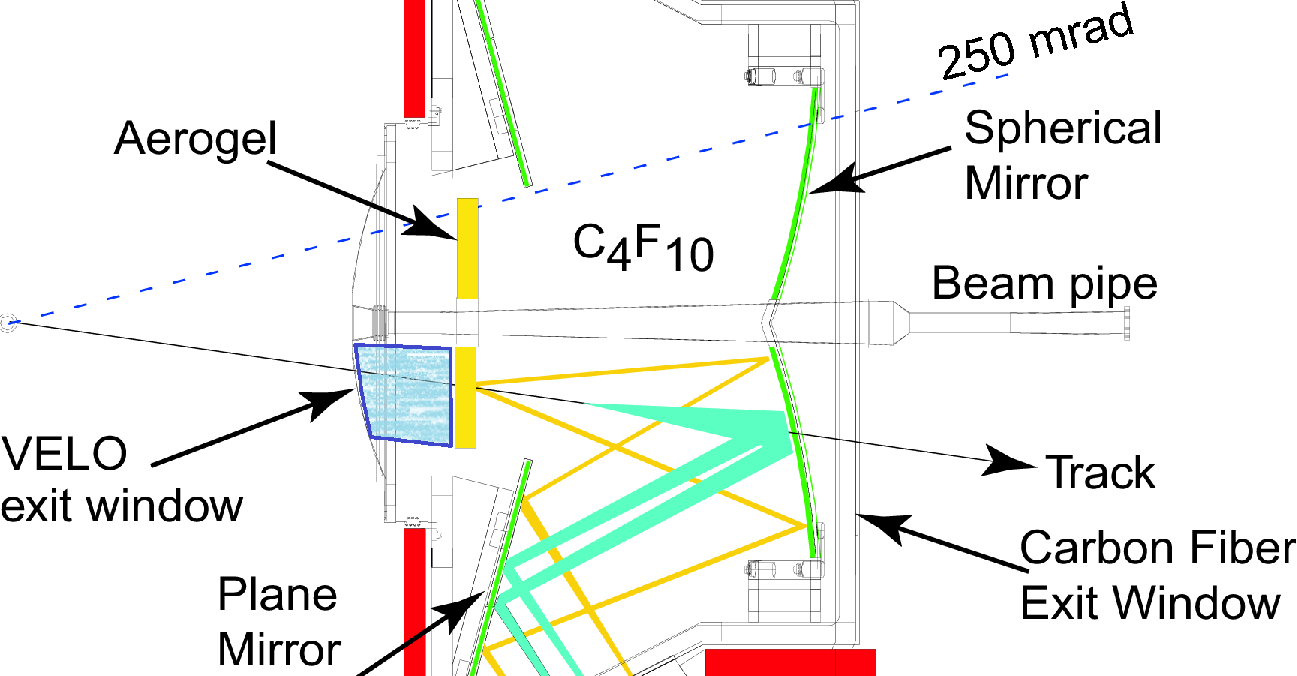}
\caption
[Sketch of \richone detector.]
{Sketch of \richone detector~\cite{Papanestis:2017zcj}.} 
\label{fig:rich1}
\end{figure}

For both \richone and \richtwo, the spherical mirrors are used to focus the light onto the flat mirror, which then projects the light onto the plane of Hybrid Photodetectors (HPDs). The use of flat mirror allows to reduce the geometrical size of detectors and to locate the photodetector outside the \lhcb acceptance. The optical systems are split into two halves: top-bottom for \richone and left-right for \richtwo. 
The optical system is shown on Fig.~\ref{fig:rich1} at the example of \richone.

The HPD is a hybrid of PMT and silicon pixel detectors. Firstly, photons produce photoelectrons from the photocathode, then the electrons are accelerated by the electric field of 16 kV in the vacuum tube and then are focused onto silicon pixel array providing a signal multiplication in one step. The pixel size of HPD is $2.5~\mm \times 2.5~\mm$.

\begin{figure}[t]
\centering{
  \subfigure[Kaon PID efficieny (red) and $\pi \to\ K$ misidentification (black).]{ 
    \protect\protect\protect\includegraphics[width=0.45\textwidth]{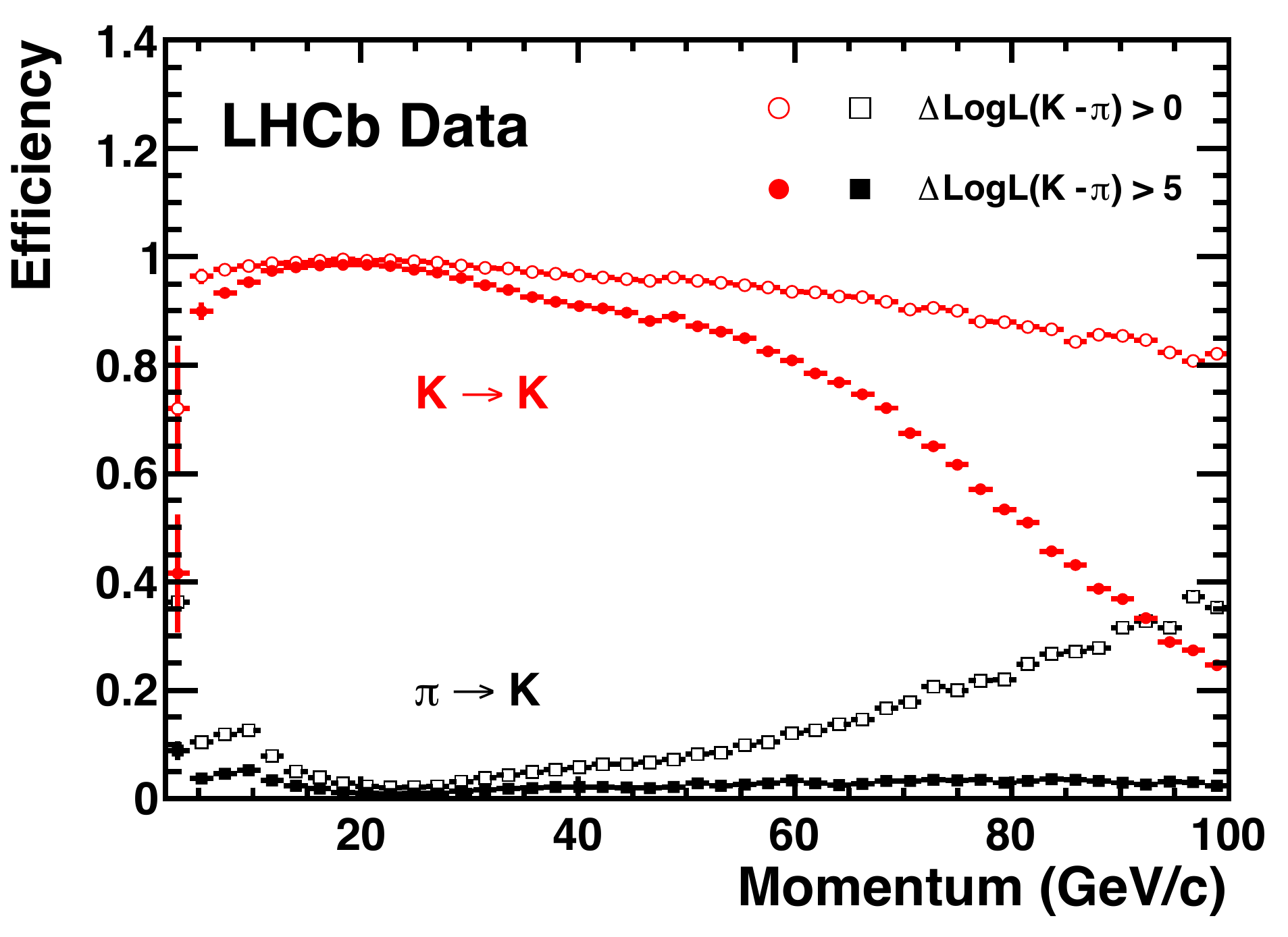}}
  \quad
  \subfigure[Proton PID efficieny (red) and $K \to p$ misidentification (black).]{ 
    \protect\protect\protect\includegraphics[width=0.45\textwidth]{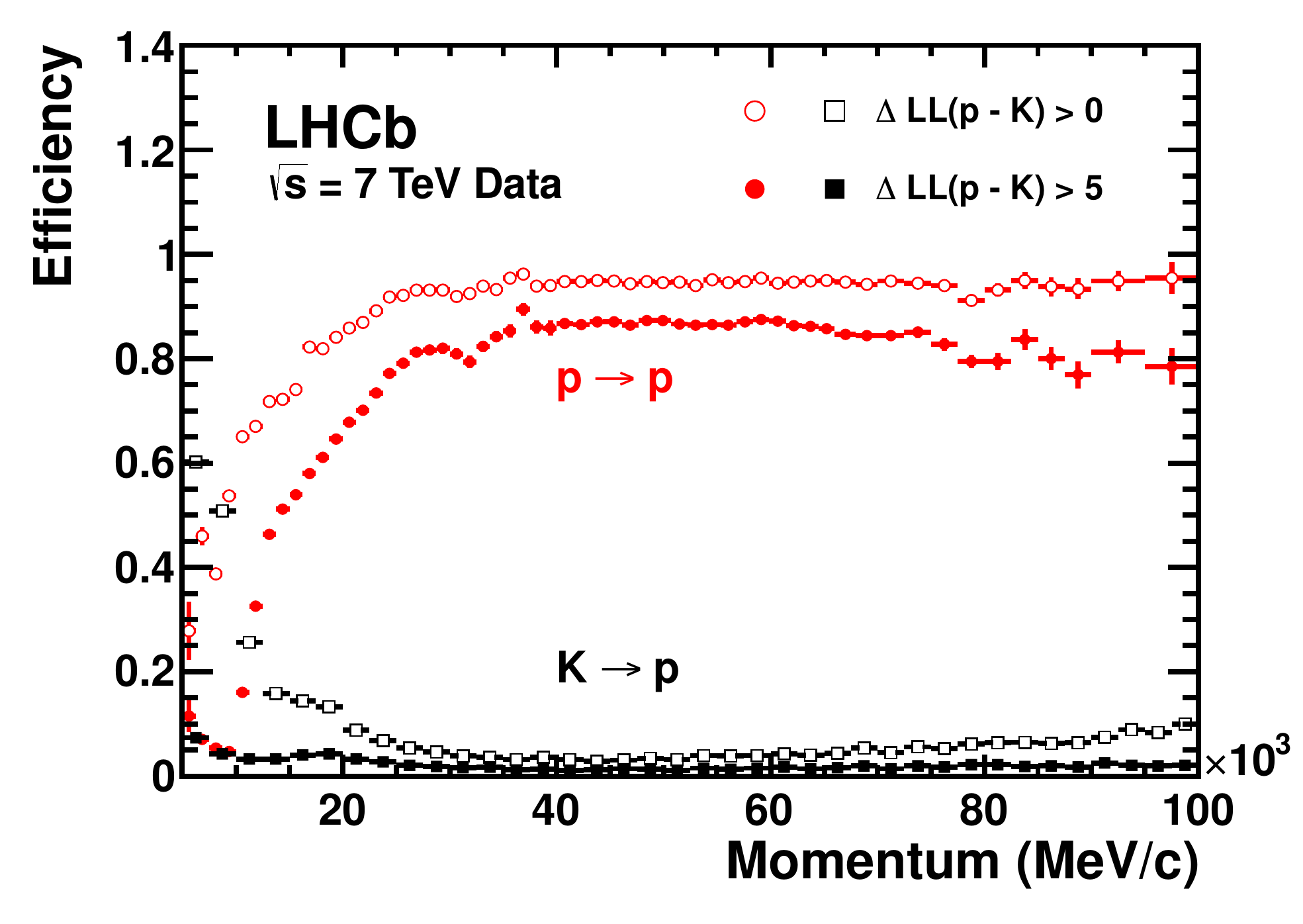}}
  \quad
  \subfigure[Proton PID efficieny (red) and $\pi \to p$ misidentification (black).]{
    \protect\protect\protect\includegraphics[width=0.45\textwidth]{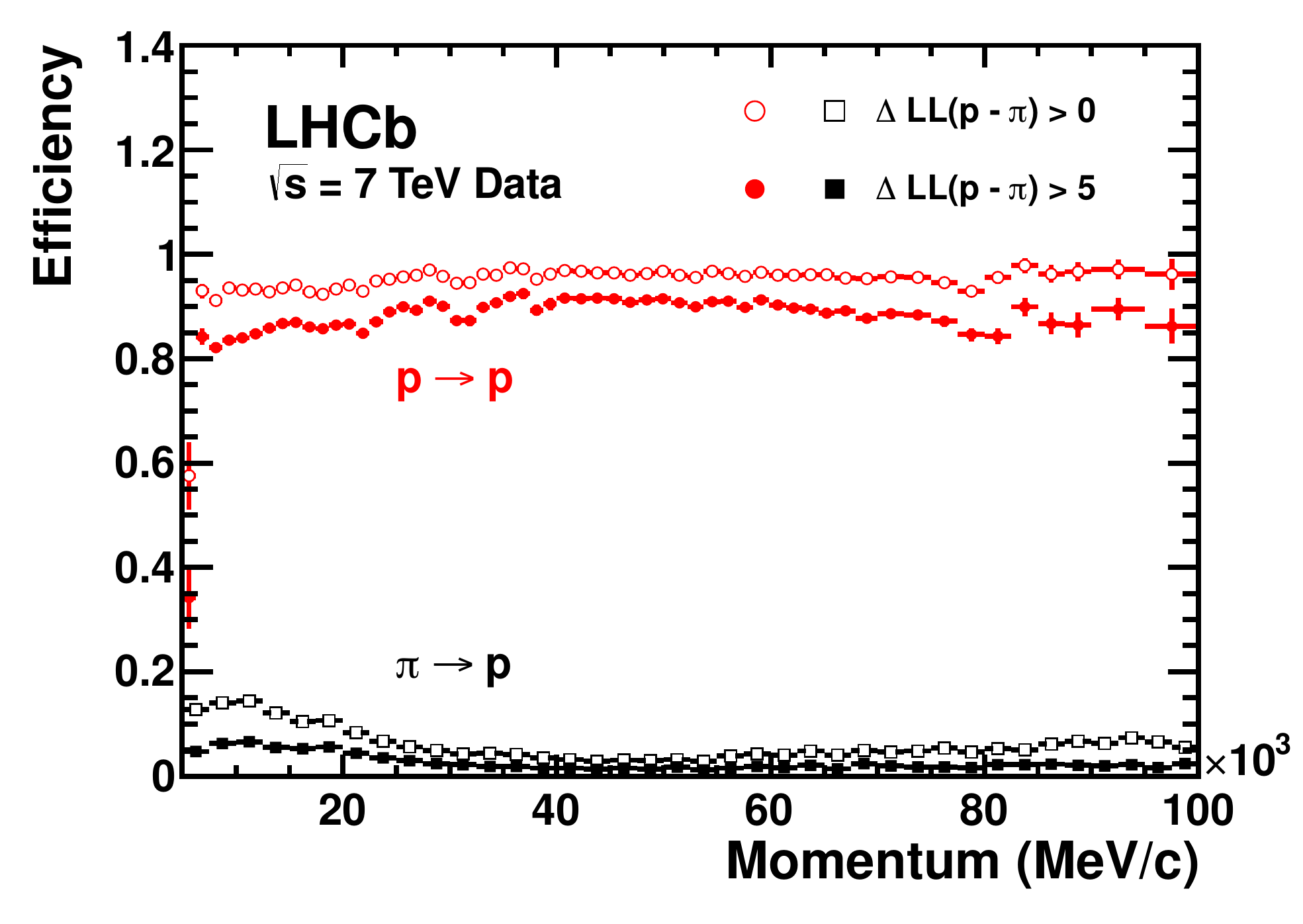}}
}
\caption
[Performance of charged hadron ID.]
{Performance of charged hadron ID~\cite{Papanestis:2017zcj}.}
\label{fig:misid} 
\end{figure}

For PID a global likelihood variable using information from all PID detectors is used. This variable is a product of the likelihoods from individual detectors. For the case of \rich detectors, the reconstructed rings are compared with the ring expected from the measured track momentum with different charged hadron hypotheses. The combination of constructed likelihoods together with the information from calorimeters and muon detector yields a global likelihood value. The $PIDp$ variable represents the likelihood of proton hypothesis, $PIK$ - kaon hypothesis, etc. Another technique used in the \lhcb is based on the multivariate classification by the neural network, which yields alternative \texttt{ProbNN} variable. The latter approach shows a better separation power due to taking into account possible correlations between signals in all detector systems.
A performance of charged hadron ID is illustrated on Fig.~\ref{fig:misid}. Typically, \rich detectors provide good proton ID for large proton momenta (above 30 \gev), while kaon PID is performant at lower momentum range up to 60 \gev. 

To improve the accuracy of PID simulation, the \lhcb uses calibration samples of well-known decays. The requirement on the decays used for PID calibration is that they can be selected using only kinematical requirements. For example, decays $\decay{\Lambda}{p\pim}$, $\decay{\Lambda_c}{p \Kp \pim}$, $\decay{\KS}{\pip\pim}$ and 
$D^{∗+}\to(\D^0\to\Km\pip) \pip$ are used to extract calibration samples for protons, kaons and pions. The $PIDCalib$ package makes use of these calibration samples and provides efficiency tables of the PID requirement as a function of kinematical and multiplicity variables. Often, these tables are used in data analyses to estimate total PID efficiency and correct simulation samples.

\begin{figure}[h]
\centering
  \subfigure[The \pip\pim invariant mass spectrum without \rich information.]{ 
    \centering
    \protect\protect\protect\includegraphics[width=0.465\textwidth]{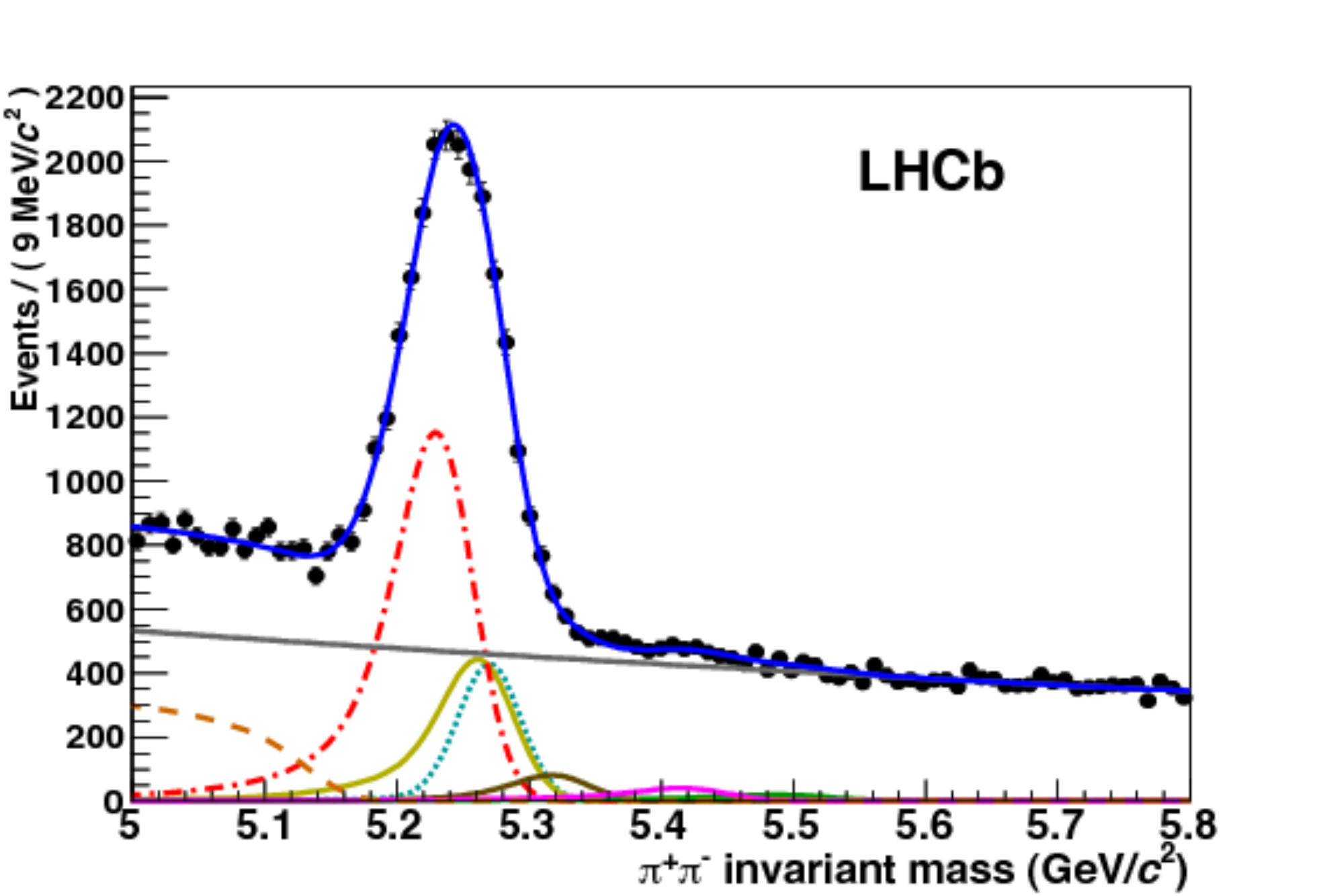}
    \label{fig:woRICH}
  }
\quad
  \subfigure[The \pip\pim invariant mass spectrum with \rich information.]{
    \centering
    \protect\protect\protect\includegraphics[width=0.465\textwidth]{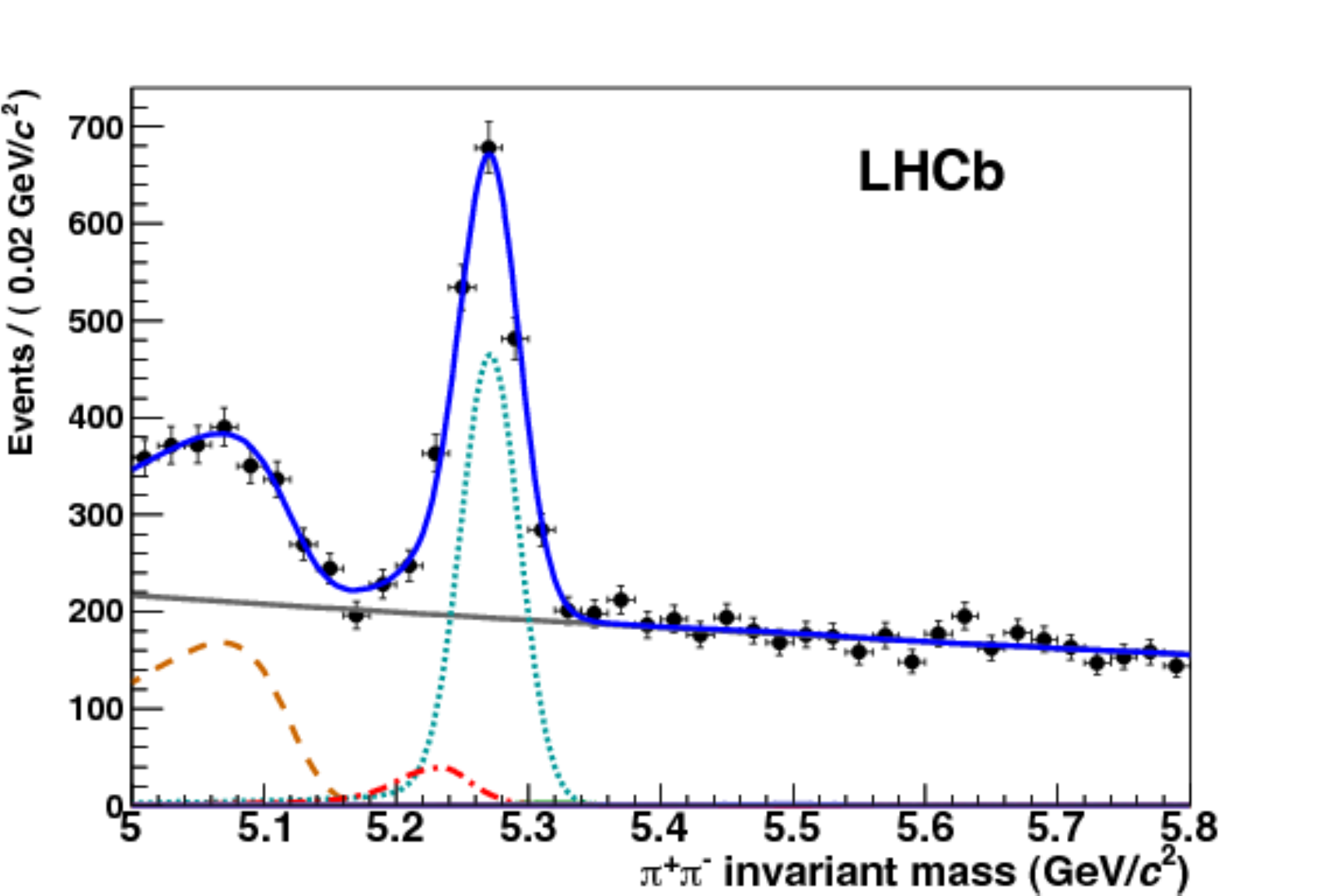}
    \label{fig:wRICH}
  }
\caption
[Comparison of the $\Bz\to\pip\pim$ candidates invariant mass distribution with and without information from \rich detectors.]
{Comparison of the $\Bz\to\pip\pim$ candidates invariant mass distribution with (right) and without (left) information from \rich detectors~\cite{Adinolfi:2012qfa}. The contributions from different \bquark-hadron decay modes ($\decay{\Bz}{\Kp\pim}$ red dashed-dotted line, three body \Bz decays orange dashed-dashed, $\decay{\Bs}{\Kp\Km}$ yellow line, $\decay{\Bs}{\Kp\pim}$ brown line, $\decay{\Lb}{p \Km}$ purple line, $\decay{\Lb}{p \pim}$ green line), are eliminated by requiring a positive identification of pions, kaons and protons and only the signal and two background contributions remain visible in the plot on the right. The grey solid line is the combinatorial background.} 
 \label{fig:B2pipi}
\end{figure}

A performance of charged hadron ID can also be illustrated with the following example. The performance of charged hadron ID is illustrated in Fig.~\ref{fig:B2pipi}, which compares the invariant mass spectra of $\Bz\to\pip\pim$ with and without  information from the \rich detectors. If the \rich information is not used, the observed peak is a sum of different \bquark-hadron decay modes to two charged hadrons (left plot), and only \rich PID is providing relatively clean and narrow $\Bz\to\pip\pim$ sample (right plot).
\clearpage

\subsection{Muon detector}
The muon detector is designed to provide muon particle identification. 
For studies of charmonium states via their decays to hadrons, the information from the muon detector is not used. However, a robust muon identification is crucial for analyses involving, for example, $\decay{\jpsi}{\mup\mu^{-}}$ or $\decay{\psitwos}{\mup\mu^{-}}$ decays. 

\begin{figure}[ht]
\centering
\protect\protect\includegraphics[width=0.9\textwidth]{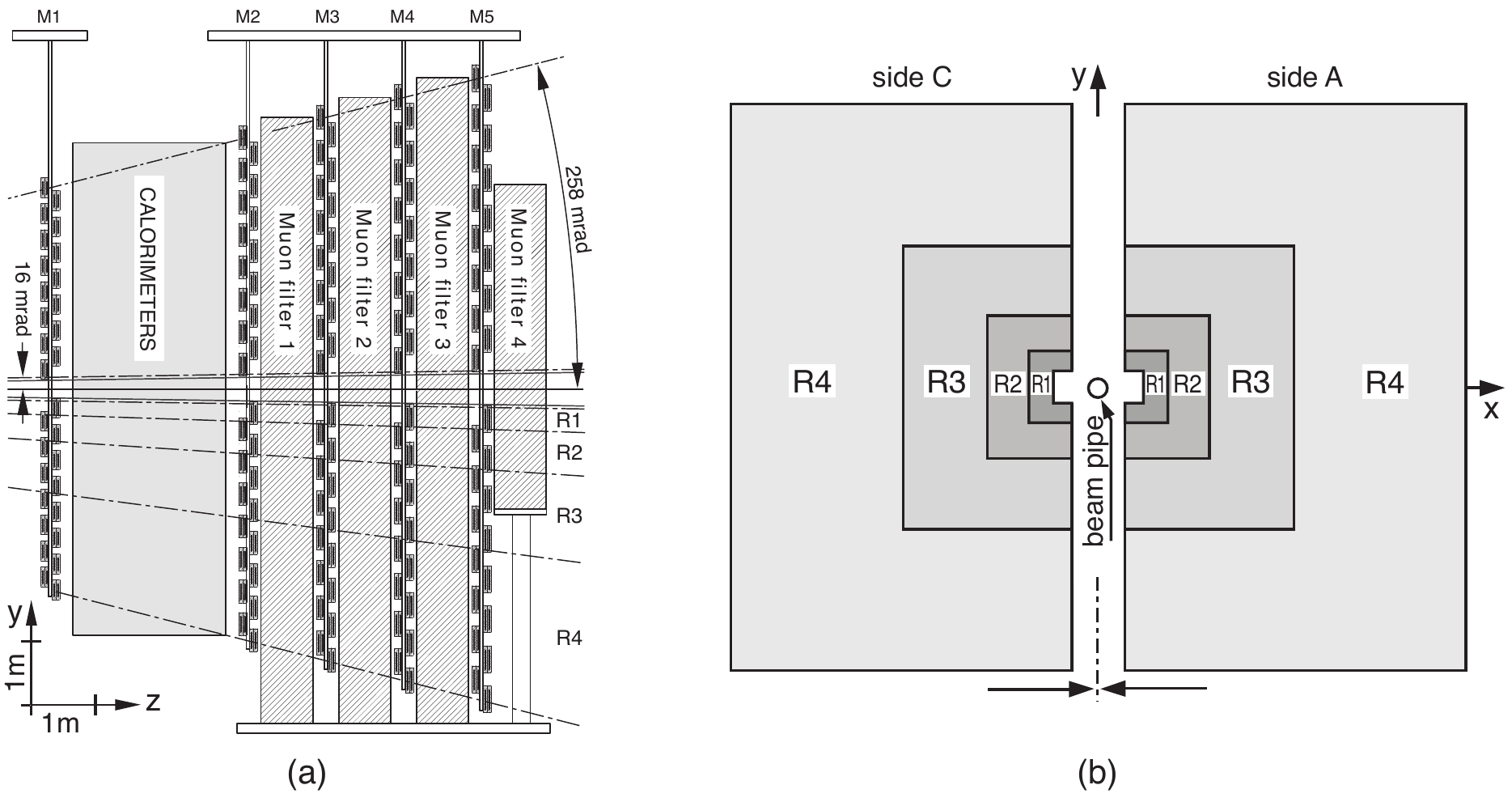}
\caption
[The \lhcb muon detector.]
{The \lhcb muon detector~\cite{LHCb:2001ab}.} 
\label{fig:muon}
\end{figure}

The muon detector consists of five stations M1-5. The first station M1 is located between the \richtwo detector and the calorimeters to improve track matching between tracking and muon detectors, while the stations M2-5 are located downstream the HCAL. The size of stations is increasing with increasing the distance from the interaction point. The sketch of the muon detector is shown on Fig.~\ref{fig:muon}.
The stations are divided into four regions R1-4. The linear scale of the next region is twice bigger than the scale of the previous one. The R2-4 regions are made of Multi-Wire Proportional Chambers (MWPC), the central R1 region is made of Gas Electron Multipliers (GEM) due to large particle fluxes in the central region and higher radiation resistance of GEM detectors.
The chambers are composed of logical pads of different dimensions, depending on the distance from the beam axis and the from interaction point. 
The sizes of chambers in the inner regions vary between $6.3\mm \times 31.3\mm$ in M2 station and $31~\mm \times 39~\mm$ in the last station, in order to maintain the occupancy similar in each region.
The MWPCs uses a mixture of $Ar$, $CO_2$ and $CF_4$ gases.

Between the M2-5 stations, iron absorber of thickness of  80~\cm thick are placed, which depth corresponds to $15 \lambda_I$. The acceptance of the muon detector is similar to that of the trackers. 
The muons stations are also used in the online (\hltone) and offline (\hlttwo) trigger for fast muon identification. Note, that only muon identification is used in the online trigger.
The \lhcb muon system is described in more details in Ref.~\cite{LHCb:2001ab}.

\section{Trigger and data processing}
\label{sec:trigger}
A flexible trigger system of \lhcb consists of low-level L0, HLT1 software and off-line HLT2 levels~\cite{LHCb-DP-2012-004}. 
The 40 MHz rate of \lhc bunch crossings corresponds to roughly 10 MHz of interactions visible by \lhcb. The hardware trigger L0 reduces this rate to about 1 MHz. Then the online software trigger selects events with a rate of above 100 kHz. Finally, the output rate of the offline HLT2 trigger decisions is about 10 kHz, which is a storable event rate.

\begin{figure}[ht]
\centering
\protect\protect\includegraphics[width=0.4\textwidth]{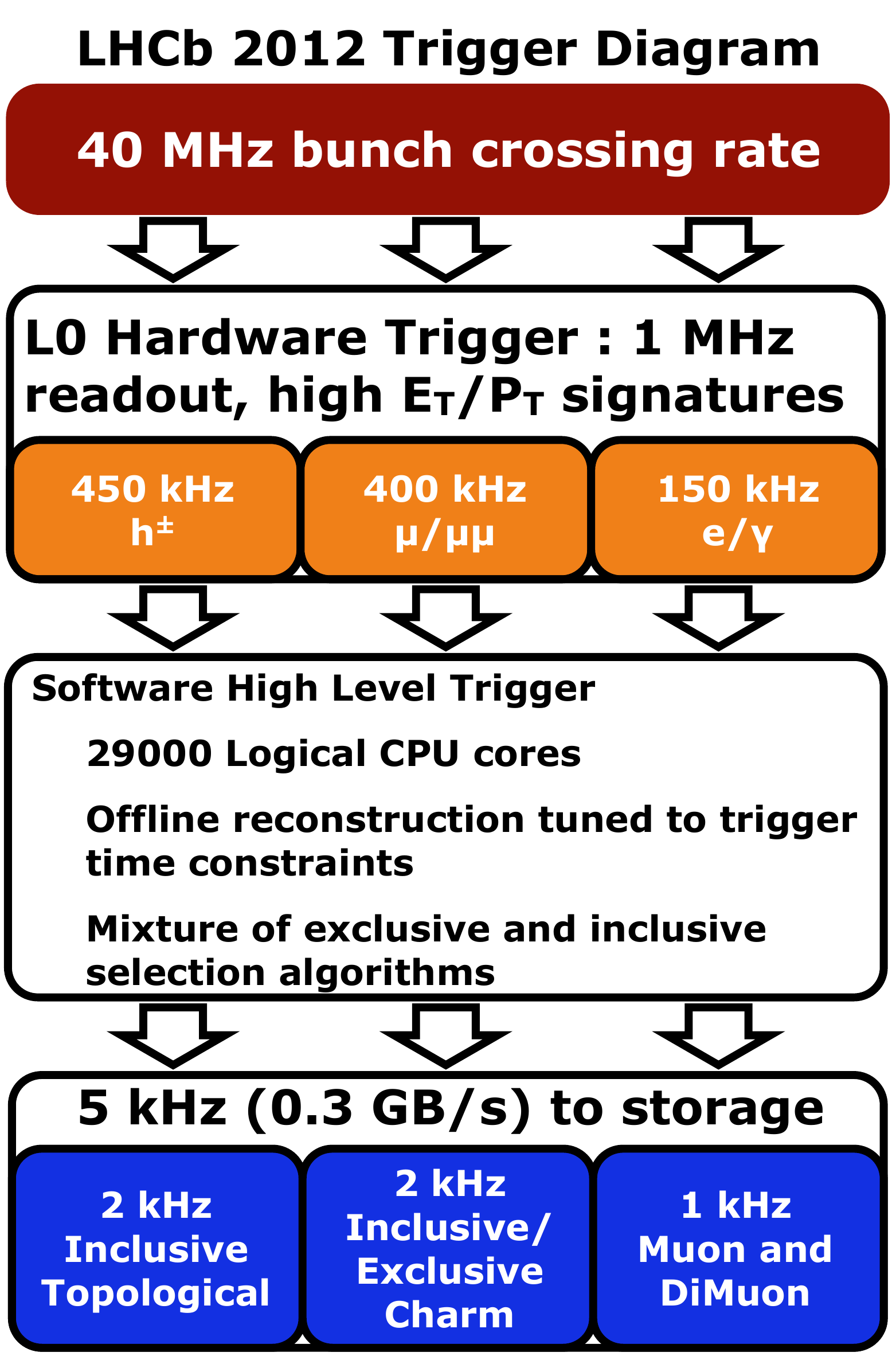}
\protect\protect\includegraphics[width=0.4\textwidth]{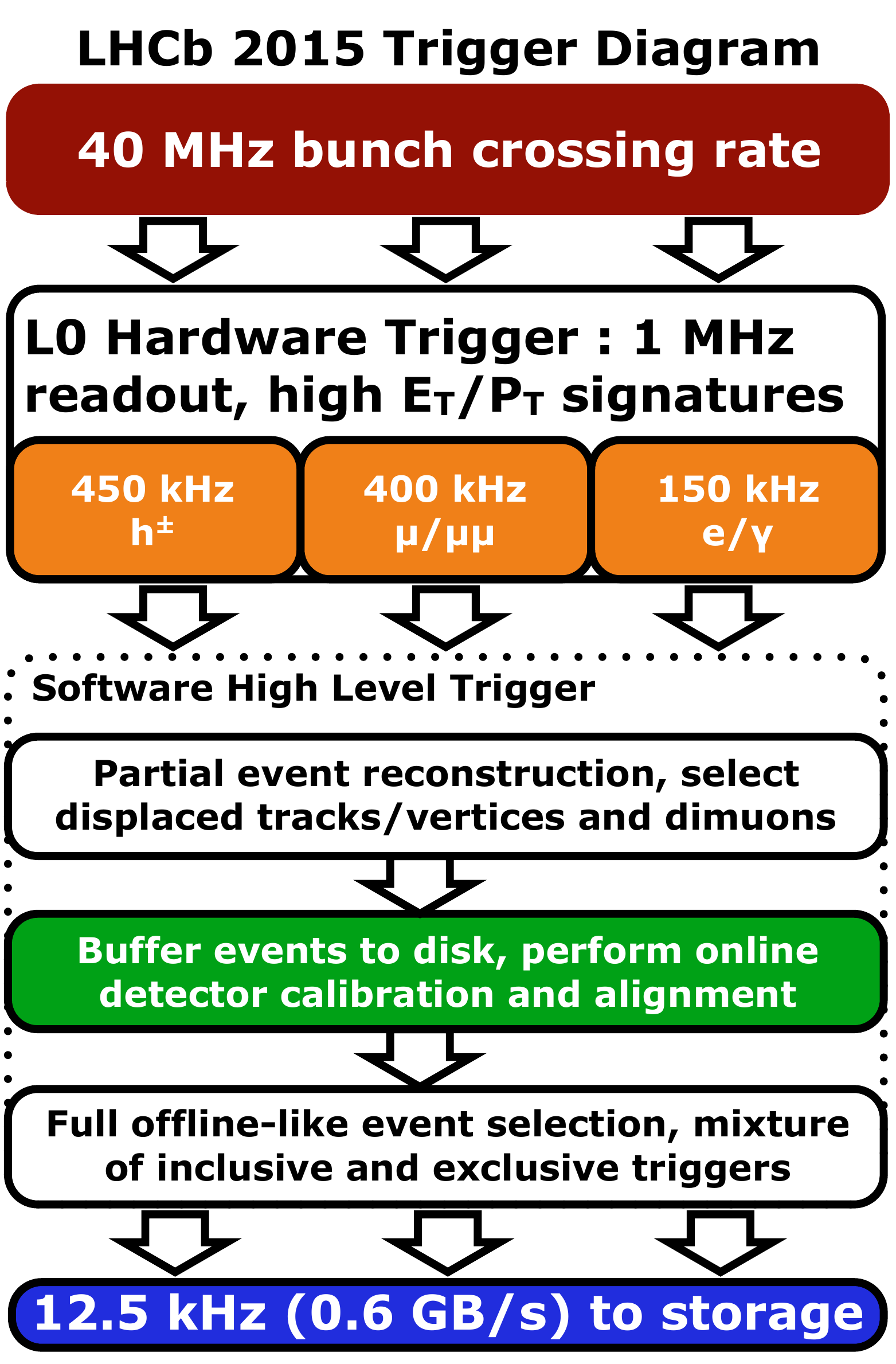}
\caption
[The scheme of the \lhcb trigger for Run I and Run II.]
{The scheme of the \lhcb trigger for (left) Run I and (right) Run II.} 
\label{fig:trigger}
\end{figure}

The trigger schemes used during the \lhcb Run I and II are shown on Fig.~\ref{fig:trigger}.

The lowest level L0 trigger uses a fast information from calorimeter, muon and VELO detectors. Thanks to L0 trigger, the event rate is reduced to a level appropriate for a fast analysis of the information from other detector systems.
The \hltone trigger performs a partial event reconstruction including simplified tracks and vertex reconstruction further reducing the rate. 
The \hlttwo trigger performs a full event reconstruction including a complex particle identification needed to perform a selection of certain exclusive and inclusive \bquark or \cquark decays. The events or candidates passing \hlttwo trigger are then stored to the disk. The bandwidth of the software trigger is limited by the available computing resources.
The trigger requirements are adjusted in order to split the available bandwidth between different physics cases in an optimal way. 

The following positive trigger decisions are defined:
\begin{itemize}
\item Trigger On Signal (TOS) - the final reconstructed candidate is the one satisfying trigger requirement;
\item Trigger Independent from Signal (TIS) - another candidate in the event triggered the decision;
\item Trigger Decision (DEC) - a logical sum of TIS and TOS.
\end{itemize}

\subsection{Hardware L0 trigger}
The L0 trigger is synchronized with the \lhc bunch crossing. Depending on the state of the Front-End electronics, the L0 can either pass or throttle the event satisfying the trigger requirement.
The L0 trigger comprises three different types of trigger decisions.

The \texttt{L0\_Muon} and \texttt{L0\_DiMuon} are based on the information from muon detector to select events containing muons with sufficient transverse momentum. 
The trigger based on the information from calorimeters are \texttt{L0\_Hadron}, \texttt{L0\_Electron} or \texttt{L0\_Photon}. The \texttt{L0\_PileUp} is used for luminosity measurement.

The Muon trigger uses L0 processors connected to each quadrant of the muon detector. Processors perform a search among the tracks with $\pt > 500~\mev$ and identify two tracks with largest \pt in the corresponding quadrant.
The \texttt{L0\_Muon} sets a threshold on the minimum transverse momentum of the track with a typical value of about 1.5~\gev, while the \texttt{L0\_DiMuon} sets a threshold on the minimal product of two muon tracks \pt with a typical value of about $(1.3~\gev)^2$. The trigger also sets a threshold on the maximal number of hits in the SPD detector to reject events producing an excessively high level of combinatorial background.
The transverse momentum is measured using Muon detector only, which provides a \pt resolution of about 20\%. All M1-5 stations are required to have track hits. The presence of M1 station, located upstream the calorimeters, is essential for the \pt measurement at the L0 trigger level. 
This trigger is especially important for studies involving reconstruction of resonances decaying into a pair of muon (e.g. \jpsi, \psitwos, \OneS, \TwoS). Notably, this trigger is used for \jpsi and \psitwos production and polarisation measurements.

The \texttt{L0\_PileUp} trigger uses the information from two $r$-sensors of the VELO detector. This trigger identifies events with single and multiple interactions.

The Calorimeter trigger is based on the transverse energy, $E_T$, deposit calculation in the ECAL or HCAL. The transverse energy is computed from clusters of $2\times2$ cells located in the same zone. Each of the calorimeters front-end board selects the highest $E_T$ among 32 clusters.
The \texttt{L0\_Photon} requires a presence of $E_T$ deposit in ECAL above the threshold of about 2.5~\gev with a matching signal from PS and no signal from the corresponding cells of SPD. 
The \texttt{L0\_Electron} requirement is similar, but contrary to \texttt{L0\_Photon}, at least one SPD cell hit should be present in a region corresponding to PS cells hits.
The \texttt{L0\_Hadron} requires a presence of $E_T$ deposit in HCAL and the matching cluster of ECAL  higher than the threshold of about 3.5~\gev. 

\begin{figure}[ht]
\centering
\protect\protect\includegraphics[width=0.6\textwidth]{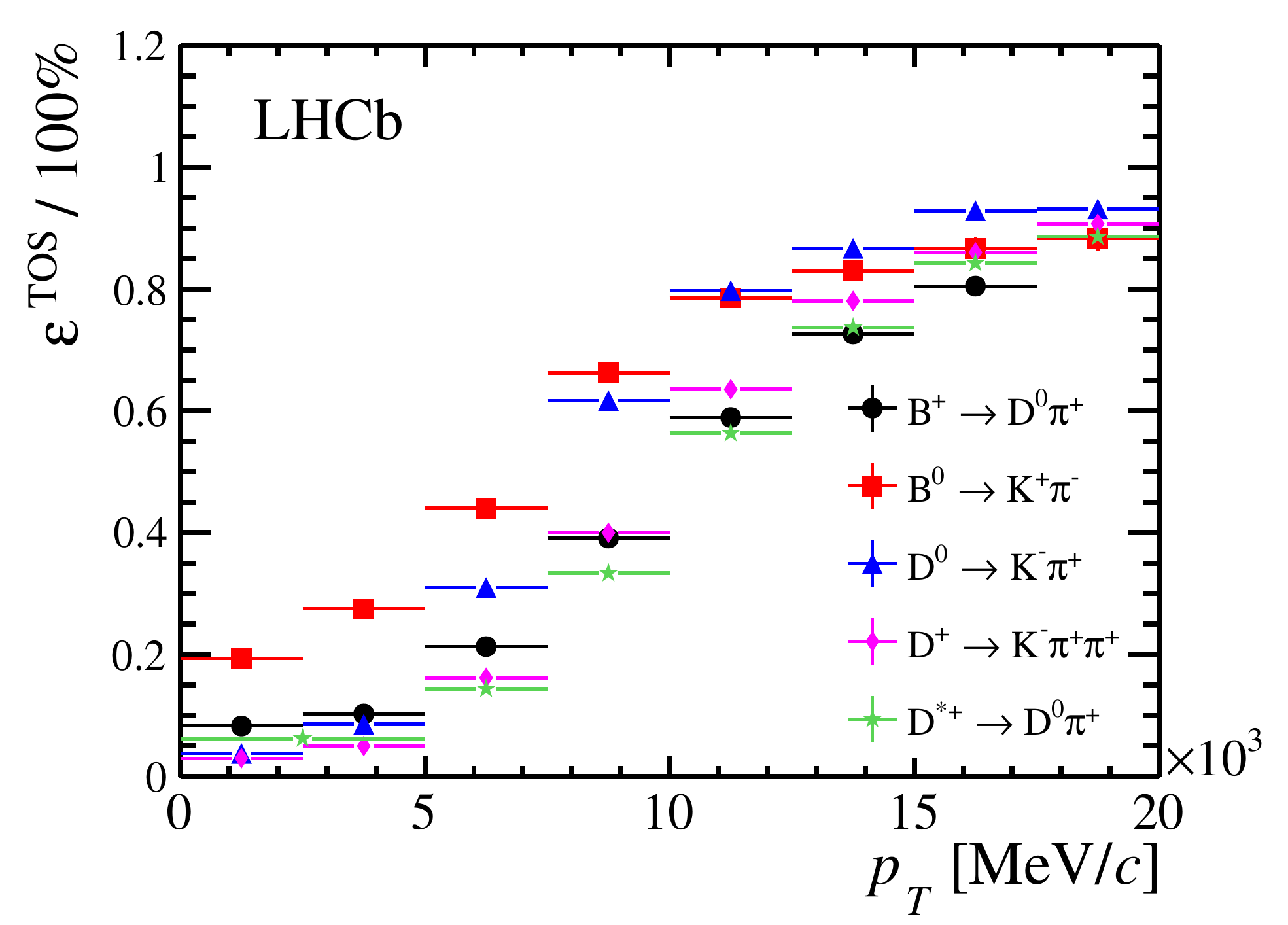}
\caption{The TOS efficiency of \texttt{L0\_Hadron} trigger for $\decay{\Bp}{\Dz\pip}$, $\decay{\Bz}{\Kp\pim}$, $\decay{\Dz}{\Kp\pim}$, $\decay{\Dp}{\Km\pip\pip}$ and $\decay{\D^{*+}}{\Dz\pip}$ decays.} 
\label{fig:l0Had}
\end{figure}

The \texttt{L0\_Hadron} trigger is used in the analyses described in Chapters~\ref{ch:ppbar} and~\ref{ch:phiphi}.
The efficiency of \texttt{L0\_Hadron} trigger for several hadronic $B$- and $D$-meson decay modes is shown on Fig.~\ref{fig:l0Had}. As expected, the efficiency is increasing with the transverse momentum.

\subsection{Software trigger and stripping} 
\subsubsection{\hltone}
The event passing the decision of any of the L0 triggers, it is transferred to the Event Filter Farm, which is used to execute the \hltone applications.
At the level of \hltone, the track and vertex reconstruction is performed without a full event reconstruction.
The idea of \hltone is to find tracks with high $p$ and \pt and to search for vertices with a reasonable fit quality. In addition, the tracks with large IP significance are selected to identify \bquark- and \cquark-decay candidates.
Besides, the track and vertex reconstruction, a fast muon PID is performed at the \hltone level. This is done by extrapolating VELO tracks to muon stations and comparing the signal from muon detector. The information from TT is used for better determination of track \pt. The track reconstruction is optimised in order to achieve a fast execution.

Charged hadron ID using \rich detectors requires a reconstruction of the rings associated with tracks. The timing of the rings reconstruction is similar to that needed for tracking. However, fast algorithms of rings reconstruction are available. Unfortunately, the PID using \rich information is not performed at the \hltone trigger level. The studies of any prompt decays to hadrons would benefit from the PID at the \hltone level with a dramatic decrease of the \hltone bandwidth. Due to limited timing of the \hltone, the PID would require the \rich calibration constants to be available online, which is not the case for the existing implementation of the \hltone trigger.

A similar situation takes place for the reconstruction of the downstream tracks. The studies of long-lived particles and long-lived baryons such as $\Lambda$ and $\Xi$ would benefit from the downstream track reconstruction with increased total efficiency.

There are two \hltone trigger lines dedicated to prompt charmonium studies using decays to hadrons. 
The \texttt{Hlt1DiProtonDecision} aims at selecting prompt and non-prompt \ppbar pairs. 
This trigger line selects hadron tracks with large \pt, which form a good quality vertex. The \pt of the proton-antiproton system is required to be larger that 6.5~\gev. More details about the \texttt{Hlt1DiProton} requirements are given in Chapter~\ref{ch:ppbar}. Initially, this trigger line was created to study prompt $\etac(1S)$ mesons. For the data taking in 2018, I suggested the line splitting into two lines
\texttt{Hlt1DiProtonLow} covering \ppbar invariant mass range of 2.8-3.3~\gev and \texttt{Hlt1DiProtonHigh} covering \ppbar invariant mass range of 2.8-3.3~\gev. Besides, the requirement on the minimum transverse momentum of the proton-antiproton system was reduced to 5.5~\gev. This change was implemented in order to search for prompt \etactwos mesons with a possible prompt production measurement predicted by theorists. The total rate of \texttt{Hlt1DiProton} lines in 2018 was about 10 $kHz$.

Another trigger line prepared for 2018 data taking is devoted to prompt charmonium decays to \phiphi (\texttt{Hlt1Ccbar2PhiPhi}). This line selects four hadrons with a large \pt. 
Pairs of hadrons with a kaon mass hypothesis are restricted to have a good vertex with an invariant mass compatible with that of $\phi(1020)$ meson. The $\phi\phi$ system is also required to form a good quality common vertex. The effective requirement on the $\phi\phi$ system \pt is about $\pt>5~\gev$. Following the observation of the $\decay{\etactwos}{\phi\phi}$ described in Chapter~\ref{ch:phiphi}, this line is also designed for studies of prompt charmonium decays to $\phi\phi$. The \etac, \etactwos and \chiczero states are targeted.
The total rate of \texttt{Hlt1Ccbar2PhiPhi} lines in 2018 was about 1 $kHz$.

The \texttt{Hlt1TrackAllL0} trigger line is meant to be “universal” for most analyses of \bquark-hadron decays. The outcome of this trigger line is registered for all L0 trigger decisions. The selection requires to have a track with IP larger than 0.1~\mm and transverse momentum larger than $1.6~\gev$. In the \lhcb Run II this line was split into two: \texttt{Hlt1OneTrackMVA} and \texttt{Hlt1TwoTrackMVA}, where a multivariate classifier was used in order to distinguish events with one and two displaced tracks, respectively.

\subsubsection{\hlttwo}
The offline trigger \hlttwo performs a more complete event reconstruction. 
Events passing the \hltone decision are stored in the buffer for further execution of the \hlttwo algorithms execution. Information from all detector systems available at the \hlttwo level.

At this stage specialised trigger selections for a number of inclusive and exclusive final states are applied. They are meant to include all types of events of interest for \lhcb.  Events passing the \hlttwo decision are then stored.

The \hlttwo uses more accurate track and vertex reconstruction compared to the one of \hltone also using information from online alignment and detector calibration. The selections applied at the \hlttwo level are more complex and are targetting specifical cases.
The deferred trigger, developed for RunII, allows HLT an overcommitment of 20-30\%.
Using a 1 PB storage at the farm, the deferred trigger then runs between the LHC fills. Using a deferred trigger made it possible to lower track reconstruction thresholds.
The \hlttwo in the configuration applied in Run I writes about 5 kHz to the storage, including about 2 kHz of inclusive \bquark-hadron candidates, about 2 kHz of inclusive charm candidates and about 1 kHz decay signatures with muons. 

Trigger lines selecting prompt \ppbar and $\phi\phi$ decays of charmonia are applied similarly as for \hltone. In addition to kinematic restrictions, the PID using \rich information for protons and kaons is required to reject specific decays producing a hadronic background. The trigger line selecting prompt \ppbar pairs will be used in the analysis described in Chapter~\ref{ch:ppbar}.

A complex universal \hlttwo trigger lines select multibody decays of \bquark-hadrons. These lines use the topology of event and are named \texttt{Hlt2Topo(2,3,4)BodyBBDT}. The selection relies on the presence of high-\pt displaced track, which contributed to a good quality vertex significantly displaced from PV. The remaining tracks of the vertex are required to have a large \pt sum.
Then a sample of simulated \bquark-hadron inclusive decays was used to traine the multivariate classification based on a Bonsai Boosted Decision Tree~\cite{Gligorov:2012qt}. The most powerful variable used in the classification is the corrected mass defined as $M_{corr}=\sqrt{M^2+\pt_{miss}^2}+\pt_{miss}$, where $M$ is an invariant corresponding to a vertex and $\pt_{miss}^2$ is an observed missing momentum due to mismatching between the direction of reconstructed momentum of decaying particle and the direction defined by PV and decay vertex.
In the thesis, this \hlttwo trigger line is used for a precision $\etac(1S)$ mass determination discussed in Chapter~\ref{ch:mass} and also in the analysis described in Chapter~\ref{ch:phiphi}.

\subsubsection{Data processing and stripping}
The total amount of raw data recorded by \lhcb corresponds to approximately 1 Tb/s, which is impossible to store with available computing resources and techniques. Therefore, raw events passing trigger requirements are stored in terms of reconstructed charged and neutral tracks, PID information in Data Summary Tape (DST) files. The obtained DST files are stored and reduced into reduced DST (rDST) files by eliminating unnecessary information from the event. 
The data stored in rDST format allows to measure momentum of tracks, positions of PV and decay vertices, etc.
The rDST files are then reprocessed using a set of preselection criteria (Stripping lines), which further reduces the amount of data. In order to take into account correlations between different stripping lines, during the stripping, the data is grouped into several streams according to the event topology, final states, PID, etc. The stripped data is then saved and replicated at special GRID storages available for \lhcb users. 
In \lhcb Run II, a part of the data passing \hlttwo decision is stored directly without offline stripping procedure to the disk into so-called Turbo streams.

\begin{singlespace}
\chapter{Study of $\etac(1S)$ production using its decay to \ppbar at \sqs=13 TeV}
\label{ch:ppbar}
\end{singlespace}
The pioneering \lhcb measurement of the \etac at \sqs=7 and 8~\tev has been performed using the \ppbar channel.
Due to the challenging background level and limited trigger bandwidth, the precision of the measurement and available fiducial region were limited. Despite that, the obtained \etac production measurement is a perfect example of how even imprecise measurement can challenge the theory.
Besides the physical result, it demonstrated the accessibility of the promptly and non-promptly produced \etac mesons by the \lhcb experiment. 
To date, the \ppbar decay channel is the most popular for studies requiring the \etac reconstruction at \lhcb. 
This chapter documents the measurement of the \etac production cross-section in \proton\proton collisions at \sqs=13~\tev using the \decay{\etac}{\ppbar} decay. 
The analysis validates the \etac production measurement at \sqs=7 and 8~\tev and yields the first \etac prompt production measurement at \sqs=13~\tev. The obtained result is more precise than the one obtained at \sqs=7 and 8~\tev.
At this point, the measurements of the \etac production using its decay to \ppbar at \lhcb remain the only available \etac production measurements at hadron colliders.

After the analysis setup introduced in Section~\ref{sec:xsecDetermination}, the data and simulation samples are discussed in Section~\ref{sec:data}. Selection criteria and signal efficiencies are addressed in Section~\ref{sec:select}. The two analysis techniques are described in Sections~\ref{sec:signal_extraction} and~\ref{sec:signal_extraction_runI}. The results are discussed in Section~\ref{sec:results}.
\newpage
\section{Analysis setup}
\protect\label{sec:xsecDetermination}
For a measurement of the \etac prompt production, 
a pseudo-proper lifetime is calculated and modeled similarly to the analysis
described in Ref.~\cite{LHCb-PAPER-2015-037} in order to distinguish between prompt production and \bquark-decays production (\tzfit).
The production of the \etac in $\bquark$-hadron decays is performed using \tzcut used in Ref.~\cite{LHCb-PAPER-2014-029}. The ratio of the \etac and \jpsi production cross-sections is measured in bins of \pt and then the \etac production is derived using measured \jpsi production at \lhcb~\cite{LHCb-PAPER-2015-037}.
The ratio of the \etac and \jpsi production can be expressed as
\begin{equation}
\begin{aligned}
\frac{\sigma^{prompt}_{\etac}}
{\sigma^{prompt}_{\jpsi}} &= \frac{N^{prompt}_{\etac}}{N^{prompt}_{\jpsi}}\times \frac{\epsilon_{\jpsi}}{\epsilon_{\etac}}\times \frac{\BR_{\JpsiToPpbar}}{\BR_{\EtacToPpbar}}, \\\
\frac{\sigma^{\bquark}_{\etac}}{\sigma^{\bquark}_{\jpsi}} = \frac{\BR_{\bToEtacX}}{\BR_{\bToJpsiX}} &= \frac{N^{\bquark}_{\etac}}{N^{\bquark}_{\jpsi}}\times \frac{\epsilon_{\jpsi}}{\epsilon_{\etac}}\times \frac{\BR_{\JpsiToPpbar}}{\BR_{\EtacToPpbar}},
\end{aligned}
\end{equation}
where 
$\sigma^{prompt}_{\etac(\jpsi)}$ is the prompt production cross-section of \etac(\jpsi), 
$\sigma^{\bquark}_{\etac(\jpsi)}$ is the production cross-section of \etac(\jpsi) in inclusive \bquark-decays, 
$\BR_{\bquark\to\etac(\jpsi)X}$ is the inclusive branching fraction of the \bquark-quark decay to \etac(\jpsi), 
$N^{prompt}_{\etac(\jpsi)}$ is the yield of prompt \etac(\jpsi) candidates, 
$N^{\bquark}_{\etac(\jpsi)}$ is the yield of \etac(\jpsi) from inclusive \bquark-decays, 
$\frac{\epsilon_{\jpsi}}{\epsilon_{\etac}}$ is the ratio of total efficiencies to reconstruct, trigger and select \JpsiToPpbar and \EtacToPpbar decays, 
$\BR_{\JpsiToPpbar}$ and $\BR_{\EtacToPpbar}$ are the branching fractions of the \JpsiToPpbar and \EtacToPpbar decays from Ref.~\cite{PDG2017}, respectively.

Using the measurement of \jpsi prompt differential production cross-section and the \jpsi production in inclusive \bquark-decays performed using the \decay{\jpsi}{\mu\mu} decay from Refs.~\cite{LHCb-PAPER-2015-037,PDG2017}, the \etac production can be extracted as
\begin{equation}
\begin{aligned}
\sigma^{prompt}_{\etac} &= \sigma^{prompt}_{\jpsi}\times \frac{N^{prompt}_{\etac}}{N^{prompt}_{\jpsi}}\times \frac{\epsilon_{\jpsi}}{\epsilon_{\etac}}\times \frac{\BR_{\JpsiToPpbar}}{\BR_{\EtacToPpbar}}, \\\
\sigma^{\bquark}_{\etac} &= \sigma^{\bquark-decays}_{\jpsi}\times \frac{N^{\bquark}_{\etac}}{N^{\bquark}_{\jpsi}}\times \frac{\epsilon_{\jpsi}}{\epsilon_{\etac}}\times \frac{\BR_{\JpsiToPpbar}}{\BR_{\EtacToPpbar}}, \\\
\BR_{\bToEtacX} &= \BR_{\bToJpsiX} \times \frac{N^{\bquark}_{\etac}}{N^{\bquark}_{\jpsi}}\times \frac{\epsilon_{\jpsi}}{\epsilon_{\etac}}\times \frac{\BR_{\JpsiToPpbar}}{\BR_{\EtacToPpbar}}.
\end{aligned}
\end{equation}
Both prompt production and production in \bquark-decays are measured using two different techniques: \tzfit and \tzcut.
The ratio of the signal event yield for the \etac prompt production measurement is quoted using \tzfit technique, while the \tzcut is used for a cross-check. The \etac production in inclusive \bquark-hadron decays is quoted by \tzcut and \tzfit is used for a cross-check.

\section{Data sample, trigger and simulation} 
\protect\label{sec:data}
This analysis uses the $\proton\proton$ collision data recorded by the LHCb experiment at $\sqs=13$ TeV with an integrated luminosity $\int\lum dt \approx 2.0 \invfb$ accumulated in 2015 and 2016. All detector subsystems were stable and fully operational during the data taking period corresponding to the present analysis.
For data processing, the reconstruction version Reco15a (Reco16), and stripping version Stripping24 (Stripping28) were used for 2015 (2016) data. 

The basic level L0 Hadron (\texttt{L0HadronDecision$\_$TOS}) trigger is applied.
The candidates are required to be selected (\texttt{TOS}) by dedicated trigger lines of HLT1 and HLT2, \texttt{Hlt1DiProtonDecision$\_$TOS}, \texttt{Hlt2DiProtonDecision$\_$TOS} (\texttt{Hlt2CcDiHadronDiProtonDecision$\_$TOS}) are used for the analysis of charmonium production for both 2015 (2016) data.

For the \etac mass measurement, a low-background data sample with larger statistics selecting $b\to \etac$X is used.
In the data sample, the basic level \texttt{L0 Hadron decision} (L0HadronDecision$\_$TOS) trigger is applied. The trigger lines \texttt{TOS} of HLT1, \texttt{Hlt1(Two)TrackMVADecision$\_$TOS}, and HLT2, \texttt{Hlt2Topo(2,3,4)BodyDecision$\_$TOS} are used.


The Monte Carlo (MC) samples used to study the \etac and the \jpsi mass resolution, 
as well as the background contribution from the $\JpsiToPpbarPiz$ channel are summarised in the Table~\ref{tab:MC}. In the simulation, \proton\proton collisions are generated using
\pythia~\cite{Sjostrand:2006za, Sjostrand:2007gs} 
 with a specific \lhcb configuration~\cite{LHCb-PROC-2010-056}.  Decays of hadronic particles
are described by \evtgen~\cite{Lange:2001uf}, in which final-state
radiation is generated using \photos~\cite{Golonka:2005pn}. The
interaction of the generated particles with the detector, and its response,
are implemented using the \geant
toolkit~\cite{Allison:2006ve, Agostinelli:2002hh} as described in
Ref.~\cite{LHCb-PROC-2011-006}.
\begin{table}[ht] 
\begin{center}
{\small{
\centering
\begin{tabular}{l|l} 
 Sample                       & Sample size     \\ \hline \hline
prompt \etac                  & 2015: 0.62 M    \\
                              & 2016: 2.40 M    \\ \hline
\etac from-\bquark            & 2015: 0.26 M    \\
                              & 2016: 1.01 M    \\ \hline
prompt \jpsi                  & 2015: 0.67 M    \\
                              & 2016: 2.41 M    \\ \hline
\jpsi from-\bquark            & 2015: 0.19 M    \\
                              & 2016: 0.60 M    \\ \hline
$\JpsiToPpbarPiz$             & 2015: 0.80 M    \\
                              & 2016: 3.01 M     
\end{tabular}
}}
\protect\caption{Simulated samples.}
\protect\label{tab:MC}
\end{center}
\end{table}
For all simulation samples a phase-space decay model is used, the daughter proton and antiproton are required to flight into detector's acceptance and to have transverse momentum of $\pt(\proton)>0.9 \gev$ to speed-up MC production. For MC samples of prompt \jpsi, \jpsi from \bquark-decays and $\JpsiToPpbarPiz$, the \jpsi meson was generated without polarisation. The prompt \etac mesons are generated as \jpsi with modified mass and width according to known values from Ref.~\cite{PDG2017}. The latter is done in order to optimize MC samples generations since generation of promptly produced \etac mesons is much slower compared to that of \jpsi.
For all MC samples reconstructed signal candidates and their daughter particles are required to match the generated ones.

\section{Event selection} 
\protect\label{sec:select}
Due to large number of random \ppbar combinations originated from PV, the background conditions and the limited trigger bandwidth complicate the analysis. 
In order to achieve a tolerable trigger rate, strongly selective requirements, including proton identification, are applied already at the trigger level. 

The \etac and \jpsi candidates are reconstructed from a pair of oppositely charged tracks identified as protons by the \lhcb detector. Both proton track candidates are required to have a good quality of track reconstruction, $\chisqndf<2.5$ and probability that track consists of random hit combinations (ghost probability) less than 0.2. In order to suppress combinatorial background and reduce the trigger bandwidth, the proton tracks are required to have transverse momenta larger than 1.9~\gev and momenta larger than 12.5~\gev. The distance of closest approach between two tracks is required to be less than 0.1\mm.
The transverse momentum of the proton-antiproton system is required to be higher than 6.5 \gev, and charmonium candidate vertex quality $\chisqndf < 4$. Trigger specifically rejects high multiplicity events, causing excessively high combinatorial background, by requiring the SPD multiplicity to be less than 300.

A sample enriched in true protons have to be selected already at the trigger level.
For that, the information from \rich detectors is used at the trigger level to separate protons from pions and kaons. 
The proton identification requirements $\Delta\log\mathcal{L}^{p-K}>10$ and $\Delta\log\mathcal{L}^{p-\pi}>20$ are used at the level of the \hlttwo.  

Further selection performed by a dedicated stripping line (\texttt{StrippingCcbar2PpbarLineDecision}) applies almost the same requirements as in the trigger selection. The only exception is a more tight PID requirement of $\Delta\log\mathcal{L}^{p-K}>15$. Trigger settings of dedicated \texttt{Hlt1DiProton} line were tightened before 2016 data taking. When combining 2015 and 2016 data samples, more tight cuts, matching trigger requirements from 2016 settings, are used in the offline selection. Additional cuts are applied to cut off tails of distributions created by mismatching between HLT1 and HLT2 requirements to avoid unnecessary edge effects.

The set of selection criteria used in the trigger, stripping and offline selection are almost identical, as illustrated in Table~\ref{tab:triggerCuts}.
\begin{table}[t] 
{\small{
\centering
\begin{tabular}{l|l|l|l|l|l} 
          & 	Variable                     & \hltone            & \hlttwo         & Stripping     & Offline  \\
          & 	                             &                 &              &               & selection\\ \hline
Trigger   &   & L0Hadron   & Hlt1DiProton  & $-$ & \footnotesize{L0Hadron$\_$TOS} \\
          &   &    &   &                         & \footnotesize{Hlt1DiProton$\_$TOS} \\
          &   &    &   &                         & \footnotesize{Hlt2DiProton$\_$TOS} \\ \hline
Protons	  & \pt, \gev                     & $>1.9$   & $>1.9$    & $>1.95$& $>2.0$ \\
  		    & \ptot, \gev                   & $>12.5$  &           & $>10.0$& $>12.5$\\
  		    & $\pt/\ptot$                    & $>0.0366$&           &        & $>0.0366$ \\
  		    & Track $\chi^{2}/\mathrm{NDF}$  & $<2.5$   & $<3.0$    & $<4.0$ & $<2.5$ \\
   		    & Ghost probability              & $<0.2$         &           &        & $<0.2$ \\
     		  & $\Delta\log\mathcal{L}^{p-\pi}$& $-$      & $>20$     &$>20$   & $>20$  \\
     		  & $\Delta\log\mathcal{L}^{p-K}$  & $-$      & $>10$     &$>15$   & $>15$  \\ \hline

\ppbar& \pt, \gev                     & $>6.5$   & $>6.5$    & $>6.0$      & $>6.5$ \\ 
  		    & Vertex $\chisqndf$             & $<4$     & $<9$      &         & $<4.0$ \\ 
  		    & Vertex DOCA, \mm               & $<0.1$   &           &         & $<0.1$\\ 
  		    & Mass, \gevc                   & $2.8-3.3$& $2.8-4.0$ &$2.8-4.0$& $2.85-3.25$ \\ \hline
          & SPD multiplicity           & $<300$   & $<300$    &$<300$   & $<300$ 
\end{tabular}
}}
\protect\caption{Trigger, stripping and offline selection criteria.}
\protect\label{tab:triggerCuts}
\end{table}

\begin{figure}[b]
\centering{
        \subfigure[The efficiency ratio for prompt production.]{ 
      \protect\protect\protect\includegraphics[width=0.465\textwidth]{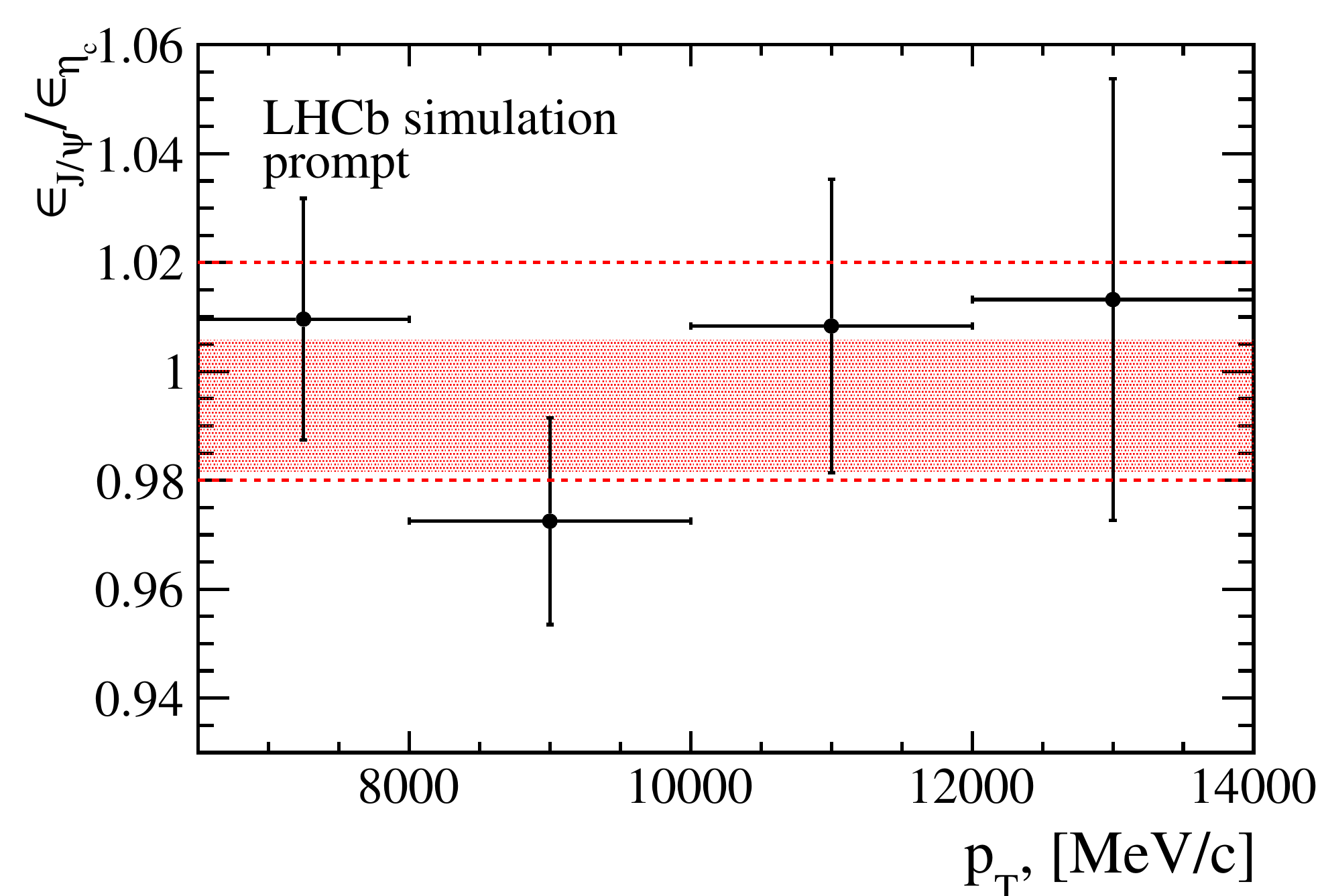}
      \put(-214,95){\rotatebox{90}{\colorbox{shadecolor}{$\epsilon_{\jpsi}/\epsilon_{\etac}$}}}
      \put(-57,3){\colorbox{shadecolor}{\pt, \mev}}
      \protect\label{fig:effRatioPrompt}
      }
\quad
        \subfigure[The efficiency ratio for production in \bquark-decays.]{ 
       \protect\protect\protect\includegraphics[width=0.465\textwidth]{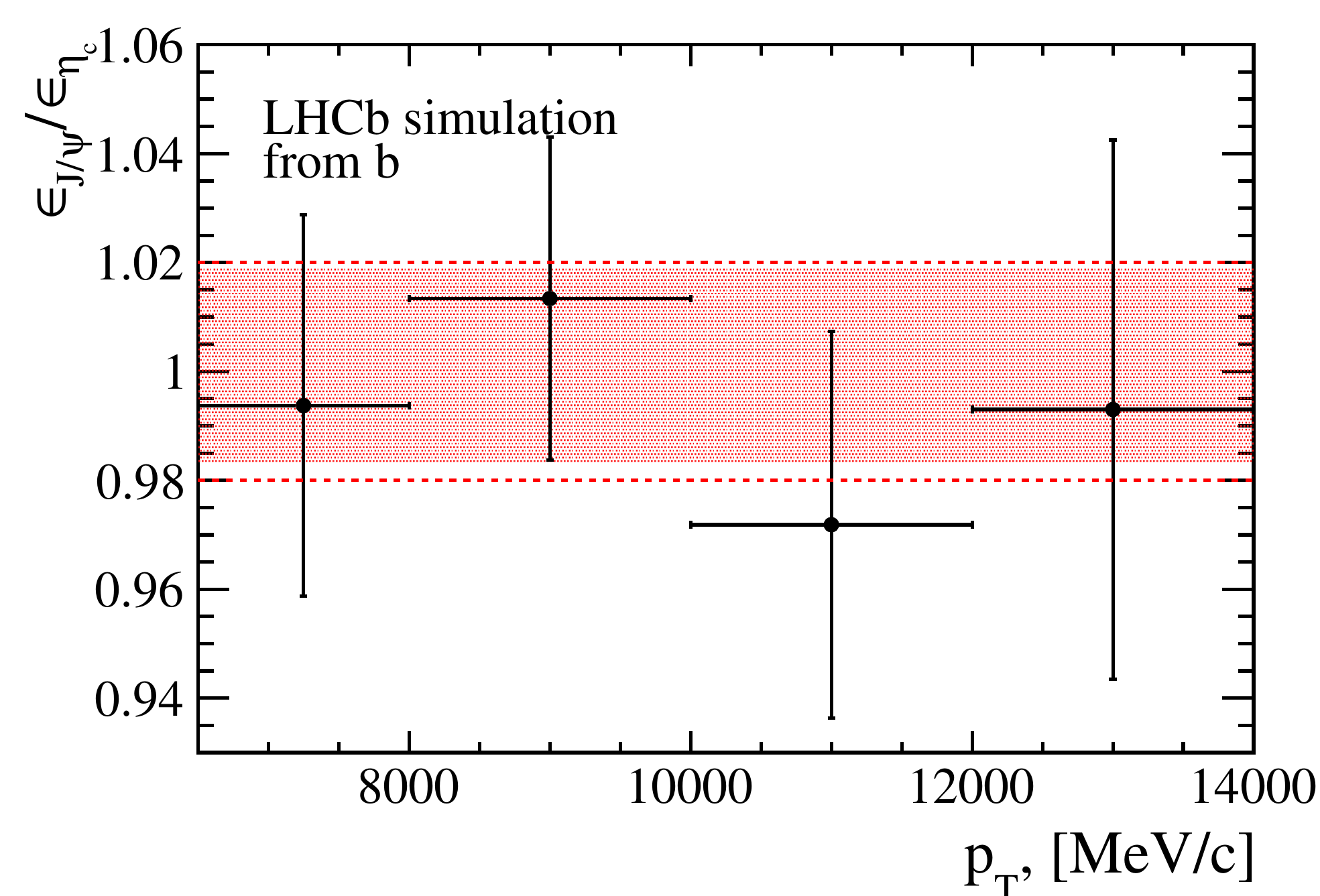}
      \put(-214,95){\rotatebox{90}{\colorbox{shadecolor}{$\epsilon_{\jpsi}/\epsilon_{\etac}$}}}
      \put(-57,3){\colorbox{shadecolor}{\pt, \mev}}
      \protect\label{fig:effRatioFromB}
      }
 }
\protect\caption
[The \jpsi to \etac total efficiency ratio in bins of \pt.]
{The \jpsi to \etac total efficiency ratio in bins of \pt. Red boxes show the total efficiency ratio. Red dashed lines illustrate the efficiency ratio uncertainty considered in the analysis.} 
\protect\label{fig:effRatio}
\end{figure}

Since the masses of \etac and \jpsi states are close to each other and kinematic distributions in \JpsiToPpbar and \EtacToPpbar decays are similar, one expects similar reconstruction, trigger and stripping efficiencies.
The efficiency ratio of \JpsiToPpbar and \EtacToPpbar is determined using simulation samples to be
\begin{equation}
  \frac{\epsilon_{\jpsi}}{\epsilon_{\etac}} = 1.00\pm0.02,
\end{equation}
where the uncertainty is due to MC sample sizes. Note that, uncertainty on efficiency ratio gives a negligible contribution to a total systematic uncertainty (see Sections~\ref{sec:syst} and~\ref{sec:fitRunI}). Effect of the \jpsi meson polarisation is taken into account in the evaluation of systematic uncertainty.
The efficiency ratio is also extracted in bins of \pt with no significant deviation from unity observed. No significant difference is found between efficiencies of prompt charmonia production and charmonia production in inclusive \bquark-decays. The efficiency ratio for prompt and for inclusive \bquark-decays production in bins of \pt is shown on Fig.\ref{fig:effRatio}.

The PID efficiency ratio for \etac and \jpsi has been cross-checked by applying tighter PID cuts on MC. For this cross-check the selection requrements 
$[PIDp>25\:\&\:(PIDp-PIDK)>15]$, $[PIDp>20\:\&\:(PIDp-PIDK)>20]$ and 
$[PIDp>25\:\&\:(PIDp-PIDK)>20]$ are used. 
Comparison of efficiency ratios for different PID selection requirements is shown in Table~\ref{tab:PIDCutsCheck}. No significant effect is observed, and the result is considered to be stable against PID requirement variations.
\begin{table}[h]
\centering
\small
\begin{tabular}{cc|c}
         & PID requirement & $\epsilon_{\jpsi}/\epsilon_{\etac}$ \\ \hline 
(nominal)& $PIDp>20\:\&\:(PIDp-PIDK)>15$ & $1.00\pm0.02$ \\  \hline
         & $PIDp>25\:\&\:(PIDp-PIDK)>15$ & $0.99\pm0.02$ \\  
         & $PIDp>20\:\&\:(PIDp-PIDK)>20$ & $1.00\pm0.02$ \\
         & $PIDp>25\:\&\:(PIDp-PIDK)>20$ & $0.99\pm0.02$
\end{tabular} 
\protect\caption{The \protect\jpsi and \protect\etac efficiency ratio from MC for different PID requirements.}
\protect\label{tab:PIDCutsCheck}
\end{table} 

Another cross-check of PID efficiency is done by estimating PID efficiency ratio using the PID calibration samples within \textit{PIDCalib} package. The PID efficiency map is extracted using calibration samples of \decay{\Lambda}{\proton\pim} for \lhcb Run II data and then applied to the \etac and \jpsi MC samples. The extracted PID efficiency ratio is compared to PID efficiency ratio extracted from MC samples as
\begin{equation*}
\protect\label{eq:pidcalib}
\begin{aligned}
\epsilon_{\jpsi}^{PID,PIDCalib}/\epsilon_{\etac}^{PID,PIDCalib} &= 0.98\pm0.04, \\
\epsilon_{\jpsi}^{PID,MC}/\epsilon_{\etac}^{PID,MC}             &= 0.99\pm0.04
\end{aligned}
\end{equation*}
No significant difference of \jpsi and \etac efficiency ratio from unity is observed for both performed cross-checks.

Below, two different techniques are employed to measure the \etac production cross-section.

\clearpage
\section{$t_z$-fit technique}
\label{sec:signal_extraction}
In order to distinguish between promptly produced charmonium candidates 
and charmonium candidates from \bquark-hadron decays, the yields of \jpsi and \etac are extracted in bins of pseudo-proper lifetime $t_z$. The $t_z$ value is defined as
 \begin{equation}
    t_z = \frac{(z_{d} - z_{p})M_{\ppbar}}{p_{z}},
 \end{equation}
where $z_{p}$ and $z_{d}$ are the $z$-coordinates of PV and charmonium candidate decay vertices, respectively, 
$M_{\ppbar}$ is the reconstructed charmonium mass and ${p_{z}}$ is the longitudinal component of its momentum. 

The yields of \jpsi and \etac candidates are determined from simultaneous extended binned  maximum-likelihood fit to the $M(\ppbar)$ distribution. 
Fit of the invariant mass is performed simultaneously in 28 bins of [\pt; $t_z$]. 
The bin edges of charmonium \pt are [6.5, 8.0, 10.0, 12.0, 14.0] expressed in \gev and the $t_z$ bin edges are [-10.0, -0.125, -0.025, 0., 0.2, 2., 4., 10.] expressed in \ps. In the simultaneous fit, the masses of \jpsi and \etac mesons and the resolution parameter, described below, are common free fit parameters throughout all 28 bins.
 
The extracted yields in bins [\pt; $t_z$] together with their statistical uncertainties are fitted to $t_z$ in 4 bins of \pt to distinguish promptly produced charmonia and charmonia produced in inclusive \bquark-decays.
For that the simultaneous integral \chisq fit was used, which finds the bin centre-of-mass according to the shape of the fit function. The latter is important for sharp functions as it is the case for the fit to $t_z$.
From  the fit to $t_z$ distribution, the ratios of prompt \etac and prompt \jpsi yields $\frac{N^{prompt}_{\etac}}{N^{prompt}_{\jpsi}}$, and \etac yields from \bquark-decays and \jpsi yields from \bquark-decays $\frac{N^{\bquark}_{\etac}}{N^{\bquark}_{\jpsi}}$, are extracted together with their uncertainties in four bins of charmonium \pt.

To measure total \etac production cross-section the same procedure is implemented with \pt integrated over the range $6.5 \gev<\pt<14.0 \gev$.

\subsection{Fit to the invariant mass}
\label{sec:massFit}
The signal shape is defined by the detector resolution and the natural width in the case of the \etac resonance. 
The detector resolution effect on invariant mass distribution is described by a double Gaussian function. 
Parameters of double Gaussian are extracted from simultaneous unbinned maximum-likelihood fit of four MC samples (prompt \etac, prompt \jpsi, \etac from \bquark-decays and \jpsi from \bquark-decays) to $M_{\ppbar} - M^{Gen}_{\ppbar}$, where $M_{\ppbar}$ is the reconstructed mass and $M^{Gen}_{\ppbar}$ is the generated mass. The resolution ratio for the \etac and \jpsi peaks is fixed to the value from simulation. Corresponding systematic uncertainties are estimated in section~\ref{sec:syst}.

\begin{figure}[ht]
\centering
\protect\protect\protect\includegraphics[width=0.95\linewidth]{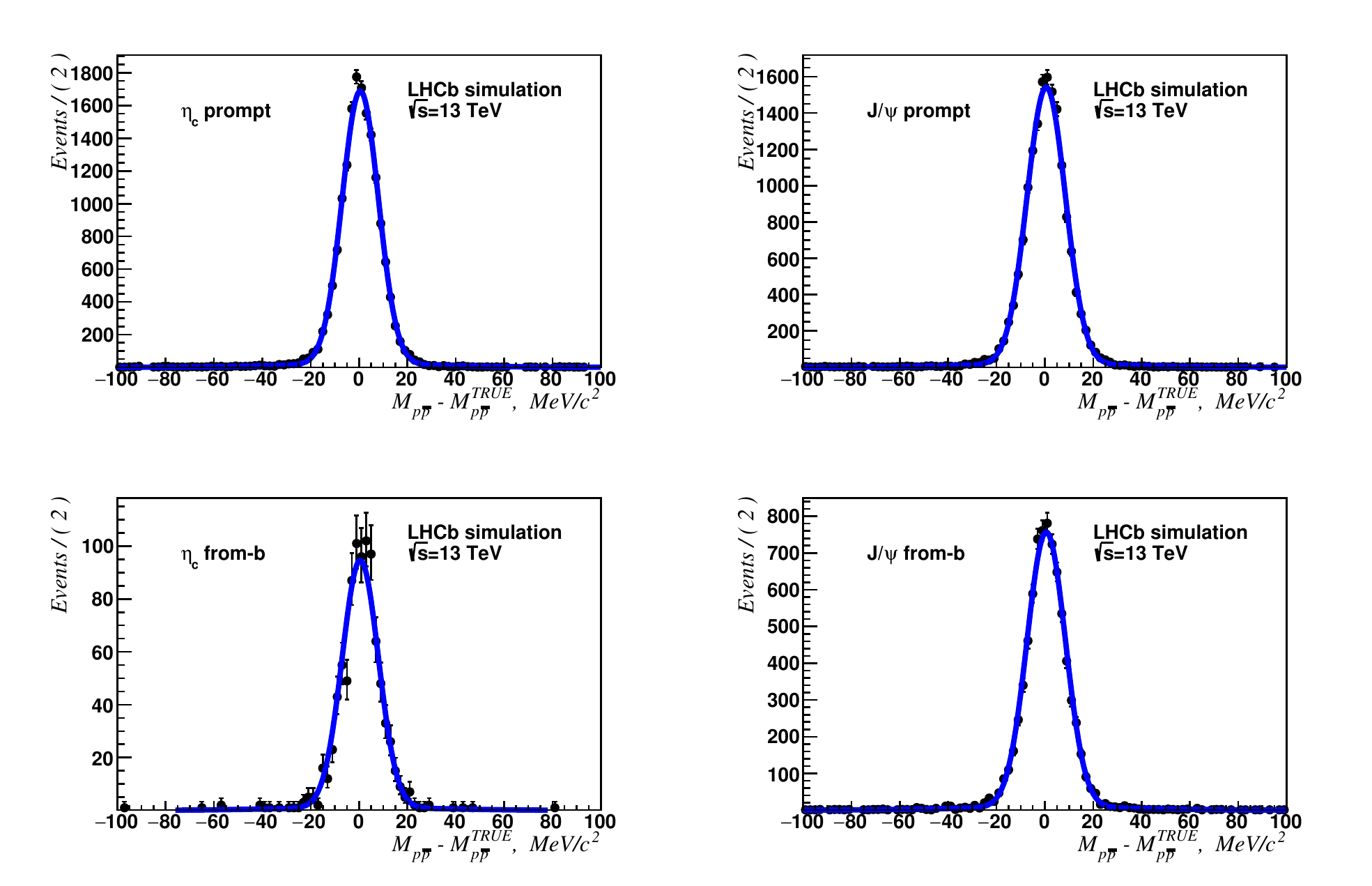}
\put(-420,185){\rotatebox{90}{\colorbox{shadecolor}{\scriptsize Candidates / 2 \mev}}}
\put(-315,146){\colorbox{shadecolor}{\scriptsize $M_{\ppbar} - M^{Gen}_{\ppbar}$, \mev}}
\put(-420,45){\rotatebox{90}{\colorbox{shadecolor}{\scriptsize Candidates / 2 \mev}}}
\put(-315,8){\colorbox{shadecolor}{\scriptsize $M_{\ppbar} - M^{Gen}_{\ppbar}$, \mev}}
\put(-205,185){\rotatebox{90}{\colorbox{shadecolor}{\scriptsize Candidates / 2 \mev}}}
\put(-100,146){\colorbox{shadecolor}{\scriptsize $M_{\ppbar} - M^{Gen}_{\ppbar}$, \mev}}
\put(-205,45){\rotatebox{90}{\colorbox{shadecolor}{\scriptsize Candidates / 2 \mev}}}
\put(-100,8){\colorbox{shadecolor}{\scriptsize $M_{\ppbar} - M^{Gen}_{\ppbar}$, \mev}}
\caption
[Distribution of the $M_{\ppbar} - M^{Gen}_{\ppbar}$ value in the MC samples.]
{Distribution of the $M_{\ppbar} - M^{Gen}_{\ppbar}$ value in the MC samples: prompt \etac (top left), prompt \jpsi (top right), \etac from \bquark-decays (bottom left) and \jpsi from \bquark-decays (bottom right). The solid blue lines represent a simultaneous fit by a double Gaussian function to all four MC samples.} 
\label{fig:MassResoMCFit}
\end{figure}

The $M_{\ppbar} - M^{Gen}_{\ppbar}$ distribution for all MC samples together with the fit curve are shown on the Fig.~\ref{fig:MassResoMCFit}. In the fit to MC samples the same resolution models for prompt \etac(\jpsi) and \etac(\jpsi) from \bquark-decays are used. The ratio of \jpsi and \etac resolutions is introduced as a ratio of \etac and \jpsi narrow gaussian widths $\sigma_n(\etac)/\sigma_n(\jpsi)$. The mean values of \etac and \jpsi double Gaussians are different independent fit parameters. The ratio of double Gaussian width $\sigma_n/\sigma_w$ and the fraction of narrow gaussian component $f_n$ are common fit parameters for all four MC samples. 
Simultaneous fit shows good description of $M_{\ppbar} - M^{Gen}_{\ppbar}$ for all samples. 

The fit yields the ratio of \jpsi and \etac resolutions to be $\sigma_{\etac}/\sigma_{\jpsi} = 0.94 \pm 0.07$, 
the ratio of the two Gaussian widths to be $\sigma_n / \sigma_w = 0.21 \pm 0.01$ 
and the fraction of the narrow Gaussian component $f_n$ to be about $95\%$. Note, that only the ratio of the \etac and \jpsi resolutions is taken from MC when fitting data, while the absolute values are constrained by the narrow and significant \jpsi peak when fit to data is performed.

\begin{figure}[ht]
\centering
\protect\protect\protect\includegraphics[width=0.95\linewidth]{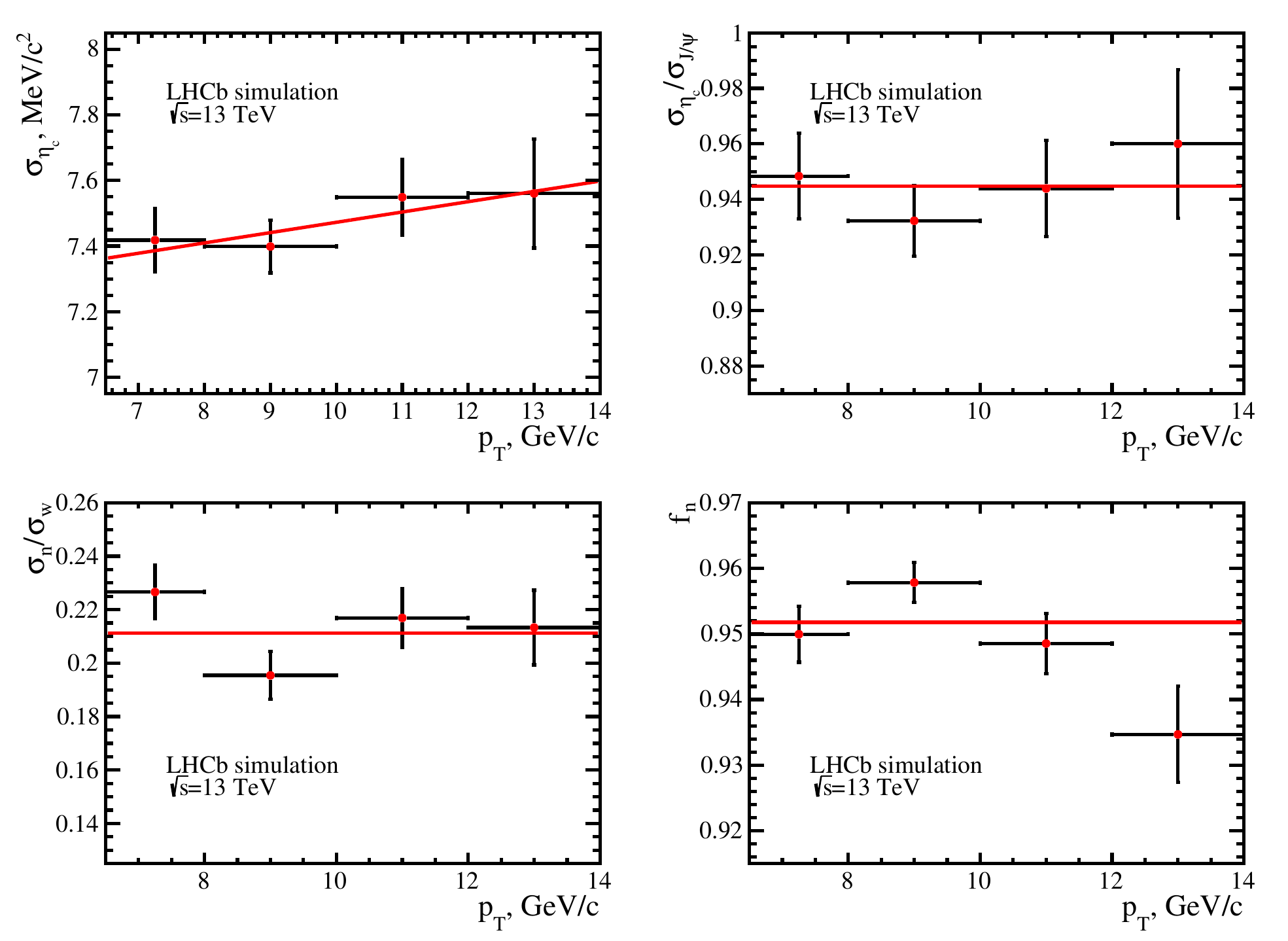}
\put(-427,255){\rotatebox{90}{\colorbox{shadecolor}{$\sigma_n$, \mev}}}
\put(-275,161){\colorbox{shadecolor}{\pt, \gev}}
\put(-430,113){\rotatebox{90}{\colorbox{shadecolor}{$\sigma_n/\sigma_w$}}}
\put(-275,5){\colorbox{shadecolor}{\pt, \gev}}
\put(-216,260){\rotatebox{90}{\colorbox{shadecolor}{$\sigma_{\etac}/\sigma_{\jpsi}$}}}
\put(-60,161){\colorbox{shadecolor}{\pt, \gev}}
\put(-215,133){\rotatebox{90}{\colorbox{shadecolor}{$f_n$}}}
\put(-60,5){\colorbox{shadecolor}{\pt, \gev}}
\caption
[The \pt dependences of double Gaussian parameters  of invariant mass resolution obtained from simultaneous fit to all four MC samples.]
{The \pt dependences of double Gaussian parameters  of invariant mass resolution obtained from simultaneous fit to all four MC samples. Red lines represent \pt-dependences used in the fit to data.} 
\label{fig:PT_dist_mass}
\end{figure}
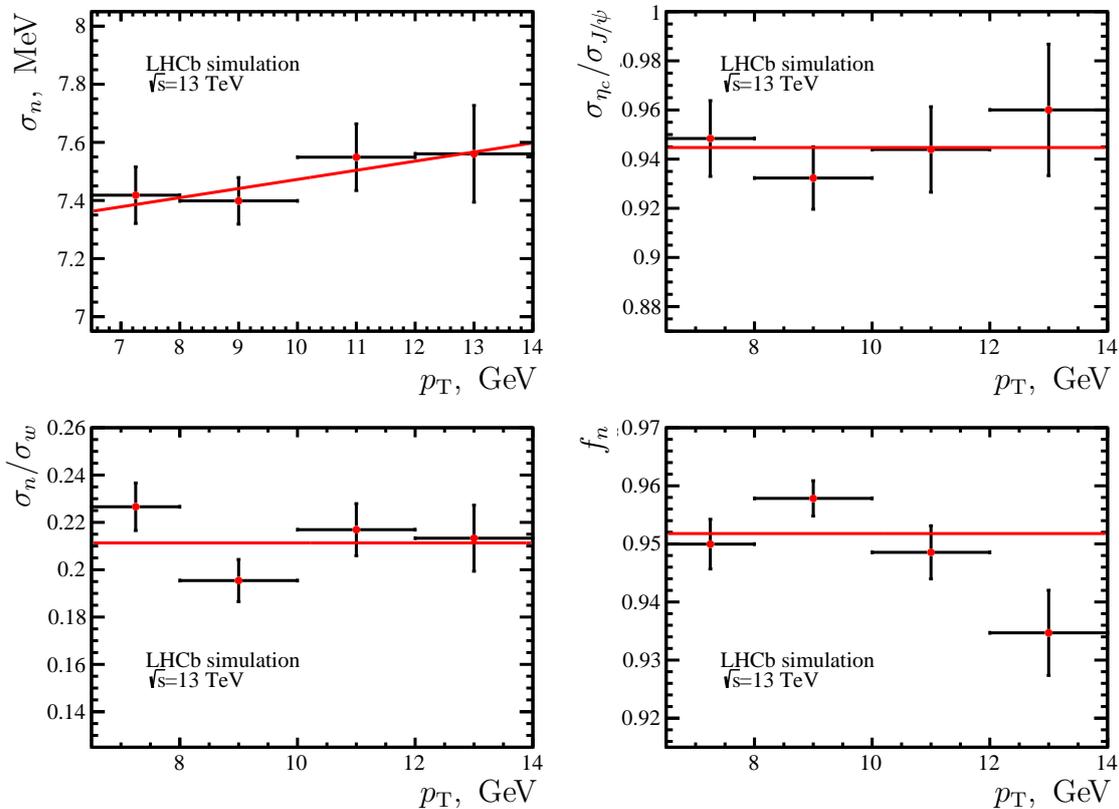

In order to study a dependence of the invariant mass resolution model as a function of charmonium transverse momentum, the same fit is performed in bins of \pt using simulation samples. The corresponding dependences of $\sigma_{\etac}/\sigma_{\jpsi}$, $f_n$, $\sigma_n / \sigma_w$ are shown on Fig.~\ref{fig:PT_dist_mass}. No significant \pt-dependence is observed for $\sigma_{\etac}/\sigma_{\jpsi}$, $f_n$ and $\sigma_n / \sigma_w$, hence no \pt-dependence is assumed in the nominal fit to data. The linear slope of \pt-dependence of $\sigma_n$ is extracted from simulation and is then used in the fit to data for differential production cross-section measurement. The value of the slope is extracted to be $a_{\sigma_n^{MC}}=(3.1\pm2.9)\times 10^{-2}$. The slope is cross-checked using \bToEtacX data sample; the fit to data yields the slope value of $a_{\sigma_n^{data}}=(9.4\pm5.6)\times 10^{-2}$ and is consistent with simulation. 

A dependence of reconstructed charmonium mass as a function of $t_z$ is considered. Figure~\ref{fig:massFitwoDeltaM} shows the curve of simultaneous invariant mass fit to data in seven $t_z$ bins for a \pt-integrated data sample. In the fit model on Fig.~\ref{fig:massFitwoDeltaM}, peak positions are assumed to be the same in all $t_z$ bins. Pull distributions show clear shifts of peak positions in several bins of $t_z$. The most notable shifts are observed in the second and fourth $t_z$ bins.

\afterpage{%
\begin{figure}[t]
\centering
\protect\protect\protect\includegraphics[angle=90, width=0.6\linewidth]{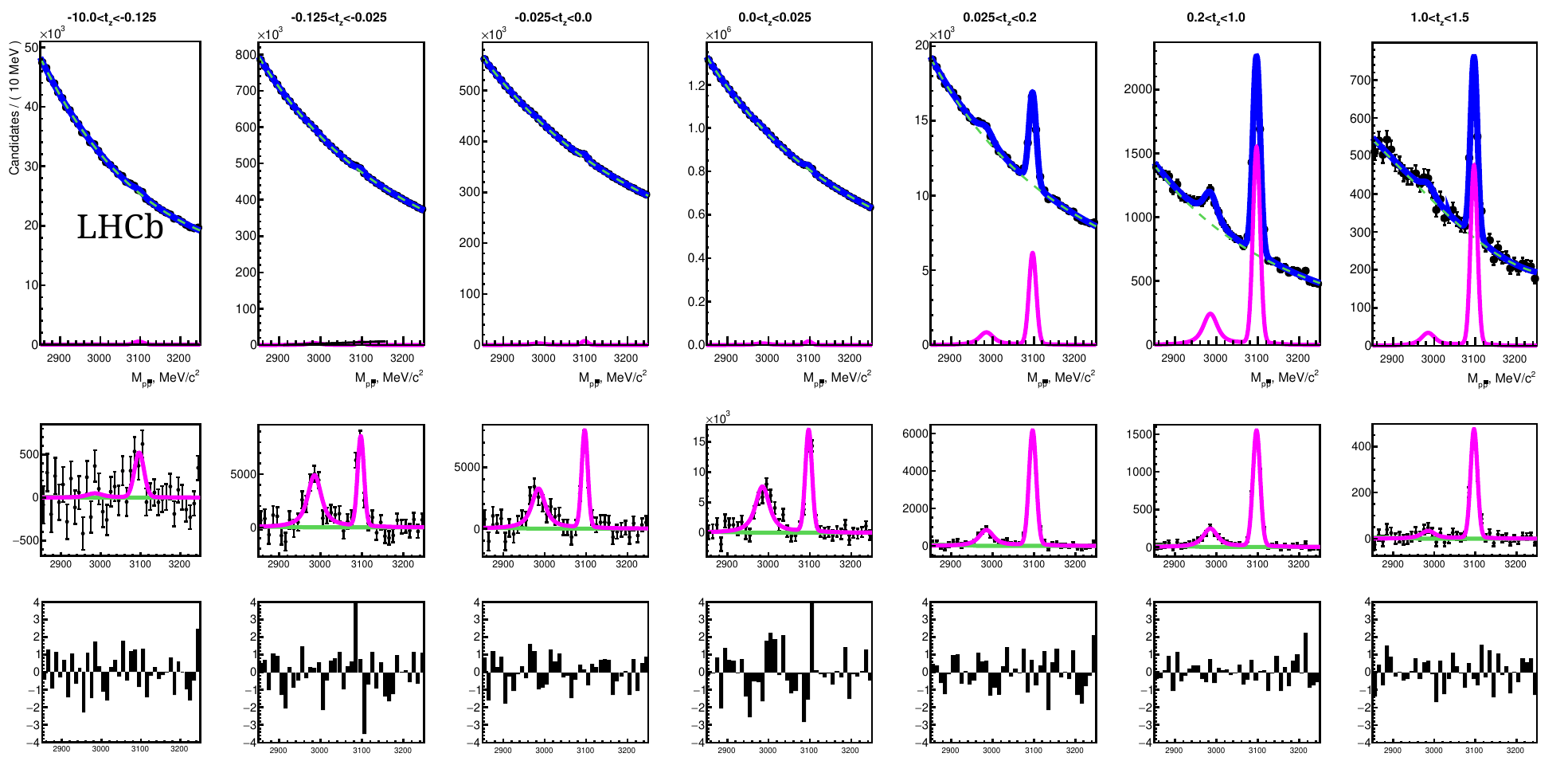}
\put(-180,20){\rotatebox{90}{\small\textbf{-ANA-2018-035}}}
\put(-142,30){\rotatebox{90}{\colorbox{shadecolor}{\scriptsize$M_{\ppbar}, \mev$}}}
\put(-142,110){\rotatebox{90}{\colorbox{shadecolor}{\scriptsize$M_{\ppbar}, \mev$}}}
\put(-142,190){\rotatebox{90}{\colorbox{shadecolor}{\scriptsize$M_{\ppbar}, \mev$}}}
\put(-142,270){\rotatebox{90}{\colorbox{shadecolor}{\scriptsize$M_{\ppbar}, \mev$}}}
\put(-142,350){\rotatebox{90}{\colorbox{shadecolor}{\scriptsize$M_{\ppbar}, \mev$}}}
\put(-142,430){\rotatebox{90}{\colorbox{shadecolor}{\scriptsize$M_{\ppbar}, \mev$}}}
\put(-142,510){\rotatebox{90}{\colorbox{shadecolor}{\scriptsize$M_{\ppbar}, \mev$}}}
\caption
[The $M_{\ppbar}$ distribution for seven bins of $t_z$ in the \pt-integrated sample ($6.5 \gev<\pt<14.0 \gev$). The solid blue lines represent the total fit result assuming the same peak positions in $t_z$ bins.]
{The $M_{\ppbar}$ distribution for seven bins of $t_z$ in the \pt-integrated sample ($6.5 \gev<\pt<14.0 \gev$). The solid blue lines represent the total fit result \textbf{assuming the same peak positions in $t_z$ bins}. Magenta and green lines show the signal and background components, respectively. The corresponding residual and pull distributions are shown on second and third line plots.} 
\label{fig:massFitwoDeltaM}
\end{figure}
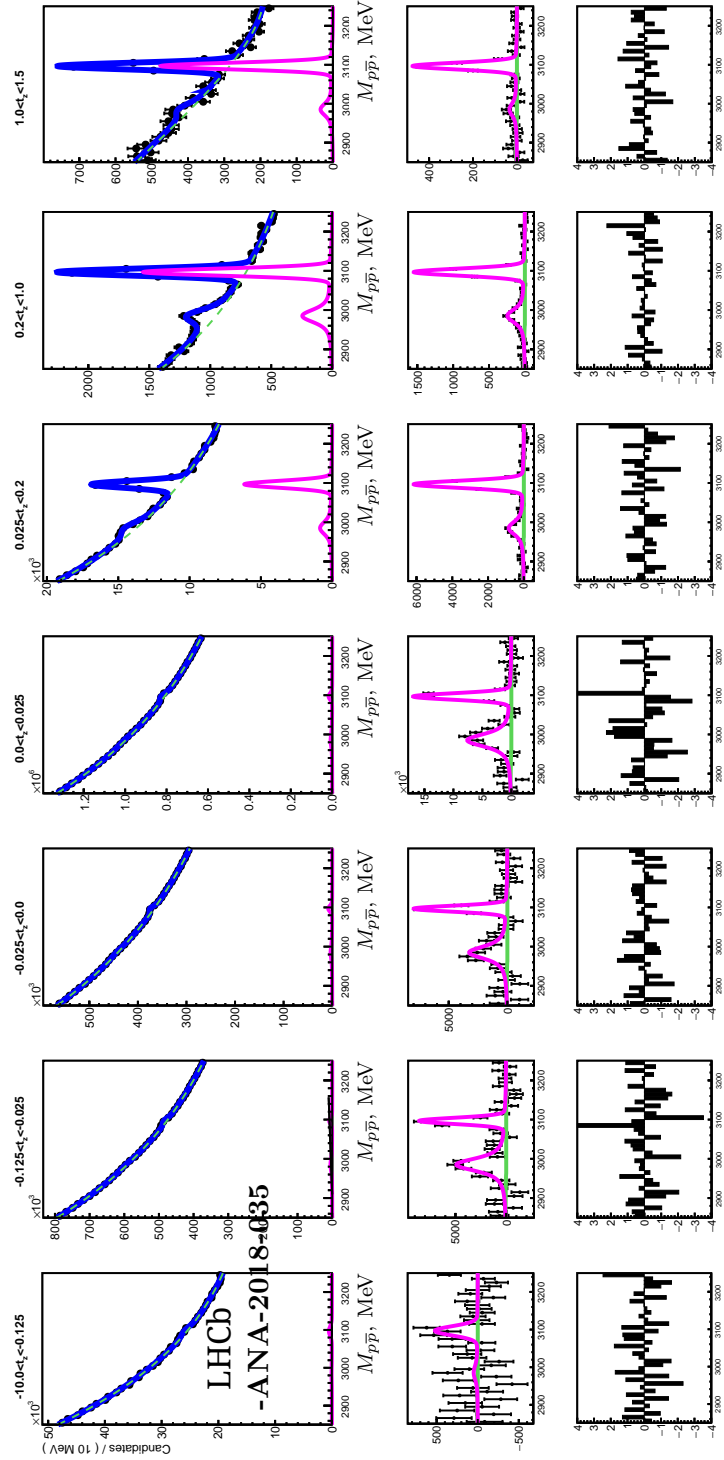
\clearpage
}

\begin{figure}[htb]
\centering
  \subfigure[Deviations of reconstructed mass, $\Delta m_{t_z}$.]{ 
    \centering
    \protect\protect\protect\includegraphics[width=0.465\textwidth]{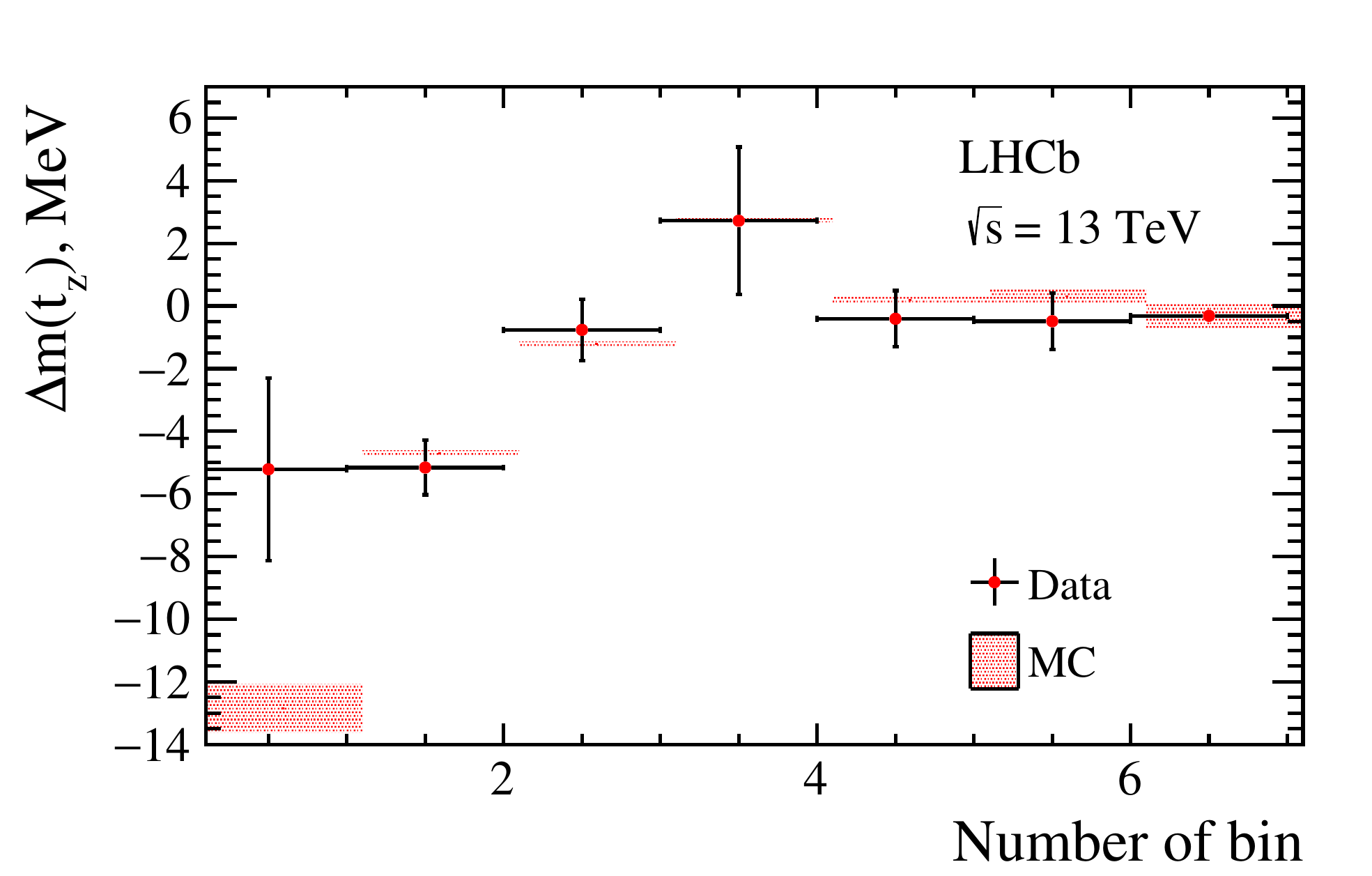}
    \label{fig:shiftsMass}
  }
\quad
  \subfigure[Correction factors of mass resolution, $\alpha_{t_z}$.]{
    \centering
    \protect\protect\protect\includegraphics[width=0.465\textwidth]{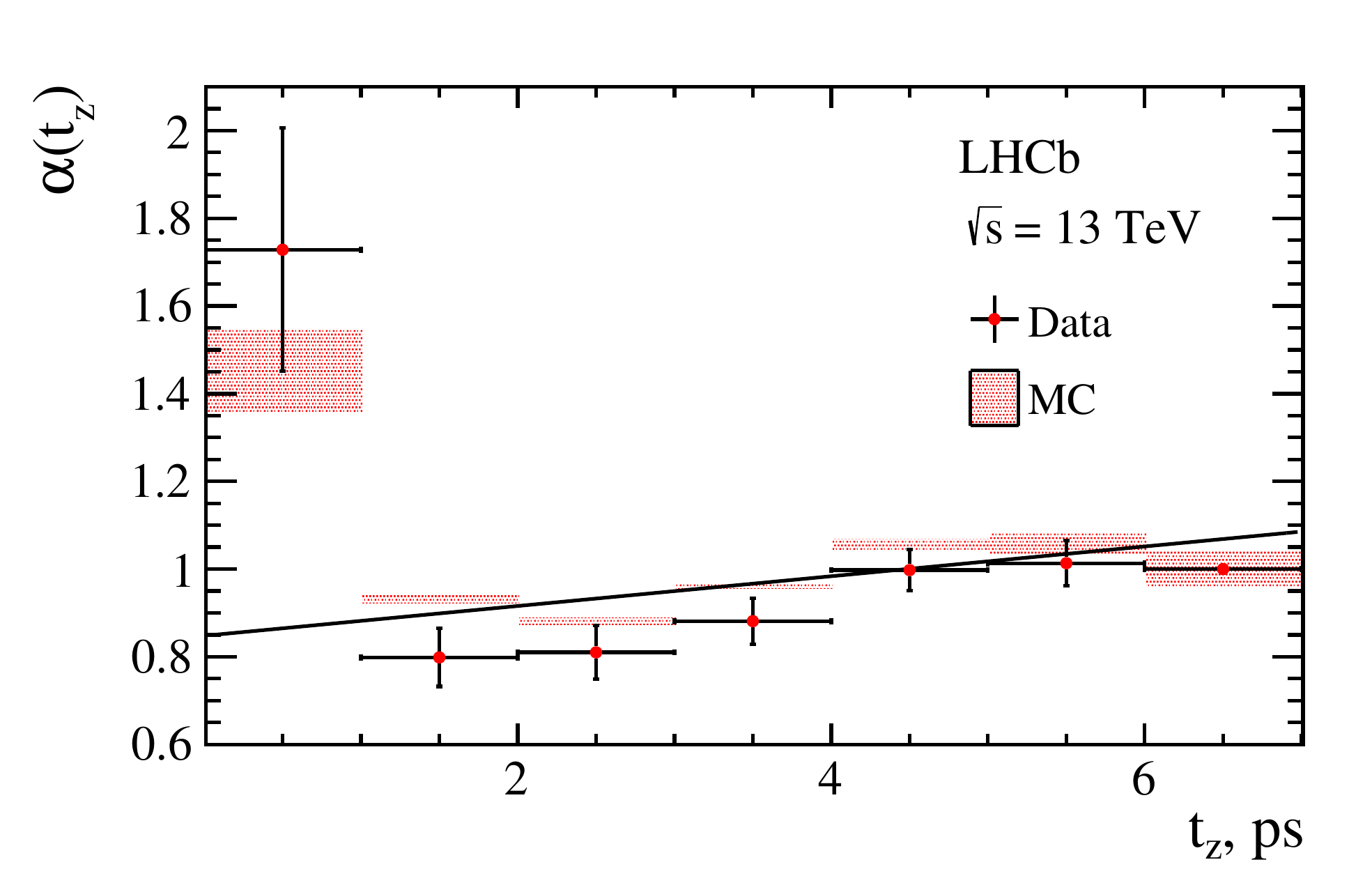}
    \label{fig:shiftsReso}
  }
 \caption
 [Mass deviations and correction factors of mass resolution as a function of $t_z$ bin mumber from simultaneous fit to the \etac and \jpsi invariant masses in the MC samples in bins of $t_z$ and from fit to total \pt-integrated data sample.]
 {Mass deviations~\subref{fig:shiftsMass} and correction factors of mass resolution~\subref{fig:shiftsReso} as a function of $t_z$ bin mumber from simultaneous fit to the \etac and \jpsi invariant masses in the MC samples in bins of $t_z$ (red boxes) and from fit to total \pt-integrated data sample (black points with error bars).} 
\label{fig:shifts}
\end{figure}

Reconstructed charmonium mass as a function of $t_z$ is studied using simulation samples. Simultaneous unbinned maximum-likelihood fit to $M_{\ppbar} - M^{Gen}_{\ppbar}$ in bins of $t_z$ is performed using \jpsi and \etac MC samples. The fit model assumes the shifts of \etac and \jpsi peak positions to be the same, while the resolution is described by the double Gaussian function as described above. Deviations with respect to the peak position in the last $t_z$ bin ($\Delta m_{t_z}$) are shown on Figure~\ref{fig:shiftsMass} in bins of $t_z$. The deviations vary by up to 4 \mev, which can cause a substantial bias when extracting signal yields in bins of $t_z$. Hence the corrections on peak positions in bins of $t_z$ are applied, while the difference of the \jpsi and \etac masses is kept constant throughout $t_z$ bins.

Similar effect is observed for the invariant mass resolution. Using MC samples, the correction factors, $\alpha_{t_z}$, of mass resolution parameter $\sigma_{\etac}$ in $t_z$ bins are extracted from the fits. The $\alpha_{t_z}$ is the ratio of the resolution in a given $t_z$ bin to that in the last $t_z$ bin. The obtained values of $\alpha_{t_z}$ in bins of $t_z$ are shown on Figure~\ref{fig:shiftsReso}. This effect is taken into account by introducing $\alpha_{t_z}$ parameters in the fit model.

Alternatively, as a cross-check, mass shifts and mass resolution correction factors are extracted from data by performing invariant mass fits in bins of $t_z$ for a total \pt-integrated data sample (Figs.~\ref{fig:shiftsMass} and ~\ref{fig:shiftsReso}). Corresponding systematic uncertainty is estimated in Section~\ref{sec:syst}.

\begin{figure}[t]
\centering
\protect\protect\protect\includegraphics[width=0.6\textwidth]{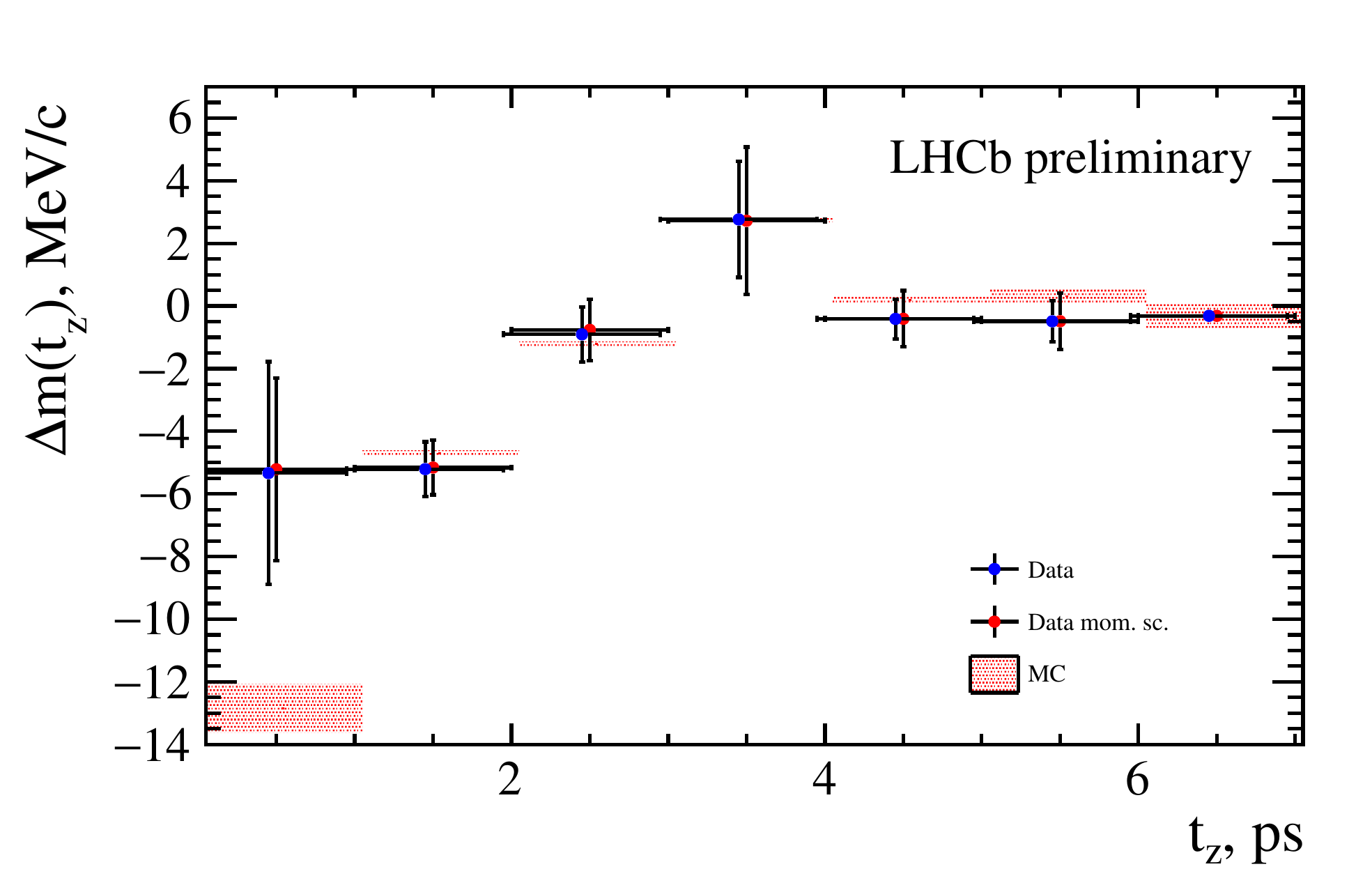}
\put(-270,80){\rotatebox{90}{\colorbox{shadecolor}{$\Delta m(t_z), \mev$}}}
\put(-120,140){\colorbox{shadecolor}{\footnotesize{\lhcb-ANA-2018-035}}}
\put(-80,5){\colorbox{shadecolor}{\small{$t_z$ bin number}}}
\caption
[Peak position shifts from \pt-integrated data sample with applying momentum scale calibration, without applying momentum scale calibration and from MC as function of $t_z$ bin number.]
{Peak position shifts from \pt-integrated data sample with applying momentum scale calibration (blue points), without applying momentum scale calibration (red points) and from MC (red boxes) as function of $t_z$ bin number.} 
\label{fig:shifts_vs_m_scaled}
\end{figure}

Peak position shifts extracted from data with and without implementing momentum scale calibration are compared on Fig.~\ref{fig:shifts_vs_m_scaled}. The shifts from data are extracted from simultaneous fit in 7 bins of $t_z$. No significant effect of momentum scale calibration is observed.

\begin{figure}[t]
\centering
\protect\protect\protect\includegraphics[width=0.6\textwidth]{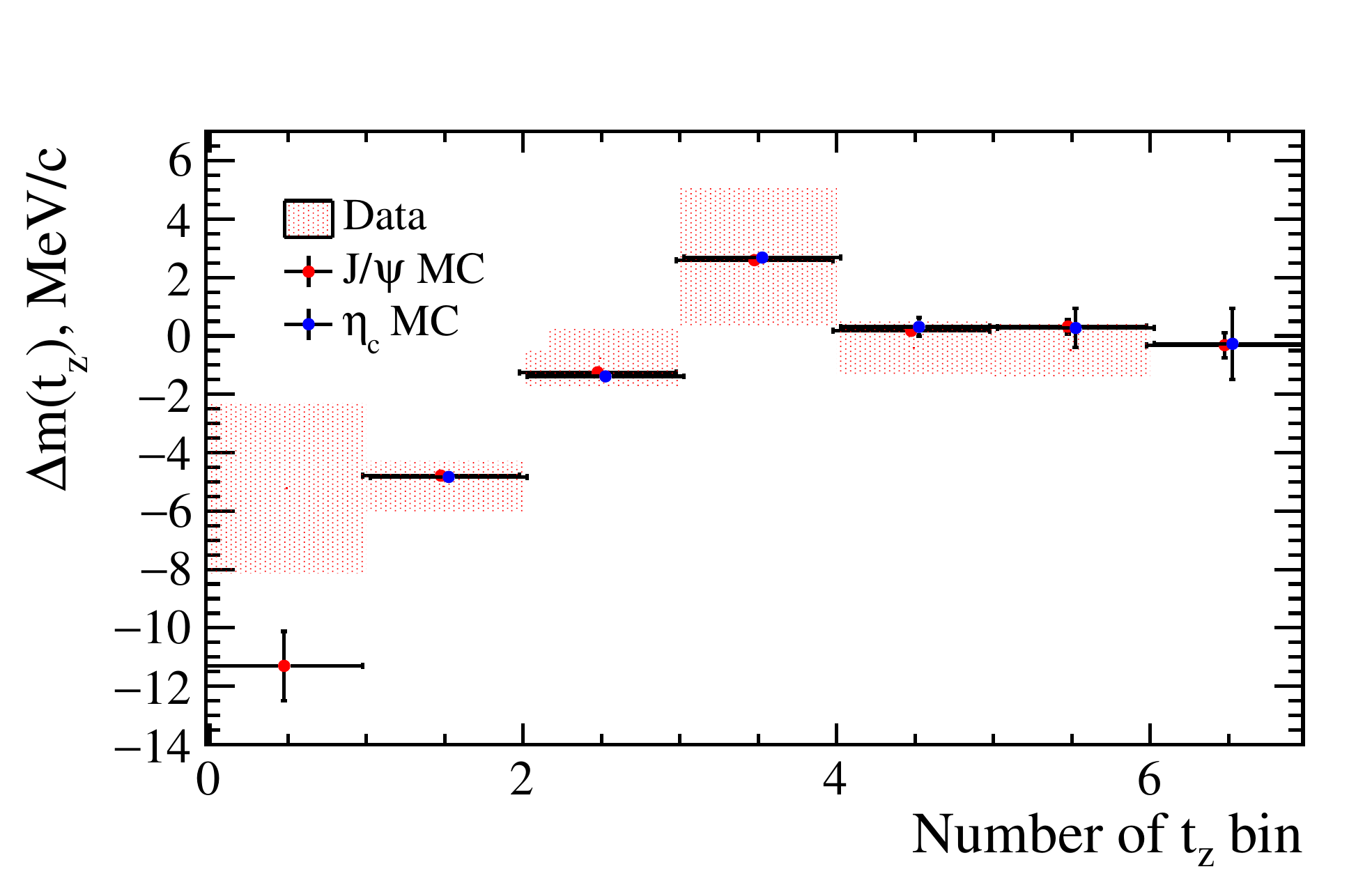}
\put(-270,70){\rotatebox{90}{\colorbox{shadecolor}{$\Delta m(t_z), \mev$}}}
\put(-120,130){\footnotesize{\lhcb-ANA-2018-035}}
\caption
[Peaks positions shifts for \etac and \jpsi separately from \pt-integrated MC sample.]
{Peaks positions shifts for \etac (blue points) and \jpsi (red points) separately from \pt-integrated MC sample.} 
\label{fig:shifts_etac_vs_jpsi}
\end{figure}

The comparison of peak position shifts for \etac and \jpsi obtained from simulation is shown on Fig.~\ref{fig:shifts_etac_vs_jpsi}. No significant difference between \etac and \jpsi shifts is observed.

\begin{figure}[t]
\centering
\protect\protect\protect\includegraphics[width=0.7\textwidth]{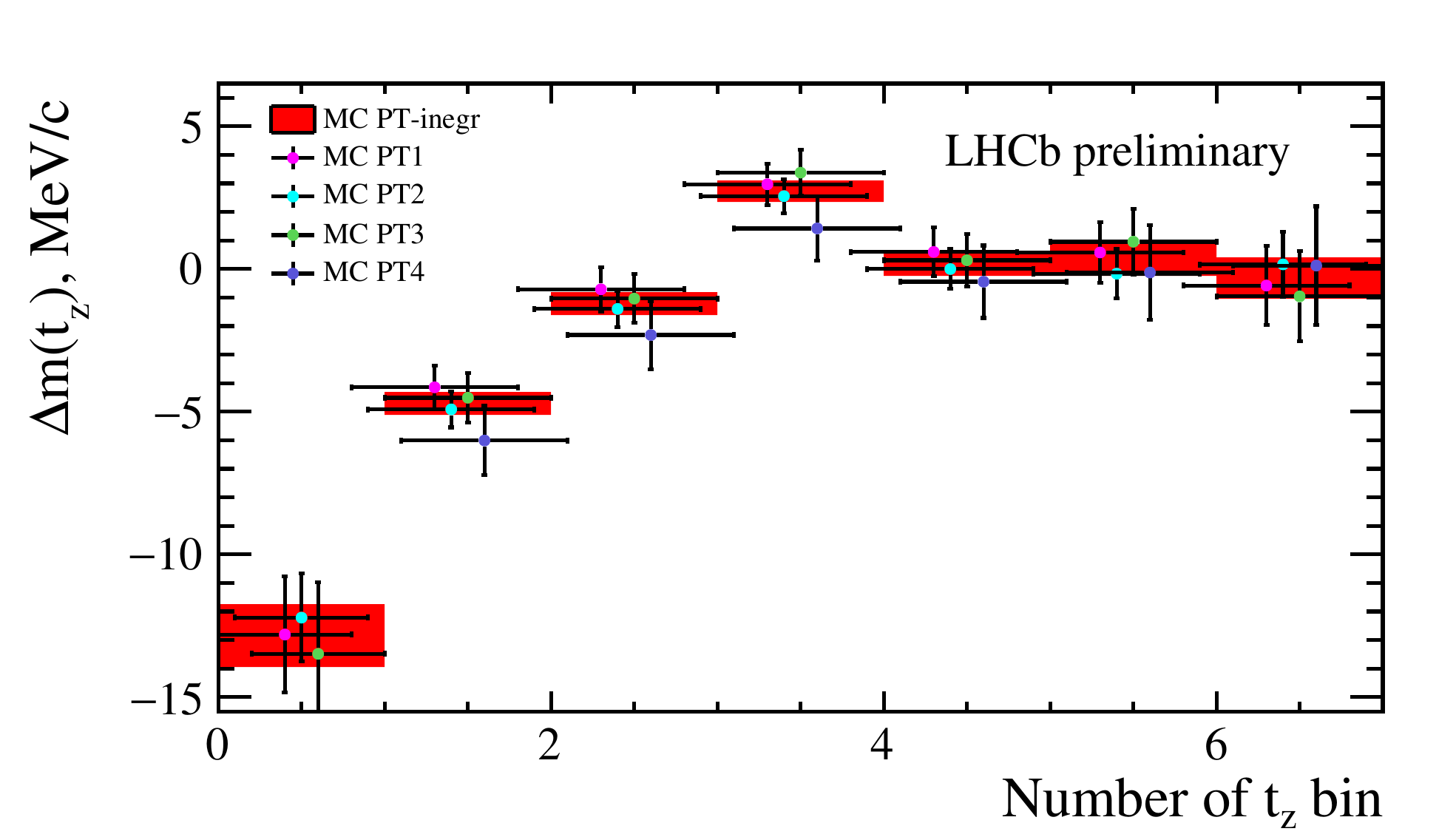}
\put(-315,85){\rotatebox{90}{\colorbox{shadecolor}{$\Delta m(t_z), \mev$}}}
\put(-120,146){\colorbox{shadecolor}{\footnotesize{\lhcb-ANA-2018-035}}}
\caption
[Peak position shifts in \pt-bins from binned and from \pt-integrated MC samples.]
{Peak position shifts in \pt-bins (points) from binned and from \pt-integrated MC samples (red boxes).}
\label{fig:shifts_vs_PT}
\end{figure}

The comparison of peak position shifts obtained with simulation in bins of \pt is shown on Fig.~\ref{fig:shifts_vs_PT}. No significant \pt-dependence of shifts is observed.

The effect of \etac natural width $\Gamma_{\etac}$ exceeds that of the detector resolution $\sigma_{n}$ in signal shape model. 
The \etac peak is thus described using a convolution of double Gaussian (DG) and relativistic Breit-Wigner (RBW) functions in the fit to data, while \jpsi peak is described by a double Gaussian function.

Complete signal shape model used in the fit to data is summarised by Eq.~\ref{eq:massSigModel}.
\begin{equation}
\label{eq:massSigModel}
\begin{aligned}
S_{\etac}(M_{\ppbar}) &\propto RBW(M_{\ppbar},m_{\etac}+\Delta m_{t_z},\Gamma_{\etac},J_{\etac}=0) \otimes DG(\sigma_n\times\alpha_{t_z}, \sigma_{n}/\sigma_{w},f_{n}) \\
S_{\jpsi}(M_{\ppbar}) &\propto \delta(M_{\ppbar}-m_{\jpsi}-\Delta m_{t_z}) \otimes DG(\sigma_n\times\frac{\sigma_{\jpsi}}{\sigma_{\etac}}\times\alpha_{t_z}, \sigma_{n}/\sigma_{w},f_{n}),
\end{aligned}
\end{equation}
where $J_{\etac}=0$ is the spin of \etac; 
$\sigma_n$, $\sigma_{\jpsi}/\sigma_{\etac}$, $\sigma_{n}/\sigma_{w}$ and $f_n$  are resolution parameters as discussed above;
$\Delta m_{t_z}$ and $\alpha_{t_z}$ are the peak position and resolution corrections in bins of $t_z$. The summary of signal shape parametrisation in the fit is given in Table~\ref{tab:massSigModel}.
\begin{table}[ht] 
\small{
\centering
\begin{tabular}{l|l} 
 Parameter & Comment \\ \hline \hline
 $\sigma_n / \sigma_w$	            & Fixed from MC  \\	\hline 
 $f_n$	                            & Fixed from MC  \\ \hline
 $\sigma_{\etac}/\sigma_{\jpsi}$	  & Fixed from MC  \\ \hline
 $\sigma_n$                         & Common free parameter, \\ &
                                     linear slope of \pt-dependence extracted from MC \\ &
                                     for differential production measurement \\ \hline
 $m_{\jpsi} - m_{\etac}$            & Common free parameter for all fits in bins of $t_z$ and $\pt$ \\ \hline             
 $m_{\jpsi}$                        & Common free parameter for all fits in bins of $t_z$ and $\pt$ \\ \hline
 $\Gamma_{\etac}$                   & Fixed to world average value~\cite{PDG2017} (31.8 \mev) \\ \hline
$\Delta m_{t_z}$                  & Fixed from MC in each $t_z$ bin \\ \hline
 $\alpha_{t_z}$                   & Fixed from MC in each $t_z$ bin
\end{tabular}
\caption{Summary of signal parametrisation in the simultaneous invariant mass fit.} 
\label{tab:massSigModel}
}
\end{table}

\clearpage
The combinatorial background composed of random combinations of charged hadrons passing proton identification hypothesis is parametrised using an exponential function multiplied by a second order polynomial function.

\begin{figure}[ht]
\centering
\protect\protect\protect\includegraphics[width=0.53\linewidth]{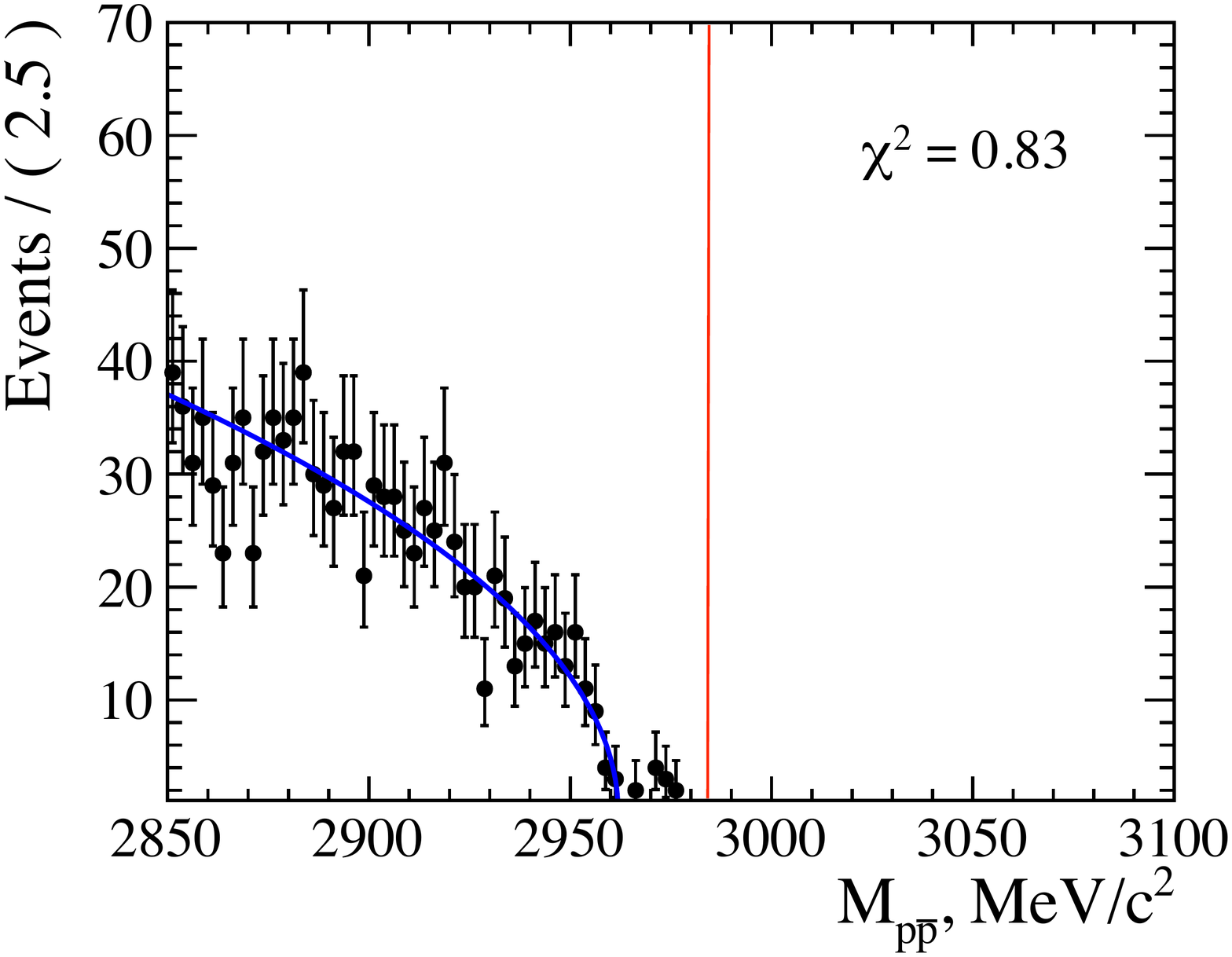}
\put(-190,145){\scriptsize{\lhcb simulation}}
\put(-225,60){\rotatebox{90}{\colorbox{shadecolor}{\small Candidates / 2~\mev}}}
\put(-83,8){\colorbox{shadecolor}{\small $M_{\ppbar}, \mev$}}
\caption
[The $M_{\ppbar}$ distribution from the simulated $\JpsiToPpbarPiz$ sample.]
{The $M_{\ppbar}$ distribution from the simulated $\JpsiToPpbarPiz$ sample. The solid blue line represents the fit by the square root function from Eq.~\ref{eq:pppi0}. The \etac mass value is indicated by red solid line.} 
\label{fig:pppi0}
\end{figure}

Besides the pure combinatorial background, proton-antiproton pairs from higher mass charmonium states decays to three or more particles can produce wide structures in the $\ppbar$ invariant mass spectrum. The only notable partially reconstructed background is that from the $\JpsiToPpbarPiz$ decays with the contribution in the range below $M_{\jpsi} - M_{\piz} = 3096.9 - 135.0 = 2961.9$~\mev, which can potentially affects the $\etac$ region description. This process is specifically included in the fit model.
Its contribution to the $\ppbar$ invariant mass spectrum around the threshold region is parametrised by a square-root shape as in Ref.~\cite{LHCb-PAPER-2014-029}:
\begin{equation} 
\label{eq:pppi0}
 f_{\JpsiToPpbarPiz} (M) \propto 
\begin{cases}
   \sqrt{M_{\jpsi} - M_{\piz} - M_{\ppbar}} & \text{if } M_{\ppbar} \leq M_{\jpsi} - M_{\piz} \ , \\
   0                                                    & \text{if } M_{\ppbar} > M_{\jpsi} - M_{\piz}.
 \end{cases}
\end{equation} 
This shape contains no free parameters. 
Applicability of the shape from equation~(\ref{eq:pppi0}) is verified using the MC sample, as shown in Fig.~\ref{fig:pppi0}. The suggested model shows a good agreement with MC yielding a good fit quality, $\chisqndf<1$.
Using branching fractions and the efficiencies in considered invariant mass range for $\JpsiToPpbar$ and $\JpsiToPpbarPiz$ channels, the contribution from $\JpsiToPpbarPiz$ is normalised as
\begin{equation}
 n_{\JpsiToPpbarPiz} = n_{\JpsiToPpbar}\times
 \frac{\epsilon_{\JpsiToPpbarPiz}}{\epsilon_{\JpsiToPpbar}}
 \times\frac{\BR_{\JpsiToPpbarPiz}}{\BR_{\JpsiToPpbar}} \ .
\end{equation}
Using the ratio of branching fractions $\BR_{\JpsiToPpbarPiz} / \BR_{\JpsiToPpbar} = \brRatioJpsipppizJpsipp$ from Ref.~\cite{PDG2017}, 
and the ratio of efficiencies $\epsilon_{\JpsiToPpbarPiz}/\epsilon_{\JpsiToPpbar} = \epsRatioJpsipppizJpsipp$ from simulation. The efficiency ratio is small due to mass fit region limit.
One can conclude that the $\JpsiToPpbarPiz$ channel produces a non-peaking contribution 
to the $\ppbar$ invariant mass spectrum that amounts to about 3\% of the $\JpsiToPpbar$ signal. 
In the fit to the invariant mass spectra, the amount of contribution from \JpsiToPpbarPiz is bound to the observed yields of \JpsiToPpbar.

\subsubsection{Fit to data}
\begin{figure}[t]
\centering
\protect\protect\protect\includegraphics[angle=90, width=0.6\textwidth]{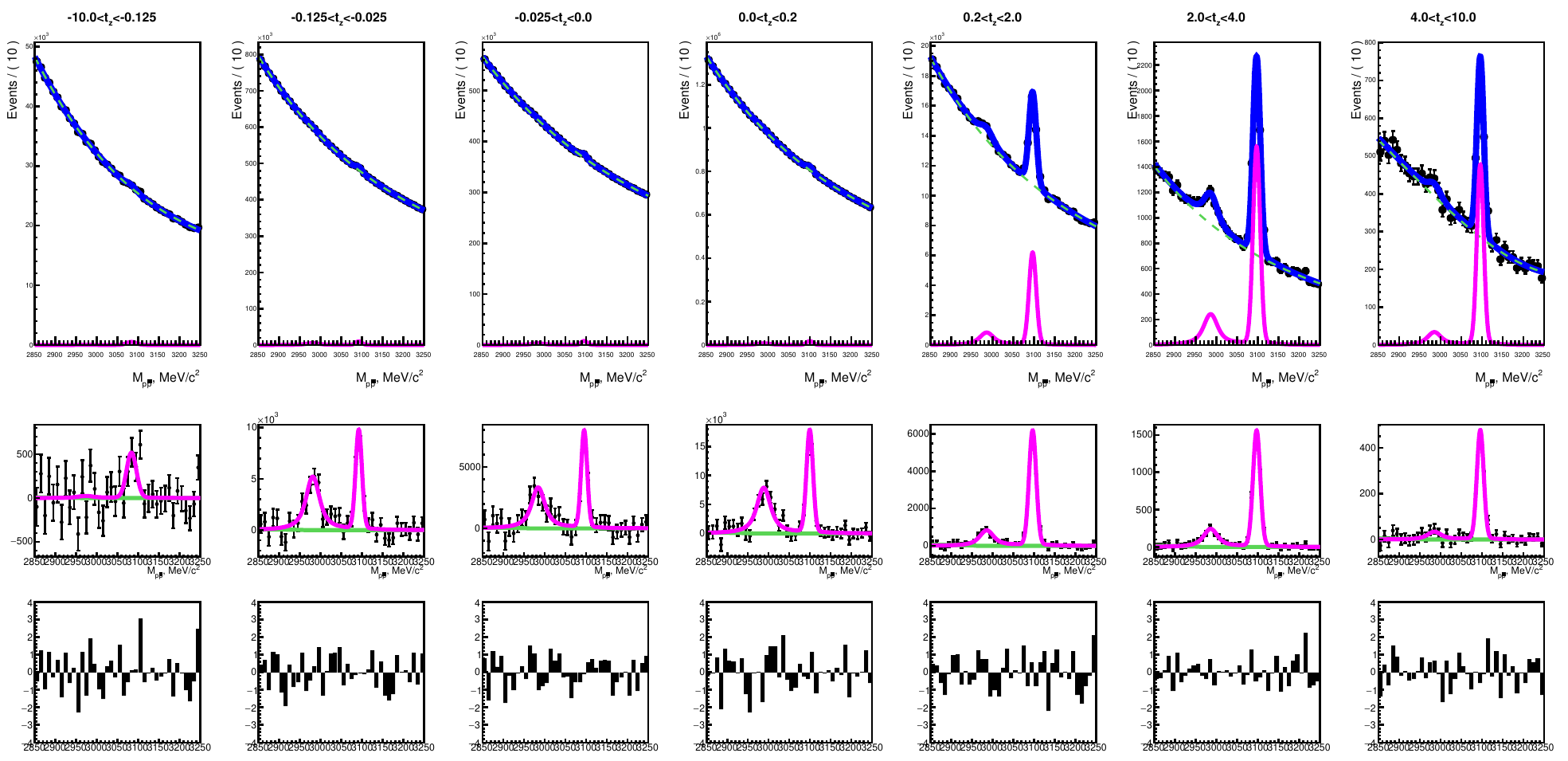}
\put(-170,20){\rotatebox{90}{\small\textbf{\lhcb-ANA-2018-035}}}
\put(-142,30){\rotatebox{90}{\colorbox{shadecolor}{\scriptsize$M_{\ppbar}, \mev$}}}
\put(-142,110){\rotatebox{90}{\colorbox{shadecolor}{\scriptsize$M_{\ppbar}, \mev$}}}
\put(-142,190){\rotatebox{90}{\colorbox{shadecolor}{\scriptsize$M_{\ppbar}, \mev$}}}
\put(-142,270){\rotatebox{90}{\colorbox{shadecolor}{\scriptsize$M_{\ppbar}, \mev$}}}
\put(-142,350){\rotatebox{90}{\colorbox{shadecolor}{\scriptsize$M_{\ppbar}, \mev$}}}
\put(-142,430){\rotatebox{90}{\colorbox{shadecolor}{\scriptsize$M_{\ppbar}, \mev$}}}
\put(-142,510){\rotatebox{90}{\colorbox{shadecolor}{\scriptsize$M_{\ppbar}, \mev$}}}
\put(-73,0){\colorbox{shadecolor}{\makebox(5,560){\textcolor{white}{a}}}}
\put(-6.5,0){\colorbox{shadecolor}{\makebox(5,560){\textcolor{white}{a}}}}
\caption
[The $M_{\ppbar}$ distribution for seven bins of $t_z$ for \pt-integrated sample $6.5 \gev<\pt<14.0 \gev$. The solid blue lines represent the total fit result.]
{The $M_{\ppbar}$ distribution for seven bins of $t_z$ for \pt-integrated sample $6.5 \gev<\pt<14.0 \gev$. The solid blue lines represent the total fit result. Magenta and green lines show the signal and background components, respectively. The corresponding residual and pull distributions are shown below.} 
\label{fig:massFitInt}
\end{figure}

Projections of simultaneous fit for the entire \pt-range $6.5 \gev<\pt<14.0 \gev$ are shown on Fig.~\ref{fig:massFitInt}. The residual and pull distributions are displayed below the corresponding projections. In general, fit yields a good description of $M_{\ppbar}$ distributions in all $t_z$ bins.

\begin{figure}[ht]
\centering
\protect\protect\protect\includegraphics[angle=90, width=0.6\textwidth]{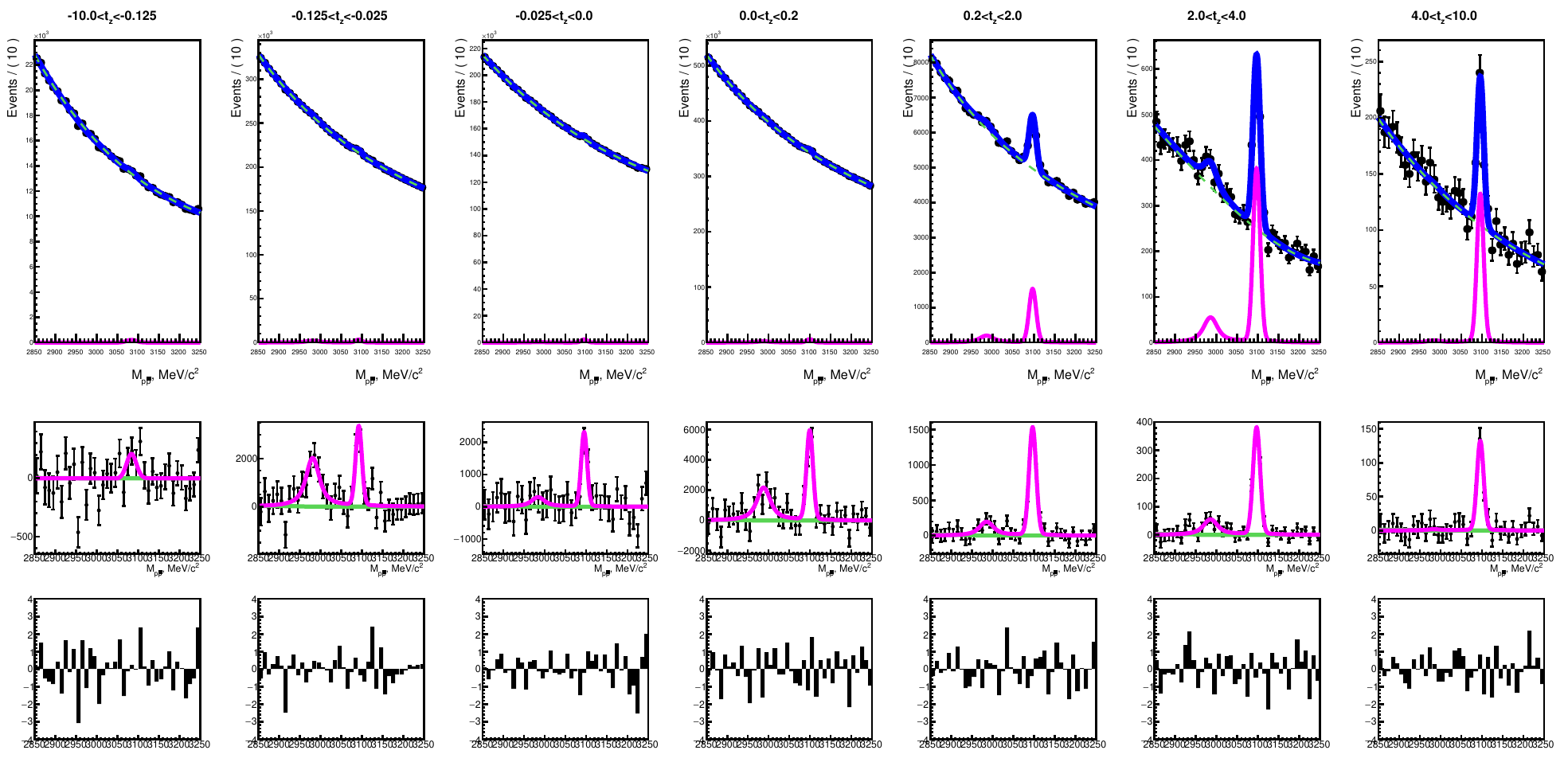}
\put(-170,20){\rotatebox{90}{\small\textbf{\lhcb-ANA-2018-035}}}
\put(-142,30){\rotatebox{90}{\colorbox{shadecolor}{\scriptsize$M_{\ppbar}, \mev$}}}
\put(-142,110){\rotatebox{90}{\colorbox{shadecolor}{\scriptsize$M_{\ppbar}, \mev$}}}
\put(-142,190){\rotatebox{90}{\colorbox{shadecolor}{\scriptsize$M_{\ppbar}, \mev$}}}
\put(-142,270){\rotatebox{90}{\colorbox{shadecolor}{\scriptsize$M_{\ppbar}, \mev$}}}
\put(-142,350){\rotatebox{90}{\colorbox{shadecolor}{\scriptsize$M_{\ppbar}, \mev$}}}
\put(-142,430){\rotatebox{90}{\colorbox{shadecolor}{\scriptsize$M_{\ppbar}, \mev$}}}
\put(-142,510){\rotatebox{90}{\colorbox{shadecolor}{\scriptsize$M_{\ppbar}, \mev$}}}
\put(-73,0){\colorbox{shadecolor}{\makebox(5,560){\textcolor{white}{a}}}}
\put(-6.5,0){\colorbox{shadecolor}{\makebox(5,560){\textcolor{white}{a}}}}
\caption
[The $M_{\ppbar}$ distribution for seven bins of $t_z$ for $6.5 \gev<\pt<8.0 \gev$.]
{The $M_{\ppbar}$ distribution for seven bins of $t_z$ for $6.5 \gev<\pt<8.0 \gev$. The solid blue lines represent the total fit result. Magenta and green lines show the signal and background components, respectively. The corresponding residual and pull distributions are shown below.} 
\label{fig:massFitPT1}
\end{figure}

\begin{figure}[ht]
\centering
\protect\protect\protect\includegraphics[angle=90, width=0.6\textwidth]{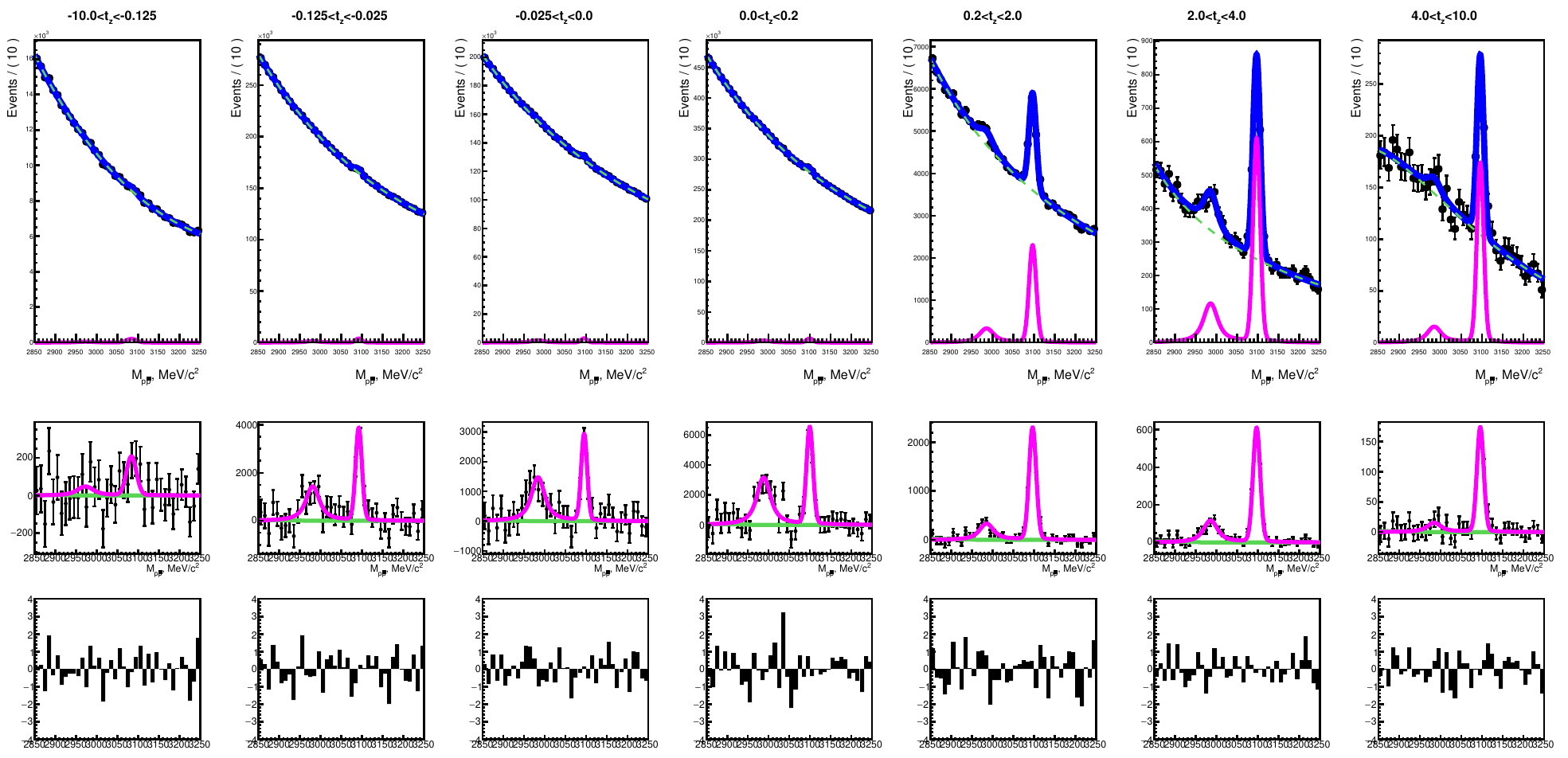}
\put(-170,20){\rotatebox{90}{\small\textbf{\lhcb-ANA-2018-035}}}
\put(-142,30){\rotatebox{90}{\colorbox{shadecolor}{\scriptsize$M_{\ppbar}, \mev$}}}
\put(-142,110){\rotatebox{90}{\colorbox{shadecolor}{\scriptsize$M_{\ppbar}, \mev$}}}
\put(-142,190){\rotatebox{90}{\colorbox{shadecolor}{\scriptsize$M_{\ppbar}, \mev$}}}
\put(-142,270){\rotatebox{90}{\colorbox{shadecolor}{\scriptsize$M_{\ppbar}, \mev$}}}
\put(-142,350){\rotatebox{90}{\colorbox{shadecolor}{\scriptsize$M_{\ppbar}, \mev$}}}
\put(-142,430){\rotatebox{90}{\colorbox{shadecolor}{\scriptsize$M_{\ppbar}, \mev$}}}
\put(-142,510){\rotatebox{90}{\colorbox{shadecolor}{\scriptsize$M_{\ppbar}, \mev$}}}
\put(-73,0){\colorbox{shadecolor}{\makebox(5,560){\textcolor{white}{a}}}}
\put(-6.5,0){\colorbox{shadecolor}{\makebox(5,560){\textcolor{white}{a}}}}
\caption
[The $M_{\ppbar}$ distribution for seven bins of $t_z$ for $8.0 \gev<\pt<10.0 \gev$.]
{The $M_{\ppbar}$ distribution for seven bins of $t_z$ for $8.0 \gev<\pt<10.0 \gev$. The solid blue lines represent the total fit result. Magenta and green lines show the signal and background components, respectively. The corresponding residual and pull distributions are shown below.} 
\label{fig:massFitPT2}
\end{figure}

\begin{figure}[ht]
\centering
\protect\protect\protect\includegraphics[angle=90, width=0.6\textwidth]{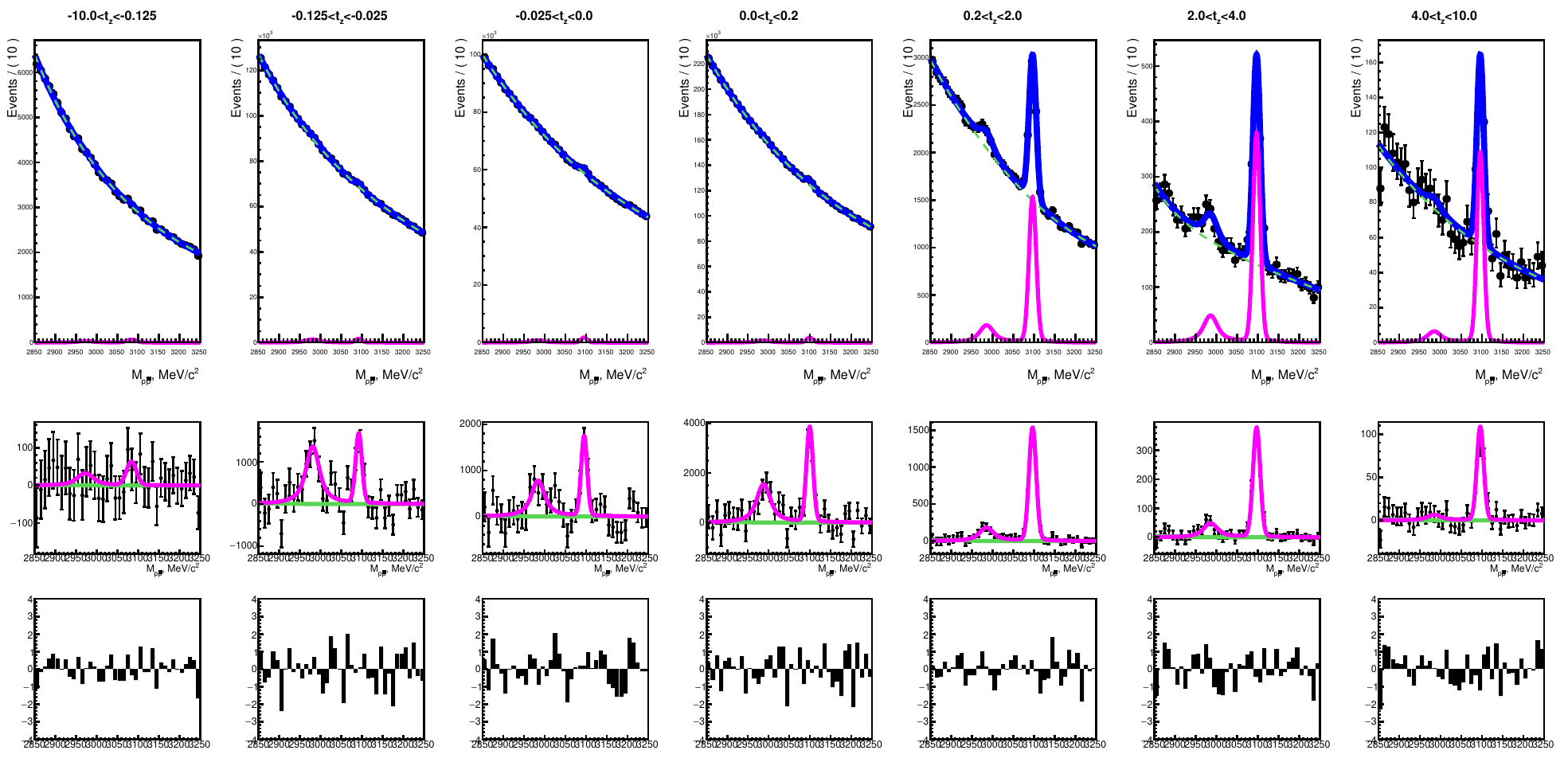}
\put(-170,20){\rotatebox{90}{\small\textbf{\lhcb-ANA-2018-035}}}
\put(-142,30){\rotatebox{90}{\colorbox{shadecolor}{\scriptsize$M_{\ppbar}, \mev$}}}
\put(-142,110){\rotatebox{90}{\colorbox{shadecolor}{\scriptsize$M_{\ppbar}, \mev$}}}
\put(-142,190){\rotatebox{90}{\colorbox{shadecolor}{\scriptsize$M_{\ppbar}, \mev$}}}
\put(-142,270){\rotatebox{90}{\colorbox{shadecolor}{\scriptsize$M_{\ppbar}, \mev$}}}
\put(-142,350){\rotatebox{90}{\colorbox{shadecolor}{\scriptsize$M_{\ppbar}, \mev$}}}
\put(-142,430){\rotatebox{90}{\colorbox{shadecolor}{\scriptsize$M_{\ppbar}, \mev$}}}
\put(-142,510){\rotatebox{90}{\colorbox{shadecolor}{\scriptsize$M_{\ppbar}, \mev$}}}
\put(-73,0){\colorbox{shadecolor}{\makebox(5,560){\textcolor{white}{a}}}}
\put(-6.5,0){\colorbox{shadecolor}{\makebox(5,560){\textcolor{white}{a}}}}
\caption
[The $M_{\ppbar}$ distribution for seven bins of $t_z$ for ($10.0 \gev<\pt<12.0 \gev$.]
{The $M_{\ppbar}$ distribution for seven bins of $t_z$ for ($10.0 \gev<\pt<12.0 \gev$. The solid blue lines represent the total fit result. Magenta and green lines show the signal and background components, respectively. The corresponding residual and pull distributions are shown below.} 
\label{fig:massFitPT3}
\end{figure}

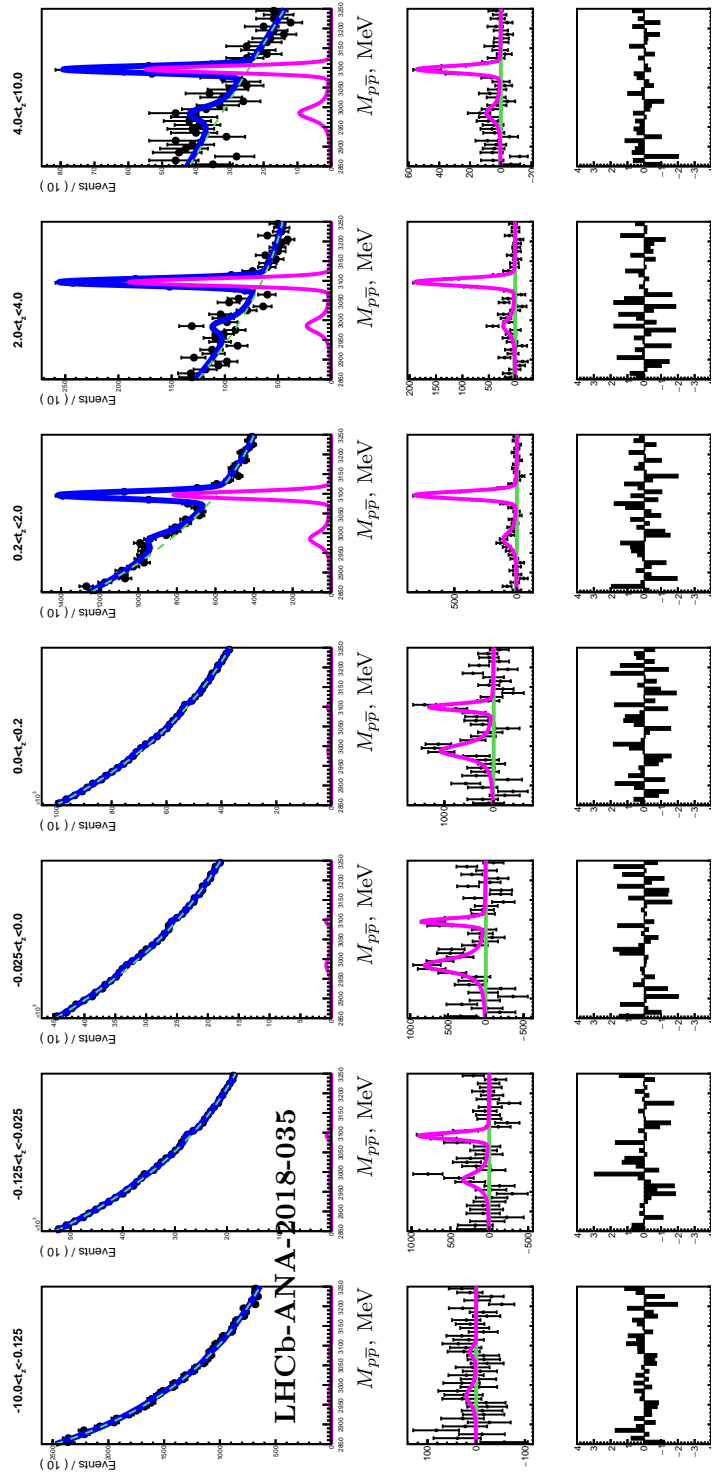
\begin{figure}[ht]
\centering
\protect\protect\protect\includegraphics[angle=90, width=0.6\textwidth]{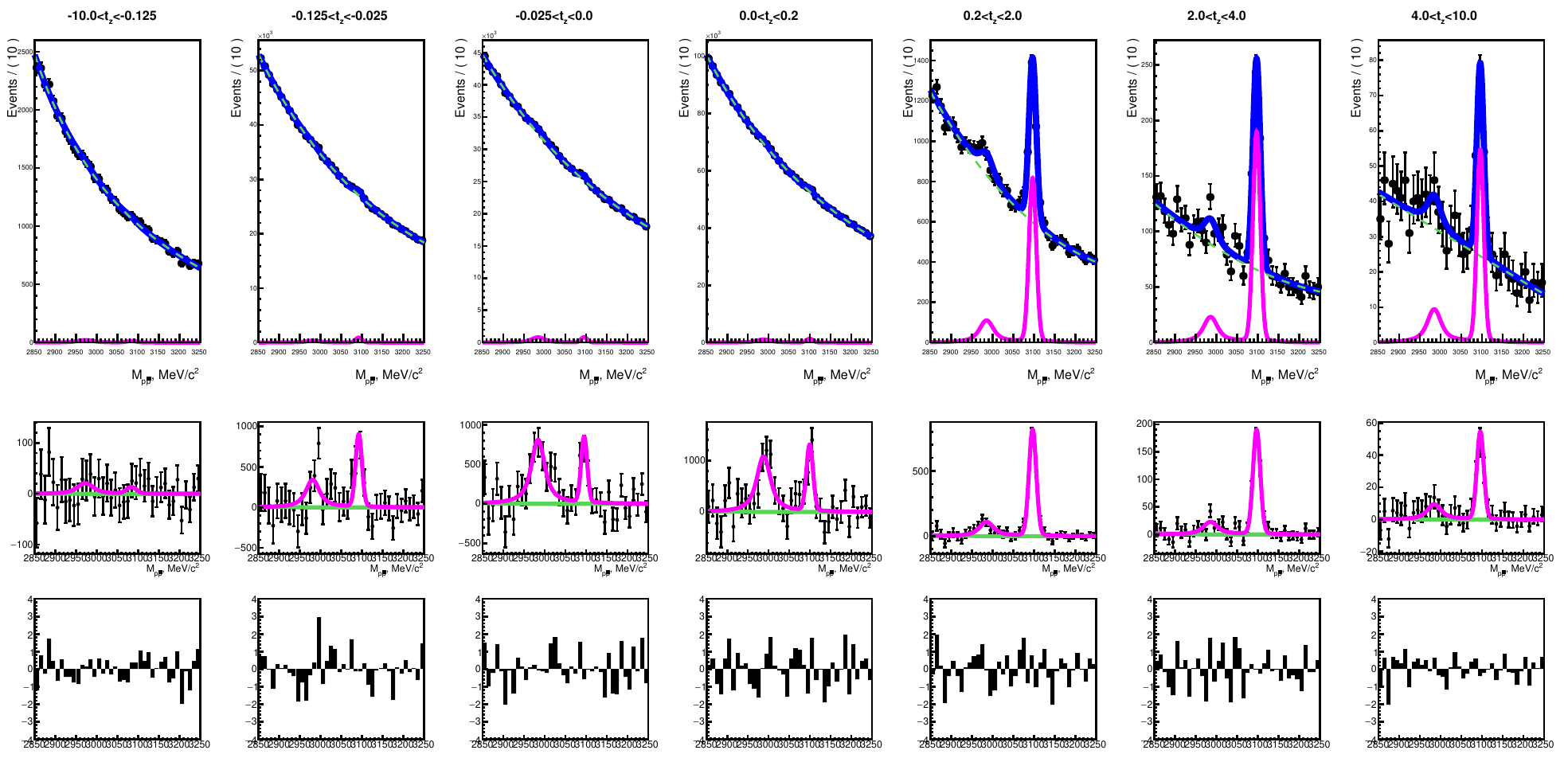}
\put(-170,20){\rotatebox{90}{\small\textbf{\lhcb-ANA-2018-035}}}
\put(-142,30){\rotatebox{90}{\colorbox{shadecolor}{\scriptsize$M_{\ppbar}, \mev$}}}
\put(-142,110){\rotatebox{90}{\colorbox{shadecolor}{\scriptsize$M_{\ppbar}, \mev$}}}
\put(-142,190){\rotatebox{90}{\colorbox{shadecolor}{\scriptsize$M_{\ppbar}, \mev$}}}
\put(-142,270){\rotatebox{90}{\colorbox{shadecolor}{\scriptsize$M_{\ppbar}, \mev$}}}
\put(-142,350){\rotatebox{90}{\colorbox{shadecolor}{\scriptsize$M_{\ppbar}, \mev$}}}
\put(-142,430){\rotatebox{90}{\colorbox{shadecolor}{\scriptsize$M_{\ppbar}, \mev$}}}
\put(-142,510){\rotatebox{90}{\colorbox{shadecolor}{\scriptsize$M_{\ppbar}, \mev$}}}
\put(-73,0){\colorbox{shadecolor}{\makebox(5,560){\textcolor{white}{a}}}}
\put(-6.5,0){\colorbox{shadecolor}{\makebox(5,560){\textcolor{white}{a}}}}
\caption
[The $M_{\ppbar}$ distribution for seven bins of $t_z$ for $12.0 \gev<\pt<14.0 \gev$.]
{The $M_{\ppbar}$ distribution for seven bins of $t_z$ for $12.0 \gev<\pt<14.0 \gev$. The solid blue lines represent the total fit result. Magenta and green lines show the signal and background components, respectively. The corresponding residual and pull distributions are shown below.} 
\label{fig:massFitPT4}
\end{figure}

Projections of simultaneous fit in the \pt bins are shown on Figs.~\ref{fig:massFitPT1},~\ref{fig:massFitPT2},~\ref{fig:massFitPT3} and~\ref{fig:massFitPT4}. The residual and pull distributions are displayed below the corresponding projections. Also in the projections the fit reproduces well the observed $M_{\ppbar}$ distributions.

The value of detector resolution parameter from the simultaneous fit is $\sigma_{\etac}=\resoSim \mev$ is reasonably comparable to the value from simulation $\sigma_{\etac}^{MC}=7.50\pm0.05 \mev$.
Simultaneous fit yields the following values of \jpsi mass $M_{\jpsi}=\jpsiMassTzFitSim$ and \jpsi and \etac mass difference $\Delta M _{\jpsi , \, \etac} = \etacMassDiffTzFitSim$, where the uncertainties are statistical only. These values agree with the world average values $M_{\jpsi}^{PDG}=\jpsiMassPDG$ and $\Delta M _{\jpsi , \, \etac}^{PDG} = \etacMassDiffPDG$~\cite{PDG2017} once systematic uncertainty is considered (the systematic uncertainty on $M_{\jpsi}$ is about $0.6 \mev$). 
Detailed consideration of the \etac mass and measurements is discussed in Chapter~\ref{ch:mass}.

\newpage

\clearpage

\subsection{Fit to the $t_z$ distribution}
\label{sec:fit2tz}
The $t_z$ resolution is studied using simulation samples similarly to the invariant mass resolution. 
Events with wrongly assigned primary vertex were excluded from the $t_z$ resolution study, therefore obtained resolution model is not distorted by these events.

\begin{figure}[b]
\centering
\protect\protect\protect\includegraphics[width=0.95\linewidth]{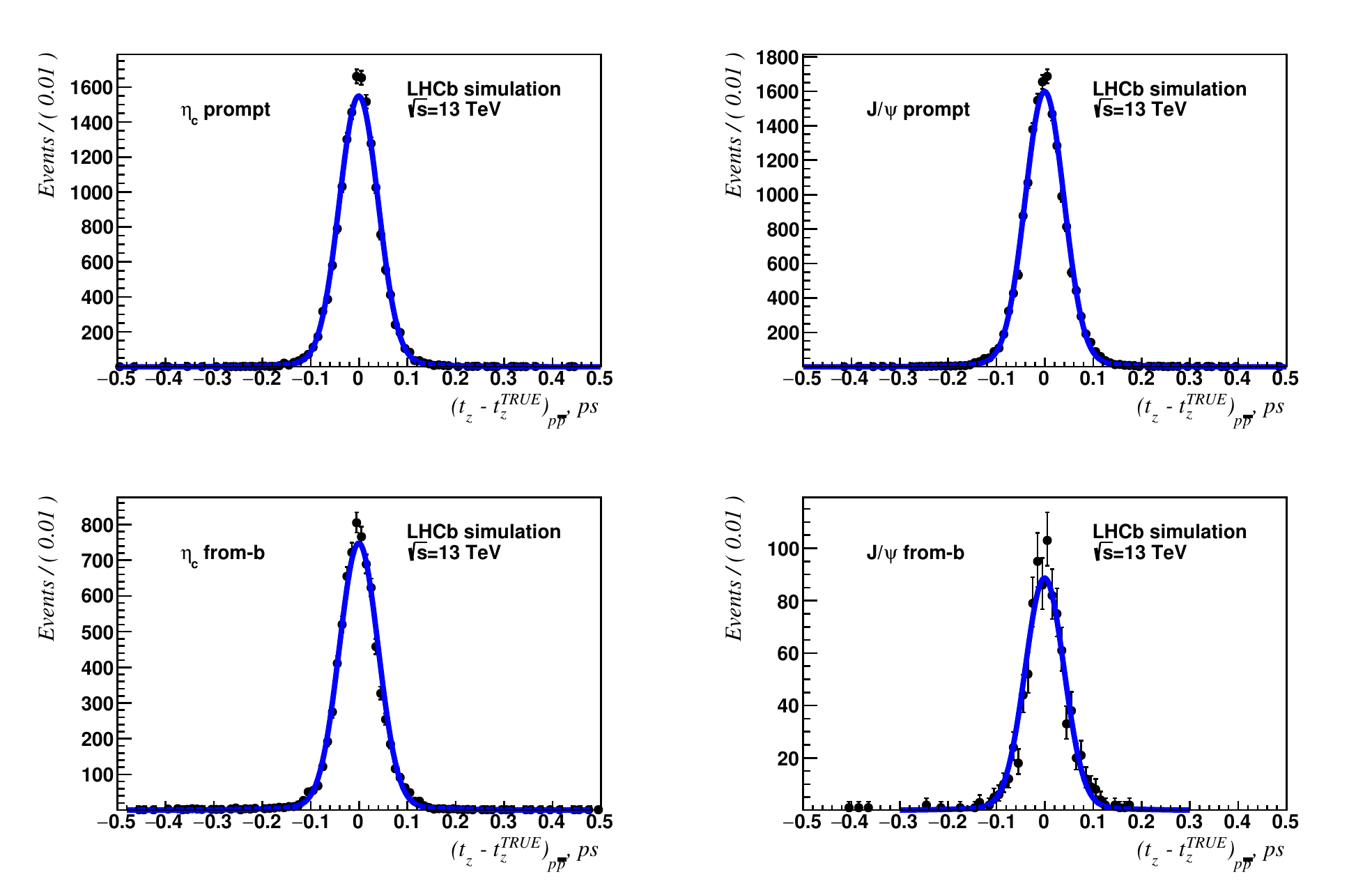}
\put(-420,180){\rotatebox{90}{\colorbox{shadecolor}{\scriptsize Candidates / 0.01 \ps}}}
\put(-295,146){\colorbox{shadecolor}{\scriptsize $t_z - t_z^{Gen}$, \ps}}
\put(-420,40){\rotatebox{90}{\colorbox{shadecolor}{\scriptsize Candidates / 0.01 \ps}}}
\put(-295,8){\colorbox{shadecolor}{\scriptsize $t_z - t_z^{Gen}$, \ps}}
\put(-205,180){\rotatebox{90}{\colorbox{shadecolor}{\scriptsize Candidates / 0.01 \ps}}}
\put(-80,146){\colorbox{shadecolor}{\scriptsize $t_z - t_z^{Gen}$, \ps}}
\put(-205,40){\rotatebox{90}{\colorbox{shadecolor}{\scriptsize Candidates / 0.01 \ps}}}
\put(-80,8){\colorbox{shadecolor}{\scriptsize $t_z - t_z^{Gen}$, \ps}}
\caption
[Distribution of the $t_z - t_z^{Gen}$ in the MC samples: prompt \etac, prompt \jpsi, \etac from \bquark-decays and \jpsi from \bquark-decays.]
{Distribution of the $t_z - t_z^{Gen}$ in the MC samples: prompt \etac (top left), prompt \jpsi (top right), \etac from \bquark-decays (bottom left) and \jpsi from \bquark-decays (bottom right). The solid blue lines represent the result of the simultaneous fit by a double Gaussian function to all four MC samples.}
\label{fig:fitMC}
\end{figure}
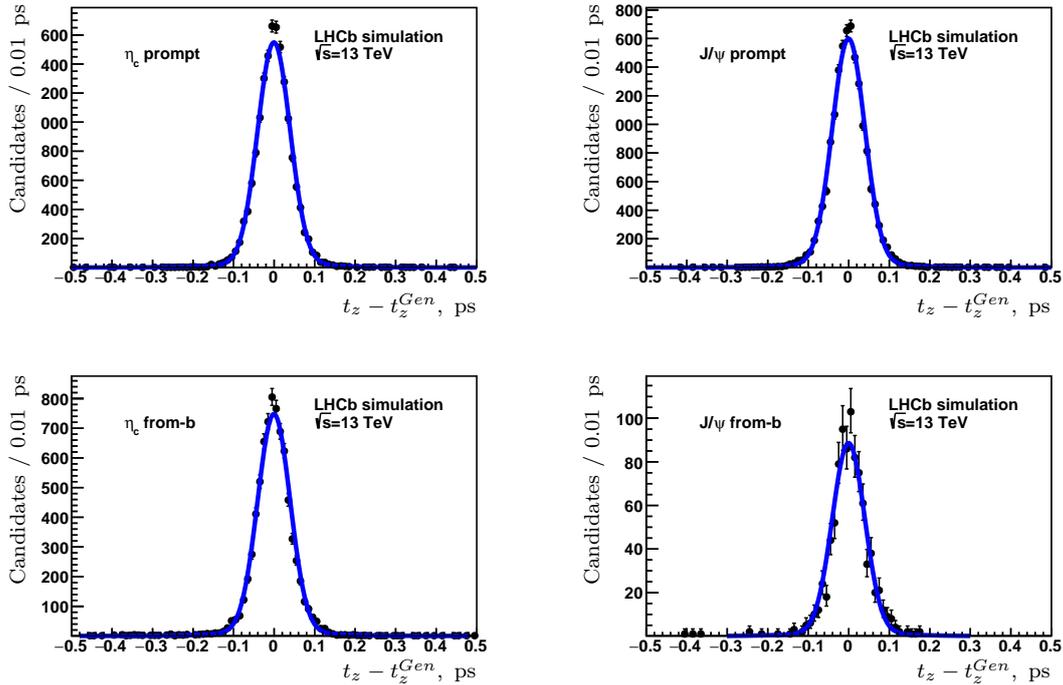

The $t_z$ resolution is described by a double Gaussian function. The resolution model for \etac and \jpsi signals is assumed to be the same, which is confirmed by the fit described below.
Parameters of double Gaussian are extracted from simultaneous unbinned maximum-likelihood fit of four MC samples (prompt \etac, prompt \jpsi, \etac from \bquark-decays and \jpsi from \bquark-decays) to $t_z - t_z^{Gen}$, where $t_z$ is the reconstructed pseudo-proper lifetime and $t_z^{Gen}$ is the pseudo-proper lifetime at generator level (Fig~\ref{fig:fitMC}). 
The ratio of the two Gaussian widths is determined to be $S_w / S_n = 3.10 \pm 0.09$ 
and the fraction of the narrow Gaussian component $\beta$ to be about $95\%$. 

\begin{figure}[t]
\centering
\protect\protect\protect\includegraphics[width=0.95\linewidth]{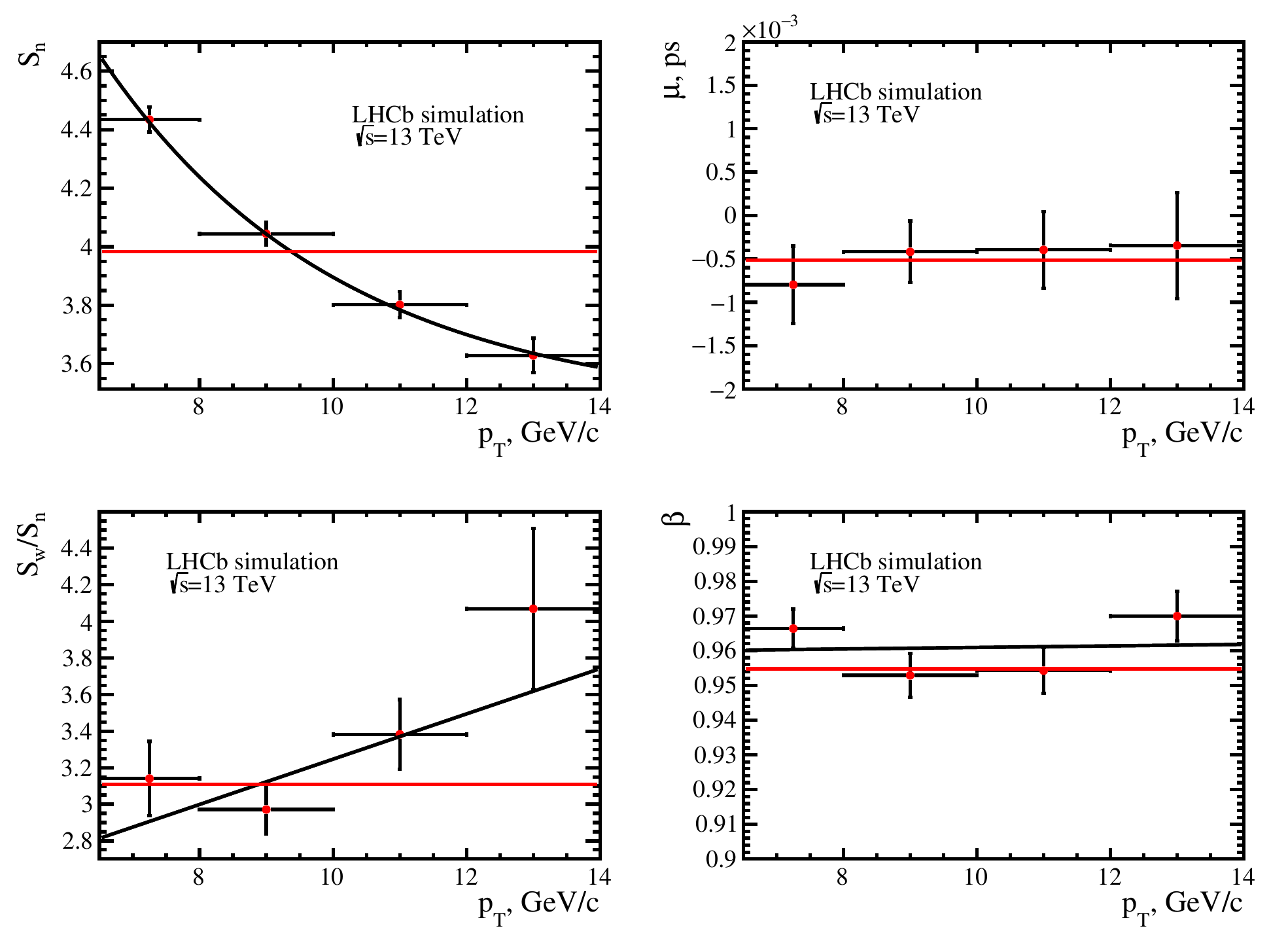}
\put(-275,163){\colorbox{shadecolor}{\pt, \gev}}
%
\put(-275,5){\colorbox{shadecolor}{\pt, \gev}}
%
\put(-60,163){\colorbox{shadecolor}{\pt, \gev}}
%
\put(-60,5){\colorbox{shadecolor}{\pt, \gev}}
\caption
[The \pt-dependences of double Gaussian parameters $S_{\etac}$, $\mu$, $S_w/S_n$ and $\beta$ of the $t_z$ resolution model, as obtained from the simultaneous fit to all four MC samples.]
{The \pt-dependences of double Gaussian parameters $S_{\etac}$ (top left), $\mu$ (to right), $S_w/S_n$ (bottom left) and $\beta$ (bottom right) of the $t_z$ resolution model, as obtained from the simultaneous fit to all four MC samples. Red solid horizontal lines represent values from the fit to the \pt-integrated MC samples. The black line shows the result of the fit for the $S_n$ dependence on $t_z$ (top left).} 
\label{fig:PT_dist_tz}
\end{figure}
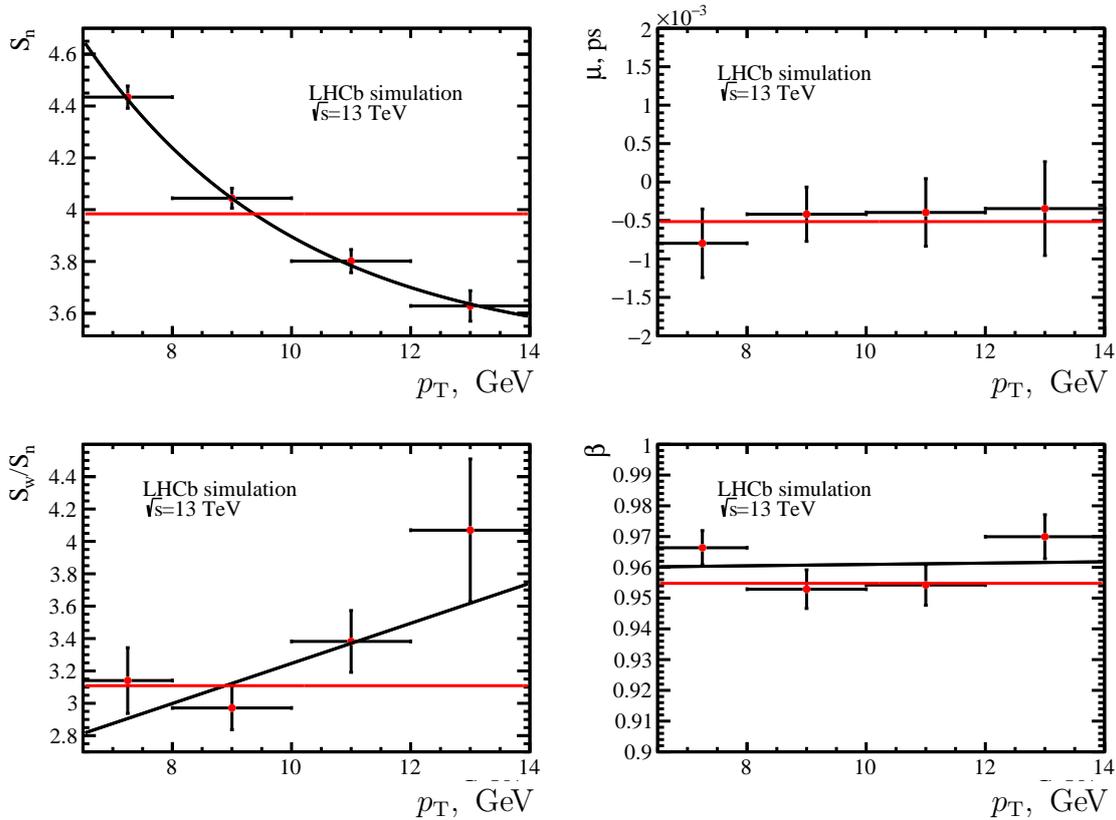

In order to study dependence of $t_z$ resolution model on charmonium transverse momentum, the same fit is performed in bins of $\pt(\ppbar)$. The dependences of $t_z$ bias $\mu$, $S_w / S_n$, $\beta$ on \pt are shown on Fig.~\ref{fig:PT_dist_tz}. No significant \pt-dependence is observed for $\beta$, $S_w / S_n$ and $\mu$, hence no \pt-dependence of these parameters is assumed in the nominal fit to data. 
A notable \pt-dependence of $S_n$ is observed and is described by the sum of an exponential function and a constant. The obtained dependence is used in the $t_z$-fit to data.
The values of double Gaussian parameters from the fit to the \pt-integrated ($6.5 \gev < \pt < 14.0 \gev$) MC samples are shown by red solid horizontal lines.

Signal model includes two components: prompt charmonia, which is parametrised as a $\delta$-function convoluted with a resolution function; and charmonia produced in inclusive \bquark-decays, which is parametrised by a decay function convoluted with a resolution function.

\begin{figure}[t]
\centering
\protect\protect\protect\includegraphics[width=0.85\linewidth]{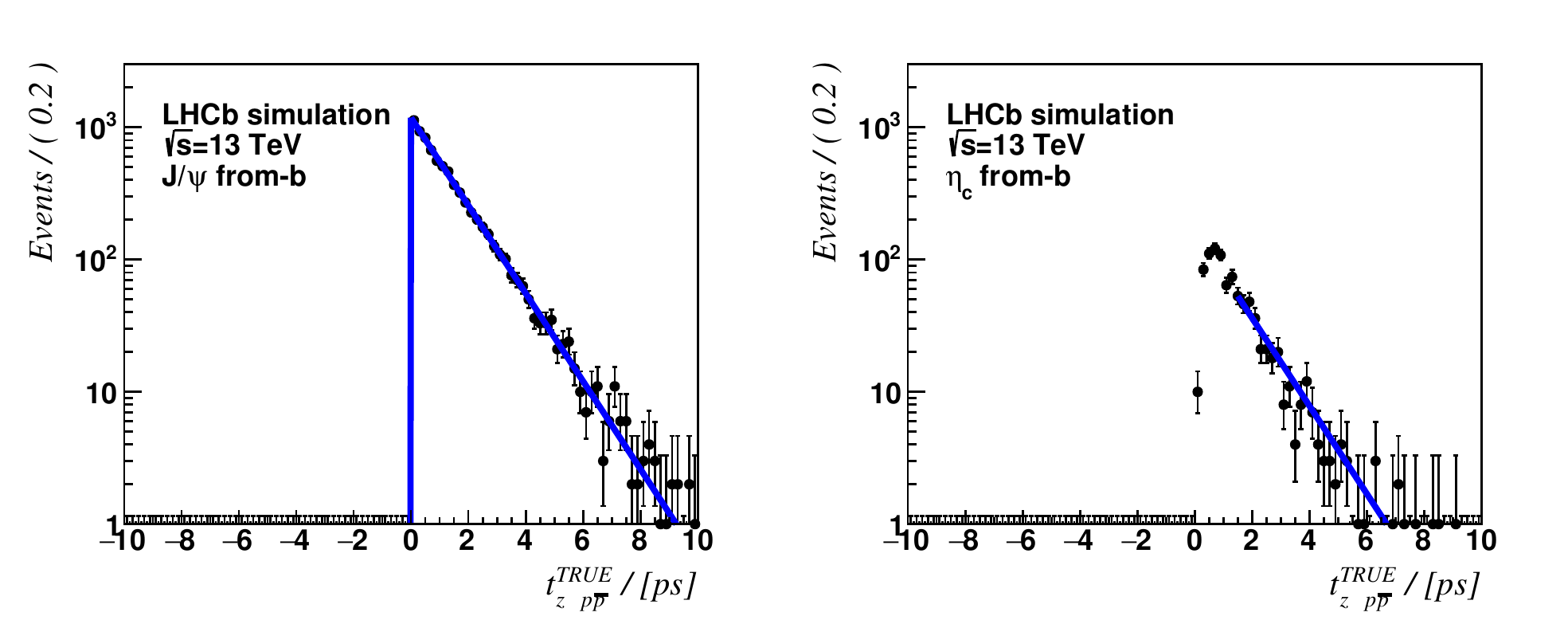}
\put(-384,45){\rotatebox{90}{\colorbox{shadecolor}{\footnotesize Candidates / 0.2 \ps}}}
\put(-253,5){\colorbox{shadecolor}{\footnotesize $t_z^{Gen}$, \ps}} 
\put(-190,45){\rotatebox{90}{\colorbox{shadecolor}{\footnotesize Candidates / 0.2 \ps}}}
\put(-60,5){\colorbox{shadecolor}{\footnotesize $t_z^{Gen}$, \ps}}
\caption
[Distribution of the $t_z^{Gen}$ in the MC samples comprising \jpsi from \bquark-decays and \etac from \bquark-decays.]
{Distribution of the $t_z^{Gen}$ in the MC samples comprising \jpsi from \bquark-decays (left) and \etac from \bquark-decays (right). The solid blue lines represent the result of the simultaneous fit by a decay function to both MC samples.} 
\label{fig:MC_TRUE_Tz}
\end{figure}

\begin{figure}[ht].
\centering
\protect\protect\protect\includegraphics[width=0.5\linewidth]{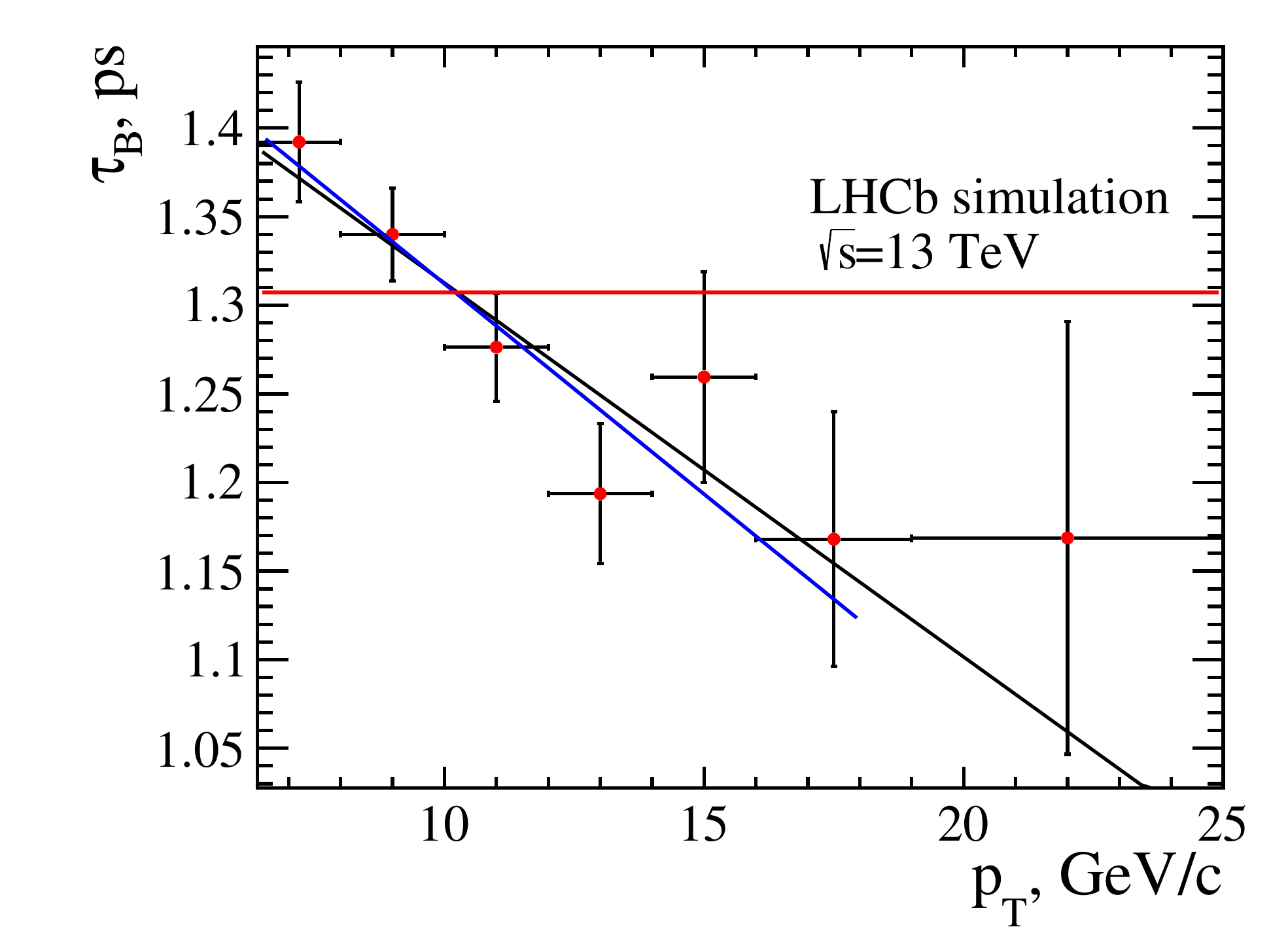}
\put(-213,120){\rotatebox{90}{\colorbox{shadecolor}{$\tau_b$, \ps}}}
\put(-57,3){\colorbox{shadecolor}{\small \pt, \gev}}
\caption
[The \pt dependence of of $\tau_b$ from simultaneous fit to both \etac from \bquark-decays and \jpsi from \bquark-decays MC samples in four \pt bins.]
{The \pt dependence of of $\tau_b$ from simultaneous fit to both \etac from \bquark-decays and \jpsi from \bquark-decays MC samples in four \pt bins. The black and blue lines represent the results of the fit by a linear function in different fit ranges. The red horizontal line shows the value of $\tau_b$ from the fit to \pt-integrated MC samples.} 
\label{fig:tau_vs_pt}
\end{figure}

The \pt-dependence of the exponential slope $\tau_b$ of the decay function for charmonia produced in inclusive \bquark-decays is studied using MC samples of \etac from and \jpsi mesons originating from \bquark-decays. Simultaneous extended maximum-likelihood fit is performed to the $t_z^{Gen}$ distributions. Since the \etac mesons from \bquark-decays are restricted to be produced in the decays of long-lived (with lifetime more than 1 \ps) \bquark-hadrons at the generator level, the fit range for the \etac sample starts at 1.5 \ps. The example of the fit to $t_z^{Gen}$ for \pt-integrated MC samples is shown on Fig.\ref{fig:MC_TRUE_Tz}. The values of $\tau_b$ extracted from the fits to MC in bins of \pt are shown on Fig.\ref{fig:tau_vs_pt}. The \pt-dependence of $\tau_b$ is approximated by a linear function; the shape of this dependence is used in the following fit to data.

\begin{figure}[ht]
\centering
\protect\protect\protect\includegraphics[width=0.5\linewidth]{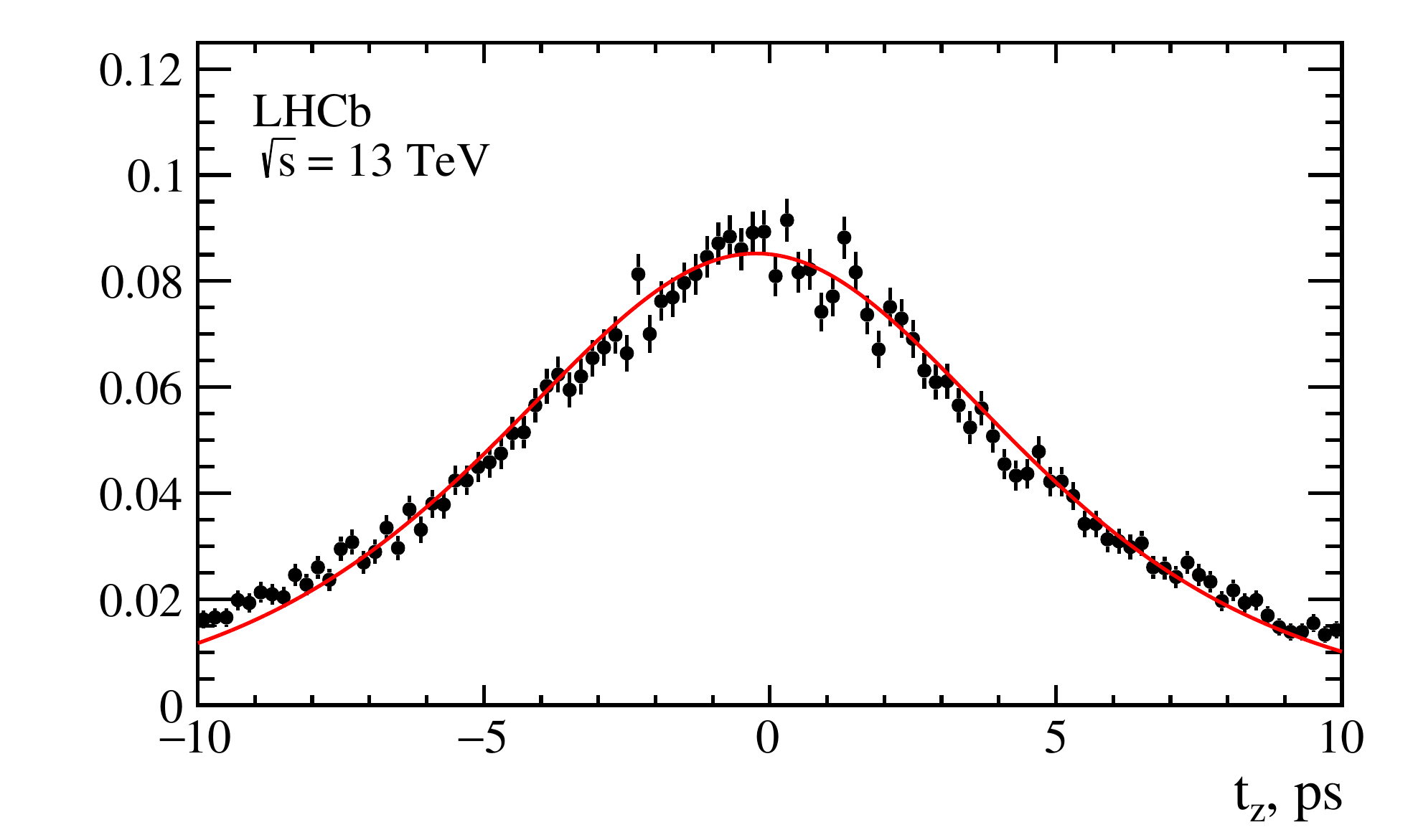}
\put(-190,117){\colorbox{shadecolor}{\scriptsize{\lhcb-ANA-2018-035}}}
\caption
[The $t_z$-distribution for the entire \pt-range from the next-event method.]
{The $t_z$-distribution for the entire \pt-range from the next-event method. The red line represents the fit using non-parametric kernel estimated function.} 
\label{fig:mismatch}
\end{figure}

\clearpage
Events with wrongly assigned PV are taken into account in the fit to data. The shape of the $t_z$-distribution for such events is extracted from data using the next-event method as explained by Eq.~\ref{eq:nextEvent}:
\begin{equation}
t_z^{next} = \frac{(z_{\ppbar}-z_{PV}^{next})\times M_{\ppbar}}{\ptot_{z}},
\label{eq:nextEvent}
\end{equation}
where $z_{\ppbar}$ is the coordinate of \bquark-decay vertex and $z_{PV}^{next}$ is the primary vertex from the next event with the smallest impact parameter with respect to the \bquark-decay vertex of the considered event. The shape is extracted using kernel estimated function in each bin of transverse momentum separately. The example of this shape as obtained for the entire \pt-range is shown on Fig.~\ref{fig:mismatch}. In the fit model to data it is assumed that the shapes and the fractions of the events with wrongly assigned primary vertex are the same for the \etac and \jpsi signals.

Complete description of the $t_z$-fit model is thus given by Eq.~\ref{eq:tzModel}
\begin{equation}
\begin{aligned}
& F^{\etac}(t_z) &= \Big(N_p^{\etac}\delta(t_z)+\frac{N_b^{\etac}}{\tau_b}e^{-t_z/\tau_b}\Big)&\otimes DG(\mu,S_n,S_n/S_w,\beta) + N_m^{\etac} f_{m}(t_z),\\
& F^{\jpsi}(t_z) &= \Big(N_p^{\jpsi}\delta(t_z)+\frac{N_b^{\jpsi}}{\tau_b}e^{-t_z/\tau_b}\Big)&\otimes DG(\mu,S_n,S_n/S_w,\beta) + N_m^{\jpsi} f_{m}(t_z),
\end{aligned}
\label{eq:tzModel}
\end{equation}
where $N_p^{\etac(\jpsi)}$ is the yield of prompt \etac(\jpsi), $N_b^{\etac(\jpsi)}$ is the yield of the \etac(\jpsi) from \bquark-decays, $N_m^{\etac(\jpsi)}$ is the yield of \etac(\jpsi) from events with wrongly assigned primary vertex, $f_{m}(t_z)$ denotes the shape of events with wrongly assigned primary vertex.
The fractions of \jpsi and \etac candidates from events with wrongly assigned PV are assumed to be equal as
\begin{equation}
\frac{N_m^{\jpsi}}{N_b^{\jpsi}+N_p^{\jpsi}}\equiv\frac{N_m^{\etac}}{N_b^{\etac}+N_p^{\etac}}.
\end{equation}

The summary of the $t_z$ fit  parametrisation is given in Table~\ref{tab:tzModel}.
\begin{table}[h!] 
\centering
\small{
\begin{tabular}{l|l} 
 Parameter &  Comment \\ \hline \hline
 $\mu$	                            & Common free parameter \\ \hline	 
 $S_n/S_w$	                        & Fixed from MC \\ \hline
 $\beta$	                          & Fixed from MC \\ \hline
 $S_n$                              & Common free parameter, 
                        \\ & shape of \pt-dependence extracted from MC 
                        \\ & for differential production measurement \\ \hline
 $\tau_b$                           & Common free parameter of average value $\langle \tau_b \rangle$, \\ & shape of \pt-dependence extracted from MC 
              \\ & for differential production measurement \\ \hline
 $N_b^{\jpsi}, N_p^{\jpsi}$         & Free fit parameters \\ \hline
 $N_b^{\etac}/N_b^{\jpsi}, N_p^{\etac}/N_p^{\jpsi}$ & Free fit parameters \\ \hline
 $N_m^{\jpsi}/(N_b^{\jpsi}+N_p^{\jpsi})$ & \\ 
 $N_m^{\etac}/(N_b^{\etac}+N_p^{\etac})$ & Free fit parameters, required to be the same for \jpsi and \etac
\end{tabular}
\caption{Summary on $t_z$-fit parametrisation used in the fit to data.} 
\label{tab:tzModel}
}
\end{table}

\begin{figure}[ht]
\centering
\protect\protect\protect\includegraphics[width=1.0\textwidth]{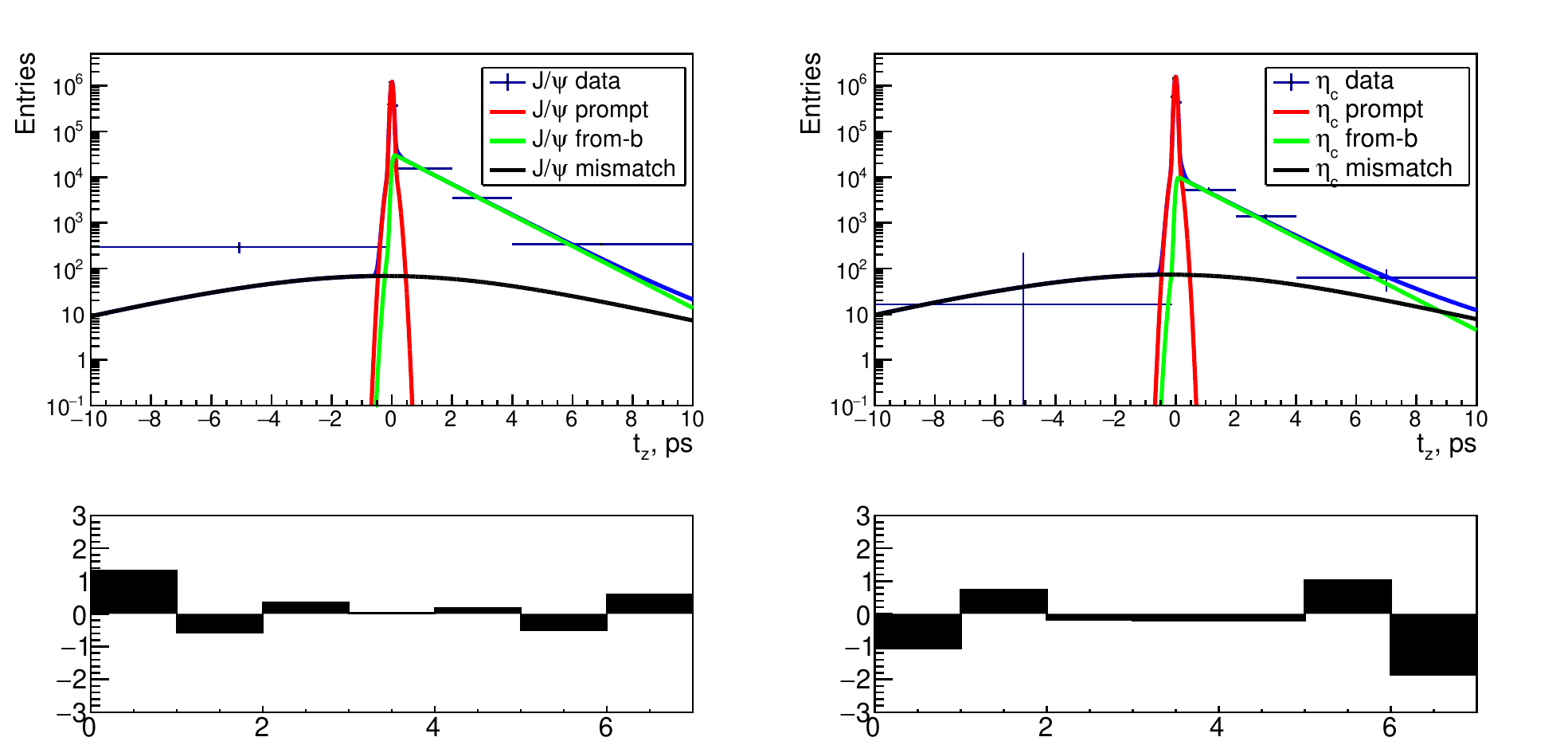}
\put(-46,-7){\scriptsize{$N(t_z)$}}
\put(-274,-7){\scriptsize{$N(t_z)$}}
\put(-192,185){\scriptsize{\lhcb-ANA-2018-035}}
\put(-420,185){\scriptsize{\lhcb-ANA-2018-035}}
\caption
[The $t_z$ distribution for \jpsi and $\etac$ for entire \pt-range $6.5 \gev<\pt<14.0 \gev$ and the result of simultaneous integral \chisq fit. ]
{The $t_z$ distribution for \jpsi (left) and $\etac$ (right) for entire \pt-range $6.5 \gev<\pt<14.0 \gev$ and the result of simultaneous integral \chisq fit. Red lines show prompt components, green lines show $\etac$ and \jpsi from inclusive \bquark-decays, black lines show contributions from the events with wrongly associated PV. The corresponding pull distributions are shown below, where the $N(t_z)$ on the pull distributions denotes the $t_z$ bin number.} 
\label{fig:tzFitInt}
\end{figure}

\begin{figure}[ht]
\centering
\protect\protect\protect\includegraphics[width=0.8\textwidth]{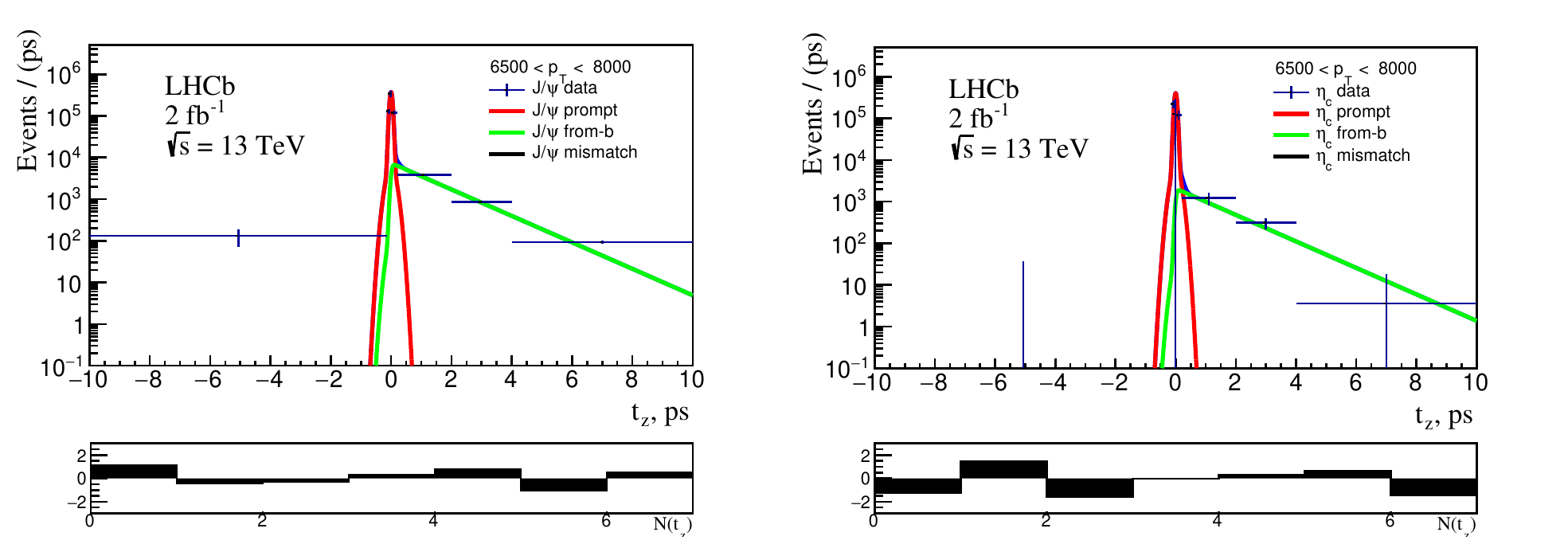}
\vspace{0.3cm}
\protect\protect\protect\includegraphics[width=0.8\textwidth]{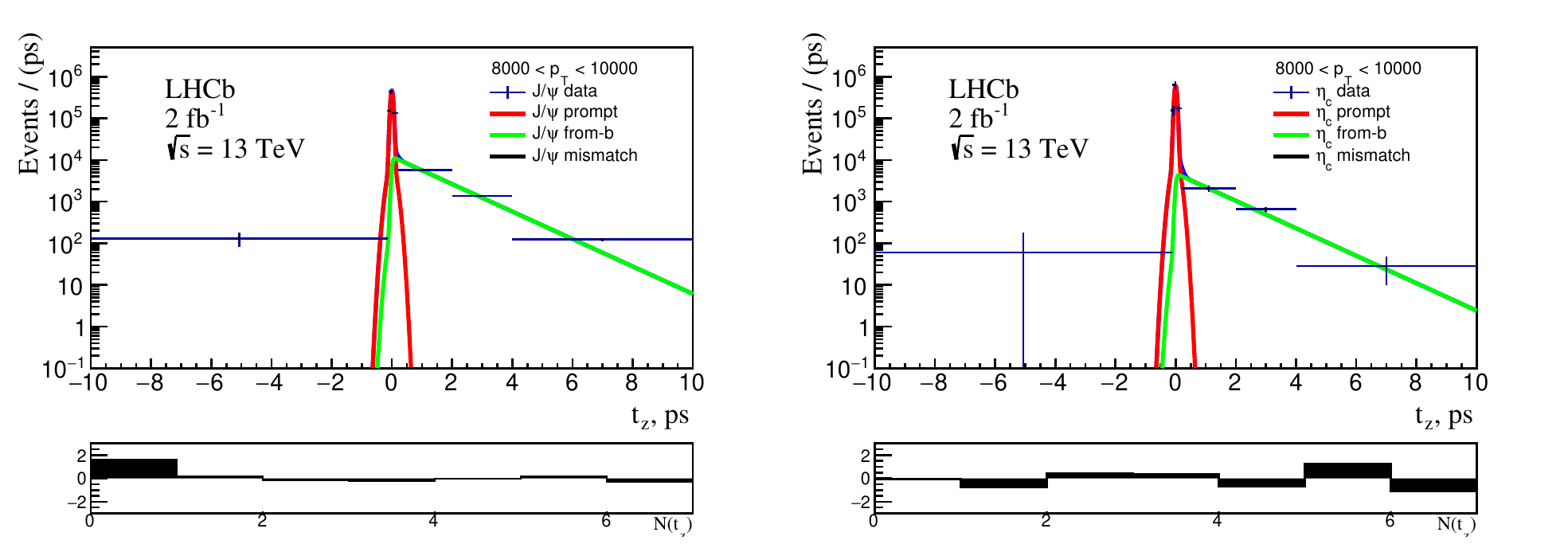}
\vspace{0.3cm}
\protect\protect\protect\includegraphics[width=0.8\textwidth]{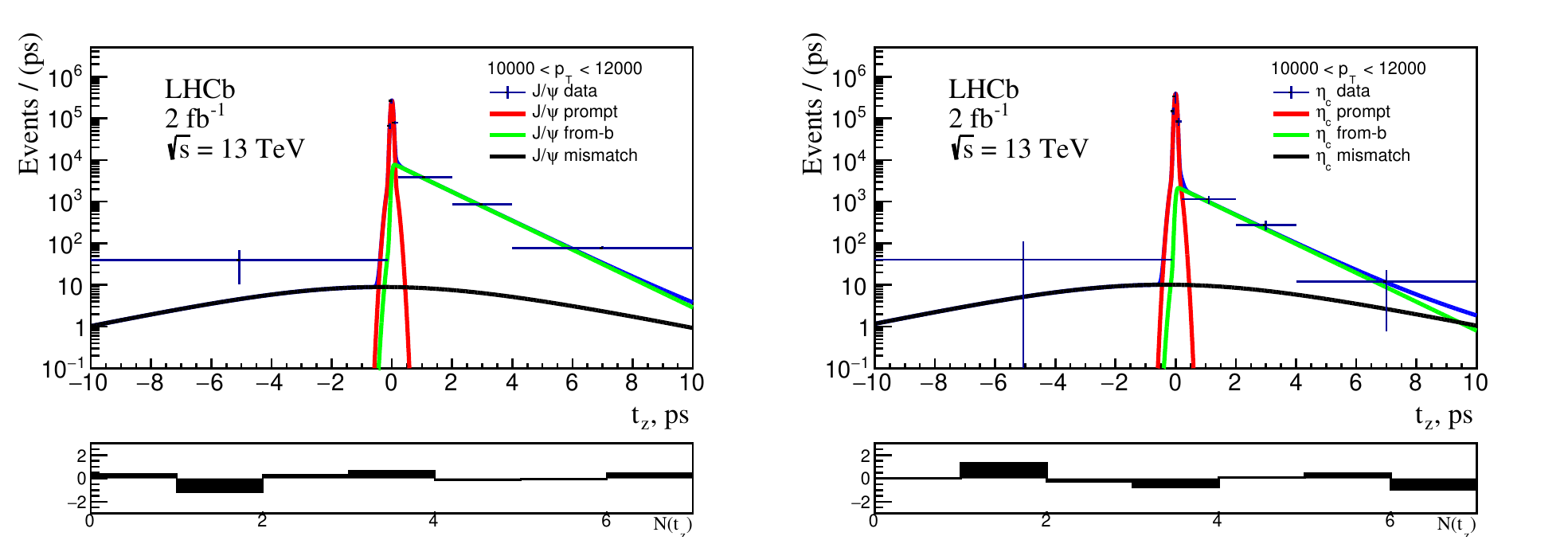}
\vspace{0.3cm}
\protect\protect\protect\includegraphics[width=0.8\textwidth]{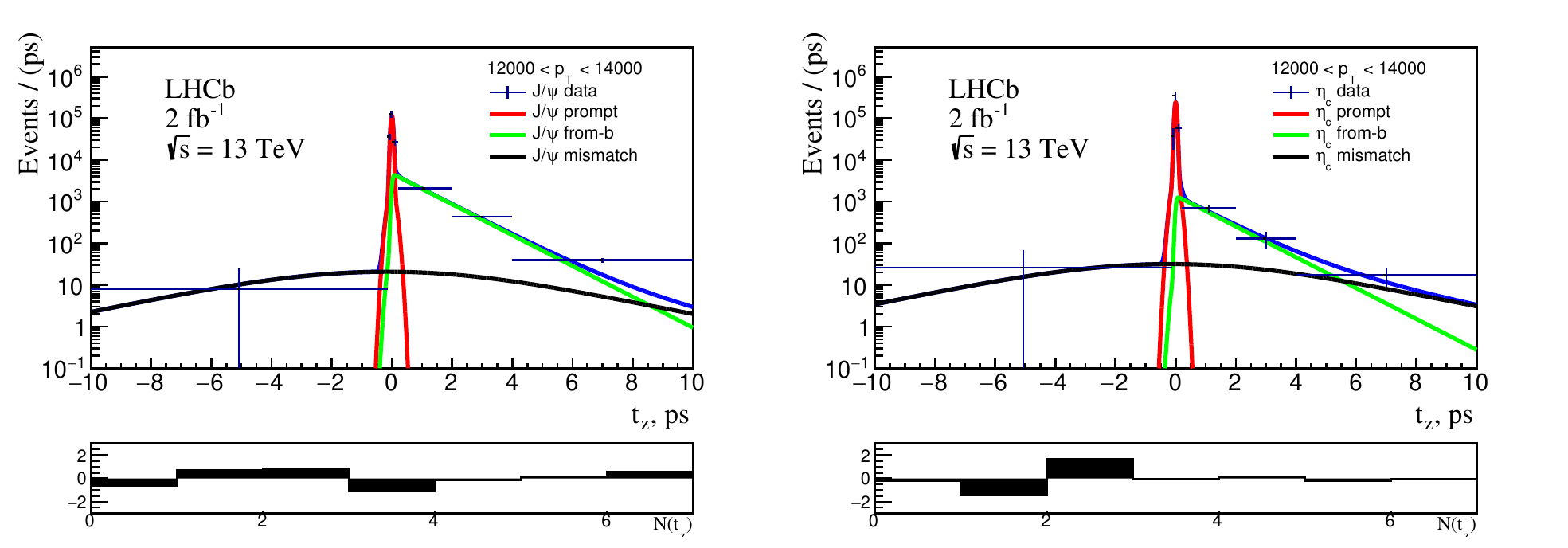}
\caption
[The $t_z$ distribution for \jpsi (left) and $\etac$ (right) for all \pt-bins and the result of simultaneous integral \chisq fit.]
{The $t_z$ distribution for \jpsi (left) and $\etac$ (right) for all \pt-bins and the result of simultaneous integral \chisq fit. Red lines show prompt components; green lines show $\etac$ and \jpsi from inclusive \bquark-decays; black lines show contributions from the events with wrongly associated PV. The corresponding pull distributions are shown below, where the $N(t_z)$ on the pull distributions denotes the $t_z$ bin number} 
\label{fig:tzFitAll}
\end{figure}

Results of the simultaneous fit to $t_z$ for entire \pt-range are shown on Fig.~\ref{fig:tzFitInt}. Results of simultaneous fit to $t_z$ in each bin of \pt are shown on Fig.~\ref{fig:tzFitAll}. Note that pull distributions take into account the integral option of the fit and the centre-of-mass of each bin is evaluated according to the shape of the fit function. Since the fit function is very sharp, the centre-of-masses of bins significantly differs from bins centers.

\newpage
\clearpage

Simultaneous fit yields the following values for the parameters, which are common across the \pt-bins:
\begin{equation}
\begin{aligned}
\mu  &= (-1.3\pm1.8)\times 10^{-3} \ps,\\ 
S_n  &= (4.28\pm0.28)\times 10^{-2} \ps, \\
\langle \tau_b \rangle &= 1.28\pm0.02 \ps.
\end{aligned}
\label{eq:tzSharedValues}
\end{equation}
The value of $\mu$ is in good agreement with the value obtained from MC, $\mu^{MC}  = (-0.5\pm0.2)\times 10^{-3} \ps$, as well as the value of $\langle \tau_b \rangle$, $\langle \tau_b^{MC} \rangle = 1.31\pm0.01\ps$. The value of $S_n$ is also in a good agreement with simulation.

Values of $\tau_b$ parameter are expected to be the same for \etac and \jpsi. This is in agreement with the fact that simultaneous fit with common $\tau_b$ parameter for \etac and \jpsi well describes $t_z^{Gen}$ distributions in the MC samples.

In general, \Bz, \Bs and $\Lambda_b$ lifetimes are close to 1.5 \ps within 2\% accuracy, which is better than the statistical precision of the $\tau_b$ parameter from the fit to the $t_z$ distribution in data. However the \Bp meson has significantly different lifetime of 1.64 \ps. A systematic shift can then impact the results only if the \Bp fraction contributing to the observed inclusive \bquark-decays to \etac is significantly different from that to \jpsi. These considerations allow to perform the following very conservative reasoning. The upper limit of q possible impact corresponds to the difference between the $\tau_b$ for \etac and \jpsi of a value 5\%, which is estimated from the lifetimes of different \bquark-species and the corresponding fragmentation fractions. 

Also, a fit to data with two different $\tau_b$ free parameters for \etac and for \jpsi is consistent with the same $\tau_b$ values for \etac and \jpsi, and also consistent to the fit assuming a 5\% difference in $\tau_b$ values within a large uncertainty of the $\tau_b$ value for \etac: 
\begin{equation}
\begin{aligned}
\tau_b^{\jpsi}   &=  1.28 \pm  0.03 \ps, \\
\tau_b^{\etac}   &=  1.19 \pm  0.12 \ps. 
\end{aligned}
\label{eq:differentTaubFit}
\end{equation}
A cross-check  fit to $t_z$ was performed using the \etac and \jpsi $\tau_b$ values different by 5\%. The results on the yield ratios in \pt bins are shown in Tab.~\ref{tab:taubXcheckPrompt} for  prompt production and in Tab.~\ref{tab:taubXcheckFromB} for production in \bquark-decays. They are consistent with the baseline fit well within statistical uncertainties, where the baseline fit assumes equal $\tau_b$ values for \etac and \jpsi as described above. 
\begin{table}[b]
\centering
\small
\begin{tabular}{c|c|c} 
\pt-range             & Baseline fit  & Fit with 5\% difference between       \\
                      &               & $\tau_b^{\etac}$ and $\tau_b^{\jpsi}$ \\ \hline
$6.5 \gev < \pt < 8.0  \gev$  & $1.08 \pm 0.21$ & $1.08 \pm 0.21$ \\
$8.0 \gev < \pt < 10.0 \gev$  & $1.29 \pm 0.18$ & $1.29 \pm 0.18$ \\
$10.0\gev < \pt < 12.0 \gev$  & $1.46 \pm 0.23$ & $1.47 \pm 0.23$ \\
$12.0\gev < \pt < 14.0 \gev$  & $2.13 \pm 0.40$ & $2.12 \pm 0.40$
\end{tabular} 
\caption{Yield ratio $N_{\etac}^{prompt}/N_{\jpsi}^{prompt}$ of prompt charmonia in bins of \pt for baseline fit and for the fit assuming 5\% difference between $\tau_b^{\etac}$ and $\tau_b^{\jpsi}$.}
\label{tab:taubXcheckPrompt}
\end{table}

\begin{table}[t]
\centering
\small
\begin{tabular}{c|c|c} 
\pt-range             & Baseline fit    & Fit with 5\% difference between       \\
                      &                 & $\tau_b^{\etac}$ and $\tau_b^{\jpsi}$ \\ \hline
$6.5 \gev < \pt < 8.0  \gev$  & $0.281 \pm 0.071$ & $0.263 \pm 0.068$ \\
$8.0 \gev < \pt < 10.0 \gev$  & $0.396 \pm 0.047$ & $0.383 \pm 0.046$ \\
$10.0\gev < \pt < 12.0 \gev$  & $0.277 \pm 0.052$ & $0.270 \pm 0.051$ \\
$12.0\gev < \pt < 14.0 \gev$  & $0.293 \pm 0.073$ & $0.289 \pm 0.073$
\end{tabular} 
\caption{Yield ratio $N_{\etac}^{b-decays}/N_{\jpsi}^{b-decays}$ of charmonia from \bquark-decays in bins of \pt for baseline fit and for the fit assuming 5\% difference between $\tau_b^{\etac}$ and $\tau_b^{\jpsi}$.}
\label{tab:taubXcheckFromB}
\end{table}

The yields of prompt charmonia and charmonia from \bquark-decays in \pt-bins from baseline simultaneous fit result and yields from fit to \pt-integrated data sample are summarised in Table~\ref{tab:yieldsSim}.
\begin{table}[t]
\centering
\small
\begin{tabular}{c|c|c|c|c} 
\pt-range & $N^{prompt}_{\jpsi}$ & $N^{\bquark-decays}_{\jpsi}$ & $\frac{N^{prompt}_{\etac}}{N^{prompt}_{\jpsi}}$ & $\frac{N^{\bquark-decays}_{\etac}}{N^{\bquark-decays}_{\jpsi}}$  \\ \hline
$6.5 \gev < \pt < 8.0  \gev$  & $22650\pm1658$ & $5050\pm182$  & $1.082\pm0.212$ & $0.281\pm0.071$ \\
$8.0 \gev < \pt < 10.0 \gev$  & $25675\pm1494$ & $7943\pm197$  & $1.291\pm0.177$ & $0.396\pm0.047$ \\
$10.0\gev < \pt < 12.0 \gev$  & $13817\pm995$ & $5296\pm152$  & $1.463\pm0.229$ & $0.277\pm0.052$ \\
$12.0\gev < \pt < 14.0 \gev$  & $5712\pm644$   & $2789\pm101$  & $2.124\pm0.401$ & $0.293\pm0.074$ \\ \hline
$6.5 \gev < \pt < 14.0 \gev$  & $68298\pm2545$ & $21282\pm357$ & $1.316\pm0.113$ & $0.331\pm0.030$  \\
\end{tabular} 
\caption
[Yields of prompt charmonia and charmonia from \bquark-decays in \pt-bins from baseline simultaneous fit to $t_z$ and for \pt-integrated sample.]
{Yields of prompt charmonia and charmonia from \bquark-decays in \pt-bins from baseline simultaneous fit to $t_z$ and for \pt-integrated sample.}
\label{tab:yieldsSim}
\end{table}

\clearpage
\subsection{Systematic uncertainties} 
\label{sec:syst}
The systematic uncertainties due to following reasons are considered.
\begin{itemize}
\item Signal description in simultaneous fit to invariant mass:
  \begin{itemize}
  \item Knowledge of the \etac natural width, $\Gamma_{\etac}$;
  \item Invariant mass resolution mismodeling;
  \item \pt-dependence of \etac and \jpsi resolution ratio,
   $\sigma_{\etac}/\sigma_{\jpsi}$
  \\ (relevant for differential cross-section measurement);
  \item Resolution correction factors $\alpha_{t_z}$ in bins of $t_z$;
  \end{itemize}
\item Background description in simultaneous fit to invariant mass:
  \begin{itemize}
  \item Combinatorial background description;
  \item Description of the feed-down from the \JpsiToPpbarPiz decay;
  \end{itemize}
\item Signal description in simultaneous fit to $t_z$:
  \begin{itemize}
  \item Bias $\mu$;
  \item The $t_z$-resolution mismodeling;
  \item Mismodeling of \pt-dependence for $t_z$ resolution;
  \\ (relevant for differential cross-section measurement);
  \item Mismodeling of \pt-dependence of $\tau_b$ distribution 
  \\ (relevant for differential cross-section measurement);
  \end{itemize}
\item The \etac and \jpsi efficiency ratio;
\item Non-zero \jpsi polarisation;
\item Uncertainties on $\BR_{\JpsiToPpbar}$ and $\BR_{\EtacToPpbar}$;
\item Uncertainties on \jpsi production (for absolute \etac production cross-section determination).
\end{itemize}
Systematic uncertainties on \etac production corresponding to signal and background description in fits to the invariant mass and $t_z$ are estimated using alternative fit  parameterisations. Each uncertainty is estimated as a difference between the nominal fit result and the alternative fit result. Bin-to-bin variations of uncorrelated systematic uncertainties remain small compared to statistical uncertainty. Since there are no physics reasons for these variations, they are interpreted as fluctuations and therefore bin contents are smoothed in order to reduce the effect of fluctuations.

\subsubsection{Uncertainties related to signal and background shape description in the fit to invariant mass}
The uncertainty corresponding to the knowledge of the \etac natural width is estimated by comparing the results of the simultaneous fit to invariant mass for the \pt-integrated data sample, when $\Gamma_{\etac}$ is set to the world average value of 31.8\mev~\cite{PDG2017} and when $\Gamma_{\etac}$ is set to the value of 34.0 \mev from the analysis of $\decay{\Bp}{\ppbar \Kp}$~\cite{LHCb-PAPER-2016-016}.
This uncertainty is correlated between \pt-bins. Therefore the relative systematic uncertainty obtained from \pt-integrated data sample is taken as an estimate of a relative systematic uncertainty in each bin.

The uncertainty corresponding a mismodeling of the invariant mass resolution is estimated by alternatively describing the detector resolution using a symmetric double-sided Crystal Ball function. The tail parameters and the ratio of \etac and \jpsi resolution parameters are extracted from the fit to MC samples and fixed in bins of \pt. The \pt dependence of the resolution parameter for the \etac signal is extracted from MC similarly to that is done for nominal fit parametrisation. This uncertainty is correlated between \pt-bins. Therefore the relative systematic uncertainty obtained from \pt-integrated data sample is taken as an estimate of a relative systematic uncertainty in each bin.

\begin{figure}[t]
\centering
\protect\protect\protect\includegraphics[width=0.8\linewidth]{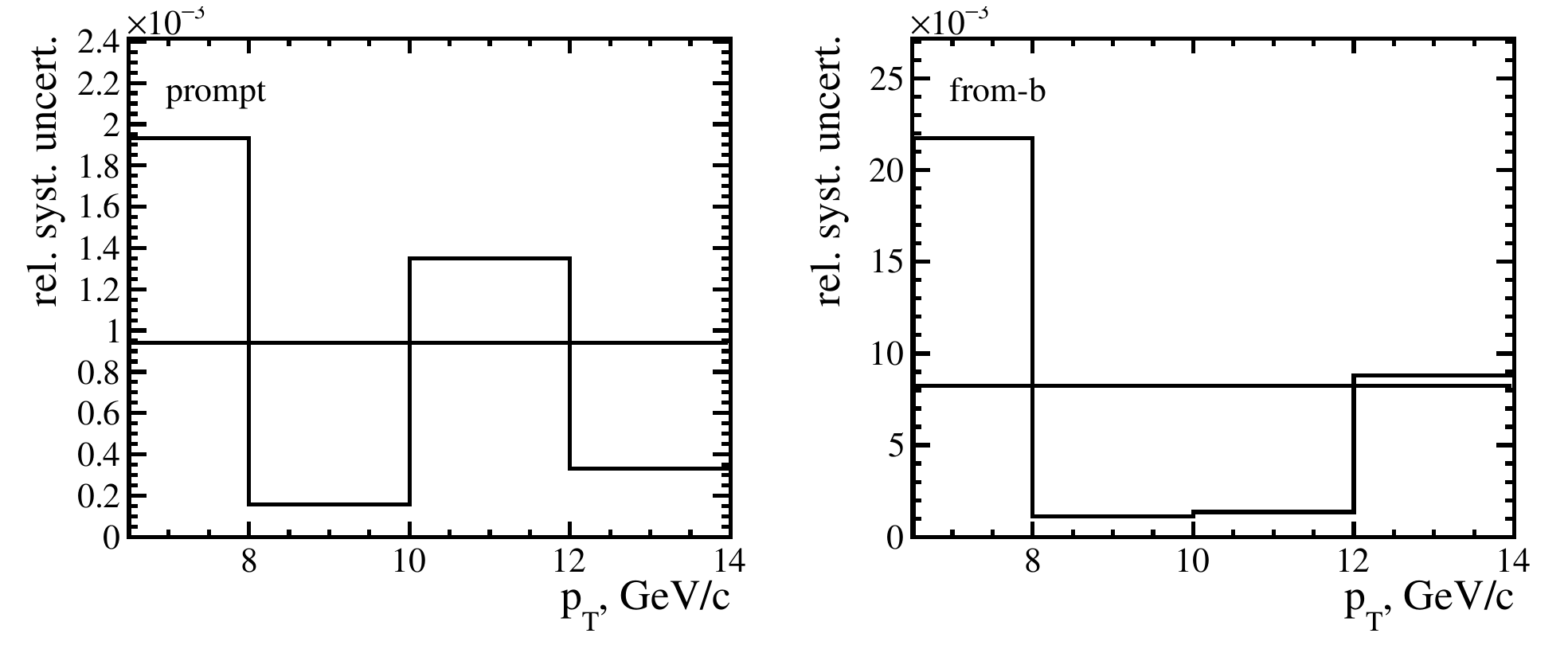}
\put(-95,129){\scriptsize{\lhcb-ANA-2018-035}}
\put(-276,129){\scriptsize{\lhcb-ANA-2018-035}}
\caption
[Relative systematic uncertainty due to the \pt-dependence of \etac and \jpsi resolution ratio in bins of \pt.]
{Relative systematic uncertainty due to the \pt-dependence of \etac and \jpsi resolution ratio in bins of \pt. The solid black line shows a smoothing curve.} 
\label{fig:smoothSystResoRatio}
\end{figure}

The uncertainty related to the \pt-dependence of the \etac and \jpsi resolution ratio is estimated by introducing a linear dependence  of $\sigma_{\etac}/\sigma_{\jpsi}$ as a function of \pt. The slope of the linear function is extracted from MC.
This systematic effect is relevant for differential cross-section measurement.
This uncertainty is parametrised as a constant in all bins as shown on Fig.~\ref{fig:smoothSystResoRatio}.



\begin{figure}[t]
\centering
\protect\protect\protect\includegraphics[width=0.8\linewidth]{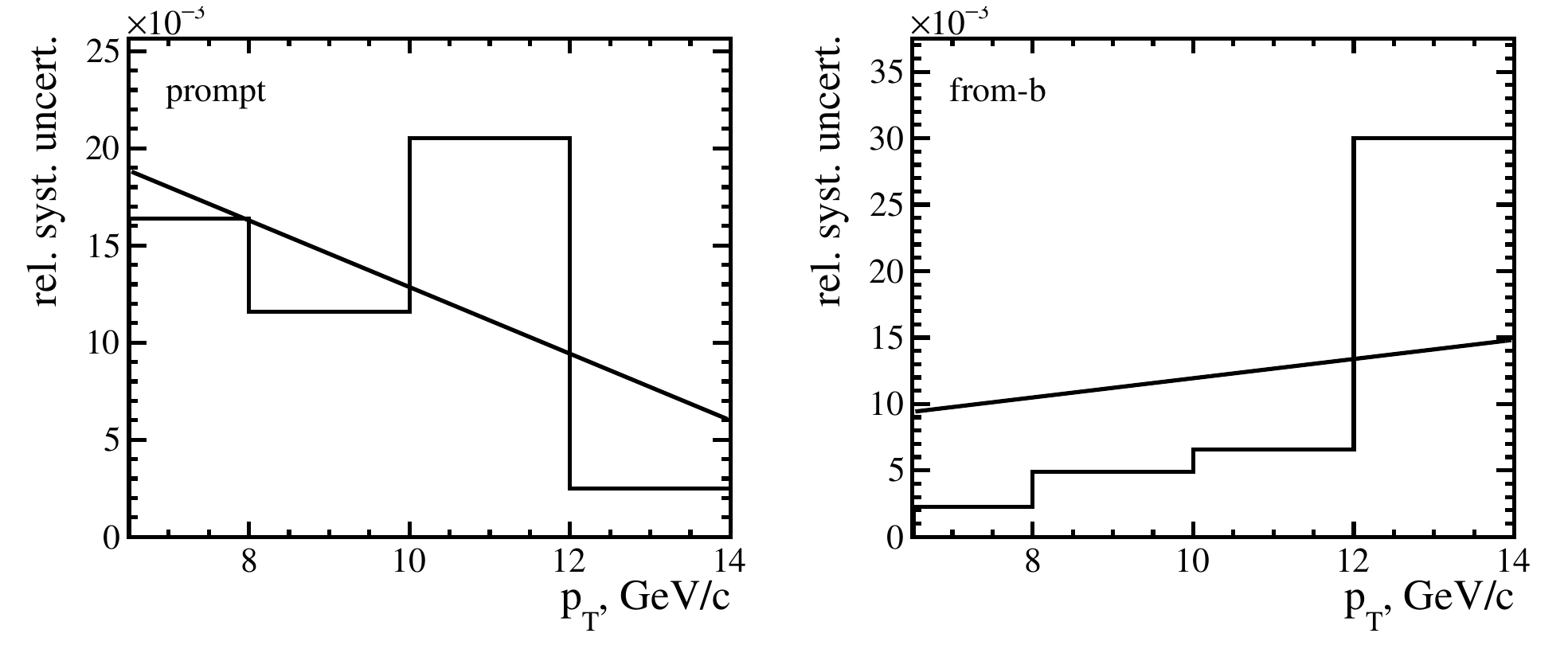}
\put(-95,129){\scriptsize{\lhcb-ANA-2018-035}}
\put(-276,129){\scriptsize{\lhcb-ANA-2018-035}}
\caption
[Relative systematic uncertainty due to resolution correction factors $\alpha_{t_z}$ in bins of \pt.]
{Relative systematic uncertainty due to resolution correction factors $\alpha_{t_z}$ in bins of \pt. The solid black line shows a smoothing curve.} 
\label{fig:smoothSystAlphaTz}
\end{figure}

The uncertainty corresponding to the resolution correction factors $\alpha_{t_z}$ is estimated by parametrising $t_z$-dependence of $\alpha_{t_z}$ by a linear function. Parameters of the linear function are extracted from the fit to MC. This uncertainty is parametrised to be linearly dependent on the bin as shown on Fig.~\ref{fig:smoothSystAlphaTz}.

\begin{figure}[b]
\centering
\protect\protect\protect\includegraphics[width=0.8\linewidth]{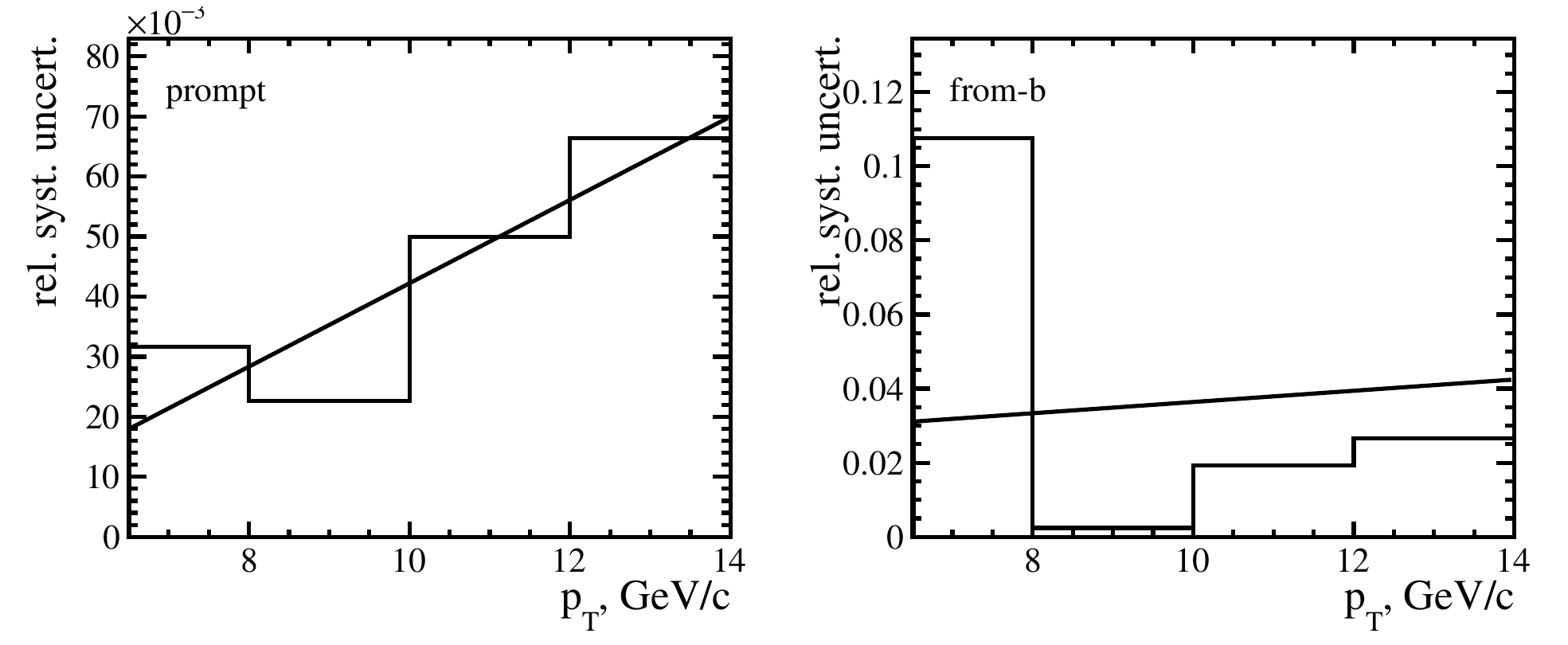}
\put(-95,129){\scriptsize{\lhcb-ANA-2018-035}}
\put(-276,129){\scriptsize{\lhcb-ANA-2018-035}}
\caption
[Relative systematic uncertainty due to combinatorial background description in bins of \pt.]
{Relative systematic uncertainty due to combinatorial background description in bins of \pt. The solid black line shows a smoothing curve.} 
\label{fig:smoothSystCombBkg}
\end{figure}

The uncertainty corresponding to combinatorial background description is estimated via  an alternative combinatorial background parametrisation with a third-order polynomial function. This uncertainty is parametrised to be linearly dependent on the bin as shown on Fig.~\ref{fig:smoothSystCombBkg}.

\begin{figure}[t]
\centering
\protect\protect\protect\includegraphics[width=0.8\linewidth]{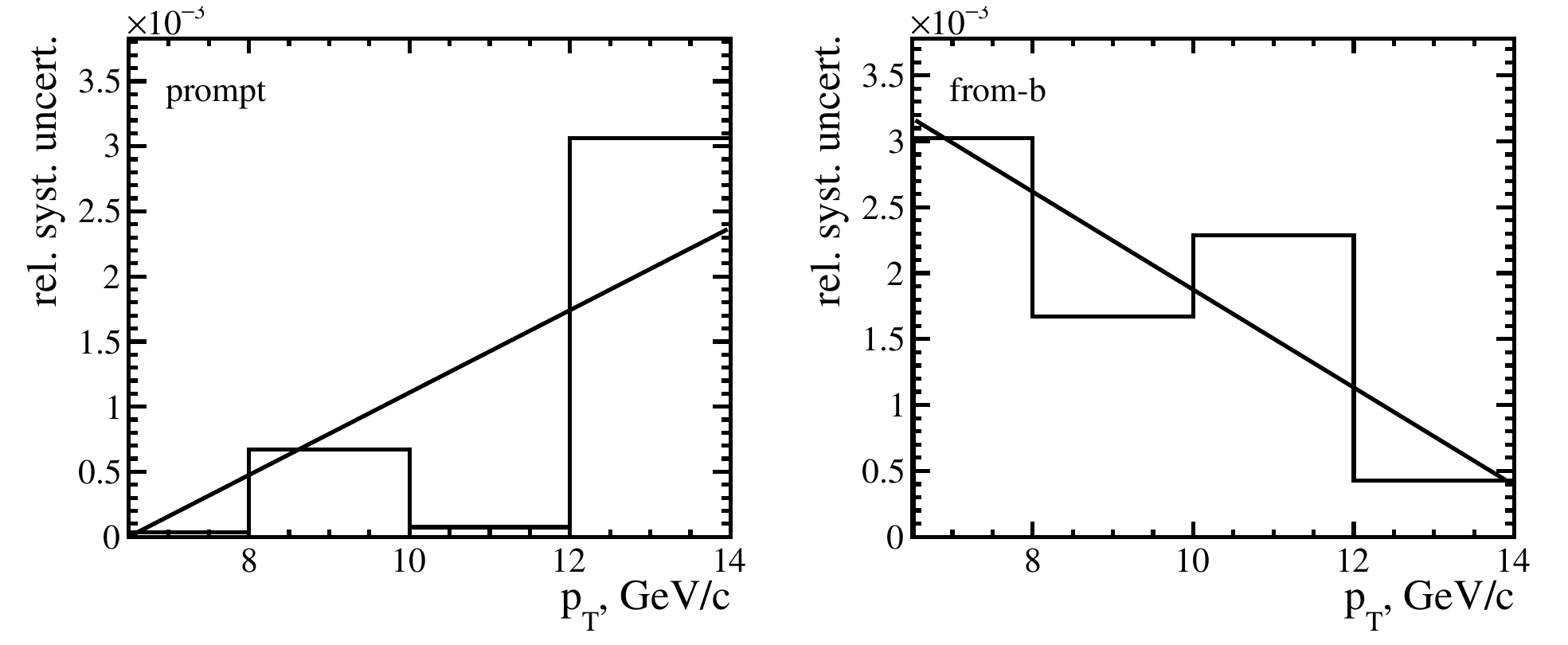}
\put(-95,129){\scriptsize{\lhcb-ANA-2018-035}}
\put(-276,129){\scriptsize{\lhcb-ANA-2018-035}}
\caption
[Relative systematic uncertainty due to description of the feed-down from the \JpsiToPpbarPiz decay in bins of \pt.]
{Relative systematic uncertainty due to description of the feed-down from the \JpsiToPpbarPiz decay in bins of \pt. The solid black line shows a smoothing curve.} 
\label{fig:smoothSystPPPi0}
\end{figure}

The uncertainty corresponding to the description of the feed-down from the \JpsiToPpbarPiz decay is estimated by shifting the value of the efficiency ratio $\epsilon_{\JpsiToPpbarPiz}/\epsilon_{\JpsiToPpbar}=\epsRatioJpsipppizJpsipp$ by its standard deviation and by shifting the value of the branching fraction ratio $\BR_{\JpsiToPpbar}/\BR_{\JpsiToPpbarPiz}$ according to the uncertainty from Ref.~\cite{PDG2017}. This uncertainty is parametrised to be linearly dependent on the bin as shown on Fig.~\ref{fig:smoothSystPPPi0}.

\subsubsection{Uncertainties related to signal description in the fit to $t_z$}
\begin{figure}[t]
\centering
\protect\protect\protect\includegraphics[width=0.8\linewidth]{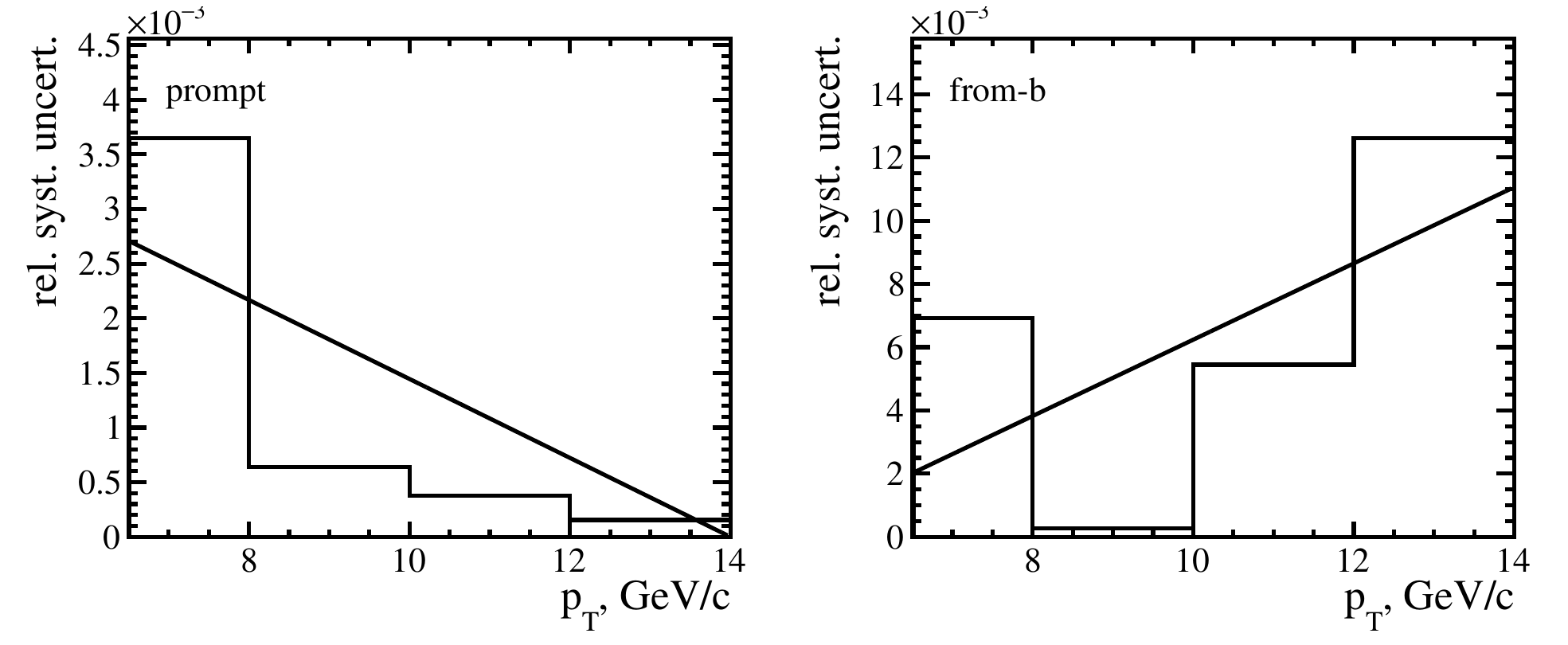}
\put(-95,129){\scriptsize{\lhcb-ANA-2018-035}}
\put(-276,129){\scriptsize{\lhcb-ANA-2018-035}}
\caption
[Relative systematic uncertainty due to $t_z$-resolution mismodeling in bins of \pt.]
{Relative systematic uncertainty due to $t_z$-resolution mismodeling in bins of \pt. The solid black line shows a smoothing curve.} 
\label{fig:smoothSystTzResoModel}
\end{figure}

The uncertainty corresponding to the $t_z$ resolution mismodeling is estimated by introducing a linear \pt-dependence of $S_w/S_n$ and $\beta$ parameters extracted from simulation. This uncertainty is parametrised to be linearly dependent on the bin as shown on Fig.~\ref{fig:smoothSystTzResoModel}.

\begin{figure}[b]
\centering
\protect\protect\protect\includegraphics[width=0.8\linewidth]{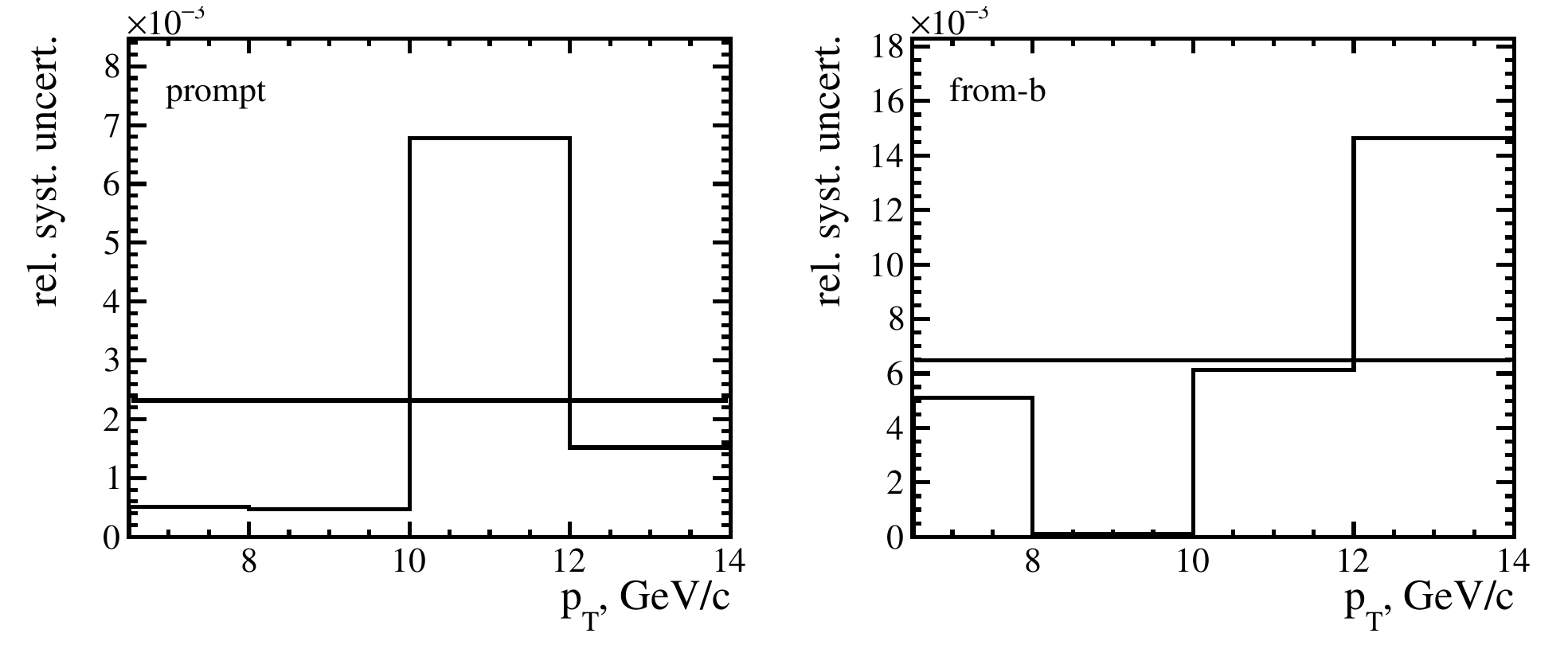}
\put(-95,129){\scriptsize{\lhcb-ANA-2018-035}}
\put(-276,129){\scriptsize{\lhcb-ANA-2018-035}}
\caption
[Relative systematic uncertainty due to bias $\mu$ in bins of \pt.]
{Relative systematic uncertainty due to bias $\mu$ in bins of \pt. The solid black line shows a smoothing curve.} 
\label{fig:smoothSystMu}
\end{figure}

The uncertainty corresponding to the bias $\mu$ is estimated by alternatively setting $\mu\equiv0$ in the fit to $t_z$. This uncertainty is parametrised as a constant in all bins as shown on Fig.~\ref{fig:smoothSystMu}.

\begin{figure}[t]
\centering
\protect\protect\protect\includegraphics[width=0.8\linewidth]{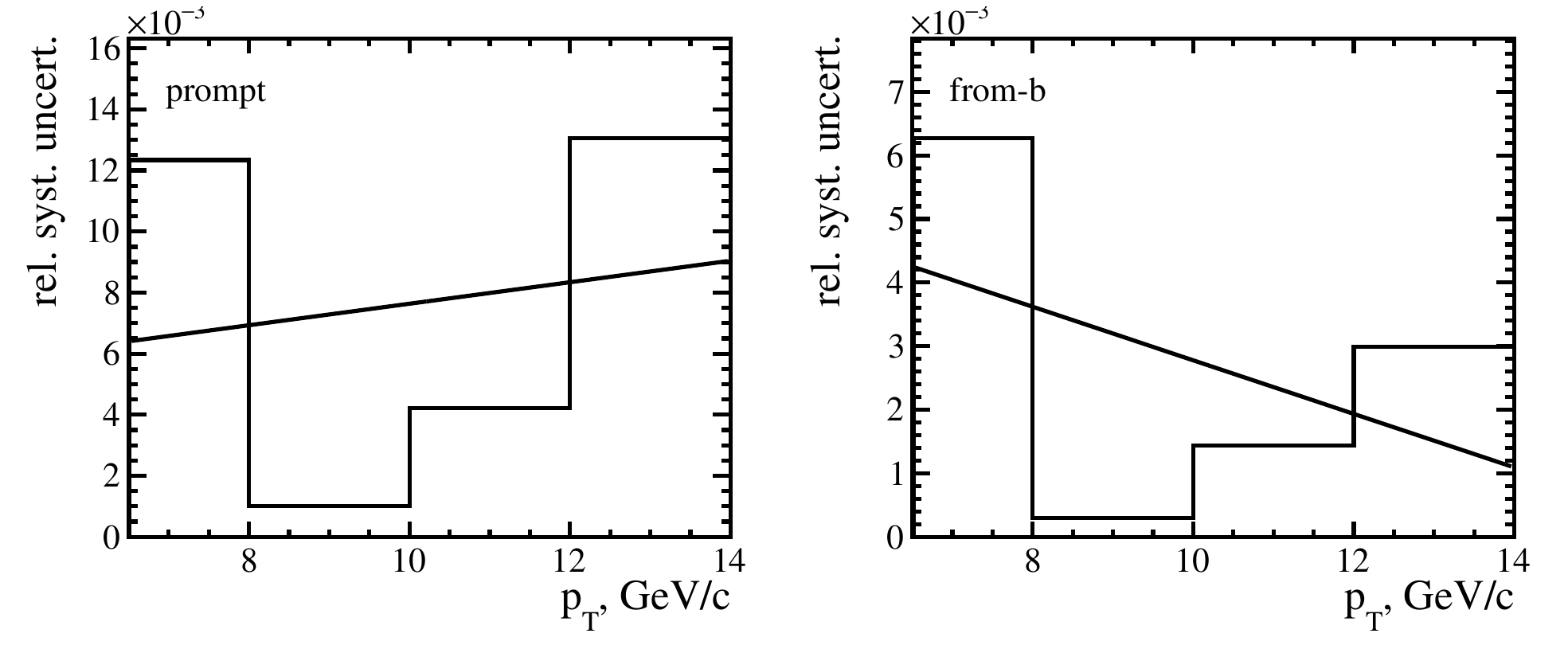}
\put(-95,129){\scriptsize{\lhcb-ANA-2018-035}}
\put(-276,129){\scriptsize{\lhcb-ANA-2018-035}}
\caption
[Relative systematic uncertainty due to mismodeling of \pt-dependence of $t_z$ resolution in bins of \pt.]
{Relative systematic uncertainty due to mismodeling of \pt-dependence of $t_z$ resolution in bins of \pt. The solid black line shows a smoothing curve.} 
\label{fig:smoothSystTzResoPt}
\end{figure}

The uncertainty corresponding to mismodeling of the \pt-dependence of $t_z$ resolution is estimated by alternatively parametrising its shape as a sum of two exponential functions. Parameters of this shape are extracted from the fit to simulation. This uncertainty is relevant for the differential cross-section measurement. This uncertainty is parametrised to be linearly dependent on the bin as shown on Fig.~\ref{fig:smoothSystTzResoPt}.

\begin{figure}[t]
\centering
\protect\protect\protect\includegraphics[width=0.8\linewidth]{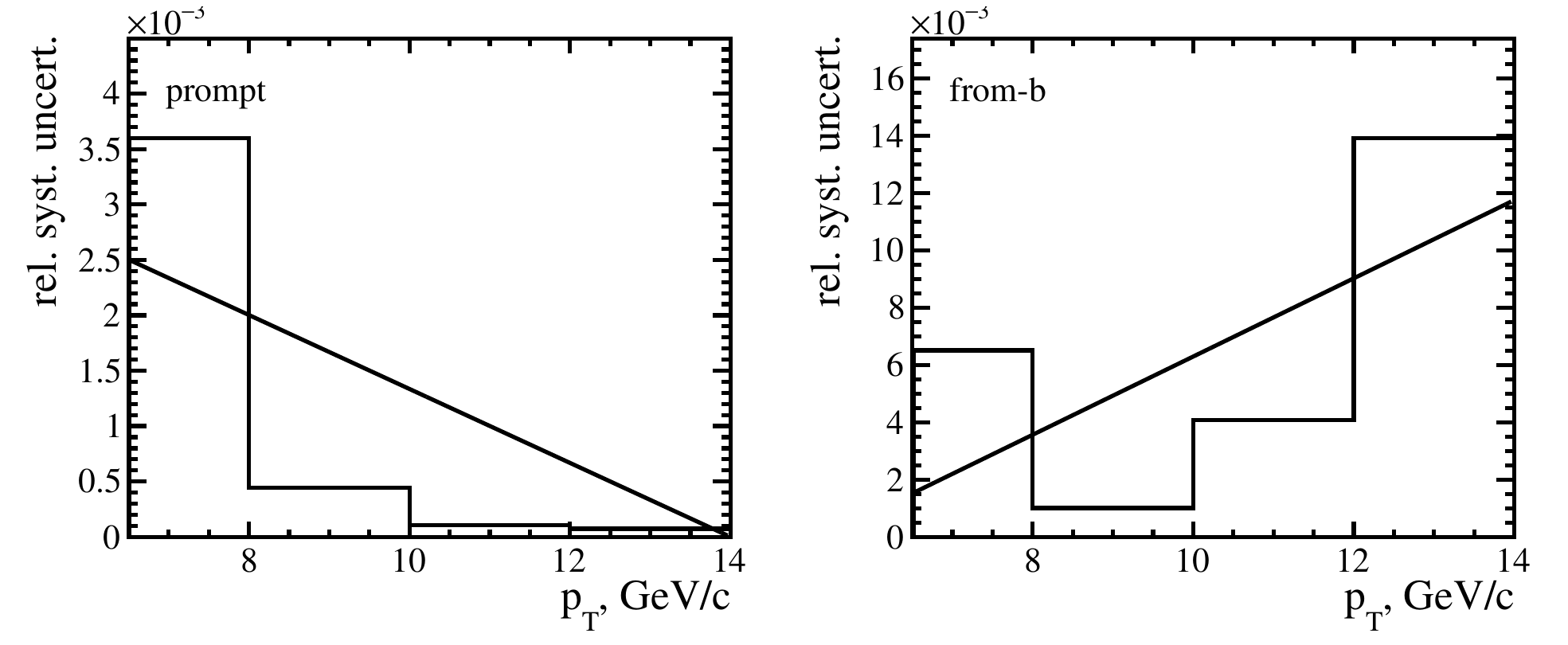}
\put(-95,129){\scriptsize{\lhcb-ANA-2018-035}}
\put(-276,129){\scriptsize{\lhcb-ANA-2018-035}}
\caption
[Relative systematic uncertainty due to mismodeling of \pt-dependence of $\tau_b$ distribution in bins of \pt.]
{Relative systematic uncertainty due to mismodeling of \pt-dependence of $\tau_b$ distribution in bins of \pt. The solid black line shows a smoothing curve.}
\label{fig:smoothSystTau}
\end{figure}

The uncertainty corresponding to mismodeling of the \pt-dependence of $\tau_b$ is 
estimated by parametrising its shape using a linear function extracted from the fit to simulation in the extended fit range. This uncertainty is relevant for the differential cross-section measurement. This uncertainty is parametrised to be linearly dependent on the bin as shown on Fig.~\ref{fig:smoothSystTau}.

The uncertainty corresponding to the \etac and \jpsi efficiency ratio is estimated via changing the \etac and \jpsi efficiency ratio by the uncertainty corresponding to the MC sample sizes.

Possible non-zero polarisation of prompt \jpsi mesons affects their reconstruction efficiency. The \jpsi polarisation has not been measured yet by \lhcb at the \sqs=13~\tev centre-of-mass energy. The \lhcb experiment studied \jpsi polarisation at \sqs=7~\tev~\cite{LHCb-PAPER-2013-008}. Small non-zero longitudinal polarisation was measured with no significant polarisation dependence on transverse momentum or rapidity observed. Small polarisation was also observed by \alice experiment in the forward kinematical regime at \sqs=7~\tev~\cite{Abelev:2011md}. The \cms experiment measured the \jpsi polarisation to be small for $|y|<1.2$ at \sqs=7~\tev~\cite{Chatrchyan:2013cla}. 
The uncertainty of the present measurement is estimated by reweighting prompt \jpsi simulation sample using the following weights:
\begin{equation}
\frac{3}{ 4 \pi \times ( 3 - \lambda_{\theta} )} \times ( 1-\lambda_{\theta}\cos^{2}\theta ) \ ,
\end{equation}
where $\theta$ is the angle between the proton direction in the \jpsi rest frame and the \jpsi boost axis and $\lambda_{\theta}$ is the polarisation parameter.
To estimate the systematic uncertainty the MC sample is reweighted using a typical value of $\lambda_{\theta}=\pm0.1$, as suggested by \jpsi production cross-section measurement at the \sqs=13 \tev~\cite{LHCb-PAPER-2015-037}. This uncertainty is correlated between \pt-bins.

Systematic uncertainties on relative \etac production measurement in the entire $6.5 \gev<\pt<14.0 \gev$ range are shown in Table~\ref{tab:systTotal}.
Detailed tables of systematic uncertainties for each bin of \pt are given in Tables~\ref{tab:systPT1},~\ref{tab:systPT2},~\ref{tab:systPT3} and ~\ref{tab:systPT4}.

The dominant source of uncorrelated systematic uncertainty for prompt \etac production
is related to combinatorial background description. The dominant sources of uncorrelated systematic uncertainties on \etac production in \bquark-decays are related to combinatorial background description and the \pt-dependence of the \etac and \jpsi resolution ratio. The dominant source of correlated systematic uncertainties on both prompt \etac production and \etac production in \bquark-decays is related to knowledge of \etac natural width and the invariant mass resolution model.

Uncertainties on the branching fractions of the \JpsiToPpbar 
and \EtacToPpbar  decay modes are taken into account to estimate corresponding uncertainties in the production cross-section measurements. They are combined in a separate systematic uncertainty, correlated between the bins of transverse momentum. The uncertainty consists of two separate uncertainties on $\BR_{\JpsiToPpbar}=\brJpsipp$ and $\BR_{\EtacToPpbar}=\brEtacpp$~\cite{PDG2017} and amounts to about 10\%.

When extracting the absolute \etac production cross-section values, the uncertainties on the measured \jpsi production cross-section~\cite{LHCb-PAPER-2015-037} are also taken into account.

\clearpage
\newpage

\clearpage

\clearpage

\clearpage

\begin{table}[t]
\centering
\small
\begin{tabular}{c|c|c} 
& $N^{prompt}_{\etac}/N^{prompt}_{\jpsi}$ & $N^{\bquark-decays}_{\etac}/N^{\bquark-decays}_{\jpsi}$  \\ \hline \hline
Mean value                    & 1.316 & 0.331 \\ \hline \hline
Stat. uncertainty             & 8.6   & 9.2 \\ \hline  
$\alpha_{t_z}$ corrections    & 1.7   & 0.3 \\   
Mass resolution model         & 3.0   & 3.8 \\  
Comb. bkg. description        & 3.4   & 1.7 \\
Variation of $\Gamma(\etac)$  & 5.2   & 5.1 \\  
Contribution from \JpsiToPpbarPiz     & $<0.1$ & 0.7 \\
\hline     
Bias $\mu$                         & 0.7    & 0.2  \\   
$t_z$-resolution model        & $<0.1$ & $<0.1$  \\   
\hline          
\jpsi polarisation  & 1.8 & $-$ \\   
\hline \hline  
Total systematic            & 7.4 & 6.6 \\    
\hline \hline  
\end{tabular} 
\caption
{Mean values and relative uncertainties (in \%) in the \etac and \jpsi yield ratios for \pt-integrated $6.5 \gev<\pt<14.0 \gev$ data sample.}
\label{tab:systTotal}
\end{table} 
\clearpage

\begin{table}[t]
\centering
\small
\begin{tabular}{c|c|c} 
& $N^{prompt}_{\etac}/N^{prompt}_{\jpsi}$ & $N^{\bquark-decays}_{\etac}/N^{\bquark-decays}_{\jpsi}$  \\ \hline \hline
Mean value                              & 1.082 & 0.281  \\ \hline \hline
Stat. uncertainty                       & 19.6  & 25.4  \\ \hline     
$\alpha_{t_z}$ corrections              & 1.8 & 1.0   \\   
\pt-dependence of $\sigma_{\etac}/\sigma_{\jpsi}$ & 0.1 & 0.8  \\   
Comb. bkg. description                                & 2.3   & 3.2  \\ 
Contribution from \JpsiToPpbarPiz                     &$<0.1$ & 0.3  \\  
\hline  
\pt-dependence of $t_z$ resolution                    & 0.7 & 0.4  \\   
\pt-dependence of $\tau_{B}$                          & 0.2 & 0.3  \\   
Bias $\mu$                                                 & 0.3 & 0.2  \\   
$t_z$-resolution model                                & 0.2 & 0.3  \\   
\hline  
\hline  
Total systematic uncorrelated                                       & 3.0 & 3.6  \\   
\hline  
\hline  
\jpsi polarisation                                    & 2.1 & $-$  \\   
Mass resolution model                                 & 3.0 & 3.8  \\   
Variation of $\Gamma_{\etac}$                         & 5.2 & 5.1  \\   
\hline  
\hline  
Total systematic correlated                                         & 6.4 & 6.4  \\   
\hline  
\hline  
Total systematic                                              & 7.0 & 7.3 \\ 
\hline  
\hline     
\end{tabular} 
\caption{Mean values and relative uncertainties (in \%) in the \etac and \jpsi yield ratios for $6.5 \gev<\pt<8.0 \gev$.}
\label{tab:systPT1}
\end{table}

\begin{table}[b]
\centering
\small
\begin{tabular}{c|c|c} 
& $N^{prompt}_{\etac}/N^{prompt}_{\jpsi}$ & $N^{\bquark-decays}_{\etac}/N^{\bquark-decays}_{\jpsi}$  \\ \hline \hline
Mean value                              & 1.291 & 0.396  \\ \hline \hline
Stat. uncertainty                       & 13.7  & 12.0  \\ \hline  
$\alpha_{t_z}$ corrections              & 1.5   & 1.1   \\   
\pt-dependence of $\sigma_{\etac}/\sigma_{\jpsi}$ & 0.1 & 0.8  \\   
Comb. bkg. description                                & 3.5 & 3.5  \\
Contribution from \JpsiToPpbarPiz                     & 0.1 & 0.2  \\    
\hline  
\pt-dependence of $t_z$ resolution                    & 0.7 & 0.3  \\   
\pt-dependence of $\tau_{B}$                          & 0.2 & 0.5  \\   
Bias $\mu$                                                 & 0.3 & 0.2  \\   
$t_z$-resolution model                                & 0.2 & 0.5  \\   
\hline  
\hline  
Total systematic uncorrelated                                       & 3.9 & 3.9  \\   
\hline  
\hline  
\jpsi polarisation                                    & 1.8 & $-$  \\   
Mass resolution model                                 & 3.0 & 3.8  \\   
Variation of $\Gamma_{\etac}$                         & 5.2 & 5.1  \\   
\hline  
\hline  
Total systematic correlated                                         & 6.3 & 6.4  \\   
\hline  
\hline  
Total systematic                                              & 7.4 & 7.5 \\ 
\hline  
\hline     
\end{tabular} 
\caption{Mean values and relative uncertainties (in \%) in the \etac and \jpsi yield ratios for $8.0 \gev<\pt<10.0 \gev$.}
\label{tab:systPT2}
\end{table}

\begin{table}[t]
\centering
\small
\begin{tabular}{c|c|c} 
& $N^{prompt}_{\etac}/N^{prompt}_{\jpsi}$ & $N^{\bquark-decays}_{\etac}/N^{\bquark-decays}_{\jpsi}$  \\ \hline \hline
Mean value                               & 1.463 & 0.277  \\ \hline \hline
Stat. uncertainty                        & 15.6  & 18.9  \\ \hline  
 $\alpha_{t_z}$ corrections              & 1.1   & 1.3   \\   
\pt-dependence of $\sigma_{\etac}/\sigma_{\jpsi}$     & 0.1 & 0.8  \\   
Comb. bkg. description                                & 4.9 & 3.8  \\ 
Contribution from \JpsiToPpbarPiz                     & 0.1 & 0.2  \\    
\hline  
\pt-dependence of $t_z$ resolution                    & 0.8 & 0.2  \\   
\pt-dependence of $\tau_{B}$                          & 0.1 & 0.8  \\   
Bias $\mu$                                                 & 0.3 & 0.2  \\   
$t_z$-resolution model                                & 0.1 & 0.7  \\   
\hline  
\hline  
Total systematic uncorrelated                                       & 5.1 & 4.3  \\   
\hline  
\hline  
\jpsi polarisation                                    & 1.6 & $-$  \\   
Mass resolution model                                 & 3.0 & 3.8  \\   
Variation of $\Gamma_{\etac}$                         & 5.2 & 5.1  \\   
\hline  
\hline  
Total systematic correlated                                         & 6.2 & 6.4  \\   
\hline  
\hline  
Total systematic                                              & 8.1 & 7.7 \\ 
\hline  
\hline     
\end{tabular} 
\caption{Mean values and relative uncertainties (in \%) in the \etac and \jpsi yield ratios for $10.0 \gev<\pt<12.0 \gev$.}
\label{tab:systPT3}
\end{table}

\begin{table}[b]
\centering
\small
\begin{tabular}{c|c|c} 
& $N^{prompt}_{\etac}/N^{prompt}_{\jpsi}$ & $N^{\bquark-decays}_{\etac}/N^{\bquark-decays}_{\jpsi}$  \\ \hline \hline
Mean value                               & 2.125 & 0.293  \\ \hline \hline 
Stat. uncertainty                        & 18.9  & 25.2  \\ \hline  
 $\alpha_{t_z}$ corrections              & 0.8   & 1.4   \\   
\pt-dependence of $\sigma_{\etac}/\sigma_{\jpsi}$     & 0.1 & 0.8  \\   
Comb. bkg. description                                & 6.3 & 4.1  \\ 
Contribution from \JpsiToPpbarPiz                     & 0.2 & 0.1  \\    
\hline  
\pt-dependence of $t_z$ resolution                    & 0.9   & 0.2  \\   
\pt-dependence of $\tau_{B}$                          &$<0.1$ & 1.0  \\   
Bias $\mu$                                                 & 0.3   & 0.2  \\   
$t_z$-resolution model                                & $<0.1$ & 1.0  \\   
\hline  
\hline  
Total systematic uncorrelated                                       & 6.4 & 4.7  \\   
\hline  
\hline  
\jpsi polarisation                                    & 1.6 & $-$  \\   
Mass resolution model                                 & 3.0 & 3.8  \\   
Variation of $\Gamma_{\etac}$                         & 5.2 & 5.1  \\   
\hline  
\hline  
Total systematic correlated                                         & 6.2 & 6.4  \\   
\hline  
\hline  
Total systematic                                              & 8.9 & 7.9 \\ 
\hline  
\hline     
\end{tabular} 
\caption{Mean values and relative uncertainties (in \%) in the \etac and \jpsi yield ratios for $12.0 \gev<\pt<14.0 \gev$.}
\label{tab:systPT4}
\end{table}  
\clearpage

\newpage
\section{Separation technique}
\label{sec:signal_extraction_runI}

In the analysis of \etac production with \lhcb Run I data, the data sample was split into prompt sample and \bquark-decays sample. The charmonia signal in each sample is dominated by the corresponding production process.
A separation between promptly produced charmonia decaying at PV and charmonia produced in \bquark-decays with large typically $t_z$ values is performed using $t_z$ requirement. 
To select charmonia from \bquark-decays, an additional requirement on impact parameter significance \chisqip
\footnote{The \chisqip is defined as the \chisq difference of the PV reconstructed with and without considered track.} of both proton and antiproton candidates is applied.
The efficiencies and cross-feed between the two samples are obtained from simulation to extract prompt and non-prompt production. 
Below, the same analysis technique is applied for data collected at \sqs=13~\tev. 

Signal selection criteria are the same as discussed in Section~\ref{sec:select}.
The separation between samples is achieved by applying the requirement $t_z< 80 \fs$ to select the prompt sample and the requirements $t_z> 80 \fs$ and $\chisqip > 16$ to select the \\bquark-decays sample.

The number of observed \etac candidates in the prompt and \\bquark-decays sample can be written in the following way
\begin{equation} \label{eq:runIraw}
 \left\{\begin{aligned}
  &n^{p}_{\etac}  &= \epsilon^{P \to P} \boldsymbol{N^{P}_{\etac}} + \epsilon^{b \to P} \boldsymbol{N^{b}_{\etac}} \\ 
  &n^{b}_{\etac}  &= \epsilon^{b \to b} \boldsymbol{N^{b}_{\etac}} + \epsilon^{P \to b} \boldsymbol{N^{P}_{\etac}},
\end{aligned}
\right.
\end{equation}
where $n^{p}_{\etac}$ and $n^{b}_{\etac}$ are \etac yields in the prompt sample and in \bquark-decays sample from simultaneous fit of the invariant mass of the two samples, respectively; 
$N^{P}_{\etac}$ is the number of promptly produced \etac; 
$N^{b}_{\etac}$ is the number of \etac produced in \bquark-decays;
$\epsilon^{P \to P}$ is the separating requirement efficiency for selecting promptly produced \etac using selection criteria of prompt sample; 
$\epsilon^{P \to b}$ is separating requirement efficiency efficiency for selecting promptly produced \etac using selection of \bquark-decays sample; 
and the efficiencies $\epsilon^{b \to b}$ and $\epsilon^{b \to P}$ for \etac produced in \bquark-hadron decays are defined in similar way. 
Similar definitions for \jpsi equally apply.

Solving equations (\ref{eq:runIraw}), the number of promptly produced \etac meson is 
\begin{equation}
  N^{P}_{\etac} = \frac{\epsilon^{b \to b} n^{p}_{\etac}  - \epsilon^{b \to P} n^{b}_{\etac}}
  {\epsilon^{P \to P}\epsilon^{b \to b} - \epsilon^{P \to b}\epsilon^{b \to P}},
\end{equation}
and the number of \etac produced in \bquark-hadron decays is
\begin{equation}
  N^{b}_{\etac} = \frac{\epsilon^{P \to P} n^{b}_{\etac}  - \epsilon^{P \to b} n^{p}_{\etac}}
  {\epsilon^{P \to P}\epsilon^{b \to b} - \epsilon^{P \to b}\epsilon^{b \to P}}.
\end{equation}

Hence, the relative \etac and \jpsi production can be expressed in the following way.
\begin{equation}
\begin{aligned}
\frac{\sigma^{prompt}_{\etac}}{\sigma^{prompt}_{\jpsi}} &= \frac{\epsilon^{b \to b} n^{p}_{\etac} - \epsilon^{b \to P} n^{b}_{\etac}}{\epsilon^{b \to b} n^{p}_{\jpsi} - \epsilon^{b \to P} n^{b}_{\jpsi}} \times \frac{\epsilon_{\jpsi}}{\epsilon_{\etac}}\times \frac{\BR_{\JpsiToPpbar}}{\BR_{\EtacToPpbar}} \\
\frac{\sigma^{\bquark-decays}_{\etac}}{\sigma^{\bquark-decays}_{\jpsi}} = \frac{\BR_{\bToEtacX}}{\BR_{\bToJpsiX}} &= \frac{\epsilon^{P \to P} n^{b}_{\etac} - \epsilon^{P \to b} n^{p}_{\etac}}{\epsilon^{P \to P} n^{b}_{\jpsi} - \epsilon^{P \to b} n^{p}_{\jpsi}}\times \frac{\epsilon_{\jpsi}}{\epsilon_{\etac}}\times \frac{\BR_{\JpsiToPpbar}}{\BR_{\EtacToPpbar}},
\end{aligned}
\end{equation}
where $\epsilon_{\etac}$ and $\epsilon_{\jpsi}$ are the total reconstruction and selection efficiencies before applying separating requirement for \etac and \jpsi respectively.

\subsection{Separating requirements efficiencies}
Efficiencies of the separating requirements are extracted from MC samples to evaluate the cross-feed between the samples. Values of the efficiencies are listed in Table~\ref{tab:crossTalk}. No significant difference between the \etac and \jpsi separating requirement efficiencies is observed. 
\begin{table}[ht]
\begin{center} 
\begin{tabular}{l|l} 
  $\epsilon^{P \to P}$  &  $0.964\pm0.011$    \\ \hline 
  $\epsilon^{b \to b}$  &  $0.692\pm0.013$    \\ \hline
  $\epsilon^{P \to b}$  &  $0.0007\pm0.0002$  \\ \hline
  $\epsilon^{b \to P}$  &  $0.064\pm0.003$    \\ 
\end{tabular}
\caption{Cross-feed efficiencies between prompt and \bquark-decays samples for \tzcut.} 
\label{tab:crossTalk}
\end{center}
\end{table}

A good agreement between data and MC in $t_z$ distribution is observed 
for all parameters describing $t_z$-resolution model and $\tau_b$, which leads to good agreement of $t_z$ requirement efficiency between data and MC.
To extract the efficiency of $t_z<80 \fs$ $(>80 \fs)$ requirement from data one can integrate the curve of the fit to $t_z$ obtained in Section~\ref{sec:fit2tz}. The comparison of this requirement efficiency estimated from data and from MC is shown in Table~\ref{tab:tzcutEffTab}. The values are well consistent within the uncertainty due to MC sample sizes.
\begin{table}[ht]
\centering
\small
\begin{tabular}{c|c|c} 
                          & from \tzfit                            & MC                \\ \hline
$\eps^{\jpsi, prompt  }(t_z<80 \fs)$  & $0.955 \pm 0.003_{(stat. uncorrelated)}$ & $0.964 \pm 0.011$ \\
$\eps^{\jpsi, b-decays}(t_z>80 \fs)$  & $0.938 \pm 0.002_{(stat. uncorrelated)}$ & $0.936 \pm 0.016$
\end{tabular} 
\caption{Comparison of the $t_z<80 \fs$ $(>80 \fs)$ requirement efficiency as estimated in data and MC for prompt charmonia and charmonia from \bquark-decays.}
\label{tab:tzcutEffTab}
\end{table}

The requirement on proton \chisqip is also used in the analysis with \tzcut. Here, its efficiency enters the definition of $\eps^{P \to b}$ and $\eps^{b \to b}$.
The \chisqip variable is proved to be well described by MC. The cross-feed described by $\eps^{P \to b}$ is small and the contamination of \bquark-decays sample by prompt \etac is about 1.5\% (about 0.4\% for \jpsi), which is estimated using known \etac prompt yield from \tzfit. The effect of this cross-feed is also checked by setting $\eps^{P \to b}\equiv 0$, which does not lead to a significant change in $N^{prompt}_{\etac}/N^{prompt}_{\jpsi}$ and $N^{b-decays}_{\etac}/N^{b-decays}_{\jpsi}$ within 1\% level.

Since the contamination of \bquark-decays sample by prompt candidates is small, one can evaluate the value of $\eps^{b \to b}$ from data by comparing a signal yield in \bquark-decays sample and the total number of signal candidates from \bquark-decays obtained from \tzfit. The comparison between data and MC values is shown in Table~\ref{tab:tzcutEffb2bTab}.
\begin{table}[ht]
\centering
\small
\begin{tabular}{c|c|c} 
             & from \tzfit and \tzcut      & MC                \\ \hline
$\eps^{b \to b}$  & $0.699 \pm 0.076_{(stat)}$  & $0.692 \pm 0.013$ \\
\end{tabular} 
\caption{Comparison of the $\eps^{b \to b}$ requirement efficiency from data and MC.}
\label{tab:tzcutEffb2bTab}
\end{table}

From all above one can conclude that all possible systematic effects due to the cross-feed efficiencies are well within the actual estimate of the cross-feed uncertainty.

\subsection{Fit to the invariant mass}
\label{sec:fitRunI}
The same signal and background parameterisations as in Section ~\ref{sec:massFit} are used to describe the invariant mass distribution. No impact of separating requirement on the resolution model is observed using MC samples; hence, the resolution model is kept identical to that used in Section~\ref{sec:syst}. The only difference is that the main resolution parameters $\sigma_n$ in \pt-bins are free fit parameters. 
No significant peak position shifts are observed between prompt and \bquark-decays samples. Therefore peak positions are required to be identical for prompt and \bquark-decays samples.
The mass of \jpsi meson and the \jpsi and \etac mass difference are free fit parameters in the fit to the invariant mass for \pt-integrated sample. Obtained values and statistical uncertainties of the \jpsi mass and the \jpsi and the \etac mass difference are then used in the gaussian constraints imposed for the fits in bins of \pt. The applicability of constraints on masses of \jpsi and \etac are cross-checked by performing individual fits to prompt and \bquark-decays samples in \pt-bins.
 
A summary of signal shape parametrisation in simultaneous fit to both prompt and \bquark-decay samples is shown in Table~\ref{tab:massSigModel_RunI}.
\begin{table}[ht] 
\small{
\centering
\begin{tabular}{l|l} 
 Parameter & \\ \hline \hline
 $\sigma_n / \sigma_w$              & Fixed from MC  \\ \hline 
 $f_n$                              & Fixed from MC  \\ \hline
 $\sigma_{\etac}/\sigma_{\jpsi}$    & Fixed from MC  \\ \hline
 $\sigma_{\etac}$                   & Individual parameter for each \pt-bin \\ \hline
 $m_{\jpsi} - m_{\etac}$            & Common free parameter for both samples, \\ 
                                    & gaussian constraint for fits in bins of \pt \\ \hline        
 $m_{\jpsi}$                        & Common free parameter for both samples, \\
                                    & gaussian constraint for fits in bins of \pt \\ \hline
 $\Gamma_{\etac}$                   & Fixed to the world average from Ref.~\cite{PDG2017} (31.8 \mev)
\end{tabular}
\caption{Summary of signal parametrisation in the simultaneous invariant mass fit.} 
\label{tab:massSigModel_RunI}
}
\end{table}

\begin{figure}[ht]
\centering
\protect\protect\protect\includegraphics[width=1.0\textwidth]{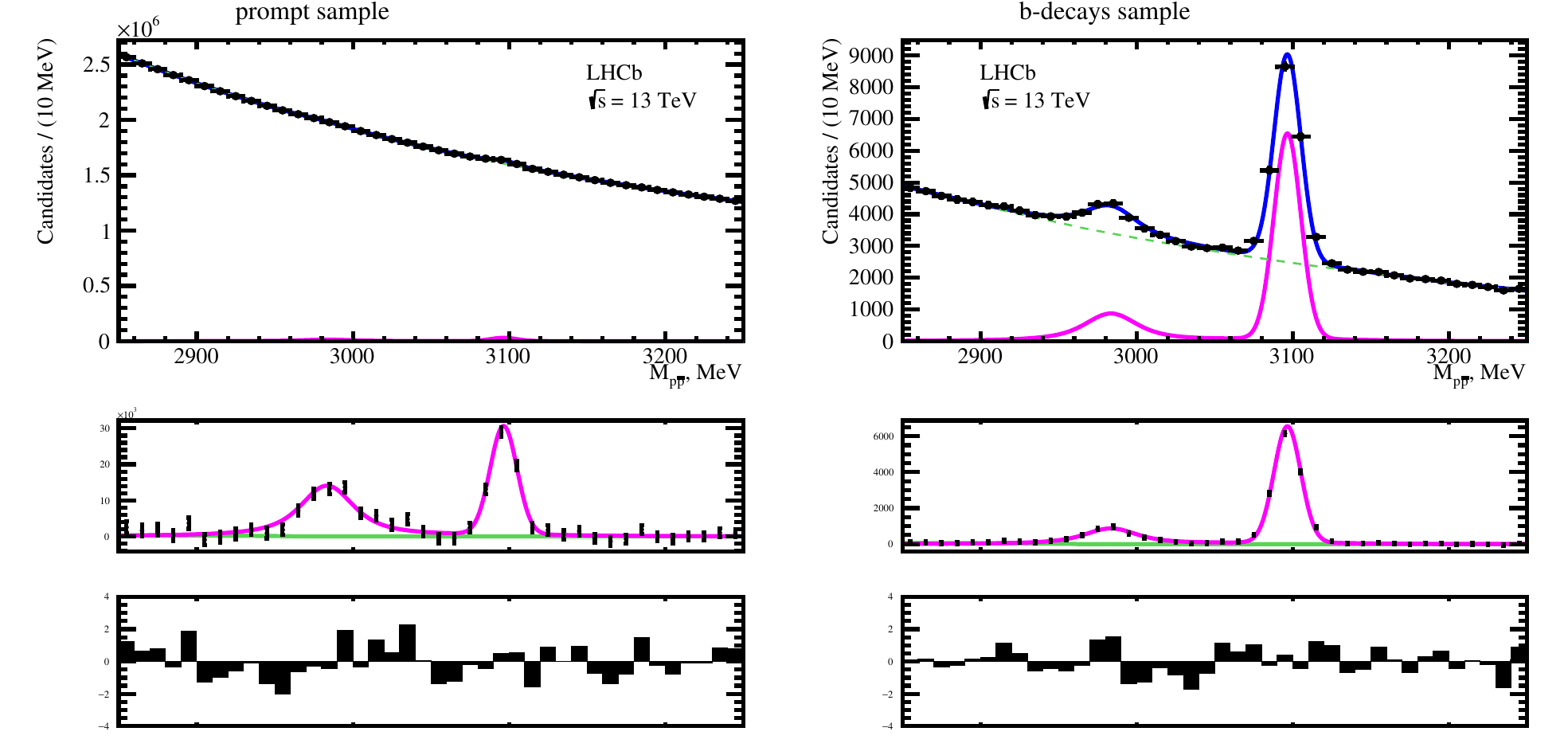}
\caption
[The $M_{\ppbar}$ distribution for prompt and \bquark-decays \pt-integrated samples $6.5 \gev<\pt<14.0 \gev$.]
{The $M_{\ppbar}$ distribution for prompt (left) and \bquark-decays (right) \pt-integrated samples $6.5 \gev<\pt<14.0 \gev$. The solid blue lines represent the total fit result. Magenta and green lines show the signal and background components, respectively. The corresponding residual and pull distributions are shown below.} 
\label{fig:massFitRunIInt}
\end{figure}

Projections of the simultaneous fit for the entire \pt-range $6.5 \gev<\pt<14.0 \gev$ are shown on Fig.~\ref{fig:massFitRunIInt}. The residual and pull distributions are displayed below the corresponding projections. In general, fit yields a good description of both $M_{\ppbar}$ distributions. Simultaneous fit yields the following values of the \jpsi mass $M_{\jpsi}=\etacMassDiffTzFitSim$ and the \jpsi and \etac mass difference $\Delta M _{\jpsi , \, \etac} = \jpsiMassTzCutInt$. These values agree with the world average values $M_{\jpsi}^{PDG}=\jpsiMassPDG$ and $\Delta M _{\jpsi , \, \etac}^{PDG} = \etacMassDiffPDG$~\cite{PDG2017}. 

\begin{figure}[ht]
\centering
\protect\protect\protect\includegraphics[width=1.0\textwidth]{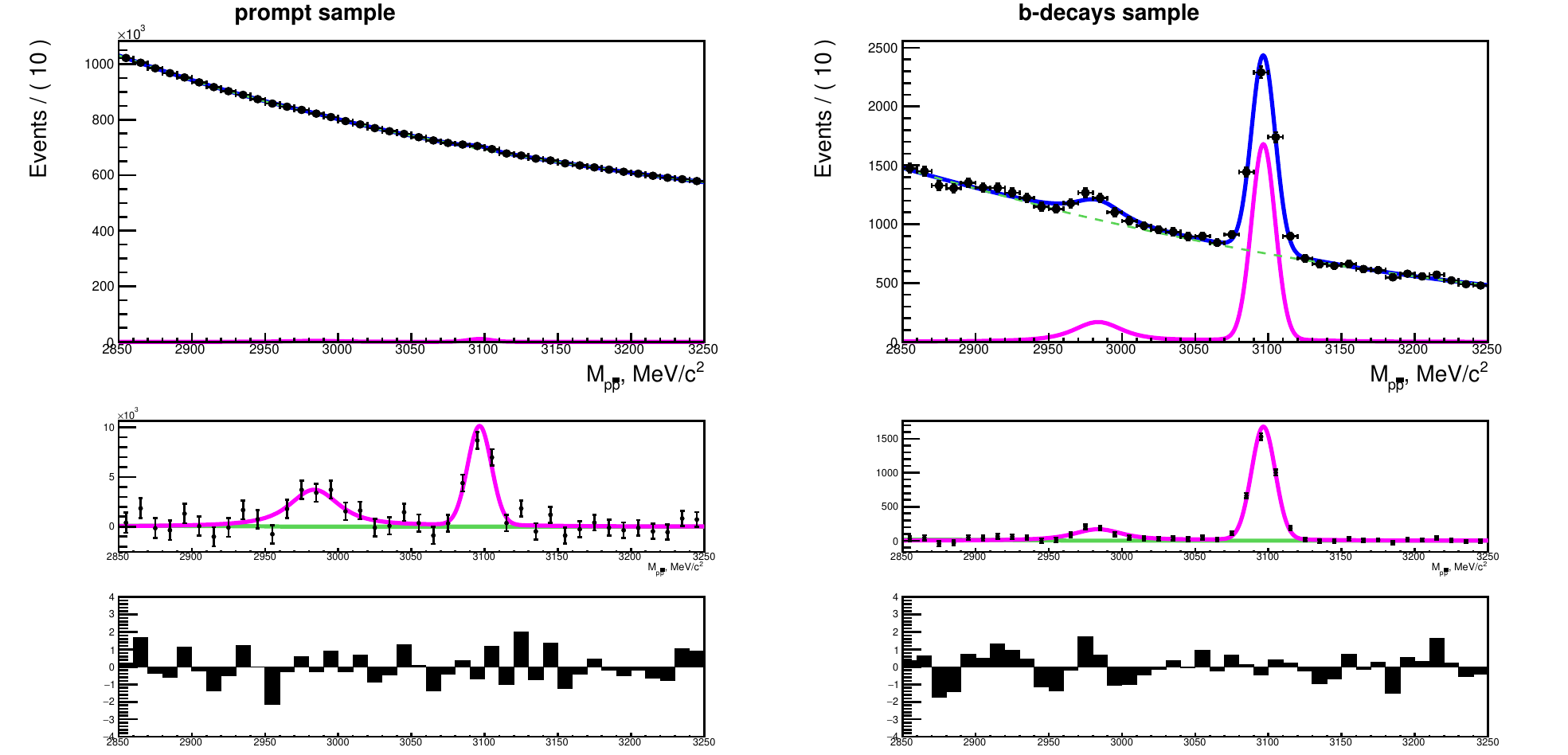}
\put(-185,198){\scriptsize{\lhcb-ANA-2018-035}}
\put(-334,198){\scriptsize{\lhcb-ANA-2018-035}}
\caption
[The $M_{\ppbar}$ distribution for prompt and \bquark-decays for $6.5 \gev<\pt<8.0 \gev$.]
{The $M_{\ppbar}$ distribution for prompt (left) and \bquark-decays (right) for $6.5 \gev<\pt<8.0 \gev$. The solid blue lines represent the fit result. Magenta and green lines show the signal and background components, respectively. The corresponding residual and pull distributions are shown below.} 
\label{fig:massFitRunIPT1}
\end{figure}

\begin{figure}[ht]
\centering
\protect\protect\protect\includegraphics[width=1.0\textwidth]{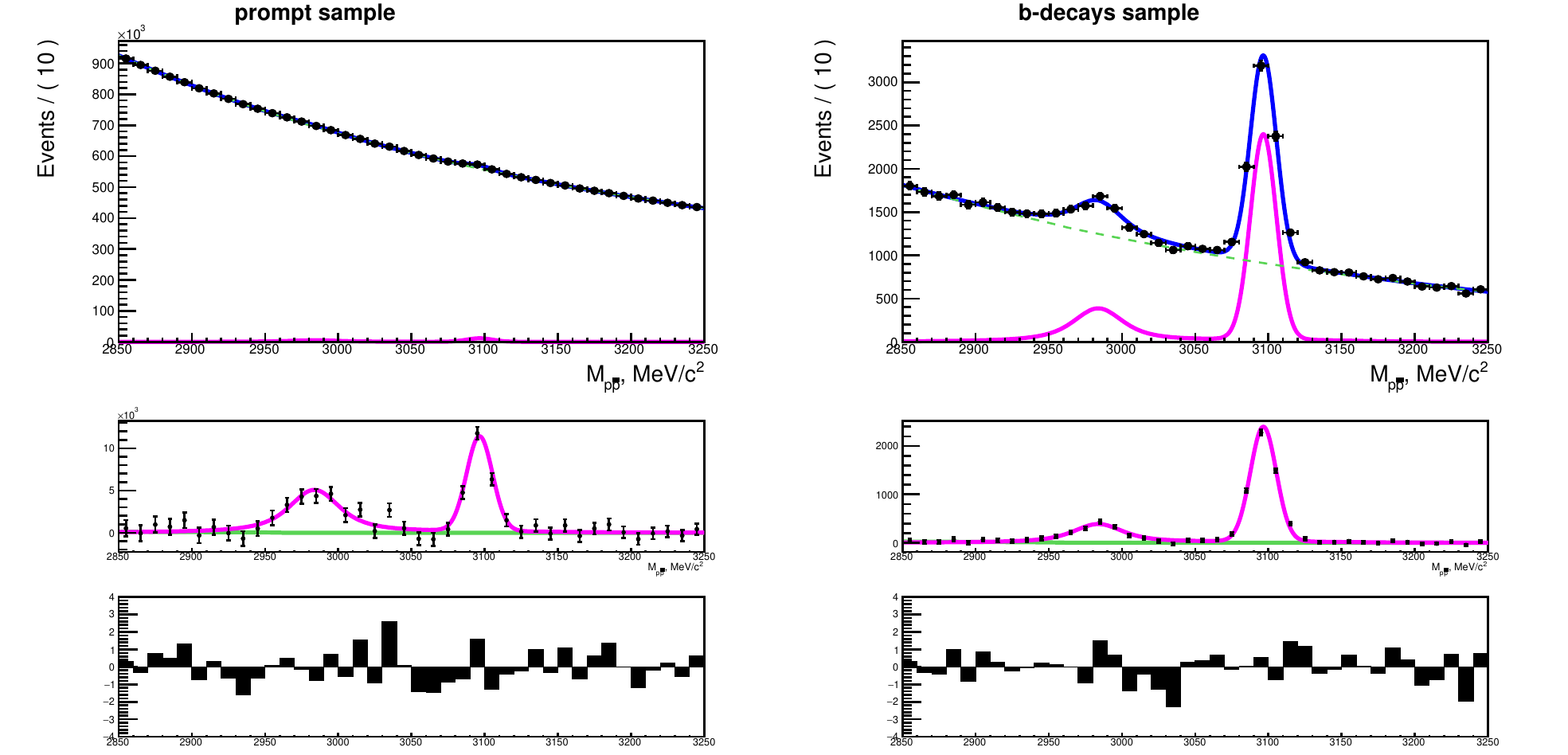}
\put(-185,198){\scriptsize{\lhcb-ANA-2018-035}}
\put(-334,198){\scriptsize{\lhcb-ANA-2018-035}}
\caption
[The $M_{\ppbar}$ distribution for prompt and \bquark-decays for $8.0 \gev<\pt<10.0 \gev$.]
{The $M_{\ppbar}$ distribution for prompt (left) and \bquark-decays (right) for $8.0 \gev<\pt<10.0 \gev$. The solid blue lines represent the fit result. Magenta and green lines show the signal and background components, respectively. The corresponding residual and pull distributions are shown below.} 
\label{fig:massFitRunIPT2}
\end{figure}

\begin{figure}[ht]
\centering
\protect\protect\protect\includegraphics[width=1.0\textwidth]{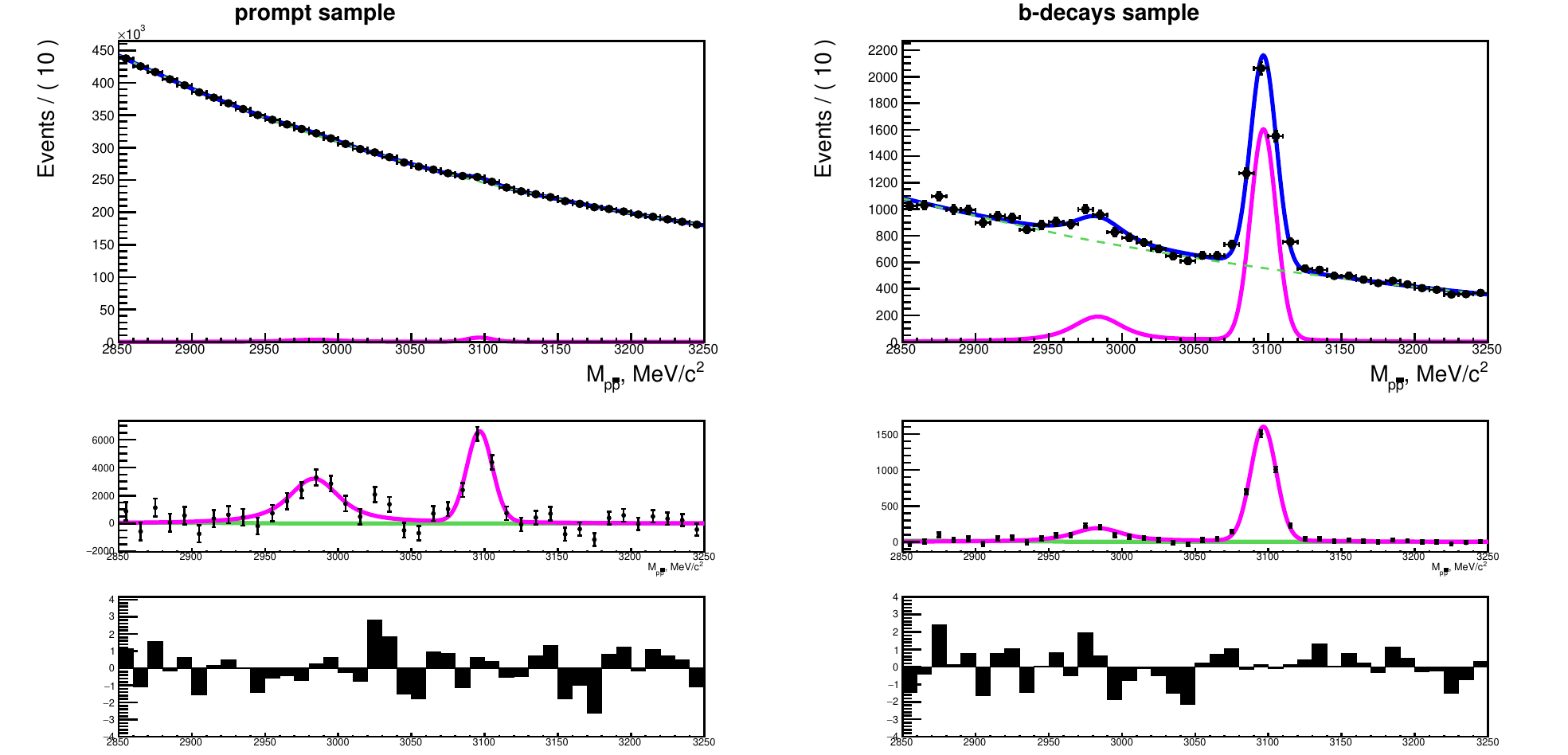}
\put(-185,198){\scriptsize{\lhcb-ANA-2018-035}}
\put(-334,198){\scriptsize{\lhcb-ANA-2018-035}}
\caption
[The $M_{\ppbar}$ distribution for prompt and \bquark-decays for ($10.0 \gev<\pt<12.0 \gev$.]
{The $M_{\ppbar}$ distribution for prompt (left) and \bquark-decays (right) for ($10.0 \gev<\pt<12.0 \gev$. The solid blue lines represent the fit result. Magenta and green lines show the signal and background components, respectively. The corresponding residual and pull distributions are shown below.} 
\label{fig:massFitRunIPT3}
\end{figure}

\begin{figure}[ht]
\centering
\protect\protect\protect\includegraphics[width=1.0\textwidth]{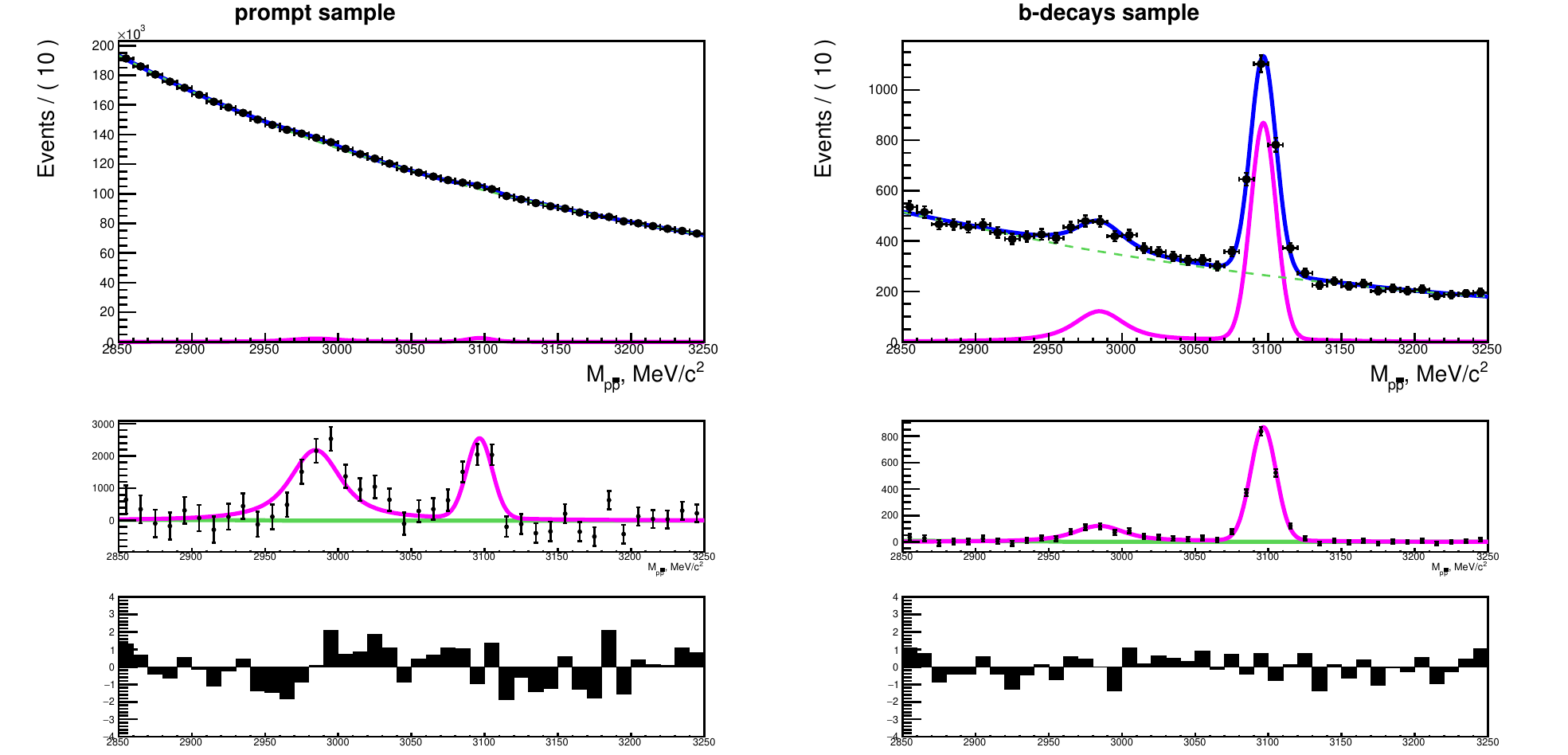}
\put(-185,198){\scriptsize{\lhcb-ANA-2018-035}}
\put(-334,198){\scriptsize{\lhcb-ANA-2018-035}}
\caption
[The $M_{\ppbar}$ distribution for prompt and \bquark-decays for $12.0 \gev<\pt<14.0 \gev$.]
{The $M_{\ppbar}$ distribution for prompt (left) and \bquark-decays (right) for $12.0 \gev<\pt<14.0 \gev$. The solid blue lines represent the fit result. Magenta and green lines show the signal and background components, respectively. The corresponding residual and pull distributions are shown below.} 
\label{fig:massFitRunIPT4}
\end{figure}

Projections of the simultaneous fits to prompt and \bquark-decays samples in \pt bins are shown on Figs.~\ref{fig:massFitRunIPT1},~\ref{fig:massFitRunIPT2},~\ref{fig:massFitRunIPT3} and~\ref{fig:massFitRunIPT4}. In general, fit yields a good description of all $M_{\ppbar}$ distributions.

\clearpage 

\clearpage

\subsection{Systematic uncertainties}
\begin{figure}[ht]
\centering
\protect\protect\protect\includegraphics[width=0.78\linewidth]{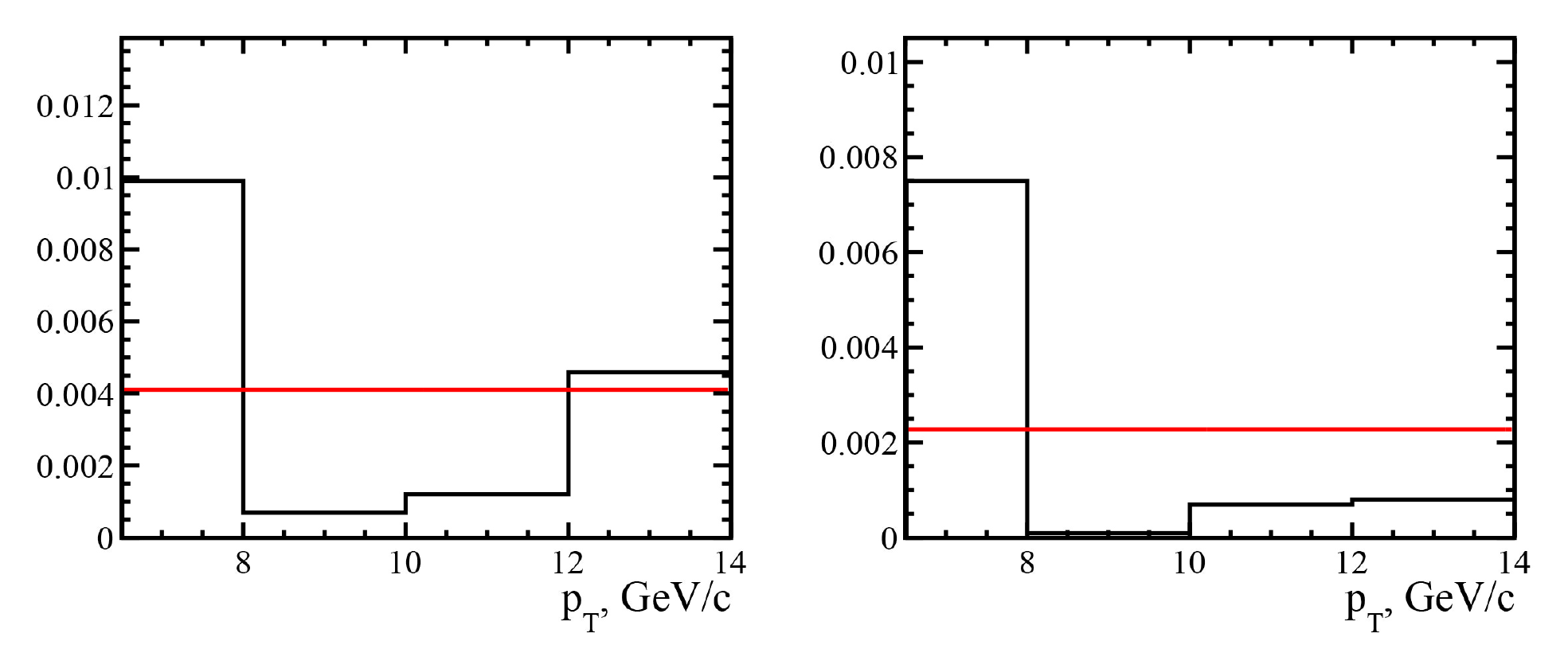}
\put(-95,126){\scriptsize{\lhcb-ANA-2018-035}}
\put(-276,126){\scriptsize{\lhcb-ANA-2018-035}}
\put(-356,60){\rotatebox{90}{\small{rel. syst. uncert.}}}
\put(-179,60){\rotatebox{90}{\small{rel. syst. uncert.}}}
\caption
[The relative systematic uncertainty due to the \pt-dependence of \etac and \jpsi resolution ratio in bins of \pt.]
{The relative systematic uncertainty due to the \pt-dependence of \etac and \jpsi resolution ratio in bins of \pt. The solid red line shows a smoothing curve.} 
\label{fig:smoothSystResoRatio_RunI}
\end{figure}

\begin{figure}[ht]
\centering
\protect\protect\protect\includegraphics[width=0.78\linewidth]{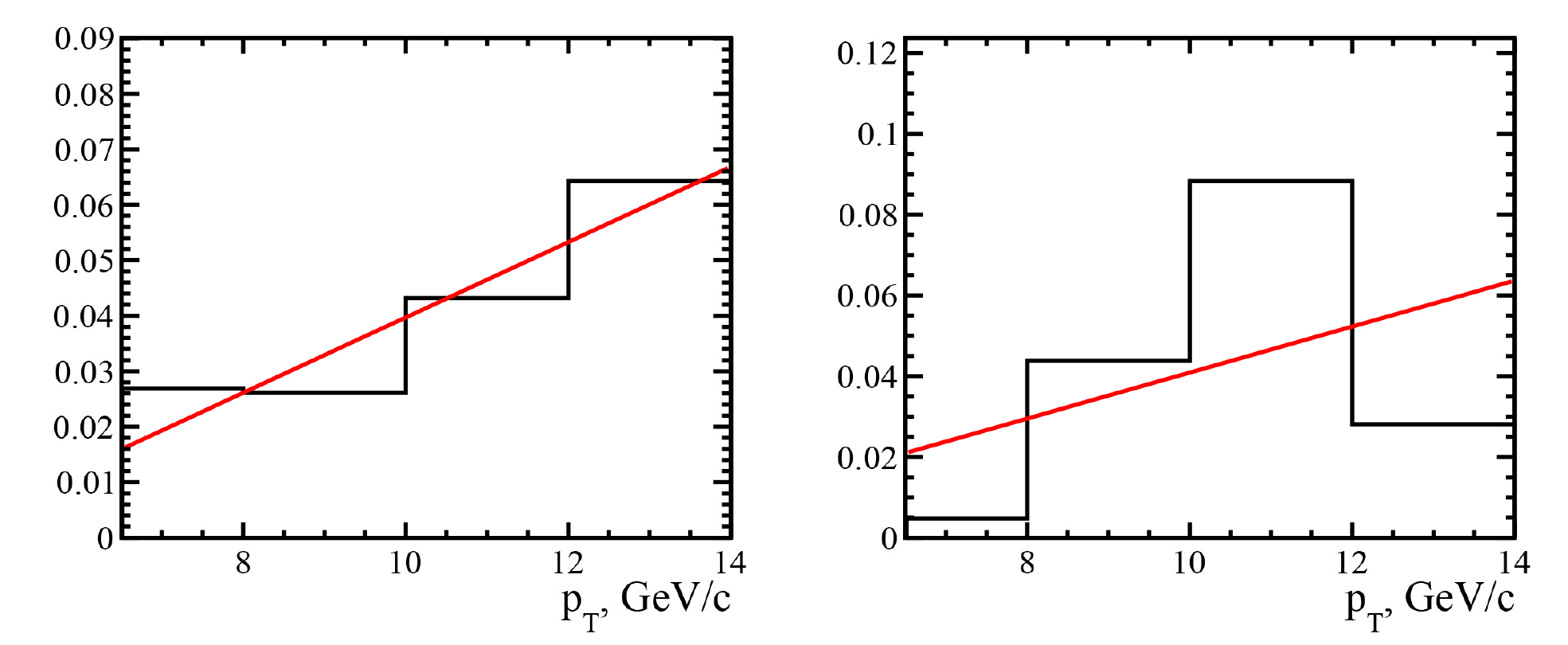}
\put(-95,126){\scriptsize{\lhcb-ANA-2018-035}}
\put(-276,126){\scriptsize{\lhcb-ANA-2018-035}}
\put(-356,60){\rotatebox{90}{\small{rel. syst. uncert.}}}
\put(-179,60){\rotatebox{90}{\small{rel. syst. uncert.}}}
\caption
[The relative systematic uncertainty due to combinatorial background description in bins of \pt.]
{The relative systematic uncertainty due to combinatorial background description in bins of \pt. The solid red line shows a smoothing curve.} 
\label{fig:smoothSystCombBkg_RunI}
\end{figure}

\begin{figure}[ht]
\centering
\protect\protect\protect\includegraphics[width=0.78\linewidth]{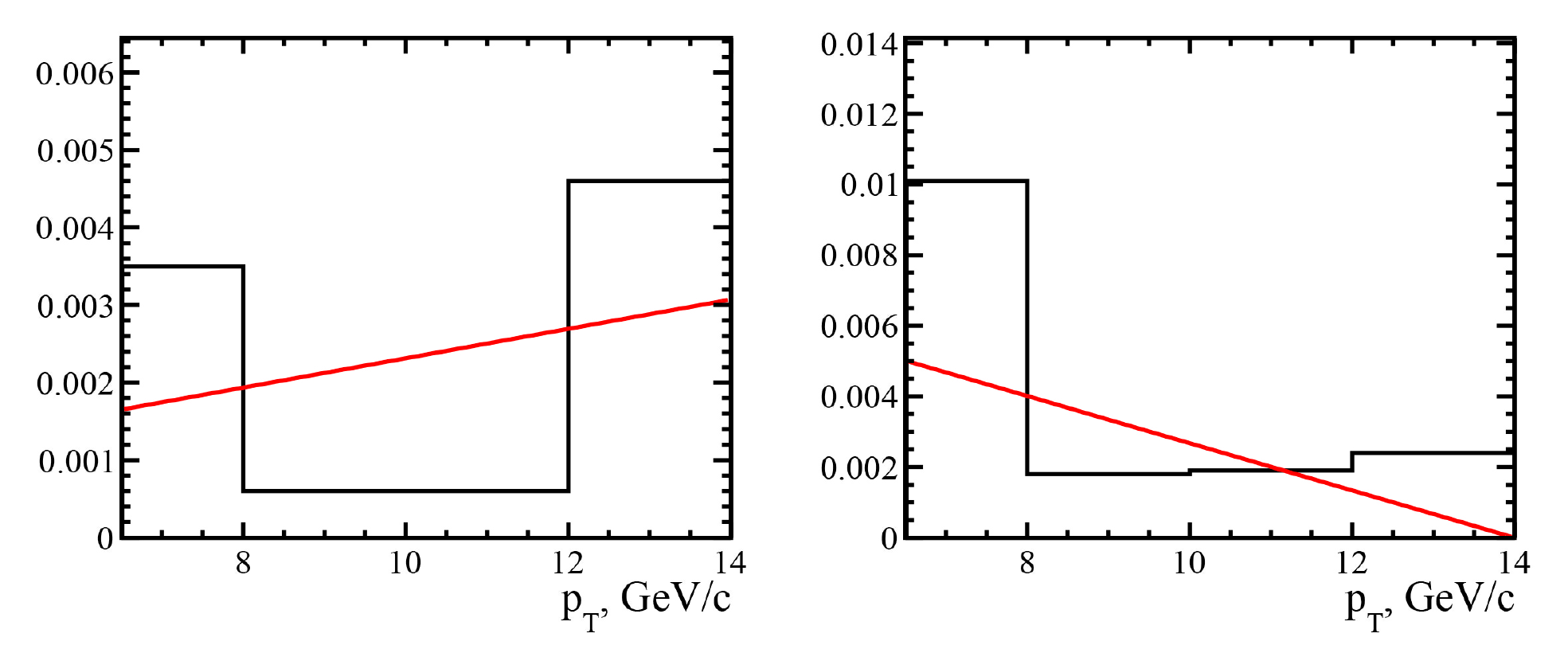}
\put(-95,126){\scriptsize{\lhcb-ANA-2018-035}}
\put(-276,126){\scriptsize{\lhcb-ANA-2018-035}}
\put(-356,60){\rotatebox{90}{\small{rel. syst. uncert.}}}
\put(-179,60){\rotatebox{90}{\small{rel. syst. uncert.}}}
\caption
[The relative systematic uncertainty due to description of the feed-down from the \JpsiToPpbarPiz decay in bins of \pt.]
{The relative systematic uncertainty due to description of the feed-down from the \JpsiToPpbarPiz decay in bins of \pt. The solid red line shows a smoothing curve.} 
\label{fig:smoothSystPPPi0_RunI}
\end{figure}

The following list of systematic uncertainties is identical for both \tzfit and \tzcut:
\begin{itemize}
\item Signal description in simultaneous fit to invariant mass distributions:
  \begin{itemize}
  \item Knowledge of the \etac natural width, $\Gamma_{\etac}$;
  \item Invariant mass resolution mismodeling;
  \item \pt-dependence of the \etac and \jpsi resolution ratio, $\sigma_{\etac}/\sigma_{\jpsi}$ 
  \\ (relevant for differential cross-section measurement);
  \end{itemize}
\item Background description in simultaneous fit to invariant mass distributions:
  \begin{itemize}
  \item Combinatorial background description;
  \item Description of the feed-down from the \JpsiToPpbarPiz decay;
  \end{itemize}
\item The \etac and \jpsi efficiency ratio;
\item Non-zero \jpsi polarisation;
\item Uncertainties on $\BR_{\JpsiToPpbar}$ and $\BR_{\EtacToPpbar}$;
\item Uncertainties on \jpsi production for absolute \etac production cross-section measurement.
\end{itemize}
The estimation of each of these uncertainties is performed in the same way as in \etac production analysis using \tzfit discussed in Section~\ref{sec:syst}. The uncertainties related to the \etac and \jpsi efficiency ratio, combinatorial background description and the description of the feed-down from the \JpsiToPpbarPiz decay 
are parametrised as shown on Figs.~\ref{fig:smoothSystResoRatio_RunI},~\ref{fig:smoothSystCombBkg_RunI} and~\ref{fig:smoothSystPPPi0_RunI} respectively.

The only additional systematic uncertainty is related to evaluation of the cross-feed. This uncertainty is estimated by modifying the efficiency values of the separating requirement by their uncertainties. Efficiencies of separation requirements are in good agreement between data and MC 
and possible discrepancies are well below the uncertainty due to MC sample sizes.

The dominant source of uncorrelated systematic uncertainty for prompt \etac production
is related to combinatorial background description. The dominant sources of uncorrelated systematic uncertainties on \etac production in \bquark-decays are related to combinatorial background description and the \pt-dependence of the \etac and \jpsi resolution ratio. The dominant source of correlated systematic uncertainties on both prompt \etac production and \etac production in \bquark-decays is related to the knowledge of the \etac natural width and the invariant mass resolution model.

Systematic uncertainties on relative \etac production measurement in the entire $6.5 \gev<\pt<14.0 \gev$ range are shown in Table~\ref{tab:systRunITotal}.
Detailed tables of systematic uncertainties for each bin of \pt are given in Tables~\ref{tab:systRunIPT1},~\ref{tab:systRunIPT2},~\ref{tab:systRunIPT3} and ~\ref{tab:systRunIPT4}.

Uncertainties on the branching fractions of the \JpsiToPpbar 
and \EtacToPpbar  decay modes are combined in a separate systematic uncertainty as in Section~\ref{sec:syst}.

\clearpage
\clearpage

\begin{table}[ht]
\centering
\small
\begin{tabular}{c|c|c} 
& $N^{p}_{\etac}/N^{p}_{\jpsi}$ & $N^{b}_{\etac}/N^{b}_{\jpsi}$  \\ \hline \hline
Mean value                              & 1.183 & 0.333  \\ \hline \hline  
Stat. uncertainty                       & 8.8   & 5.8  \\ \hline  
\pt-dependence of $\sigma_{\etac}/\sigma_{\jpsi}$     & 0.2   & 0.1  \\ 
Comb. bkg. description                                & 2.0   & 2.3  \\
Contribution from \JpsiToPpbarPiz                     &$<0.1$ & 0.2  \\     
Cross-feed                                            & 0.9   & 0.8  \\     
Mass resolution model                                 & 2.7   & 3.1  \\   
Variation of $\Gamma_{\etac}$                         & 4.8   & 3.6  \\ 
\jpsi polarisation                                    & 1.8   & $-$  \\  
\hline  
\hline  
Total systematic                                              & 6.2   & 5.4 \\ 
\hline  
\hline     
\end{tabular} 
\caption{Mean values and relative uncertainties (in \%) in the \etac and \jpsi yields for \pt-integrated data sample $6.5 \gev<\pt<14.0 \gev$}
\label{tab:systRunITotal}
\end{table} 
\clearpage

\begin{table}[ht]
\centering
\small
\begin{tabular}{c|c|c} 
& $N^{p}_{\etac}/N^{p}_{\jpsi}$ & $N^{b}_{\etac}/N^{b}_{\jpsi}$  \\ \hline \hline
Mean value                              & 0.984 & 0.263  \\ \hline \hline
Stat. uncertainty                       & 22.7  & 15.4  \\ \hline 
\pt-dependence of $\sigma_{\etac}/\sigma_{\jpsi}$     & 0.4   & 0.2  \\  
Comb. bkg. description                                & 2.1   & 2.5  \\ 
Contribution from \JpsiToPpbarPiz                     & 0.2   & 0.7  \\    
Cross-feed                                            & 1.9   & 1.4  \\   
\hline  
\hline  
Total systematic uncorrelated                                       & 2.9   & 3.0  \\   
\hline  
\hline  
Mass resolution model                                 & 2.7 & 3.1  \\   
Variation of $\Gamma_{\etac}$                         & 4.8 & 3.6  \\ 
\jpsi polarisation                                    & 2.1 & $-$  \\    
\hline  
\hline  
Total systematic correlated                                         & 5.8 & 4.8  \\   
\hline  
\hline  
Total systematic                                              & 6.5 & 5.6 \\ 
\hline  
\hline     
\end{tabular} 
\caption{Mean values and relative uncertainties (in \%) in the \etac and \jpsi yields for the first \pt bin $6.5 \gev<\pt<8.0 \gev$}
\label{tab:systRunIPT1}
\end{table}

\begin{table}[ht]
\centering
\small
\begin{tabular}{c|c|c} 
& $N^{p}_{\etac}/N^{p}_{\jpsi}$ & $N^{b}_{\etac}/N^{b}_{\jpsi}$  \\ \hline \hline
Mean value                              & 1.118 & 0.395  \\ \hline \hline 
Stat. uncertainty                       & 16.1  & 8.2  \\ \hline  
\pt-dependence of $\sigma_{\etac}/\sigma_{\jpsi}$     & 0.4 & 0.2  \\ 
Comb. bkg. description                                & 3.3 & 3.5  \\
Contribution from \JpsiToPpbarPiz                     & 0.2 & 0.5  \\     
Cross-feed                                            & 1.1 & 1.3  \\   
\hline  
\hline  
Total systematic uncorrelated                                       & 3.5 & 3.8  \\   
\hline  
\hline  
Mass resolution model                                 & 2.7 & 3.1  \\   
Variation of $\Gamma_{\etac}$                         & 4.8 & 3.6  \\ 
\jpsi polarisation                                    & 1.8 & $-$  \\    
\hline  
\hline  
Total systematic correlated                                         & 5.7 & 4.8  \\   
\hline  
\hline  
Total systematic                                              & 6.7 & 6.1 \\ 
\hline  
\hline     
\end{tabular}
\caption{Mean values and relative uncertainties (in \%) in the \etac and \jpsi yields for the first \pt bin $8.0 \gev<\pt<10.0 \gev$}
\label{tab:systRunIPT2}
\end{table}

\begin{table}[ht]
\centering
\small
\begin{tabular}{c|c|c} 
& $N^{p}_{\etac}/N^{p}_{\jpsi}$ & $N^{b}_{\etac}/N^{b}_{\jpsi}$  \\ \hline \hline
Mean value                              & 1.241 & 0.290  \\ \hline \hline
Stat. uncertainty                       & 16.9  & 12.8  \\ \hline  
\pt-dependence of $\sigma_{\etac}/\sigma_{\jpsi}$     & 0.4 & 0.2  \\ 
Comb. bkg. description                                & 4.6 & 4.7  \\
Contribution from \JpsiToPpbarPiz                     & 0.3 & 0.3  \\     
Cross-feed                                            & 1.2 & 1.7  \\   
\hline  
\hline  
Total systematic uncorrelated                                       & 4.8 & 5.0  \\   
\hline  
\hline  
Mass resolution model                                 & 2.7 & 3.1  \\   
Variation of $\Gamma_{\etac}$                         & 4.8 & 3.6  \\ 
\jpsi polarisation                                    & 1.6 & $-$  \\    
\hline  
\hline  
Total systematic correlated                                         & 5.7 & 4.8  \\   
\hline  
\hline  
Total systematic                                              & 7.5 & 6.9 \\ 
\hline  
\hline     
\end{tabular}
\caption{Mean values and relative uncertainties (in \%) in the \etac and \jpsi yields for the first \pt bin $10.0 \gev<\pt<12.0 \gev$}
\label{tab:systRunIPT3}
\end{table}

\begin{table}[ht]
\centering
\small
\begin{tabular}{c|c|c} 
& $N^{p}_{\etac}/N^{p}_{\jpsi}$ & $N^{b}_{\etac}/N^{b}_{\jpsi}$  \\ \hline \hline
Mean value                              & 2.238 & 0.348  \\ \hline  \hline
Stat. uncertainty                       & 18.3  & 13.4  \\ \hline 
\pt-dependence of $\sigma_{\etac}/\sigma_{\jpsi}$     & 0.4 & 0.2  \\ 
Comb. bkg. description                                & 6.0 & 5.8  \\
Contribution from \JpsiToPpbarPiz                     & 0.3 & 0.1  \\   
Cross-feed                                            & 1.4 & 1.0  \\   
\hline  
\hline  
Total systematic uncorrelated                                       & 6.2 & 5.9  \\   
\hline  
\hline  
Mass resolution model                                 & 2.7 & 3.1  \\   
Variation of $\Gamma_{\etac}$                         & 4.8 & 3.6  \\ 
\jpsi polarisation                                    & 1.6 & $-$  \\    
\hline  
\hline  
Total systematic correlated                                         & 5.7 & 4.8  \\   
\hline  
\hline  
Total systematic                                              & 8.4 & 7.6 \\ 
\hline  
\hline     
\end{tabular}
\caption{Mean values and relative uncertainties (in \%) in the \etac and \jpsi yields for the first \pt bin $12.0 \gev<\pt<14.0 \gev$}
\label{tab:systRunIPT4}
\end{table}  

\clearpage
\section{Summary and discussion}
\label{sec:results}

The \pt-differential production of the \etac meson for both prompt charmonium production and production in inclusive \bquark-hadron decays is obtained below.

The ratios of \etac and \jpsi differential production cross-sections obtained with the \tzfit
 are shown on Fig.~\ref{fig:sigmaRelPrompt} for prompt \etac and on Fig.~\ref{fig:sigmaRelFromB} for \etac produced in inclusive \bquark-decays. 
The relative \etac prompt production is similar to those measured at \sqs=7 and 8 \tev~\cite{LHCb-PAPER-2014-029}. The linear slope of \pt-dependece of relative \etac to \jpsi prompt production is obtained to be $0.23\pm0.11\, \gev^{-1}$ and is not significantly different from zero. The relative \etac production in inclusive \bquark-decays is consistent with those measured at \sqs=7 and 8 \tev~\cite{LHCb-PAPER-2014-029}. 
\begin{figure}[htb]
\centering{
        \subfigure[Prompt-production. The result of the fit by a linear function is overlaid.]{ 
          \protect\protect\protect\includegraphics[width=0.465\textwidth]{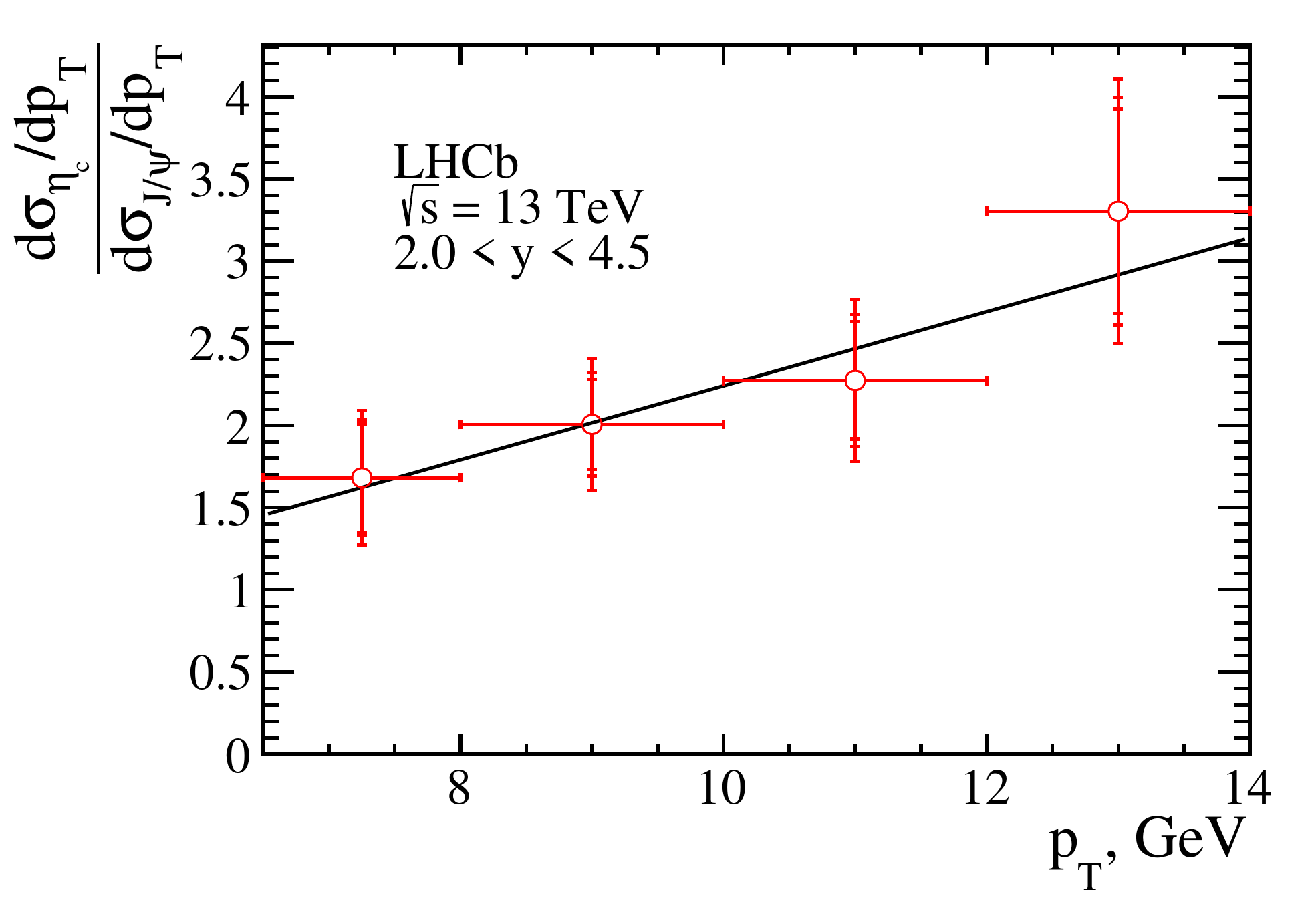}
          \label{fig:sigmaRelPrompt}}
\quad
        \subfigure[Production in inclusive \bquark-decays.]{
          \protect\protect\protect\includegraphics[width=0.465\textwidth]{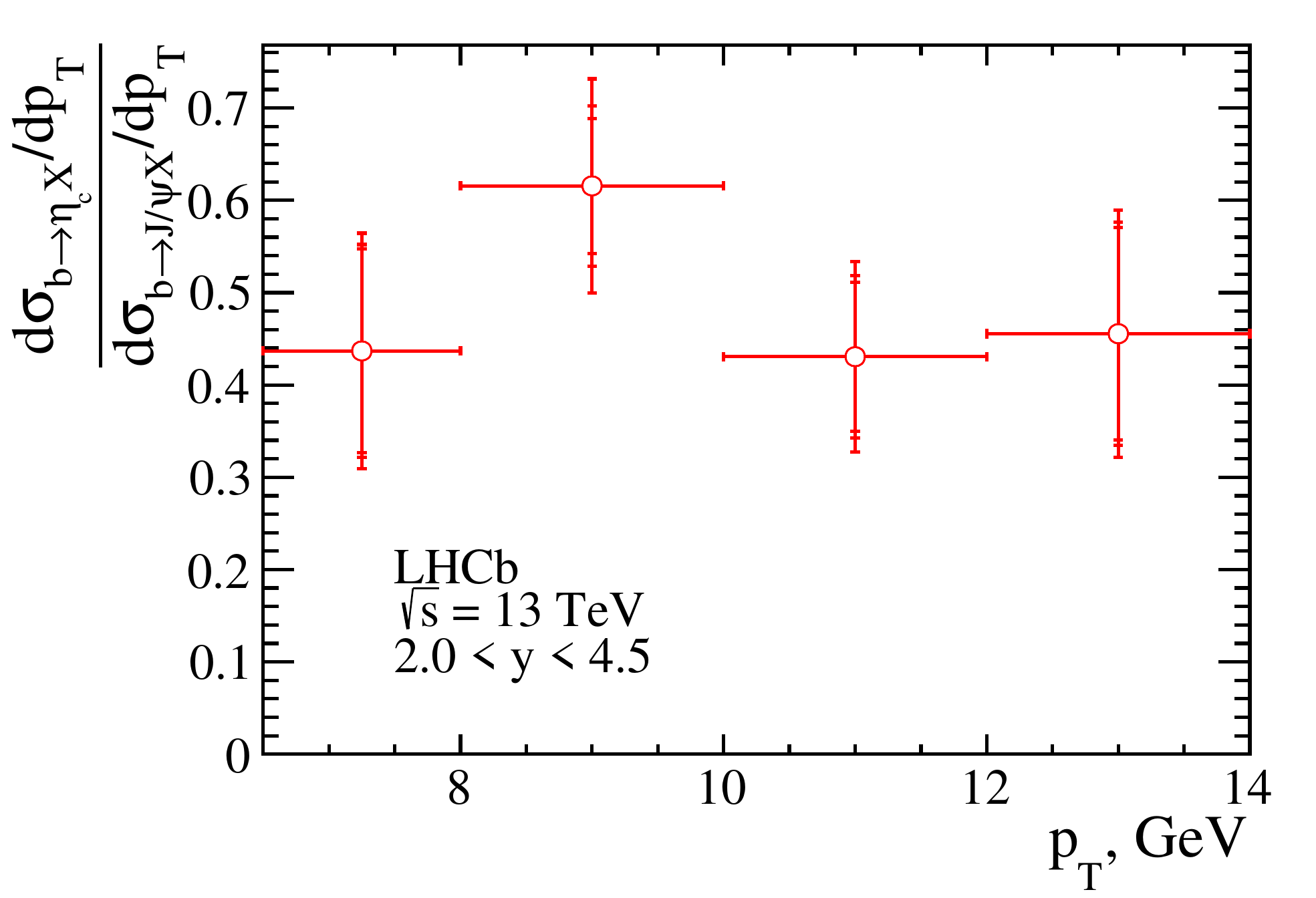}
          \label{fig:sigmaRelFromB}}
 }
\caption
[The ratios of \etac and \jpsi differential production cross-sections.]
{The ratios of \etac and \jpsi differential production cross-sections. The uncertainties shown are statistical, systematic, and the uncertainty due to the \JpsiToPpbar and \JpsiToPpbar branching fractions.} 
\label{fig:sigmaRel}
\end{figure}

\begin{figure}[h]
\centering{
        \subfigure[Prompt production cross section.]{ 
          \protect\protect\protect\includegraphics[width=0.465\textwidth]{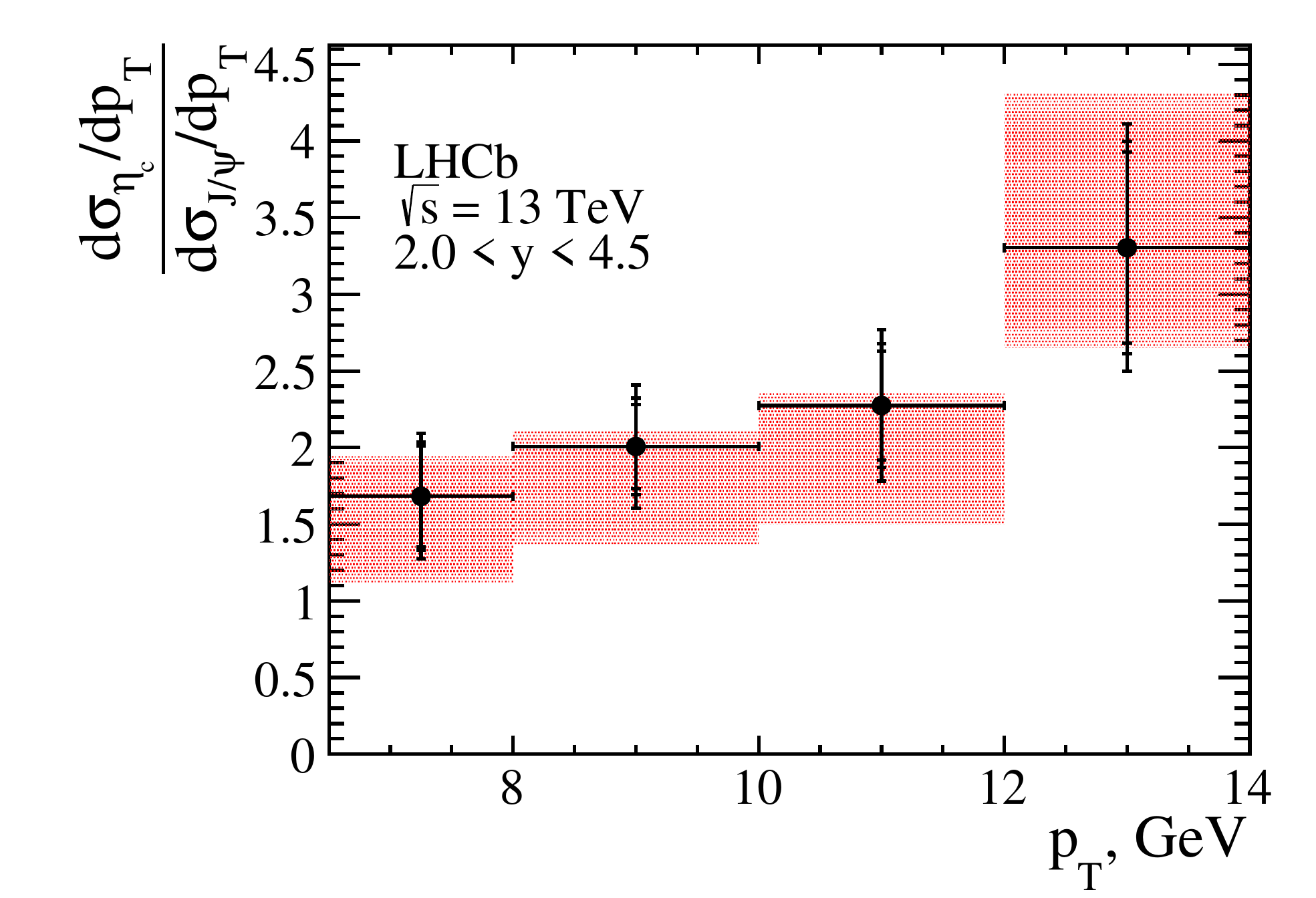}
          \label{fig:sigmaRelPrompt_vsRunI}}
\quad
        \subfigure[Production in inclusive \bquark-decays.]{ 
        \protect\protect\protect\includegraphics[width=0.465\textwidth]{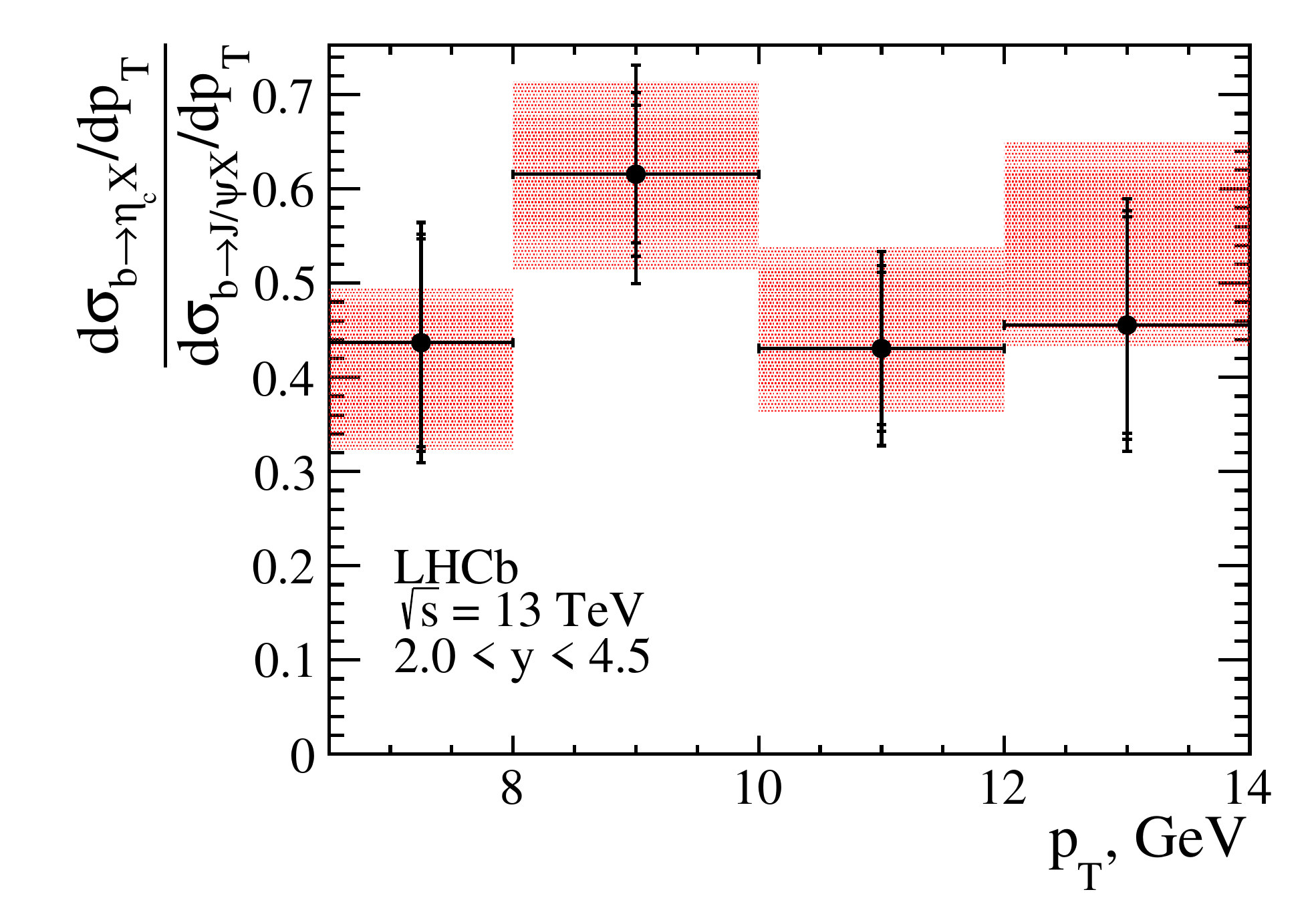}
        \label{fig:sigmaRelFromB_vsRunI}}
 }
\caption
[The ratios of \etac and \jpsi differential production cross-sections compared to result obtained with \tzcut.]
{The ratios of \etac and \jpsi differential production cross-sections (points) compared to result obtained with \tzcut (red boxes). The uncertainties shown are statistical, systematic, and the uncertainty 
    due to the \JpsiToPpbar and \EtacToPpbar branching fractions.} 
\label{fig:vsRunI}
\end{figure}

\clearpage
The comparison of the measurements using the \tzcut and the \tzfit is shown on Fig.~\ref{fig:vsRunI}.
Both measurements give consistent results in all \pt-bins for both prompt production and production in \bquark-decays. 
The two measurements are strongly correlated. For the prompt production the measurement using \tzfit is more robust and is chosen as a final result.
For the measurement of the production in \bquark-decays the \tzcut gives a more precise result which is retained. The obtained values of the relative differential cross-sections using both analysis techniques listed in Tables~\ref{tab:relPromptTable} and~\ref{tab:relSecondaryTable}. 
\begin{table}[b]
\centering
\small
\begin{tabular}{c|c|c}
\pt, \gev      & \multicolumn{2}{c}{$d\sigma^{prompt}_{\etac}/d\sigma^{prompt}_{\jpsi}$} \\ \hline
   & \tzfit & \tzcut \\ \hline
6.5 - 8.0     & 1.68 $\pm$ 0.33 $\pm$ 0.06 $\pm$ 0.11 $\pm$ 0.21 & 1.53 $\pm$ 0.35 $\pm$ 0.05 $\pm$ 0.09 $\pm$ 0.19 \\ 
8.0 - 10.0    & 2.01 $\pm$ 0.28 $\pm$ 0.09 $\pm$ 0.13 $\pm$ 0.25 & 1.74 $\pm$ 0.28 $\pm$ 0.07 $\pm$ 0.10 $\pm$ 0.22   \\  
10.0 - 12.0   & 2.27 $\pm$ 0.36 $\pm$ 0.13 $\pm$ 0.14 $\pm$ 0.28 & 1.93 $\pm$ 0.33 $\pm$ 0.10 $\pm$ 0.11 $\pm$ 0.24   \\ 
12.0 - 14.0     & 3.30 $\pm$ 0.62 $\pm$ 0.22 $\pm$ 0.21 $\pm$ 0.41 & 3.48 $\pm$ 0.64 $\pm$ 0.23 $\pm$ 0.20 $\pm$ 0.43 \\
\end{tabular} 
\caption{The \pt-differential ratios of \etac and \jpsi differential prompt production cross-sections.}
\label{tab:relPromptTable}
\centering
\small
\begin{tabular}{c|c|c}
\pt, \gev     & \multicolumn{2}{c}{$d\sigma^{\bquark-decays}_{\etac}/d\sigma^{\bquark-decays}_{\jpsi}$} \\ \hline 
    & \tzfit & \tzcut \\ \hline
6.5 - 8.0     & 0.44 $\pm$ 0.11 $\pm$ 0.02 $\pm$ 0.03 $\pm$ 0.05 & 0.41 $\pm$ 0.06 $\pm$ 0.01 $\pm$ 0.02 $\pm$ 0.05 \\ 
8.0 - 10.0    & 0.62 $\pm$ 0.07 $\pm$ 0.03 $\pm$ 0.04 $\pm$ 0.08 & 0.61 $\pm$ 0.05 $\pm$ 0.03 $\pm$ 0.03 $\pm$ 0.08 \\  
10.0 - 12.0   & 0.43 $\pm$ 0.08 $\pm$ 0.02 $\pm$ 0.03 $\pm$ 0.05 & 0.45 $\pm$ 0.06 $\pm$ 0.02 $\pm$ 0.02 $\pm$ 0.06  \\ 
12.0 - 14.0     & 0.46 $\pm$ 0.12 $\pm$ 0.02 $\pm$ 0.03 $\pm$ 0.06 & 0.54 $\pm$ 0.07 $\pm$ 0.03 $\pm$ 0.02 $\pm$ 0.07  \\   
\end{tabular} 
\caption{The \pt-differential ratios of \etac and \jpsi differential production cross-sections in inclusive \bquark-decays.}
\label{tab:relSecondaryTable}
\end{table}  

\begin{figure}[h]
\centering{
        \subfigure[Prompt-production cross section from \tzfit. The results of the fits by exponential functions are overlaid.]{ 
      \protect\protect\protect\includegraphics[width=0.465\textwidth]{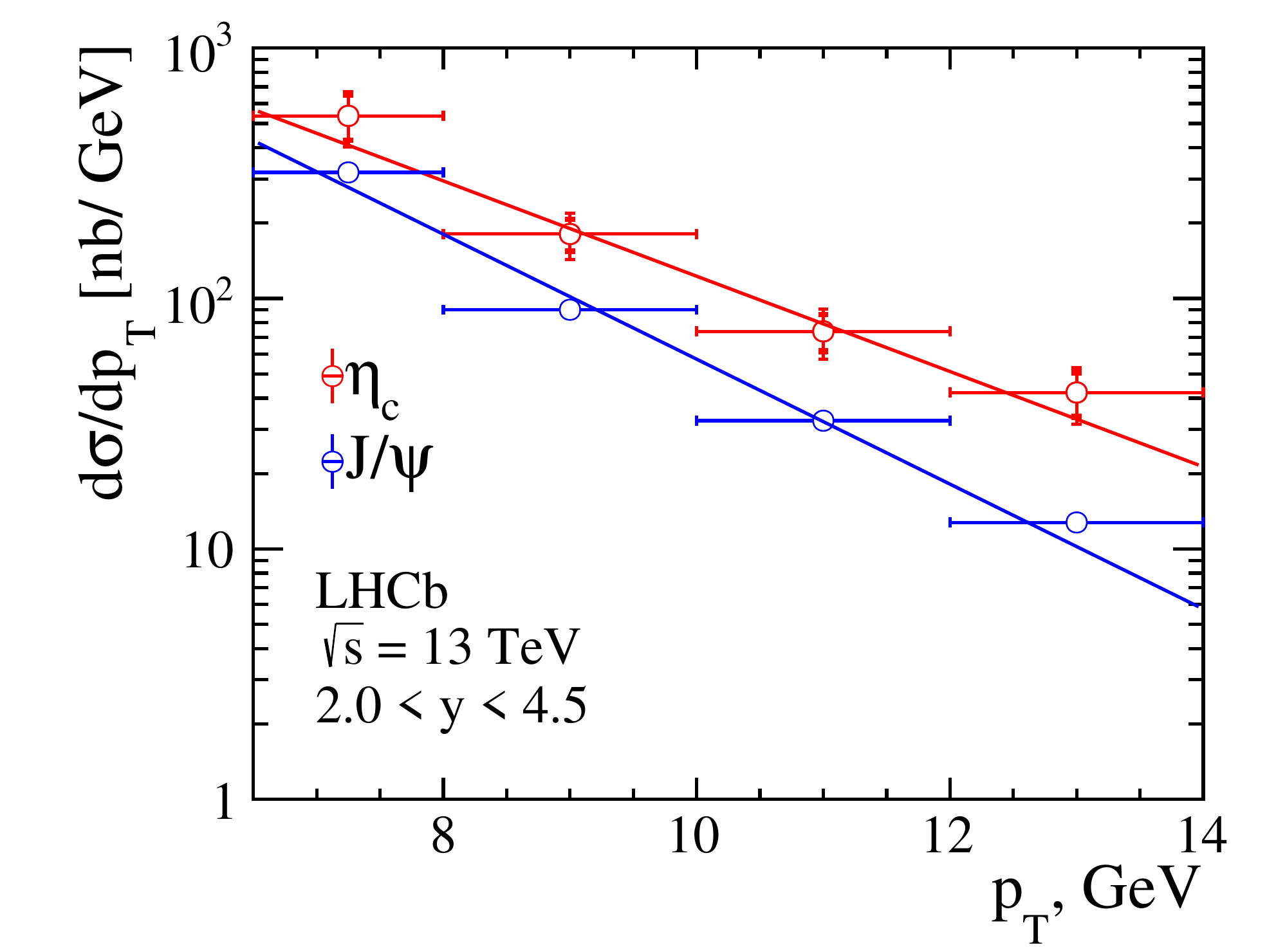}
      \label{fig:sigmaAbsPrompt}
      }
\quad
        \subfigure[Production in inclusive \bquark-decays from \tzcut.]{ 
       \protect\protect\protect\includegraphics[width=0.465\textwidth]{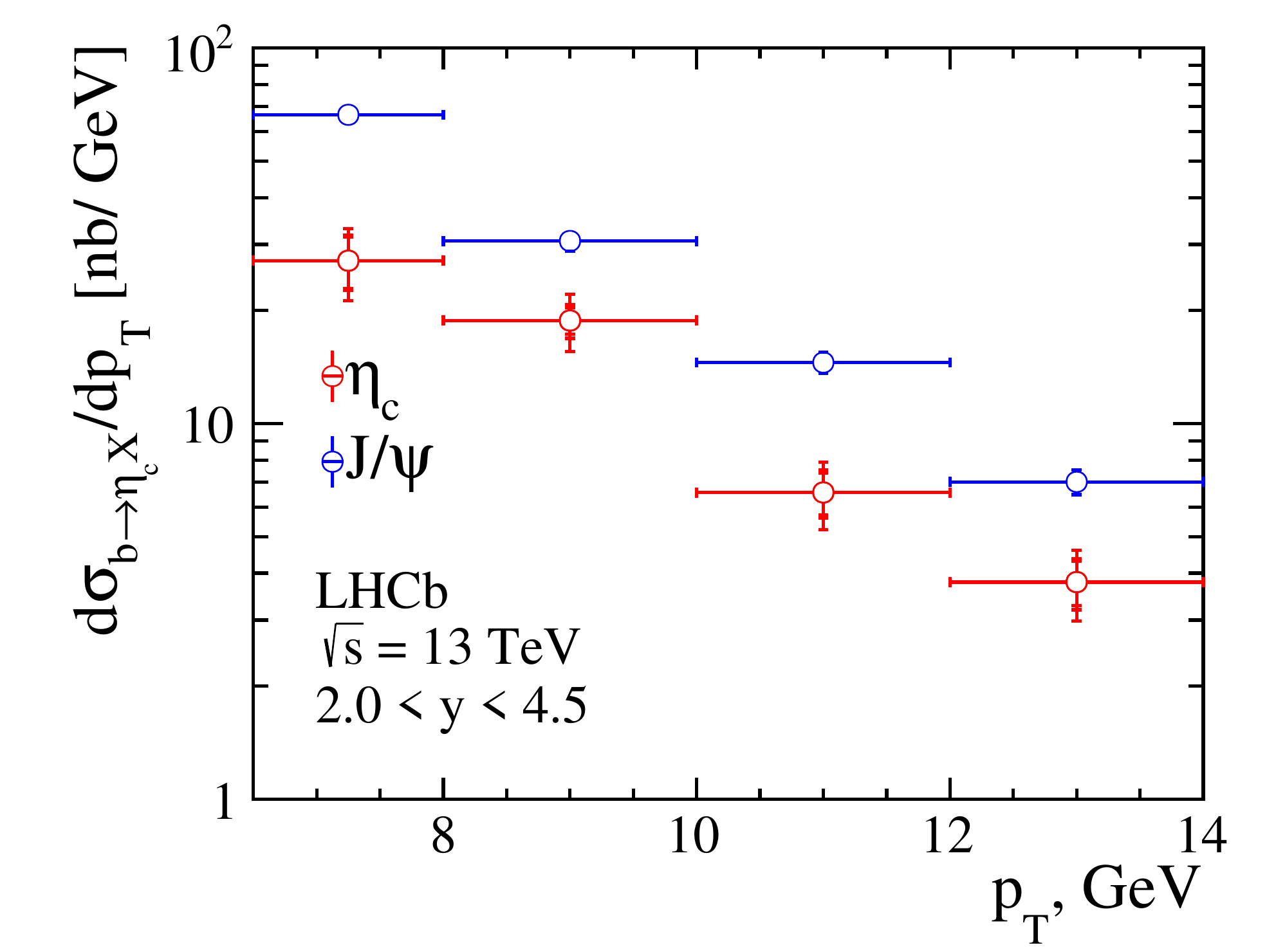}
      \label{fig:sigmaAbsFromB}
      }
 }
\caption
[The \etac and \jpsi \pt-differential production cross section in inclusive \bquark-decays from \tzcut.]
{The \etac (red) and \jpsi (blue) \pt-differential production cross section in inclusive \bquark-decays from \tzcut. 
The shown uncertainties for \etac production are statistical, systematic, and the uncertainty due to the \JpsiToPpbar and \EtacToPpbar branching fractions and \jpsi production cross-section.} 
\label{fig:sigmaAbs}
\end{figure}

\clearpage
\newpage
The absolute \etac differential production cross-sections are shown on Fig.~\ref{fig:sigmaAbs} for both prompt \etac and \etac produced in inclusive \bquark-decays. The obtained values of the absolute differential cross-sections using both analysis techniques are listed in Tables~\ref{tab:PromptTable} and~\ref{tab:SecondaryTable}. 
This is the first \pt-differential cross-section measurement of the \etac prompt production at \sqs=13~\tev. For illustative reasons, the exponential slopes of \pt-dependences of \etac and \jpsi prompt production are obtained to be $e_{\etac}=0.44\pm0.06 \gev^{-1}$ and $e_{\jpsi}=0.57\pm0.01 \gev^{-1}$, respectively. 
Contrary to NRQCD expectations, the \lhcb result indicates a steeper dependence of differential cross-section for \jpsi compared to that of \etac. 
It is important to confirm and possibly measure more accurately the difference in the \pt-slope by extending the measurement to larger \pt values. A value of the \pt-slope larger than prediction from Ref.~\cite{Feng:2019zmn}could be an indication of a possible color octet contribution. 
\clearpage

\begin{table}[t]
\centering
\footnotesize
\begin{tabular}{c|c|c}
\pt, \gev      & \multicolumn{2}{c}{$d\sigma^{prompt}_{\etac}/d\pt$, \nb/\gev}  \\ \hline
  & \tzfit & \tzcut \\ \hline
6.5 - 8.0     & 536.09 $\pm$ 105.04 $\pm$ 19.61 $\pm$ 34.19 $\pm$ 70.67 & 487.53$\pm$110.49$\pm$17.01$\pm$28.24$\pm$64.27  \\ 
8.0 - 10.0    & 180.92 $\pm$ 24.81 $\pm$ 7.90 $\pm$ 11.35 $\pm$ 24.97 & 156.68$\pm$25.23$\pm$ 6.32$\pm$ 8.97$\pm$21.62 \\  
10.0 - 12.0   & 73.92 $\pm$ 11.57 $\pm$ 4.07 $\pm$ 4.60 $\pm$ 10.32 & 62.70$\pm$10.61$\pm$ 3.29$\pm$ 3.54$\pm$ 8.75 \\ 
12.0 - 14.0     & 42.12 $\pm$ 7.95 $\pm$ 2.83 $\pm$ 2.62 $\pm$ 6.01 & 44.36$\pm$ 8.13$\pm$ 2.88$\pm$ 2.52$\pm$ 6.33 \\ 
\end{tabular} 
\caption{The \pt-differential \etac prompt production.}
\label{tab:PromptTable}
\footnotesize
\centering
\small
\begin{tabular}{c|c|c}
\pt, \gev     & \multicolumn{2}{c}{$d\sigma^{\bquark-decays}_{\etac}/d\pt$, \nb/\gev}     \\ \hline 
   & \tzfit & \tzcut \\ \hline
6.5 - 8.0     & 29.02 $\pm$ 7.33 $\pm$ 1.19 $\pm$ 1.86 $\pm$ 3.99 & 27.16$\pm$ 4.23$\pm$ 0.99$\pm$ 1.34$\pm$ 3.74 \\ 
8.0 - 10.0    & 18.87 $\pm$ 2.24 $\pm$ 0.81 $\pm$ 1.19 $\pm$ 2.62 & 18.82$\pm$ 1.52$\pm$ 0.81$\pm$ 0.91$\pm$ 2.61 \\  
10.0 - 12.0   & 6.27 $\pm$ 1.18 $\pm$ 0.30 $\pm$ 0.41 $\pm$ 0.88 & 6.56$\pm$ 0.84$\pm$ 0.34$\pm$ 0.32$\pm$ 0.93 \\ 
12.0 - 14.0     & 3.19 $\pm$ 0.81 $\pm$ 0.17 $\pm$ 0.21 $\pm$ 0.47 & 3.79$\pm$ 0.51$\pm$ 0.23$\pm$ 0.18$\pm$ 0.55 \\  
\end{tabular} 
\caption{The \pt-differential \etac production cross-section in inclusive \bquark-decays.}
\label{tab:SecondaryTable}
\end{table}

The relative prompt production rates of the \etac and \jpsi states in the LHCb fiducial region is measured to be 
\begin{align*}
\left( \sigma_{\etac}/\sigma_{\jpsi} \right)_{13 \tev}^{6.5 \gev < \pt < 14.0 \gev, 2.0<y<4.5} 
  &= \etacPromptRelativeXsec. 
\end{align*}
using \tzfit.
The \etac prompt production cross section in the LHCb fiducial region is then derived to be 
\begin{align*}
\left( \sigma_{\etac} \right)_{13 \tev}^{6.5 \gev < \pt < 14.0 \gev, 2.0<y<4.5}  
  &= \etacPromptAbsoluteXsec.
\end{align*}
For comparison, according to Ref.~\cite{LHCb-PAPER-2015-037}, the \jpsi production at the same kinematic regime was measured to be:
\begin{align*}
\left( \sigma_{\jpsi} \right)_{13 \tev}^{6.5 \gev < \pt < 14.0 \gev, 2.0<y<4.5}  
  &= \jpsiPromptAbsoluteXsec.
\end{align*}
This is the first measurement of the \etac production at \sqs=13~\tev. The obtained result supports the first conclusions from Ref.~\cite{LHCb-PAPER-2014-029} on more prolific \etac production compared to \jpsi.
The obtained \etac prompt production cross-section 
is in a good agreement with color singlet model prediction of $1.56^{+0.83}_{-0.49}\,^{+0.38}_{-0.17}~\mub$, where the first and second uncertainties are due to scale and PDF (CT14NLO), respectively~\cite{Feng:2019zmn}. This leaves a limited room for a potential color octet contribution. This confirms the conclusion~\cite{Han:2014jya,Bodwin:2014gia,Zhang:2014ybe,Butenschoen:2014dra} from \etac production studies at \sqs=7 \tev and \sqs=8 \tev~\cite{LHCb-PAPER-2014-029}.

The \etac inclusive branching fraction from \bquark-hadron decays is measured to be:
\begin{align*}
\BR_{\bToEtacX}/\BR_{\bToJpsiX} &= \etacSecondaryRelativeBR 
\end{align*}
and
\begin{align*}
\BR_{\bToEtacX} &= \etacSecondaryAbsoluteBR. 
\end{align*}
using more precise \tzcut.

\begin{figure}[htb]
\centering{
        \subfigure[Relative \etac prompt production cross section.]{
        \protect\protect\protect\includegraphics[width=0.465\textwidth]{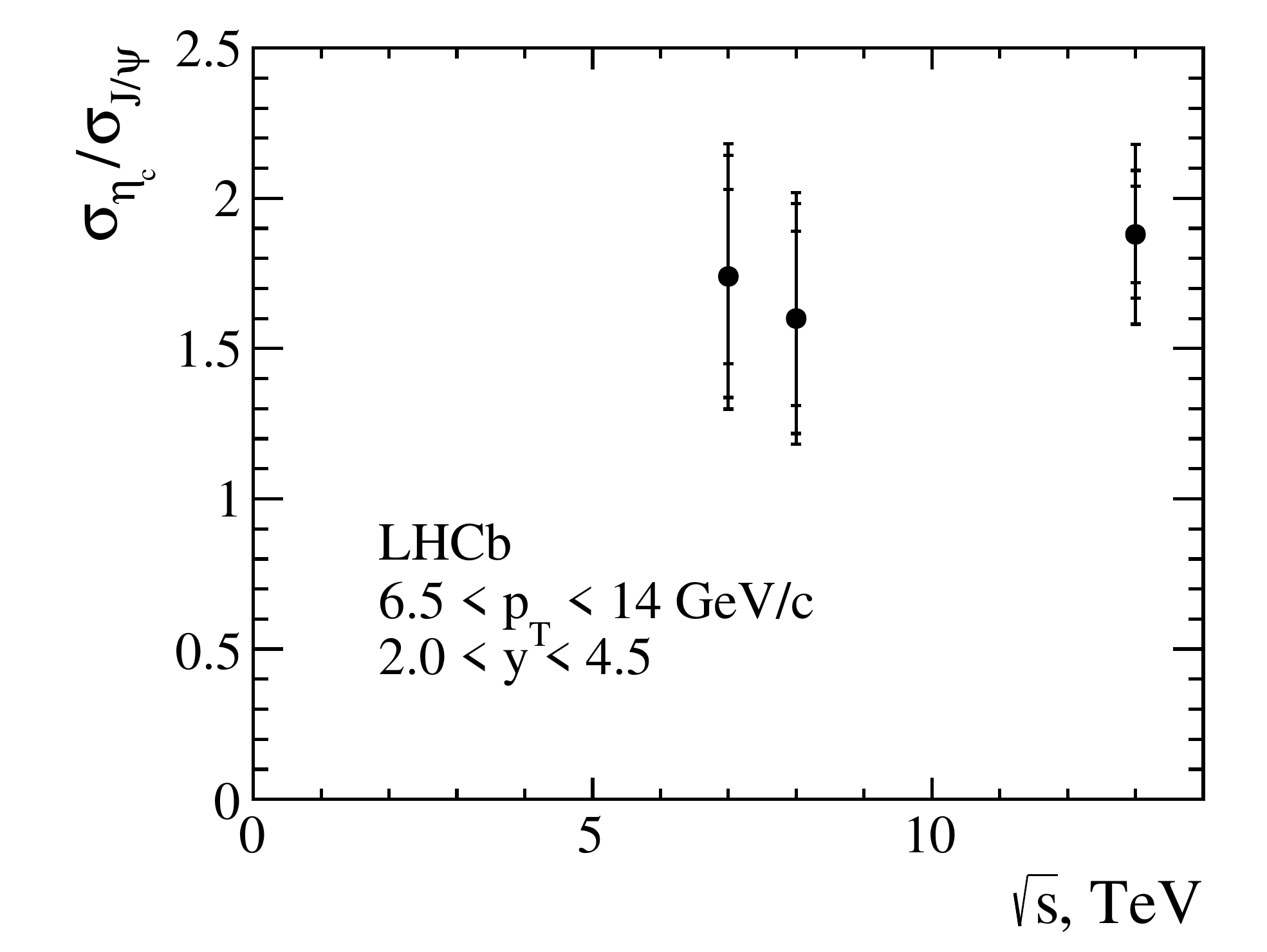}
      \label{fig:sigmaAbsPrompt}
          }
\quad
        \subfigure[Absolute \etac (black points) and \jpsi (blue points) prompt production cross section.]{ 
        \protect\protect\protect\includegraphics[width=0.465\textwidth]{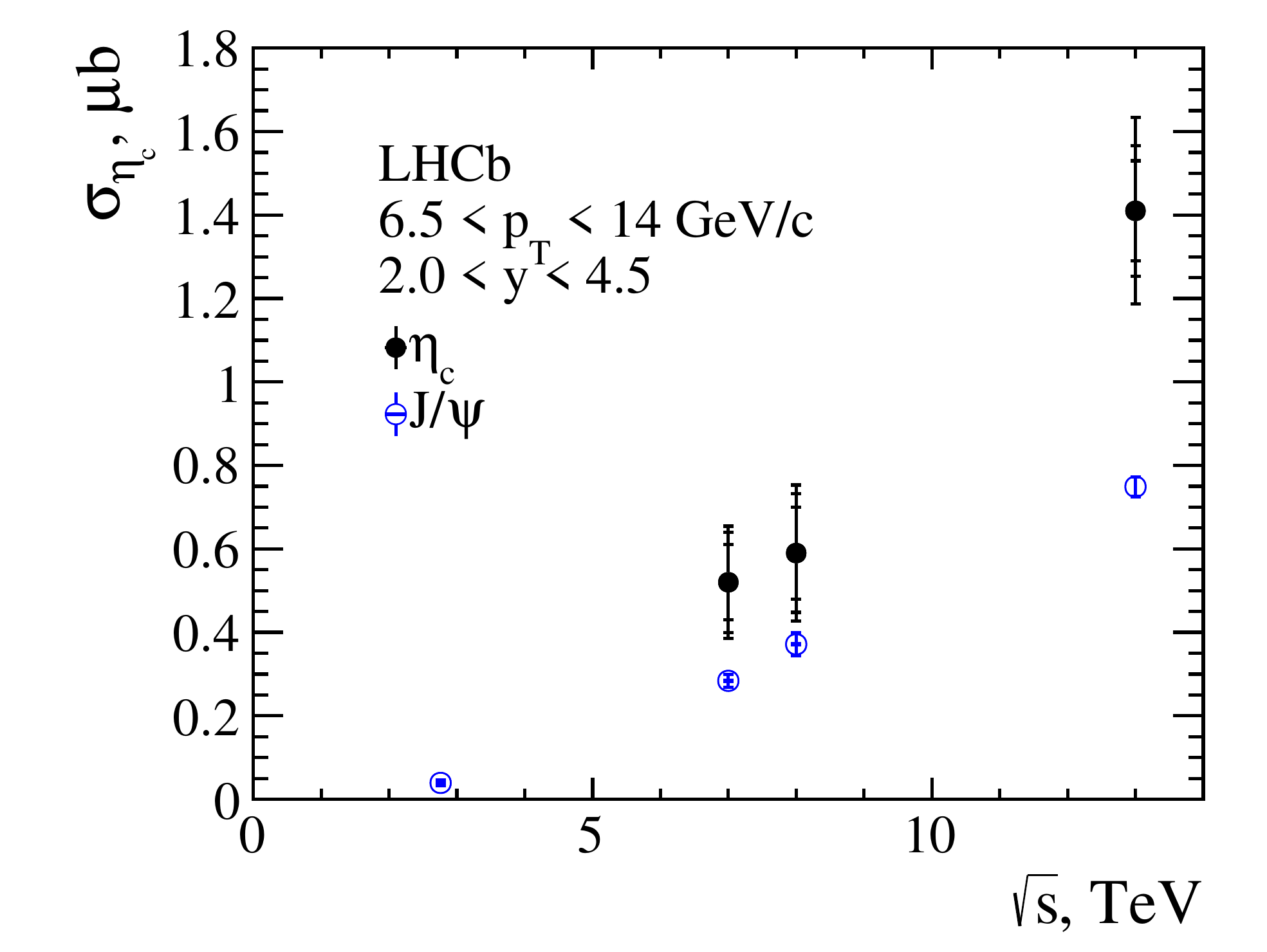}
      \label{fig:sigmaAbsFromB}
          }
 }
\caption{The \etac production as a function of centre-of-mass energy.} 
\label{fig:Xsec_vs_s}
\end{figure}

\begin{figure}[htb]
\centering{
        \subfigure[Prompt production.]{
        \protect\protect\protect\includegraphics[width=0.465\textwidth]{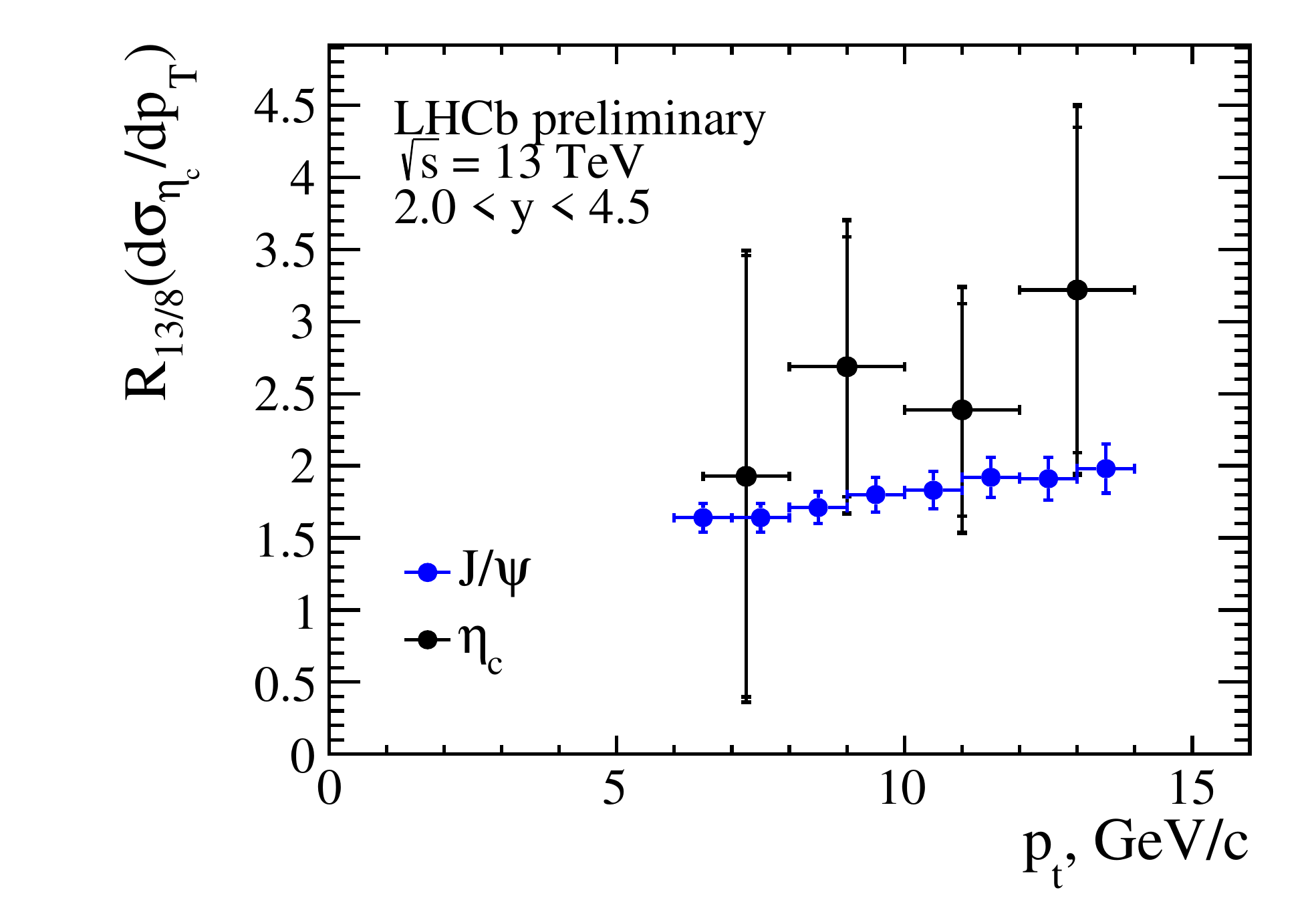}
         \label{fig:13/8Prompt}
          }
\quad
        \subfigure[Production in inclusive \bquark-decays.]{
        \protect\protect\protect\includegraphics[width=0.465\textwidth]{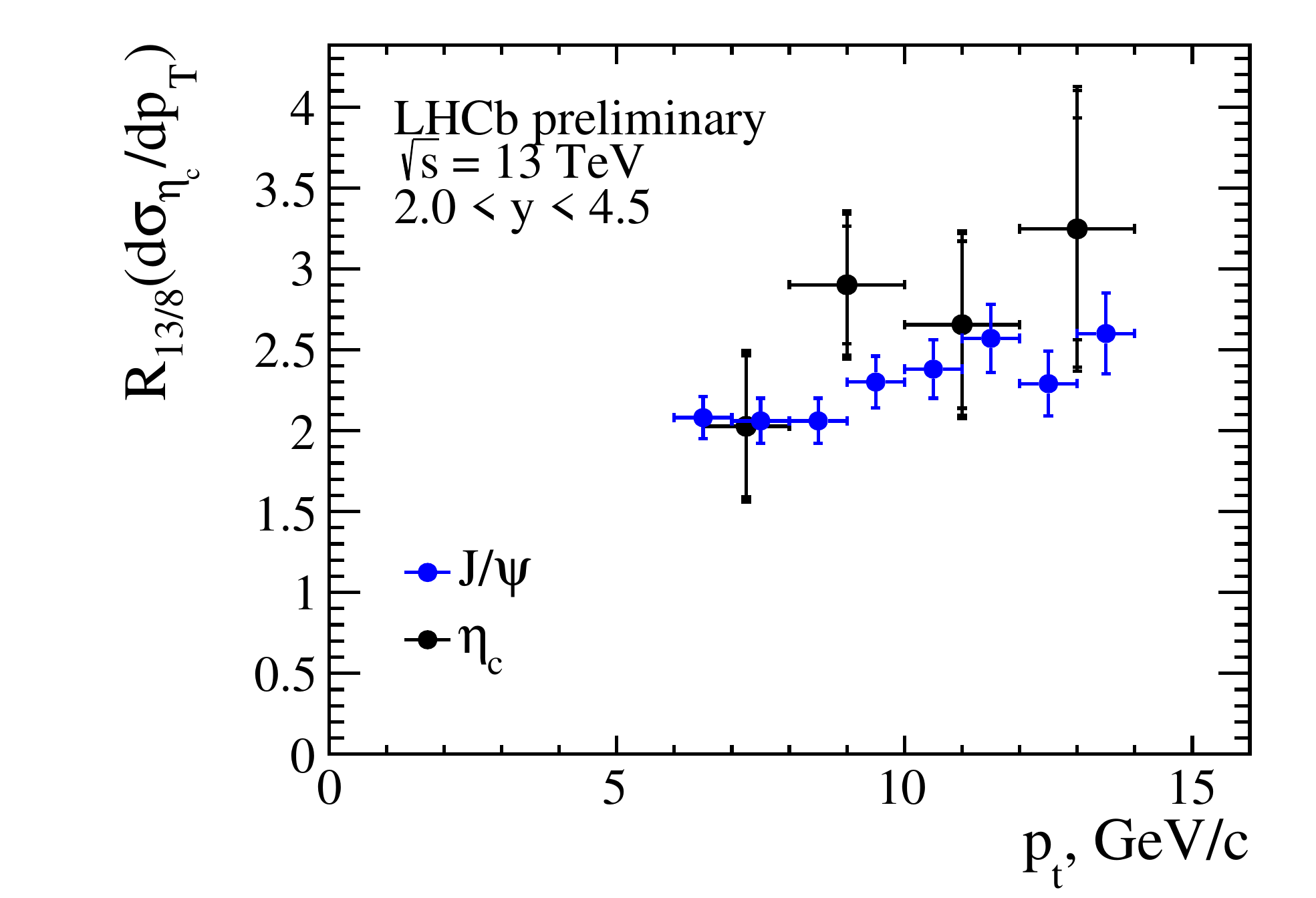}
        \label{fig:13/8FromB}
          }
 }
\caption
[Ratios of the \etac and \jpsi differential production cross-sections at \sqs=13 and 8 TeV.]
{Ratios of the \etac (black points) and \jpsi (blue points) differential production cross-sections at \sqs=13 and 8 TeV. The uncertainties shown are statistical, systematic and due to \jpsi production cross-section ratio.} 
\label{fig:13/8}
\end{figure}

Using also the \lhcb measurement of the \etac prompt production at the \sqs=7 \tev and \sqs=8 \tev~\cite{LHCb-PAPER-2014-029}, the prompt \etac production cross-section is shown as a function of the centre-of-mass energy on Fig.~\ref{fig:Xsec_vs_s}. Using the \jpsi production cross-section ratio from Ref.~\cite{LHCb-PAPER-2015-037}, the ratio of \etac production at \sqs=13 \tev and \sqs=8 \tev is shown on Fig~\ref{fig:13/8}. The corresponding \jpsi production ratio in the same kinematic regime from Ref.~\cite{LHCb-PAPER-2015-037} is shown for comparison. Note, that the uncertainty on the \etac ratio is strongly dominated by the statistical uncertainties of both measurements at 8 and 13~\tev and hence the ratio is less precise than the absolute cross-section measurement.

\begin{singlespace}
\chapter{Study of charmonium states production using decays to $\phi\phi$}
\end{singlespace}
\label{ch:phiphi}
The charmonium decays to $\phi\phi$ is promising to access non-$1^{-}$ states.
It requires a reconstruction of four kaon tracks, which leads to smaller reconstruction efficiency compared to charmonium decays to \ppbar with only two tracks in the final state. Nevertheless, it is not a priori clear, which channel is more advantageous to measure charmonium production.
In proton-proton collisions, the number of produced $\phi\phi$ pairs is smaller than the number of \ppbar pairs since four \squark quarks have to be created to produce $\phi\phi$ combination. In addition, the narrow $\phi$ resonance is situated near the $\Kp\Km$ threshold, and hence the background level under the $\phi$ signal is limited. Besides, the branching fractions of decays of excited charmonium states to \ppbar are typically smaller than the ones for $\phi\phi$ decays.

This chapter summarises studies of charmonium states $\etac(1S)$, \chiczero, \chicone, \chictwo and \etactwos production in \bquark-hadron inclusive decays using charmonia decays to $\phi \phi$ with the \lhcb experiment. Within this analysis, the \chiczero and \etactwos unambiguous signals are reconstructed, which is already essential to test theoretical predictions. 
Due to a much smaller number of $\phi\phi$ combinations produced at PV, the trigger aiming at the reconstruction of prompt $\phi\phi$ pairs would require much smaller trigger bandwidth compared to the trigger selecting \ppbar combinations.
Therefore, this analysis can also be considered as a first step to measuring prompt production of charmonium states, which are not accessible using other decay channels (e.g. \etactwos and \chiczero).

This chapter is organised in the following way. After the analysis setup described in Section~\ref{sec:phiphiSetup}, the data and simulation samples are discussed in Section~\ref{sec:dataPhiPhi}. The selection criteria and signal efficiencies are shown in Section~\ref{sec:sel}. The results on the charmonium production in \bquark-hadron decays are presented in Section~\ref{sec:ccsection}. Section~\ref{sec:limits} stands for a search for production of charmonium-like states. Finally, the summary is given in Section~\ref{sec:sumPhiPhi}.
\clearpage
\section{Analysis setup}
\label{sec:phiphiSetup} 
The main target of present analysis is to measure production of \chic and \etactwos states in inclusive \bquark-hadron decays.
Since strong decays of $J^{PC} = 1^{- -}$ (\jpsi, \psitwos) states to two $\phi$ are forbidden, the decay mode $\etac(1S) \to \phi \phi$ is used as normalisation. 
The branching fraction of inclusive \bquark decays to \etac meson was measured at \lhcb~\cite{LHCb-PAPER-2014-029} to be 
$\BR(\bquark \to \etac(1S) X) = (4.88 \pm 0.64 \pm 0.29 \pm 0.67_{\BR}) \times 10^{-3}$, 
where the third uncertainty is due to uncertainties 
on the \jpsi inclusive branching fraction from \bquark-hadron decays and branching fractions 
of the decays of \jpsi and $\etac(1S)$ to the \proton\antiproton final state.

A relative production of charmonium states A and B in the inclusive $b$-hadron decays is calculated from the ratio of observed event yields, efficiency ratio 
and ratio of branching fractions of A and B decays to $\phi \phi$, 
\begin{equation}
\frac{\BR ( b \to A X ) \times \BR ( A \to \phi \phi )}{\BR ( b \to B X ) \times \BR ( B \to \phi \phi )} 
= \frac{N_A}{N_B} \times \frac{ \varepsilon_B }{ \varepsilon_A } \ ,
\label{eq:odin}
\end{equation}
where $N_A$($N_B$) are the observed yields of A(B) state;
$\varepsilon_{A , B}$ are the corresponding total efficiencies to reconstruct, trigger and select $A \to \phi \phi$ and $B \to \phi \phi$ decays. 
For the states with similar kinematics - a good example is the ratio of the production of \chic states - efficincies are similar, with their ratio close to unity.

\section{Data sample, trigger and simulation} 
\label{sec:dataPhiPhi}
The present analysis uses the \proton\proton collision data recorded by the LHCb experiment at $\sqrt{s} = 7 \tev$ in 2011 and at $\sqrt{s} = 8 \tev$ in 2012. 
The analysis is based on an integrated luminosity $\int\mathcal{L}dt \approx 1.0$ fb$^{-1}$ accumulated in 2011 
and an integrated luminosity of $\int\mathcal{L}dt \approx 2.0$ fb$^{-1}$ accumulated in 2012. 
For data processing, the reconstruction version 14 (Reco14), is used. 

The same trigger lines as for the $\Bs \to \phi \phi$ study in Ref.~\cite{LHCb-PAPER-2014-026} were used. 
The L0 Hadron decision \texttt{L0HadronDecision$\_$TOS} or \texttt{L0Global$\_$TIS} are applied at L0 trigger level. 
At the level of the HLT1, \texttt{HLT1TrackAllL0Decision$\_$TOS} was used. 
At the level of the HLT2, \texttt{HLT2Topo(2,3,4)BodyBBDTDecision$\_$TOS} or 
\texttt{HLT2IncPhiDecision$\_$TOS} lines were used. 
The dedicated stripping lines (\texttt{StrippingCcbar2PhiPhiDetachedLine}, 
version 20r1 (Stripping20r1) are used.

The simulated events for this analysis are obtained using the \texttt{Pythia} (version 6 and 8) event generator and the \texttt{GEANT4} package. 
The following MC samples are used to study the \etac, \chic and \etactwos mass resolutions and efficiencies:
\begin{table}[ht] 
\begin{center}
{\small{
\centering
\begin{tabular}{l|l|l} 
 Sample                    & Event type & Sample size  \\ \hline \hline
$\etac (1S)\to \phi \phi$  & 18104060   & 2.0 M         \\
$\chiczero \to \phi \phi$  & 18104030   & 1.1 M         \\
$\chicone  \to \phi \phi$  & 18104040   & 1.1 M         \\
$\chictwo  \to \phi \phi$  & 18104050   & 1.1 M         \\
$\etac (2S)\to \phi \phi$  & 18104080   & 1.1 M     
\end{tabular}
}}
\caption{Simulation samples.}
\label{tab:MCphi}
\end{center}
\end{table}
In the simulation charmonium states are required to be produced in the decays of long-lived \bquark-hadrons. 
Charmonia decays as well proceed via phase space decay model. At the generator level, all daughter particles are required to fly into \lhcb acceptance. Reconstructed signal candidates and their daughter particles are required to match the generated ones.

\section{Event selection} 
\label{sec:sel}
Selection aims at distinguishing pure $\phi$ candidates from the background 
by using charged kaon identification, 
narrow $\phi$ signal and at a later stage employing 2D fit procedure to select true \phiphi combinations (Section~\ref{sec:procedure}).
In order to select \bquark-hadrons, 
flying on average about $1 \cm$ in the \lhcb detector before their decay,
and suppress combinatorial background associated to the PV, 
measurements of impact parameter of daughter kaons and a distance between \bquark-production and \bquark-decay vertices are used. 

The $\phi$ candidates are reconstructed from oppositely charged particles identified as kaons by the \lhcb detector, $ProbNNk > 0.1$. 
Both kaon track candidates are required to have a good quality of track reconstruction, $\chisqndf<3$.
In order to suppress combinatorial background, the kaon tracks are required 
to have transverse momenta larger than $0.5 \gev$. 
Since decays of \bquark-hadrons are searched for, kaon tracks consistent with originating from PV 
are eliminated from the analysis by requiring large $\chi^2$ value of the correcponding IP with respect to any PV, $\chisqip > 4$. 

The $K^+ K^-$ pairs forming $\phi$ candidates are required to have a good quality common vertex, $\chisqndf < 25$. 
The $K^+ K^-$ invariant mass is required to be within $\pm 12 \mev$ from the known $\phi$ mass~\cite{PDG2016}. 

Two $\phi$ candidates are required to form good quality common vertex, $\chisqndf < 9$. 
Finally, in order to further suppress combinatorial background associated with the tracks coming from PV, the common $\phi\phi$ vertex is required to be well-separated from 
the corresponding collision vertex with a flight distance significance of $\chi^2 > 100$. 

Table \ref{tab:cuts} summarizes selection criteria for charmonia and \Bs meson decays to $\phi \phi$. 
\begin{table}[t] 
{\small{
\centering
\begin{tabular}{l|l|l|l} 
   & Variable	&  Denotion & Requirement \\  \hline

  Kaons	& Track quality &  \chisqndf & $<3$ \\
        & Impact parameter to primary vertex & $\chi^{2}_{IP}$ & $>4$ \\
  	& Transverse momentum & \pt, \gev & $>0.5$ \\
  	& Identification & ProbNNk  & $>0.1$ \\  \hline

 $\phi$	& Vertex quality & $\chi^2$ & $<25$ \\  \hline
 	& Invariant mass & $| M_{K^+ K^-} - M_{\phi} |$, \mev & $<12$ \\  \hline

 $\phi\phi$ & Vertex quality &  \chisqndf & $<9$ \\  
            & Distance between the decay vertex & $\chi^2$ & $>100$ \\  
            & and the primary vertex & & \\  \hline
\end{tabular}
}}
\caption{Selection criteria for charmonia decays to $\phi \phi$.} 
\label{tab:cuts}
\end{table}
Almost all selection had to be fixed already at the trigger/stripping level to limit the corresponding bandwidth. 
Exceptions are kaon identification and distance between primary and secondary vertices.
The method applied to extract the signal does not seem to require strong kaon identification, since narrow $\phi$
peaks are selected. The charmonium yields are checked for stability 
against variantions in the PID requirement with no significant difference observed.

In order to obtain ratios of the branching fractions, 
efficiency ratios are determined using simulation samples to be
\begin{align*}
\frac{\varepsilon ( \chiczero \to \phi \phi )}{\varepsilon ( \etac (1S) \to \phi \phi )} &= 0.98 \pm 0.02 \ , \\
\frac{\varepsilon ( \chicone \to \phi \phi )}{\varepsilon ( \etac (1S) \to \phi \phi )} &= 1.04 \pm 0.02 \ , \\
\frac{\varepsilon ( \chictwo \to \phi \phi )}{\varepsilon ( \etac (1S) \to \phi \phi )} &= 1.16 \pm 0.03 \ , \\
\frac{\varepsilon ( \etac (2S) \to \phi \phi )}{\varepsilon ( \etac (1S) \to \phi \phi )} &= 1.40 \pm 0.04 \ ,
\end{align*}
where the uncertainties reflect the MC sample sizes. 

\begin{figure}[h]
\centering
\protect\protect\includegraphics[width=0.6\linewidth]{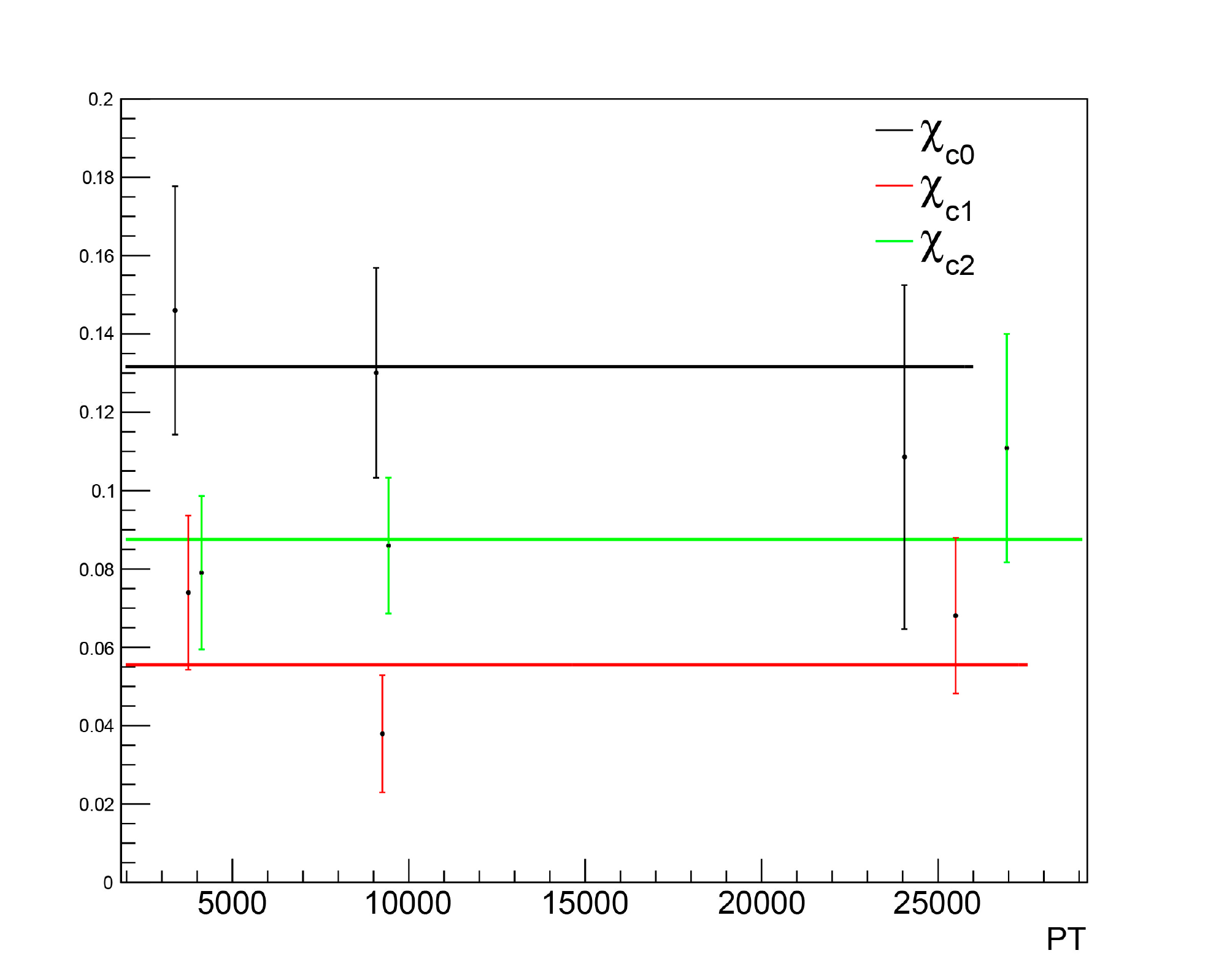}
\put(-280,80){\rotatebox{90}{$\frac{\BR(\bquark\to\chic X)\cdot \BR(\chic\to \phi\phi)}{\BR(\bquark\to\etac X) \cdot \BR(\etac\to \phi\phi)}$}}
\put(-240,185) {\footnotesize \lhcb-ANA-2015-038}
\put(-95,3){\colorbox{shadecolor}{\small $\pt(\phi\phi)$ $[\gev]$}}
\caption
[Ratio of the \chic states production in \bquark-hadron decays to that of $\etac(1S)$ in three bins of the \pt.]
{Ratio of the \chic states production in \bquark-hadron decays to that of $\etac(1S)$ 
in three bins of the \pt. 
Only statistical uncertainties are shown. 
Horizontal line corresponds to the fit result.} 
\label{fig:fitpt}
\end{figure}

\begin{figure}[h]
\centering
\protect\protect\includegraphics[width=0.6\linewidth]{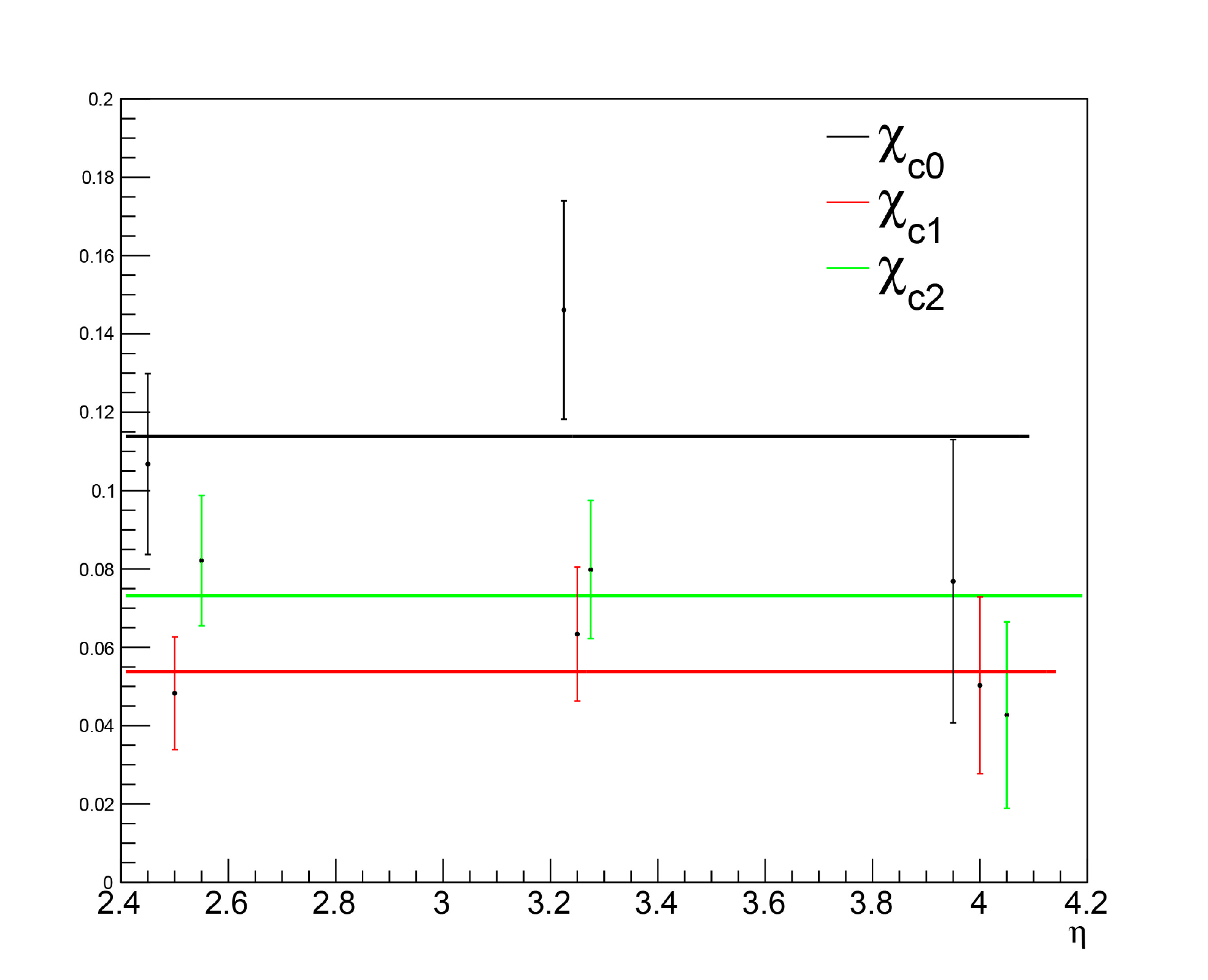}
\put(-280,80){\rotatebox{90}{$\frac{\BR(\bquark\to\chic X)\cdot \BR(\chic\to \phi\phi)}{\BR(\bquark\to\etac X) \cdot \BR(\etac\to \phi\phi)}$}}
\put(-230,185){\small \lhcb-ANA-2015-038}
\put(-40,3){\colorbox{shadecolor}{\small $\eta$ \phantom{111}}}
\caption
[Ratio of the \chic states production in \bquark-hadron decays to that of $\etac(1S)$ in three bins of pseudo-rapidity.]
{Ratio of the \chic states production in \bquark-hadron decays to that of $\etac(1S)$ in three bins of pseudo-rapidity. 
Only statistical uncertainties are shown. 
Horizontal line corresponds to the fit result.} \label{fig:fiteta}
\end{figure}

\begin{figure}[h]
\centering
\protect\protect\includegraphics[width=0.6\linewidth]{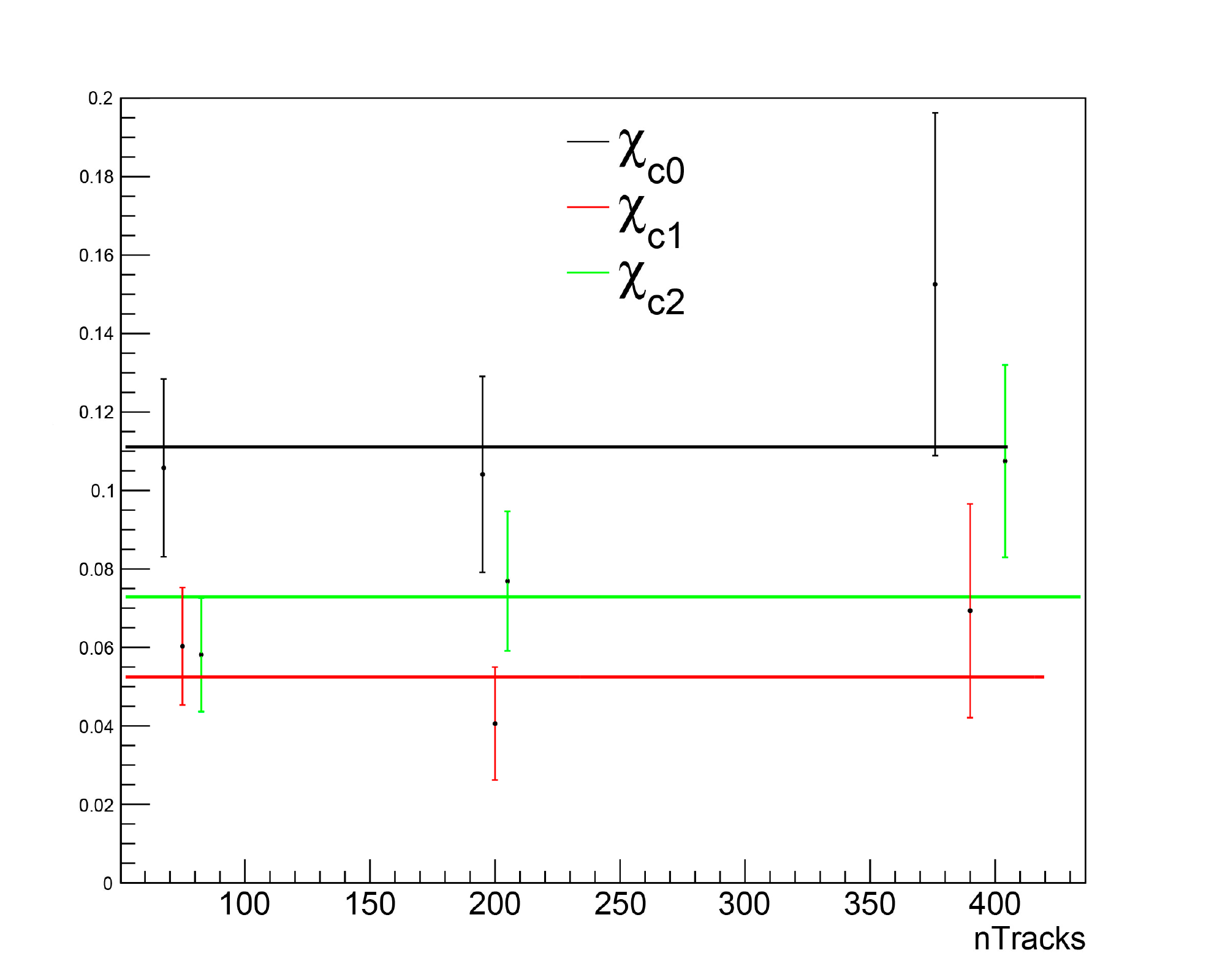}
\put(-280,80){\rotatebox{90}{$\frac{\BR(\bquark\to\chic X)\cdot \BR(\chic\to \phi\phi)}{\BR(\bquark\to\etac X) \cdot \BR(\etac\to \phi\phi)}$}}
\put(-240,185){\footnotesize \lhcb-ANA-2015-038}
\put(-115,3){\colorbox{shadecolor}{\small Number of tracks}}
\caption
[Ratio of the \chic states production in \bquark-hadron decays to that of $\etac(1S)$ in three bins of the event multiplicity.]
{Ratio of the \chic states production in \bquark-hadron decays to that of $\etac(1S)$ in three bins of the event multiplicity. 
Only statistical uncertainties are shown. 
Horizontal line corresponds to the fit result.} 
\label{fig:fitmult}
\end{figure}

\begin{figure}[h]
\centering
\protect\protect\includegraphics[width=0.6\linewidth]{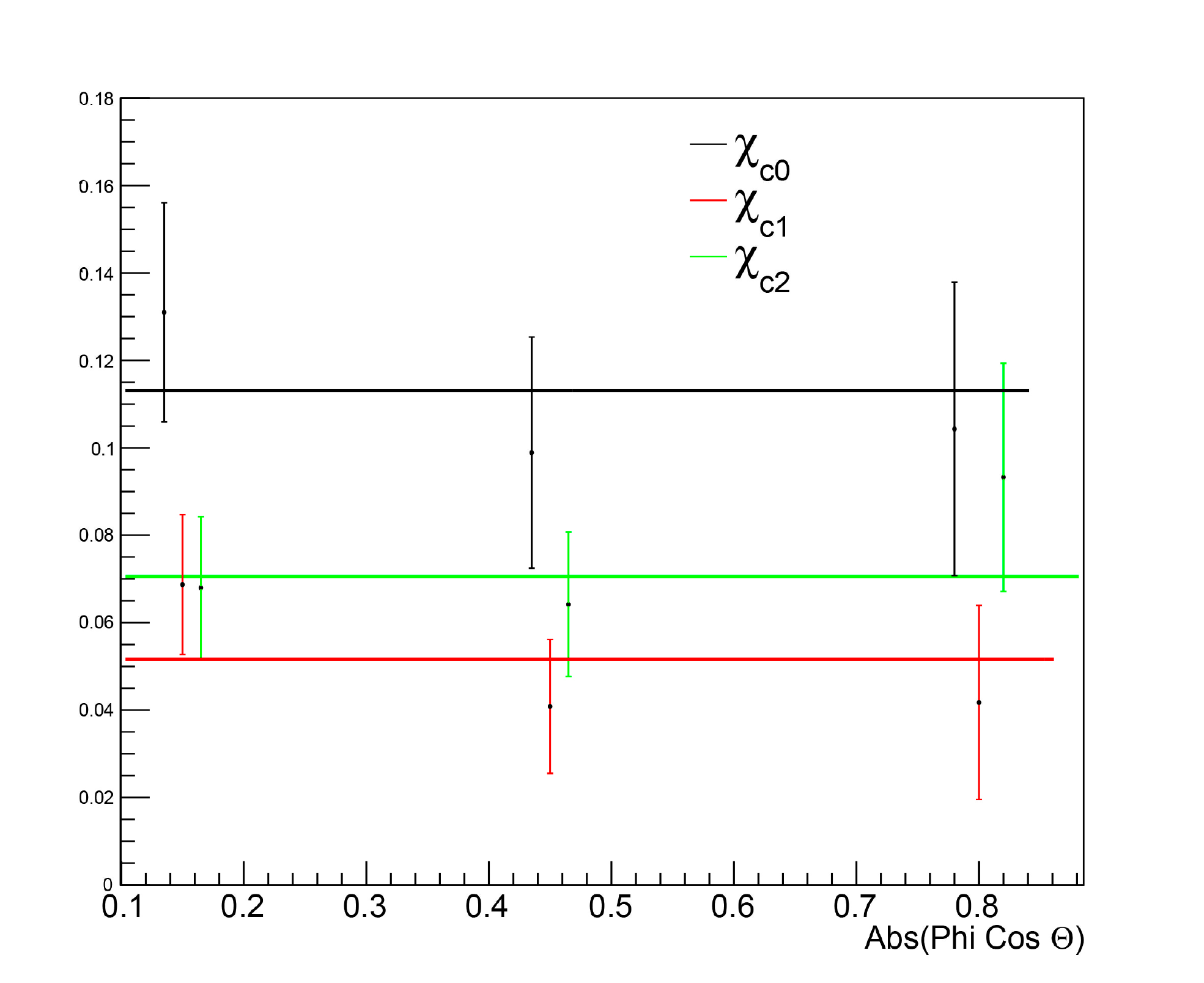}
\put(-280,80){\rotatebox{90}{$\frac{\BR(\bquark\to\chic X)\cdot \BR(\chic\to \phi\phi)}{\BR(\bquark\to\etac X) \cdot \BR(\etac\to \phi\phi)}$}}
\put(-240,185) {\footnotesize \lhcb-ANA-2015-038}
\put(-85,3){\colorbox{shadecolor}{\small \phantom{111} $|\cos \theta_\phi|$}}
\caption
[Ratio of the \chic states production in \bquark-hadron decays to that of $\etac(1S)$ in three bins of the absolute value of the $\cos \theta_\phi$, where $\theta_\phi$ is the flight angle of the $\phi$ 
meson in the charmonium rest frame with respect to the charmonium boost direction.]
{Ratio of the \chic states production in \bquark-hadron decays to that of $\etac(1S)$ 
in three bins of the absolute value of the $\cos \theta_\phi$, where $\theta_\phi$ is the flight angle of the $\phi$ 
meson in the charmonium rest frame with respect to the charmonium boost direction. 
Only statistical uncertainties are shown. 
Horizontal line corresponds to the fit result.} 
\label{fig:fitpol}
\end{figure}

Potential difference in the MC description of basic event properties and kinematics distributions 
could influence the efficiency ratios. 
A data-based cross-check for the distributions in \pt, pseudo-rapidity, event multiplicity, and polarization have been performed.
Figures~\ref{fig:fitpt}, ~\ref{fig:fiteta}, ~\ref{fig:fitmult}, and ~\ref{fig:fitpol} show ratios of the \chic states production in \bquark-hadron decays to that of the $\etac(1S)$, 
in three bins of \pt, pseudo-rapidity, event multiplicity, and polarization, respectively. The potential impact on the efficiencies are compared to the corresponding statistical uncertainties 
in Table~\ref{tab:diffs}. 
\begin{table}[t]
\centering
\begin{tabular}{l|c|c|c|c}
        & $\etac (1S)$ & \chiczero & \chicone & \chictwo \\ \hline 
Statistical uncertainty & $5 \%$ & $15 \%$ & $18 \%$ & $18 \%$ \\ \hline 
$\cos \theta_\phi$ & $+1 \%$ & $-4 \%$ & $-8 \%$ & $+6 \%$ \\ 
Event multiplicity & $-2 \%$ & $-8 \%$ & $< 1 \%$ & $-7 \%$ \\ 
Pseudo-rapidity & $-1 \%$ & $-8 \%$ & $-4 \%$ & $-2 \%$ \\ 
\end{tabular}
\caption{Effect on the efficiencies from potential differences 
in pseudo-rapidity, event multiplicity, and polarization, 
for considered charmonium states.
\label{tab:diffs}}
\end{table}
No significant dependence within statistical uncertainties is observed in any bin of each variable considered. 
The efficiencies have nevertheless been corrected to the central values of the observed difference in \pt distribution. 
Given similar quantum numbers of the \etac states and a small-size sample of the reconstructed $\etac(2S)$ 
candidates, the ratio of the corresponding efficiencies from the simulation is used. 

\clearpage
\clearpage

\clearpage
\section{Extraction of $\phi\phi$ signal yield}
\label{sec:procedure}
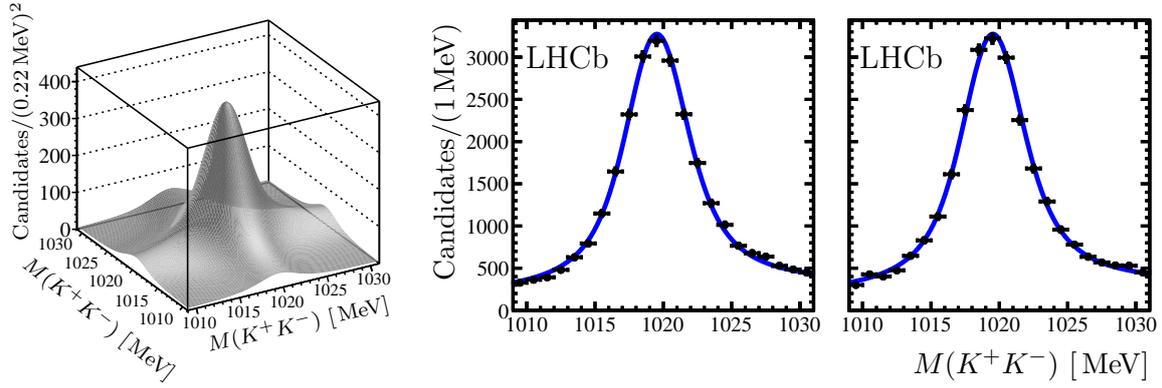
\begin{figure}[t]
\centering
\begin{picture}(450,150)
\put(15,6){\protect\protect\includegraphics[width=450px]{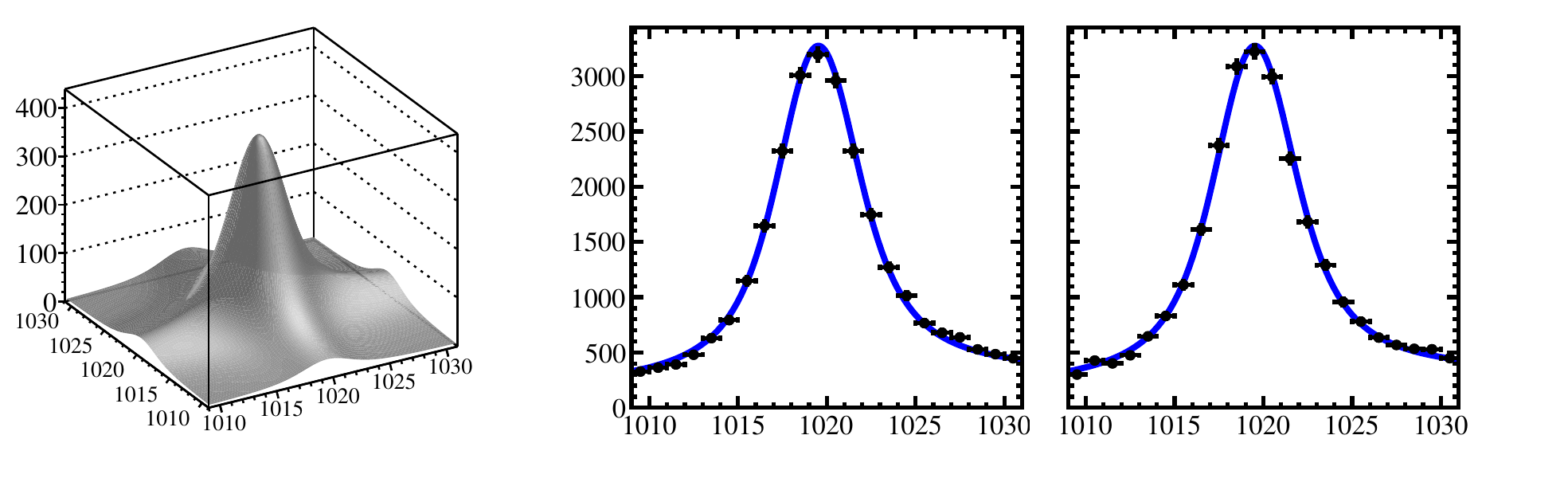}
                \put(-300,35){\rotatebox{90}{\small{Candidates/$( 1 \mev )$}}}
                \put(-458,50){\rotatebox{90}{\scriptsize{Candidates/$( 0.22 \mev )^2$}}}
                \put(-265,115){\lhcb}
                \put(-140,115){\lhcb}
                \put(-120,0) {\small{$M( \Kp \Km ) \ [ \mev ]$}}
                \put(-385,8) {\rotatebox{13}{\scriptsize{$M( \Kp \Km ) \ [ \mev ]$}}}
                \put(-455,40) {\rotatebox{322}{\scriptsize{$M( \Kp \Km ) \ [ \mev ]$}}}}
\end{picture}
\protect\caption{Result of the 2D fit to the $2 ( K^+ K^- )$ invariant mass distribution 
along with the projections to the $\Kp \Km$ invariant mass axes in the $\etac (1S)$ signal region.} 
\label{fig:phiproj}
\end{figure}

In order to extract a pure \phiphi component, 
the two-dimensional unbinned maximum likelihood fit corresponding to the two $K^+ K^-$ combinations, 
in the bins of the $K^+ K^- K^+ K^-$ invariant mass, was performed. 
Each of the two $K^+ K^-$ combinations is randomly assigned as the first or the second $\phi$ candidates.  
The two-dimensional fit accounts for the $\phi \phi$, $\phi K^+ K^-$ and $K^+ K^- K^+ K^-$ components, 
taking into account also the threshold factor. 
In the 2D fit of the $K^+ K^-$ invariant masses, $\phi$ signal is described by the convolution 
of the Breit-Wigner function
to describe natural width of the $\phi$ resonance, 
and double Gaussian function to describe the effect of detector resolution. 
The ratio of the two Gaussian widths $\sigma_1 / \sigma_2$ of $0.41 \pm 0.01$ and the fraction of narrow 
Gaussian $N_1 / (N_1 + N_2)$ of $0.87 \pm 0.01$ are taken from simulation. 
Combinatorial background is described by the first order polynomial. 
A threshold factor $\sqrt{x} = \sqrt{m_{KK} - 2 m_K}$ to describe phase space difference 
is introduced in both signal and combinatorial background shapes. 
The complete description function is written as 
\begin{align*}
F (x_1 , x_2) = & N_{\phi \phi} \times S_1 \times S_2 \ + \\
& N_{\phi K K} \times ( S_1 \times k_2 \times \sqrt{x_2} + 
S_2 \times k_1 \times \sqrt{x_1} ) \ + \\
& N_{K K K K} \times ( k_3 \times \sqrt{x_1} \times \sqrt{x_2} ) \ , 
\end{align*}
where signal functions $S_1$ and $S_2$ correspond to the PDF of the two $\phi$ candidates, 
and $k_i$ are normalization coefficients. The fit shape accounts for $\phi \phi$, $\phi \Kp \Km$ and $\Kp \Km \Kp \Km$ 
contributions and takes into account the available phase space. 
The two-dimensional fit function as well as the projections on the two axes, for the complete event 
sample are shown on Figure~\ref{fig:phiproj}, respectively. 

The $\phi\phi$ mass spectrum is obtained from the $\phi\phi$ signal yield determined from the 2D-fit in each bin of invariant mass. 
The obtained sample contains true two-$\phi$ combinations, that are either random combinations 
or originate from the decay of common mother particles. 

No clear contribution from the $f_0 (980)$ resonance is seen in the \Kp\Km  invariant mass distribution. 
However a potential effect due to $f_0 (980)$ 
is estimated in the following as a potential source of systematic uncertainty. 

In the following production ratios are extracted from signal event yields 
obtained from the fit to the pure $\phi \phi$ invariant mass spectra. 
The invariant mass spectrum of pure $\phi \phi$ combinations is used to study 
charmonia production in inclusive \bquark-hadron decays, 
study $\etac$ and $\chic$ masses and the natural width of the $\etac(1S)$, 
and measure $\BR (\Bs \to \phi \phi)$. 

In Section~\ref{sec:ccyield}, the 2D fit is performed in bins of the \Kp\Km\Kp\Km invariant mass 
using the technique discussed in this section to construct the invariant mass distribution 
of the di-$\phi$ candidates, which is subsequently fit to extract charmonia yields.
\clearpage
\section{Production of \chic and \etactwos in inclusive \bquark-decays}
\protect\label{sec:ccsection}
\subsection{Fit to the invariant mass of \phiphi}
\protect\label{sec:ccyield}
Using the technique discussed in Section~\protect\ref{sec:procedure} the invariant mass spectrum of the pure \phiphi pairs is constructed and is fit to extract the yields of charmonia decaying to \phiphi. 

The invariant mass distribution of the \phiphi candidates 
is fit to the sum of the signal shapes for the \etac family, $\etac (1S)$ and $\etactwos$, 
and \chic family, \chiczero, \chicone and \chictwo, 
and the background shape. 
Each of the above charmonium states is described by the convolution of the relativistic 
Breit-Wigner function (RBW) to account for the natural width of the resonances, 
\protect\begin{align*}
RBW = & \frac{x \cdot \Gamma_f}{(M^2 - x^2)^2 + M^2 \cdot \Gamma_f^2} \ , \\
 & \Gamma_f = \Gamma \times \left( \frac{K(x)}{K(M)} \right) ^{2J + 1} \times \left( \frac{F(rK(x))}{F(rK(M))} \right) \times \frac{M}{x} \ , \\
 & K(y) = \frac{\sqrt{( y^2 -2 m_{\phi}^2 )^2 - 4 m_{\phi}^4}}{2y} \ , \\
 & F(y) = \left\{ \protect\begin{array}{ll} 1, & J = 0 \\ \frac{1}{1 + y^2} , & J = 1 \\ \frac{1}{9 + 3y^2 + y^4} , & J = 2 \ , \protect\end{array} \right. 
\protect\end{align*}
and a double Gaussian function (DG) to account for detector resolution. 
In the above expression $x$ and $y$ are the decay products centre-of-mass energies, 
$M$ and $\Gamma$ are the resonance mass and natural width, respectively, 
$J$ is the total angular momentum and $r$ is the radial parameter 
of the decaying meson~\cite{VonHippel:1972fg}. 
Natural width of the \etac(1S) state is left a free parameter in the fit, 
while natural widths of the \chiczero, \chicone, \chictwo and \etac(2S) are fixed to their world average values~\cite{PDG2016}. 

\begin{figure}[h]
\centering
\protect\includegraphics[width=0.75\linewidth]{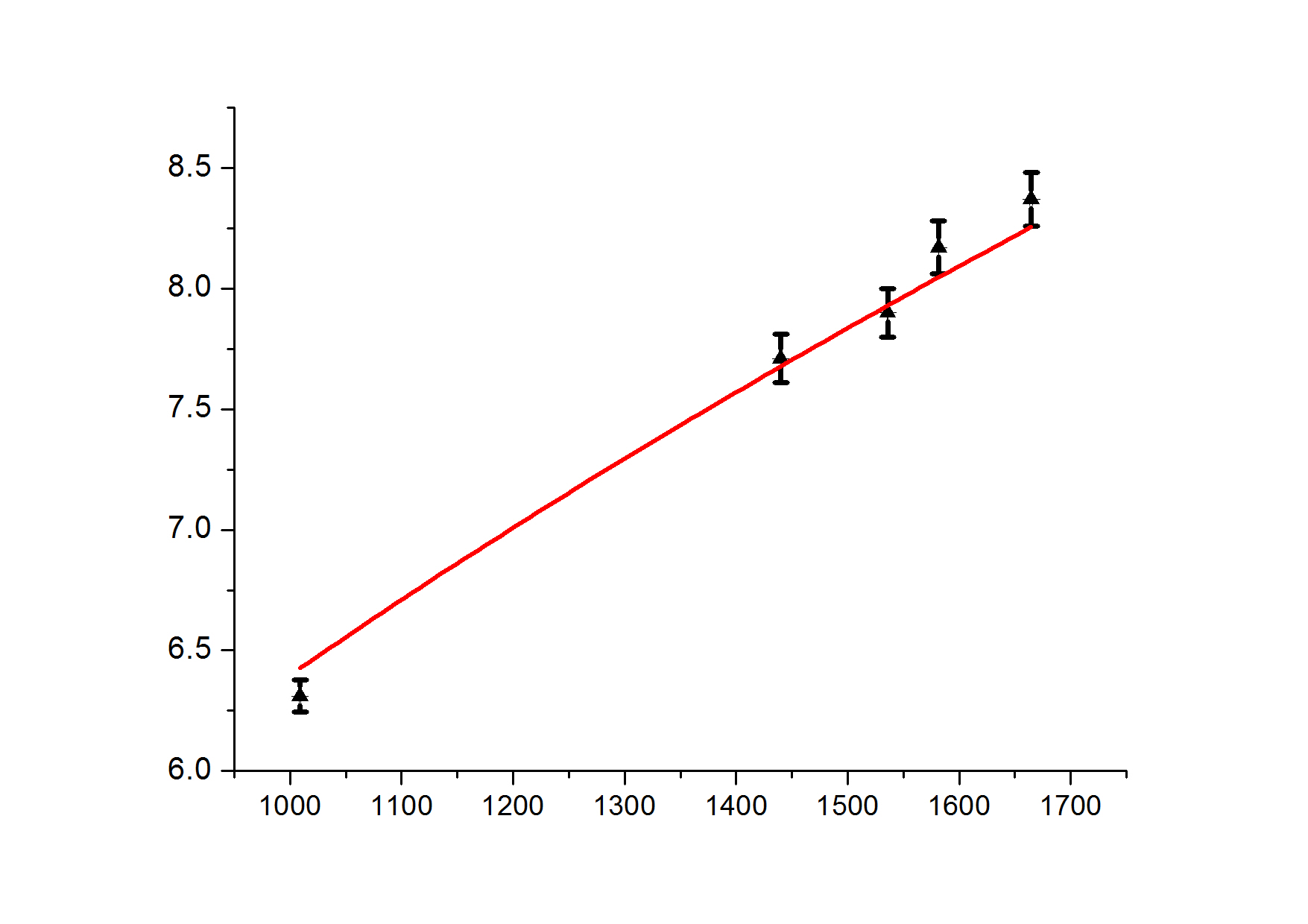}
                \put(-315,180){\rotatebox{90}{\small{\mev}}}
                \put(-315,110){\rotatebox{90}{\small{$\sigma_{narrow}$}}}
                \put(-190,8) {\small{Energy release}}
                \put(-260,190) {\small{\lhcb simulation}}
\protect\caption{Resolution obtained on the simulated samples depending on the energy release. 
The \etac and \chic states are shown on the plot. 
Fit using the function $k \cdot \sqrt{x - 4m_K}$.}
\protect\label{fig:sigmascale}
\protect\end{figure}

The values of the ratio of two Gaussian widths 
and the fraction of the narrow Gaussian are fixed to the values determined from simulation - $\sigma_2 / \sigma_1 = 2.16$  and $f_1=0.81$, respectively.
Resolution effect is scaled according to the energy release in agreement with the MC based expectations 
(Fig.~\protect\ref{fig:sigmascale}). 
\protect
In total, one free parameter in the \phiphi invariant mass fit accounts for the detector resolution effect. 
Resolution obtained in the simulation is compared to that from data in Table~\protect\ref{tab:sigmas}. 
\protect\begin{table}[h]
\centering
\protect\begin{tabular}{r|c|c}
 Resonance & Simulation & Data  \\ 
\hline
\etac(1S)      & $6.3 \pm 0.1$ & $7.4 \pm 0.6$ \\ 
\chiczero(1P)  & $7.7 \pm 0.1$ & $8.8 \pm 0.8$  \\
\chicone(1P)   & $7.9 \pm 0.1$ & $9.1 \pm 0.8$  \\ 
\chictwo(1P)   & $8.2 \pm 0.1$ & $9.2 \pm 0.8$ \\
\etac(2S)      & $8.4 \pm 0.1$ & $9.5 \pm 0.8$  \\
\protect\end{tabular}
\protect\caption{Resolution (narrow Gaussian $\sigma$) as obtained from simulation and data samples.
\protect\label{tab:sigmas}}
\protect\end{table}
The resolution values in the table are obtained from using a single free fit parameter both with 
the data and the simulation samples. 
The simulation values are obtained from the simultaneous fit to the simulated signal samples. 
The data values are obtained from the nominal fit to the combined data sample. 
Therefore in both cases the correlation is present. 
Resolution dependence on energy release is consistent for simulation and data samples wint a trend for a simulation to underestimate the resolution, as for other \lhcb analyses.

The charmonium-like $X(3872)$ and $X(3915)$ (or $X(3915)$) states are taken into account in the fit, 
in order to evaluate systematic uncertainty of the 
main fit results, 
as well as to obtain upper limits on the probabilities 
$\BR (\bquark \to X(3872) X) \times \BR ( X(3872) \to \phi \phi)$ 
and $\BR (\bquark \to X(3915) X) \times \BR ( X(3915) \to \phi \phi)$. 
The upper limits of charonium-like states production is given in Section~\protect\ref{sec:limits}.

Natural width of the \etactwos meson is fixed to the central value of $\Gamma_{\etac (2S)} = 11.3 ^{+ 3.2} _{- 2.9} \mev$ from Ref.~\cite{PDG2016}. 
Possible variations are taken into account by providing the results as a function of the $\etactwos$ natural width. 

The combinatorial background, i.e. a contribution from random \phiphi combinations, 
is described by the product of a 
first-order polynomial, exponential function and a factor to account for the available 
phase space: 
\[
BGR = \sqrt{z} \cdot \exp ( A z ) \cdot (1 + B z) \ , 
\]
where $z = M (K^+ K^- K^+ K^-) - 2 M (\phi)$. 

The complete description can then be denoted as 
\protect\begin{align*}
PDF & = \ RBW ( M_{\etac (1S)} , \Gamma_{\etac (1S)} , J = 0, r = 1.5 \gev^{-1}) \otimes DG (M_{\etac (1S)}) \ + \\
& + \ RBW ( M_{\chiczero} , \Gamma_{\chiczero} = const , J = 0, r = 1.5 \gev^{-1}) \otimes DG (M_{\chiczero}) \ + \\
& + \ RBW ( M_{\chicone} , \Gamma_{\chicone} = const , J = 1, r = 1.5 \gev^{-1}) \otimes DG (M_{\chicone}) \ + \\
& + \ RBW ( M_{\chictwo} , \Gamma_{\chictwo} = const , J = 2, r = 1.5 \gev^{-1}) \otimes DG (M_{\chictwo}) \ + \\
& + \ RBW ( M_{\etac (2S)} , \Gamma_{\etac (2S)} = const , J = 0, r = 1.5 \gev^{-1}) \otimes DG (M_{\etac (2S)}) \ + \\
& + \ BGR \ .
\protect\end{align*}
Free parameters in the fit are  yields and masses of the resonances, the $\etac (1S)$ natural width, 
one resolution parameter $k$, $A$ and $B$ background description parameters. 

\begin{figure}[h]
\centering
\protect\includegraphics[width=1.\linewidth,height=10.cm]{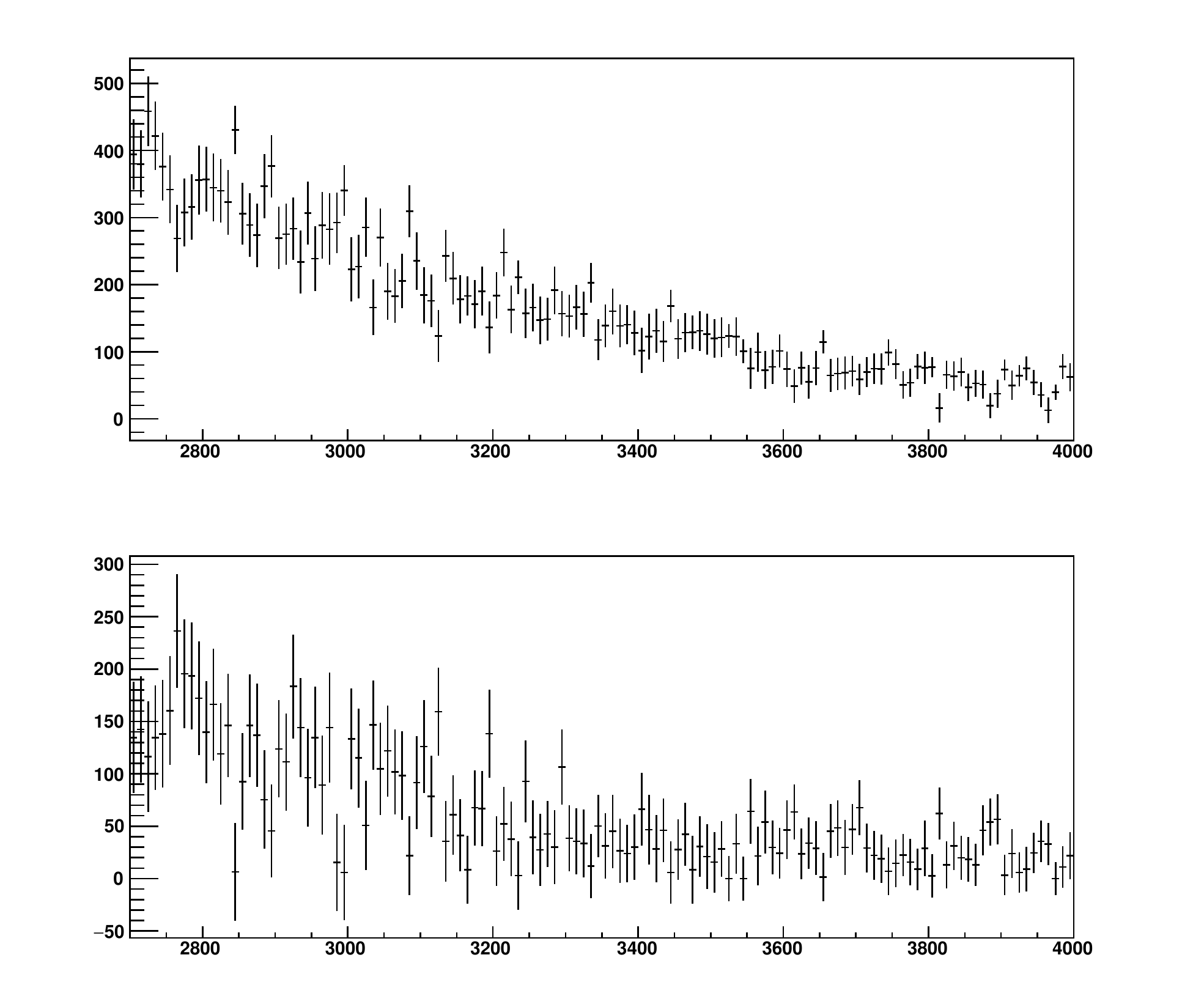}
                \put(-440,170){\rotatebox{90}{Events/10 \mev}}
                \put(-85,-2){\mev}
                \put(-165,250){\lhcb-ANA-2015-038}
                \put(-150,235){3 fb$^{-1}$}
                \put(-155,140){$M(\phi \Kp \Km )$, \mev}
                \put(-180,-2){$M(\Kp \Km \Kp \Km)$,}
\protect\protect\caption[Distribution of the $\phi \Kp \Km$ and $\Kp \Km \Kp \Km$ invariant mass for combined data sample.]{Distribution of the $\phi \Kp \Km$ (top) and $\Kp \Km \Kp \Km$ (bottom) invariant mass for combined data sample, accumulated at $\protect\sqs = 7 \protect\tev$ and $\protect\sqs = 8 \protect\tev$. Data points are the results of the 2D fit.}
\protect\label{fig:multi}
\protect\end{figure}
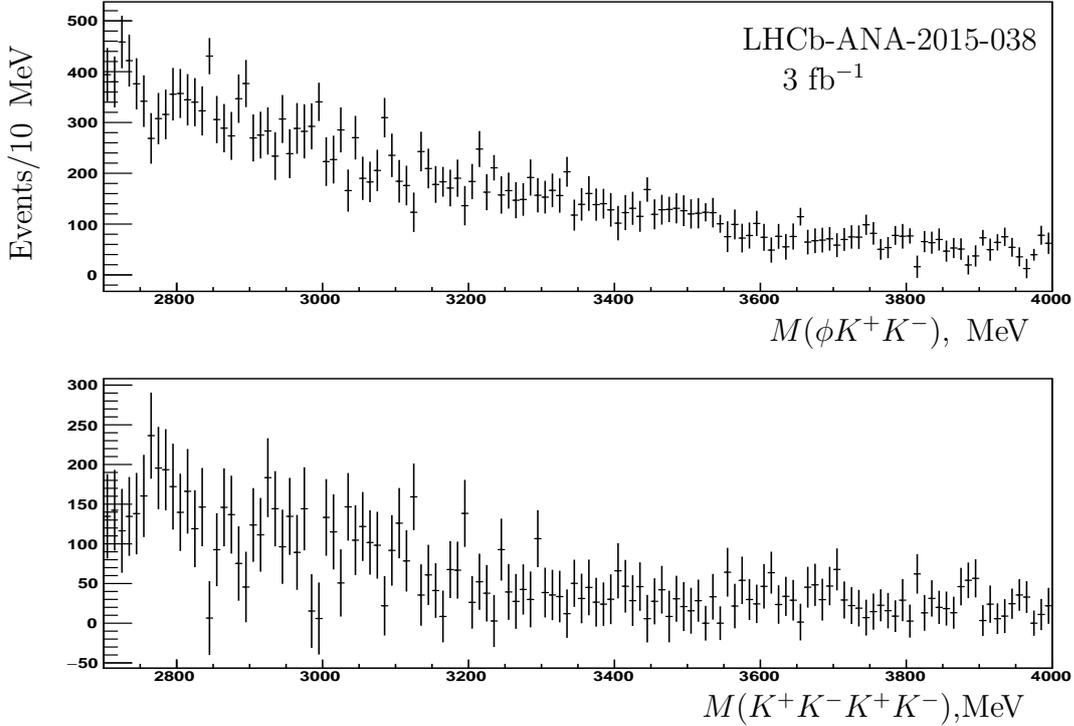

A binned $\chi^2$ fit is performed on the \phiphi invariant mass distribution taking into 
account the fact that the error bars reflect the 2D fit results, so the error on the yield is the one obtained from 2D fit but not Poisson error. 
Figure~\protect\ref{fig:ccphiphi} shows the fit to the spectrum of the invariant mass of \phiphi combinations,
for combined data sample. 
As explained before, each bin of the invariant mass distribution shown on Fig.~\protect\ref{fig:ccphiphi} 
is a result of the 2D fit as described in Section~\protect\ref{sec:procedure}. 
Signals from \etac(1S), \chiczero, \chicone, \chictwo and \etac(2S) decays into \phiphi are clearly visible. 
For illustration, Fig.~\protect\ref{fig:ccall} shows invariant mass spectra for charmonia decays to $\phi \phi$ before performing 2D fit procedure. The background level on Fig.\protect\ref{fig:ccphiphi} is almost twice smaller to that on Fig.\protect\ref{fig:ccall}. The later is due to statistical unfolding of $\phi\phi$. However, the statistical errors on Fig.\protect\ref{fig:ccphiphi} obtained from 2D fit are larger than poisson errors.
\afterpage{
\protect\begin{figure}[h!]
\centering
\protect\begin{picture}(430,260)
\put(0,10){\protect\includegraphics[width=450px]{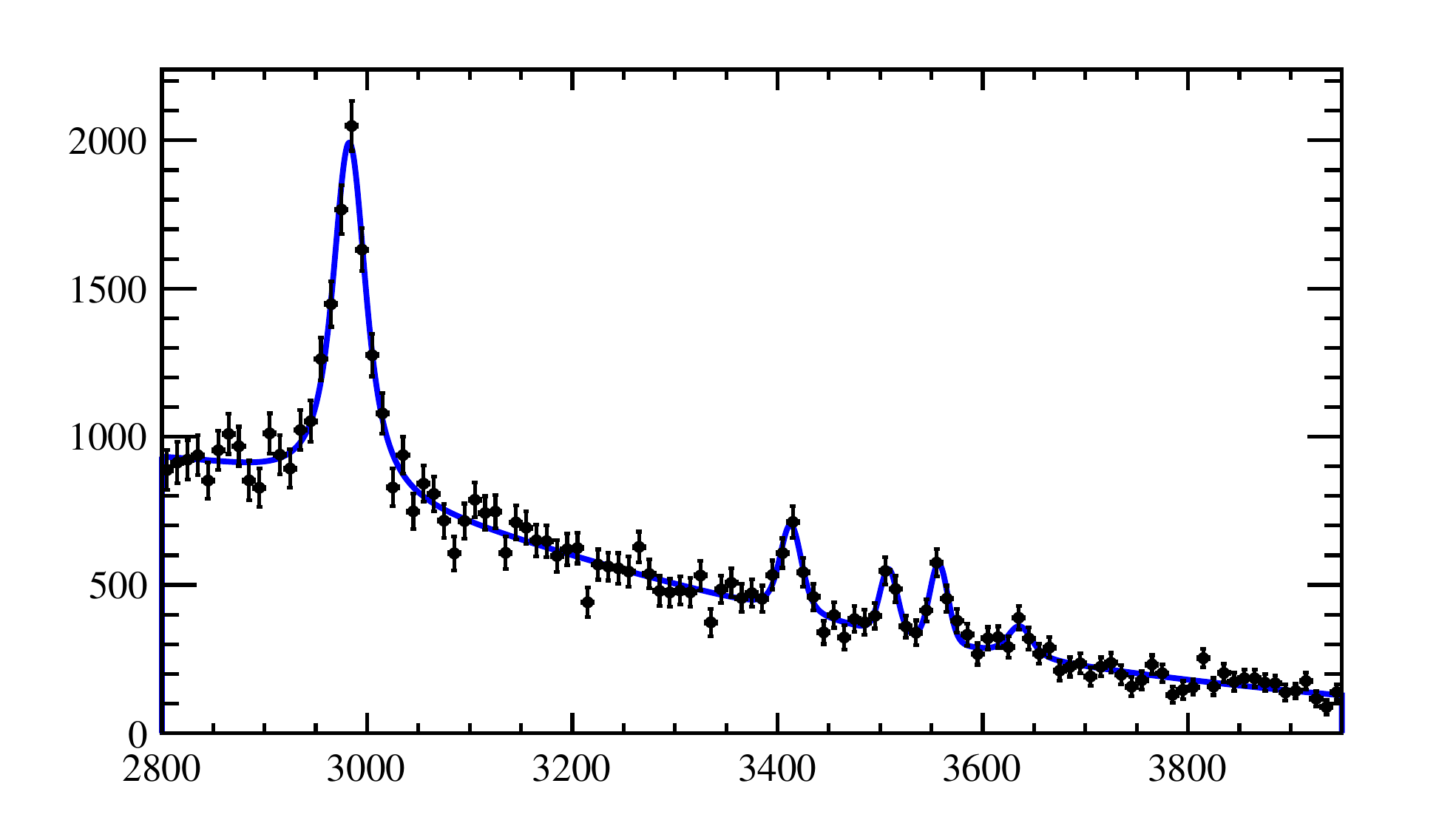}
                \put(-447,125){\rotatebox{90}{{Candidates/$( 10 \mev )$}}}
                \put(-100,210) {\lhcb}
                \put(-335,220) {$\etac(1S)$}
                \put(-210,115) {\chiczero}
                \put(-185,100) {\chicone}
                \put(-160,95) {\chictwo}
                \put(-140,80) {$\etac(2S)$}
                \put(-110,0) {$M(\phi\phi), \mev $}}
\protect\end{picture}
\protect\caption
[Distribution of the invariant mass of $\phi \phi$ combinations. 
The number of candidates in each bin is obtained from the corresponding 2D fit.]
{Distribution of the invariant mass of $\phi \phi$ combinations. 
The number of candidates in each bin is obtained from the corresponding 2D fit. 
The peaks corresponding to the \ccbar resonances are marked on the plot.} 
\protect\label{fig:ccphiphi}
\protect\begin{picture}(430,260)
\put(10,0){\protect\includegraphics[width=450px]{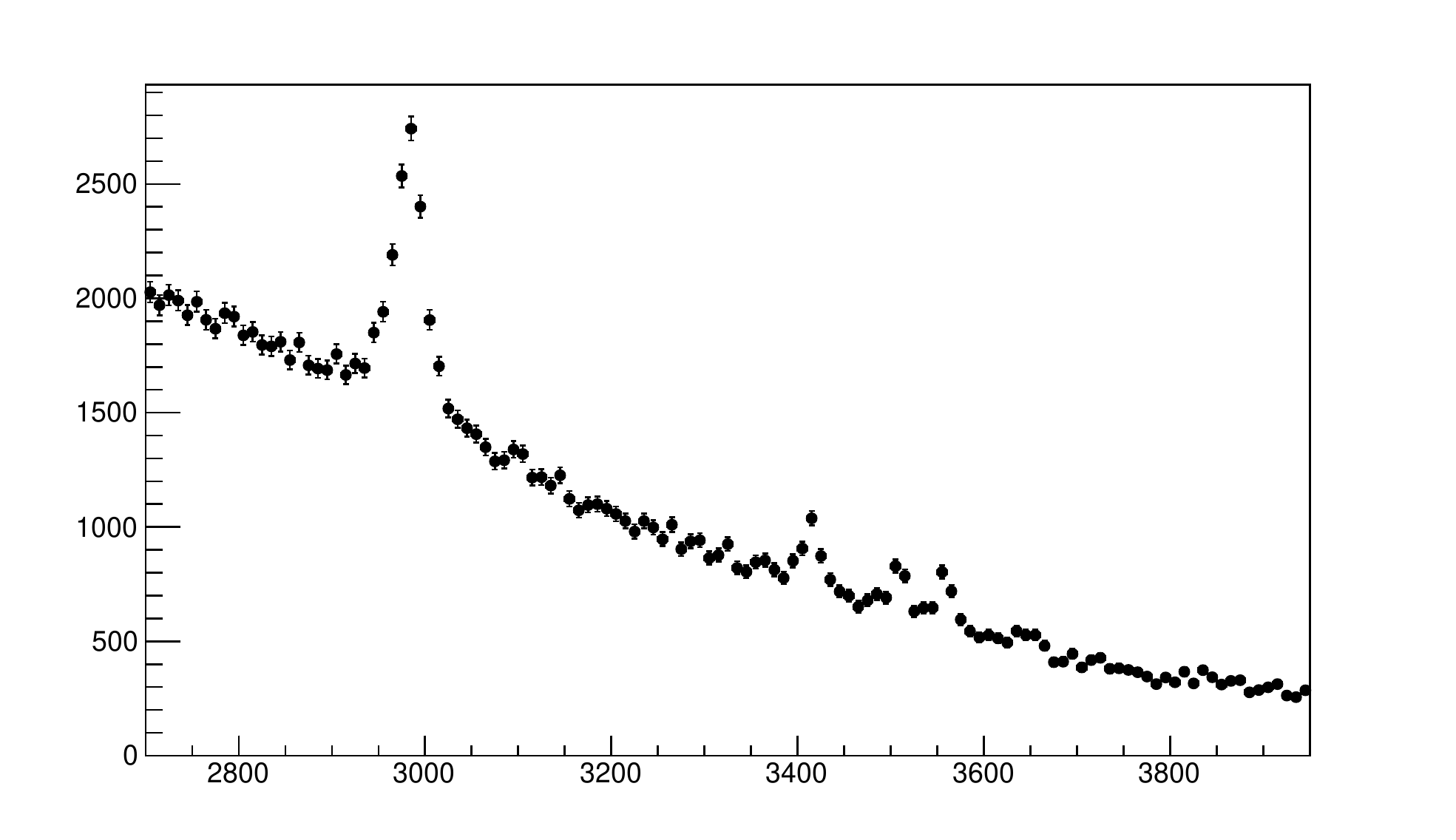}}
\protect\end{picture}
            \put(-410,115){\rotatebox{90}{{Candidates/$( 10 \mev )$}}}
            \put(-160,210){\lhcb-ANA-2015-038}
            \put(-160,195){3 fb$^{-1}$}
            \put(-280,220){$\etac(1S)$}
            \put(-175,115){\chiczero}
            \put(-150,100){\chicone}
            \put(-130,95){\chictwo}
            \put(-105,75){$\etac(2S)$}
            \put(-145,0){$M(\Kp\Km\Kp\Km)$, \mev}
\protect\caption{Distribution of the $\phi \phi$ invariant mass for combined data sample. No 2D fit is performed.}
\protect\label{fig:ccall}
\protect\end{figure}
}
\clearpage
Figure~\protect\ref{fig:multi} shows invariant mass spectra for 
the $\phi \Kp \Km$ and  $\Kp \Km \Kp \Km$ combinations plotted using the results of the 2D fit. No significant resonance contributions are observed in the $\phi \Kp \Km$ and $\Kp \Km \Kp \Km$ invariant mass distributions.
\protect

Table~\protect\ref{tab:evtsyears} compares the 
event yields at a centre-of-mass energy of \protect\sqs = 7~\protect\tev and \protect\sqs = 8~\protect\tev for the considered charmonum states. 
\protect\begin{table}[h]
\centering
\protect\begin{tabular}{c|c|c|c|c|c}
                 & $\etac (1S)$   & \protect\chiczero & \chicone & \chictwo &  $\etac (2S)$ \\ \hline
 $\protect\sqs = 7 \protect\tev$ & $2008 \pm 215$ & $289 \pm 74$ & $141 \pm 47$ & $168 \pm 52$ & $20 \pm 53$ \\ 
 $\protect\sqs = 8 \protect\tev$ & $4440 \pm 350$ & $619 \pm 107$ & $314 \pm 72$ & $431 \pm 85$ & $336 \pm 86$ \\ 
 All data & $6476 \pm 418$ & $933 \pm 128$ & $460 \pm 89$ & $611 \pm 97$ & $365 \pm 100$ 
\protect\end{tabular}
\protect\protect\caption{Event yields at a centre-of-mass energy of \protect\sqs = 7 \protect\tev and \protect\sqs = 8 \protect\tev and for the combined data sample, for the considered charmonia states.}
\protect\label{tab:evtsyears}
\protect\end{table}
Scaling the $\etac (2S)$ yield from the $\protect\sqs = 7 \protect\tev$ data sample using the central values of the 
\chictwo yields from the $\protect\sqs = 7 \protect\tev$ and $\protect\sqs = 8 \protect\tev$ data samples (not taking into account their uncertainties), 
a difference between the $\etac (2S)$ yields for the $\protect\sqs = 7 \protect\tev$ and $\protect\sqs = 8 \protect\tev$ data samples 
as estimated from statistical uncertainties only is about $2 \protect\sigma$.
Numbers of other signal candidates are consistent between the data samples collected 
at a centre-of-mass energy of $\protect\sqs = 7 \protect\tev$ and $\protect\sqs = 8 \protect\tev$, 
and with the combined data sample. 

\begin{figure}[b]
\centering
\protect\protect\includegraphics[width=0.75\linewidth]{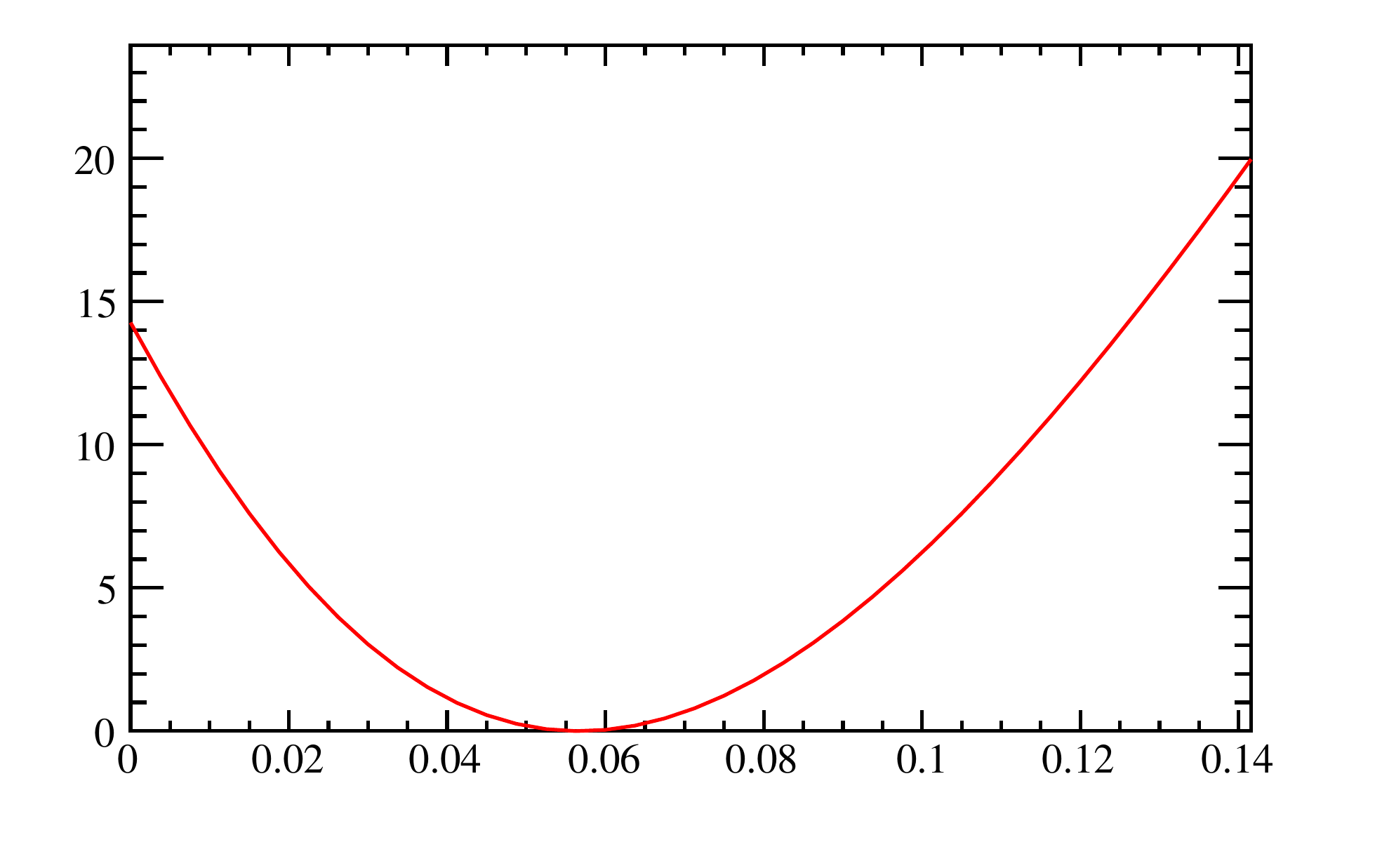}
                \put(-335,130){\rotatebox{90}{{$\Delta \chisq$}}}
                \put(-120,180) {\lhcb}
                \put(-120,165) {3 fb$^{-1}$}
                \put(-170,5) {{$\frac{N ( \etac (2S) )}{N ( \etac (1S) )}$}}
\protect\caption{Difference of \chisq of the fit as a function of the ratio of the \etactwos and $\etac(1S)$ 
event yields.} \protect\label{fig:sigetac}
\protect\end{figure}

The ratios of resonance yields from the fit are summarized in Table~\protect\ref{tab:ccc} 
for the ratios inside the family, 
and in Table~\protect\ref{tab:cce} for the ratios with respect to the decays to $\bquark\to\etac(1S)X$.  
\protect\begin{table}[h]
\centering
\protect\begin{tabular}{r|c}
 Resonance & Event yield ratio \\ 
\hline 
$N_{\chicone} / N_{\chiczero}$  & $0.494 \pm 0.107 \pm 0.012$  \\ 
$N_{\chictwo} / N_{\chiczero}$  & $0.656 \pm 0.121 \pm 0.015$  \\ 
$N_{\etac(2S)} / N_{\etac(1S)}$  & $0.056 \pm 0.016 \pm 0.005$  \\ 
\protect\end{tabular}
\protect\caption{Charmonium event yield ratios inside families from the fit to \phiphi invariant mass spectrum.
\protect\label{tab:ccc}}
\protect\end{table}
\protect\begin{table}[h]
\centering
\protect\begin{tabular}{r|c}
 Resonance & Event yield ratio \\ 
\hline
$N_{\chiczero} / N_{\etac(1S)}$  & $0.144 \pm 0.022 \pm 0.011$  \\ 
$N_{\chicone} / N_{\etac(1S)}$  & $0.071 \pm 0.015 \pm 0.006$  \\ 
$N_{\chictwo} / N_{\etac(1S)}$  & $0.094 \pm 0.016 \pm 0.006$  
\protect\end{tabular}
\protect\caption{Charmonium event yield ratios with respect to decays to $\etac (1S)$ from the fit to \phiphi invariant mass spectrum.
\protect\label{tab:cce}}
\protect\end{table}
Significance of the \etactwos to $\etac(1S)$ event yield ratio is illustrated on Fig.~\protect\ref{fig:sigetac}. The statistical significance, not including systematic, for the $N_{\etac (2S)}$ signal is estimated from the \chisq-profile 
to be 3.7 standard deviations.  
\protect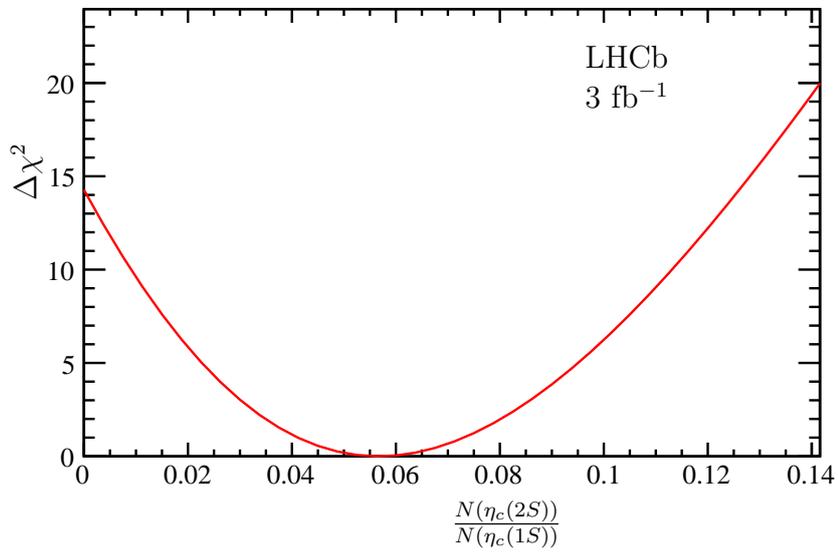

\clearpage
\subsection{Systematic uncertainties}
\protect\label{sec:ccsyst}
Systematic uncertainties are obtained by including potential contribution 
from other resonances, varying detector resolution, varying fit range, 
implementing alternative background parametrization, 
accounting for potential contribution from the $f_0 (980)$ state 
to the 2D fit, and fixing masses of the \chic states to the known values~\cite{PDG2016}. 
In order to evaluate systematic uncertainty related to potential contribution
from other resonances, contributions from $X (3872)$, $X(3915)$ and $\chictwo (3930)$
are included in the fit. 
Systematic uncertainties related to detector resolution are conservatively estimated by using 
the $\etac (1S)$ resolution as obtained from the simulation. 
Systematic uncertainties associated to the impact of the detector resolution description 
on the signal shapes are estimated by comparing the nominal fit results to those obtained 
using a single Gaussian instead of double Gaussian shape. 
The uncertainty associated to the description using the Breit-Wigner shape is estimated 
by varying radial parameter $r$ between $0.3 \gev^{-1}$ and $5 \gev^{-1}$. 
In order to estimate uncertainty related to the natural width of $\etactwos$ the $\Gamma_{\etac (2S)}$ 
value is varied within the PDG~\cite{PDG2016} uncertainties. 
The systematic associated to the fit range is estimated by restricting the fit to the \chic and $\etactwos$ region ($3.15 \gevc - 3.95 \gevc$) 
is used to estimate the corresponding systematic uncertainty. 
Alternative background parametrization using a parabola function is used for the corresponding 
systematic uncertainty estimate. 
Systematic uncertainty associated to the background parametrization in the 2D fit is estimated 
by adding slope parameters for the $\phi \Kp \Km$ and $\Kp \Km \Kp \Km$ description. 

Effect of potential contribution from the $f_0 (980)$ state to the 2D fit
is estimated by including the $f_0 (980)$ contribution 
following the example from Ref.~\cite{LHCb-PAPER-2011-002}.  
In order to evaluate potential contribution from the $f_0 (980)$ state to the 2D fit, 
the signal regions for each considered resonance is fit including the term describing 
the $f_0 (980)$ contribution, and varying the $f_0 (980)$ parameters. 
Using the $f_0 (980)$ description with the Breit-Wigner function and varying parameters 
according to the uncertainties from Ref.~\cite{PDG2016}, 
the obtained results are shown in Table~\protect\ref{tab:fnol}. 
\protect\begin{table}[h]
\centering
\protect\begin{tabular}{l|c|c|c|c|c}
                                     &  $\etac(1S)$ & $\chiczero$ & \chicone & \chictwo & $\etac(2S)$   \\ \hline 
$M = 990 \mev$, $\Gamma = 70 \mev$ &  $< 1$       & $1$         & $2$      & $< 1$    & $3$           \\ \hline
$M = 990 \mev$, $\Gamma = 40 \mev$ &  $< 1$       & $3$         & $3$      & $< 1$    & $6$           \\ \hline
$M = 990 \mev$, $\Gamma = 100 \mev$ &  $1$       & $< 1$         & $1$      & $< 1$    & $< 1$           \\ \hline
$M = 970 \mev$, $\Gamma = 70 \mev$ &  $< 1$       & $1$         & $< 1$      & $< 1$    & $3$           \\ \hline
$M = 1010 \mev$, $\Gamma = 70 \mev$ &  $< 1$       & $6$         & $< 1$      & $1$    & $< 1$           \\ \hline
$\Delta N$-max                        &  $1$       & $6$         & $3$      & $1$    & $6$           \\ \hline
\protect\end{tabular}
\protect\caption
[Estimated difference $\Delta N_{\phi \phi}$ in the region of the 
$\etac (1S)$, $\chiczero$, $\chicone$, $\chictwo$ and $\etactwos$ resonances originated from accounting for the contribution from the $f_0 (980)$ state in the 2D fit.]
{Estimated difference $\Delta N_{\phi \phi}$ in the region of the 
$\etac (1S)$ ($2920 - 3050 \mev$), 
$\chiczero$ ($3370 - 3460 \mev$), 
$\chicone$ ($3460 - 3530 \mev$), 
$\chictwo$ ($3530 - 3600 \mev$) and 
$\etactwos$ ($3600 - 3660 \mev$) resonances 
originated from accounting for the contribution from the $f_0 (980)$ state in the 2D fit. 
Parameters of the $f_0 (980)$ state are varied according to the uncertainties from the Ref.~\cite{PDG2016}.
\protect\label{tab:fnol}}
\protect\end{table}
Using the $f_0 (980)$ description with the Flatte function~\cite{Flatte:1976xu} and varying parameters 
following the example from Ref.~\cite{LHCb-PAPER-2011-002}, 
the obtained results are shown in Table~\protect\ref{tab:flatte}. 
\protect\begin{table}[h]
\centering
\protect\begin{tabular}{l|c|c|c|c|c}
                                     &  $\etac(1S)$ & $\chiczero$ & \chicone & \chictwo & $\etac(2S)$   \\ \hline 
$M = 990 \mev$, $g_2 / g_1 = 4.12$ &  $+ 1$       & $- 1$         & $< 1$      & $< 1$    & $- 2$           \\ \hline
$M = 990 \mev$, $g_2 / g_1 = 3.80$ &  $- 1$       & $< 1$         & $< 1$      & $< 1$    & $- 2$           \\ \hline
$M = 990 \mev$, $g_2 / g_1 = 4.44$ &  $- 2$       & $< 1$         & $< 1$      & $< 1$    & $- 3$           \\ \hline
$M = 970 \mev$, $g_2 / g_1 = 4.12$ &  $+ 2$       & $- 1$         & $< 1$      & $< 1$    & $- 2$           \\ \hline
$M = 1100 \mev$, $g_2 / g_1 = 4.12$ &  $- 2$       & $+ 1$         & $< 1$      & $< 1$    & $- 1$           \\ \hline
$\Delta N$-max                        &  $2$       & $1$         & $< 1$      & $< 1$    & $3$           \\ \hline
\protect\end{tabular}
\protect\caption
[Estimated difference $\Delta N_{\phi \phi}$ in the region of the 
$\etac (1S)$, $\chiczero$, $\chicone$, $\chictwo$ and 
$\etactwos$ resonances 
originated from accounting for the contribution from the $f_0 (980)$ state in the 2D fit with different Flatte parametrisations of the $f_0 (980)$ resonance.]
{Estimated difference $\Delta N_{\phi \phi}$ in the region of the 
$\etac (1S)$ ($2920 - 3050 \mev$), 
$\chiczero$ ($3370 - 3460 \mev$), 
$\chicone$ ($3460 - 3530 \mev$), 
$\chictwo$ ($3530 - 3600 \mev$) and 
$\etactwos$ ($3600 - 3660 \mev$) resonances 
originated from accounting for the contribution from the $f_0 (980)$ state in the 2D fit with different Flatte parametrisations of the $f_0 (980)$ resonance. 
\protect\label{tab:flatte}}
\vspace{-0.5cm}
\protect\end{table}
Maximum differences ($\Delta N$-max) over the two $f_0 (980)$ parametrizations 
are conservatively attributed as an estimate of the corresponding 
source of systematic uncertainty for each charmonium state. 

Uncertainty associated to the description of the $\phi$ signal peak resolution is estimated 
by fixing the resolution in the 2D fit at the value suggested by simulation. 
Uncertainty on the description of the \chic signal peaks is estimated by fixing the \chic masses 
at their nominal values. 
Uncertainty related to momentum scale calibration is negligible and is not accounted in the evaluation 
of the systematic uncertainty on the yield ratios. 
Combined systematic uncertainty is obtained as a quadratic sum 
of the individual systematic contributions. 
Background description as well as potential contribution from other resonances 
dominate combined systematic uncertainties. 
In the yield ratios systematic uncertainty is smaller or comparable to the statistical one. 

The details of systematic uncertainty estimates for the ratios of charmonia yields are shown in Tables~\protect\ref{tab:cccsyst}
and~\protect\ref{tab:ccesyst}.  
\protect\begin{table}[h]
\centering
{\small{
\protect\begin{tabular}{l|c|c|c}
 & $N_{\chicone} / N_{\chiczero}$ & $N_{\chictwo} / N_{\chiczero}$ & $N_{\etac(2S)} / N_{\etac(1S)}$ \\ \hline 
Including $X(3872)$, $X(3915)$, $\chictwo (3930)$ & $ 0.006$ & $ 0.008$ & $0.003$ \\
Fix $\etac(1S)$ resolution                   & & & \\
to MC value                                  & $ 0.001$ & $ 0.001$  & $< 0.001$ \\
Resolution described                         &  &  &  \\
with a single Gaussian                       & $< 0.001$ & $< 0.001$  & $- 0.002$ \\
Varying $r$ parameter                        &  &  &  \\
  between $0.5$ and $3 \gev^{-1}$           & $< 0.001$ & $< 0.001$  & $< 0.001$ \\
Varying $\Gamma_{\etac (2S)}$                & $< 0.001$ & $ 0.001$  & $- 0.003$ \\
Fit \chic and \etactwos region only          & 0.001    & $- 0.004$  & -  \\
Alternative bgrd parametrization             & 0.002    & 0.011    &$< 0.001$ \\ 
Accounting for $f_0 (980)$ in 2D fit         & $ 0.005$ & $ 0.005$ & $ 0.001$ \\ 
Fix \chic masses at nominal values           & $- 0.010$ & $- 0.002$ & $< 0.001$ \\ 
Fix resolution in 2D fit                     & & & \\
at MC value                                  & $< 0.001$ & $- 0.001$ & $< 0.001$ \\ 
Add slope parameter                          & & & \\ 
for the $\phi \Kp \Km$ component             & $< 0.001$ & $ 0.001$ & $< 0.001$ \\ 
in 2D fit                                    & & & \\ 
Add slope parameter                          & & & \\ 
for the $\Kp \Km \Kp \Km$ component          & $< 0.001$ & $< 0.001$ & $< 0.001$ \\ 
in 2D fit                                    & & & \\ \hline
Combined systematic uncertainty              & $0.012$   & $0.015$ & $0.005$ \\
\protect\end{tabular}
}}
\protect\caption{Systematic uncertainty of the obtained charmonium event yield ratios
within families.
\protect\label{tab:cccsyst}}
\protect\end{table}\protect\begin{table}[h]
\centering
{\small{
\protect\begin{tabular}{l|c|c|c}
 & $N_{\chiczero} / N_{\etac(1S)}$ & $N_{\chicone} / N_{\etac(1S)}$ & $N_{\chictwo} / N_{\etac(1S)}$  \\ \hline 
Including $X(3872)$, $X(3915)$, $\chictwo (3930)$ & 0.004     & $ 0.003$ & 0.003  \\
Fix $\etac(1S)$ resolution to MC value       & $< 0.001$ & $< 0.001$ & $< 0.001$ \\
Resolution described with a single Gaussian  & $< 0.001$ & $< 0.001$ & $< 0.001$ \\
Varying $r$ parameter                        & & & \\ 
  between $0.5 \gev^{-1}$ and $3 \gev^{-1}$ & $< 0.001$ & $< 0.001$ & $< 0.001$ \\
Varying $\Gamma_{\etac (2S)}$                  & $< 0.001$ & $< 0.001$ & $< 0.001$ \\
Alternative bgrd parametrization             & $- 0.010$ & $- 0.005$ & $- 0.005$ \\ 
Accounting for $f_0 (980)$ in 2D fit         & $ 0.001$  & $< 0.001$ & $< 0.001$ \\ 
Fix \chic masses at nominal values        & $- 0.002$ & $- 0.002$ & $- 0.001$  \\ 
Fix resolution in 2D fit at MC value         & $< 0.001$ & $< 0.001$ & $< 0.001$ \\ 
Add slope parameter                          & & & \\ 
for $\phi \Kp \Km$ component in 2D fit       & $- 0.002$ & $< 0.001$ & $ 0.001$ \\ 
Add slope parameter                          & & & \\ 
for $\Kp \Km \Kp \Km$ component in 2D fit    & $< 0.001$ & $< 0.001$ & $< 0.001$ \\ \hline
Combined systematic uncertainty              & $0.011$ & $0.006$ & $0.006$  \\
\protect\end{tabular}
}}
\protect\caption{Systematic uncertainty of the obtained charmonium event yield ratios
with respect to the decays with $\etac (1S)$.
\protect\label{tab:ccesyst}}
\protect\end{table}

Stability of the obtained results are checked by shifting the \phiphi invariant mass distribution by half a bin.
Table~\protect\ref{tab:bincheckr} compares the results for yield ratios
to those obtained with the shift by half a bin of the invariant mass 
distribution. 
\protect\begin{table}[t]
\centering
\protect\begin{tabular}{l|c|c}
                            & Measured value             & Shift with respect    \\ 
                            &                            & to the measured value \\ \hline
$N_{\chiczero} / N_{\etac(1S)}$ & $0.144 \pm 0.022 \pm 0.011$ & $0.006$ \\ \hline
$N_{\chicone} / N_{\etac(1S)}$  & $0.071 \pm 0.015 \pm 0.006$ & $0.004$ \\ \hline
$N_{\chictwo} / N_{\etac(1S)}$  & $0.094 \pm 0.016 \pm 0.007$ & $0.007$ \\ \hline
$N_{\etac(2S)} / N_{\etac(1S)}$ & $0.056 \pm 0.016 \pm 0.005$ & $0.003$ \\ \hline
$N_{\chicone} / N_{\chiczero}$  & $0.494 \pm 0.107 \pm 0.012$ & $0.005$ \\ \hline
$N_{\chictwo} / N_{\chiczero}$  & $0.656 \pm 0.121 \pm 0.014$ & $0.022$ 
\protect\end{tabular}
\protect\caption{Cross-check for charmonia yield ratios against a shift by half a bin of the invariant mass 
distribution.\protect\label{tab:bincheckr}}
\protect\end{table}
 
Another cross-check has been performed by using \sPlot technique instead of the 2D fit procedure. No significant deviations from nominal result is observed. However, the \sPlot technique is not strict enough for unfolding true \phiphi contributions in wide range of $M(\phi\phi)$ due to the correlation of background parameters and $M(\phi\phi)$.

\protect\clearpage
\subsection{Results and discussion}
\protect\label{sec:ccprod}
\subsubsection{Branching fractions of inclusive charmonia production in b-decays}
The double ratios of the inclusive branching fractions constitute the main results of the section since the branching fractions of charmonium states to \phiphi are not well measured. Hence, in the double ratios the related systematic uncertainties partially cancel. In addition to that, the PDG average and PDG fit values of the $\decay{\etac}{\phi\phi}$ significantly differ, which is adressed in Section~\protect\ref{sec:brEtac2PhiPhi}. 

In order to exctract simple ratios or absolute branching fractions further input is needed.
In the following the $\etac (1S)$ production rate in \bquark-hadron decays and branching fractions of the charmonia decays to \phiphi are used. 
The $\etac (1S)$ inclusive production in \bquark-decays was measured by \lhcb using decays to \proton\antiproton, 
$\BR ( b \to \etac (1S) X ) = ( 4.88 \pm 0.97 ) \times 10^{-3}$~\cite{LHCb-PAPER-2014-029}. 
Branching fractions of the charmonia decays to \phiphi from Ref.~\cite{PDG2016} are used. 
However, the measured \chic production shows a disagreement when measured using $\etac (1S)$ production for normalization 
and when measured without a normalization. 
In addition, Ref.~\cite{PDG2016} indicates a tension for the $\BR ( \etac (1S) \to \phi \phi )$ value when comparing 
a direct determination and a fit including all available measurements. 
Therefore, an average of the results from \belle~\cite{Huang:2003dr} and \babar~\cite{Aubert:2004gc} using \Bp decays to $\phi \phi \Kp$, 
$\BR ( \etac (1S) \to \phi \phi ) = ( 3.21 \pm 0.72 ) \times 10^{-3}$, is used below. 
The uncertainty of this average dominates a majority of the further results in this section, and improvement of the $\BR ( \etac (1S) \to \phi \phi )$ 
knowledge is critical to reduce the uncertainties of the related results. 
The values $\BR ( \chiczero \to \phi \phi ) = (7.7 \pm 0.7)  \times 10^{-4}$, $\BR ( \chicone \to \phi \phi ) = (4.2 \pm 0.5)  \times 10^{-4}$, 
and $\BR ( \chictwo \to \phi \phi ) = (1.12 \pm 0.10) \times 10^{-3}$, are used for the \chic decays. 

Accounting for small differences in the trigger, reconstruction and selection efficiency 
for decays of the \chic states into \phiphi, 
relative yields of the \chic states in \bquark-hadron inclusive decays are derived as (Eq.~\protect\ref{eq:odin})
\protect\begin{align*}
\frac{\BR ( b \to \chicone X ) \times \BR ( \chicone \to \phi \phi )}{\BR ( b \to \chiczero X ) \times \BR ( \chiczero \to \phi \phi )} 
 &= 0.50 \pm 0.11 \pm 0.01 \ , \\
\frac{\BR ( b \to \chictwo X ) \times \BR ( \chictwo \to \phi \phi )}{\BR ( b \to \chiczero X ) \times \BR ( \chiczero \to \phi \phi )} 
 &= 0.56 \pm 0.10 \pm 0.01 \ . 
\protect\end{align*}
Dominant contribution to the systematic uncertainty 
comes from accounting for possible other resonances and using known \chic mass values~\cite{PDG2016}. 
The systematic uncertainty is smaller than the statistical one,
so that precision will improve with more data accumulated by \lhcb. 

Using branching fractions of the \chic decays to \phiphi from Ref.~\cite{PDG2016}, 
relative branching fractions of \bquark-hadron decays to \chic states can be derived as 
\protect\begin{align*}
\frac{\BR ( b \to \chicone X )}{\BR ( b \to \chiczero X )}  &= 0.92 \pm 0.20 \pm 0.02 \pm 0.14 \ , \\
\frac{\BR ( b \to \chictwo X )}{\BR ( b \to \chiczero X )}  &= 0.38 \pm 0.07 \pm 0.01 \pm 0.05 \ , 
\protect\end{align*}
where the first uncertainty is statistical, the second one is systematic 
and the third one is due to the branching fractions $\BR ( \chi_c \to \phi \phi )$. 

This is the first (\chiczero and \chictwo) or most precise (\chicone) determination of the \chic relative yields 
in \bquark-hadron decays. 
These results are compared to the PDG average values~\cite{PDG2016} 
for the $\Bz / \Bp$ branching fractions into \chicone and \chictwo mesons measured by 
\cleo~\cite{Chen:2000ri,Anderson:2002md}, \belle~\cite{Abe:2002wp} and \babar~\cite{Aubert:2002hc} experiments. In order to make the qualitative comparison, one need to add assumptions about the fraction of charmonium originating from decays of different \bquark-hadrons. 
The average value for the branching fraction 
$\BR ( B \to \chictwo X) = (1.4 \pm 0.4) \times 10^{-3}$~\cite{PDG2016} 
has limited precision and is different from zero by a three standard deviations. 
This is a consequence of a descrepancy between the results of 
the \belle~\cite{Abe:2002wp} and \babar~\cite{Aubert:2002hc} experiments on one side 
and the \cleo result~\cite{Chen:2000ri} on the other side, 
which calls for another measurement.  
The obtained result for relative \chicone and \chictwo production in \bquark-hadron decays 
reproduces the same ratio from  $\Bz / \Bp$ production~\cite{PDG2016}. 

To derive absolute values of the \chic yields from \bquark-hadron decays, 
the result of the \etac inclusive yield measured using decay to \proton\antiproton~\cite{LHCb-PAPER-2014-029} 
is used. 
Taking into account the difference in trigger, reconstruction and selection efficiencies 
for \etac and \chic mesons, 
$\varepsilon_{\chi_c} / \varepsilon_{\etac}$, 
the yield ratios relative to the \etac yield are constructed as 
\protect\begin{align*}
\frac{\BR ( b \to \chiczero X ) \times \BR ( \chiczero \to \phi \phi )}{\BR ( b \to \etac X ) \times \BR ( \etac \to \phi \phi )} 
 &= 0.147 \pm 0.023 \pm 0.011 \ , \\
\frac{\BR ( b \to \chicone X ) \times \BR ( \chicone \to \phi \phi )}{\BR ( b \to \etac X ) \times \BR ( \etac \to \phi \phi )} 
 &= 0.073 \pm 0.016 \pm 0.006 \ , \\
\frac{\BR ( b \to \chictwo X ) \times \BR ( \chictwo \to \phi \phi )}{\BR ( b \to \etac X ) \times \BR ( \etac \to \phi \phi )} 
 &= 0.081 \pm 0.013 \pm 0.005 \ . 
\protect\end{align*}

Relative branching fractions of \bquark-hadron decays to \chic states can be derived as 
\protect\begin{align*}
\frac{\BR ( b \to \chiczero X )}{\BR ( b \to \etac X )} &= 0.615 \pm 0.095 \pm 0.047 \pm 0.149 \ , \\
\frac{\BR ( b \to \chicone X )}{\BR ( b \to \etac X )}  &= 0.562 \pm 0.119 \pm 0.047 \pm 0.131 \ , \\
\frac{\BR ( b \to \chictwo X )}{\BR ( b \to \etac X )}  &= 0.234 \pm 0.038 \pm 0.015 \pm 0.057 \ , 
\protect\end{align*}
where last uncertainties are due to branching fractions $\BR ( \etac , \chi_c \to \phi \phi )$ and are larger than the systematic ones. 

With the branching fraction of \etac production in \bquark-hadron decays 
$\BR ( b \to \etac X ) = ( 4.88 \pm 0.97 ) \times 10^{-3}$~\cite{LHCb-PAPER-2014-029}, 
the absolute branching fractions of \chic production in \bquark-hadron decays 
are obtained as 
\protect\begin{align*}
\BR ( b \to \chiczero X ) &= ( 3.02 \pm 0.47 \pm 0.23 \pm 0.94 ) \times 10^{-3} \ , \\
\BR ( b \to \chicone X )  &= ( 2.76 \pm 0.59 \pm 0.23 \pm 0.89 ) \times 10^{-3} \ , \\
\BR ( b \to \chictwo X )  &= ( 1.15 \pm 0.20 \pm 0.07 \pm 0.36 ) \times 10^{-3} \ , 
\protect\end{align*}
where the third uncertainty is due to the uncertainties on the branching fractions
of the \bquark-hadron decays to \etac meson $\BR ( \bquark \to \etac X )$ 
and $\etac (1S)$ and \chic decays to \phiphi. 

The branching fraction of \bquark-hadron decays into \chiczero is measured for the first time, 
and is larger than the values predicted in Ref.~\cite{Beneke:1998ks}. 

The result for \bquark-decays into \chicone is the most precise measurement 
for the mixture of \Bz, \Bp, \Bs and \bquark-baryons. 
The central value of the result for \bquark-decays into \chicone is lower than the value 
measured by DELPHI~\cite{Abreu:1994rk} and L3~\cite{Adriani:1993ta} experiments at LEP, 
$0.0113 ^{+ 0.0058} _{- 0.0050} \pm 0.0004$ and $0.019 \pm 0.007 \pm 0.001$, respectively. 
However, taking into account the LEP results limited precision, 
the \lhcb result is consistent with them. 
It must also be  noticed that the mixture of \bquark-hadrons is slightly different for \lep and \lhc, thus the $\BR(\bquark\to(\ccbar)X)$ measured in each case is not expected to be exactly the same.
However the difference in the \bquark-hadron cocktail between \lep and \lhc is small compare to the precisions of present measurement. 
The value obtained is also lower than than the branching fraction of \bquark-decays into \chicone 
measured by \cleo~\cite{Anderson:2002md}, \belle~\cite{Bhardwaj:2015rju} and \babar~\cite{Aubert:2002hc}, which however only refers to a Bz and \Bp mixture, 
$0.00435 \pm 0.00029 \pm 0.00040$, 
$0.00363 \pm 0.00022 \pm 0.00034$, and  
$0.00333 \pm 0.00005 \pm 0.00024$, respectively. 
Finally, the \lhcb result for \bquark-decays into \chicone is consistent with 
the prediction in Ref.~\cite{Beneke:1998ks}. 

The branching fraction of \bquark-hadron decays into \chictwo is measured for the first time 
with the \Bz, \Bp, \Bs and \bquark-baryons mixture.  
The result is consistent with the average, corresponding 
to the \Bz, \Bp mixture, from Ref.~\cite{PDG2016}, given large PDG uncertainty. 
The obtained value has higher precision than
the results from \cleo~\cite{Chen:2000ri} and \babar~\cite{Aubert:2002hc}, less precise than recently updated \belle measurement $(0.98 \pm 0.06 \pm 0.10) \times 10^{-3}$~\cite{Bhardwaj:2015rju}
is close to the \cleo result of $(0.67 \pm 0.34 \pm 0.03) \times 10^{-3}$ 
and is different by more than $2 \sigma$ 
and \babar, $(2.10 \pm 0.45 \pm 0.31) \times 10^{-3}$. 
The comparison of the obtained results with theory prediction~\cite{Beneke:1998ks} is given in Section~\protect\ref{ch:pheno}. 

It should be mentioned, that the measured branching fractions of \bquark-hadron 
decays to charmonia comprise also decays via intermediate higher-mass charmonium resonances, 
contrary to the theory calculations, which consider only direct \bquark-hadron transitions to 
the considered charmonium state. 

\begin{figure}[h]
\centering
\protect\includegraphics[width=0.75\linewidth]{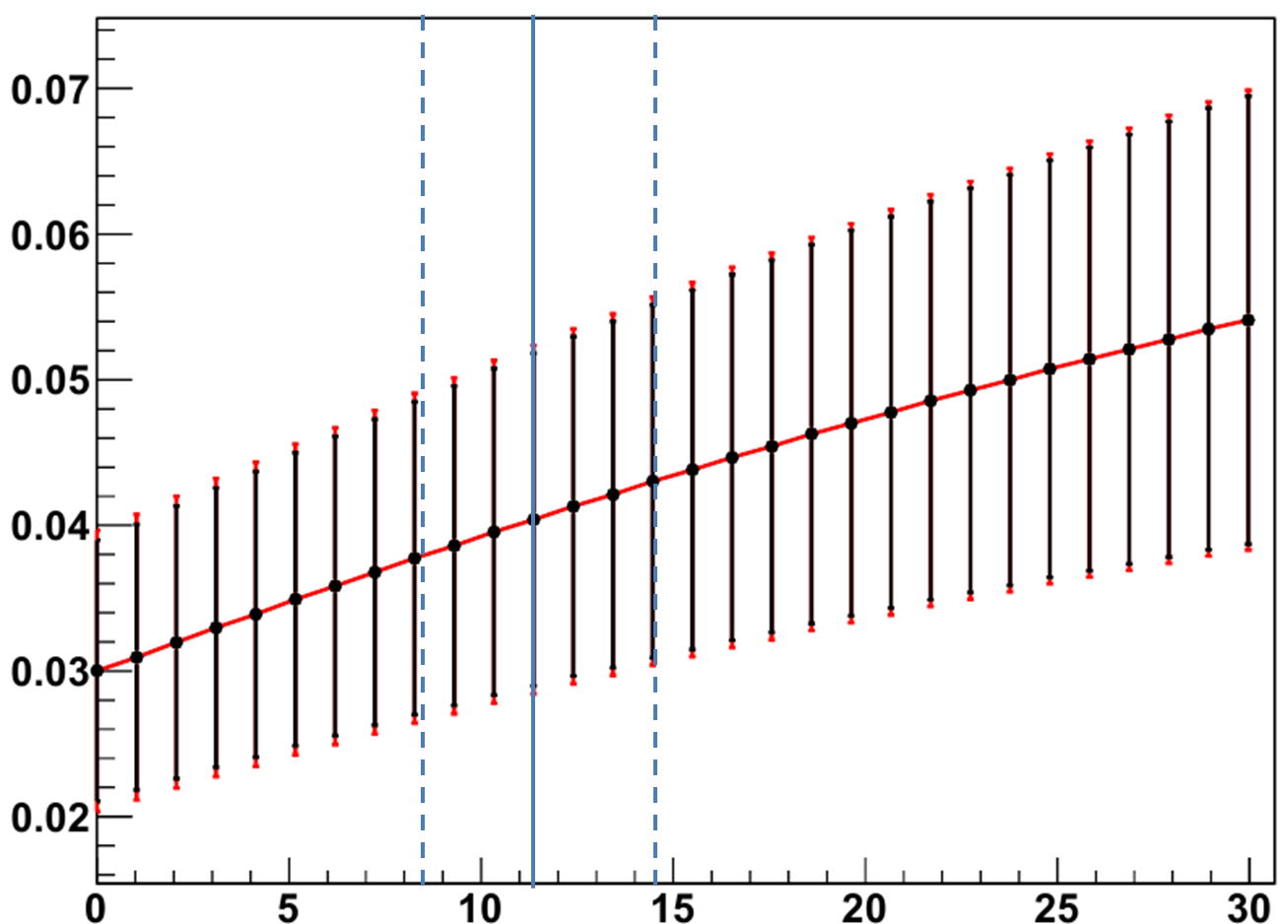}
                \put(-360,60){\rotatebox{90}{{$\frac{\BR ( b \to \etac (2S) X ) \times \BR ( \etac (2S) \to \phi \phi )}{\BR ( b \to \etac (1S) X ) \times \BR ( \etac (1S) \to \phi \phi )}$}}}
                \put(-300,220) {\lhcb}
                \put(-300,200) {3 fb$^{-1}$}
                \put(-190,-12) {{$\Gamma ( \etac (2S) )$, \mev}}
\protect\caption
[Obtained ratio of the $\etac(1S)$ and \etactwos inclusive yields 
$\frac{\BR ( b \to \etac (2S) X ) \times \BR ( \etac (2S) \to \phi \phi )}{\BR ( b \to \etac (1S) X ) \times \BR ( \etac (1S) \to \phi \phi )}$
depending on the \etactwos natural width.]
{Obtained ratio of the $\etac(1S)$ and \etactwos inclusive yields 
$\frac{\BR ( b \to \etac (2S) X ) \times \BR ( \etac (2S) \to \phi \phi )}{\BR ( b \to \etac (1S) X ) \times \BR ( \etac (1S) \to \phi \phi )}$
depending on the \etactwos natural width. 
Statistical and total uncertainties for each point are shown separately. 
The \etactwos natural width from Ref.~\cite{PDG2016} is shown as a vertical solid line, 
while dashed lines correspond to the Ref.~\cite{PDG2016} uncertainty.
} \protect\label{fig:etacvsgamma}
\protect\end{figure}

Another goal is to quantify the observed signal of $365 \pm 100$ $\etactwos$ meson candidates 
in \bquark-hadron inclusive decays. 
Taking into account the difference in trigger, reconstruction and selection efficiencies 
for $\etac(1S)$ and \etactwos mesons, 
the yield ratio relative to the $\etac (1S)$ yield was constructed as 
\protect\begin{align*}
\frac{\BR ( b \to \etac (2S) X ) \times \BR ( \etac (2S) \to \phi \phi )}{\BR ( b \to \etac (1S) X ) \times \BR ( \etac (1S) \to \phi \phi )} 
 = 0.040 \pm 0.011 \pm 0.004 \ , 
\protect\end{align*}
where systematic uncertainty is dominated by possible contributions possible contributions of other resonances 
and variation of the $\etactwos$ natural width.
The dependence of this ratio on the \etactwos natural width is shown in Fig.~\protect\ref{fig:etacvsgamma}. 
\protect

Since the decay of \etac(2S) meson to \phiphi had not been previously observed, only 
the product of the branching fraction of \bquark-hadron decays to \etac(2S) 
and the branching fraction of the $\etac (2S) \to \phi \phi$ decay mode is determined as 
\protect\begin{align*}
\BR ( b \to \etac (2S) X ) \times \BR ( \etac (2S) \to \phi \phi )
 &= ( 6.34 \pm 1.81 \pm 0.57 \pm 1.89 ) \times 10^{-7} \ , 
\protect\end{align*}
where systematic uncertainty is dominated by the uncertainty of the \etac production 
in \bquark-decays. 
This is the first indication of the \etactwos production in \bquark-decays, 
as well as the decay of \etactwos meson to the \phiphi pair. 

\newpage
\subsubsection{The \pt-differential $\chic$ and $\etac(1S)$ 
production in \bquark-decays}
\protect\label{sec:pt}

\begin{figure}[b]
\centering
\protect\includegraphics[width=1.0\linewidth]{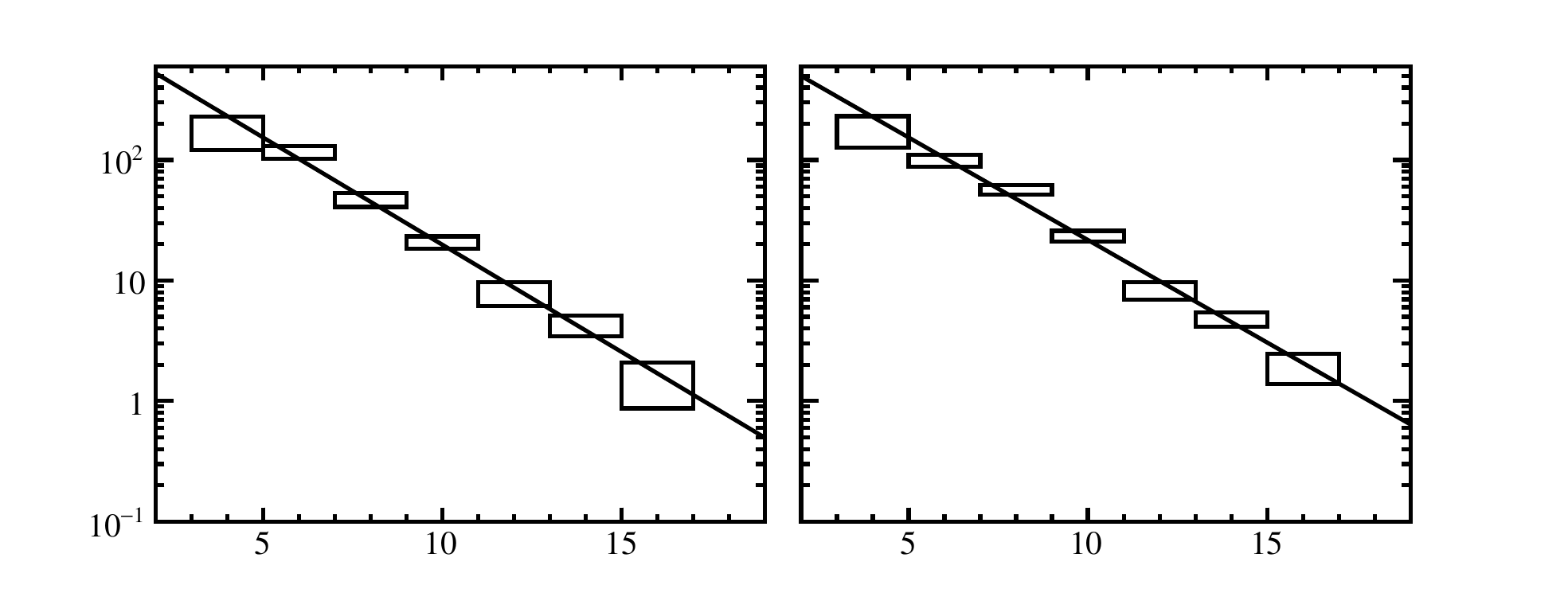}
                \put(-440,90){\rotatebox{90}{{$\frac{d \sigma}{d \pt}$ ($\frac{\nb}{\gev}$)}}}
                \put(-280,130){\lhcb}
                \put(-280,110){$1 \invfb$}
                \put(-90,130){\lhcb}
                \put(-90,110){$2 \invfb$}
                \put(-100,-3){{\pt, \gev}}
\protect\protect\caption
[Differential production cross-section of the $\etac (1S)$ state for the $\protect\sqs = 7 \protect\tev$ 
and $\protect\sqs = 8 \protect\tev$ data samples.]
{Differential production cross-section of the $\etac (1S)$ state for the $\protect\sqs = 7 \protect\tev$ (left) 
and $\protect\sqs = 8 \protect\tev$ (right) data samples. Fits (integral) to an exponential function are overlaid.}
\protect\label{fig:ptetac}
\protect\end{figure}
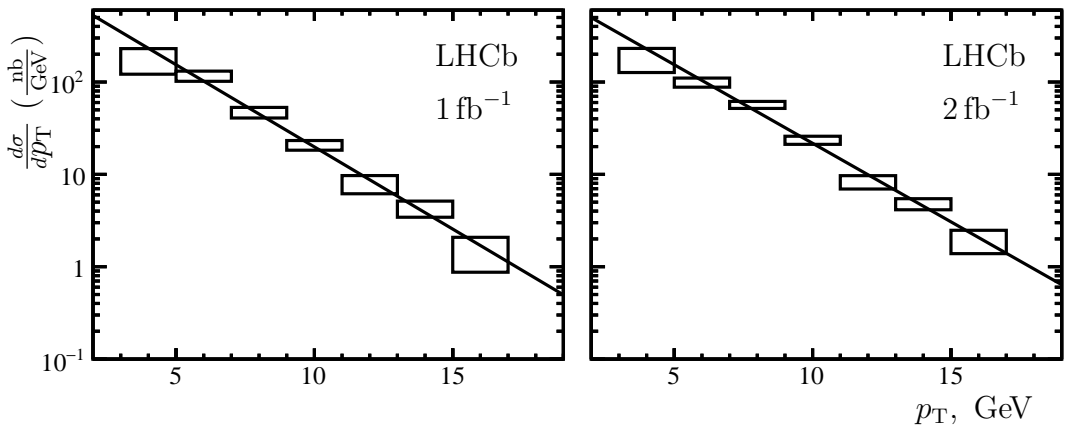

The shapes of the differential production cross-sections as a function of transverse momentum are studied 
in the \lhcb acceptance ($2 < \eta < 5$) and for $3 < \pt < 17 \gev$ and $2 < \pt < 19 \gev$ 
for the $\etac (1S)$ and \chic states, respectively. 
Figure~\protect\ref{fig:ptetac} shows the differential cross-section of the $\etac (1S)$ production at $\protect\sqs = 7 \protect\tev$ and $8 \protect\tev$. 
\protect
Only statistical and uncorrelated systematic uncertainties are taken into account. 
The distributions are fit to the exponential function. 
Dependence of the  $\etac (1S)$ production on \pt is found to be similar in the studied kinematical regime for the two centre-of-mass energies (Table~\protect\ref{tab:ptetac}). 
\protect\begin{table}[b]
\centering
\protect\begin{tabular}{l|c|c}
Data sample & Exponential slope &  \chisqndf \\ \hline
$\protect\sqs = 7 \protect\tev$ & $0.41 \pm 0.02$ & $0.41$ \\ \hline
$\protect\sqs = 8 \protect\tev$ & $0.39 \pm 0.02$ & $1.12$
\protect\end{tabular}
\protect\protect\caption{Results of the fit to the $\etac (1S)$ differential cross-section data 
for the $\protect\sqs = 7 \protect\tev$ and $8 \protect\tev$ data samples. 
\protect\label{tab:ptetac}}
\protect\end{table}
As a cross-check the corresponding  \chisqndf values are obtained using only statistical uncertainties, which also shows a good fit quality.

\begin{figure}[h]
\centering
\protect\includegraphics[width=1.0\linewidth]{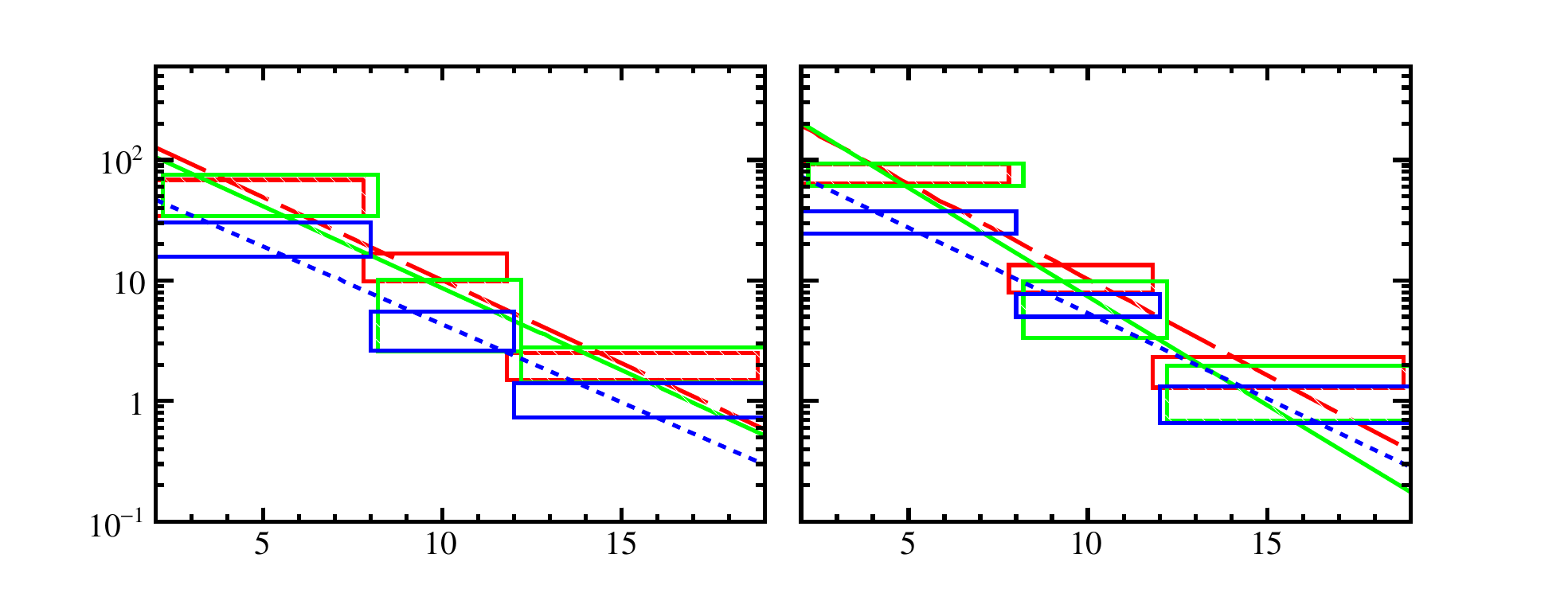}
                \put(-440,90){\rotatebox{90}{{$\frac{d \sigma}{d \pt}$ ($\frac{\nb}{\gev}$)}}}
                \put(-280,130){\lhcb}
                \put(-280,110){$1 \invfb$}
                \put(-90,130){\lhcb}
                \put(-90,110){$2 \invfb$}
                \put(-100,-3){{\pt, \gev}}
\protect\protect\caption
[Differential production cross-section of the $\protect\chiczero$ (red), $\chicone$ (green) 
and $\chictwo$ (blue) states for the $\protect\sqs = 7 \protect\tev$ and $\protect\sqs = 8 \protect\tev$ data samples.]
{Differential production cross-section of the $\protect\chiczero$ (red), $\chicone$ (green) 
and $\chictwo$ (blue) states for the $\protect\sqs = 7 \protect\tev$ (left) and $\protect\sqs = 8 \protect\tev$ (right)
data samples.
Fits (integral) to an exponential function are overlaid.}
\protect\label{fig:ptchic}
\protect\end{figure}
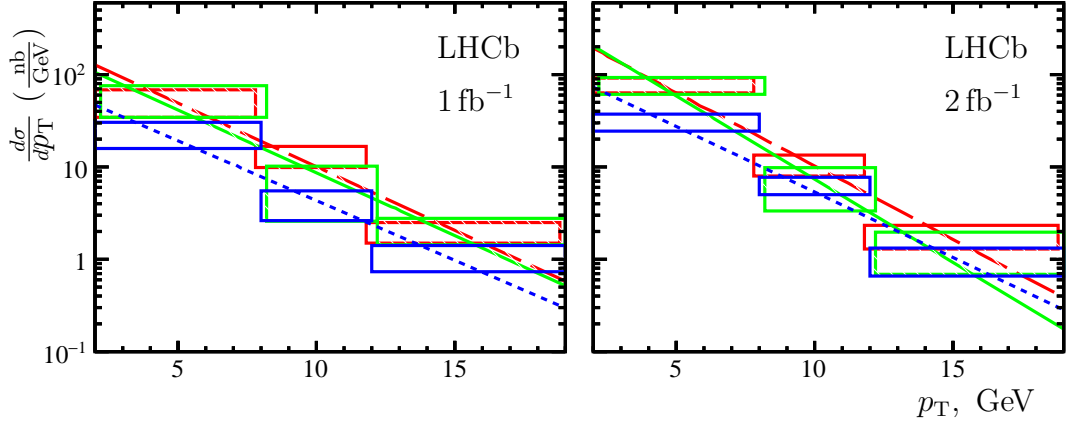

Figure~\protect\ref{fig:ptchic} shows differential production cross-sections of the \chic states for $\protect\sqs = 7 \protect\tev$ and $8 \protect\tev$ data samples. 
\protect
Only statistical and uncorrelated systematic uncertainties are taken into account. 
The fits of the numbers of \chic states in \pt bins are performed simultaneously with the integral fit. The result of the fit is given in Table~\protect\ref{tab:ptchic}. 
\protect\begin{table}[h]
\centering
\protect\protect\caption{Results of the fit to the \chic differential cross-section data 
for the $\protect\sqs = 7 \protect\tev$ and $8 \protect\tev$ data samples. 
\protect\label{tab:ptchic}}
\protect\begin{tabular}{l|c|c|c|c|c|c}
Data  & \multicolumn{3}{|c|}{Slope} & \multicolumn{3}{c}{\chisqndf} \\ \cline{2-7}
sample & \chiczero & \chicone & \chictwo & \chiczero & \chicone & \chictwo \\ \hline
$\protect\sqs = 7 \protect\tev$ & $0.32 \pm 0.04$ & $0.31 \pm 0.06$ & $0.30 \pm 0.05$ & 0.61 & 0.69 & 0.17 \\ \hline
$\protect\sqs = 8 \protect\tev$ & $0.37 \pm 0.04$ & $0.41 \pm 0.06$ & $0.33 \pm 0.04$ & 0.03 & 0.42 & 0.27 
\protect\end{tabular}
\protect\end{table}

\begin{figure}[t]
\centering
\protect\includegraphics[width=1.0\linewidth]{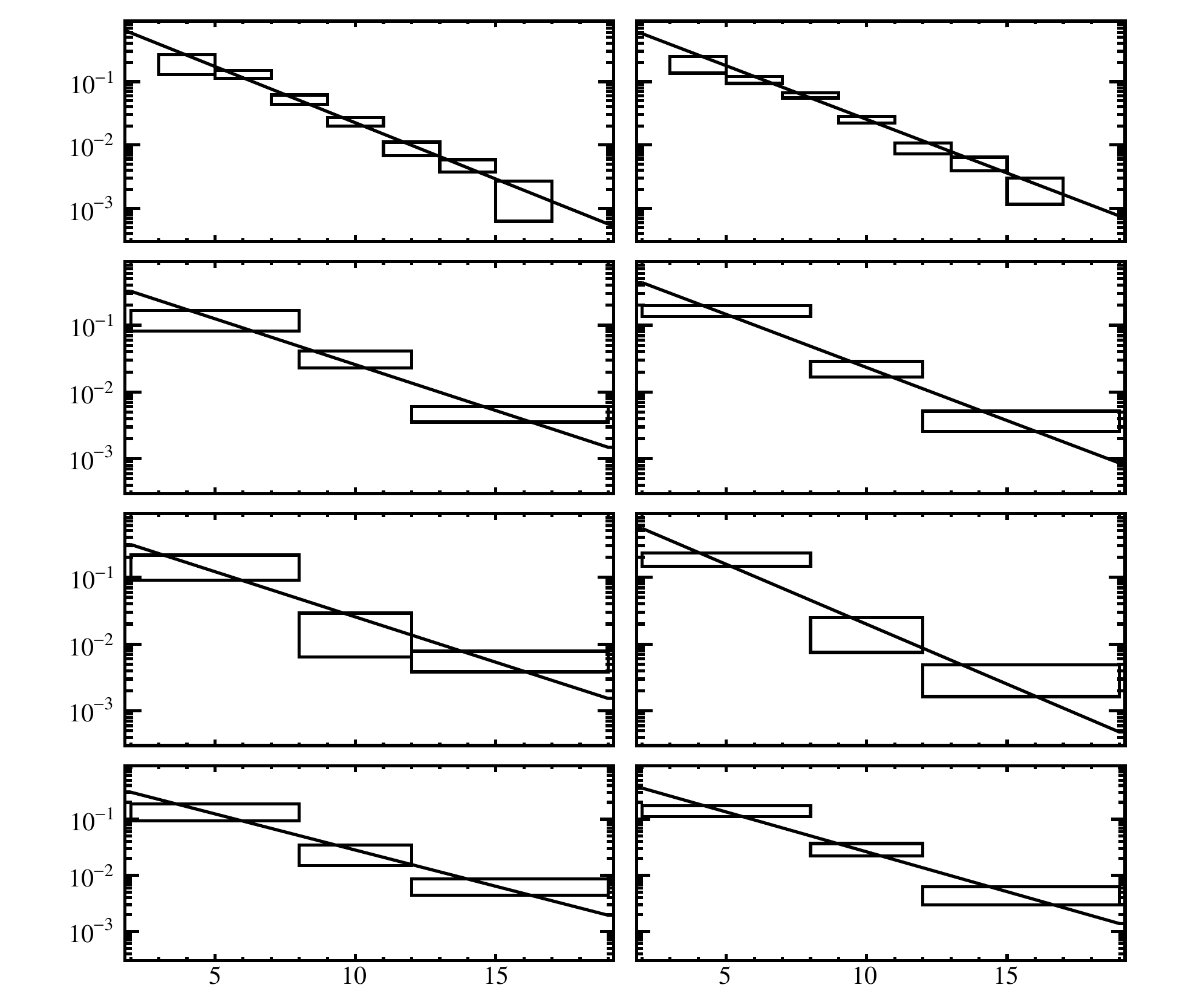}
                \put(-450,300){\rotatebox{90}{{$\frac{d \sigma}{\sigma^* dp_{\textrm{\tiny{$_{\mbox{T}}$}}}} \ [ \gev^{-1} ]$}}}
                \put(-395,300){a) $\etac (1S)$}
                \put(-395,205){c) \chiczero}
                \put(-395,110){e) \chicone}
                \put(-395,30){g) \chictwo}
                \put(-200,300){b) $\etac (1S)$}
                \put(-200,205){d) \chiczero}
                \put(-200,110){f) \chicone}
                \put(-200,30){h) \chictwo}
                \put(-270,360){\lhcb}
                \put(-270,265){\lhcb}
                \put(-270,170){\lhcb}
                \put(-270,75){\lhcb}
                \put(-270,345){$7 \protect\tev$}
                \put(-270,250){$7 \protect\tev$}
                \put(-270,155){$7 \protect\tev$}
                \put(-270,60){$7 \protect\tev$}
                \put(-70,360){\lhcb}
                \put(-70,265){\lhcb}
                \put(-70,170){\lhcb}
                \put(-70,75){\lhcb}
                \put(-70,345){$8 \protect\tev$}
                \put(-70,250){$8 \protect\tev$}
                \put(-70,155){$8 \protect\tev$}
                \put(-70,60){$8 \protect\tev$}
                \put(-75,-3){{$\pt \ [ \gev ]$}}
\protect\protect\caption
[Differential cross-sections normalized to the production cross-section integrated over the studied region, $\sigma^*$, 
of the $\etac (1S)$, \chiczero, \chicone and \chictwo states 
for the $\protect\sqs = 7 \protect\tev$ and the $\protect\sqs = 8 \protect\tev$ data samples.]
{Differential cross-sections normalized to the production cross-section integrated over the studied region, $\sigma^*$, 
of the (top to bottom) $\etac (1S)$, \chiczero, \chicone and \chictwo states 
for the (left) $\protect\sqs = 7 \protect\tev$ and the (right) $\protect\sqs = 8 \protect\tev$ data samples. 
The horizontal and vertical size of the boxes reflect the size of the \pt bins and the statistical and uncorrelated systematic uncertainties 
of the differential production cross-sections added in quadrature. 
Fits by an exponential function are overlaid.}
\protect\label{fig:ptccbar}
\protect\end{figure}

Below, each differential production cross-section is normalized to the production cross-section integrated over the studied \pt region. 
Figure~\protect\ref{fig:ptccbar} shows the normalized differential cross-sections of $\etac (1S)$, \chiczero, \chicone and \chictwo production at $\protect\sqs = 7$ and $8 \protect\tev$. 
\protect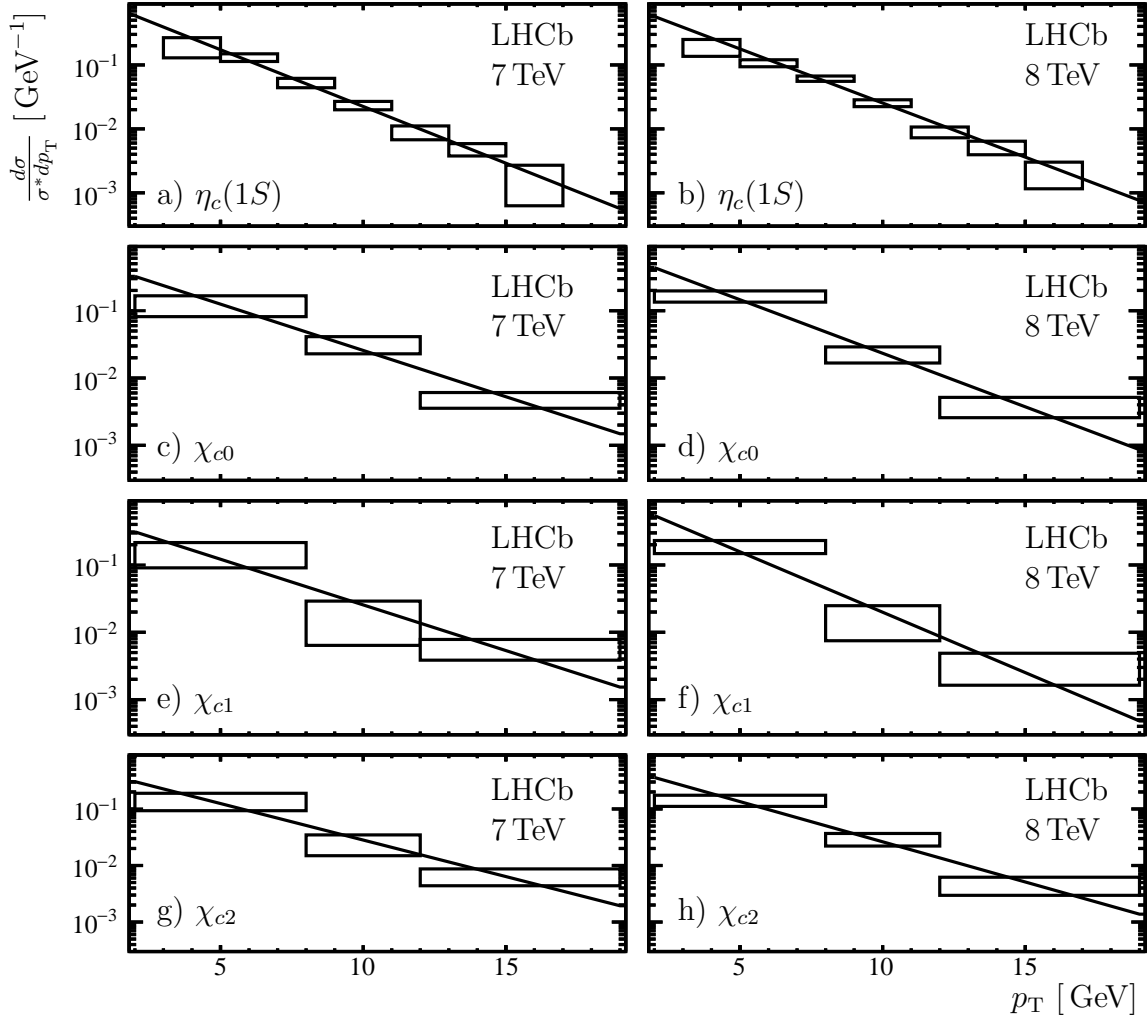

An exponential function proportional to $\exp (- \alpha \, \pt )$ is fitted to the distributions. 
The results for the slope parameters $\alpha$ are given in Table~\protect\ref{tab:ptall}.
\protect\begin{table}[t]
\centering
\protect\protect\caption{Exponential slope parameter in units of $\gev^{-1}$ from a fit to the \pt spectra of 
$\etac (1S)$, \chiczero, \chicone and \chictwo mesons. 
\protect\label{tab:ptall}}
\protect\begin{tabular}{l|c|c|c|c}
              & $\etac (1S)$ & \chiczero & \chicone & \chictwo \\ \hline
$\protect\sqs = 7 \protect\tev$ & $0.41 \pm 0.02$ & $0.32 \pm 0.04$ & $0.31 \pm 0.06$ & $0.30 \pm 0.05$  \\ 
$\protect\sqs = 8 \protect\tev$ & $0.39 \pm 0.02$ & $0.37 \pm 0.04$ & $0.41 \pm 0.06$ & $0.33 \pm 0.04$ 
\protect\end{tabular}
\protect\end{table}
Production dependence on \pt is found to be similar in the studied kinematical regime within uncertainties, 
for the two centre-of-mass energies (Table~\protect\ref{tab:ptall}). 
For \chicone and \chictwo production in \bquark-hadron decays these results extend the \atlas experiment studies~\cite{ATLAS:2014ala} 
in \pt and rapidity.
As a cross-check the corresponding \chisqndf values are obtained using only statistical uncertainties to check the effect of the systematic uncertainty, which remains negligible.

\clearpage
\section{Search for production of the $X(3872)$, $X(3915)$, and $\chictwo (3930)$ states}
\label{sec:limits}
Observation of the $X(3915)$ and $\chictwo (3930)$ states in \bquark-decays, 
or the $X(3872)$ decaying to the $\phi \phi$ pair, 
would provide interesting information on the properties of these states. 
For example, none of $X(3872)$ decays to light hadrons has been observed. Hence, the observation of the $\decay{X(3872)}{\phi\phi}$ would lead to the estimation of fraction of the charmonium component in $X(3872)$ if one considers that $X(3872)$ is a mixed state of charmonium and hadronic molecule. Similar considerations apply for $X(3915)$ and $\chictwo (3930)$ (named $X(3927)$ at the time of this analysis release).

Figure~\ref{fig:ccphiphi} shows no indication of signal from the 
$X(3872)$, $X(3915)$, and $\chictwo (3930)$ states. 
From that upper limits are obtained, relative to the observed states 
with similar quantum numbers. 

\begin{figure}[p]
\centering
\protect\protect\includegraphics[width=0.75\linewidth,height=8.0cm]{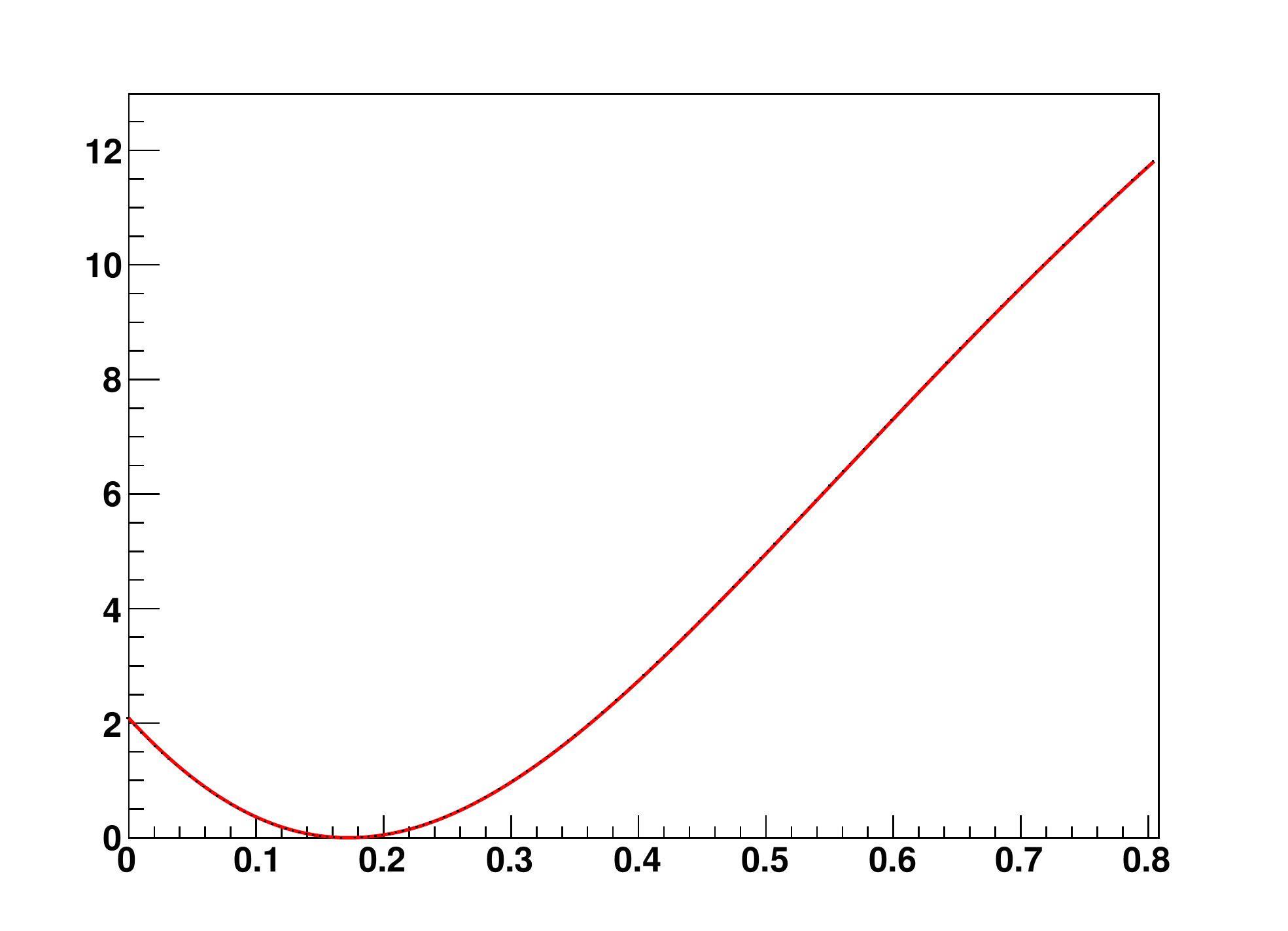}
                \put(-150,180) {\lhcb}
                \put(-150,165) {3 fb$^{-1}$}
                \put(-350,110){\rotatebox{90}{{$\Delta_{\chisq}$}}}
\linebreak
\protect\protect\includegraphics[width=0.75\linewidth,height=8.0cm]{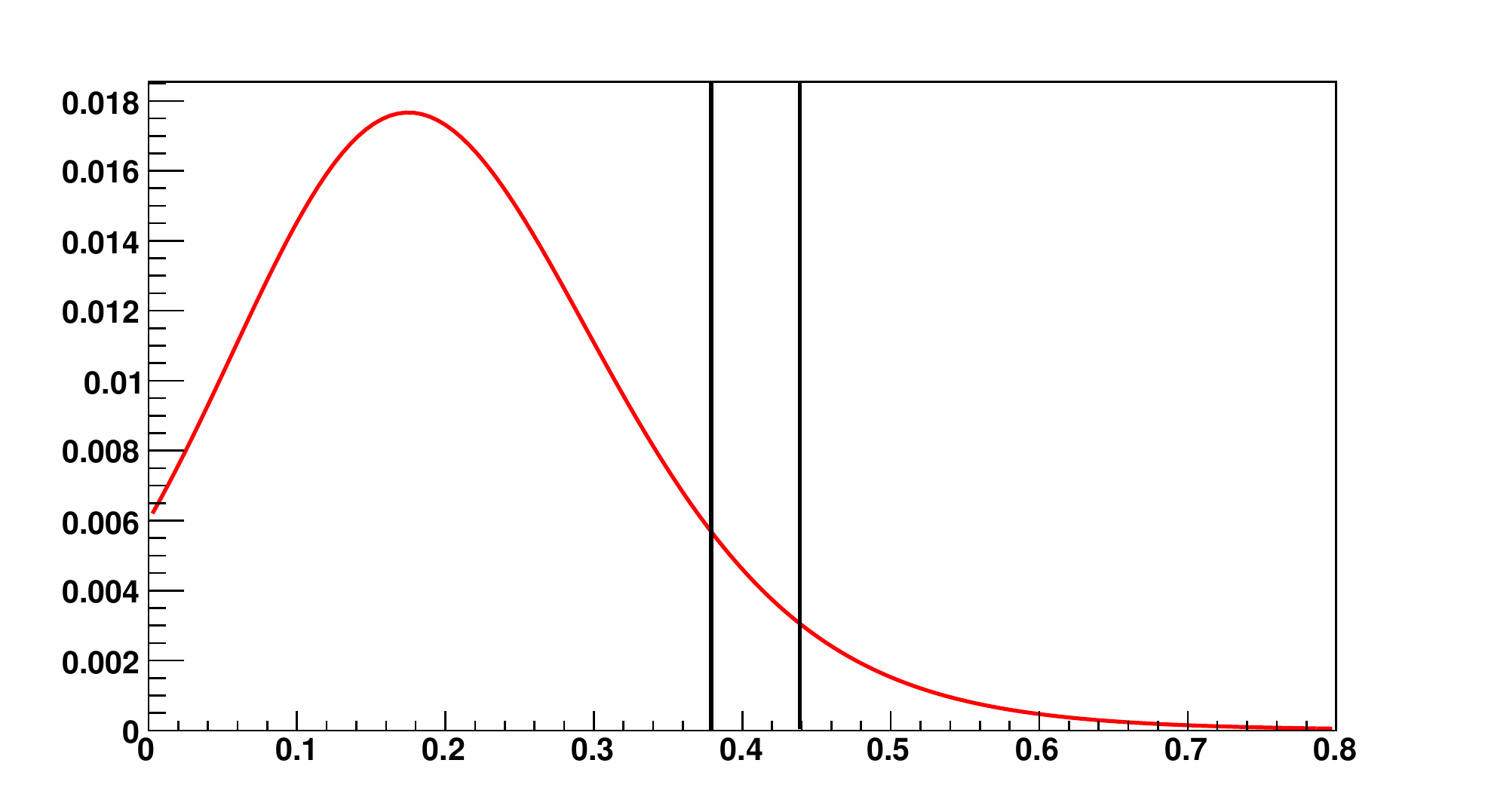}
                \put(-150,180) {\lhcb}
                \put(-150,165) {3 fb$^{-1}$}
                \put(-350,110){\rotatebox{90}{{\PDF}}}
                \put(-200,-5) {{$\frac{N ( X(3872) )}{N ( \chicone )}$}}
\caption
[The $\Delta_{\chisq}$ and \PDF distributions for the $\frac{N(X(3872))}{N(\chicone)}$ ratio.]
{The $\Delta_{\chisq}$ and \PDF distributions for the $\frac{N(X(3872))}{N(\chicone)}$ ratio. Vertical lines correspond to 
$90 \, \%$ CL and $95 \, \%$ CL upper limits.} \label{fig:ulx}
\end{figure}

\begin{figure}[p]
\centering
\protect\protect\includegraphics[width=0.75\linewidth,height=8.0cm]{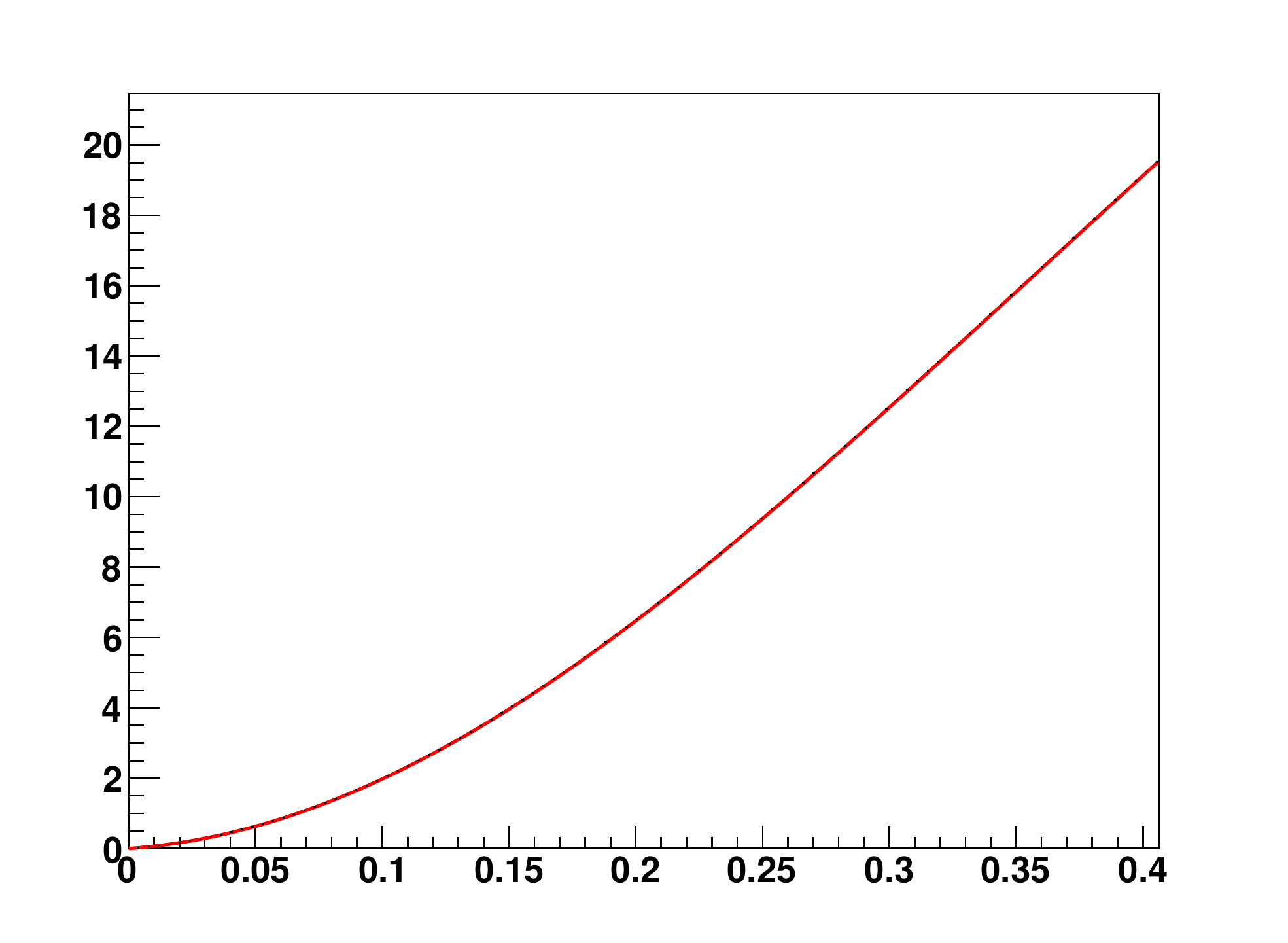}
                \put(-150,180) {\lhcb}
                \put(-150,165) {3 fb$^{-1}$}
                \put(-350,110){\rotatebox{90}{{$\Delta_{\chisq}$}}}
\linebreak
\protect\protect\includegraphics[width=0.75\linewidth,height=8.0cm]{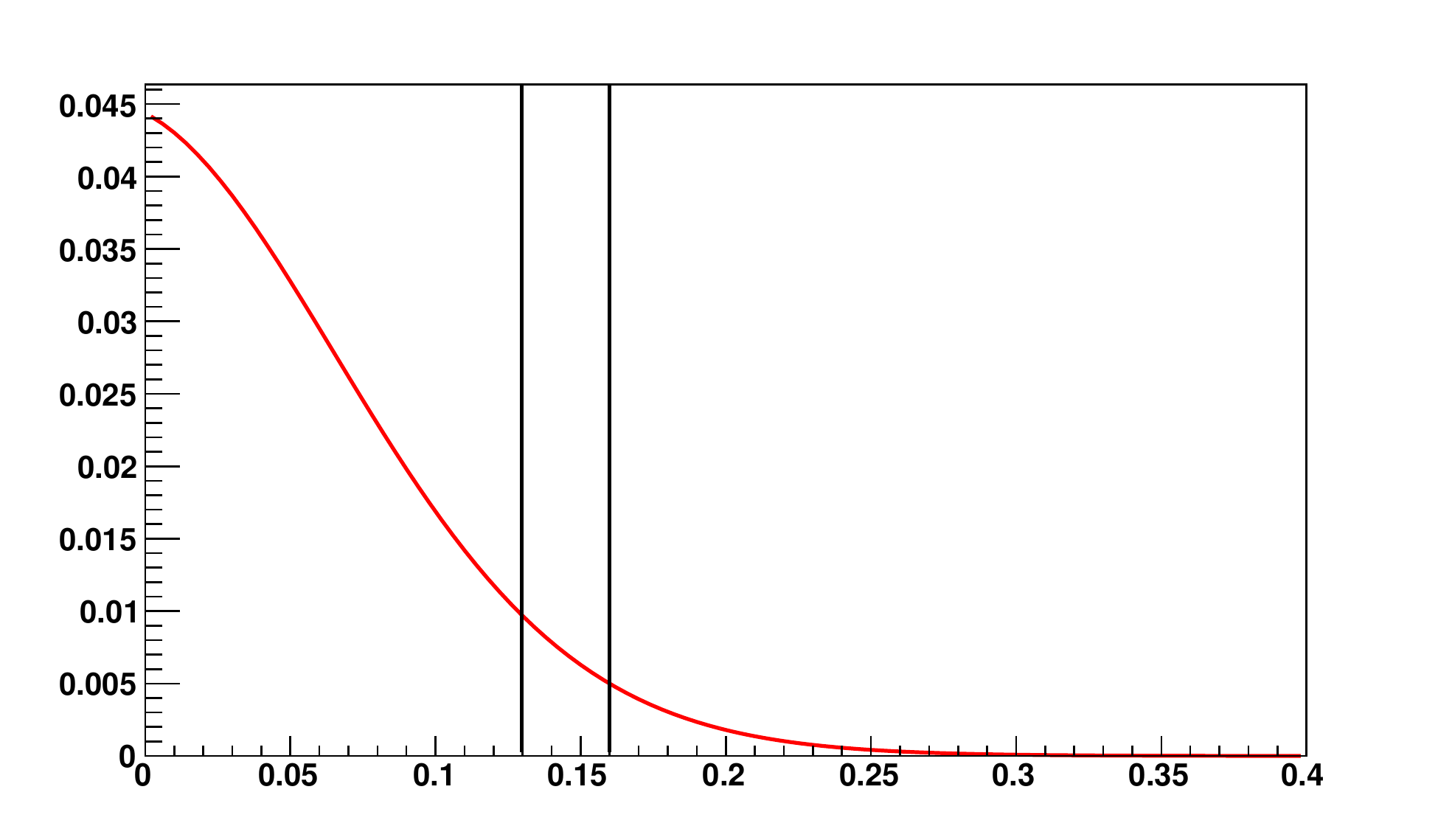}
                \put(-150,180) {\lhcb}
                \put(-150,165) {3 fb$^{-1}$}
                \put(-350,110){\rotatebox{90}{{\PDF}}}
                \put(-200,-5) {{$\frac{N ( X(3915) )}{N ( \chiczero )}$}}
\caption
[The $\Delta_{\chisq}$ and \PDF distributions for the 
$\frac{N ( X(3915) )}{N ( \chiczero )}$ 
ratio.]
{The $\Delta_{\chisq}$ and \PDF distributions for the 
$\frac{N ( X(3915) )}{N ( \chiczero )}$ 
ratio. 
Vertical lines correspond to $90 \, \%$ CL and $95 \, \%$ CL upper limit.} \label{fig:ulchinol}
\end{figure}

\begin{figure}[p]
\centering
\protect\protect\includegraphics[width=0.75\linewidth,height=8.0cm]{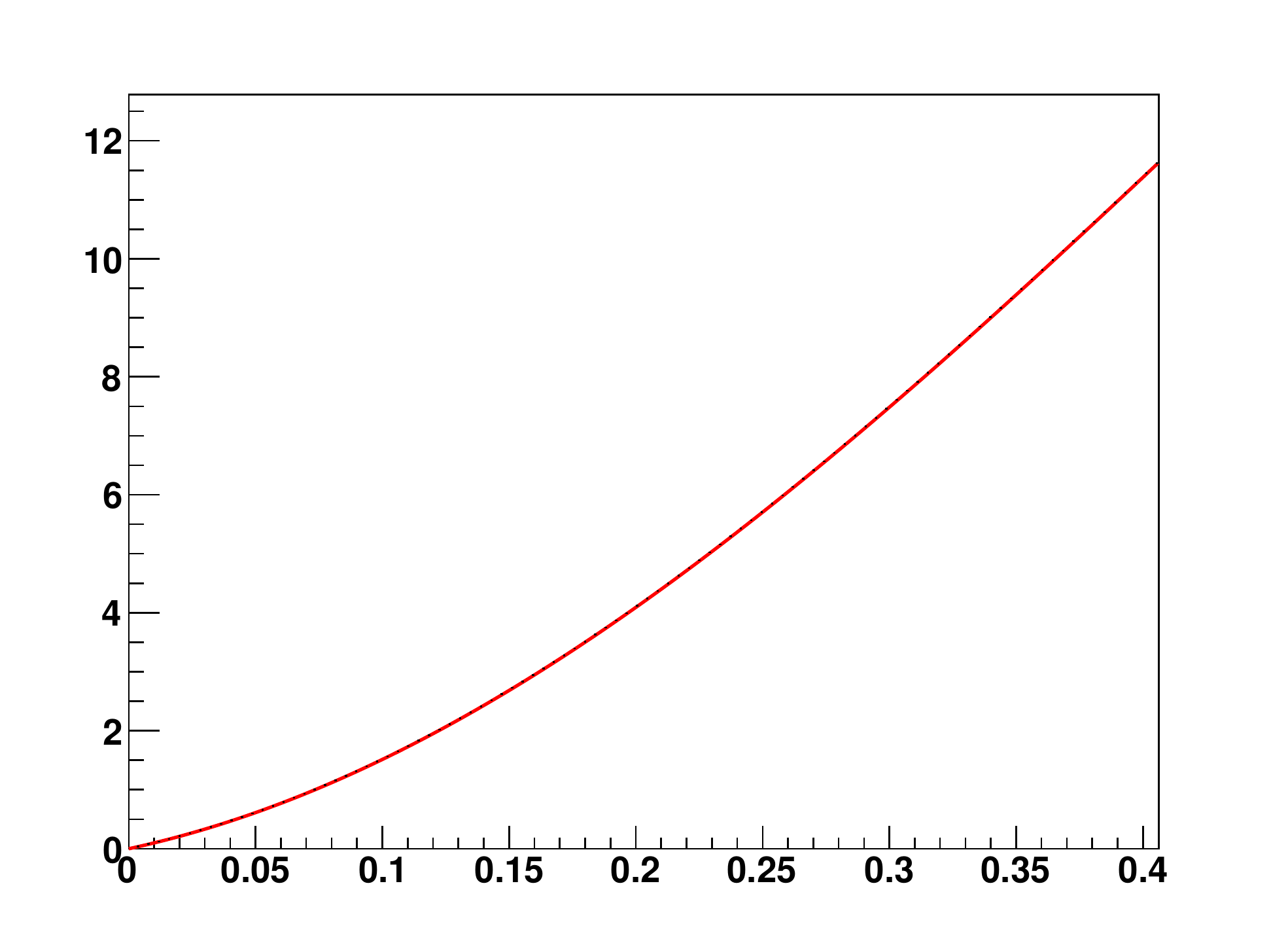}
                \put(-150,180) {\lhcb}
                \put(-150,165) {3 fb$^{-1}$}
                \put(-350,110){\rotatebox{90}{{$\Delta_{\chisq}$}}}
\linebreak
\protect\protect\includegraphics[width=0.75\linewidth,height=8.0cm]{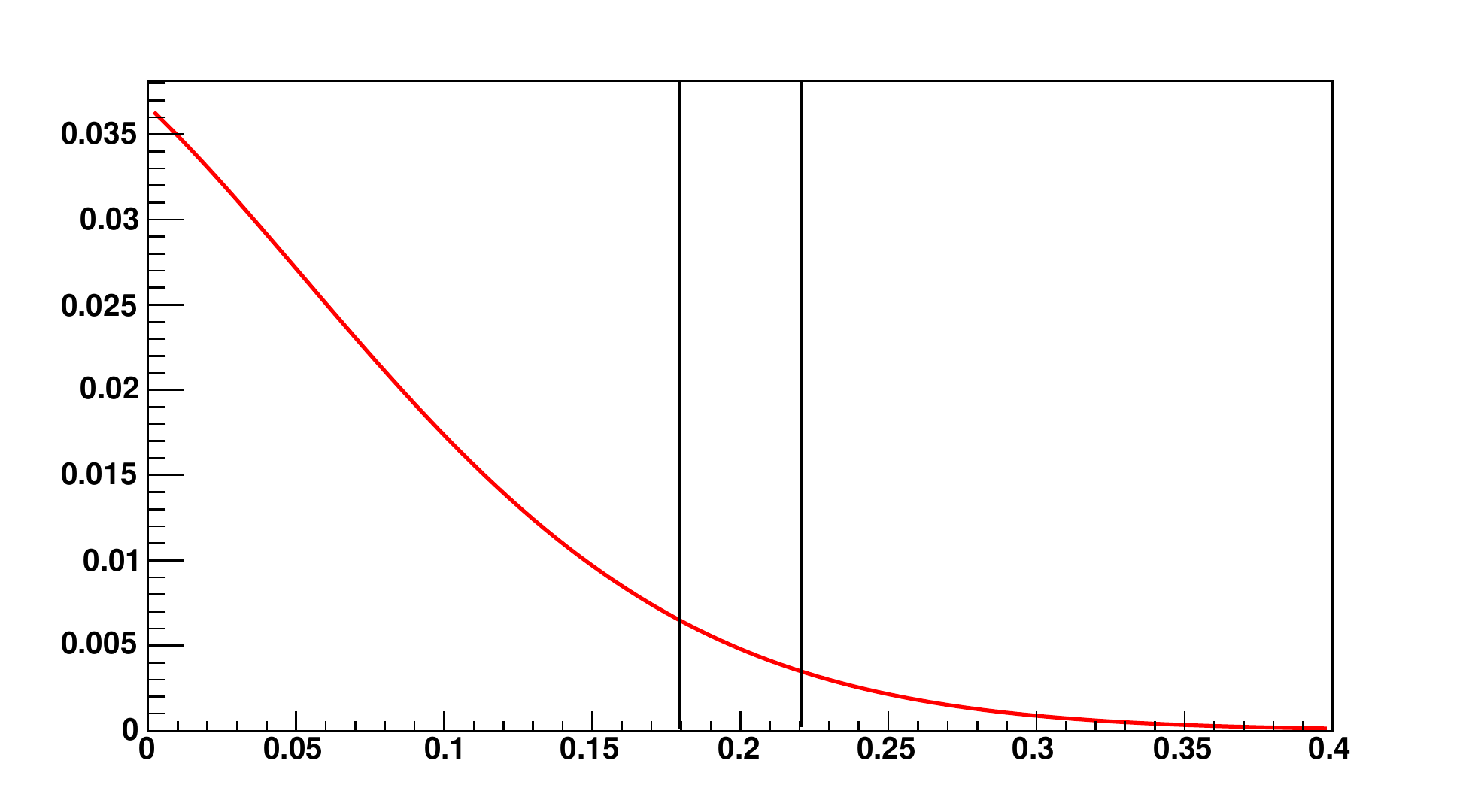}
                \put(-150,180) {\lhcb}
                \put(-150,165) {3 fb$^{-1}$}
                \put(-350,110){\rotatebox{90}{{\PDF}}}
                \put(-200,-5) {{$\frac{N ( \chictwo (3930) )}{N ( \chictwo )}$}}
\caption
[The $\Delta_{\chisq}$ and \PDF distributions for the 
$\frac{N ( \chictwo (3930) )}{N ( \chictwo )}$
ratio.]
{The $\Delta_{\chisq}$ and \PDF distributions for the 
$\frac{N ( \chictwo (3930) )}{N ( \chictwo )}$
ratio. 
Vertical lines correspond to $90 \, \%$ CL and $95 \, \%$ CL upper limit.} \label{fig:ulchidva}
\end{figure}

Figures~\ref{fig:ulx}, \ref{fig:ulchinol}, and \ref{fig:ulchidva}
show the $\Delta_{\chisq}$ and \PDF distributions for the 
$\frac{N ( X(3872) )}{N ( \chicone )}$, 
$\frac{N ( X(3915) )}{N ( \chiczero )}$, and 
$\frac{N ( \chictwo (3930) )}{N ( \chictwo )}$, 
respectively. The obtained \PDF distributions take into accound possible systematic effects. For that, the fit likelihood has been convolved with a gaussian, representing total systematic uncertainty.
The Bayessian upper limits at $90 \, \%$ and $95 \, \%$ confidence level (CL) are then extracted for the first time.

Vertical lines correspond to the $90 \%$ and $95 \%$ CL upper limits. 

Using the efficiency ratios 
$\frac{\varepsilon ( X (3872) )}{\varepsilon ( \chicone )} = 1.11$, 
$\frac{\varepsilon ( X(3915) )}{\varepsilon ( \chiczero )} = 1.16$, and 
$\frac{\varepsilon ( \chictwo (3930) )}{\varepsilon ( \chictwo )} = 1.12$, 
the following upper limits on the ratios of inclusive branching fractions are obtained:
\begin{align}
\frac{\BR ( \bquark \ra X(3872) X ) \times \BR ( X(3872) \ra \phi \phi )}{\BR ( \bquark \ra \chicone X ) \times \BR ( \chicone \ra \phi \phi )} & < 0.39 (0.34) , \\
\frac{\BR ( \bquark \ra X(3915) X ) \times \BR ( X(3915) \ra \phi \phi )}{\BR ( \bquark \ra \chiczero X ) \times \BR ( \chiczero \ra \phi \phi )} & < 0.14 (0.12) , \\
\frac{\BR ( \bquark \ra \chictwo (3930) X ) \times \BR ( \chictwo (3930) \ra \phi \phi )}{\BR ( \bquark \ra \chictwo X ) \times \BR ( \chictwo \ra \phi \phi )} & < 0.20 (0.16) 
\label{eq:res1}
\end{align}
at the $95 \%$ ($90 \%$) CL. 

Using the branching fractions for the \chiczero, \chicone, and \chictwo decays to the $\phi \phi$ final state~\cite{PDG2016}, 
observed signals of these states on Fig.~\ref{fig:ccphiphi}, the $\etac (1S)$ state production from Ref.~\cite{LHCb-PAPER-2014-029}
and efficiency ratios from the simulation, the upper limits at $95\%$ ($90 \%$) CL on
production of the $X(3872)$, $X(3915)$, and $\chictwo (3930)$ states are obtained as: 
\begin{align}
\BR ( \bquark \ra X(3872) X ) \times \BR ( X(3872) \ra \phi \phi ) & < 4.5 (3.9) \times 10^{-7} , \\
\BR ( \bquark \ra X(3915) X ) \times \BR ( X(3915) \ra \phi \phi ) &  < 3.1 (2.7) \times 10^{-7} , \\
\BR ( \bquark \ra \chictwo (3930) X ) \times \BR ( \chictwo (3930) \ra \phi \phi ) &  < 2.8 (2.3)  \times 10^{-7} . 
\end{align}

\clearpage

\section{Summary and discussion}
\label{sec:sumPhiPhi}
In summary, charmonia production in \bquark-hadron inclusive decays is studied
with the integrated luminosity of $3 \, fb^{-1}$, using charmonia decays 
to $\phi \phi$ pairs. 
These studies are performed using pure $\phi \phi$ yields from determined using a 2D-fit technique. 

Inclusive production of the $\chi_c$ states in \bquark-hadron decays are measured to be 
\begin{align*}
\BR ( b \to \chiczero X ) &= ( 3.02 \pm 0.47 \pm 0.23 \pm 0.94 ) \times 10^{-3} \ , \\
\BR ( b \to \chicone X )  &= ( 2.76 \pm 0.59 \pm 0.23 \pm 0.89 ) \times 10^{-3} \ , \\
\BR ( b \to \chictwo X )  &= ( 1.15 \pm 0.20 \pm 0.07 \pm 0.36 ) \times 10^{-3} \ . 
\end{align*}
These results will be used in the phenomenological analysis described in Chapter~\ref{ch:pheno}.

\begin{figure}[h]
\centering
\protect\protect\includegraphics[width=1.0\linewidth]{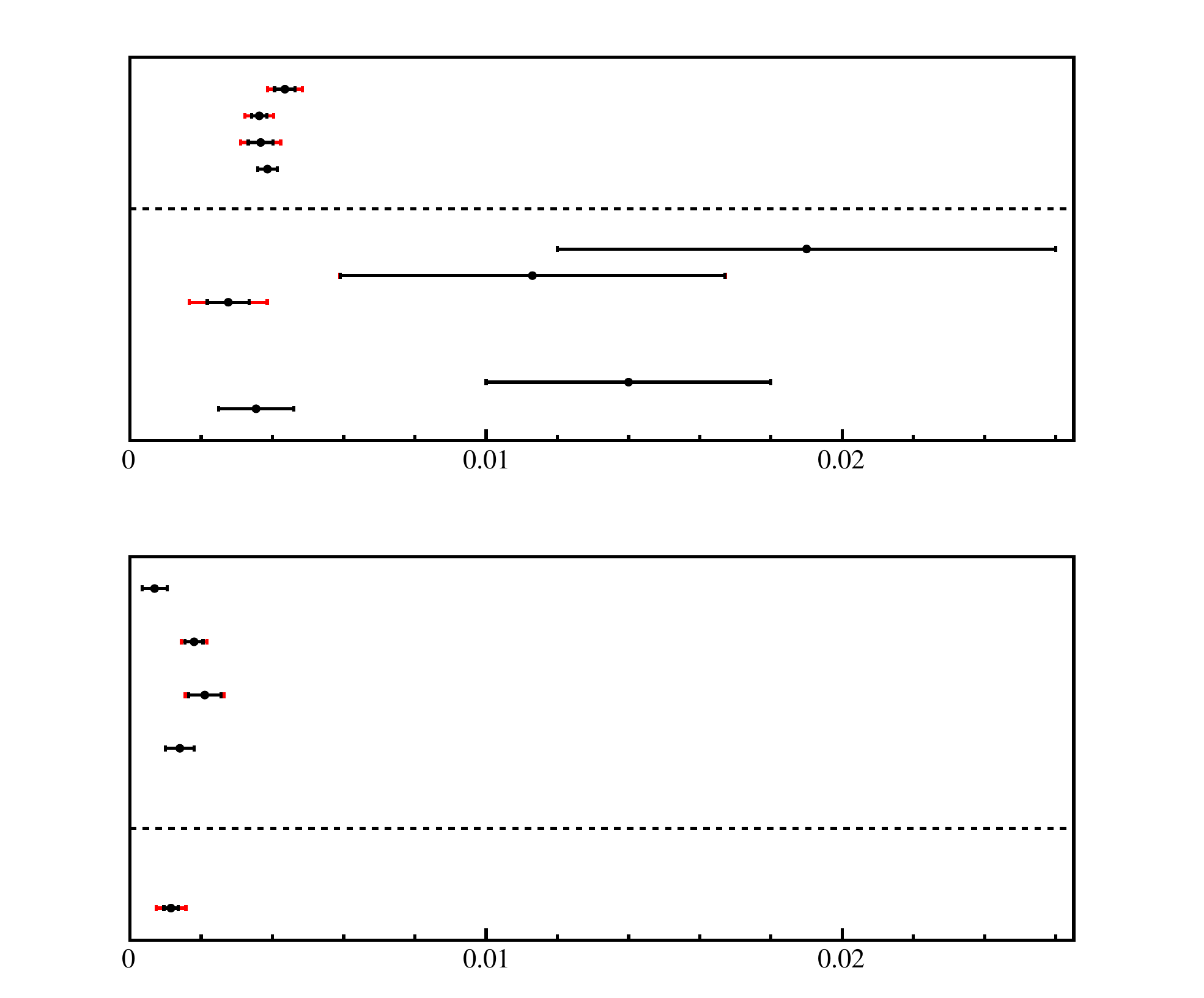}
                \put(-120,310) {\B \to \chicone X}
                \put(-330,339) {\small{\cleo2~\cite{Anderson:2002md}}}
                \put(-330,328) {\small{\belle~\cite{Abe:2002wp}}}
                \put(-330,317) {\small{\babar~\cite{Aubert:2002hc}}}
                \put(-330,306) {\small{PDG2016~\cite{PDG2016}}}
                \put(-120,225) {\bquark \to \chicone X}
                \put(-240,287) {\small{\lthree~\cite{Adriani:1993ta}}}
                \put(-323,278) {\small{\delphi~\cite{Abreu:1994rk}}}
                \put(-345,258) {\small{\lhcb, this paper}}
                \put(-265,237) {\small{PDG2016~\cite{PDG2016}}}
                \put(-375,227) {\small{Our average}}
                \put(-120,76) {\B \to \chictwo X}
                \put(-360,151) {\small{\cleo2~\cite{Chen:2000ri}}}
                \put(-360,131) {\small{\belle~\cite{Abe:2002wp}}}
                \put(-360,111) {\small{\babar~\cite{Aubert:2002hc}}}
                \put(-360,90) {\small{PDG2016~\cite{PDG2016}}}
                \put(-120,33) {\bquark \to \chictwo X}
                \put(-370,31) {\small{\lhcb, this paper}}
                \put(-120,0) {$\BR ( \B , \bquark \to \chi_c X)$}
\caption
[Summary of the branching fraction measurements for inclusive decays of light \B mesons, $\BR (\B \to \chi_c X)$, 
and of all \bquark hadrons, $\BR (\bquark \to \chi_c X)$.]
{Summary of the branching fraction measurements for inclusive decays of light \B mesons, $\BR (\B \to \chi_c X)$, 
and of all \bquark hadrons, $\BR (\bquark \to \chi_c X)$, shown in each plot above and below the dashed line, respectively. 
The branching fractions for the decays to \chicone and \chictwo are shown in the top and bottom plots, respectively. 
The world averages noted ``PDG2016'' do not include the \lhcb results. 
} \label{fig:chicdata}
\end{figure}
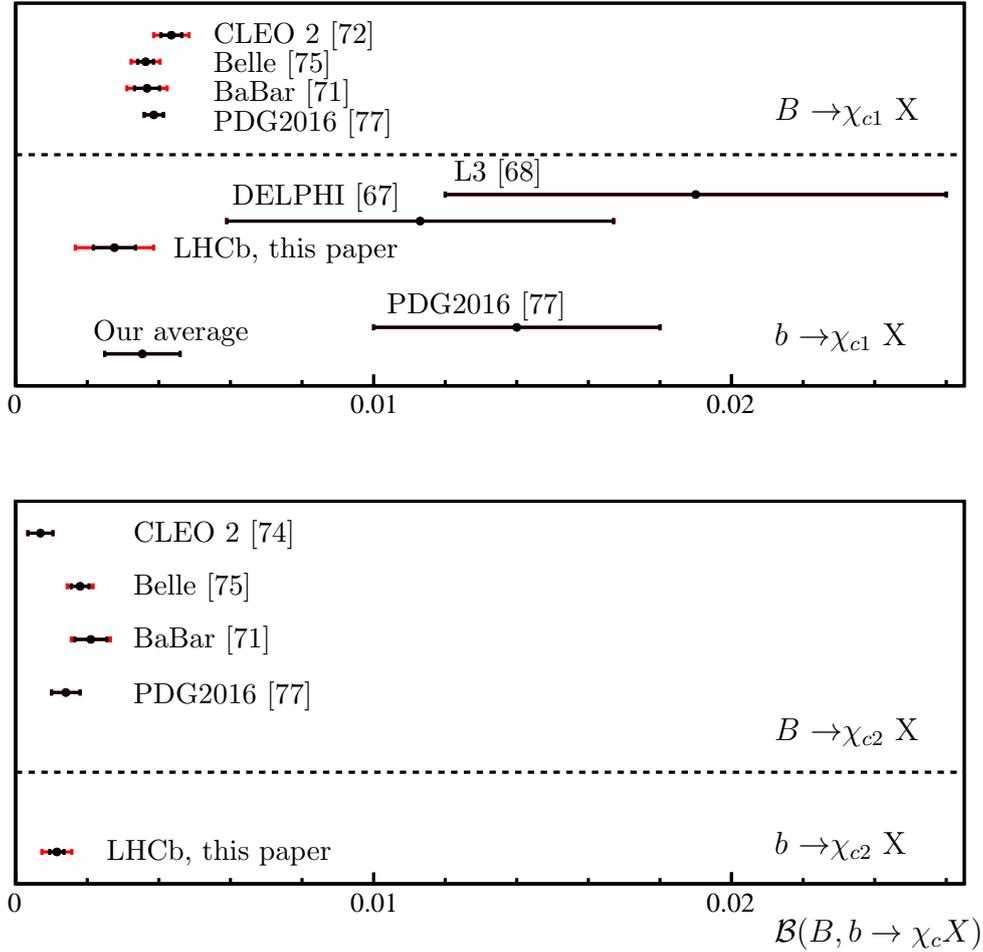

Figure~\ref{fig:chicdata} 
shows a summary of the branching fraction measurements for inclusive decays of light \B-mesons, $\BR (\B \to \chi_c X)$, 
and of mixtures of all \bquark hadrons, $\BR (\bquark \to \chi_c X)$. Note that the mixture of \bquark-hadrons is different for \lep and \lhc. 
Also indicated are the PDG averages and averages including the results from this paper. Note, that the recent update from \belle experiment~\cite{Bhardwaj:2015rju} was released after this analysis and did not enter the plot on Fig.~\ref{fig:chicdata}. The \belle measurements are the most precise among all measurements at B-factories and are shown below.
\begin{align*}
\BR(\B \to \chicone X)^{Belle} &= (3.33 \pm 0.05 \pm 0.24)\times 10^{-3}, \\
\BR(\B \to \chictwo X)^{Belle} &= (9.8 \pm 0.6 \pm 1.0)\times 10^{-4}.
\end{align*}
The \lhcb result for \bquark-hadron decays to \chiczero is the only available result and is not shown in the figure. 
Note, that no indirect contribution to the production rate is subtracted. 
However, limited contribution from \psitwos decays to the $\chi_c$ states is present. 
Relations between the $\chi_c$ branching fractions are not consistent with those 
predicted in Ref.~\cite{Beneke:1998ks}. 
The branching fraction $\BR ( b \to \chiczero X )$ is measured for the first time. 
The result for \bquark-decays into \chicone is the most precise measurement 
for the admixture of \Bz, \Bp, \Bs and \bquark-baryons. 
Central value of the result for \bquark-decays into \chicone is lower than the value 
measured by DELPHI~\cite{Abreu:1994rk} and L3~\cite{Adriani:1993ta} experiments at LEP, 
however, taking into account the LEP results limited precision, 
the \lhcb result is consistent with them. 
The obtained value is lower than the branching fraction of \bquark-decays into \chicone 
measured by \cleo~\cite{Anderson:2002md}, \belle~\cite{Abe:2002wp} and \babar~\cite{Aubert:2002hc} 
using the admixture of \Bz and \Bp. 
The branching fraction of \bquark-hadron decays into \chictwo is measured for the first time 
with the \Bz, \Bp, \Bs and \bquark-baryons admixture.  
The result is consistent with the average, corresponding 
to the \Bz, \Bp admixture, from Ref.~\cite{PDG2016}, given large PDG uncertainty. 
The obtained value has higher precision than
the results from \cleo~\cite{Chen:2000ri}, \belle~\cite{Abe:2002wp} and \babar~\cite{Aubert:2002hc}, 
is close to the \cleo result of $(0.67 \pm 0.34 \pm 0.03) \times 10^{-3}$ 
and is different by more than $2 \sigma$ 
from the results of \belle, $(1.80 ^{+ 0.23} _{- 0.28} \pm 0.26) \times 10^{-3}$, 
and \babar, $(2.10 \pm 0.45 \pm 0.31) \times 10^{-3}$. 

Transverse momentum \pt dependence of charmonia production in \bquark-decays
is studied for the $\etac (1S)$ and $\chi_c$ states in the \lhcb acceptance 
and for $\pt > 4 \gevc$. 
Precision of about 15\% for $\etac (1S)$ and between 20\% and 30\% for 
the $\chi_c$ states is achieved. 
The NLO calculations of the \pt dependence of the \etac and $\chi_c$ production in \bquark-decays 
are important to translate the obtained results to the conclusions on the production mechanisms. 

Production of the $\etac (2S)$ state in \bquark-decays was determined to be  
\begin{align*}
\BR ( b \to \etac (2S) X ) \times \BR ( \etac (2S) \to \phi \phi )
 &= ( 6.34 \pm 1.81 \pm 0.57 \pm 1.89 ) \times 10^{-7} \ .
\end{align*}
This is the first indication of the $\etac (2S)$ production in \bquark-decays, 
as well as the decay of $\etac (2S)$ meson to the $\phi \phi$ pair. 

These are the first $\chi_c$ and $\etac (2S)$ inclusive production measurements, using charmonia decays 
to hadronic final state, in the high-multiplicity environment of a hadron machine. 

In addition upper limits at $95 \%$ ( $90 \%$) CL on
the production of the $X(3872)$, $X(3915)$, and $\chictwo (3930)$, states 
in \bquark-decays are obtained as: 
\begin{align*}
\BR ( \bquark \ra X(3872) X ) \times \BR ( X(3872) \ra \phi \phi ) & < 4.5 (3.9) \times 10^{-7} , \\
\BR ( \bquark \ra X(3915) X ) \times \BR ( X(3915) \ra \phi \phi ) &  < 3.1 (2.7) \times 10^{-7} , \\
\BR ( \bquark \ra \chictwo (3930) X ) \times \BR ( \chictwo (3930) \ra \phi \phi ) &  < 2.8 (2.3)  \times 10^{-7} .
\end{align*}

One can qualitatively estimate an upper limit on the branching fraction of the $X(3872)$ decay to $\phi\phi$. Since the $X(3872)$ has likely $\chicone(2P)$ charmonium component, the radial suppresion of \bquark-hadron decays should work similarly for different charmonium states. In other words:
\begin{equation}
\frac{\BR(\bquark \to X(3872) X)}{\BR(\bquark \to \chicone X)} \sim \frac{\BR(\bquark \to \psitwos X)}{\BR(\bquark \to \jpsi X)}.
\label{eq:reas1}
\end{equation}
The right part of this equation can be extracted from Ref.~\cite{PDG2018} to be
\begin{equation}
\frac{\BR(\bquark \to \psitwos X)}{\BR(\bquark \to \jpsi X)} = 0.25 \pm 0.03.
\label{eq:reas2}
\end{equation}
Then using Eqs.\ref{eq:reas1}~\ref{eq:reas2} and~\ref{eq:res1} and the branching fraction of the $\chicone\to\phi\phi$ decay, one can estimate at a qualitative level that
\begin{align}
\frac{\BR ( X(3872) \ra \phi \phi )}{\BR ( \chicone \ra \phi \phi )} & < 1.6, \\
\BR ( X(3872) \ra \phi \phi ) < 7.5\times 10^{-4}.
\end{align}

\begin{singlespace}
\chapter{Phenomenological analysis of charmonium production}
\label{ch:pheno}
\end{singlespace}
\vspace{-0.7cm}
The first measurement of the $\etac(1S)$ prompt production and production in \bquark-hadron inclusive decays performed at \lhcb~\cite{ppbar} triggered an intention to simultaneously use all available experimental information on $S$-wave charmonium production to constrain involved LDMEs.

The \chicone and \chictwo inclusive production in \bquark-hadron decays were measured at B-factories and \lep. 
The results discussed in Chapter~\ref{ch:phiphi} provide the first measurement of branching fraction of \chiczero inclusive production in 
\bquark-hadron decays.
The relative \chicone-to-\chiczero, \chictwo-to-\chiczero and \chictwo-to-\chicone production rates have also been reported.
These measurements provide a powerful test of theoretical predictions. 
While correlated experimental and theoretical uncertainties cancel in ratios, the implication of absolute branching fractions provides a more extensive number of independent measurements, i.e. larger number of constraints. 

In this chapter the \lhcb results on the absolute branching fractions $\BR(\bquark \to \etac(1S) X)$ and $\BR(\bquark \to \chic_J X)$ and their ratios are compared to
predictions from Ref.~\cite{Beneke:1998ks} using a fit technique to quantify the agreement. 
For \jpsi and \etac mesons, a simultaneous study of prompt and \bquark-decays production is performed thanks to the theory prediction provided by H.-S. Shao~\cite{ShaoPriv}. For the analysis presented in this chapter, I would like to acknowledge fruitful discussions with E. Kou, J.-P. Lansberg and H.-S. Shao.

This chapter is organised as follows. Experimental input
 of charmonium production in \bquark-hadron decays is presented in Section~\ref{sec:ccsection}. The NRQCD prediction for inclusive charmonium production in \bquark-hadron decays used for further fits to theory is described in Section~\ref{sec:ccprod}. A comparison of $S$-wave charmonium hadroproduction and production in \bquark-hadron decays to theory together with simultaneous fit to both observables is shown in Section~\ref{sec:Swave}. 
In Section~\ref{sec:Pwave} theory predictions are compared to the measurements of $P$-wave charmonium production in \bquark-hadron decays presented in Chapter~\ref{ch:phiphi}. Finally the results are summarised in Section~\ref{sec:ThFitSummary}.
\clearpage
\section{Experimental input, feed-down subtraction}
\label{sec:ccsection}
While the \jpsi state production in \bquark-hadron inclusive decays is well measured, 
$\BR = (1.094 \pm 0.032) \%$, 
the only measurement of the $\etac(1S)$ production in \bquark-hadron inclusive decays has been performed by the \lhcb 
experiment~\cite{ppbar}. 
LHCb measured the relative $\etac(1S)$ to \jpsi production in \bquark-hadron inclusive decays~\cite{ppbar} to be 
\begin{align}
\frac{\BR ( b \to \etac(1S) X )}{\BR ( b \to \jpsi X )}  &= 0.424 \pm 0.055 \pm 0.021 \pm 0.045_{\BR}, \label{etacRel}
\end{align}
where the first uncertainty is statistical, the second is systematic, the third uncertainty is due to those on the branching fractions $\BR(\etac(1S)\to p\overline{p})$ and $\BR(\jpsi\to p\overline{p})$~\cite{PDG2016}. 
Measurements of the \chicone and \chictwo inclusive production in \bquark-hadron decays from the \epem experiments, where only light \bquark-hadrons - 
\Bp and \Bz mesons - are produced, 
were performed at \cleo~\cite{CLEO_chic1,CLEO_chic2}, \belle~\cite{BELLE_chic1,Bhardwaj:2015rju} and \babar~\cite{BABAR_chic1}. 
The world average values for branching fractions of light $B$-meson inclusive decays to the charmonium states of interest~\cite{PDG2016} are given in Table~\ref{tabBfactories}.
\begin{table}[h]
\begin{center}
\caption
[Branching fractions of $B$-meson inclusive decays to charmonium states.]
{Branching fractions of $B$-meson inclusive decays to charmonium states~\cite{PDG2016}.}
\label{tabBfactories}
\begin{tabular}{ c | c | c | c |  c | c}
 	& \etac(1S) 	    & \jpsi  		      & \chiczero 	& \chicone         & \chictwo	 	      \\ \hline
$\BR(\B \to \ccbar X) , \ \times10^{-3}$  & $<9$   & $10.94\pm0.32$  & -  		& $3.55\pm0.27$  & $1.00\pm0.17$  
\end{tabular}
\end{center}
\end{table}

The inclusive production of the \chicone and \chictwo states in \bquark-decays involving all \bquark-hadrons (\Bp, \Bz, \Bs, \Bcp and \bquark-baryons) has been studied at \lthree~\cite{L3_chic1} and \delphi~\cite{DELPHI_chic1} experiments. 
Recently, \lhcb reported the most precise \chicone and \chictwo and the first \chiczero production measurements in \bquark-hadron inclusive decays~\cite{phiphi} as well as the corresponding \chicone-to-\chiczero 
and \chictwo-to-\chiczero production ratios.
LHCb measured the branching fractions of \bquark-hadron inclusive decays into \chic states to be 
\begin{align}
\BR ( b \to \chiczero X ) &= ( 3.02 \pm 0.47 \pm 0.23 \pm 0.94_{\BR} ) \times 10^{-3} , \label{chic0Abs} \\ 
\BR ( b \to \chicone X )  &= ( 2.76 \pm 0.59 \pm 0.23 \pm 0.89_{\BR} ) \times 10^{-3} , \label{chic1Abs} \\ 
\BR ( b \to \chictwo X )  &= ( 1.15 \pm 0.20 \pm 0.07 \pm 0.36_{\BR} ) \times 10^{-3} , \label{chic2Abs}
\end{align}
where the third uncertainty is due to the uncertainties on the branching fractions of the \bquark-hadron inclusive decays 
to the $\etac (1S)$ meson, 
$\BR ( \bquark \to \etac (1S) X )$~\cite{ppbar}, and of the $\etac (1S)$ and \chic decays to a pair of $\phi$ mesons~\cite{PDG2016}. 
The relative  branching fractions are determined to be 
\begin{align}
\frac{\BR ( b \to \chicone X )}{\BR ( b \to \chiczero X )}  &= 0.92 \pm 0.20 \pm 0.02 \pm 0.14_{\BR} , \label{chic1Rel} \\ 
\frac{\BR ( b \to \chictwo X )}{\BR ( b \to \chiczero X )}  &= 0.38 \pm 0.07 \pm 0.01 \pm 0.05_{\BR} , \label{chic2Rel}
\end{align}
where the third uncertainty is due to the uncertainties on the branching fractions $\BR ( \chic \to \phi \phi )$~\cite{PDG2016}. 
The results for the \chicone and \chictwo production in \bquark-hadron inclusive decays are close to those in \Bz and \Bp inclusive decays~\cite{PDG2016}. 

The mixture of \bquark-hadrons in the \lhcb measurements consists of about 76\% of light $B$-mesons, 10\% of \Bs 
and 14\% of \Lb~\cite{fs,fLb}, while other contributions are considered to be negligible. 
The branching fraction $\BR ( \Lb \to (\ccbar) X )$ is assumed to be small compared to  $\BR ( B \to (\ccbar) X )$, while $\BR ( \Bs \to (\ccbar) X )$ is assumed to be of the same value as $\BR ( B \to (\ccbar) X )$. Hence, no significant difference in the ratio of branching fractions of inclusive decays to \chic states is expected between all \bquark-hadron and light $B$-meson systems.

The dominant feed-down contributions to \jpsi production originate from the 
$\psitwos \to \jpsi X$, $\chicone \to \jpsi \gamma$ and $\chictwo \to \jpsi \gamma$ transitions. The feed-down contributions to the $\etac(1S)$ yield 
originate from the $h_c$ and \chic decays.
The feed-down to the $\etac(1S)$ sample is expected to be small and is not taken into account, 
so that it is assumed that $\BR ( b \to \etac(1S) X ) = \BR ( b \to \etac(1S)^{direct} X ) $.
The feed-down subtracted $\etac(1S)$ to \jpsi relative production in \bquark-inclusive decays is obtained in the following way
\begin{align}
\frac{\BR ( b \to \etac(1S)^{direct} X )}{\BR ( b \to \jpsi^{direct} X )} &=\frac{\BR ( b \to \etac(1S) X )}{\BR ( b \to \jpsi X )} \times \\ \nonumber
\times \Big[1&-\frac{\BR ( b \to \chicone X )}{\BR ( b \to \etac(1S) X )}\frac{\BR ( b \to \etac(1S) X )}{\BR ( b \to \jpsi X )}\BR ( \chicone \to \jpsi \gamma) \\ \nonumber
&-\frac{\BR ( b \to \chictwo X )}{\BR ( b \to \etac(1S) X )}\frac{\BR ( b \to \etac(1S) X )}{\BR ( b \to \jpsi X )}\BR ( \chictwo \to \jpsi \gamma) \\ \nonumber
&-\frac{\BR ( b \to \psitwos X )}{\BR ( b \to \jpsi X )}\BR ( \psitwos \to \jpsi X) \Big]^{-1}.
\end{align}

Using the measurements (\ref{etacRel}), (\ref{chic0Abs}), (\ref{chic1Abs}) and (\ref{chic2Abs}) 
and the values of $\BR ( b \to \psitwos X )$, $\BR (\psitwos \to  \jpsi X)$ and $\BR (\chic \to  \jpsi \gamma)$ 
from Ref.~\cite{PDG2016}, the branching fraction 
of the direct $\etac(1S)$ production in \bquark-hadron decays relative to that of the \jpsi meson, is calculated to be
\begin{align}
\frac{\BR ( b \to \etac(1S)^{direct} X )}{\BR ( b \to \jpsi^{direct} X )}  & = 0.691 \pm 0.090 \pm 0.024 \pm 0.103 , \label{b2EtacDirect}
\end{align}
where the third uncertainty is due to that on the branching fractions involved in the calculation.

The dominant feed-down contribution to the \chic yield is from the $\psitwos \to \chic \gamma$ transition and is measured 
to be around 10\% for each of the  \chiczero, \chicone and \chictwo states~\cite{PDG2016}. 
The feed-down contributions to the \chic production is subtracted in the following way
\begin{align*}
\BR ( b \to \chiczero^{direct} X ) &= \BR ( b \to \chiczero X ) - \BR ( b \to \psitwos X )\cdot \BR (\psitwos \to  \chiczero \gamma), \\
\BR ( b \to \chicone^{direct} X )  &= \BR ( b \to \chicone X ) - \BR ( b \to \psitwos X )\cdot \BR (\psitwos \to  \chicone \gamma) , \\
\BR ( b \to \chictwo^{direct} X )  &= \BR ( b \to \chictwo X ) - \BR ( b \to \psitwos X )\cdot \BR (\psitwos \to  \chictwo \gamma) ,
\end{align*}
\begin{align*}
\frac{\BR ( b \to \chicone^{direct} X )}{\BR ( b \to \chiczero^{direct} X )}  &= \frac{\BR ( b \to \chicone X )}{\BR ( b \to \chiczero X )}
\times\frac{1-\frac{\BR ( b \to \psitwos X )\cdot \BR (\psitwos \to  \chicone \gamma)}{\BR ( b \to \chicone X )}}{1-\frac{\BR ( b \to \psitwos X )\cdot \BR (\psitwos \to  \chiczero \gamma)}{\BR ( b \to \chicone X )}}, \\
\frac{\BR ( b \to \chictwo^{direct} X )}{\BR ( b \to \chiczero^{direct} X )}  &= \frac{\BR ( b \to \chicone X )}{\BR ( b \to \chiczero X )}
\times\frac{1-\frac{\BR ( b \to \psitwos X )\cdot \BR (\psitwos \to  \chictwo \gamma)}{\BR ( b \to \chicone X )}}{1-\frac{\BR ( b \to \psitwos X )\cdot \BR (\psitwos \to  \chiczero \gamma)}{\BR ( b \to \chicone X )}}.
\end{align*}

Using measurements (\ref{chic0Abs}), (\ref{chic1Abs}), (\ref{chic2Abs}), (\ref{chic1Rel}) and (\ref{chic2Rel}) and the values of 
$\BR ( b \to \psitwos X )$ and $BR (\psitwos \to  \chic \gamma)$ from Ref.~\cite{PDG2016}, the direct \chic production rates 
in \bquark-hadron decays are calculated to be
\begin{align}
\BR ( b \to \chiczero^{direct} X ) &= ( 2.74 \pm 0.47 \pm 0.23 \pm 0.94_{\BR} ) \times 10^{-3}, \label{chic0AbsDirect} \\ 
\BR ( b \to \chicone^{direct} X )  &= ( 2.49 \pm 0.59 \pm 0.23 \pm 0.89_{\BR} ) \times 10^{-3}, \label{chic1AbsDirect} \\ 
\BR ( b \to \chictwo^{direct} X )  &= ( 0.89 \pm 0.20 \pm 0.07 \pm 0.36_{\BR} ) \times 10^{-3}, \label{chic2AbsDirect}
\end{align}
where the third uncertainty is due to the uncertainties on the branching fractions of the \bquark-hadron decays to the $\etac (1S)$ meson, $\BR ( \bquark \to \etac (1S) X )$~\cite{ppbar}, $\etac (1S)$ and \chic decays to $\phi\phi$, $\BR ( \etac \to \phi\phi)$ and $\BR ( \chic_J \to \phi\phi)$~\cite{PDG2016} 
and due to the feed-down contribution uncertainties. 

The relative direct \chic production rates in \bquark-hadron decays are calculated to be
\begin{align}
\frac{\BR ( b \to \chicone^{direct} X )}{\BR ( b \to \chiczero^{direct} X )}  &= 0.91\pm0.20\pm0.02\pm0.15_{\BR}, \label{chic1RelDirect} \\ 
\frac{\BR ( b \to \chictwo^{direct} X )}{\BR ( b \to \chiczero^{direct} X )}  &= 0.34\pm0.06\pm0.01\pm0.05_{\BR}, \label{chic2RelDirect}
\end{align}
where the third uncertainty is due to those on the branching fractions $\BR ( \chic_J \to \phi \phi )$~\cite{PDG2016} and due to the uncertaities
of the branching fractions of the decays contributing to the feed-down. 
Correlations between the uncertainties of the values of $\BR ( b \to \chiczero X )$, $\BR ( b \to \chicone X )$ and $\BR ( b \to \chictwo X )$ 
are not taken into account in the feed-down contribution uncertainty estimation, because the correlation effect is small compared to other uncertainties.

Finally, the relative branching fractions of the $\bquark \to \chic X$ inclusive decays to the measured relative 
branching fractions of exclusive $B$-meson decays to the \chic states are compared. 
A selection of the measured exclusive branching fractions from Ref.~\cite{PDG2016} is listed in Table~\ref{tabExcl}. 
All these branching fractions show suppression of the decays to the \chictwo state compared to the decays to the \chicone and \chiczero states. 
The branching fractions of the exclusive \bquark-hadron decays to the \chiczero state are smaller than the branching fractions of decays to the \chicone state.
The values of the branching fractions $\BR(B\to\chic K)$ are similar to those of $\BR(B\to\chic \Kstar)$. 
\begin{table}[h]
\begin{center}
\caption
[Branching fractions of exclusive $B$-meson decays to \chic states.]
{Branching fractions of exclusive $B$-meson decays to \chic states~\cite{PDG2016}.}
\label{tabExcl}
\begin{tabular}{ c | c | c | c  }
 				    		 & \chiczero 				    	   & \chicone                   		     & \chictwo	 	     		     \\ \hline
$\BR(\Bp\to\chic\Kp)$  	 & $(1.50\pm0.15)\times10^{-4}$  & $(4.79\pm0.23)\times10^{-4}$  & $(1.1\pm0.4)\times10^{-5}$  \\ \hline
$\BR(\Bd\to\chic\Kz)$ 	& $(1.47\pm0.27)\times10^{-4}$  & $(3.93\pm0.27)\times10^{-4}$  & $< 1.5\times10^{-5}$ 	      \\ \hline
$\BR(\Bp\to\chic\Kstarp)$   & $< 2.1\times10^{-4}$		   & $(3.0\pm0.6)\times10^{-4}$      & $< 1.52\times10^{-4}$  	      \\ \hline
$\BR(\Bd\to\chic\Kstarz)$   & $(1.7\pm0.4)\times10^{-4}$       & $(2.39\pm0.19)\times10^{-4}$   &$(4.9\pm1.2)\times10^{-5}$   \\ \hline
$\BR(\Bp\to\chic\pip)$  	& $< 1\times10^{-7}$  	    		   & $(2.2\pm0.5)\times10^{-5}$	      & $< 1\times10^{-7}$			 \\ \hline
$\BR(\Bd\to\chic\piz)$  	& $-$					   	   & $(1.12\pm0.28)\times10^{-5}$   & $-$						 \\ \hline
$\BR(\Bd\to\chic\Km\pip)$  & $-$					    	   & $(3.8\pm0.4)\times10^{-4}$       & $-$						 \\ \hline
$\BR(\Bs\to\chic\phiz)$  	& $-$					          & $(2.03\pm0.29)\times10^{-4}$   & $-$						 \\ 
\end{tabular}
\end{center}
\end{table}
\vspace{0.4cm}

\clearpage
\section{Predictions for charmonium yields in $B$-meson decays}
\label{sec:ccprod}
A theoretical description of the inclusive \bquark-hadron decays to S-wave and in particular to P-wave charmonium states is challenging. 
Despite the fact that many problems have been recognised, no clear solutions have been identified yet. 
More theoretical efforts are certainly called for. 

Authors of Ref.~\cite{Beneke:1998ks} consider two mechanisms - CS and CO - of charmonia production in $B$-meson decays. 
A negative NLO correction has been pointed out for the CS contribution, which makes it difficult to deliver a precise 
theoretical prediction.  

For S-wave charmonium the four Fock states are expected to be dominating, namely $O_1^{\jpsi}(^3S_1)$, $O_8^{\jpsi}(^3S_1)$, $O_8^{\jpsi}(^1S_0)$ and $O_8^{\jpsi}(^3P_0)$ for \jpsi and $O_1^{\etac}(^1S_0)$, $O_8^{\etac}(^1S_0)$, $O_8^{\etac}(^3S_1)$ and $O_8^{\etac}(^1P_1)$ for $\etac(1S)$.
According to the proposed formalism, the \jpsi and the $\etac(1S)$ production rates in inclusive \bquark-decays are expressed as 
\footnote{Since the present discussion is qualitative, 
I simply quote the central values ignoring the theoretical uncertainties related to the charm mass, 
renormalisation scale etc.}:
\begin{align}
\BR ( B \to \jpsi X ) =  0.754\cdot10^{-3} \langle O_1^{\jpsi}(^3S_1) \rangle + 0.195 \langle O_8^{\jpsi}(^3S_1) \rangle + \nonumber \\
					0.342 \Big[ \langle O_8^{\jpsi}(^1S_0) \rangle +   \frac{3.10}{m_c^2}\langle O_8^{\jpsi}(^3P_0) \rangle \Big], \\
\BR ( B \to \etac(1S) X ) =  2.500\cdot10^{-3} \langle O_1^{\etac}(^1S_0) \rangle + 0.342 \langle O_8^{\etac}(^1S_0) \rangle + \nonumber \\
					0.195 \Big[ \langle O_8^{\etac}(^3S_1) \rangle -   \frac{0.240}{m_c^2}\langle O_8^{\etac}(^1P_1) \rangle \Big] , \label{etacPred}
\end{align}
where $m_c$ is the mass of the \cquark-quark.
In Eq.~\ref{etacPred} the coefficient of the colour  singlet contribution suffers from large theoretical uncertainties. 
However, in this discussion we quote the so-called {\it improved} value, which is the NLO calculation with one term from NNLO level (see Ref.~\cite{Beneke:1998ks} for details).

Within the same formalism, the two Fock states, $O_1^{\chic_J}(^3P_J)$ and $O_8^{\chic_J}(^3S_1)$, are expected to be dominating in the description of the $\B \to \chic_J X$ decays as discussed in Chapter~\ref{ch:theory}.
The branching fractions of the \chic production in inclusive $B$-meson decays are then expressed as:
\begin{align}
\BR ( B \to \chiczero X ) &=  \frac{-0.0148}{m_c^2} \langle O_1^{\chiczero}(^3P_0) \rangle + 0.195 \langle O_8^{\chiczero}(^3S_1) \rangle , \label{b2chic0TheoryNSR} \\ 
\BR ( B \to \chicone X ) &=  \frac{-0.00783}{m_c^2} \langle O_1^{\chicone}(^3P_1) \rangle + 0.195 \langle O_8^{\chicone}(^3S_1) \rangle , \label{b2chic1TheoryNSR} \\ 
\BR ( B \to \chictwo X ) &=  \frac{-0.0120}{m_c^2} \langle O_1^{\chictwo}(^3P_2) \rangle + 0.195 \langle O_8^{\chictwo}(^3S_1) \rangle . \label{b2chic2TheoryNSR}
\end{align}
Here again I quote only {\it improved} values for the singlet contribution. 

The LDMEs are linked by the spin relations. For the \jpsi and $\etac(1S)$ meson production, this gives:
\begin{align*}
\langle O_1^{\etac}(^1S_0) \rangle =  \frac{1}{3} \langle O_1^{\jpsi}(^3S_1) \rangle, \\
\langle O_8^{\etac}(^1S_0) \rangle =  \frac{1}{3} \langle O_8^{\jpsi}(^3S_1) \rangle, \\
\langle O_8^{\etac}(^3S_1) \rangle =  \langle O_8^{\jpsi}(^1S_0) \rangle, \\
\langle O_8^{\etac}(^1P_1) \rangle =  3 \langle O_8^{\jpsi}(^3P_0) \rangle.
\end{align*}
Hence, both $\BR ( B \to \etac(1S) X )$ and $\BR ( B \to \jpsi X )$ can be expressed as a function of only four LDMEs:
\begin{align}
\BR ( B \to \jpsi X ) =  7.54\cdot10^{-4} \langle O_1^{\jpsi}(^3S_1) \rangle + 0.195 \langle O_8^{\jpsi}(^3S_1) \rangle + \nonumber \\ \label{b2jpsiTheory}
					       0.342 \Big[ \langle O_8^{\jpsi}(^1S_0) \rangle +   \frac{3.10}{m_c^2}\langle O_8^{\jpsi}(^3P_0) \rangle \Big],\\ 
\BR ( B \to \etac(1S) X ) =  8.33\cdot10^{-4} \langle O_1^{\jpsi}(^3S_1) \rangle + 0.114 \langle O_8^{\jpsi}(^3S_1) \rangle + \nonumber \\ \label{b2etacTheory}
					       0.195 \Big[ \langle O_8^{\jpsi}(^1S_0) \rangle -   \frac{0.720}{m_c^2}\langle O_8^{\jpsi}(^3P_0) \rangle \Big].
\end{align}

The spin relations for the \chic states production yield 
\begin{align*}
O_1 \equiv \langle O_1^{\chiczero}(^3P_0)\rangle /m_c^2,\\
O_8 \equiv \langle O_8^{\chiczero}(^3S_1) \rangle, \\
\langle O_1^{\chic_J}(^3P_J) \rangle/m_c^2 =  (2J+1) O_1, \\
\langle O_8^{\chic_J}(^3S_1) \rangle =  (2J+1) O_8.
\end{align*}
Thus, three branching fractions, $\BR ( B \to \chiczero X ), \BR ( B \to \chicone X )$ and $\BR ( B \to \chictwo X )$, are expressed as a function of only two LDMEs:
\begin{align}
\BR ( B \to \chiczero X ) =  -0.0148\ O_1 + 0.195\ O_8, \label{b2chic0Theory} \\ 
\BR ( B \to \chicone X )  =  -0.0234\ O_1 + 0.585\ O_8, \label{b2chic1Theory} \\
\BR ( B \to \chictwo X )  =  -0.0600\ O_1 + 0.975\ O_8. \label{b2chic2Theory}
\end{align}
Therefore, a measurement of the three $\BR ( B \to \chic_J X )$ values would in principle overconstrain the model and provide a crucial consistency check.

The problem of the description of the $P$-wave states production has already been pointed out in the same paper~\cite{Beneke:1998ks}. 
If the $O_1$ value is computed using the potential model~\cite{Bushm}, 
\begin{align*}
O_1 & = 4.8 \times 10^{-2} \gev^3, 
\end{align*}
and the $O_8$ value is adjusted to reproduce the \chictwo meson production rate measured at the \cleo experiment~\cite{CLEO_chic2}, 
\begin{align*}
O_8 & = 4.5 - 6.5 \times 10^{-3} \gev^3, 
\end{align*}
the \chicone state production rate is predicted to be in the range
\begin{align*}
\BR ( \bquark \to \chicone X ) & = (0.15 - 0.27) \%, 
\end{align*}
 which is below the value measured by CLEO even after taking into account large uncertainties. 

In addition, authors of Ref.~\cite{ind} extracted $O_8$ from the simultaneous fit of the \chic hadroproduction measurements at 
\cms~\cite{Chatrchyan:2012ub}, 
\lhcb~\cite{Aaij:2013dja}, 
\atlas~\cite{ATLAS:2014ala} 
and \cdf~\cite{Abulencia:2007bra} to be
\begin{align*}
O_8 & = (11.12 \pm 0.68) \times 10^{-3} \gev^3,
\end{align*}
which exceeds the value tuned using $\BR(B \to \chictwo X)$ from Ref.~\cite{CLEO_chic2}.

The solution to the problem of $P$-wave production seems to be a combination of different contributions. 
So far it remains an unsolved puzzle in the description of charmonium production. 
In the next section, I perform the fit of LDMEs using the LHCb results (Eqs.~\ref{chic0AbsDirect}-\ref{chic2RelDirect}) 
as an attempt to pin down the origin of the problem, while in the remaining part of this section, 
I qualitatively discuss this puzzle exploiting Eqs.~\ref{b2chic0TheoryNSR}-\ref{b2chic2TheoryNSR}. 

First of all, at the LO of the singlet model, only the \chicone state can be produced since the production of the \chiczero and \chictwo states
is allowed only via non-factorisable contribution within  the $V-A$ theory. 
As in the case of the $J/\psi$ meson production, the NLO corrections to the singlet contribution are negative, which induces a large theoretical uncertainty. 
Nevertheless, we can see from Eq.~\ref{b2chic1TheoryNSR} that the NLO singlet contribution partially cancels 
the LO singlet contribution of the \chicone meson, which makes its branching ratio much too small. 

There are enough evidences that the colour octet contributions are necessary to explain the observed charmonium production in $B$ decays (see e.g.~\cite{Beneke:1999gq}). The octet contributions to the $B\to \chi_{cJ} X$ decays are the same for $J=0,1,2$ as shown in Eqs.~\ref{b2chic0TheoryNSR}-\ref{b2chic2TheoryNSR}. Thus, together with the spin relations, the octet contributions follow a simple ratio $\chiczero:\chicone:\chictwo = 1:3:5$ (see Eqs.~\ref{b2chic0Theory}-\ref{b2chic2Theory}). 
For the \chiczero and \chictwo states, as the singlet contributions also have similar coefficients, 
they follow approximately the same ratio, $\chiczero : \chictwo \sim 1 : 5$. 
Therefore, the total branching fractions would also follow such a ratio, 
while the LHCb results in Eq.~\ref{chic2Rel} suggest rather opposite, 
$\BR(\bquark \to \chiczero X) > \BR(\bquark \to \chictwo X)$.
This is a new discrepancy between experimental results and NLO calculations. 

Interestingly, for the exclusive decays $B\to \chi_{cJ}K^{(*)}$, 
we find a similar enhancement (suppression) of \chiczero (\chictwo) (c.f. Table 1). 
Possible solution has been pointed out in Ref.~\cite{Beneke_B2chicK}. 
We may resort to exclusive channels for finding the solution to the inclusive puzzle. 
However, from a comparison to the inclusive branching ratios, many more channels than $B\to \chi_{cJ}K^{(*)}$ are needed 
to fill the inclusive branching ratio, which may dilute the ratio seen in these observed channels. 
In any case, the explanation of the difference between the branching fractions of the $B\to \chi_{c0}K^{(*)}$ and $B\to \chi_{c2}K^{(*)}$ 
channels is rather complicated and it might occur that the solution to the inclusive channel puzzle 
comes from several contributions. 
\vspace{0.5cm}


\clearpage
\section{Comparison of \etac and \jpsi production to theory}
\label{sec:Swave}
\subsection{Production in \bquark-hadron decays}
The values of LDMEs for the \jpsi and $\etac(1S)$ production extracted from the fits of prompt $\jpsi$ and $\psi(2S)$ production and polarization
measurements~\cite{CDFpsi2s,Aaij:2011jh,Chatrchyan:2013cla,Chatrchyan:2011kc,Aad:2011sp,Abulencia:2007us,Affolder:2000nn} to
theoretical predictions~\cite{Butenschoen:2012px,Chao:2012iv,Gong:2012ug,Bodwin:2014gia} are summarised in Table~\ref{tableLDMEs}.
The values of the relative $\etac(1S)$ production in \bquark-hadron inclusive decays derived from these predictions (Eqs.~\ref{b2jpsiTheory} and~\ref{b2etacTheory})
and the values from Table~\ref{tableLDMEs} are shown in Table~\ref{tableEtacPredictions}.
\begin{table}[h]
\begin{center}
\caption
[Values of LDMEs calculated from the \jpsi prompt production measurements used to predict $\etac(1S)$ production in \bquark-hadron inclusive decays.]
{Values of LDMEs calculated in Refs.~\cite{Butenschoen:2012px,Chao:2012iv,Gong:2012ug,Bodwin:2014gia} from the \jpsi prompt production measurements 
used to predict $\etac(1S)$ production in \bquark-hadron inclusive decays.}
\label{tableLDMEs}
\begin{tabular}{ c | c | c | c | c }
  & M. Butenschoen,   					 & K.-T. Chao         & B. Gong 	 		 & G. T. Bodwin      \\	   
  & B. A. Kniehl~\cite{Butenschoen:2012px}   & et al.~\cite{Han:2014jya} & et al.~\cite{Gong:2012ug}	 & et al.~\cite{Bodwin:2014gia}   \\ \hline
  $\langle O_1^{\jpsi}(^3S_1) \rangle$ 		       & 1.32                           & 1.16 			   & 1.16 		  	   & - 			    \\ \hline
  $\langle O_8^{\jpsi}(^3S_1) \rangle$ 		       & $0.0017\pm0.0005$      & $0.0030\pm0.0012$ & $-0.0046\pm0.0013$ & $0.011\pm0.010$ \\ \hline
  $\langle O_8^{\jpsi}(^1S_0) \rangle$ 		       & $0.0304\pm0.0035$      & $0.0890\pm0.0098$ &  $0.097\pm0.009$     & $0.099\pm0.022$ \\ \hline
  $\langle O_8^{\jpsi}(^3P_0) \rangle / m_c^2 $     & $-0.0040\pm0.0007$       & $0.0056\pm0.0021$ & $-0.0095\pm0.0025$ & $0.0049\pm0.0044$ 
\end{tabular}
\end{center}
\end{table}
\begin{table}[h]
\begin{center}
\caption
[Predictions of $\BR(B \to \etac(1S) X)$ and the \lhcb measurement.]
{Predictions of $\BR(B \to \etac(1S) X)$ using LDMEs from Refs.~\cite{Butenschoen:2012px,Chao:2012iv,Gong:2012ug,Bodwin:2014gia} and the \lhcb measurement~\cite{ppbar}.}
\resizebox{\textwidth}{!}{%
\label{tableEtacPredictions}
\begin{tabular}{ c | c | c | c | c | c}
  &\lhcb                     & M. Butenschoen,   		 & K.-T. Chao         & B. Gong 	         & G. T. Bodwin      \\	   
  &\cite{ppbar}              & B. A. Kniehl~\cite{Butenschoen:2012px}   & et al.~\cite{Han:2014jya} & et al.~\cite{Gong:2012ug}	 & et al.~\cite{Bodwin:2014gia}   \\ \hline
 $\BR(B \to \etac(1S) X)$ & $0.69\pm0.14$ & $1.04\pm1.34$ & $0.47\pm0.07$        &  $0.90\pm0.67$             & $0.48\pm0.07$
\end{tabular}}
\end{center}
\end{table}
Theoretical predictions for $\BR(\bquark \to \etac(1S) X)$ are in general agreement with the measurement. 
The values of LDMEs from Refs.~\cite{Han:2014jya} and~\cite{Bodwin:2014gia} provide more precise prediction for the 
$\etac(1S)$ production in \bquark-hadron decays because $\langle O_8^{\jpsi}(^3P_0) \rangle$ is positive.

Using expressions (\ref{b2jpsiTheory}) and (\ref{b2etacTheory}) for $\BR ( B \to \jpsi X)$ and $\BR ( B \to \etac(1S) X)$ and the measurement (\ref{b2EtacDirect}), a fit is performed to determine the allowed regions for LDMEs. 

The reliable value $\langle O_1^{\jpsi}(^3S_1) \rangle = 1.16 \gev^3$, originally coming from Buschmuller-Tye potential model~\cite{Buchmuller:1980su}, is fixed in the fits. 
The values for other LDMEs are fixed one after another to perform a fit on the plane of two remaining LDMEs as shown in Table~\ref{tablefits2Etac}.

\clearpage 
The LDMEs from the following calculations are displayed on all following plots for comparison:
\begin{itemize}
\item NRQCD fit to \jpsi production with a constraint from \etac production~\cite{Han:2014jya},
\item Simulnateous fit to hadroproduction and photoproduction~\cite{Butenschoen:2011yh},
\item Simultaneous \kt-factorization fit to \jpsi and \etac prompt production~\cite{Baranov:2019joi}.
\end{itemize}

Using expressions (\ref{b2jpsiTheory}) and (\ref{b2etacTheory}) for $\BR(B \to \jpsi X)$ and $\BR(B \to \etac(1S) X)$, the measurements (\ref{b2EtacDirect}) and $\BR(B \to \jpsi^{direct} X)$~\cite{PDG2016} are fitted simultaneously to theory in terms of LDME parameters.

\begin{figure}[t]
\centering
\subfigure[]
{\protect\protect\includegraphics[width=0.5\linewidth]{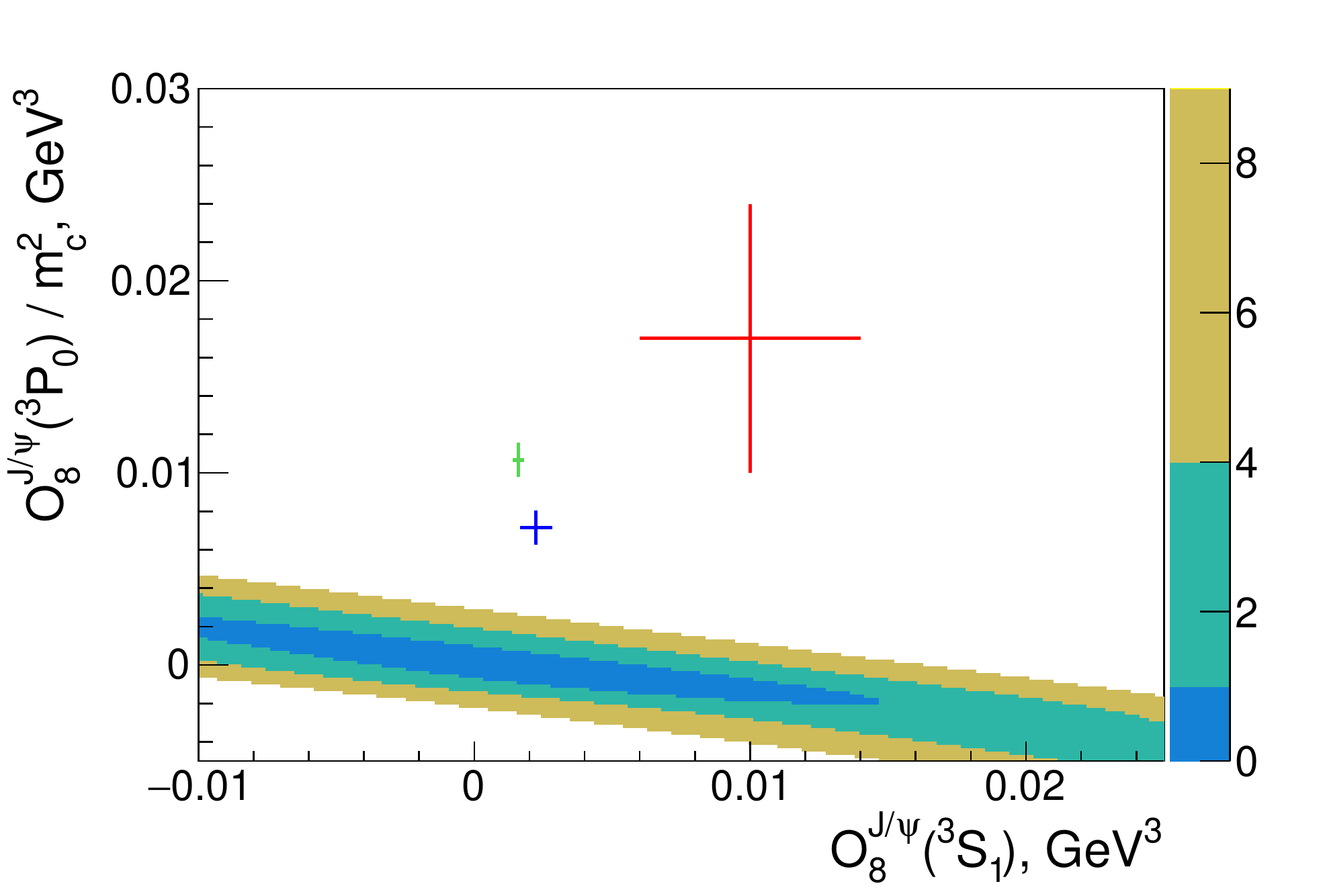}
\label{fig:etacJpsiOnly_1S0_fixed}}
\subfigure[]
{\protect\protect\includegraphics[width=0.5\linewidth]{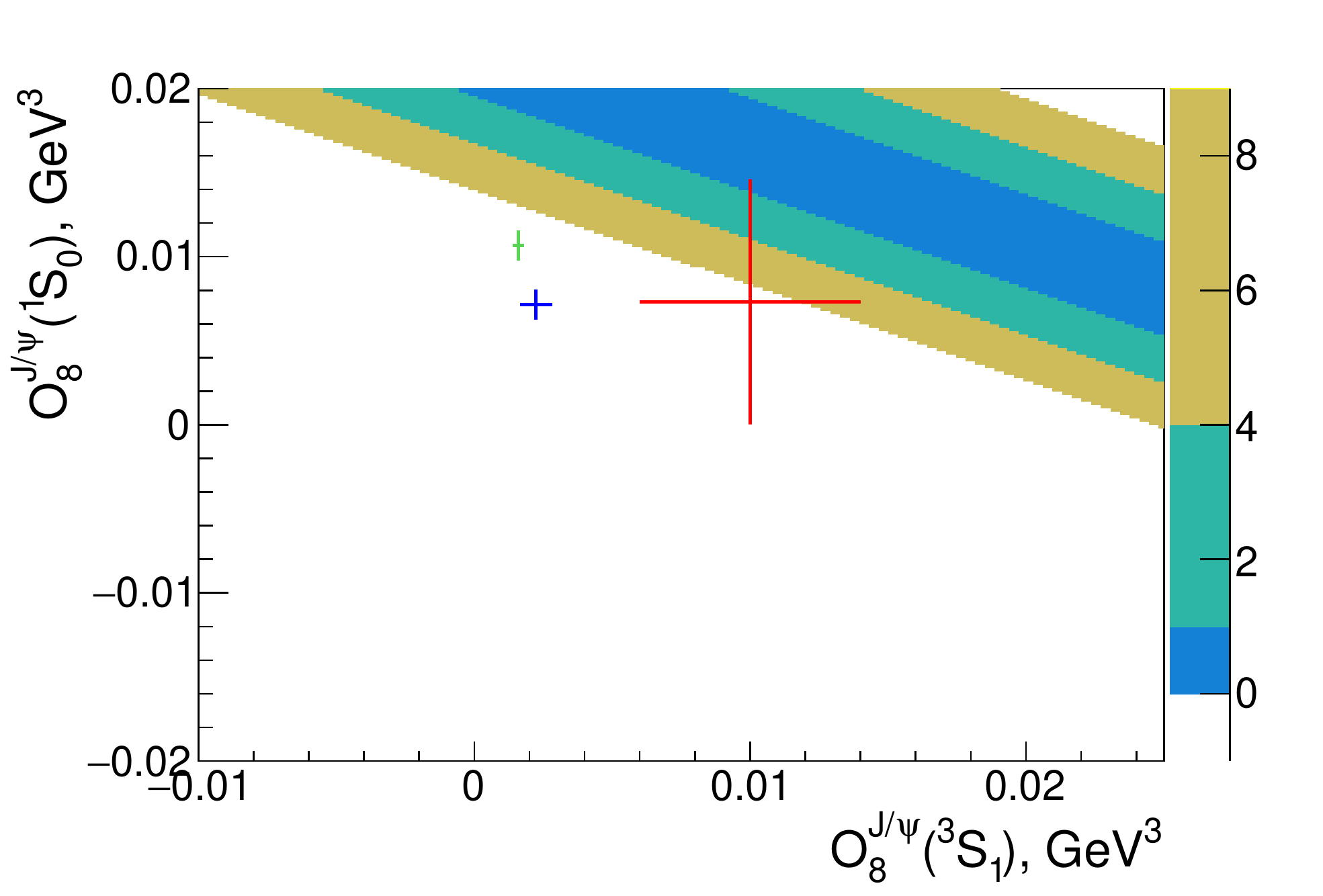}
\label{fig:etacJpsiOnly_3P0_fixed}}
\subfigure[]
{\protect\protect\includegraphics[width=0.5\linewidth]{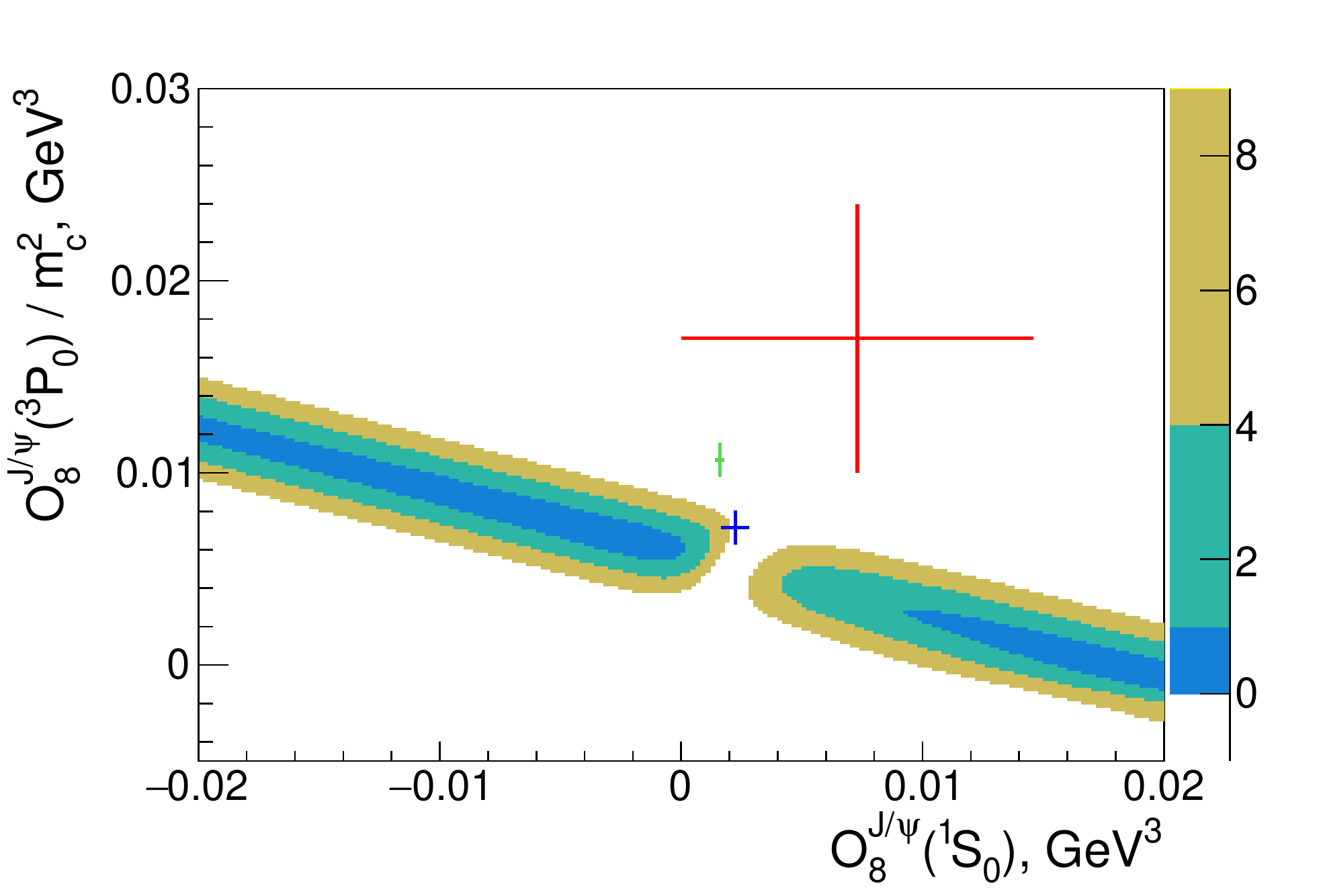}
\label{fig:etacJpsiOnly_3S1_fixed}}

\caption
[The $\Delta \chisq \rangle$ fit distribution using the measurements 
of $\frac{\BR ( b \to \etac(1S)^{direct} X )}{\BR ( b \to \jpsi^{direct} X )}$.]
{The $\Delta \chisq \rangle$ fit distribution using the measurements 
of $\frac{\BR ( b \to \etac(1S)^{direct} X )}{\BR ( b \to \jpsi^{direct} X )}$ from Eq. (\ref{b2EtacDirect}) and $\BR( B \to \jpsi^{direct} X )$~\cite{PDG2016}.
For all listed plots $\langle O_1^{\jpsi}(^3S_1)\rangle =1.16 \gev^3$ is used. 
Only area with $\Delta \chisq <9$ is shown with colour code. Red points correspond to the values from Ref.\cite{Han:2014jya}, green points - from Ref.~\cite{Baranov:2019joi}, blue points - from Ref.~\cite{Butenschoen:2011yh}.}
\label{fig:EtacAndJpsi_thUnc}
\end{figure}

Figure~\ref{fig:EtacAndJpsi_thUnc} shows the $\Delta \chisq$ of the fit on the 
$(\langle O_8^{\jpsi}(^3S_1) \rangle);\langle O_8^{\jpsi}(^1S_0) \rangle)$,
$(\langle O_8^{\jpsi}(^3S_1) \rangle);\langle O_8^{\jpsi}(^3P_0) \rangle / m_c^2 $
and
$\langle O_8^{\jpsi}(^3P_0) \rangle / m_c^2 ;\langle O_8^{\jpsi}(^1S_0) \rangle)$
planes. 
The values of LDMEs from~\cite{Han:2014jya} are overlaid. 
Total experimental uncertainties are taken into account in the fit as well as theoretical uncertainties
on the short-distance coefficient for the CS part. The correlations between the measurements are small and are therefore neglected.

Values of LDMEs, determined by the fit, for various fit options are listed in Table~\ref{tablefits2Etac}. 
The $\langle O_1^{\jpsi}(^3S_1) \rangle$	 matrix element is fixed to the value from Ref.~\cite{Han:2014jya}, the values of 
$\langle O_8^{\jpsi}(^3S_1) \rangle$, $\langle O_8^{\jpsi}(^1S_0) \rangle$ or $\langle O_8^{\jpsi}(^3P_0) \rangle / m_c^2 $ 
are fixed (A,C,E) or Gaussian constrained (B,D,F) to the values from Ref.~\cite{Han:2014jya}.
\begin{table}[h]
\begin{center}
\caption
[Results of simultaneous fits of the LDMEs to the $\BR( \bquark \to \etac(1S)^{direct} X)/\BR( \bquark \to \jpsi^{direct}  X)$.]
{Results of simultaneous fits of the LDMEs to the $\BR( \bquark \to \etac(1S)^{direct} X)/\BR( \bquark \to \jpsi^{direct}  X)$ from Eq. (\ref{b2EtacDirect}) 
and $\BR(B \to \jpsi^{direct}  X)$~\cite{PDG2016}.}
\resizebox{\textwidth}{!}{%
\label{tablefits2Etac}
\begin{tabular}{ c | c | c | c | c | c | c }
  	&   A	  &   B	  &  C	& D	  & E   	&	F \\	  \hline
$\langle O_1^{\jpsi}(^3S_1) \rangle$, $\gev^3$	& 1.16	& 1.16 	& 1.16 & 1.16 	& 1.16 	& 1.16 \\ 
 & (fixed) & (fixed) & (fixed) & (fixed)  & (fixed) & (fixed) \\ \hline
$\langle O_8^{\jpsi}(^3S_1) \rangle$ , $\gev^3$	& 0.0030 & 0.0030 & $-0.115\pm0.008$ & $-0.115\pm0.019$ & $1.65\pm0.38$	& $1.65\pm0.66$ \\ 
 & (fixed) & (constrained) &	&  & 	& 	  \\ \hline
$\langle O_8^{\jpsi}(^1S_0) \rangle$, $\gev^3$	& $0.020\pm0.005$ & $0.020\pm0.005$ & 0.089 & 0.089 & $-0.94\pm0.22$ & $-0.94\pm0.38$ \\
	& 	&	& (fixed) & (constrained) & &   \\ \hline
$\langle O_8^{\jpsi}(^3P_0) \rangle / m_c^2 $, $\gev^3$	& $-0.0006\pm0.0015$	& $-0.0006\pm0.0015$ & $-0.0011\pm0.0014$ & $-0.0011\pm0.0014$  & 0.0056 & 0.0056 \\
	& &	&				     &			         & (fixed)     	       &(constrained) \\ 
\end{tabular}}
\end{center}
\end{table}
The results show that the fit chooses different optimal values 
for $\langle O_8^{\jpsi}(^3S_1) \rangle$ and $\langle O_8^{\jpsi}(^1S_0) \rangle$ depending on the fit assumptions, which are also different from the values in Ref.~\cite{Han:2014jya}. 
The fit also chooses a value for $\langle O_8^{\jpsi}(^3P_0) \rangle / m_c^2 $ that is similar for all fit assumptions, which is however different different from the value from Ref.~\cite{Han:2014jya}. 

\clearpage
\subsection{Hadroproduction}
A theoretical description of the \etac hadroproduction using the \lhcb measurement as one of the inputs is given in Ref.~\cite{Han:2014jya}.
The authors perform a fit to the \jpsi prompt production measurement performed at \cdf. The fit reasonably describes the measured cross-sections.
The following linear combinations are defined:

\begin{equation}
\begin{aligned}
M_0 = \langle O_8^{\jpsi}(^1S_0) \rangle + r_0\cdot\langle O_8^{\jpsi}(^3P_0)/m_c^2 \rangle, \\
M_1 = \langle O_8^{\jpsi}(^3S_1) \rangle + r_1\cdot\langle O_8^{\jpsi}(^3P_0)/m_c^2 \rangle,
\end{aligned}
\end{equation}
where $r_0=-3.9$ and $r_1=-0.56$. The CO combinations $M_0$ and $M_1$ are defined in order to separate different \pt behaviour. Namely, the CO contribution related to $M_0$ behaves as $\sim \pt^{-6}$ and the contribution related to $M_1$ behaves as $\sim \pt^{-4}$, so that these contributions can be distinguished from the fit to \pt-differential cross-section.
The obtained from the fit values of $M_0$ and $M_1$ are:
\begin{equation}
\begin{aligned}
M_0 = (7.4\pm1.9)\times10^{-2} \gev^{3}, \\
M_1 = (0.05\pm0.02)\times10^{-2} \gev^{3}.
\end{aligned}
\label{eq:JpsiConstr}
\end{equation}
Note that the value of $M_1$ is very small and is consistent with zero.
In addition, uses the \lhcb measurement of \etac prompt production to further constrain CO LDMEs. By neglecting the dominant CS contribution a fit to \etac production has been performed by letting CO contributions to saturate the measured cross-section. This way the following upper limit on $O_8^{\jpsi}(^1S_0)$ CO LDME was obtained.
\begin{equation}
\langle O_8^{\jpsi}(^1S_0) \rangle < 1.46 \times 10^{-2} \gev^3
\label{eq:EtacConstr}
\end{equation}

By having the constraint from Eqs.~\ref{eq:JpsiConstr} and~\ref{eq:EtacConstr} and the fit to \etac production cross-section, one can describe the \etac and \jpsi hadroproduction simultaneously.

\begin{figure}[t]
\centering
\subfigure[]
{\protect\protect\includegraphics[width=0.5\linewidth]{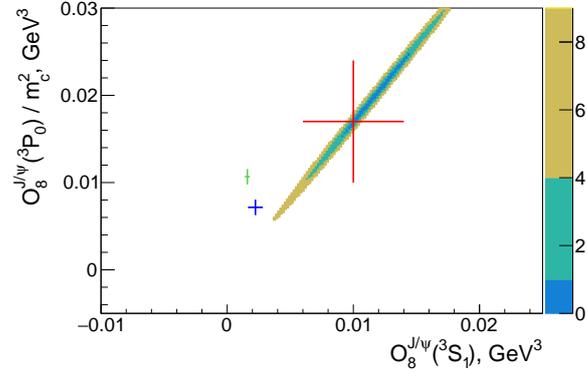}
\label{fig:etacJpsiPmt_1S0_fixed}}
\subfigure[]
{\protect\protect\includegraphics[width=0.5\linewidth]{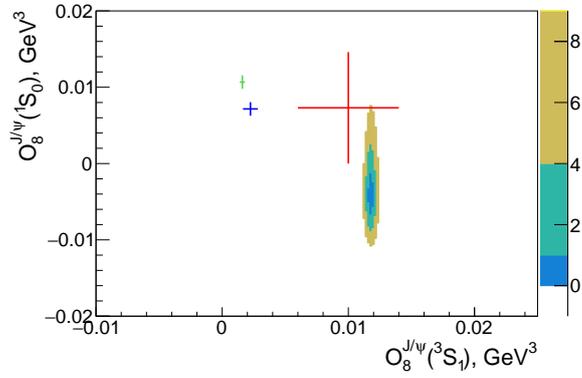}
\label{fig:etacJpsiPmt_3P0_fixed}}
\subfigure[]
{\protect\protect\includegraphics[width=0.5\linewidth]{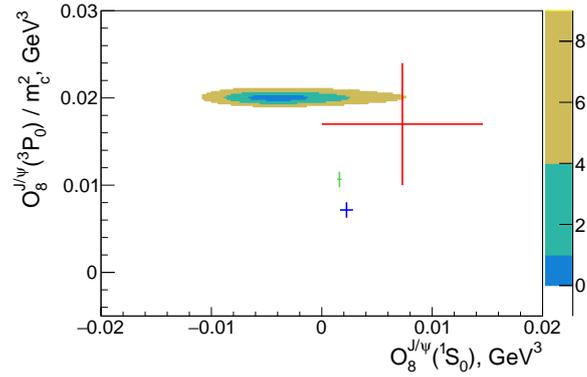}
\label{fig:etacJpsiPmt_3S1_fixed}}
\caption
[The $\Delta \chisq$ fit distribution using the \etac hadroproduction measurements and constraints from \jpsi prompt production.]
{The $\Delta \chisq$ fit distribution using the \etac hadroproduction measurements and constraints from Eq.(\ref{eq:JpsiConstr}).
For all listed plots $\langle O_1^{\jpsi}(^3S_1)\rangle =1.16 \gev^3$ is used. 
Only area with $\Delta \chisq<9$ is shown with colour code. Red points correspond to the values from Ref.\cite{Han:2014jya}, green points - from Ref.~\cite{Baranov:2019joi}, blue points - from Ref.~\cite{Butenschoen:2011yh}.}
\label{fig:EtacAndJpsiPmt_thUnc}
\end{figure}

Figure~\ref{fig:EtacAndJpsiPmt_thUnc} shows the fit $\Delta \chisq$  on the 
$(\langle O_8^{\jpsi}(^3S_1) \rangle);\langle O_8^{\jpsi}(^1S_0) \rangle)$,
$(\langle O_8^{\jpsi}(^3S_1) \rangle);\langle O_8^{\jpsi}(^3P_0) \rangle / m_c^2 $
and
$\langle O_8^{\jpsi}(^3P_0) \rangle / m_c^2 ;\langle O_8^{\jpsi}(^1S_0) \rangle)$
planes. 
The values of LDMEs from Ref.~\cite{Han:2014jya} are overlaid. 
Total experimental uncertainties are taken into account in the fit as well as theoretical uncertainties
on the short-distance coefficient for the CS part.
The dominant source of theory uncertainty is coming from the renormalisation and factorisation scales and amounts to about 35\%.
The experimental uncertainty is dominated by statistical one and amounts to about 30 to 70\%. The fit central values points are in agreement with the results from Ref.~\cite{Han:2014jya} but with significantly reduced uncertainties. The agreement is explained by the fact that the measurement of the \jpsi production is much more precise and dominates the fit. The measurement of the \etac production is not well described by the fit which causes a significant reduction in the LDMEs allowed regions. 

\clearpage
\subsection{Simultaneous study of hadroproduction and production in \bquark-hadron decays}

Similarly, one can perform a fit to \etac and \jpsi hadroproduction and production in \bquark-hadron decays simultaneously by using the same technique as in previous sections.

\begin{figure}[t]
\centering
\subfigure[]
{\protect\protect\includegraphics[width=0.5\linewidth]{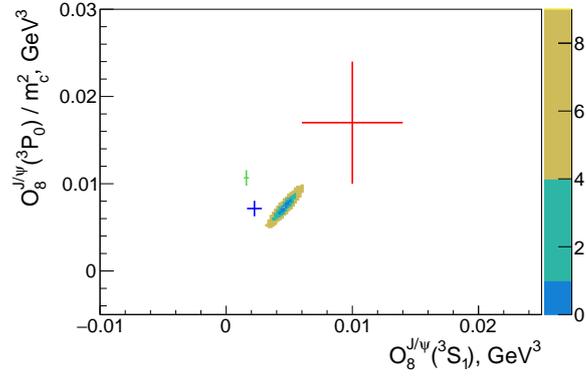}
\label{fig:etacJpsiPmt_1S0_fixed}}

\subfigure[]
{\protect\protect\includegraphics[width=0.5\linewidth]{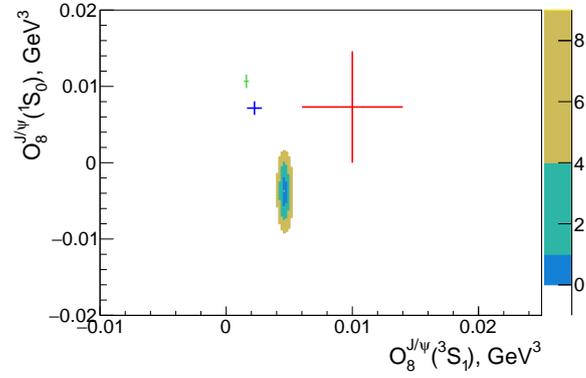}
\label{fig:etacJpsiPmt_3P0_fixed}}

\subfigure[]
{\protect\protect\includegraphics[width=0.5\linewidth]{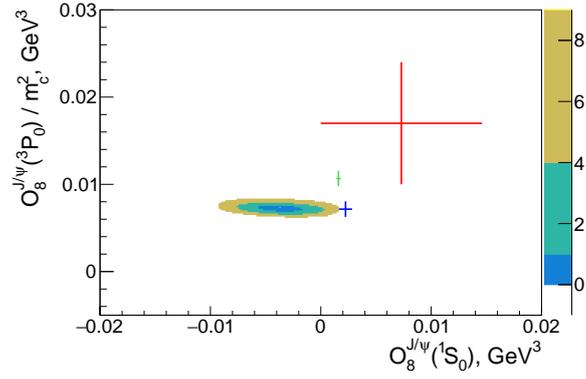}
\label{fig:etacJpsiPmt_3S1_fixed}}

\caption
[The $\Delta \chisq$ fit distribution using the \etac hadroproduction measurements, constraints from \etac and \jpsi prompt production and measurements on \etac and \jpsi production in \bquark-hadron inclusive decays.]
{The $\Delta \chisq$ fit distribution using the \etac hadroproduction measurements, constraints from Eqs.(\ref{eq:JpsiConstr}, \ref{eq:EtacConstr}) and measurements on \etac and \jpsi production in \bquark-hadron inclusive decays.
For all listed plots $\langle O_1^{\jpsi}(^3S_1)\rangle =1.16 \gev^3$ is used. 
Only area with $\Delta \chisq<9$ is shown with colour code. Red points correspond to the values from Ref.~\cite{Han:2014jya}, green points - from Ref.~\cite{Baranov:2019joi}, blue points - from Ref.~\cite{Butenschoen:2011yh}.}
\label{fig:EtacAndJpsiAll_thUnc}
\end{figure}

Figure~\ref{fig:EtacAndJpsiAll_thUnc} shows the fit $\Delta \chisq$ on the 
$(\langle O_8^{\jpsi}(^3S_1) \rangle);\langle O_8^{\jpsi}(^1S_0) \rangle)$,
$(\langle O_8^{\jpsi}(^3S_1) \rangle);\langle O_8^{\jpsi}(^3P_0) \rangle / m_c^2 $
and
$\langle O_8^{\jpsi}(^3P_0) \rangle / m_c^2 ;\langle O_8^{\jpsi}(^1S_0) \rangle)$
planes. 
The values of LDMEs from~\cite{Han:2014jya} are overlaid. 
Total experimental uncertainties are taken into account in the fit as well as theoretical uncertainties
on the short-distance coefficient for the CS part.

The goodness of fit is reasonable, $\chisqndf = 9.7/8$.
The result shows that the parameter space, which can describe all measurements is reduced. This is most remarkable for $\langle O_8^{\jpsi}(^3S_1)\rangle$;$\langle O_8^{\jpsi}(^1S_0)\rangle$ plane.
The optimal points differ from the ones obtained from the fit to prompt production only. This indicates a possible difference in LDMEs for the two production processes, contrary to basic NRQCD assumptions.

\clearpage
\section{Comparison of the \chic production in \bquark-hadron decays to theory}
\label{sec:Pwave}
\begin{figure}[t]
\captionsetup[subfigure]{singlelinecheck=false,margin=7cm}
\centering
\subfigure[]{\protect\protect\includegraphics[width=0.4\linewidth]{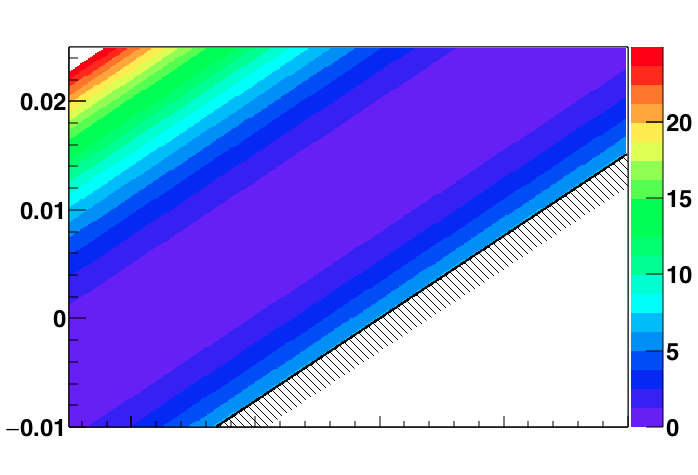}
		    \put(-18,117) {$\langle \chisq \rangle$}
		    \put(-22,-5) {$O_1$}
		    \put(-175,117) {$O_8$}
\label{fig:AbsBR_chic0}}
\quad \quad \quad \quad 
\subfigure[]{\protect\protect\includegraphics[width=0.4\linewidth]{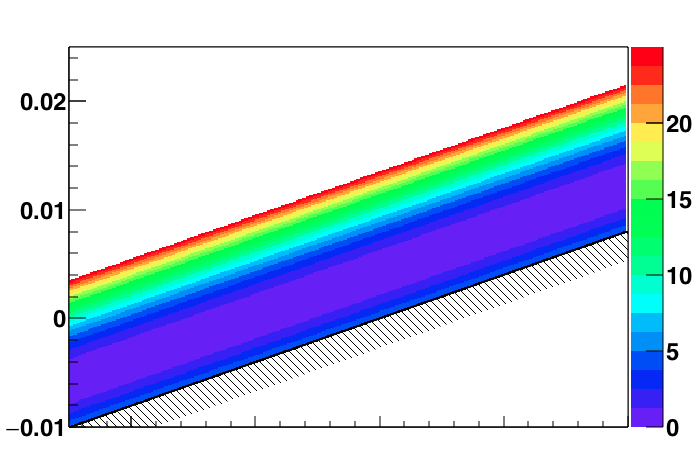}
		    \put(-18,117) {$\langle \chisq \rangle$}
		    \put(-22,-5) {$O_1$}
		    \put(-175,117) {$O_8$}
\label{fig:AbsBR_chic1}}
\quad \quad \quad \quad 
\subfigure[]{\protect\protect\includegraphics[width=0.4\linewidth]{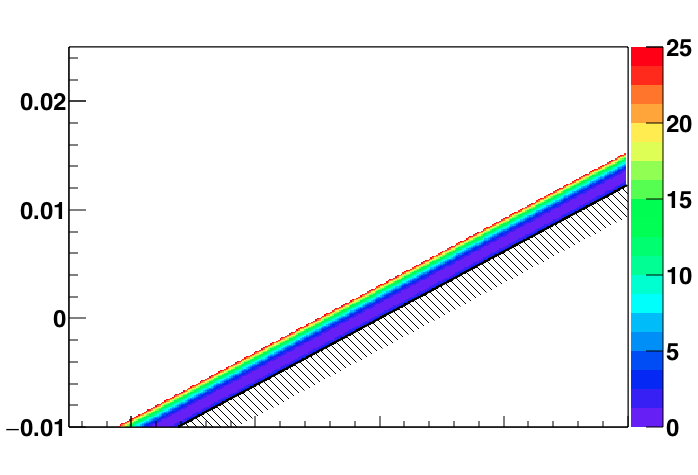}
		    \put(-18,117) {$\langle \chisq \rangle$}
		    \put(-22,-5) {$O_1$}
		    \put(-175,117) {$O_8$}
\label{fig:AbsBR_chic2}}
\quad \quad \quad \quad
\subfigure[]{\protect\protect\includegraphics[width=0.4\linewidth]{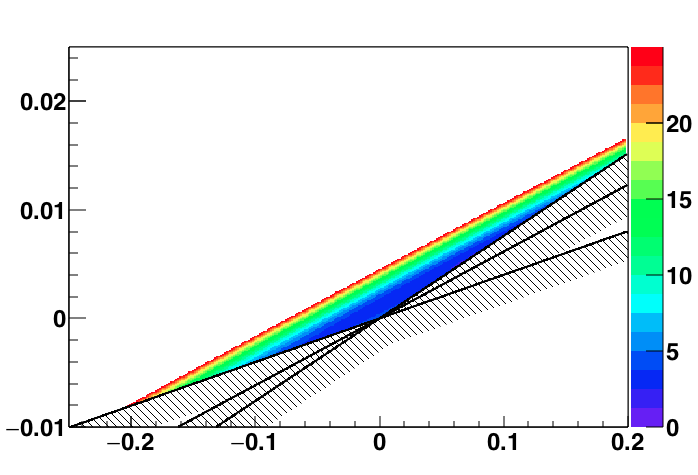}
		    \put(-18,117) {$\langle \chisq \rangle$}
		    \put(-22,-5) {$O_1$}
		    \put(-175,117) {$O_8$}
\label{fig:AbsBR_All}}
\captionsetup{singlelinecheck=off}
\caption[The $\langle \chisq \rangle$ fit distribution for the $O_1$ and $O_8$ matrix elements of \chic production in \bquark-hadron inclusive decays.]{The $\langle \chisq \rangle$ fit distribution for the $O_1$ and $O_8$ matrix elements 
using the measurement of the 
\begin{itemize}
\item \subref{fig:AbsBR_chic0} $\BR(\bquark \to \chiczero^{direct} X)$ from Eq.~\ref{chic0AbsDirect},
\item \subref{fig:AbsBR_chic1} $\BR(\bquark \to \chicone^{direct} X)$ from Eq.~\ref{chic1AbsDirect},
\item \subref{fig:AbsBR_chic2} $\BR(\bquark \to \chictwo^{direct} X)$ from Eq.~\ref{chic2AbsDirect},
\item \subref{fig:AbsBR_All} simultaneously all branching fractions $\BR(\bquark \to \chic_J^{direct} X)$.
\end{itemize}
Black lines indicate boundaries, where branching fractions become negative.
Only area with $\langle \chisq \rangle<25$ is shown with colour code.}
\label{fig:oabs}
\end{figure}

Using expressions (\ref{b2chic0Theory}), (\ref{b2chic1Theory}) and (\ref{b2chic2Theory}) for $\BR ( B \to \chic_JX)$, the measurements of $\BR(\bquark \to \chic X)$ were fitted in terms of colour singlet and colour octet matrix elements. Figure~\ref{fig:oabs} shows the $\langle \chisq \rangle$ of the fit as a function of the CS ($O_1$) and the CO ($O_8$) matrix elements. The fit was performed separately for the $\BR(\bquark \to \chiczero^{direct} X)$, $\BR(\bquark \to \chicone^{direct} X)$ and $\BR(\bquark \to \chictwo^{direct} X)$ measurements and using all three measurements simultaneously. 
The total experimental uncertainties are taken into account in the fit, while the correlations between the measurements are not taken into account.  
The fit allows to strongly restrict the allowed range for LDMEs.
The most probable values of LDMEs are determined from simultaneous fit to be
\begin{align*}
O_1^{opt} & = 0.0755 \gev^3, \\
O_8^{opt} & = 0.00575 \gev^3. 
\end{align*}

\begin{figure}[h]
\centering
\protect\protect\includegraphics[width=0.6\linewidth]{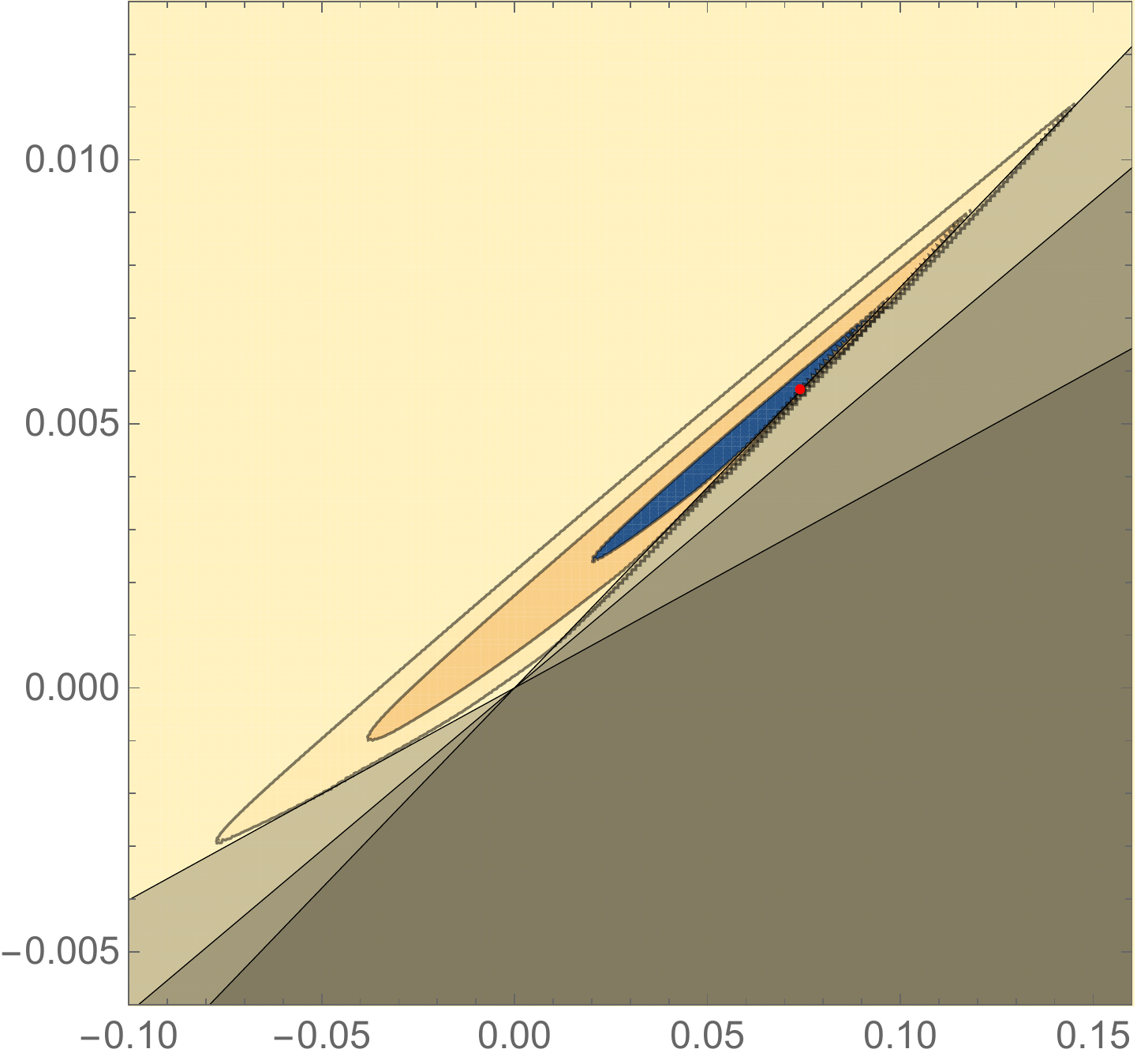}
		    \put(-20,-10) {$O_1$}
		    \put(-285,230) {$O_8$}
\caption
[The contour plot for $O_1$ and $O_8$ describing one, two and three sigma statistical contours taking into account unphysical area where at least one
of the $\BR(\bquark \to \chic_J^{direct} X)$ becomes negative.]
{The contour plot for $O_1$ and $O_8$ describing one, two and three sigma statistical contours taking into account unphysical area where at least one
of the $\BR(\bquark \to \chic_J^{direct} X)$ becomes negative. The unphysical area is filled in grey. The most probable values $(O_1^{opt},O_8^{opt})$ are shown in red.}
\label{fig:francois}
\end{figure}

As another representation of the results obtained with simultaneous fit of all $\BR(\bquark \to \chic_J^{direct} X)$ measurements,
Figure~\ref{fig:francois} shows one, two and three standard deviations contours in the ($O_1$;$O_8$) plane taking into account non-physical regions, where at least one
of the $\BR(\bquark \to \chic_J X)$ becomes negative. In order to extract the contours, the $\langle \chisq \rangle$ fit distribution from Fig.~\ref{fig:AbsBR_All} is used to generate toy 
frequency distribution $PDF(O_1,O_8)$ in the ($O_1$;$O_8$) plane. 

\begin{figure}[h]
\centering
\subfigure[]{\protect\protect\includegraphics[width=0.45\linewidth]{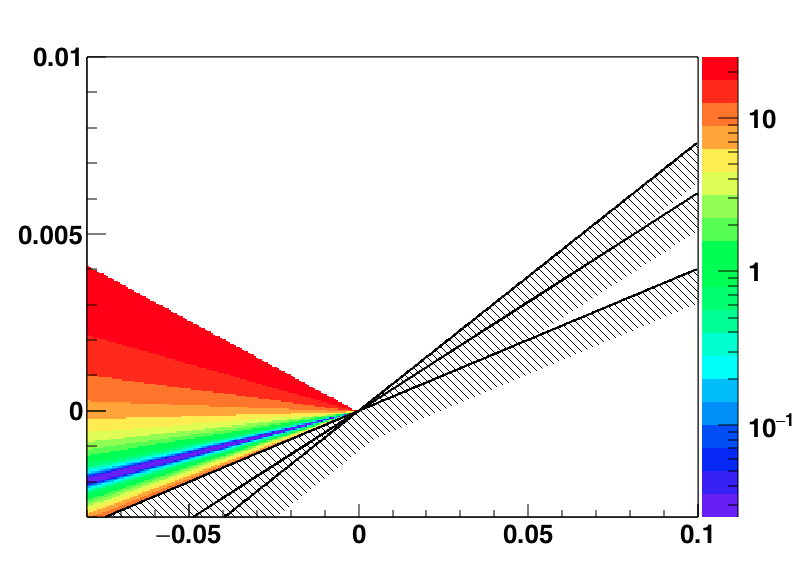}
		    \put(-25,140) {$\langle \chisq \rangle$}
		    \put(-35,-5) {$O_1$}
		    \put(-195,140) {$O_8$}
\label{fig:chi1Fit}}
\quad\quad\quad
\subfigure[]{\protect\protect\includegraphics[width=0.45\linewidth]{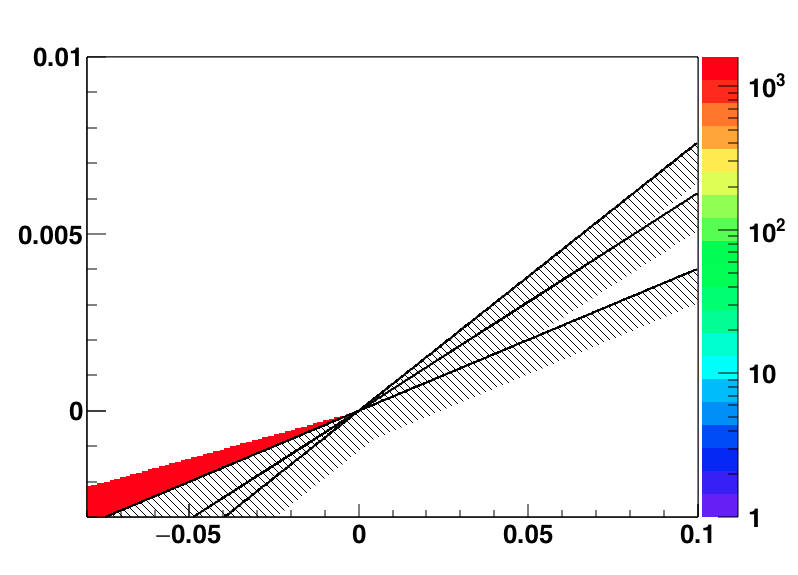}
		    \put(-25,140) {$\langle \chisq \rangle$}
		    \put(-35,-5) {$O_1$}
		    \put(-195,140) {$O_8$}
\label{fig:chi2Fit}}
\subfigure[]{\protect\protect\includegraphics[width=0.45\linewidth]{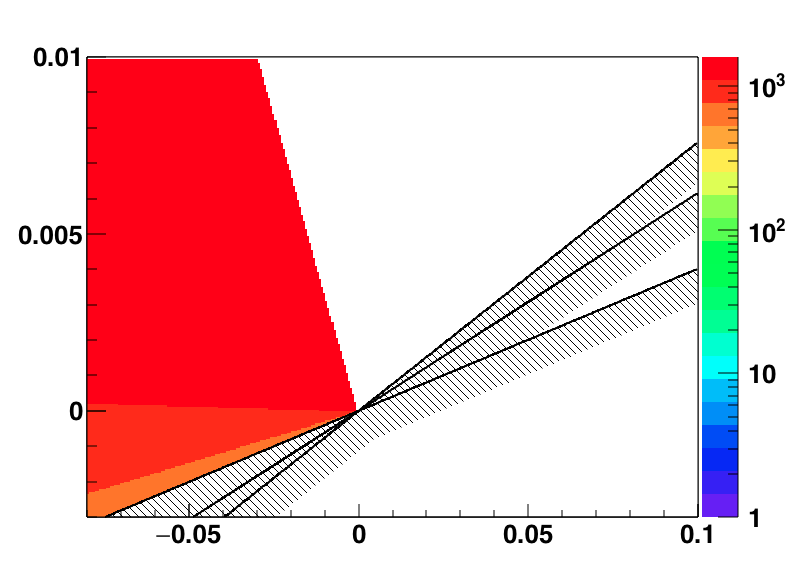}
		    \put(-25,140) {$\langle \chisq \rangle$}
		    \put(-35,-5) {$O_1$}
		    \put(-195,140) {$O_8$}
\label{fig:chi12Fit}}
\captionsetup{singlelinecheck=off}
\caption[The $\langle \chisq \rangle$ fit distribution using the measurement of the relative branching fractions of $\bquark\to\chic X$.]{The $\langle \chisq \rangle$ fit distribution using the measurement of the relative branching fractions
\begin{itemize}
\item \subref{fig:chi1Fit} $\BR(\bquark \to \chicone^{direct} X)/\BR(\bquark \to \chiczero^{direct} X)$ from Eq.~\ref{chic1RelDirect}, 
\item \subref{fig:chi2Fit} $\BR(\bquark \to \chictwo^{direct} X)/\BR(\bquark \to \chiczero^{direct} X)$ from Eq.~\ref{chic2RelDirect},
\item \subref{fig:chi12Fit} both $\BR(\bquark \to \chicone^{direct} X)/\BR(\bquark \to \chiczero^{direct} X)$ and $\BR(\bquark \to \chictwo^{direct} X)/\BR(\bquark \to \chiczero^{direct} X)$
\end{itemize}
for the $O_1$ and $O_8$ matrix elements. 
Black lines indicate boundaries, where branching fractions become negative
Only area with \subref{fig:chi1Fit} $\langle \chisq \rangle<25$ and \subref{fig:chi2Fit},\subref{fig:chi12Fit} $\langle \chisq \rangle<1600$ is shown with colour code.}
\label{fig:relfit}
\end{figure}

The $\langle \chisq \rangle$ fit
 for the  $O_1$ and $O_8$ matrix elements using the measurements of $\BR(\bquark \to \chicone^{direct} X)/\BR(\bquark \to \chiczero^{direct} X)$ and $\BR(\bquark \to \chictwo^{direct} X)/\BR(\bquark \to \chiczero^{direct} X)$ is shown on Fig.~\ref{fig:relfit}.
 The fit is performed separately for $\frac{\BR(\bquark \to \chicone^{direct} X)}{\BR(\bquark \to \chiczero^{direct} X)}$ and $\frac{\BR(\bquark \to \chictwo^{direct} X)}{\BR(\bquark \to \chiczero^{direct} X)}$ and using both measurements simultaneously. Total experimental uncertainties are taken into account in the fit, while the correlations between the measurements 
are not taken into account. Note, that in this case correlations are negligible and can be ignored.


While the fit using absolute branching fractions $\BR(\bquark \to \chic_J^{direct} X)$ can accommodate a limited range of $O_1$ and $O_8$ due to large 
experimental uncertainties, the fit to the ratio of branching fractions $\BR(\bquark \to \chicone^{direct} X)/\BR(\bquark \to \chiczero^{direct} X)$
significantly reduces the allowed $O_1$ and $O_8$ range. The \chictwo ratio of branching fractions $\BR(\bquark \to \chictwo^{direct} X)/\BR(\bquark \to \chiczero^{direct} X)$
is then not consistent with the assumed theoretical framework. 

Note that the fits to theory (Fig.~\ref{fig:relfit}) prefer negative values of $O_1$. 
This confirms the problem of unphysical negative short-distance coefficient relative to the CS LDME discussed by the authors of Ref.~\cite{Beneke:1998ks}.

The values of CS matrix elements are extracted from Eqs. (\ref{b2chic0TheoryNSR}), (\ref{b2chic1TheoryNSR}), (\ref{b2chic2TheoryNSR})
and $O_8=(11.22\pm0.68)\time10^{-3} \gev^3$~\cite{ind}
allow to extract the values of the CS matrix elements from the
$\BR(\bquark \to \chic_J^{direct} X)$ measurements 
without using spin symmetry relations to be:
\begin{align*}
\langle O_1^{\chiczero}(^3S_1) \rangle = -0.04\pm0.07 \gev^3,\\ 
\langle O_1^{\chicone}(^3S_1) \rangle = 0.51\pm0.14 \gev^3,\\ 
\langle O_1^{\chictwo}(^3S_1) \rangle = 0.83\pm0.04 \gev^3.
\end{align*}

\clearpage
\section{Summary and discussion}
\label{sec:ThFitSummary}
This chapter proposes a technique of constraining theory using simultaneously results on charmonium hadroproduction and charmonium production
in \bquark-hadron inclusive decays, under the assumption of factorization, universality of LDMEs and heavy-quark spin symmetry, where different
charmonium states are involved. 
Alternatively, when the hadroproduction and production in \bquark-hadron inclusive decays will be measured for the
charmonium states with linked LDMEs, the above assumptions can be tested in a qualitative way.

The relative $\etac(1S)$ to \jpsi production measurement is found to be in agreement with the theory prediction
when using LDMEs values from the fits~\cite{Butenschoen:2012px,Chao:2012iv,Gong:2012ug,Bodwin:2014gia} of prompt $\etac(1S)$ production measurement. 
However, matrix elements extracted from the simultaneous fit of the \jpsi and $\etac(1S)$ production in inclusive \bquark-decays slightly differ from the matrix elements, extracted using measurements of the prompt \jpsi production. At the same time, a simultaneous fit is able to describe both $S$-wave charmonium prompt production and production in \bquark-hadron inclusive decays.

The $\BR(\bquark \to \chic_J^{direct} X)$ measurements are fitted according to theoretical formalism with two free parameters representing CS and CO LDMEs linked between the decays to the \chiczero, \chicone and \chictwo charmonia. It is shown that the measurement of the ratio
of the branching fractions $\BR(\bquark \to \chicone^{direct} X)/\BR(\bquark \to \chiczero^{direct} X)$
can be accommodated by theory model and can constrain LDMEs, while the measurement of the ratio
of the branching fractions $\BR(\bquark \to \chictwo^{direct} X)/\BR(\bquark \to \chiczero^{direct} X)$
is not consistent with the theory prediction. 
Hence, calculations of the $\chic_J$ production in inclusive \bquark-hadron decays need to be revisited.

The predictions describe \bquark-decays to the $S$-wave charmonia within reduced parameter space, while the description of \bquark-decays to $P$-wave charmonia is not entirely consistent with
the \lhcb measurement. Particularly, the \chictwo production in inclusive \bquark-decays is not described by theory; the \chiczero production cannot be accomodated by theory using
prediction for CS matrix element. The problems in describing the $\chic_J$ production in inclusive \bquark decays were expected by authors of Ref.\cite{Beneke:1998ks} in the CS part. It was noted that
negative short-distance coefficient before the CS LDME is not physical. 
This would justify why the fit prefers negative values for the CS LDME.

Examining the exclusive branching fractions $\BR(B \to \chic_J K)$, authors of Ref.~\cite{Beneke_B2chicK} pointed out a potentially important
contribution of spectator scattering to the CO production.
Measurement of the $h_c$ production is important to test $P$-wave charmonia production in \bquark-hadron decays, since it is expected to be problematic similarly to \chictwo. 




\begin{singlespace}
\chapter{Measurement of charmonium resonance parameters}
\label{ch:mass}
\end{singlespace}
This chapter summarises the measurements of charmonium resonance parameters performed using \lhcb data samples of charmonia produced in \bquark-hadron inclusive decays. The \ppbar and $\phi\phi$ decays of charmonium are used similarly to production measurements described in Sections~\ref{ch:ppbar} and~\ref{ch:phiphi}. The obtained measurements of the \etac mass and potentially natural width can compete with the world average values. It proves that much larger production rate of the \etac meson at \lhcb already provides better accessibility to the \etac properties compared to that at charm and B-factories.
However, this is not yet the case for other charmonium states.

After introducing the charmonium spectroscopy in Section~\ref{sec:ccbarMass}, the measurement of the \etac mass using the decay $\decay{\etac}{\ppbar}$ is described in Section~\ref{sec:mass}. The measurement of the \etac mass and natural width using the decay $\decay{\etac}{\phi\phi}$ is described in Section~\ref{sec:masses}.
\newpage
\section{Charmonium resonance parameters}
\label{sec:ccbarMass}
The charmonium states below the \DDbar threshold are well identified as bound states of \ccbar. 
Their masses and natural widths are summarized in Table~\ref{tab:mres}. The reported average values take into account also the results from Chapter~\ref{ch:phiphi}.
\begin{table}[h]
\centering
\begin{tabular}{l|c|c}
                 & Mass, \mev            & Natural width, \mev \\ \hline
$\etac(1S)$      & $2983.9 \pm 0.5$      & $32.0 \pm 0.8$ \\ 
$\jpsi$          & $3096.900 \pm 0.006$    & $0.0929\pm0.0028$ \\
\chiczero         & $3414.71 \pm 0.30$      & $10.8 \pm 0.6$ \\  
\chicone         & $3510.67 \pm 0.05$      & $0.84 \pm 0.04$ \\
$h_c$            & $3525.38 \pm 0.11$      & $0.7 \pm 0.4$ \\
\chictwo         & $3556.17 \pm 0.07$      & $1.97 \pm 0.09$ \\
\etactwos        & $3637.6 \pm 1.2$        & $11.3^{+3.2}_{-2.9}$ \\
\psitwos         & $3686.097 \pm 0.025$    & $0.294 \pm 0.008$ 
\end{tabular}
\caption
[Charmonia masses and natural widths.]
{Charmonia masses (in \mev) and natural widths (in \mev)~\cite{PDG2018}.
\label{tab:mres}}
\end{table}
The most precise mass and width measurements of \jpsi and \psitwos have been performed by KEDR collaboration~\cite{Anashin:2015rca}. The world average values for the \jpsi and \psitwos natural width are dominated by measurements of 
CLEO~\cite{Adams:2005mp}, E835~\cite{Andreotti:2007ur} and BES~\cite{Bai:2002zn}.

The world average values of the \etac mass and width are dominated by \lhcb~\cite{Aaij:2016kxn} and BES III~\cite{BESIII:2011ab} measurements. The \lhcb measurement takes into account a possible interference between the $\decay{\Bp}{(\decay{\etac}{\ppbar})\Kp}$ and the non-resonant $\decay{\Bp}{\ppbar\Kp}$ decays. The BES III measurement required a complex description of the \etac lineshape since the \etac sample from radiative decays $\psitwos\to\etac\gamma$ was used. 
The tension of 2$\sigma$ between the two measurements of the \etac mass calls for other measurements of the \etac resonance parameters.

The world average values of the \chic resonance parameters are dominated by the BES III~\cite{Ablikim:2005yd}, E835~\cite{Andreotti:2003sk, Andreotti:2005ts}, E760~\cite{Armstrong:1991yk} measurements and the measurement of \lhcb~\cite{Aaij:2017vck} using recently discovered $\chic_{1,2}\to\jpsi\mup\mu^{-}$ decays.
Similarly, the $h_c$ mass and width world average values are dominated by the results from BES III~\cite{Ablikim:2012ur} and CLEO~\cite{Dobbs:2008ec}.

The charmonium state below the \DDbar threshold with the least studied resonance parameters is the \etactwos. The \etactwos mass is known to a precision of 1.2 \mev and is dominated by the \lhcb~\cite{Aaij:2016kxn} and \babar~\cite{delAmoSanchez:2011bt} measurements, while the most precise measurements of the \etactwos natural width have been performed by \babar~\cite{delAmoSanchez:2011bt} and BES II~\cite{Ablikim:2013gd}. This reflects a limited sample of \etactwos mesons at BES experiment.

\begin{figure}[t]
\centering
\protect\includegraphics[width=0.9\textwidth]{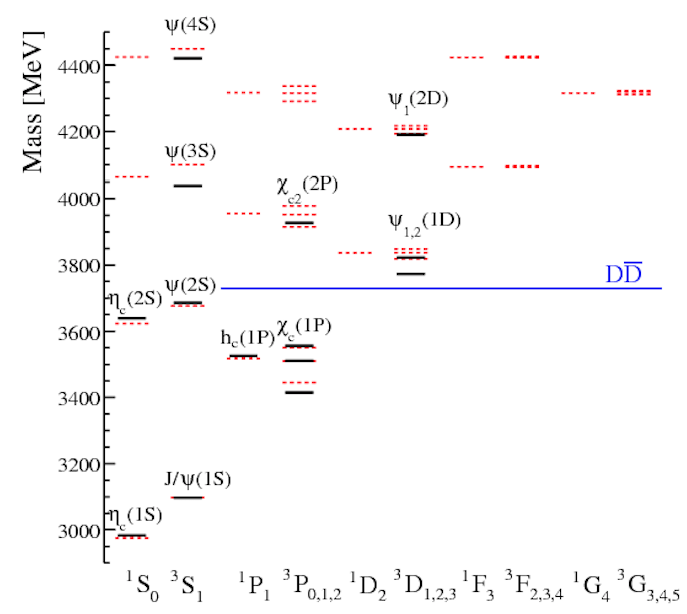}
\caption
[The comparison of charmonium spectrum to the Godfrey-Isgur model calculation.]
{The comparison of charmonium spectrum to the Godfrey-Isgur model calculation~\cite{Barnes:2005pb}. Figure is taken from Ref.~\cite{Olsen:2017bmm}.} 
\label{fig:ccSpectr}
\end{figure}

Theoretically, a spectrum of charmonium states is predicted by potential models and lattice calculations.
A comparison of the observed spectrum with the theoretical prediction by Godfrey-Isgur model~\cite{Godfrey:1985xj,Barnes:2005pb} is shown on Fig.~\ref{fig:ccSpectr}. As demostrated by Fig.~\ref{fig:ccSpectr}, the potential model provides a good spectrum description. 

While a qualitative picture of charmonium spectrum is well described by the potential model, a hyperfine splittings is another subject to be addressed by theory. 
Rigorous predictions of $P$-wave charmonium masses maybe studied using information directly from lattice QCD or by using potentials obtained from lattice and then embedded in EFTs.

For example, the splitting between the \etac and \jpsi masses, reflects the effect of relativistic spin-dependent forces. As was already stated before, the hyperfine splitting of $S$-wave quarkonium can be computed perturbatively and is a subject for precision tests.
The first precise lattice calculations of the $S$-wave quarkonium mass splitting based on NRQCD with spin-dependent terms have been performed in Ref.~\cite{Davies:1995db} followed by Ref.~\cite{Trottier:1996ce}. In the latter article, authors expect further large relativistic corrections.
As shown in Ref.~\cite{Okamoto:2001jb}, the predicted \jpsi-\etac mass difference is underestimated compared to measurements. A similar situation takes place for the mass splitting between the $\chic_J$ states. Recent results performed at the next-to-next-to-next-to-leading logarithmic ($N^{3}LL$) accuracy show a better agreement (see e.g. Ref.~\cite{Peset:2018jkf}).
The predictions are done using pNRQCD as discussed in Ref.~\cite{Pineda:2011dg}. The \etac mass and width can be also determined precisely using $\jpsi \to \etac \gamma$ transition description~\cite{Brambilla:2010ey}, namely description of the signal lineshape.
Note that the theoretical precision is worse than the experimental one. However, it has to be proven that measurements are converging to the same average and no systematic effects can cause a significant change in the world average values.

The spectroscopy of resonant charmonium states above the \DDbar threshold is more complicated due to their large natural width. There are still charmonium resonances to be discovered.
The last to date discovery of charmonium state has been performed by \lhcb~\cite{Aaij:2019evc}, where the state $X(3842)$ state has been observed. It is interpreted as the $1^3D_3$ charmonium state. This observation was performed using \DDbar spectroscopy, while decays to light hadrons do not significantly contribute to the study of charmonia above the \DDbar threshold. 

\begin{figure}[t]
\centering
\protect\protect\includegraphics[width=0.9\textwidth]{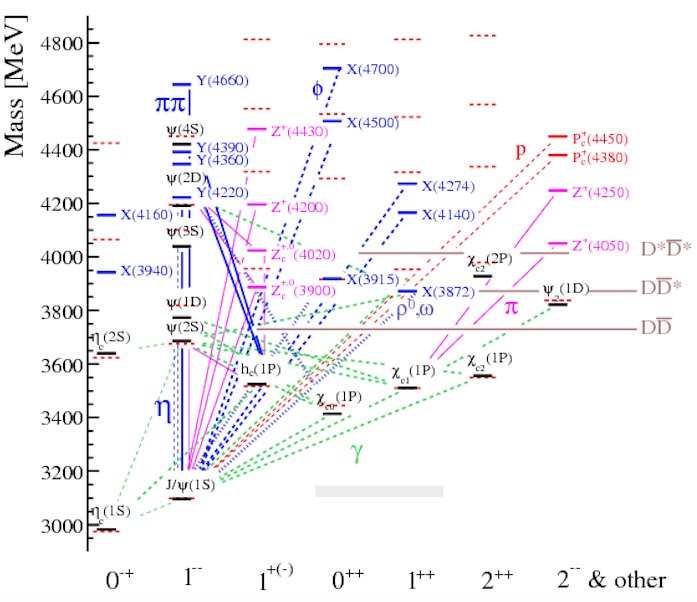}
\caption
[The spectrum of charmonium-like states.]
{The spectrum of charmonium-like states. Figure is taken from Ref.~\cite{Olsen:2017bmm}.} 
\label{fig:exot}
\end{figure}

Apart from charmonium states above the \DDbar threshold, the so-called charmonium-like exotics candidates appear in the spectrum. These states do not fit the charmonium model and hence hypotheses on their tetraquark, molecular, hybrid and adjoint charmonium are most natural to assume. Another feature of the states in this region is that they are expected to be mixed states. For example, the current understanding of the $X(3872)$ state is that it is rather a mixture of charmonium and $DD^{*}$ molecular state~\cite{Suzuki:2005ha}. The key properties of $X(3872)$ is that it is much narrower than any charmonium state expected at this mass and hence cannot be described by pure charmonium model; and that it decays to $\psitwos\gamma$ with relatively large branching fraction and hence cannot be accommodated by pure molecular model. In addition to that, the prompt production study of $X(3872)$ showed that it behaves consistently with a prediction for $\chicone(2P)$ state. All mentioned above led to the mixed interpretation of $X(3872)$ and it has been renamed as the $\chicone(3872)$ in the latest PDG release~\cite{PDG2019}. However, the available experimental inputs  and theory do not allow to establish the nature of this state. Therefore, the renaming points to ithe quantum numbers of the state without suggesting charmonium interpretation. However, a discovery of any hadronic decay of $X(3872)$ would immediately add information about its possible charmonium component.
Other exotics candidates such as $X(3915)$, $X(3832)$ and many others have been identified. The problem of these states is such that they have to be distinguished from the so-called cusp effects appearing close to the opening threshold due to a virtual hadrons loop~\cite{Blitz:2015nra,Swanson:2015bsa}.
The spectrum of charmonium-like states is shown on Fig.~\ref{fig:exot}.  

Systematic studies of exotics states are performed at B-factories and \lhcb experiment. This is, however, not the topic of this work.

\clearpage
\section{Measurement of the \jpsi and \etac mass difference using decays to \ppbar}
\label{sec:mass}
The prompt \etac production measurement requires extreme selection applied at the trigger level to compete with the challenging background conditions, at the same time avoiding biases to retain robust efficiency estimates.
Charmonia produced in \bquark-decays are reconstructed over controlled background level and are more suitable to measure \jpsi-\etac mass difference. A looser selection adjusted for the mass difference determination is chosen contrary to the production measurement, where same selection for prompt charmonium and charmonium from \bquark-hadron decays is used in order to avoid potential biases in the efficiency estimates.

Below, the \etac mass relative to the well-reconstructed and well-known \jpsi mass, 
$\Delta M _{\jpsi , \, \etac} = M_{\jpsi} - M_{\etac}$ is measured.

\subsection{Selection and optimisation}
In the data sample the basic level \texttt{L0 Hadron decision} (L0HadronDecision$\_$TOS) trigger is applied.
The trigger lines \texttt{TOS} of HLT1, \texttt{Hlt1(Two)TrackMVADecision$\_$TOS}, and HLT2, 
\texttt{Hlt2Topo(2,3,4)BodyDecision$\_$TOS} are used for mass measurement for the combined 2015 and 2016 data sample.

The set of selection criteria used in the preselection (\texttt{StrippingCcbar2PpbarDetachedLineDecision}) is summarised in 
Table~\ref{tab:offlinerequirementsTOPO}. In comparison to preselection used for the \etac production measurement in Chapter~\ref{ch:ppbar}, less tight requirements on proton and charmonium candidates \pt is used as well as less tight requirement of proton identification.
\begin{table}[ht] 
\centering
\small
\begin{tabular}{l|l|l} 
          & Variable	                      & Selection criteria 		     \\ \hline
 Trigger	&                                 & L0$\_$Hadron$\_$TOS         \\
			&                                     & Hlt1(Two)TrackMVADecision$\_$TOS\\
			&                                     & Hlt2Topo(2,3,4)BodyDecision$\_$TOS \\ \hline 
 Proton  	  & \pt, \gev                      & $>1.0  \gevc$               \\
 candidates & Track $\chi^{2}/\mathrm{NDF}$   & $<5.0$                 \\
  		      & Impact parameter \chisq         & $>9$                 \\
     		    & $\Delta\log\mathcal{L}^{p-\pi}$   & $>15$                \\
     		    & $\Delta\log\mathcal{L}^{p-K}$     & $>10$                \\ \hline

 Charmonium	& \pt, \gev                      & $-$               \\ 
 candidates & Vertex $\chi^{2}$               & $<9$                 \\ 
   		    & Flight distance \chisq          & $>25$                 \\ 
  		    & Rapidity $y$                    & $2<y<4.5$            \\ \hline

 Multiplicity & SPD multiplicity              & $<600$ \\\
\end{tabular}
\caption{Preselection criteria for the \etac and \jpsi mass difference measurement.} 
\label{tab:offlinerequirementsTOPO}
\end{table}

\begin{figure}
\centering{
        \subfigure[]
        {\protect\protect\protect\includegraphics[width=0.6\textwidth]{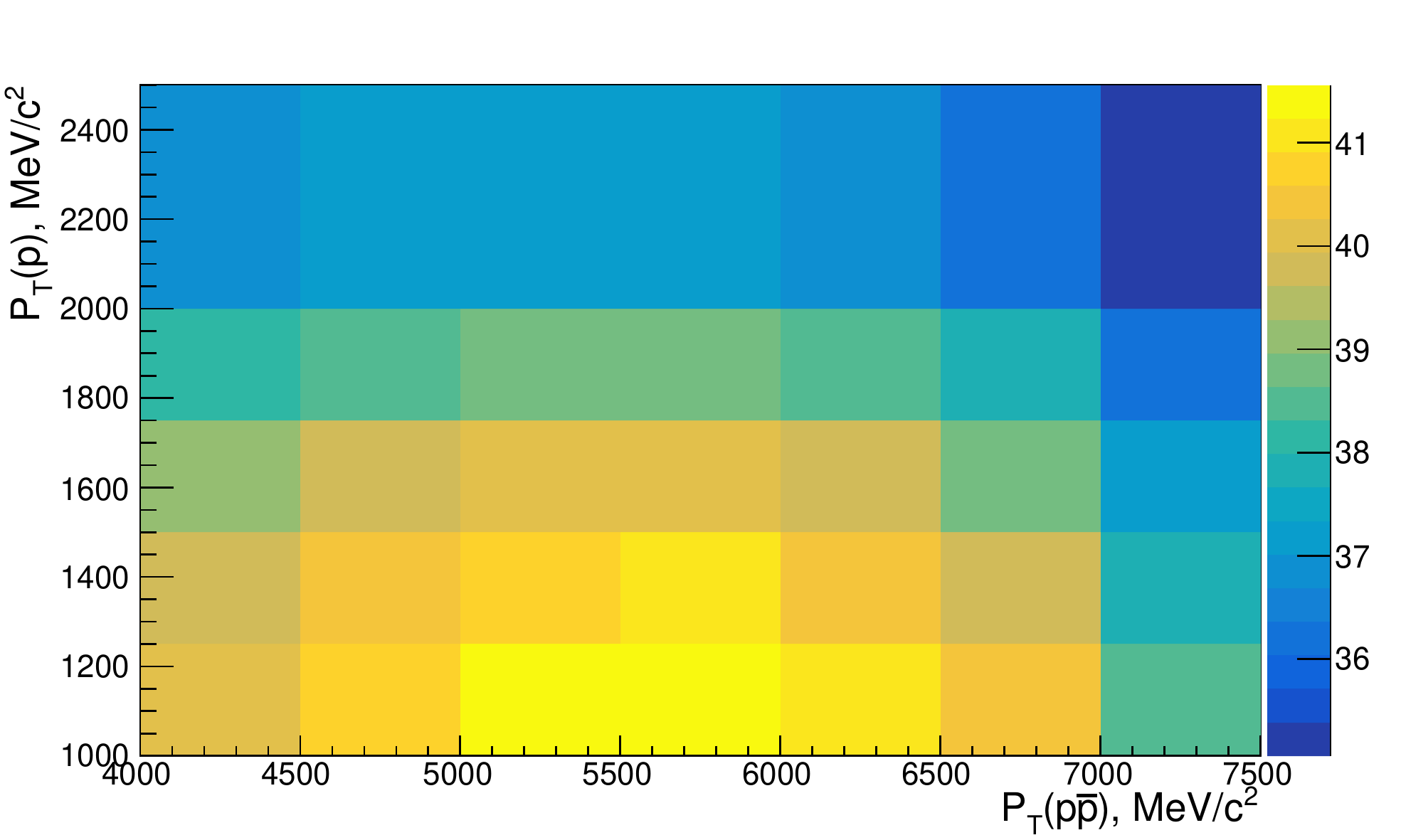}
         \put(-240,150){\footnotesize{\lhcb-ANA-2018-035}}
        \label{fig:OptimizePT}}
        \subfigure[]
        {\protect\protect\protect\includegraphics[width=0.6\textwidth]{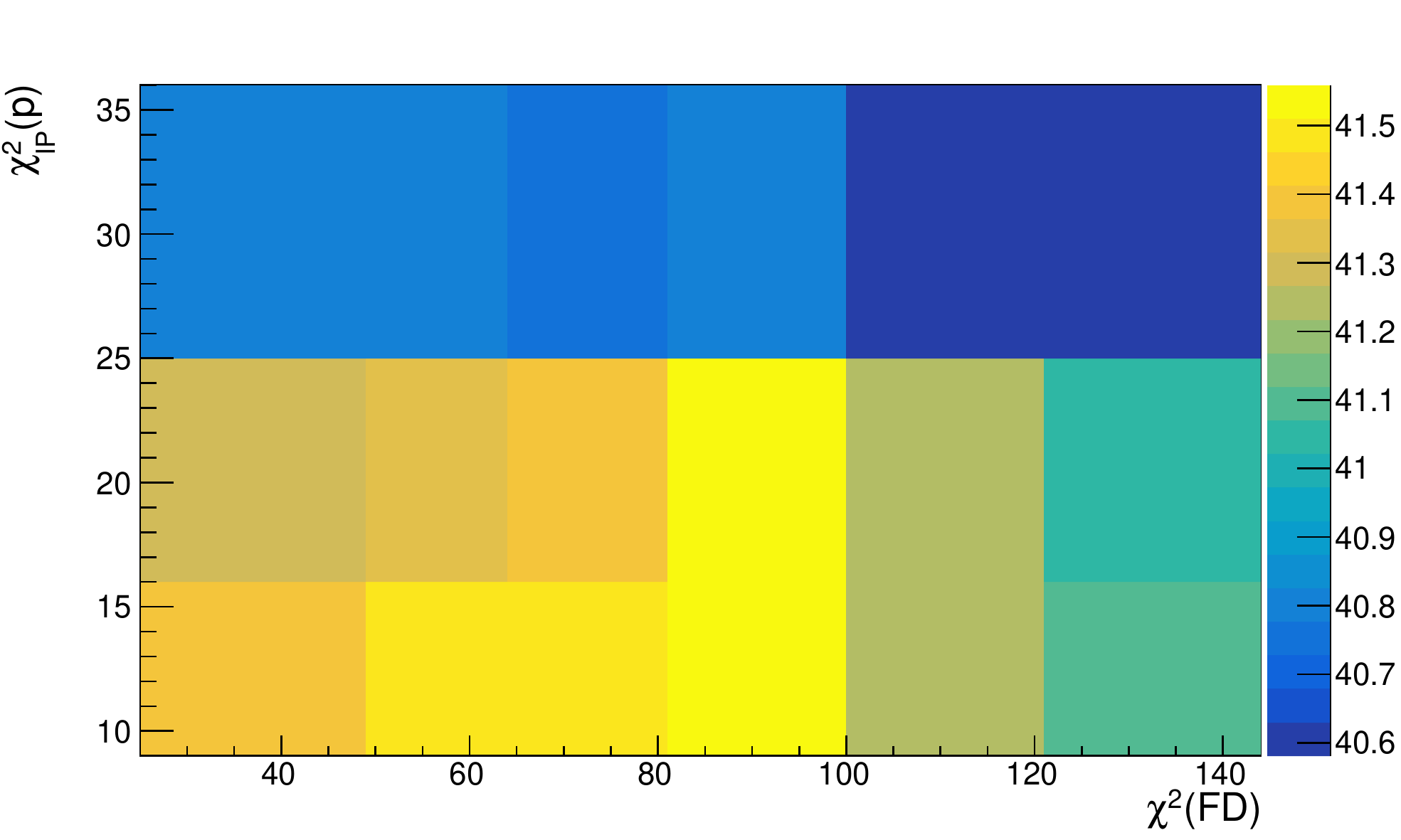}
         \put(-240,150){\footnotesize{\lhcb-ANA-2018-035}}
         \label{fig:OptimizeIP}}
 }
\caption
[Optimisation map as a function of applied requirements on proton transverse momentum \pt(\proton) and charmonium transverse momentum \pt(\ppbar), \chisq of flight distance of charmonium candidates $\chisq(FD)$ and \chisq of track impact parameter with respect to the best primary vertex $\chisq_{IP}(\proton)$.]
{Optimisation map as a function of applied requirements on proton transverse momentum \pt(\proton) and charmonium transverse momentum \pt(\ppbar)~\subref{fig:OptimizePT}, \chisq of flight distance of charmonium candidates $\chisq(FD)$ and \chisq of track impact parameter~\subref{fig:OptimizeIP} with respect to the best primary vertex $\chisq_{IP}(\proton)$. The $FoM=N_{sig}/\sqrt{N_{sig}+N_{bkg}}$ is shown with the color code. The plots are 2D projections of the 4D optimisation.} 
\label{fig:massOpt}
\end{figure}

Further optimisation of cut-based selection is performed on four basic variables: minimal \pt of charmonium, minimal \pt of both proton and antiproton, minimal \chisq of track impact parameter with respect to the closest primary vertex, minimal \chisq of flight distance of charmonium. Optimisation finds the best requirements to achieve the largest possible value of Figure-of-Merit (FoM) $N_{sig}/\sqrt{N_{sig}+N_{bkg}}$, where $N_{sig}$ is the number of signal \etac events estimated from MC simulation and scaled to the yield in data, $N_{bkg}$ is the number of background events from data sidebands ($2850 \mev < M_{\ppbar} < 2920 \mev$).
The projections of the optimisation map as a function of applied requirements are shown on Fig.\ref{fig:massOpt}.

Optimisation results suggest the optimal requirements to be $\pt(\ppbar)>5.5 \gev$ and $\chisq(FD)>81$, which are then applied in the offline analysis.
Using prompt production cross-sections of the \etac and \jpsi from Section~\ref{sec:results} and efficiencies from MC simulation, the contamination by prompt \jpsi and \etac in the data sample is estimated to be below $10^{-3}$. This number is quoted for illustration purposes.

\subsection{Fit to the invariant mass}
The mass difference $\Delta M _{\jpsi , \, \etac}$ is measured from extended maximum likelihood fit to the $M_{\ppbar}$ distribution. 
The signal and background components are modelled in the same way as discussed in Section~\ref{sec:massFit}. The \pt dependence of the $\sigma_w$, $f_n$ and $\sigma_{\etac}/\sigma_{\jpsi}$ are extracted from fits to MC simulation samples of the \jpsi from \bquark-decays and \etac from \bquark-decays in the same way as discussed in Section~\ref{sec:fitRunI}, while $\sigma_n$ is a free fit parameter. 

\begin{figure}[b]
\centering
\protect\protect\protect\includegraphics[width=1.\linewidth]{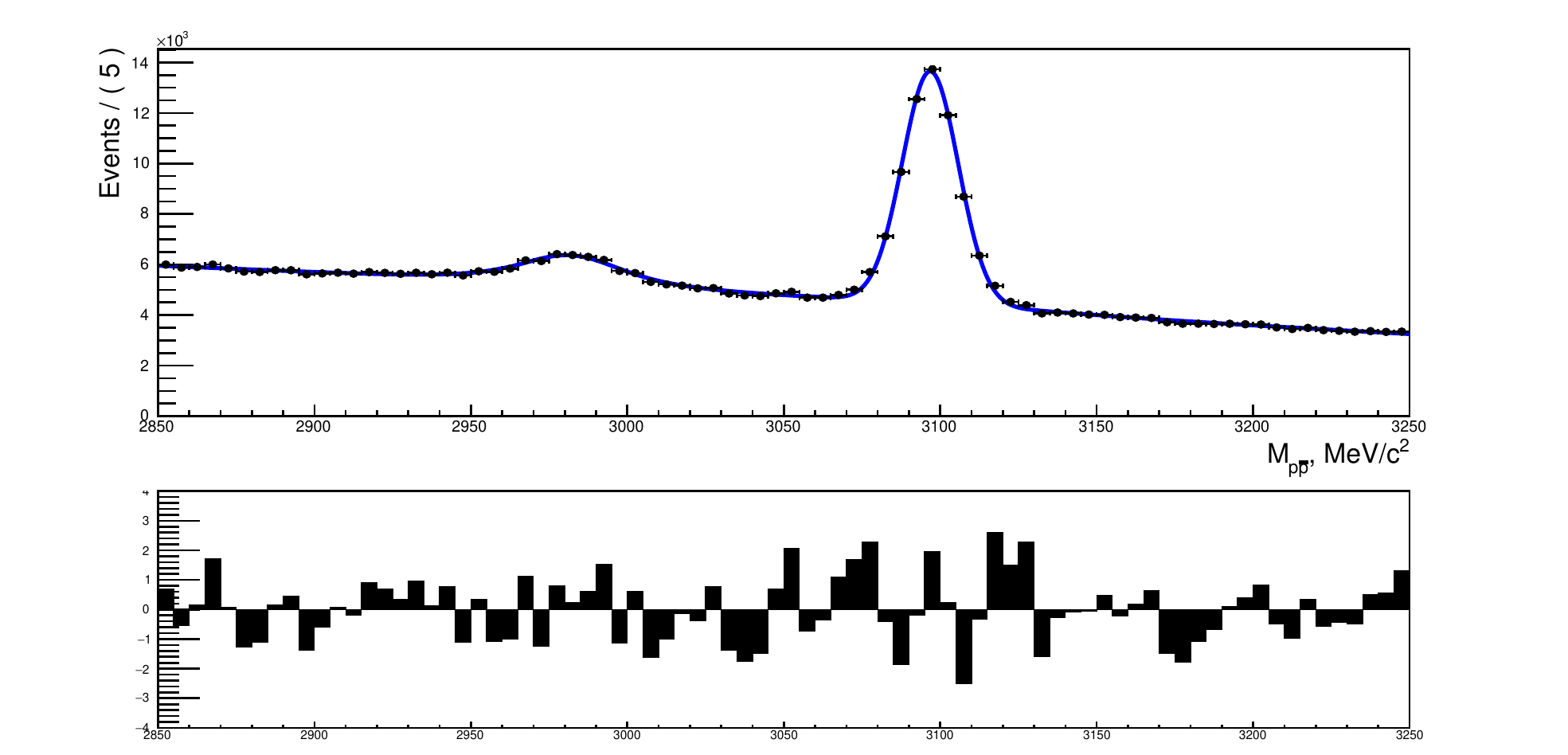}
\put(-385,190){\small{\lhcb}}
\put(-90,83){\colorbox{shadecolor}{\scriptsize $M_{\ppbar}, \mev$}}
\caption
[The distribution of $M_{\ppbar}$.]
{The distribution of $M_{\ppbar}$. The solid blues line represent the fit result. The corresponding pull distribution is shown below each plot.} 
\label{fig:massTopo}
\end{figure}

The master distribution allowing the determination of the \etac mass is shown on Fig.~\ref{fig:massTopo}.
In general, fit yields a good description of the experimental points.
The fit yields the \jpsi and \etac mass difference to be $\Delta M _{\jpsi , \, \etac} = 112.99\pm0.67 \mev$. 

This result is in agreement with the world average value $\Delta M _\jpsi , \, \etac^{PDG} = \etacMassDiffPDG$~\cite{PDG2017}.

\clearpage
\subsection{Systematic uncertainties}
The following list of systematic uncertainties is identical for both \etac production measurement analysis and the $\Delta M _{\jpsi , \, \etac}$ mass difference measurement:
\begin{itemize}
\item Signal description in simultaneous fit to the invariant mass distribution:
  \begin{itemize}
  \item Knowledge of the \etac natural width $\Gamma_{\etac}$;
  \item Invariant mass resolution mismodeling;
  \item \pt-dependence of the \etac and \jpsi resolution ratio $\sigma_{\etac}/\sigma_{\jpsi}$;
  \end{itemize}
\item Background description in simultaneous fit to the invariant mass distribution:
  \begin{itemize}
  \item Combinatorial background description;
  \item Description of the feed-down from the \JpsiToPpbarPiz decay.
  \end{itemize}
\end{itemize}
The estimation of each of these uncertainties is done in the same way as for the \etac production analysis as discussed in Section~\ref{sec:syst}. 
The systematic uncertainty related to momentum scale calibration is estimated by comparing fit result with and without momentum scale calibration applied.

The total systematic uncertainty is calculated as a quadratic sum of individual systematic uncertainties. Table~\ref{tab:deltam} summarises the systematic uncertainty estimates. The dominant source of systematic uncertainty is related to the resolution model and its \pt dependence. 
The total systematic uncertainty is smaller than the statistical one. Hence enlarging data sample by adding more data will improve the precision of this measurement. 

\begin{figure}[hb]
\centering
\protect\protect\protect\includegraphics[width=1.1\linewidth, angle=90]{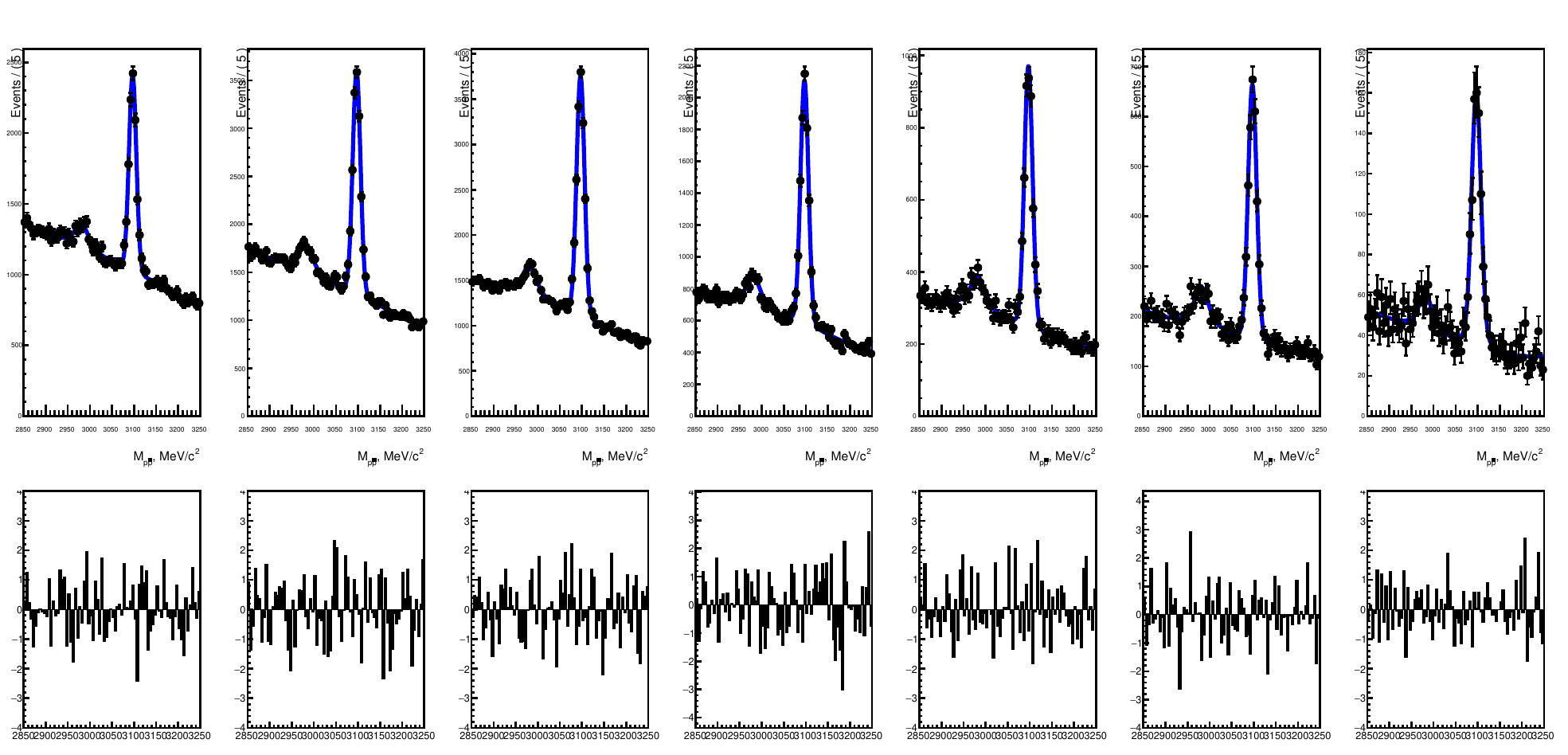}
\put(-8,0){\colorbox{shadecolor}{\makebox(5,560){\textcolor{white}{a}}}}
\caption
[Distributions of $M_{\ppbar}$ in \pt bins.]
{Distributions of $M_{\ppbar}$ in \pt bins. The solid blues line represent the simultaneous fit result. The corresponding pull distribution is shown below each plot.} 
\label{fig:massTopo7Bins}
\end{figure}

As a cross-check, the fit of the invariant mass is performed simultaneously in 7 bins of charmonium transverse momentum to take into account the dependence of resolution on charmonium \pt. 
The bin edges of charmonium $\pt$ are [5.5, 6.5, 8.0, 10.0, 12.0, 14.0, 18.0, 30.0] expressed in \gev. 
Distributions of invariant mass in each \pt-bin is shown on Fig.~\ref{fig:massTopo7Bins}.
In general, fit yields a good description of all $M_{\ppbar}$ distributions in each \pt-bin.
The fit gives the value of the \jpsi and \etac mass difference to be $\Delta M _{\jpsi , \, \etac} = \etacMassDiffInBins$, which is consistent with the nominal result. 

Since inclusive \bquark-decays comprise many exclusive decays of different long-lived \bquark-hadrons, no significant interference between non-resonant \decay{\bquark}{\proton\antiproton X} S-wave decays and \decay{\bquark}{\etac X} is expected. In the conservative estimate, the shape of the \etac peak is described together with the background by the following expression:
\begin{equation}
  f_{[\decay{\bquark}{(\decay{\etac}{\proton\antiproton})X}]+[\decay{\bquark}{\proton\antiproton X}]+interf.} = |A_{RelBW(\etac)} + e^{i\cdot\phi}A_{non.res.}|^2,
\end{equation}
where $A_{RelBW(\etac)}$ is the Relativistic Breit-Wigner amplitude for \etac, $A_{non.res.}$ in the non-resonant amplitude of the \decay{\bquark}{\proton\antiproton X} decays, $\phi$ is the phase difference between the \etac and the non-resonant amplitudes.
The non-resonant amplitude is described by the empirical expression:
\begin{equation}
  A_{non.res.}=A+B\cdot e^{i\cdot \phi_B}\cdot M_{\ppbar} + C\cdot e^{i\cdot \phi_C}\cdot M_{\ppbar}^2,
\end{equation}
where $A$, $B$, $C$, $\phi_B$ and $\phi_C$ are real free fit parameters.

\begin{figure}[h]
\centering
\protect\protect\protect\includegraphics[width=1.\linewidth]{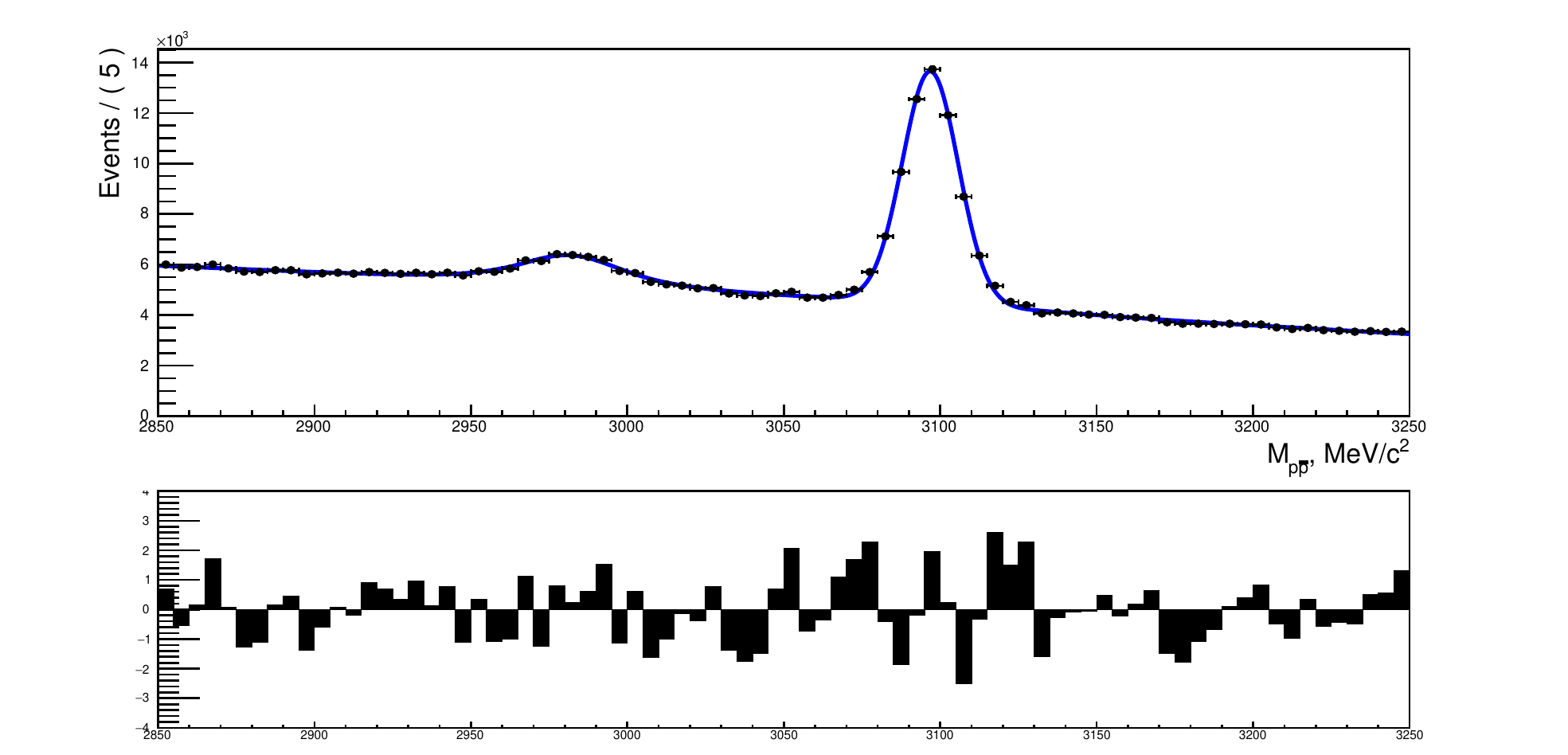}
\put(-380,190){\footnotesize{\lhcb-ANA-2018-035}}
\put(-88,85){\colorbox{shadecolor}{\tiny $M_{\ppbar}, \mev$      }}
\caption
[The distribution of $M_{\ppbar}$. The solid blues lines represent the fit result, which includes the possible interference contribution.]
{The distribution of $M_{\ppbar}$. The solid blues lines represent the fit result, which includes the possible interference contribution. The corresponding pull distributions are shown below each plot.} 
\label{fig:massTopoInterf}
\end{figure}

The result of this fit is is shown on Fig.~\ref{fig:massTopoInterf}.
The data are well described by the fit, and the mass difference is $\Delta M _\jpsi , \, \etac = 113.87\pm0.64 \mev$, consistent with the baseline fit.

\begin{table}[ht]
\begin{center} 
\begin{tabular}{l|*{1}{c}} 
 			                 				& $M_{\jpsi} - M_{\etac}, \mev$ \\ \hline 
Mean value  		         				& 112.99  		    \\ \hline 
Stat. uncertainty	     				  & 0.67 		    	\\ \hline 
Mass resolution model  	                    & 0.08	 		    \\ 
Variation of $\sigma_{\etac}/\sigma_{\jpsi}$& 0.01        \\
Variation of $\Gamma(\etac)$                & 0.04				\\
Comb. bkg. description                      & 0.03			  \\
Contribution from \JpsiToPpbarPiz           & $<0.01$      \\  
Momentum scale                              & 0.05    \\ \hline 
Total systematic uncertainty                & 0.11	 	\\ \hline 
Total uncertainty 								          & 0.68  	\\ 
\end{tabular} 
\end{center} 
 \caption{Systematic uncertainties (in \mev) for the measurement of the \jpsi and \etac mass difference.} 
\label{tab:deltam}
\end{table}

\clearpage
\section{Charmonium spectroscopy study using decays to $\phi\phi$ }
\label{sec:masses}
Masses of the $\etac$ and $\chi_c$ states and natural width of the $\etac$ state are studied below. 
Mass differences within the $\etac$ and $\chi_c$ families are specifically extracted. The measurements in this section are done using the same data sample as in Chapter~\ref{ch:phiphi}. Moreover, the same baseline fit is used.

In a preview to this section the PDG averages and the values obtained in the presented study are summarized in Table~\ref{tab:mres}. The last column represents results obtained using charmonium decays to $\phi\phi$.
\begin{table}[h]
\centering
\begin{tabular}{l|c|c|c}
                     & PDG                 & \ppbar (Section~\ref{sec:mass})& Measured value \\ \hline
$M_{\etac(1S)}$      & $2983.7 \pm 0.7$      & $2983.91\pm0.77\pm0.11$& $2982.81 \pm 0.99 \pm 0.45$ \\ \hline
$M_{\chiczero}$      & $3414.75 \pm 0.31$    && $3412.99 \pm 1.91 \pm 0.62$ \\ \hline
$M_{\chicone}$       & $3510.66 \pm 0.07$    && $3508.38 \pm 1.91 \pm 0.66$ \\ \hline
$M_{\chictwo}$       & $3556.20 \pm 0.09$    && $3557.29 \pm 1.71 \pm 0.66$ \\ \hline
$M_{\etac(2S)}$      & $3639.4 \pm 1.3$      && $3636.35 \pm 4.06 \pm 0.69$ \\ \hline
$\Gamma_{\etac(1S)}$ & $32.0 \pm 0.9$        && $31.35 \pm 3.51 \pm 2.01$ \\ \hline
$\Gamma_{\etac(2S)}$ & $11.3^{+ 3.2}_{- 2.9}$  && $ - $
\end{tabular}
\caption
[Charmonia masses and natural widths.] 
{Charmonia masses (in \mev) and natural widths (in \mev). 
\label{tab:mres}}
\end{table}

Systematic uncertainties from 
the fit to the $\phi \phi$ invariant mass spectrum including additional resonances, 
variation of detector resolution, variation of the fit range, 
variation of the background parametrization, 
uncertainties on the $\chi_c$ mass values, 
and momentum scale calibration uncertainty 
are taken into account. 
In order to evaluate systematic uncertainty related to a potential contribution
from other resonances, contributions from $X (3872)$, $\chiczero (2P)$, and $\chictwo (2P)$
are included in the fit. 
Systematic uncertainties related to detector resolution are conservatively estimated by using 
the $\etac (1S)$ resolution as obtained from the simulation. 
Fit range including only the $\chi_c$ and $\etac (2S)$ region ($3.15 \gevcc - 3.95 \gevcc$) 
and another one ($2.80 \gevcc - 3.70 \gevcc$) excluding $\etac (2S)$ region, 
are used to estimate the corresponding systematic uncertainties. 
Alternative background parametrization using a parabola function is used for the systematic 
uncertainty estimate. 
Uncertainties related to the momentum scale calibration are estimated by varying the 
calibration parameter $\alpha$ by $3 \times 10^{-4}$~\cite{Aaij:2013qja}. 
Effect of a potential contribution from the $f_0 (980)$ state to the 2D fit
is estimated by including the $f_0 (980)$ contribution with the PDG parameters. 
Varying the $f_0 (980)$ mass and natural width within the uncertainties from Ref.~\cite{PDG2016} 
is taken into account. 
The \sPlot technique also gives mass values that are consistent with those in Tab.~\ref{tab:mres} within uncertainties. 
Resulting systematic uncertainty is obtained as a quadratic sum 
of the individual contributions. 
Details of the systematic uncertainty estimate are summarized in Table~\ref{tab:msyst}. 
\begin{table}[h]
\centering
\rotatebox{90}{
\begin{tabular}{l|c|c|c|c|c|c}
    &  $M_{\etac(1S)}$ & $M_{\chiczero}$ & $M_{\chicone}$ & $M_{\chictwo}$ & $M_{\etac(2S)}$ & $\Gamma_{\etac(1S)}$ \\ \hline
With $X (3872)$, $\chiczero (2P)$, $\chictwo (2P)$ 
    & $- 0.01$ & $- 0.02$ & $< 0.01$ & $  0.08$ & $- 0.01$ & $- 0.55$ \\ 
$\etac (1S)$ resolution at MC value
    & $- 0.02$ & $- 0.02$ & $< 0.01$ & $ 0.03$ & $- 0.02$ & $ 0.64$ \\ 
Resolution described by a single Gaussian
    & $<  0.01$ & $< 0.01$ & $< 0.01$ & $< 0.01$ & $ 0.13$ & $ 0.01$ \\ 
Variation of the $r$ parameter in RBW & & & \\ 
between  $0.5 \gevc^{-1}$ and $3 \gevc^{-1}$
    & $<  0.01$ & $< 0.01$ & $< 0.01$ & $< 0.01$ & $< 0.01$ & $< 0.01$ \\ 
Variation of $\Gamma_{\etac (2S)}$ 
    & $<  0.01$ & $< 0.01$ & $< 0.01$ & $< 0.01$ & $ 0.19$ & $ 0.01$ \\ 
Fit region (3.15,3.95) \gevcc
    & $ -   $  & $  0.01$ & $-  0.02$ & $- 0.08$ & $- 0.06$ & $  -   $ \\ 
Fit region (2.80,3.70) \gevcc
    & $ - 0.02 $  & $- 0.01$ & $- 0.01$ & $< 0.01$ & $ 0.02$ & $- 0.19 $ \\ 
Background parametrization
    & $- 0.04$ & $  0.05$ & $  0.04$ & $-  0.06$ & $  0.20$ & $  1.57$ \\ 
$\chi_c$ at PDG values
    & $<  0.01$ & $  -   $ & $  -   $ & $   -  $ & $ 0.02$ & $ 0.14$ \\ 
MC resolution in 2D fit
    & $< 0.01$ & $ 0.01 $ & $- 0.04$ & $ 0.01$ & $ 0.01$ & $  0.02$ \\ 
Add slope parameter                          & & & & & & \\ 
for $\phi \Kp \Km$ component in 2D fit       
    & $- 0.11$ & $ 0.04$ & $- 0.02$ & $- 0.01$ & $ 0.02$ & $ 0.89$ \\ 
Add slope parameter                          & & & & & & \\ 
for $\Kp \Km \Kp \Km$ component in 2D fit 
    & $< 0.01$ & $ 0.03$ & $- 0.02$ & $< 0.01$ & $- 0.01$ & $< 0.01$ \\ 
Momentum scale calibration
    & $ 0.43$ & $ 0.62$ & $ 0.66$ & $ 0.66$ & $ 0.69$ & $ - $ \\ \hline
Combined systematic uncertainty
    & $  0.45$ & $  0.62$ & $  0.66$ & $  0.66$ & $  0.69$ & $  2.01$ 
\end{tabular}
}
\caption{Systematic uncertainties (deviation from the baseline value)
in the measurement of charmonia masses (in \mev) and natural widths (in \mev). 
\label{tab:msyst}}
\end{table}
\clearpage
Uncertainty related to the momentum scale calibration dominates mass determination 
for all $\etac$ and $\chi_c$ states. 
The uncertainty on $\Gamma (\etac (1S))$ measurement is dominated by the background description.

Measured charmonia masses agree with the PDG average values.
The obtained precision of the $\etac (1S)$ mass is similar to the precision 
of the PDG value, while other masses are determined with precisions below the PDG ones. 
The obtained $\etac (1S)$ mass is in agreement with the \lhcb measurement 
using decays to the $\proton \antiproton$ final states~\cite{LHCb-PAPER-2014-029}. 
The value of the $\etac (1S)$ natural width is consistent to the PDG average~\cite{PDG2016}. 

As a cross-check, a stability of the results is checked by using \sPlot 
instead of the 2D fit technique. 
Note that the amount of the pure $\phi \phi$ yield extracted using the \sPlot technique might be affected due to possible correlation of the background shape and the $\phi \phi$ invariant mass. 
Table~\ref{tab:scheckr} compares the results for yield ratios obtained in section~\ref{sec:ccyield} 
to those obtained with the \sPlot technique. 
\begin{table}[h]
\centering
\begin{tabular}{l|c|c}
                            & Measured value & Shift with respect    \\ 
                            &                & to the measured value \\ \hline
$N_{\chiczero} / N_{\etac(1S)}$ & $0.144 \pm 0.022 \pm 0.011$ & $- 0.006$ \\ \hline
$N_{\chicone} / N_{\etac(1S)}$  & $0.071 \pm 0.015 \pm 0.006$ & $- 0.002$ \\ \hline
$N_{\chictwo} / N_{\etac(1S)}$  & $0.094 \pm 0.016 \pm 0.007$ & $- 0.002$ \\ \hline
$N_{\etac(2S)} / N_{\etac(1S)}$ & $0.056 \pm 0.016 \pm 0.005$ & $- 0.007$ \\ \hline
$N_{\chicone} / N_{\chiczero}$  & $0.494 \pm 0.107 \pm 0.012$ & $- 0.006$ \\ \hline
$N_{\chictwo} / N_{\chiczero}$  & $0.656 \pm 0.121 \pm 0.014$ & $- 0.013$ 
\end{tabular}
\caption{Cross-check for charmonia yield ratios using the \sPlot technique.
\label{tab:scheckr}}
\end{table}
The weighting coefficients were obtained from the $\phi_1$ vs. $\phi_2$ fit. 
Then they were used for unbinned maximum log likelihood fit of the $M (2 \phi)$.
Table~\ref{tab:scheckm} compares the results for charmonia mass and $\Gamma_{\etac(1S)}$ values obtained 
in section~\ref{sec:masses} to those obtained with the \sPlot technique. 
\begin{table}[t!]
\centering
\begin{tabular}{l|c|c}
                             & Measured value, \mev (\mev) & Shift with respect    \\ 
                             &                & to the measured value, \mev \\ \hline
$M_{\etac(1S)}$ & $2982.81 \pm 0.99 \pm 0.45$ & $0.37$ \\ \hline
$M_{\chiczero}$ & $3412.99 \pm 1.91 \pm 0.62$ & $0.32$ \\ \hline
$M_{\chicone}$  & $3508.38 \pm 1.91 \pm 0.66$ & $0.82$ \\ \hline
$M_{\chictwo}$  & $3557.29 \pm 1.71 \pm 0.66$ & $0.33$ \\ \hline
$M_{\etac(2S)}$ & $3636.35 \pm 4.06 \pm 0.69$ & $- 1.33$ \\ \hline
$\Gamma_{\etac(1S)}$ & $31.35 \pm 3.51 \pm 2.00$ & $- 0.20$ 
\end{tabular}
\caption{Cross-check for charmonia mass and $\Gamma_{\etac(1S)}$ values using the \sPlot technique.
\label{tab:scheckm}}
\end{table}
Table~\ref{tab:scheckdm} compares the results for charmonia mass difference values obtained 
in section~\ref{sec:masses} to those obtained with the \sPlot technique. 
\begin{table}[h]
\centering
\begin{tabular}{l|c|c}
                             & Measured value, \mev & Shift with respect    \\ 
                             &                        & to the measured value, \mev \\ \hline
$M_{\chicone} - M_{\chiczero}$  & $95.38 \pm 2.71 \pm 0.11$ & $0.50$ \\ \hline
$M_{\chictwo} - M_{\chiczero}$  & $144.28 \pm 2.59 \pm 0.17$ & $0.01$ \\ \hline
$M_{\etac(2S)} - M_{\etac(1S)}$ & $653.54 \pm 4.22 \pm 0.42$ & $- 1.71$ 
\end{tabular}
\caption{Cross-check for charmonia mass difference values using the \sPlot technique.
\label{tab:scheckdm}}
\end{table}
The results are found to be stable within the statistical uncertainties. 

Charmonia mass differences within families $M_{\chicone} - M_{\chiczero}$, $M_{\chictwo} - M_{\chiczero}$, 
and $M_{\etac(2S)} - M_{\etac(1S)}$
are obtained in order to cancel part of the systematic 
uncertainty, and provide inputs for direct comparison with theory. 
Table~\ref{tab:dmres} summarizes the results for charmonia mass differences. 
\begin{table}[h]
\centering
\begin{tabular}{l|c|c}
    & PDG & Measured value \\ \hline
$M_{\chicone} - M_{\chiczero}$ & $95.91 \pm 0.83$ & $95.38 \pm 2.71 \pm 0.11$ \\ \hline
$M_{\chictwo} - M_{\chiczero}$ & $141.45 \pm 0.32$ & $144.28 \pm 2.59 \pm 0.17$ \\ \hline
$M_{\etac(2S)} - M_{\etac(1S)}$ & $655.70 \pm 1.48$ & $653.54 \pm 4.22 \pm 0.42$ 
\end{tabular}
\caption{Charmonia mass differences (in \mev). 
\label{tab:dmres}}
\end{table}
Systematic uncertainties from 
the fit to the $\phi \phi$ invariant mass spectrum including additional resonances, 
variation of detector resolution, variation of the fit range, 
variation of the background parametrization (parabola), 
momentum scale calibration uncertainty, 
and potential contribution from the $f_0 (980)$ state with a mass and natural width varied 
within uncertainties of Ref.~\cite{PDG2016} to the 2D fit technique are taken into account. 
Resulting systematic uncertainty is obtained as a quadratic sum 
of the individual contributions. 
Details of the systematic uncertainty estimate are summarized in Table~\ref{tab:dmsyst}. 
\begin{table}[h]
\centering
\begin{tabular}{l|c|c|c}
    &  $M_{\chicone} - M_{\chiczero}$ & $M_{\chictwo} - M_{\chiczero}$ & $M_{\etac(2S)} - M_{\etac(1S)}$ \\ \hline
With $X (3872)$, $\chiczero (2P)$, $\chictwo (2P)$  
        & $  0.03$ & $  0.11$ & $-  0.01$ \\ 
Masses of $\chi_c$ states & & & \\ 
at nominal values
        & $ - $ & $ - $ & $ 0.02$ \\ 
$\etac (1S)$ resolution at MC value
        & $ 0.01$ & $ 0.05$ & $- 0.04$ \\ 
Resolution described & & & \\ 
by a single Gaussian 
        & $< 0.01$ & $< 0.01$ & $< 0.01$ \\ 
Variation of $r$ parameter & & & \\ 
between  $0.5 \gevc^{-1}$ and $3 \gevc^{-1}$
        & $< 0.01$ & $< 0.01$ & $< 0.01$ \\ 
Variation of $\Gamma_{\etac (2S)}$ 
        & $ 0.01$ & $ 0.01$ & $0.19$ \\ 
Fit region (3.15,3.95) \gevcc
        & $- 0.01$ & $- 0.06$ & $ -  $ \\ 
Fit region (2.80,3.70) \gevcc
        & $0.02$ & $0.03$ & $0.02 $ \\ 
Background parametrization
        & $< 0.01$ & $- 0.08$ & $ 0.24$ \\ 
MC resolution in the 2D fit 
        & $- 0.05$ & $< 0.01$ & $- 0.01$ \\ 
Add slope parameter                          & & & \\ 
for the $\phi \Kp \Km$ component    & $- 0.06$ & $- 0.04$ & $ 0.12$ \\ 
in 2D fit       & & & \\ 
Add slope parameter                          & & & \\ 
for the $\Kp \Km \Kp \Km$ component    & $- 0.05$ & $- 0.03$ & $- 0.01$ \\ 
in 2D fit   & & &  \\ 
Momentum scale calibration
        & $  0.04$ & $ 0.04$ & $ 0.26$ \\ \hline
Combined systematic uncertainty
        & $  0.11$ & $  0.17$ & $  0.42$ 
\end{tabular}
\caption{Systematic uncertainties (deviation from the baseline value)
in the measurement of charmonia mass differences (in \mev). 
\label{tab:dmsyst}}
\end{table}
Uncertainty related to the momentum scale calibration dominates 
the $M_{\chicone} - M_{\chiczero}$ and $M_{\etac(2S)} - M_{\etac(1S)}$
mass difference measurements. 
Systematic uncertainty of the  $M_{\chictwo} - M_{\chiczero}$ measurement is dominated by the MC resolution.

The results are consistent to the central values obtained above withing statistical uncertainties. 
Measured charmonia mass differences agree with the PDG average values 
but are less precise than the world averages. 

Stability of the obtained results on the mass difference has been also cross-checked by shifting the $\phi \phi$ invariant mass distribution 
by half a bin and
and by using \sPlot technique instead of the 2D fit. 

\clearpage
\section{Summary and discussion}
\label{sec:mass_summary}

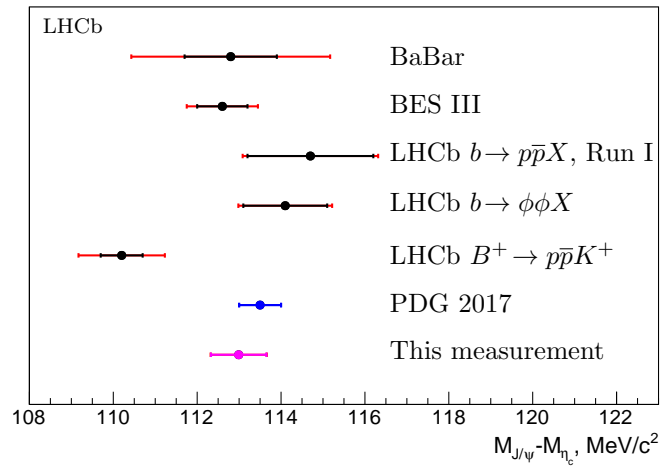
\begin{figure}[h]
\centering
\protect\protect\protect\includegraphics[width=0.65\linewidth]{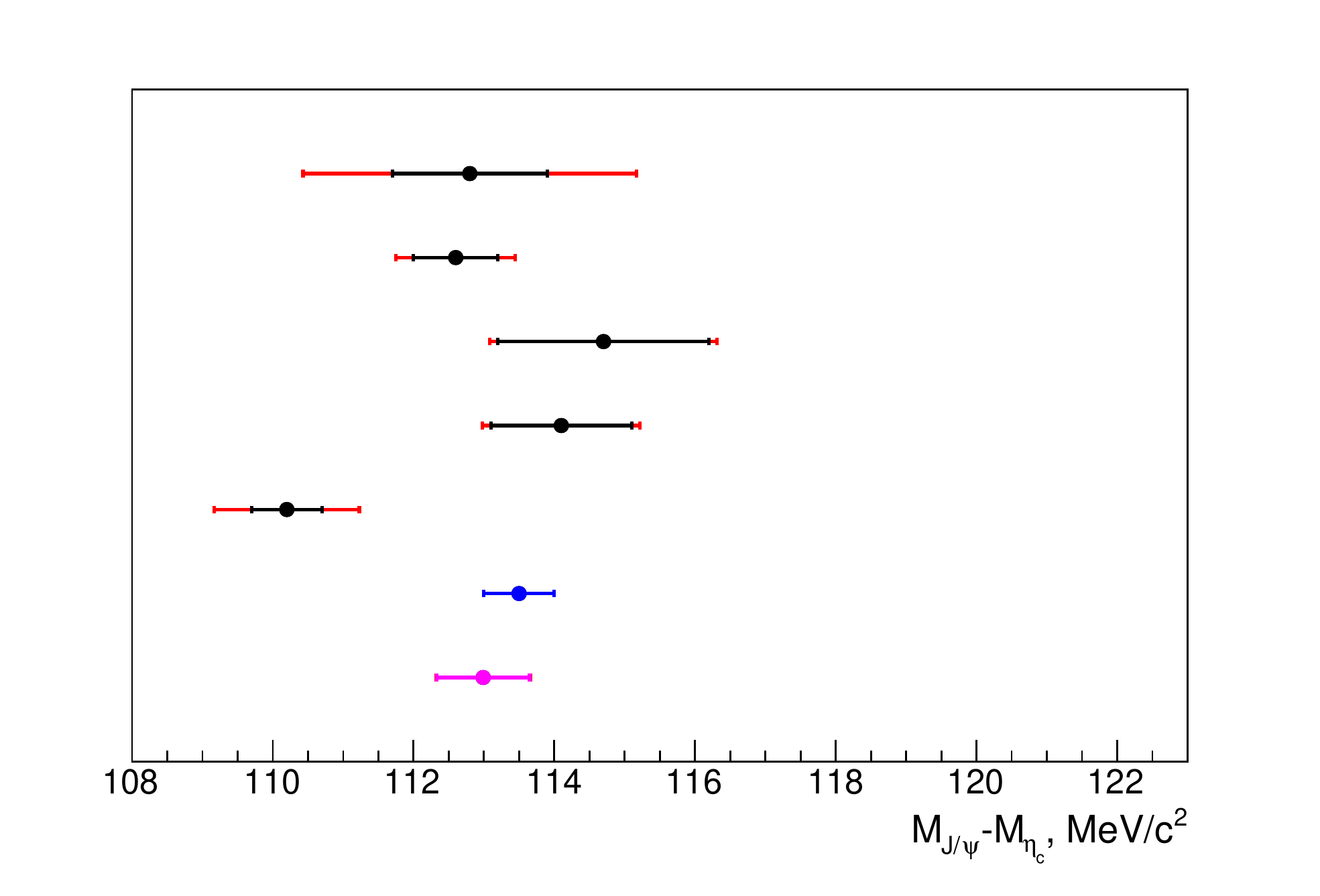}
\put(-260,170){\scriptsize{\lhcb}}
\put(-130,158){\footnotesize{\babar}}
\put(-130,140){\footnotesize{BES III}}
\put(-130,122){\footnotesize{\lhcb \decay{\bquark}{\ppbar X}}, Run I}
\put(-130,103){\footnotesize{\lhcb \decay{\bquark}{\phi\phi X}}}
\put(-130,83){\footnotesize{\lhcb \decay{\Bp}{\ppbar \Kp}}}
\put(-130,65){\footnotesize{PDG 2017}}
\put(-130,47){\footnotesize{This measurement}}
\caption
[Mass difference $M_{\jpsi}-M_{\etac}$ measurement compared to the measurements from \babar, BES III and \lhcb.]
{Mass difference $M_{\jpsi}-M_{\etac}$ measurement compared to the measurements from \babar~\cite{Lees:2014iua}, BES III~\cite{BESIII:2011ab} and \lhcb~\cite{LHCb-PAPER-2014-029,phiphi,LHCb-PAPER-2016-016}; black error bars represent statistical uncertainties, red error bars represent total uncertainties. The blue point with error bars shows the world average, the magenta point with error bars represents this measurement.} 
\label{fig:massMeasurements}
\end{figure}

Using a sample of $\decay{\bquark}{(\decay{\ccbar}{\ppbar})X}$ candidates, the \jpsi and \etac mass difference is measured. 
The obtained result, 
$\Delta M_\jpsi , \, \etac = M_{\jpsi} - M_{\etac} = \etacMassDiff$, 
is consistent with the world average value and is the most precise single \etac mass measurement to date.
The comparison of the obtained result with recent BES III result~\cite{BESIII:2011ab}, the latest $B$-factory measurement~\cite{Lees:2014iua}, the \lhcb result using decays to \ppbar within Run I data~\cite{LHCb-PAPER-2014-029}, the \lhcb measurement using \decay{\bquark}{\phi\phi X}~\cite{phiphi} and the \lhcb measurement using exclusive \decay{\Bp}{\ppbar \Kp} decays~\cite{LHCb-PAPER-2016-016} is shown on Fig.~\ref{fig:massMeasurements}. The second most precise to date measurement of \etac mass has been performed by BES III~\cite{BESIII:2011ab}, where a sample of \etac produced in $\decay{\psitwos}{\etac\gamma}$ was used. The BES III measurement is more complicated requiring a description of the corresponding tails of the \etac signal model taking place due to radiative transitions in the production process. The \lhcb measurement from Ref.~\cite{LHCb-PAPER-2016-016} takes into account interference between \decay{\Bp}{(\etac\to\ppbar) \Kp} and \decay{\Bp}{\ppbar \Kp} non-resonant decays. 
The new \lhcb result obtained here represents not only the single most precise determination of the mass splitting between the \jpsi and the \etac, but it is also free from the systematic effects which can influence other measurements of similar precision.

Masses and natural widths of the \etac and $\chi_c$ states are determined to be 
\begin{align*}
M_{\etac(1S)} &= 2982.81 \pm 0.99 \pm 0.45 \mev \ , \\ 
M_{\chiczero} &= 3412.99 \pm 1.91 \pm 0.62 \mev \ , \\
M_{\chicone}  &= 3508.38 \pm 1.91 \pm 0.66 \mev \ , \\
M_{\chictwo}  &= 3557.29 \pm 1.71 \pm 0.66 \mev \ , \\
M_{\etac(2S)} &= 3636.35 \pm 4.06 \pm 0.69 \mev \ , \\ 
\Gamma_{\etac(1S)} &= 31.35 \pm 3.51 \pm 2.00 \mev \ . 
\end{align*}
using a sample of $\decay{\bquark}{(\decay{\ccbar}{\phi\phi})X}$ decays.
Measured charmonia masses agree with the PDG average values. 
The obtained precision of the $\etac(1S)$ mass is similar to the precision 
of the PDG value, while other masses are determined with precisions below the PDG ones. 

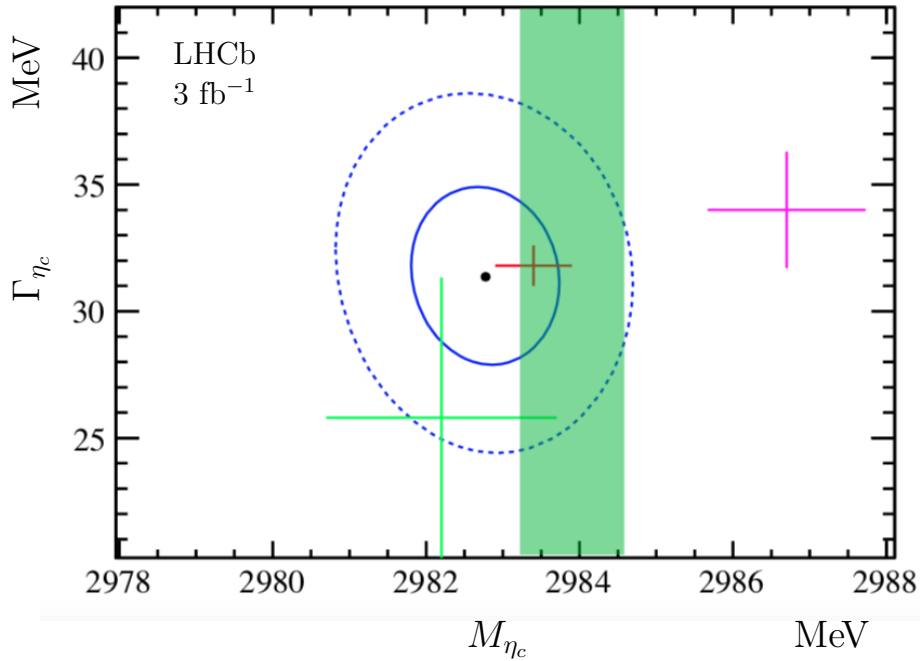
\begin{figure}[h]
\centering
                \protect\protect\includegraphics[width=0.75\textwidth]{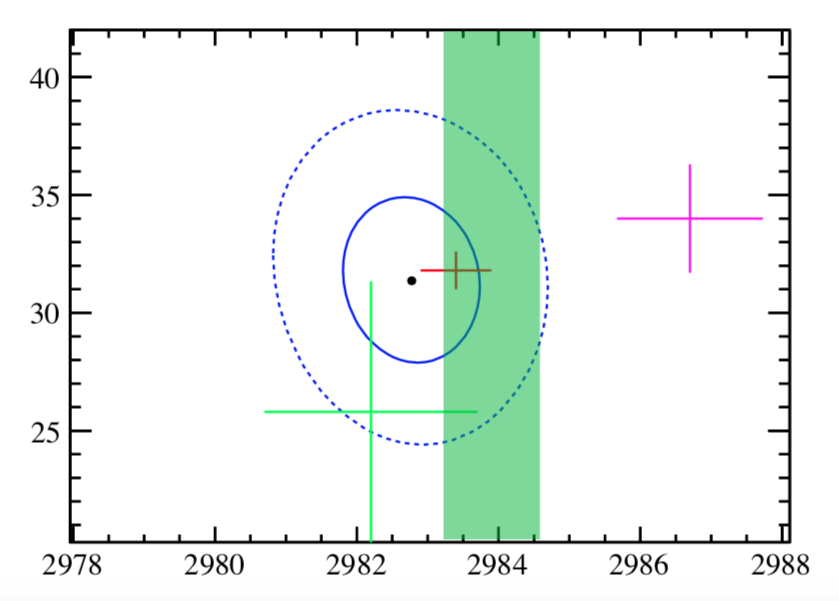}
                \put(-60,-10) {\large \mev}
                \put(-180,-10) {\large $M_{\etac}$}
                \put(-350,120){\rotatebox{90}{\large $\Gamma_{\etac}$}}
                \put(-350,190){\rotatebox{90}{\large \mev}}
                \put(-290,210) {\lhcb}
                \put(-290,195) {3 fb$^{-1}$}
\caption
[The \etac mass measurements from Section~\ref{sec:mass} and contour plot of $\Gamma_{\etac}$ and $M_{\etac}$ using $\etac \to \phi \phi$  decay for the combined data sample.]
{The \etac mass measurements from Section~\ref{sec:mass} (green band) and contour plot of $\Gamma_{\etac}$ and $M_{\etac}$ using $\etac \to \phi \phi$ (blue contour) decay for the combined data sample. The two curves indicate 68.3~C.L. (one-sigma) and 95.5~C.L. (two-sigma) contours. Only statistical uncertainties are shown. The red, green, and magenta points with error bars indicate the PDG average~\cite{PDG2016}, the result from Ref.~\cite{LHCb-PAPER-2014-029}, 
and the result from Ref.~\cite{LHCb-PAPER-2016-016}, respectively.} 
\label{fig:contb}
\end{figure}

Fig.~\ref{fig:contb} shows the $\Gamma_{\etac (1S)}, \, M_{\etac (1S)}$ contour plot, obtained from the analysis 
of \bquark-hadron decays into \etac meson, where the \etac candidates are reconstructed 
via the $\etac (1S) \to \phi \phi$ decay, for the combined data sample. 
Measurements of the \etac mass and natural width using \etac meson decays to $\phi \phi$ are consistent 
with the studies using decays to \proton\antiproton~\cite{LHCb-PAPER-2014-029} superimposed on the plot 
as a green point with error bars, 
and with the PDG average~\cite{PDG2016} superimposed on the plot as a red point with error bars. 
The measured \etac mass is below the result in Ref.~\cite{LHCb-PAPER-2016-016}. 
The obtained precision of the $\etac (1S)$ mass is similar to the precision 
of the PDG value, while 
the $\etac$ natural width measurement has a precision below that of the PDG average value.

Mass differences within charmonia families are measured to be 
\begin{align*}
M_{\chicone} - M_{\chiczero} &= 95.38 \pm 2.71 \pm 0.11 \mev \ , \\ 
M_{\chictwo} - M_{\chiczero} &= 144.28 \pm 2.59 \pm 0.17 \mev \ , \\ 
M_{\etac(2S)} - M_{\etac(1S)} &= 653.54 \pm 4.22 \pm 0.42 \mev \ .
\end{align*}
Measured charmonia mass differences agree with the PDG average values 
and have precisions below the PDG ones.

For all measurements listed in this chapter statistical uncertainty is larger that the systematic one. Therefore, measurements will benefit from larger data samples.

\chapter{Study of \Bs decays to $\phi$ mesons}
\label{ch:bs}
In addition to charmonium production measurements, signatures of multiple $\phi$ mesons can be used to study decays of the \Bs meson.
Large centre-of-mass energy together with the powerful charged hadron ID and selective trigger make the \lhcb experiment the ideal place for the measurements of \Bs decays.

Section~\ref{sec:bstwo} describes the analysis of relatively well-known $\Bs\to\phi\phi$ decay and measurement of its branching fraction, which is, however, a cross-check to another more precise \lhcb measurement~\cite{Aaij:2015cxj}. The analysis described in Section~\ref{sec:bstwo} compares the yields of observed $\decay{\etac}{\phi\phi}$ and $\Bs\to\phi\phi$ and has another normalisation channel to that of nominal \lhcb measurement, which leads to different sources of systematic uncertainties between the two measurements. In addition to that, using the measured $\BR(\Bs\to\phi\phi)$, the $\BR(\etac\to\phi\phi)$ is extracted aiming to solve a consistency problem in the corresponding world average value.

Section~\ref{sec:bsthree} describes a first evidence of the $\Bs\to\phi\phi\phi$ decay, studies of its decay model and search for intermediate resonances. Finally, the results obtained are summarised in Section~\ref{sec:BSsummary}.
\clearpage
\section{The $\Bs \ra \phi \phi$ decay}
\label{sec:bstwo}
\begin{figure}[h]
\centering
\protect\protect\includegraphics[width=0.5\linewidth]{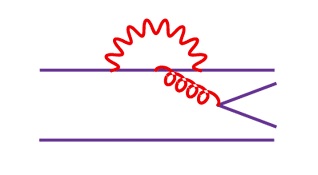}
                \put(-210,70) {{\bquarkbar}}
                \put(-20,20) {{\squark}}
                \put(-210,20) {{\squark}}
                \put(-20,30) {{\squarkbar}}
                \put(-20,60) {{\squark}}
                \put(-20,70) {{\squarkbar}}
                \put(-120,117) {{\Wp}}
                \put(-240,45) {{\Bs}}
                \put(0,25) {{$\phi$}}
                \put(0,65) {{$\phi$}}
\caption{Quark diagram describing $\Bs \to \phi \phi$ decay.} \label{fig:diagtwo}
\end{figure}

The $\Bs \ra \phi \phi$ is forbidden at the tree level in the SM and proceeds via a gluonic penguin diagram $\bquark \to \squark \squarkbar \squark$ shown on Fig.~\ref{fig:diagtwo}. This rare decay is an excellent probe of potential New Physics (NP) contributions and can be used to search for new heavy particles, which enter the penquin loop~\cite{Bartsch:2008ps,Beneke:2006hg,Cheng:2009mu}.

Measurements of the polarization amplitudes and triple product asymmetries 
in the $\Bs \ra \phi \phi$ decay mode were pointed out to provide 
important probes of the 
non-factorizable penguin-annihilation effects~\cite{Kagan:2004uw}, 
final state interactions~\cite{Datta:2007qb}, 
and NP contributions to the penguin loops~\cite{Chen:2005mka,Huang:2005qb}. 
Recently, the \lhcb experiment performed 
a measurement of the time-dependent \CP-violating asymmetry 
in $\Bs \ra \phi \phi$ decays~\cite{LHCb-PAPER-2013-007}, 
and probed the \CP-violating phase \phis for the first time. 
The branching fraction $\BR ( \Bs \ra \phi \phi )$ is calculated 
using pertubative QCD approach (Ref.~\cite{Ali:2007ff} and references therein) 
and QCD factorization (Ref.~\cite{Cheng:2009mu,Beneke:2006hg} and references therein). 
However, experimental knowledge of the branching fraction for this mode 
remains limited, with measurements from CDF~\cite{Acosta:2005eu,Aaltonen:2011rs}
and upper limit set by the SLD experiment~\cite{Abe:1999ze}. 
In the recent CDF result~\cite{Aaltonen:2011rs}, 
$\BR ( \Bs \ra \phi \phi ) = ( 17.7 \pm 2.4 ^{+ 5.7} _{- 3.2} ) \times 10^{-6}$, 
the systematic uncertainty is dominated by the precision 
of the branching fraction for the normalization channel $\Bs \to \jpsi \phi$. 
This measurement was limited by large systematic uncertainties and calls for 
the $\BR ( \Bs \ra \phi \phi )$ determination using alternative approach.
Later, \lhcb provided a new measurement using the $B\to\phi K^{*0}(892)$ decay as a normalization
$\BR ( \Bs \ra \phi \phi ) = ( 18.4 \pm 0.5 \pm 0.7 \pm 1.1_{f_s/f_d}\pm 1.2_{norm} ) \times 10^{-6}$, where the third uncertainty is due to \bquark quark fragmentation ratio $f_s/f_d$ and the last uncertainty is related to normalization and relevant branching fraction~\cite{Aaij:2015cxj}.

Reconstructing the \Bs meson via its decay to $\phi\phi$, 
and comparing the \etac and \Bs event yields, I suggested a new alternative approach to access $\BR ( \Bs \to \phi \phi )$ described below. 

\subsection{Signal extraction and systematic uncertainties}
The $\Bs \to \phi \phi$ decay mode is studied below to extract the $\BR(\Bs\to\phi\phi)/\BR(\etac\to\phi\phi)$ ratio and as a normalization mode for the 
$\BR ( \Bs \to \phi \phi \phi )$ measurement. 

The $\Bs \to \phi \phi$ candidates are reconstructed 
using selection criteria similar to those applied for charmonia reconstruction 
via decays to $\phi \phi$ in the production analysis, as discussed in section~\ref{sec:ccyield}. 
Charged kaon separation against pions, $ProbNNk > 0.1$, 
and kaon transverse momentum $\pt > 0.5 \gev$ are required. 
Kaons from each $\phi$ candidate are required to form a good quality vertex, $\chisq < 25$. 
Two $\phi$ candidates are required to also form a good quality common vertex, $\chisqndf < 9$, 
well distinguished from the corresponding primary vertex with a significant distance 
between the two vertices, $\chisq > 100$. 
A dedicated MC sample of $\Bs\to\phi\phi$ decays is used to study signal resolution and efficiency.
The decay model uses amplitudes measured by CDF. 
The following efficiency ratio for $\etac\to\phi\phi$ and $\Bs\to\phi\phi$ decays was obtained.
\begin{align*}
\frac{\varepsilon ( \etac (1S) \to \phi \phi )}{\varepsilon ( \Bs \to \phi \phi )} &= 0.31 \pm 0.01
\end{align*}

\begin{figure}[h]
\centering
\protect\protect\includegraphics[width=0.8\linewidth]{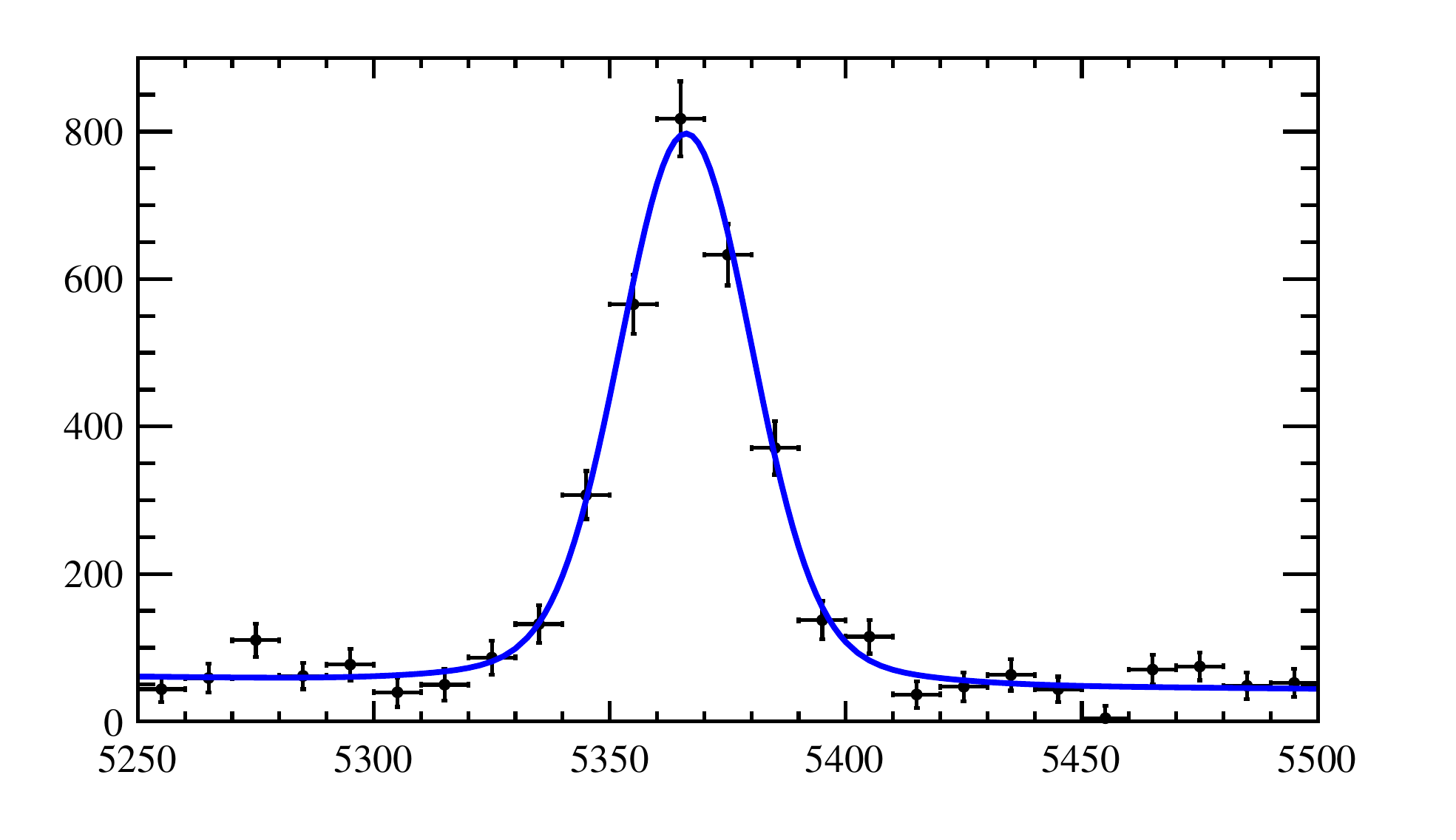}
                \put(-360,100){\rotatebox{90}{\small{Events/10 \mev}}}
                \put(-60,-5) {\small{\mev}}
                \put(-200,-5) {\small{$M( \phi \phi )$}}
                \put(-140,165) {\lhcb}
                \put(-140,150) {3 fb$^{-1}$}
\caption[Invariant mass spectrum of the $\phi \phi$ combinations in the \Bs region]{Invariant mass 
spectrum of the $\phi \phi$ combinations in the region of the \Bs mass for combined data sample. 
The number of candidates in each bin comes from the 2D fit, decribed in section~\ref{sec:procedure}.
} 
\label{fig:bsphiphi}
\end{figure}
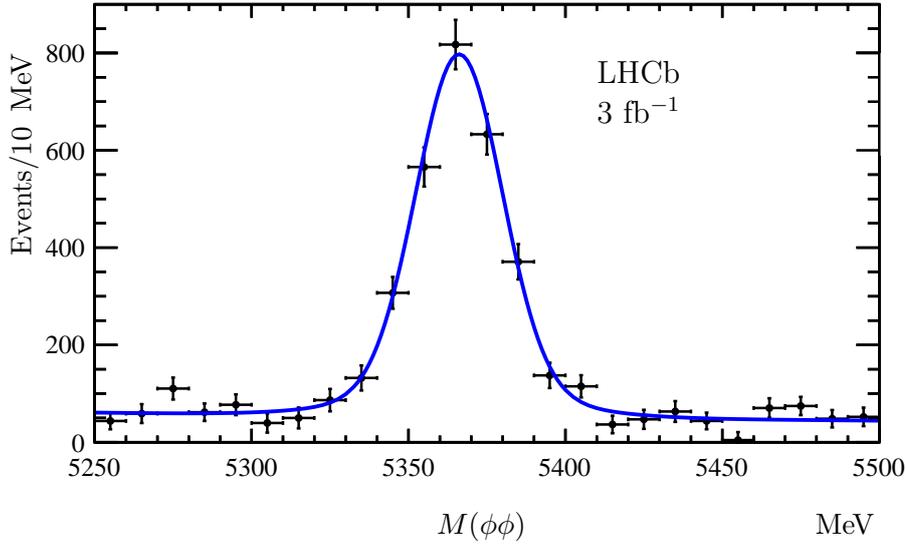

The two-dimensional fit selects pure $\phi \phi$ combinations, 
suppressing a significant reflection from $\Bz \to \phi \Kstarz$. 
Separate analysis of the data samples, corresponding to $\protect\sqs = 7 \protect\tev$ and $\protect\sqs = 8 \protect\tev$, 
shown consistent results for signal and background models and event yields, 
so that the combined data sample is considered. 
A fit to the $\phi \phi$ invariant mass spectrum in the region of the \Bs mass is shown on Fig.~\ref{fig:bsphiphi}. 

A double Gaussian function is used to describe the \Bs signal shape, while an exponential function 
modelled combinatorial background. 
The ratio of the two Gaussian widths of $0.52 \pm 0.01$ and the fraction of narrow 
Gaussian of $0.81 \pm 0.01$ are taken from simulation. 
The fit yields $2701 \pm 114 \pm 84$ candidates in the \Bs signal peak, 
and the \Bs mass value $M_{\Bs} = 5366.15 \pm 0.64 \mev$, in agreement 
with the PDG average of $5366.77 \pm 0.24 \mev$~\cite{PDG2016}. 

In the measurement of the \Bs signal yield, systematic uncertainties from the 
variation of background shape (constant), 
resolution description 
and potential $f_0 (980)$ contribution to the 2D fit technique 
are taken into account. 
Resulting systematic uncertainty is obtained as a quadratic sum 
of the individual contributions. 
Details of the systematic uncertainty estimate are summarized in Table~\ref{tab:bstwosyst}. 
\begin{table}[h]
\centering
\begin{tabular}{l|c}
    &  $N ( \Bs )$ \\ \hline
Background shape variation, $\phi \phi$ & $- 2$ \\ 
Resolution in 2D fit at MC value & $- 23$ \\ 
$f_0 (980)$ in the 2D fit & $2$ \\ 
Resolution for \Bs described by a single Gaussian & $- 81$ \\ \hline
Combined & $84$ 
\end{tabular}
\caption
[Systematic uncertainties in the measurement of the \Bs signal yield.]
{Systematic uncertainties (deviation from the baseline value)
in the measurement of the \Bs signal yield (in number of candidates) 
\label{tab:bstwosyst}}
\end{table}
Uncertainty related to the resolution description in the 2D fit and the \Bs resolution description give the largest contribution
the systematic uncertainty in the \Bs signal yield determination. 

\subsection{Extraction of $\BR ( \Bs \ra \phi \phi )$}
Relating $\Bs \to \phi \phi$ decay to the production of 
the $\etac (1S)$ state in \bquark-hadron decays provides an alternative approach 
of the $\BR ( \Bs \ra \phi \phi )$ determination. 

In the measurement of the $\Bs \to \phi \phi$ branching fraction, 
the normalization to the $\bquark \to \etac (1S) X$ is used, where the $\etac (1S)$ production in \bquark-decays was measured in Ref.~\cite{LHCb-PAPER-2014-029}. 
Thus having reconstructed the \etac and \Bs mesons in the $\etac \to \phi\phi$ and $\Bs \to \phi\phi$ 
decay modes, comparing the \etac and \Bs event yields, 
and accounting for the efficiency difference,  
the branching fraction $\BR ( \Bs \to \phi \phi )$ can be obtained. 
In the ratio of the \Bs production, where \Bs is reconstructed via the $\Bs \to \phi \phi$ decay, 
to the \etac production in \bquark-hadron inclusive decays, the \Bs fragmentation from the \bquark-quark 
has to be taken into account, 
\begin{equation}
\frac{\BR ( \bquarkbar \to \Bs ) \times \BR ( \Bs \to \phi \phi )}{\BR ( b \to \etac X ) \times \BR ( \etac \to \phi \phi )} 
= \frac{N_{\Bs}}{N_{\etac}} \times \frac{ \varepsilon_{\etac} }{ \varepsilon_{\Bs} } \ ,
\end{equation}
where $N_{\etac}$ and $N_{\Bs}$ are the event yields for \etac and \Bs signals, respectively, 
and the efficiency ratio estimated using MC.
The ratio of \Bs production to the \etac inclusive production in \bquark-hadron decays 
is thus obtained to be 
\begin{equation}
\frac{\BR ( \bquarkbar \to \Bs ) \times \BR ( \Bs \to \phi \phi )}{\BR ( b \to \etac X ) \times \BR ( \etac \to \phi \phi )} 
= 0.128 \pm 0.010 \pm 0.007 \ .
\end{equation}
Using the above value, together with the ratio 
$\BR ( \bquark \to \etac X ) / \BR ( \bquark \to \jpsi X )$ obtained in the same \pt region, 
allows to extract the branching fraction $\BR ( \Bs \to \phi \phi )$ as 
\begin{align}
\BR ( \Bs \to \phi \phi ) & = 
\frac{N_{\Bs}}{N_{\etac}} \times \frac{ \varepsilon_{\etac} }{ \varepsilon_{\Bs} } \times \\
 & \times \frac{\BR ( b \to \etac X ) \times \BR ( \etac \to \proton \antiproton )}{\BR ( b \to \jpsi X ) \times \BR ( \jpsi \to \proton \antiproton )} 
\times \\ 
 & \times \frac{\BR ( \etac \to \phi \phi )}{\BR ( \etac \to \proton \antiproton )}
\times \BR ( b \to \jpsi X ) \times \BR ( \jpsi \to \proton \antiproton ) / \BR ( \bquarkbar \to \Bs ) \ .
\label{eq:BrBs2phiphi}
\end{align}
In the above expression, 
the ratio on the second line has been measured in Ref.~\cite{LHCb-PAPER-2014-029} to be 
$\frac{\BR ( b \to \etac X ) \times \BR ( \etac \to \proton \antiproton )}{\BR ( b \to \jpsi X ) \times \BR ( \jpsi \to \proton \antiproton )} = 0.302 \pm 0.039 \pm 0.015 = 0.302 \pm 0.042$ 
for $\pt ( \etac, \jpsi ) > 6.5 \gev$, and can be used as an estimate for the present calculations; 
the ratio of the \etac branching fractions to the $\phi \phi$ and $\proton \antiproton$ final states 
$\BR ( \etac \to \phi \phi ) / \BR ( \etac \to \proton \antiproton ) = 1.17 \pm 0.18$~\cite{PDG2016}
and is dominated by the accuracy of BES measurements; 
the inclusive branching fraction of \bquark-hadrons into \jpsi, $\BR ( b \to \jpsi X ) = ( 1.16 \pm 0.10 ) \%$, 
where decays of the mixture of \Bpm, \Bz, \Bs and \bquark baryons are considered~\cite{PDG2016}; 
and the branching fraction of the \jpsi meson decay to the \proton\antiproton final state 
$\BR ( \jpsi \to \proton \antiproton ) = ( 2.120 \pm 0.029 ) \times 10^{-3}$~\cite{PDG2016}. 

The fragmentation of the \bquarkbar quark to \Bs is driven by strong dynamics
in the nonperturbative regime, and no reliable theoretical prediction can be made. 
Thus the $f_s$ is also determined experimentally. 
The \lhcb experiment determined $\frac{f_s}{f_d}$ 
via semileptonic~\cite{LHCb-PAPER-2011-018} and hadronic~\cite{LHCb-PAPER-2012-037} decays, 
and found it consistent with being independent on the rapidity and transverse momentum. 
The two measurements agree to each other, and yield the average of 
\begin{align*}
\frac{f_s}{f_d} = 0.256 \pm 0.020 \ .
\end{align*}
Though $\frac{f_s}{f_d}$ 
is not a priori a "universal" number, the \lhcb result is similar to those obtained by the experiments 
at LEP and Tevatron.
Assuming universality, 
the Ref.~\cite{PDG2016} proposes a value of $f_s = \BR ( \bquarkbar \to \Bs ) = 0.107 \pm 0.014$.

From the above input, the branching fraction $\BR ( \Bs \to \phi \phi )$ is obtained to be 
\begin{align*}
\BR ( \Bs \to \phi \phi ) = ( 2.06 \pm 0.16 \pm 0.12 \pm 0.27_{f_s} \pm 0.47_{\BR} ) \times 10^{-5} \ , 
\end{align*}
where the first uncertainty is statistical, the second is systematic, the third one is due to $f_s$ and the last one is due to normalization and especially due to $\frac{\BR ( \etac \to \phi \phi )}{\BR ( \etac \to \proton \antiproton )}$, which limits precision of this measurement. 

Alternatively, the $f_s$ can be extracted from the \lhcb results on \bquark-hadron production~\cite{LHCb-PAPER-2014-004}
yielding $f_s = 0.096 \pm 0.005$. 
Using this value, the branching fraction $\BR ( \Bs \to \phi \phi )$ is obtained to be 
\begin{align*}
\BR ( \Bs \to \phi \phi ) = ( 2.18 \pm 0.17 \pm 0.11 \pm 0.14_{f_s} \pm 0.65_{\BR} ) \times 10^{-5} \ . 
\end{align*}

The above value of 
$\BR ( \Bs \to \phi \phi )$
is measured with a different technique with respect to the previous 
results~\cite{Abe:1999ze,Acosta:2005eu,Aaltonen:2011rs}. 
The measurement is consistent with the previous CDF results and has a precision 
similar to that of the PDG value~\cite{PDG2016}. 
The measurement is consistent with the new \lhcb result~\cite{Aaij:2015cxj} 
using normalization 
to the $B \to \phi K^{*0}(892)$ decay mode, 
$\BR ( \Bs \ra \phi \phi ) = (1.84 \pm 0.05 \pm 0.07 \pm 0.11_{f_s / f_d} \pm 0.12_{norm}) \times 10^{-5}$. 
The result is also consistent with theoretical calculations~\cite{Beneke:2006hg,Ali:2007ff,Cheng:2009mu}. 

Precision of the measured branching fraction $\BR ( \Bs \to \phi \phi )$ is fully dominated 
by the systematic uncertainty, which is in turn dominated by the 
uncertainty in the ratio of the \etac branching fractions 
$\frac{\BR ( \etac \to \phi \phi )}{\BR ( \etac \to \proton \antiproton )}$, 
and the knowledge of the $f_s$ parameter. 
Averages~\cite{PDG2016} of the branching fractions 
$\BR ( \etac \to \phi \phi )$ and $\BR ( \etac \to \proton \antiproton )$ 
rely on the so far most precise measurements by DM2~\cite{Bisello:1990re} 
and BES~\cite{Bai:2003tr,Ablikim:2005yi,Ablikim:2012ur} experiments. 
Precision of all above measurements is dominated by systematic uncertainties. 
None of the two experiments directly performed a measurement of the ratio of the two branching fractions, 
which would allow partial cancellation of systematic uncertainty, and 
would consequently reduce the systematic uncertainty of the branching fraction $\BR ( \Bs \to \phi \phi )$ measured in this section. 

In summary, the branching ratio 
$\BR ( \Bs \to \phi \phi ) = ( 2.18 \pm 0.17 \pm 0.11 \pm 0.14_{f_s} \pm 0.65_{\BR} ) \times 10^{-5}$
is calculated with a different technique with respect to the previous 
results~\cite{Abe:1999ze,Acosta:2005eu,Aaltonen:2011rs}. 
The measurement is consistent with the previous CDF results and has a precision 
similar to that of the PDG value~\cite{PDG2016}. 
The result is consistent and is less precise than the new \lhcb 
result~\cite{Aaij:2015cxj}.  
The result is also consistent with theoretical calculations~\cite{Beneke:2006hg,Ali:2007ff,Cheng:2009mu}.

\subsection{Extraction of the $\BR(\etac\to\phi\phi)/\BR(\etac\to\ppbar)$}
\label{sec:brEtac2PhiPhi}
An opposite approach with respect to that discussed in Section~\ref{sec:bstwo} can be elaborated to resolve a tension between the PDG average and PDG fit values of $\BR(\etac\to\phi\phi)$~\cite{PDG2018}
\begin{equation}
\begin{aligned}
\BR(\etac\to\phi\phi)^{PDG\,fit}     &= (1.79\pm 0.20)\times 10^{-3} \\
\BR(\etac\to\phi\phi)^{PDG\,average} &= (2.8 \pm 0.4)\times 10^{-3}
\end{aligned}
\end{equation}

Using Eq.~\ref{eq:BrBs2phiphi} and external inputs, the ratio of the branching fractions 
for the $\etac (1S)$ decays to $\phi \phi$ and to $\proton \antiproton$ is determined below. 
The measured \Bs and $\etac (1S)$ yields and efficiency ratio, the branching fraction 
\mbox{$\BR ( \Bs \ra \phi \phi ) = (1.84 \pm 0.05 \pm 0.07 \pm 0.11_{f_s / f_d} \pm 0.12_{\scalebox{0.6}{norm}}) \times 10^{-5}$}~\cite{LHCb-PAPER-2015-028}, 
the \jpsi production rate in \bquark-hadron decays $\BR ( \bquark \ra \jpsi X ) = (1.16 \pm 0.10 ) \%$~\cite{PDG2016}, 
the relative production rates of $\etac (1S)$ and \jpsi in \bquark-hadron decays 
$\frac{\BR( \bquark \ra \etac (1S) X) \times \BR( \etac (1S) \ra \proton \antiproton )}{\BR( \bquark \ra \jpsi X) \times \BR( \jpsi \ra \proton \antiproton )} = 0.302 \pm 0.042$~\cite{LHCb-PAPER-2014-029}, 
the branching fraction $ \BR( \jpsi \ra \proton \antiproton ) = ( 2.120 \pm 0.029 ) \times 10^{-3}$~\cite{PDG2016}, 
the ratio of fragmentation fractions \mbox{$f_s / f_d = 0.259 \pm 0.015$}~\cite{Aaij:2013qqa}, 
and the \Lb fragmentation fraction $f_{\Lb}$ momentum dependence from Ref.~\cite{LHCb-PAPER-2014-004} are used.
The ratio of the branching fractions for the $\etac (1S)$ decays to $\phi \phi$ and to $\proton \antiproton$ is determined as 
\[
\frac{\BR ( \etac (1S) \ra \phi \phi )}{\BR ( \etac (1S) \ra \proton \antiproton )} = 1.79 \pm 0.14 \pm 0.09 \pm 0.10_{f_s / f_d} \pm 0.03_{f_{\Lb}} \pm 0.29_{\BR} , 
\]
where the third uncertainty is related to $f_s / f_d$, the fourth uncertainty is related to $f_{\Lb}$, 
and the fifth uncertainty is related to uncertainties of the production rates and decay branching fractions involved. 
This value is larger than the value computed from the world average branching fractions given in Ref.~\cite{PDG2016}, $\frac{\BR ( \etac (1S) \ra \phi \phi )}{\BR ( \etac (1S) \ra \proton \antiproton )} = 1.19 \pm 0.14$, and indicates a consistency problem.

\section{The $\Bs \ra \phi \phi \phi$ decay}
\label{sec:bsthree}
\begin{figure}[h]
\centering
\protect\protect\includegraphics[width=0.5\linewidth]{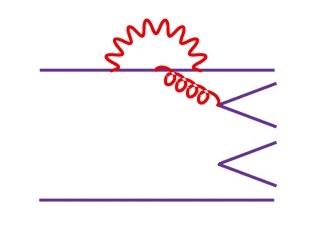}
                \put(-210,120) {{\bquarkbar}}
                \put(-20,27) {{\squark}}
                \put(-210,27) {{\squark}}
                \put(-20,37) {{\squarkbar}}
                \put(-20,67) {{\squark}}
                \put(-20,77) {{\squarkbar}}
                \put(-20,110) {{\squark}}
                \put(-20,120) {{\squarkbar}}
                \put(-120,165) {{\Wp}}
                \put(-240,75) {{\Bs}}
                \put(0,32) {{$\phi$}}
                \put(0,72) {{$\phi$}}
                \put(0,115) {{$\phi$}}
\caption{Quark diagram describing three-body $\Bs \to \phi \phi \phi$ decay.} 
\label{fig:diagthree}
\end{figure}
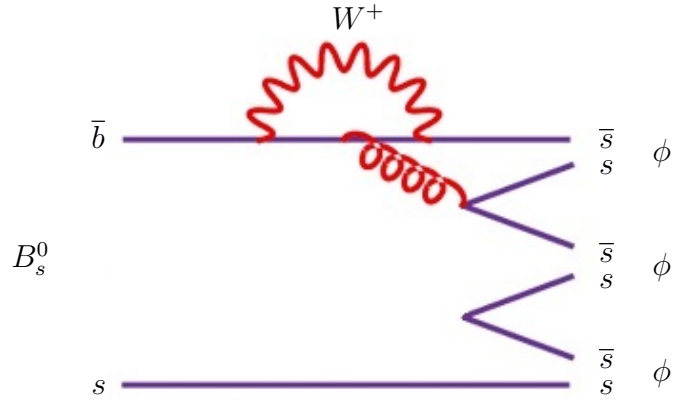

The three-body $\Bs \to \phi \phi \phi$ decay can be described by a penguin quark diagram 
shown on Fig.~\ref{fig:diagthree}. 
This diagram is similar to the diagram describing the $\Bs \to \phi \phi$ decay mode, 
(Fig.~\ref{fig:diagtwo})
and involves a creation of an additional \ssbar pair. 
The transition thus leads to the final state with six strange quarks. 

The $\Bs \to \etac \phi$ decay mode followed by the $\etac \to \phi \phi$ decay, 
is an example of the decay via intermediate resonance, yielding three-$\phi$ system. 
The $\Bs \to \etac \phi$ decay is described by an internal emission of \W boson. This decay has been observed by \lhcb recently and the branching fraction was measured to be
\begin{equation}
\BR(\Bs \to \etac \phi) = (5.01\pm0.53\pm0.27\pm0.63)\times 10^{-4},
\end{equation}
where the last uncertainty is due to the limited knowledge of the external branching fractions.
The similar $\Bs \to \jpsi \phi$ decay occurs with a branching fraction of 
$\BR ( \Bs \to \jpsi \phi ) = (1.08 \pm 0.08) \times 10^{-3}$~\cite{PDG2018} and was used as a normalization. The four decay modes were used to reconstruct the \etac meson, namely decays to \ppbar, $\Kp\Km\Kp\Km$, $\Kp\Km\pip\pim$ and $\pip\pim\pip\pim$ final states. 

The difference between the branching fractions for the $\Bs \to \etac \phi$ and 
$\Bs \to \jpsi \phi$ decays can however be expected due to the fact, that the latter decay is a $P \to VV$ transition. 
For example, branching fractions of light \B-meson decays to $\jpsi K^*$ exceed 
those of light \B-meson decays to $\etac K^*$ by a factor 2, where only neutral \B-decays 
are measured precisely enough to draw this conclusion at a quantitative level~\cite{PDG2018}. 

In addition, in order to make a comparison between the $\Bs \to \etac \phi$ and three-body contributions 
to the $\Bs \to \phi \phi \phi$ decay, the $\etac \to \phi \phi$ branching fraction should be taken into account. 
Once the $\Bs \to \phi \phi \phi$ decay is observed, studying its resonance structure can yield interesting information on the QCD contribution to weak \bquark-decays.

\protect\clearpage
\subsection{Signal extraction and systematic uncertainties}
\label{sec:triphir}
Adding another reconstructed $\phi$ candidate to the $\phi \phi$ system, 
allows to search for the $\Bs \to \phi \phi \phi$ decay. 
The $\Bs \to \phi \phi$ decay is used as normalization.

Reconstruction of the $\Bs \to \phi \phi \phi$ decay mode employs 
selection criteria, that are similar to those used for the analysis of the $\Bs \to \phi \phi$ decay. 
Table~\ref{tab:cuts} summarizes selection criteria for charmonia and \Bs meson decays to $\phi \phi$ 
and \Bs decays to $\phi \phi \phi$. 
\begin{table}[h] 
{\small{
\centering
\begin{tabular}{l|l|l|l} 
   & Variable   &  Denotion & Requirement \\  \hline

  Kaons & Track quality &  \chisqndf & $<3$ \\
        & Impact parameter to primary vertex & $\chi^{2}_{IP}$ & $>4$ \\
    & Transverse momentum & \pt, \gev & $>0.5$ \\
    & Identification & ProbNNk  & $>0.1$ \\  \hline

 $\phi$ & Vertex quality & $\chi^2$ & $<25$ \\  \hline
    & Invariant mass & $| M_{K^+ K^-} - M_{\phi} |$, \mev & $<12$ \\  \hline

 $\phi\phi$ & Vertex quality &  \chisqndf & $<9$ \\  
            & Distance between the decay vertex & $\chi^2$ & $>100$ \\  
            & and the primary vertex & & \\  \hline

 $\phi\phi\phi$ & Vertex quality &  \chisqndf & $<9$ \\  
            & Distance between the decay vertex & $\chi^2$ & $>100$ \\  
            & and the primary vertex & & 
\end{tabular}
}}
\caption{Selection criteria for charmonia and \Bs meson decays to $\phi \phi$ and \Bs decays to $\phi \phi \phi$.} \label{tab:cuts}
\end{table}

A dedicated MC sample of $\Bs \to \phi \phi \phi$ decays is used to describe detector resolution and signal efficiency.
The efficiency ratio of $\Bs \to \phi \phi \phi$ and $\Bs \to \phi \phi$ decay modes is determined to be
\begin{align*}
\frac{\varepsilon ( \Bs \to \phi \phi \phi )}{\varepsilon ( \Bs \to \phi \phi )} &= 0.26 \pm 0.01.
\end{align*}

\begin{figure}[h]
\centering
\protect\protect\includegraphics[width=0.95\linewidth]{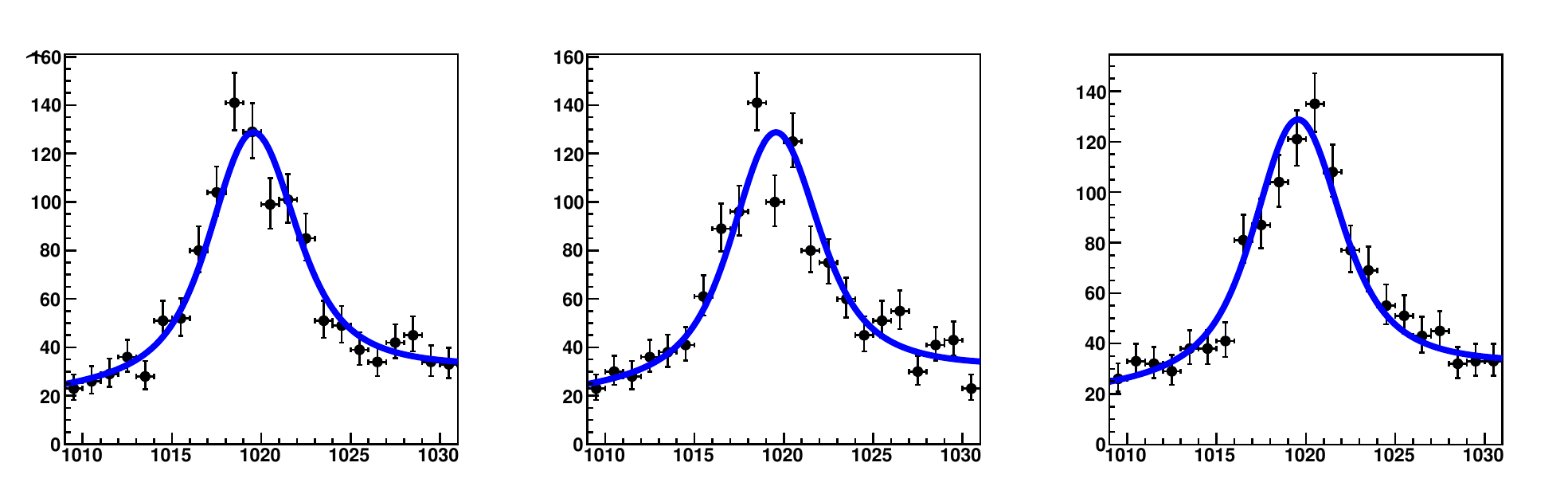}
                \put(-420,30){\rotatebox{90}{\small{Events/1 \mev}}}
                \put(-80,-5) {\scriptsize{$M(\Kp \Km)$ \mev}}
                \put(-225,-5) {\scriptsize{$M(\Kp \Km)$ \mev}}
                \put(-370,-5) {\scriptsize{$M(\Kp \Km)$ \mev}}
                \put(-120,110) {\scriptsize{\lhcb}}
                \put(-265,110) {\scriptsize{\lhcb}}
                \put(-405,110) {\scriptsize{\lhcb}}
\caption{Projections of the entire sample of the $\phi \phi \phi$ candidates 3D fit on each $\phi$ candidate.}
\label{fig:threed}
\end{figure}
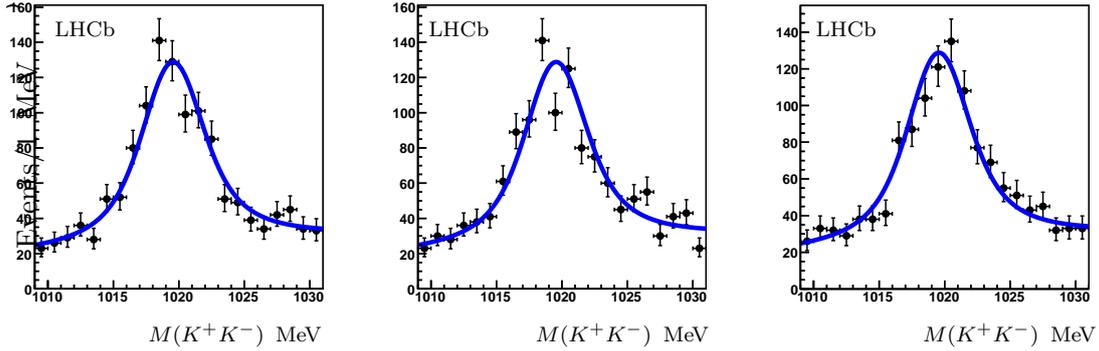

In the $\Bs \to \phi \phi \phi$ analysis, in order to extract pure $\phi \phi \phi$ combinations 
a maximum likelihood unbinned 3D fit is used, 
similar to the 2D fit used to extract pure $\phi \phi$ combinations, 
\begin{align*}
F (x_1 , x_2 , x_3) = & N_{\phi \phi \phi} \times S_1 \times S_2 \times S_3 \ + \\
& N_{\phi \phi K K} \times ( S_1 \times S_2 \times B_3 \ + \ S_1 \times B_2 \times S_3 \ + \  
B_1 \times S_2 \times S_3 ) + \\
& N_{\phi K K K K} \times ( S_1 \times B_2 \times B_3 \ + \ B_1 \times S_2 \times B_3 \ + \  
B_1 \times B_2 \times S_3 ) + \\
& N_{K K K K K K} \times B_1 \times B_2 \times B_3 \ , 
\end{align*}
where signal contributions $S_1$, $S_2$ and $S_3$ are described by the product 
of the convolution of the Breit-Wigner function and double Gaussian function 
and the square root to account for the phase space difference, 
and background contributions $B_1$, $B_2$ and $B_3$ are decribed by the square root function. 
The ratio of the two Gaussian widths $\sigma_1 / \sigma_2$ of $0.40 \pm 0.01$ and the fraction of narrow 
Gaussian $N_1 / (N_1 + N_2)$ of $0.87 \pm 0.01$ are taken from simulation. 
The fit shape accounts for $\phi \phi \phi$, $\phi \phi \Kp \Km$, $\phi \Kp \Km \Kp \Km$ 
and $\Kp \Km \Kp \Km \Kp \Km$ 
contributions and takes into account the available phase space. 
Projections of the 3D fit for the entire sample of candidates on each $\phi$ candidate are shown 
on Fig.~\ref{fig:threed}. 
No contribution from the $f_0 (980)$ resonance is seen on the plots. 
However a potential effect due to $f_0 (980)$ 
is estimated in the following as a potential source of systematic uncertainty. 

\begin{figure}[b]
\centering
\protect\protect\includegraphics[width=0.8\linewidth]{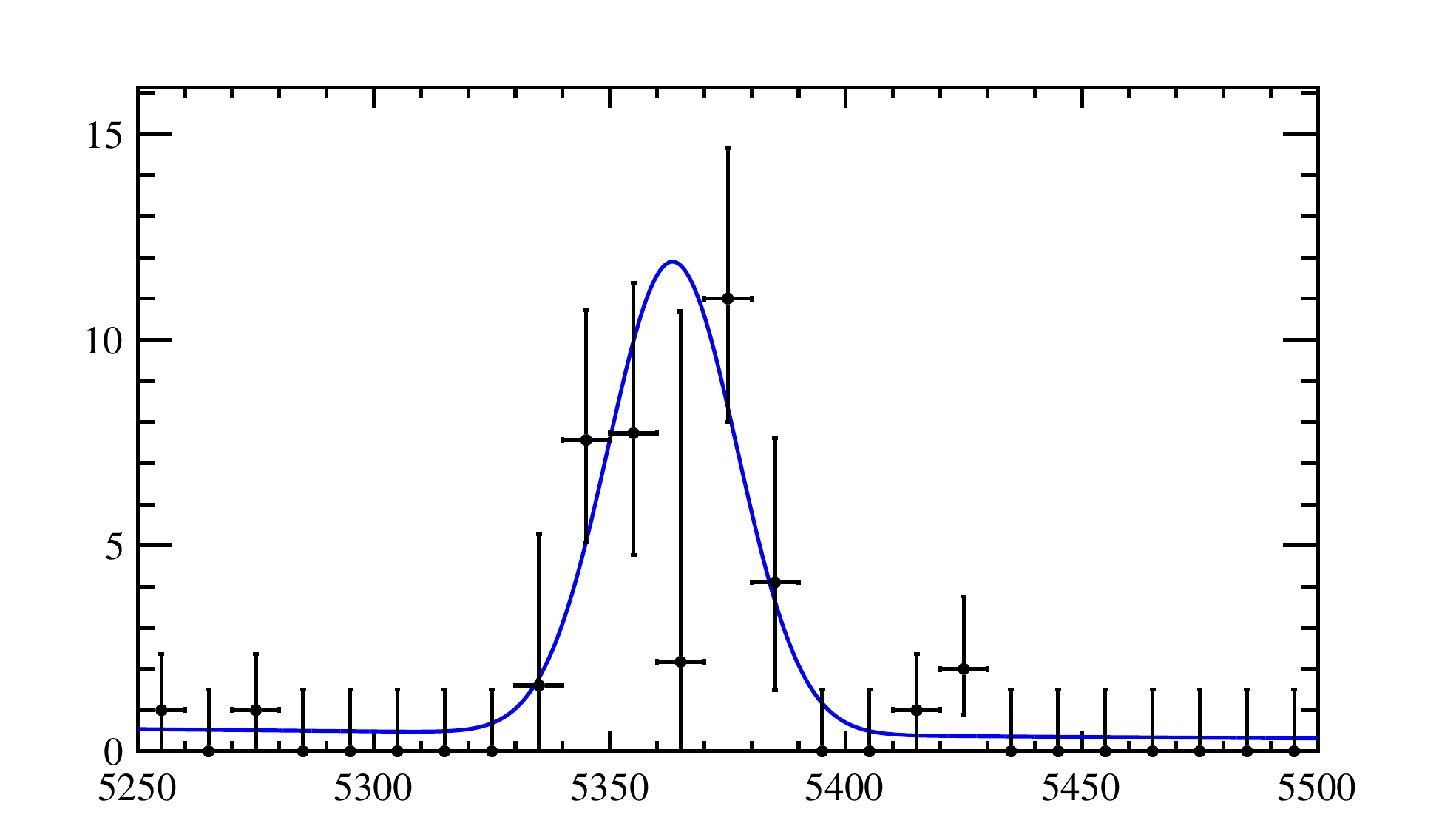}
                \put(-360,110){\rotatebox{90}{\small{Events/10 \mev}}}
                \put(-60,0) {\small{\mev}}
                \put(-200,0) {\small{$M( \phi \phi \phi)$}}
                \put(-140,165) {\lhcb}
                \put(-140,150) {3 fb$^{-1}$}
\protect\caption{Distribution of the $\phi \phi \phi$ invariant mass 
for combined data sample accumulated at $\protect\sqs = 7 \protect\tev$ and $\protect\sqs = 8 \protect\tev$.} 
\label{fig:bstriphi}
\end{figure}
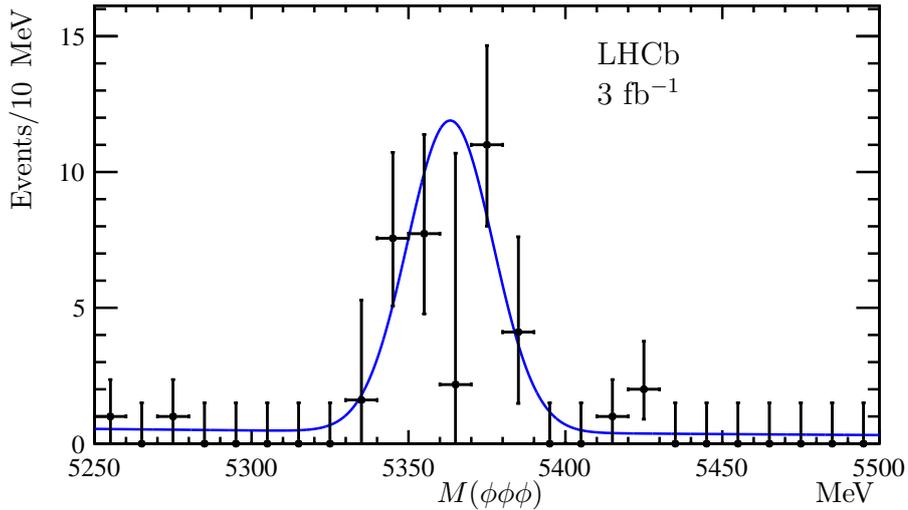

\begin{figure}[h]
\centering
\protect\protect\includegraphics[width=0.7\linewidth]{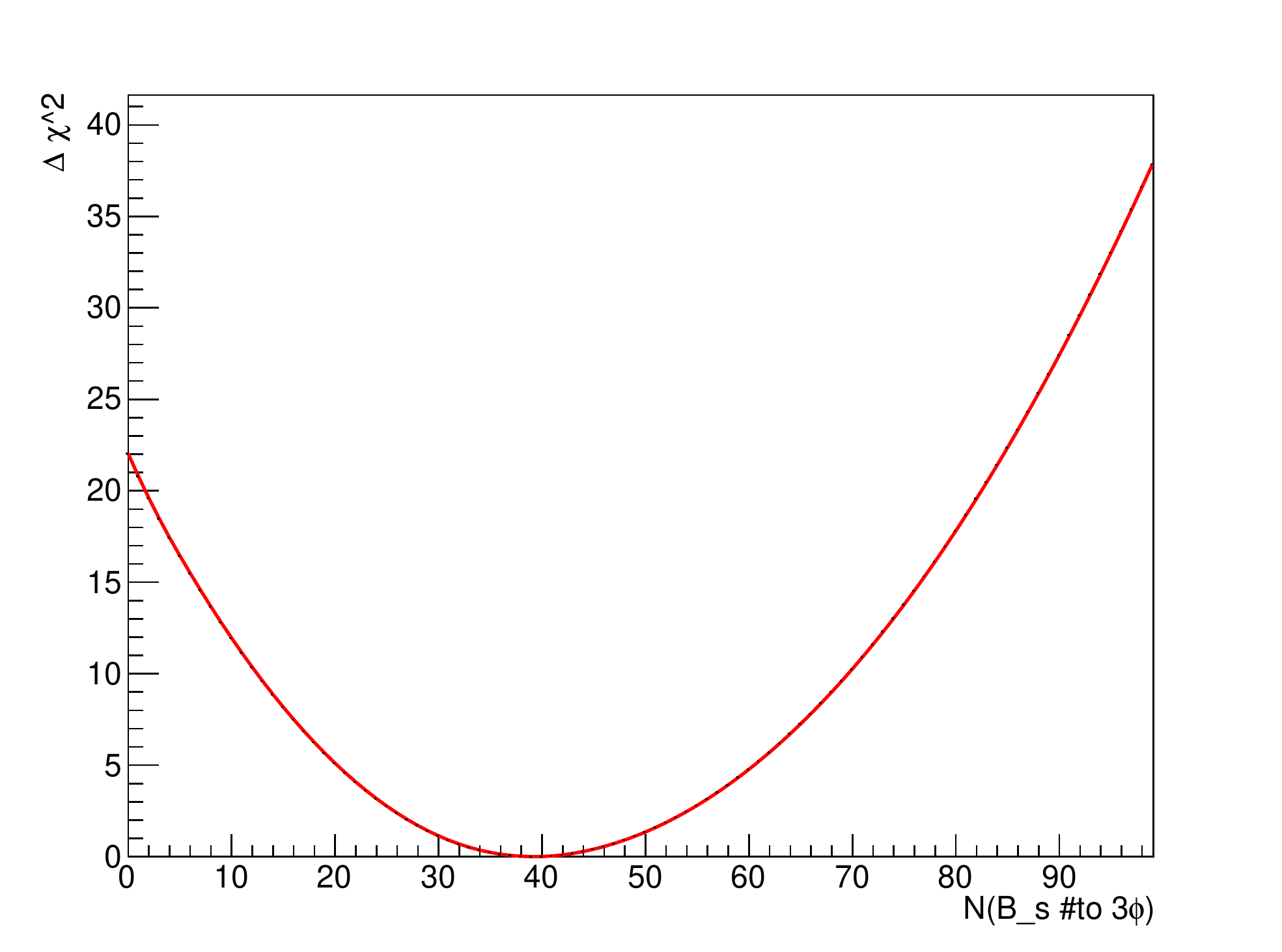}
                \put(-315,190){\rotatebox{90}{\colorbox{shadecolor}{$\Delta \chisq$}}}
                \put(-150,190) {\lhcb-ANA-2015-038}
                \put(-150,175) {3 fb$^{-1}$}
                \put(-125,2){\colorbox{shadecolor}{$N_{signal} (\Bs\to\phi\phi\phi)$           }}
\protect\caption{Fit quality of the $\Bs \to \phi \phi \phi$ signal $\Delta \chisq$ depending on the number 
of signal candidates assumed by the fit.
\protect\label{fig:signa}}
\protect\end{figure}
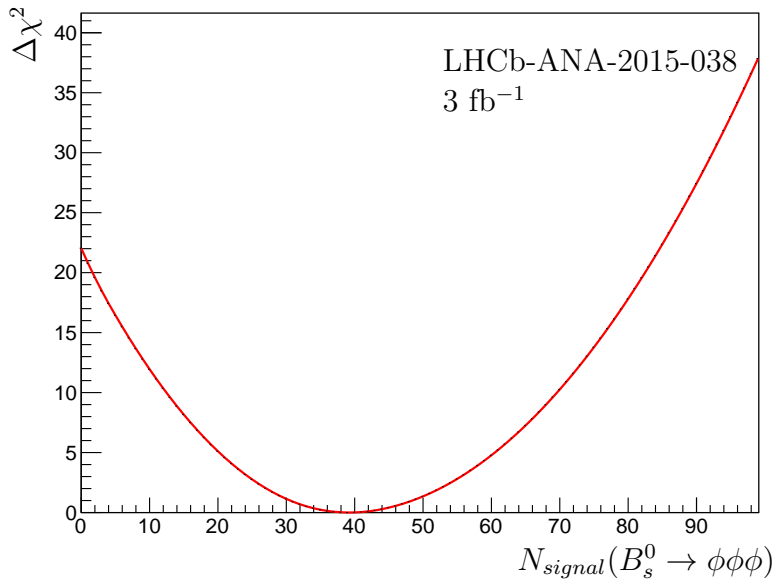

Figure ~\ref{fig:bstriphi} shows invariant mass distribution for pure $\phi \phi \phi$ combinations. 
A fit to the invariant mass spectrum, using a double Gaussian function to describe a clear \Bs signal corresponding 
to the transition $\Bs \to \phi \phi \phi$, and an exponential to describe combinatorial background, is performed. 
The ratio of the two Gaussian widths $\sigma_1 / \sigma_2$ of $0.45 \pm 0.02$ and the fraction of narrow 
Gaussian $N_1 / (N_1 + N_2)$ of $0.85 \pm 0.03$ are taken from simulation\footnote{Resolutions are 
found to be $\sigma_1 = 9.8 \pm 0.2 \mev$ (MC) and $13.2 \pm 2.9 \mev$ (data). 
The $\phi$ resolution is fixed to the value from the fit in the whole mass range.. 
One parameter of the \Bs resolution is left free in the fit, 
ratio of the two Gaussians and the fraction of the narrow Gaussian are fixed to the MC values.} 
The fit yields $41 \pm 10 \pm 5$ \Bs candidates over a low level background. 
Significance of the $\Bs \to \phi \phi \phi$ signal is estimated by judging the fit quality 
using the fit function comprising or not the signal shape. 
An estimate of about $4.7 \sigma$ is obtained from Fig.~\ref{fig:signa}. 
Here the mass and resolution of \Bs are fixed to the values calculated in the $\Bs \to \phi \phi$ analysis. 
\protect

\begin{figure}[h]
\centering
\protect\protect\includegraphics[width=0.7\linewidth]{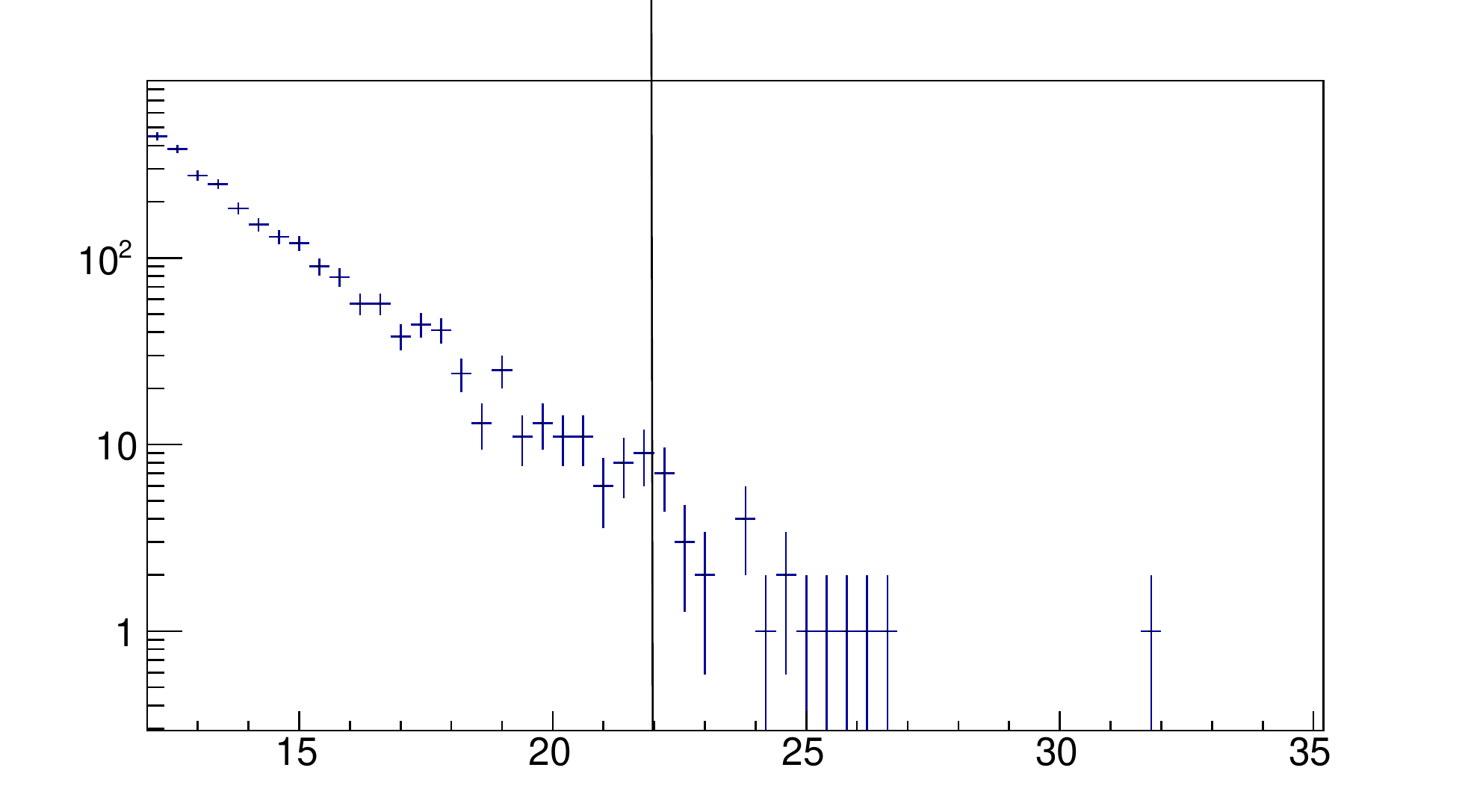}
                \put(-145,140) {\lhcb-ANA-2015-038}
                \put(-180,0) {{$\chisq_{B} - \chisq_{S+B}$}}
\protect\caption
{Difference between the  $\chisq$ values of the fit with background shape only and signal and background shapes, $\chisq_{B}$ and $\chisq_{S+B}$, for the $3 \times 10^7$ toy simulation samples generated according to the fit to data with corresponding uncertainties, excluding a signal region. 
\protect\label{fig:signb}}
\protect\end{figure}

Alternatively, the $3 \times 10^7$ toy simulation samples, 
were generated according to the fit to data with corresponding uncertainties, excluding the signal region.
These samples were fit to the background shape only and to a sum of the background and signal shapes. 
A difference between the corresponding $\chisq$ values of the fit, $\chisq_{B}$ and $\chisq_{S+B}$, is shown on Fig.~\ref{fig:signb}. 
\protect
Arrow points to the $\chisq_{B} - \chisq_{S+B}$ value of 22.0 as obtained from the fit to the data sample. 
This corresponds to the significance above $4.9 \sigma$ and $p$-value of $8.1 \times 10^{-7}$ 
for the observed signal~\cite{Cowan:2010js}.  
Details of the systematic uncertainty estimate are summarized in Table~\ref{tab:bsthrsyst}. 
\begin{table}[h]
\centering
\begin{tabular}{l|c}
    &  $N ( \Bs )$ \\ \hline
Background shape variation, $\phi \phi \phi$ & $< 1$ \\ 
Resolution at MC value in 3D fit & $- 1$ \\ 
Resolution of \Bs described by a single Gaussian & $- 2$ \\ 
$f_0 (980)$ in the 3D fit & $1$ \\ 
Decay model & $4$ \\  \hline
Combined & $5$ 
\end{tabular}
\caption{Systematic uncertainties (deviation from the baseline value)
in the measurement of the \Bs signal yield (in number of candidates). 
\label{tab:bsthrsyst}}
\end{table}
Uncertainties related to the background description in the 3D fit and to the decay model 
of $\phi$ polarisation 
dominate the systematic uncertainty in the \Bs signal yield determination. 

\subsubsection{Effect of the MC description of the \Bs \pt spectrum}
\label{sec:bspt}
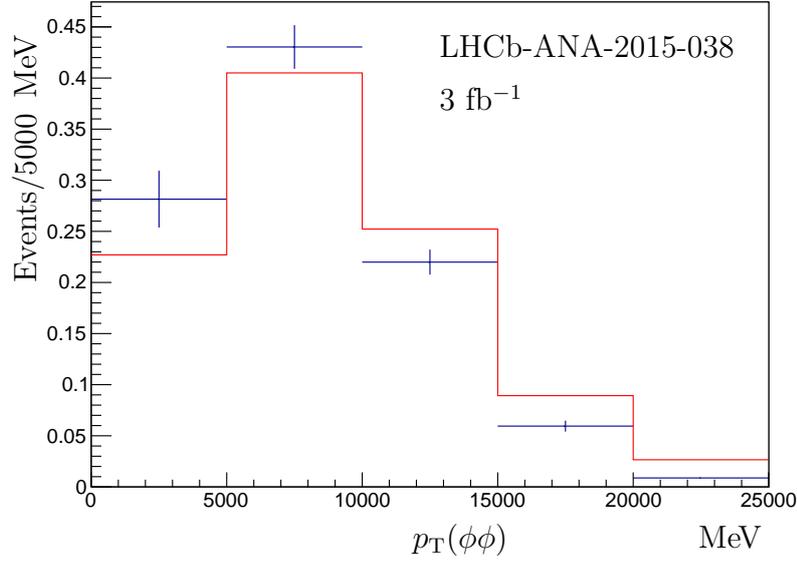
\begin{figure}[h]
\centering
\protect\protect\includegraphics[width=0.7\linewidth]{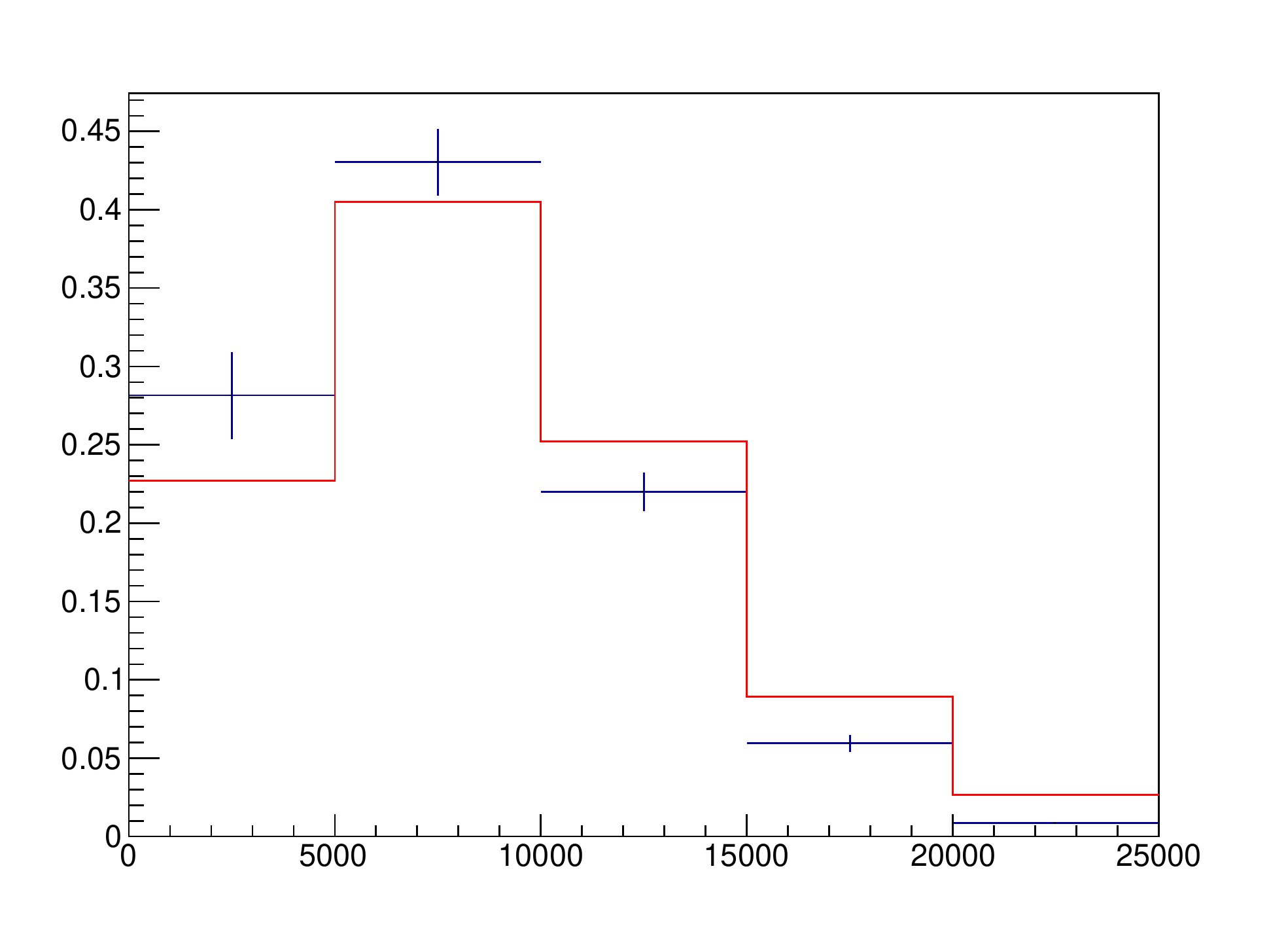}
                \put(-315,100){\rotatebox{90}{{Events/5000 \mev}}}
                \put(-60,0) {{\mev}}
                \put(-155,185) {\lhcb-ANA-2015-038}
                \put(-155,165) {3 fb$^{-1}$}
                \put(-165,0) {{$\pt ( \phi \phi )$}}
\caption
[Reconstructed \pt dependence of \Bs candidates reconstructed via the $\Bs \to \phi \phi$ decay in data and simulation.]
{Reconstructed \pt dependence of \Bs candidates reconstructed via the $\Bs \to \phi \phi$ decay in data (blue points with error bars) 
and simulation (red histogram).} \label{fig:pteff}
\end{figure}

When calculating a ratio of the branching fractions for 
the $\Bs \to \phi \phi \phi$ and $\Bs \to \phi \phi$ decays, 
knowledge of the \pt spectrum of the \Bs mesons can modify the ratio 
of corresponding efficiencies. 
The \pt dependence of \Bs candidates reconstructed via
the $\Bs \to \phi \phi$ decay in data and simulation is shown on Fig.~\ref{fig:pteff}. 
The difference between the two spectra was accounted as a correction to the efficiency ratio 
for the $\Bs \to \phi \phi \phi$ and $\Bs \to \phi \phi$ channels.


The estimated effect is found to be at the level of $8.8 \%$. 
Corresponding correction is applied. 
The corresponding contribution to the systematic uncertainty is below 1\% and is neglected.

\subsection{Measurement of the  $\BR ( \Bs \ra \phi \phi \phi )$}
\label{sec:triphi}

The $\Bs \to \phi \phi \phi$ branching fraction
is measured relatively to the $\Bs \to \phi \phi$ channel, 
\begin{equation}
\frac{\BR ( \Bs \to \phi \phi \phi )}{\BR ( \Bs \to \phi \phi )}
= \frac{N_{\Bs \to \phi \phi \phi}}{N_{\Bs \to \phi \phi}} 
\times \frac{ \varepsilon_{\Bs \to \phi \phi}}{ \varepsilon_{\Bs \to \phi \phi \phi} } 
\times \frac{1}{\BR ( \phi \to \Kp \Km )}
\ .
\end{equation}

Reconstructing the decays of \Bs meson to two $\phi$ mesons, $\Bs \to \phi \phi$, 
and three $\phi$ mesons, $\Bs \to \phi \phi \phi$, with similar selection criteria, 
allows a determination of the ratio of their branching fractions to be 
\begin{align*}
\frac{\BR ( \Bs \to \phi \phi \phi )}{\BR ( \Bs \to \phi \phi )} 
 = 
\frac{N_{\Bs \to \phi \phi \phi}}{N_{\Bs \to \phi \phi}}
 \times \frac{\varepsilon_{\Bs \to \phi \phi}}{\varepsilon_{\Bs \to \phi \phi \phi}} 
 \times \frac{1}{\BR ( \phi \to \Kp \Km )} \ .
\end{align*}
In the above expression, the event yields are measured in the present analysis, 
and the efficiency ratio is estimated using MC simulation to be 
$\varepsilon_{\Bs \to \phi \phi \phi} / \varepsilon_{\Bs \to \phi \phi} = 0.26 \pm 0.01$, 
assuming that the $\Bs \to \phi \phi \phi$ transition proceeds as a three-body decay. 
The efficiency correction of $\alpha = 0.912 \pm 0.001$ related to the MC description 
of the \Bs \pt spectrum is additionally applied. 
The ratio of the branching fraction is thus obtained as 
\begin{align*}
\frac{\BR ( \Bs \to \phi \phi \phi )}{\BR ( \Bs \to \phi \phi )} 
 = 0.117 \pm 0.030 \pm 0.015 \ ,
\end{align*}
where the systematic uncertainty is dominated by the uncertainty due to the decay model. 
Using 
$\BR ( \Bs \to \phi \phi ) = ( 1.84 \pm 0.05 \pm 0.07 \pm 0.12_{norm} \pm 0.11_{f_s / f_d} ) \times 10^{-5}$ 
from Ref.~\cite{LHCb-PAPER-2015-028},
the branching fraction for the \Bs meson decay to three $\phi$ mesons is found to be 
\begin{align*}
\BR ( \Bs \to \phi \phi \phi ) = ( 2.15 \pm 0.54 \pm 0.28 \pm 0.21 ) \times 10^{-6},
\end{align*}
where the last uncertainty is due to involved exteranl branching fractions knowledge.

\clearpage
\subsection{Decay model of the $\Bs \ra \phi \phi \phi$ decay}
\label{sec:bsreso}
\begin{figure}[h]
\protect\centering
\protect\protect\protect\includegraphics[width=0.7\linewidth]{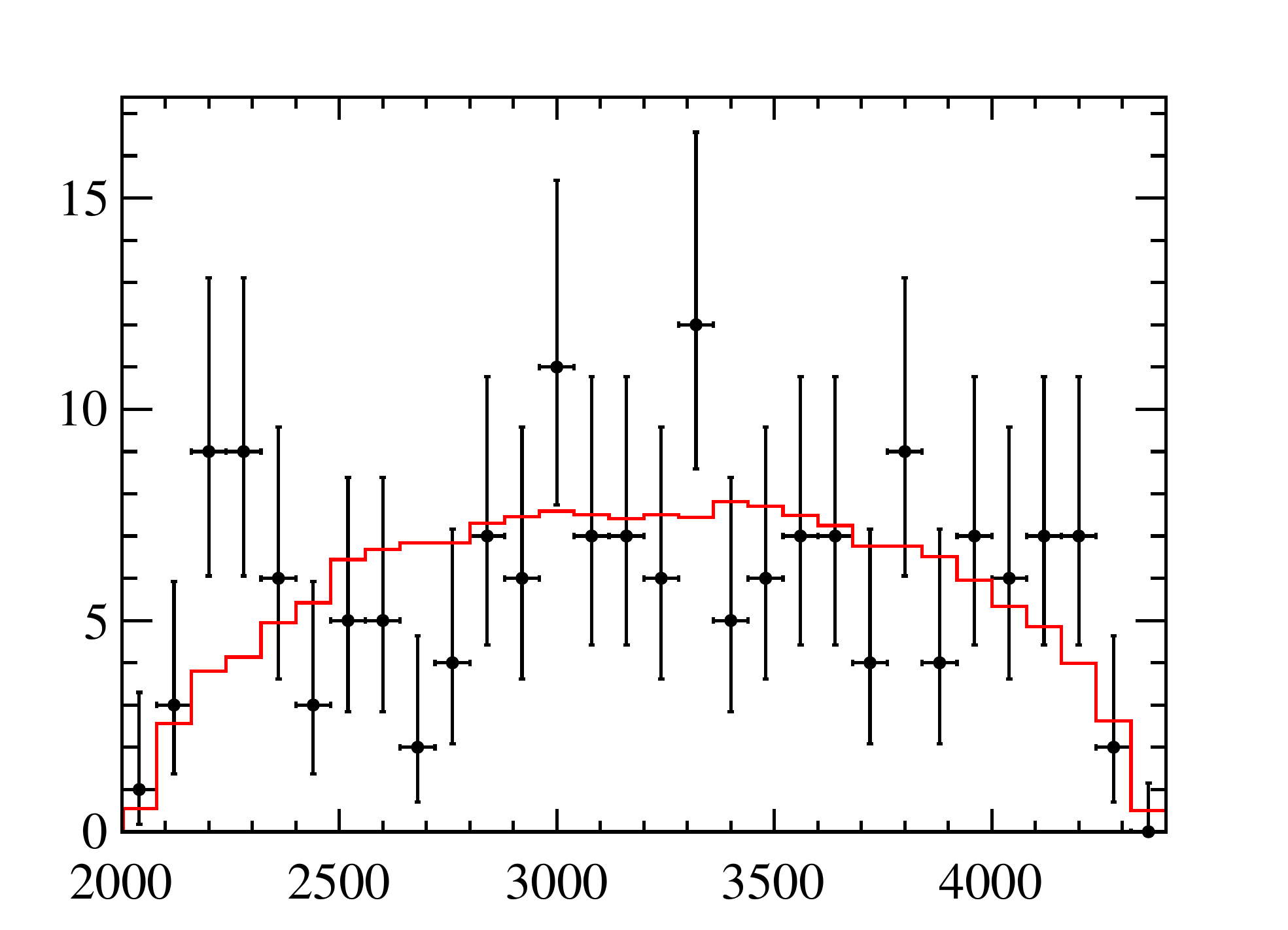}
                \protect\put(-315,110){\protect\rotatebox{90}{\protect\small{Events/80 \protect\mev}}}
                \protect\put(-60,-3) {\protect\small{\protect\mev}}
                \protect\put(-170,-3) {\protect\small{$M( \phi \phi)$}}
                \protect\put(-270,190) {\protect\lhcb}
                \protect\put(-270,170) {3 \protect\invfb}
\protect\caption
[Invariant mass distribution 
of the $\phi \phi$ pair from the $\Bs \to \phi \phi \phi$ candidates 
for combined data sample accumulated at $\protect\sqs = 7 \protect\tev$ and $\protect\sqs = 8 \protect\tev$.]
{Invariant mass distribution 
of the $\phi \phi$ pair from the $\Bs \to \phi \phi \phi$ candidates 
for combined data sample accumulated at $\protect\sqs = 7 \protect\tev$ and $\protect\sqs = 8 \protect\tev$. 
A phase space distribution as obtained from simulation (red histogram) is overlaid.} 
\protect\label{fig:twotriphi}
\protect\end{figure}
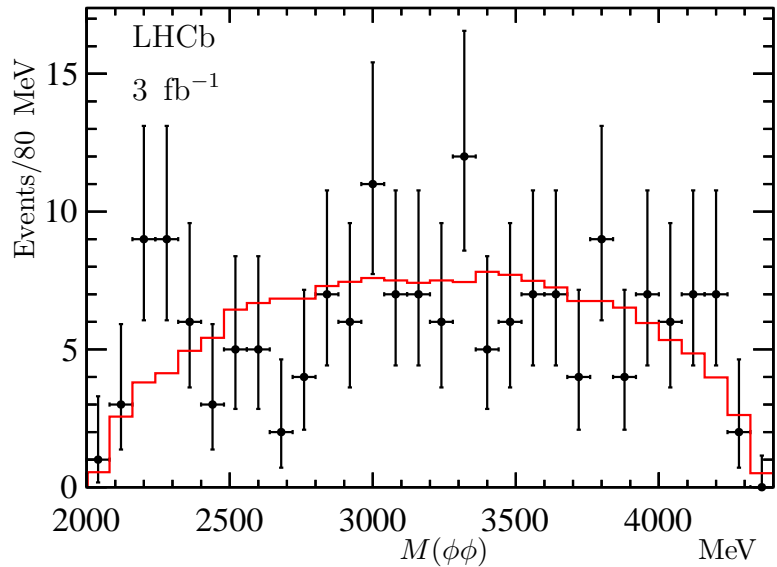

In order to search for contributions from possible intermediate resonances 
the invariant mass of $\phi \phi$ pairs is looked at. 
Fig.~\ref{fig:twotriphi} shows the invariant mass distribution of $\phi \phi$ pairs 
from the $\Bs \to \phi \phi \phi$ candidates. The candidates with invariant mass between $5.325 \gevc$ 
and $5.415 \gevc$ are considered. With the limited sample of $\Bs \to \phi \phi \phi$ candidates the 3D fit technique to remove contributions 
from $\Kp\Km$ combinations that are not from $\phi$ decays cannot be used for this measurement. 
Instead, all $\phi$ mesons contributing in the mass range of the \Bs are used, with an estimated signal purity of $71 \%$. 
\protect
A phase space distribution as obtained from simulation is overlaid for comparison. 
Though the event sample is too small to conclude, no indication 
of a dominating resonance contribution is visible 
for the $\etac(1S)$, \chiczero, \chicone, \chictwo or $\etac(2S)$ contribution. 
In addition, there is no indication of the $f_2 (2300)$ or $f_2 (2340)$ presence. 
A small increase in the number of candidates is observed around $2200 \mev$ close 
to the phase space threshold, which is however not compatible with any known 
resonant state. 
This increase cannot be attributed to a nearby $\phi (2170)$ state, 
since $\phi (2170)$ has the quantum numbers 
$J^{PC} = 1^{- -}$ incompatible with the decay to $\phi \phi$. 

\begin{figure}[h]
\protect\centering
\protect\protect\protect\includegraphics[width=0.7\linewidth]{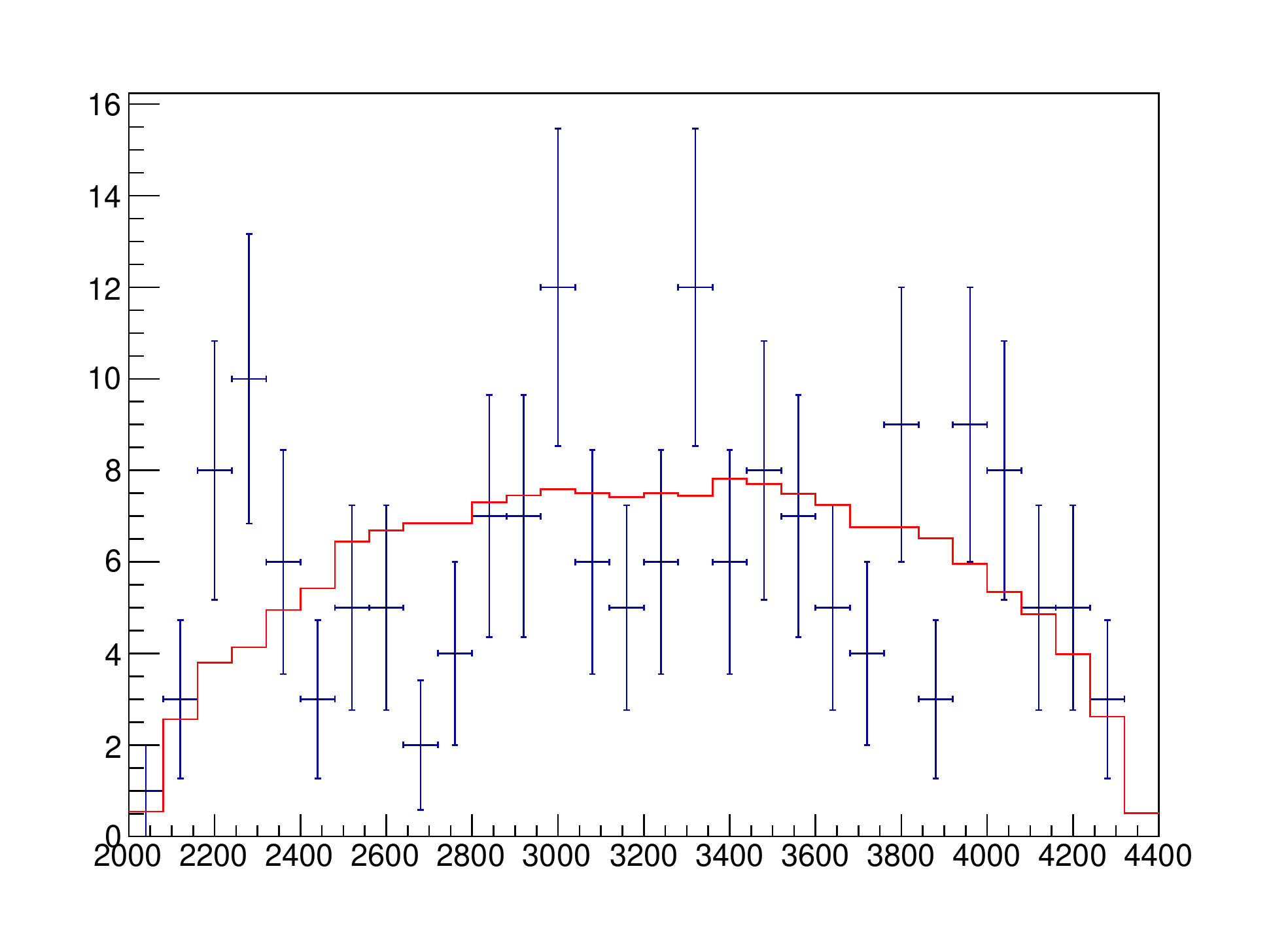}
                \put(-315,110){\rotatebox{90}{\small{Events/80 \mev}}}
                \put(-60,-3) {\small{\protect\mev}}
                \put(-170,-3) {\small{$M( \phi \phi)$}}
                \put(-270,190) {\protect \lhcb-ANA-2015-038}
\protect\caption
[Invariant mass distribution 
of the $\phi \phi$ pair from the $\Bs \to \phi \phi \phi$ candidates 
for combined data sample accumulated at $\protect\sqs = 7 \protect\tev$ and $\protect\sqs = 8 \protect\tev$, using a constraint to the \Bs mass.]
{Invariant mass distribution 
of the $\phi \phi$ pair from the $\Bs \to \phi \phi \phi$ candidates 
for combined data sample accumulated at $\protect\sqs = 7 \protect\tev$ and $\protect\sqs = 8 \protect\tev$, 
using a constraint to the \Bs mass. 
A phase space distribution as obtained from simulation (red histogram) is overlaid.}
\protect\label{fig:twophibsc}
\protect\end{figure}

As an attempt to improve the resolution, 
Fig.~\ref{fig:twophibsc} shows the invariant mass distribution of $\phi \phi$ pairs 
from the $\Bs \to \phi \phi \phi$ candidates, using a constraint to the \Bs mass.  
\protect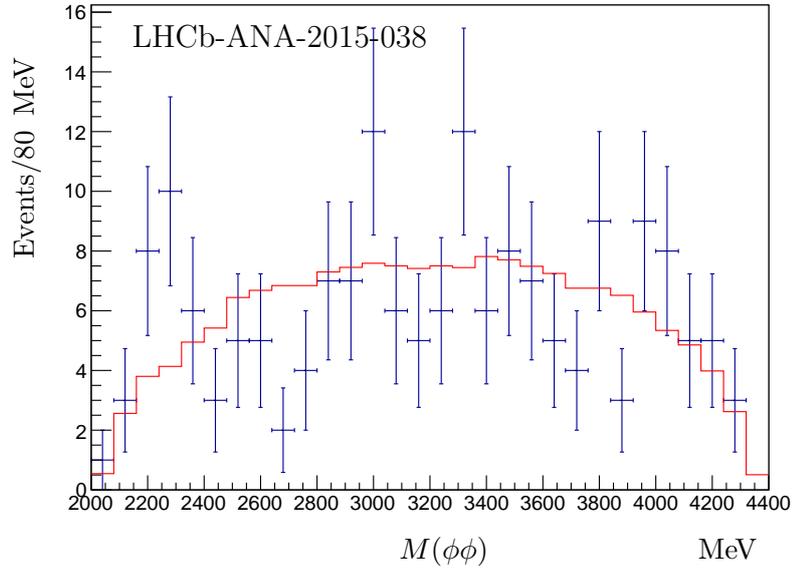
No significant improvement is observed.

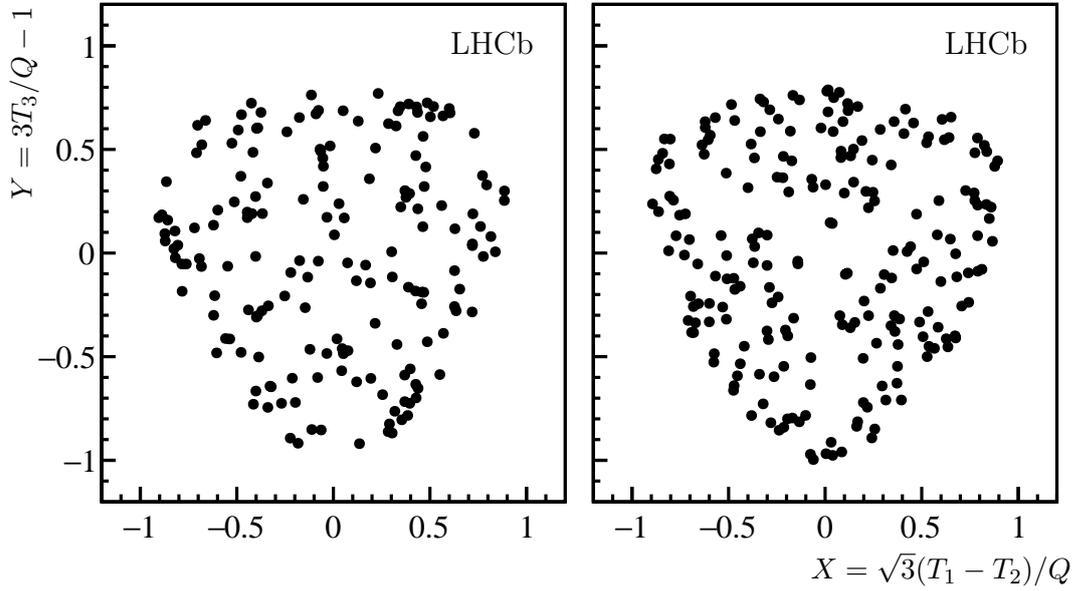
\begin{figure}[h]
\centering
\protect\protect\includegraphics[width=1.0\linewidth]{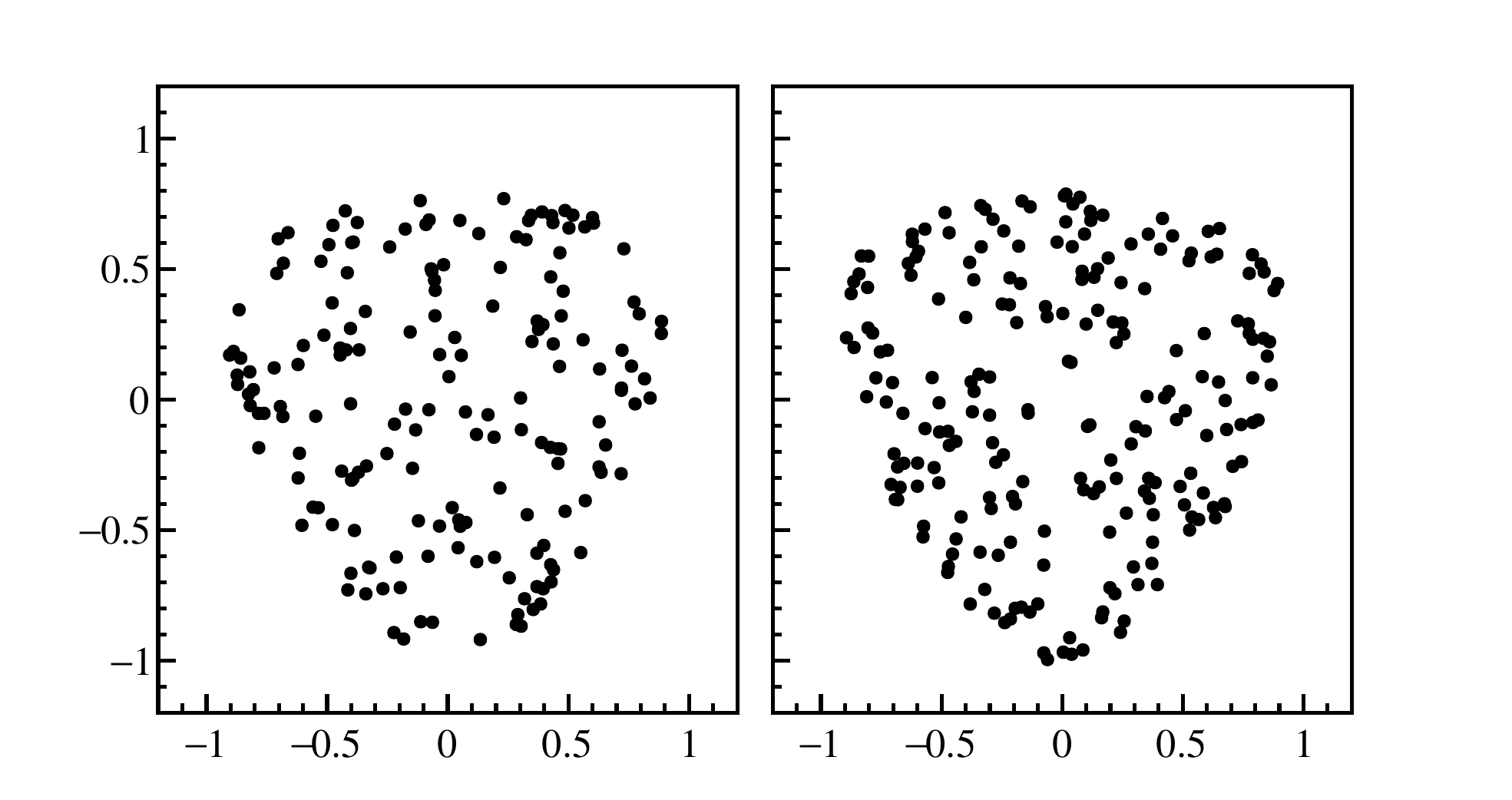}
                \put(-275,195){\lhcb}
                \put(-90,195){\lhcb}
                \put(-440,140){\rotatebox{90}{\small{$Y = 3 T_3 / Q - 1$}}}
                \put(-140,-3) {\small{$X = \sqrt{3} ( T_1 - T_2 ) / Q$}}
\protect\caption
[Symmetrized Dalitz plot for the \Bs signal and the sideband regions.
The \Bs candidates are constrained to the known \Bs mass.]
{Symmetrized Dalitz plot~\protect\cite{Adolph:2008vn} for (left) the \Bs signal and (right) the sideband regions.
The \Bs candidates are constrained to the known \Bs mass.}
\label{fig:dalitz}
\end{figure}

As another cross-check, a symmetrized Dalitz plot is constructed following Ref.~\cite{Adolph:2008vn}
for the \Bs signal region (Fig.~\ref{fig:dalitz}, left) and sidebands ($4.925 - 5.325 \gev$ and $5.415 - 5.815 \gev$, Fig.~\ref{fig:dalitz}, right), 
using the $X = \sqrt{3} ( T_1 - T_2 ) / Q$ and $Y = 3 T_3 / Q - 1$ axes, where $T_{1,2,3}$ are kinetic energies of $\phi$ mesons 
in the rest frame of \Bs and $Q$ is the energy released in the $\Bs \to \phi \phi \phi$ decay. 
The \Bs candidates are constrained to the known \Bs mass.
No evidence for resonant contributions is observed with the available statistics. 

\begin{figure}[t]
\centering
\begin{picture}(330,225)
\put(10,10){\protect\protect\includegraphics[width=315px]{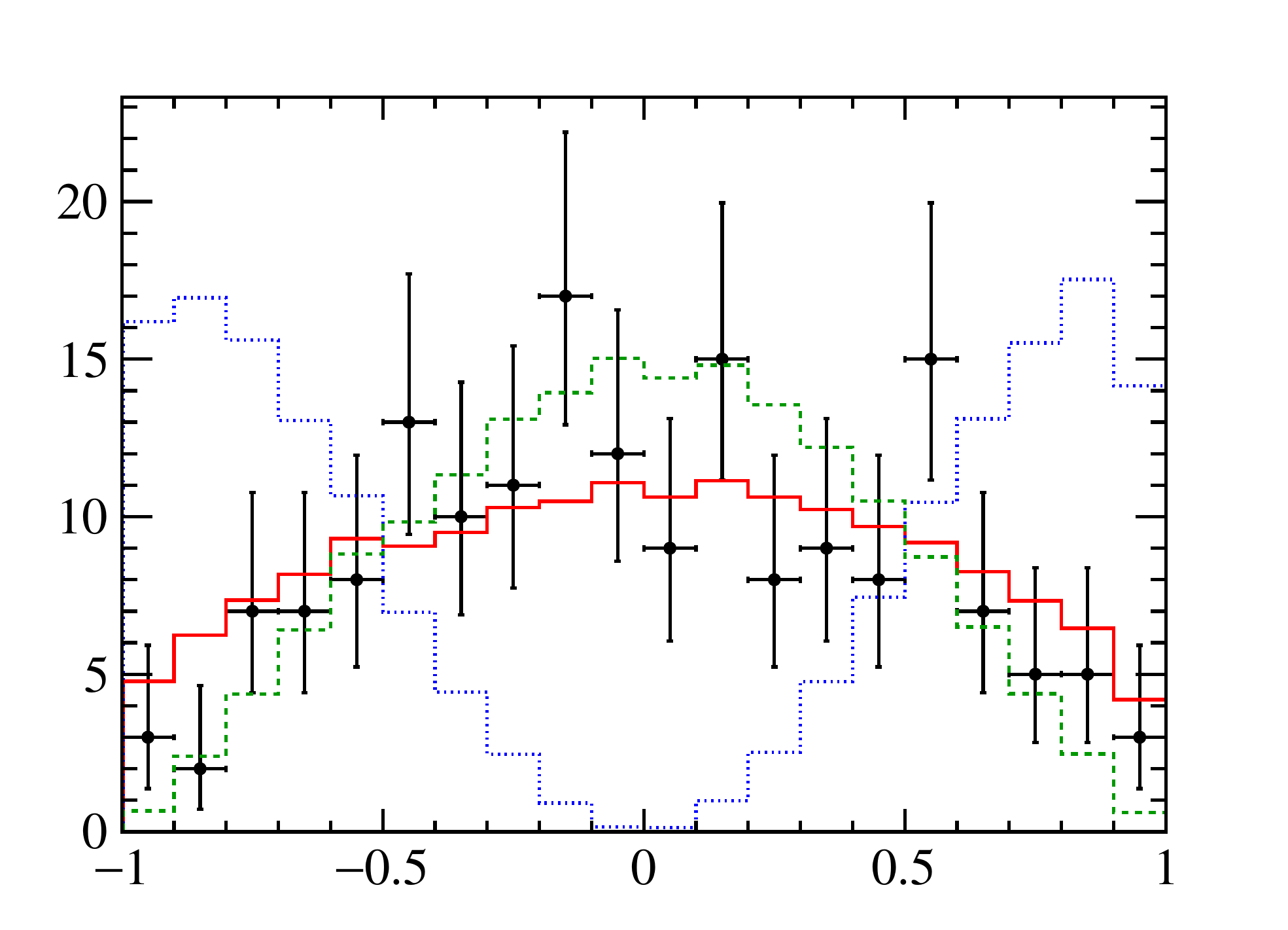}
                \put(-320,125){\rotatebox{90}{{Candidates/0.4}}}
                \put(-60,-3) {{$\cos ( \theta )$}}
                \put(-270,185) {\lhcb}}
\end{picture}
\protect\caption
[The $\phi$ meson angular distribution for the $\Bs \to \phi \phi \phi$ candidates with the overlaid distribution from the simulation with no polarization and two extreme, transverse and longitudinal, polarizations.]
{The $\phi$ meson angular distribution for the $\Bs \to \phi \phi \phi$ candidates (points with error bars)
with the overlaid distribution from the simulation with no polarization (red solid histogram) and two extreme, transverse 
(green dashed histogram) and longitudinal (blue dotted histogram), polarizations.}
\protect\label{fig:pol}
\protect\end{figure}
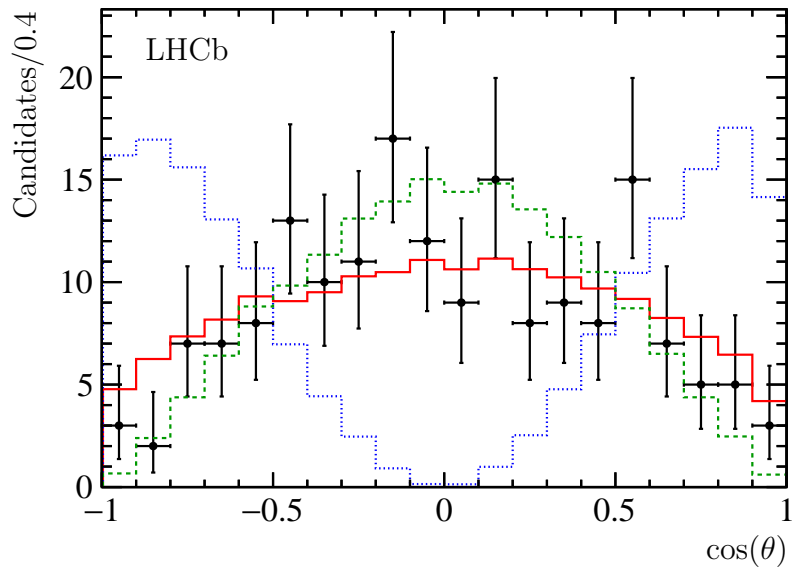

The polarization of the $\phi$ mesons is studied by means of the angle $\theta$ between the direction of flight of a $\phi$ meson 
in the \Bs rest frame and the \Bs direction in the laboratory frame.  
Figure~\ref{fig:pol} 
compares the $\cos ( \theta )$ distribution for the $\Bs \to \phi \phi \phi$ signal candidates in data with expectations from simulation
using different assumptions for the polarization. 
The purely longitudinal polarization clearly does not describe the data. 
The difference between the expectations for no polarization and purely transverse polarization is used
to estimate the corresponding systematic uncertainty in the $\BR ( \Bs \to \phi \phi \phi )$ measurement. 
\protect

\begin{figure}[h]
\centering
\protect\protect\includegraphics[width=0.7\linewidth]{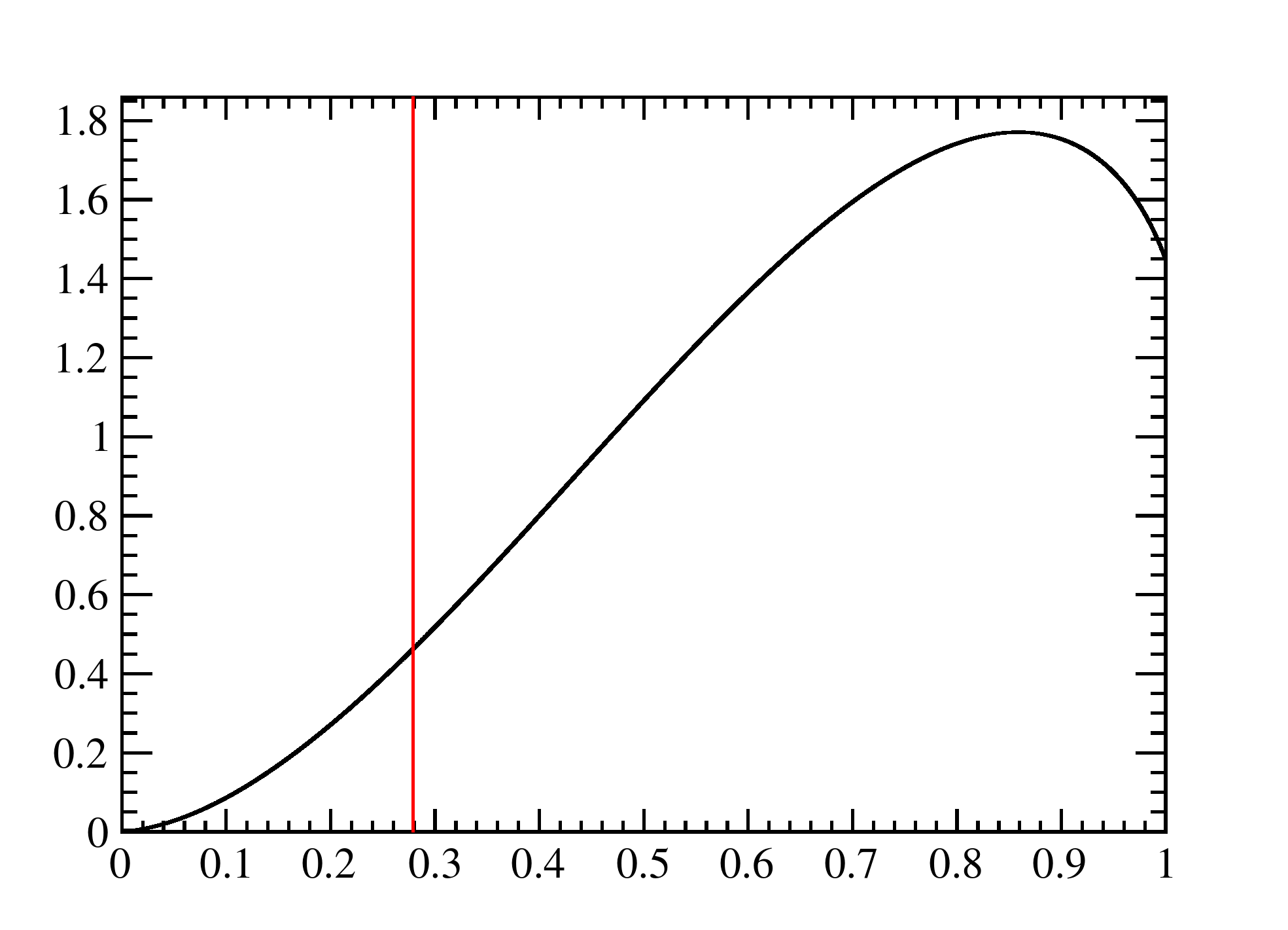}
                \put(-320,180){\rotatebox{90}{{\PDF}}}
                \put(-40,-3) {{$f_{\textrm{T}}$}}
                \put(-270,190) {\lhcb}
\protect\caption
[The \PDF for the fraction of transverse $\phi$ meson polarization $f_{\textrm{T}}$ for the $\Bs \to \phi \phi \phi$ candidates.]
{The \PDF for the fraction of transverse $\phi$ meson polarization $f_{\textrm{T}}$ for the $\Bs \to \phi \phi \phi$ candidates. 
The $95 \%$ Bayesian lower limit is shown by the red vertical line.} 
\protect\label{fig:poln}
\protect\end{figure}

To quantify the fraction of transverse polarization, $f_{\textrm{T}}$, in the data, the probability density function (\PDF) 
for $f_{\textrm{T}}$ is shown in Fig.~\ref{fig:poln}. 
The most probable value is $f_{\textrm{T}} = 0.86$. 
Assuming a uniform prior in the physically allowed range, a Bayesian lower limit of $f_{\textrm{T}} > 0.28$ at $95\%$ CL is found. 
\protect

\clearpage
\section{Summary and discussion}
\protect\label{sec:BSsummary}

To validate the analysis technique, the $\BR ( \Bs \to \phi \phi )$ is first determined to be
\begin{equation*}
\BR ( \Bs \to \phi \phi ) =  ( 2.18 \pm 0.17 \pm 0.11 \pm 0.14_{f_s} \pm 0.65_{\BR}) \times 10^{-5}
\end{equation*}
and is measured with a different technique with respect to the previous 
results~\cite{Abe:1999ze,Acosta:2005eu,Aaltonen:2011rs}. 
The result is consistent with the previous CDF and \lhcb results and has a precision worse than that of the PDG value~\cite{PDG2018}. 
The result is also consistent with theoretical calculations~\cite{Beneke:2006hg,Ali:2007ff,Cheng:2009mu}.
The ratio of the branching fractions for the $\etac (1S)$ decays to $\phi \phi$ and to $\proton \antiproton$ is determined as 
\[
\frac{\BR ( \etac (1S) \ra \phi \phi )}{\BR ( \etac (1S) \ra \proton \antiproton )} = 1.79 \pm 0.14 \pm 0.09 \pm 0.10_{f_s / f_d} \pm 0.03_{f_{\Lb}} \pm 0.29_{\BR} , 
\]
which is larger than the world average value and indicates a problem in it.

The transition $\Bs \to \phi \phi \phi$ is observed for the first time with a significance about $4 \sigma$, 
and its branching fraction is measured to be 
\begin{equation*}
\BR ( \Bs \to \phi \phi \phi ) = ( 2.15 \pm 0.54 \pm 0.28 \pm 0.21 ) \times 10^{-6}.
\end{equation*} 
No clear resonant structure is observed in the $\phi \phi$ invariant mass distribution. 
Depending on how the $\Bs \to \phi \phi \phi$ transition proceeds, 
the above value becomes an upper limit on the penguin transition and pickup of the \ssbar pair, 
with six strange quarks in the final state, 
and/or on the contribution of intermediate resonances such as 
the $\Bs \to \etac \phi$ mode. The result does not contradict to the \lhcb measurement of the $\BR (\Bs \to \etac \phi)$.  Finally, transverse polarization of $\phi$ mesons from $\Bs\to\phi\phi\phi$ decays is favoured by LHCb data. The contribution from transversly polarised $\phi$ measons is found to be larger than $f_{\textrm{T}} > 0.28$ at $95\%$ CL.

\chapter{Summary and prospects}
\label{ch:prospects}
In summary, using an integrated luminosity of 2.0 \invfb collected in 2015 and 2016, the prompt \etac production at $\sqs=13$~\tev centre-of-mass energy is measured for the first time using the decay $\decay{\etac}{\ppbar}$. 
The relative prompt production rates of the \etac and \jpsi states in the LHCb fiducial region ($2.0 < y < 4.5$, $6.5 \gevc < \pt < 14.0 \gevc$)
are measured using \tzfit and \tzcut to distinguish prompt charmonium and charmonium produced in \bquark-hadron decays. The measurement uncertainty is dominated by the statistical one, therefore it can be improved by using larger data sample. The obtained precision is better that the one of the measurement performed at \sqs=7 and 8~\tev mainly due to larger production cross-section and hence reduced statistical uncertainty. Since the precision of the measurement is already better than the theoretical one, a new possible measurement of the \etac prompt production at \sqs=13~\tev with a larger data sample will not further constrain the theory. 
However, the full Run II \lhcb data sample can potentially allow a small extension of the \pt-range studied. 

The additional high-\pt point of the differential production cross-section measurement 
can potentially separate more efficiently
CS and CO contributions. 

The branching fraction from \bquark-hadron to \etac inclusive decays is measured.
The precision of the measurement is limited by systematic uncertainty, which is dominated by that on $\BR(\etac\to\ppbar)$. Hence, further significant improvements on $\BR(\bquark\to\etac X)$ precision can come from measuring $\BR(\etac\to\ppbar)$ to a better precision at available or future charm factories.
The results of this work confirm the first measurement of the branching fraction of inclusive \bquark-decays to the \etac meson and provides the most precise measurement of $\BR(\bquark\to\etac X)$. 

A phenomenological analysis shows that the measured value of $\BR(\bquark\to\etac X)/\BR(\bquark\to\jpsi X)/$ can be accomodated by available theoretical prediction~\cite{Beneke:1998ks}. A simultaneous fit to the
\etac prompt production measurements using prediction from Ref.~\cite{Han:2014jya}; 
the \jpsi prompt production cross-section measurement in limited \pt-range;
the measurement of \etac production in \bquark-decays~\cite{Beneke:1998ks};
and the measurement of \etac inclusive production in \bquark-decays 
allows to reduce the parameter space of involved LDMEs, which provides a reasonable description of all observables. However, the latter is achieved by numerical cancelation of large CO contributions. This calls for new theoretical predictions and possibly new approaches. A good description of the \etac prompt production by \kt-factorization approach should be tested using other observables such as, for example, photoproduction and \epem production cross-sections and \jpsi polarisation.

Production of other charmonium states in \bquark-hadron inclusive decays is studied
with an integrated luminosity of $3 \, fb^{-1}$, using charmonia decays 
to $\phi \phi$ pairs. 
Inclusive production of all $\chi_c$ states in \bquark-hadron inclusive decays are measured.
The branching fraction $\BR ( b \to \chiczero X )$ is measured for the first time. 
The result for \bquark-decays into \chicone is the most precise measurement 
for the mixture of \Bz, \Bp, \Bs and \bquark-baryons. 
The branching fraction of \bquark-hadron decays into \chictwo is measured for the first time  with the \Bz, \Bp, \Bs and \bquark-baryons mixture.  
The result is consistent with the world average of the \Bz, \Bp mixture from Ref.~\cite{PDG2019}. 
The measurements can be further improved using a larger \lhcb Run II data sample.
The precision of the absolute inclusive branching fraction measurements is limited by the knowledge of the branching fractions of charmonium decays to $\phi\phi$. In this thesis it is demonstrated that the current world average value of the $\BR(\decay{\etac}{\phi\phi})$ is rather not reliable and new measurements at charm factories are called for.

The shape of transverse momentum dependence of charmonia production in \bquark-decays is studied for the $\etac (1S)$ and $\chi_c$ states in the \lhcb acceptance and for $\pt > 4 \gevc$. 
A precision of about 15\% for $\etac (1S)$ and between 20\% and 30\% for 
the $\chi_c$ states is achieved. 
.

The first evidence of the $\etactwos$ production in inclusive \bquark-decays and the first evindence of the $\etactwos \to \phi\phi$ are reported in this thesis. A measurement of $\BR(\decay{\bquark}{\etactwos X}\times \BR(\etactwos \to \phi\phi))$ has been performed. A larger data sample will improve the precision of the measurement. A future measurement of the $\BR(\etactwos \to \phi\phi)$ is needed in order to extract $\BR(\bquark\to\etactwos X)$.

A limits on the productuct of inclusive branching fraction of \bquark quark decays to $X(3872)$, $X(3915)$ and $\chictwo(3930)$ and branching fractions of the corresponding decays to $\phi\phi$ are reported.

Using a sample of $\decay{\bquark}{(\decay{\ccbar}{\ppbar})X}$ candidates, the \jpsi and \etac mass difference is measured. 
The obtained result, is consistent with the world average value and is the most precise single \etac mass measurement to date. This measurement is consistent to and is competing with the most precise individual measurements. Possibly, the entire \lhcb Run II data sample of $\decay{\bquark}{(\decay{\ccbar}{\ppbar})X}$ candidates can provide  results of \etac mass and natural width competing with the world average value.
All obrained measurements of charmonium resonance parameters agree with the corresponding world averages.

A branching fraction of the $\Bs \to \phi \phi$ decay has been determined using $\etac \to \phi \phi$ as a reference. An evidence for the $\Bs \to \phi \phi \phi$ decay has been reported at the level of about four standard deviations together with a branching fraction measurement. The resonant structure of the decay including the $\Bs \to (\etac \to \phi \phi) \phi$ contribution is expected to be seen within the entire Run II data sample. This study is an important cross-check of the $\BR(\etac\to\phi\phi)$ value.

Finally, I would like to outline the following wish list of experimental prospects at \lhcb based on charmonium decays to hadrons. 
This list reflects my personal vision and preferences with no aim of being exhaustive.
The main goal is to improve precision of charmonium production measurements, to access more charmonium states, and to search for further complementary measurements.
\begin{enumerate}
\item The natural extension of the work presented in this thesis is a measurement of \textbf{prompt \etactwos production} using its decays to \ppbar and $\phi\phi$. The corresponding trigger lines have been prepared in this work and included in the online trigger during the proton-proton collisions program of \lhcb in 2018. The data sample corresponds to about 2.2\invfb of integrated luminosity recorded at \sqs=13~\tev. 

According to preliminary studies, the expected upper limit on the \etactwos production using the \ppbar decay channel is smaller than the CO contributions projected from \psitwos production measurements. If the \etactwos production is saturated by the CS contribution (as for the \etac prompt production) the amount of data 
would be however not sufficient to perform a measurement. Only an upper limit could be set. In this case, the upper limit would anyway test of the available predictions. Moreover, this upper limit should be used in the simultaneous description of \psitwos prompt production cross-section and polarisation.

Projections for the $\phi\phi$ final states are more difficult, since it is the first data sample of prompt $\phi\phi$ decays at \lhcb and the backround properties are not studied yet.
While the branching fraction of the $\decay{\etactwos}{\ppbar}$ can be extracted using available experimental inputs, there is no experimental information on the $\BR(\decay{\etactwos}{\phi\phi})$. That is why this branching fraction can be extracted from theory only. Potentially, signals from \chic states can be seen in the prompt invariant mass spectrum as well. Especially, a measurement of \chiczero prompt production would be extremely important since it has not been measured yet.

The analysis of the \etactwos production using both decay channels has been started. However, no judgements can be done before the corresponding simulation samples will be produced.

\item Probes of prompt charmonium production using \textbf{other hadronic decays with available Run II (and Run I) data}.
Most of prompt charmonium studies require a dedicated trigger operating during the data taking. Final states including long-lived particles can be triggered using universal \lhcb trigger lines requiring a presence of track(s) displaced from PV. However, a common problem of such decays is that the signal efficiency is reduced due to, particularly, the absence of downstream track reconstruction at the \hltone trigger level.
Nevertheless, I suggest the two following final states.
\begin{itemize}
	\item \textbf{The $\KS K \pi$ final state}. The advantage is that the typical branching fraction of charmonium decays to $\KS K \pi$ is at least an order of magnitude larger than the branching fractions of decays to \ppbar and $\phi\phi$. On the other hand, a high level of combinatorial background is expected since most of the hadrons created at PV are pions. Besides, four tracks have to be reconstructed. 
    The $\KS K \pi$ decay channel is  promising for the \etac mass and width measurement, since a sample $\ccbar \to \KS K \pi$ of exclusive \bquark-decays can potentially provide the cleanest \etac signal at \lhcb. This work has been started at \lhcb.
    The preselection for charmonium decays to $\KS K \pi$ has been written and the entire \lhcb data set for this analysis will be available by the end of 2019.
	\item \textbf{The $\Lambda \bar{\Lambda}$ final state}. This final state requires a reconstruction of two long-lived baryons and therefore the efficiency will be even smaller than for the decay to $\KS K \pi$ channel. Another disadvantage is that the branching fractions are much smaller than the ones of decays to $\KS K \pi$.
	Since the mass of the $\Lambda$ baryon is close to the $\proton \pi$ threshold, this decay channel can be compared to $\phi\phi$. A larger production rate of $\Lambda \bar{\Lambda}$ combinations can be naively expected compared to that of $\phi\phi$. A random $\phi\phi$ combination at PV is more rarely produced because it requires a creation of four \squark-quarks in PV compared to only two \squark quarks for $\Lambda \bar{\Lambda}$. On the other hand, the selection can benefit from variables describing $\Lambda$ lifetime or its vertex displacement from PV. 
	Therefore charmonium decays to $\Lambda \bar{\Lambda}$ are rather less promising than to $\KS K \pi$. The preselection of charmonium decays to $\Lambda \bar{\Lambda}$ have been prepared and the entire \lhcb data for the analysis set will be available by the end of 2019.
\end{itemize}

\item \textbf{The $\eta_b$ production}. I would like to particularly stress this item. All other items in the list appeared to be within experimental possibilities and expectations, while the $\eta_b$ production provides crucial observables for NRQCD and is an essential extension of this work. The measurement can provide an answer, whether the simultaneous description of $S$-wave bottomonium states production has the same complications as the ones observed for charmonium. It can also shed the light in the CO and CS interplay in bottomonium production.   

This task is challenging due to a poor knowledge of the $\eta_b$ meson. No exclusive $\eta_b$ decay branching fraction has been measured yet. The studies of $\eta_b$ will also require a dedicated trigger line in future. Compared to charmonium case, the combinatorial background level in the bottomonium mass region is much reduced. Hence, a smaller bandwdith will be required to reconstruct $\eta_b$ using the same decay channel. I would suggest to develop additional trigger lines for the $\eta_b$ decays to two and four stable hadrons and to $\KS K \pi$. 
A trigger line for bottomonium decays to $K^{*}K^{*}$ was active during the \lhcb Run II, but because of four tracks reconstruction its efficiency is small.

\item \textbf{The $h_c$ study} at \lhcb. The $h_c$ meson has not been seen at \lhcb yet. The study of prompt $h_c$ production is well motivated by theory since it provides a new obervable, which has never been measured at hadron machines. The same apply for production in \bquark-hadron decays
\begin{itemize}
 \item The first promising channel to access $h_c\to\KS K \pi$ discussed above. The data is already available and the analysis has been started.
 \item Alternatively, the $h_c$ can be resonstructed using decays to $\etac \gamma$ and $\etac \mup \mu^{-}$ final states. This, however, requires a reconstruction of the \etac state. The trigger line aiming at prompt $\etac \mup \mu^{-}$ selection was operational during the data taking in 2018.
 \item Following the observation of the $h_c$ decay to $\ppbar \pip \pim$, this decay channel can be used to access $h_c$ produced in \bquark-hadron decays.
\end{itemize}

\item \textbf{Central exclusive production} (CEP) of \etac. This topic requires a relatively easy analysis, The observation of the \etac produced in CEP will be directly interpreted as the odderon discovery. Different decay channels can be considered, while the data is available only for the \ppbar final state. 

\item Similarly, studies of the \textbf{\etac production in ion-ion collisions} can be studied. In this case, dedicated trigger lines are not needed since the \lhcb stores all information from ion-ion collisions. On the other hand, relatively low luminosity limits the measurement.

\item \textbf{Other hadronic decays of charmonium} to study charmonium from \bquark-decays. Among possible final states I would highlight $\Lambda \bar{\Lambda(1520)}$ and $\Lambda(1520) \bar{\Lambda}$. The studies of these decays can yield their branching fraction measurements with the data already available.

\item Studies involving \etac reconstruction. This subject is not related to charmonium production but the experience on \etac production measurements can be transferred to other studies involving \etac meson in the final state.
\begin{itemize}
	\item Study of \textbf{semileptonic decays $\Bc\to\etac\mu\nu$ and possibly $\Bc\to\etac\tau\nu$}. This subject is interesting for future lepton universality measurement of the ratio $\BR(\Bc\to\etac\tau\nu)/\BR(\Bc\to\etac\mu\nu)$. The $\etac\to\ppbar$ and $\etac\to\KS K \pi$ decay channels can be used. Due to a small lifetime of \Bc meson, the universal \bquark-hadron trigger lines are not optimal. Therefore, a dedicated trigger line should be developed in future. Nevertheless, the situation is not that much extreme as for prompt charmonium case, so that the studies have been started with already available data. The corresponding preselection has been prepared and the entire \lhcb data set will be available by the end of 2019.
	\item Another example of non-production measurements is \textbf{a search for hadron exotic candidates decaying to the \etac meson}. One of such studies has been already performed, finding evidence of $Z(4010)\to \etac \pi$. Similar studies can be performed for other decay channels involving \etac, to (in)validate charmonium-like states previously observed via decays to \jpsi and \psitwos. 
\end{itemize}
\end{enumerate}

\chapter*{Main thesis results}
\addcontentsline{toc}{chapter}{Main thesis results}  
Main results obtained in the thesis are listed below.

\begin{enumerate}
\item \lhcb operation
	\begin{itemize}
	\item \textbf{Trigger development} for prompt charmonium decays to $\ppbar$ and $\phi\phi$ final states
	\item \textbf{Preparation of data sets} (stripping) for charmonium produced in \bquark-hadron decays using a list of charmonium decays to hadrons
	\end{itemize}
\item Physics analysis
	\begin{itemize}
		\item First measurement of \textbf{prompt \etac production at \sqs=13 TeV}
		\item First or most precise measurements of \textbf{production of $\boldsymbol{\etac}$, $\boldsymbol{\chiczero}$, $\boldsymbol{\chicone}$, $\boldsymbol{\chictwo}$ and $\boldsymbol{\etactwos}$ in \bquark-hadron inclusive decays}
		\item Search for $X(3872)$, $X(3915)$, $\chictwo(2P)$ production in \bquark-hadron inclusive decays
		\item Measurement of the \textbf{$\boldsymbol{\etac}$ resonance parameters}
		\item Measurement of the branching fraction of the $\decay{\boldsymbol{\etac}}{\boldsymbol{\phi}\boldsymbol{\phi}}$ decay 
		\item Measurement of the branching fraction of the $\boldsymbol\Bs\to\boldsymbol{\phi}\boldsymbol{\phi}$ decay
		\item First evidence of the $\mathbf{\Bs}\boldsymbol{\to}\boldsymbol{\phi}\boldsymbol{\phi}\boldsymbol{\phi}$ decay
	\end{itemize}
\item Phenomenological analysis
	\begin{itemize}
		\item Proposal of a \textbf{new technique of simultaneous fit to long-distance matrix elements}
		\item Simultaneous study of the \etac and \jpsi prompt production and production in \bquark-hadron inclusive decays
		\item Simultaneous study of $\chic_J$ states production in \bquark-hadron inclusive decays
	\end{itemize}
\end{enumerate}

\chapter*{List of measurements}
\addcontentsline{toc}{chapter}{List of measurements}  
A list of the most important measurements performed in this thesis is summarized below.
\vspace{0.5cm}

\textbf{\etac prompt production at \sqs=13 \tev}

Integral prompt production cross-section, relative and absolute measurements:
\begin{align*}
(\sigma_{\etac}/\sigma_{\jpsi})_{13 \tev}^{6.5 \gevc < p_T < 14.0 \gevc, 2.0<y<4.5}  
	&= \etacPromptRelativeXsec, 
\end{align*}
\begin{align*}
(\sigma_{\etac})_{13 \tev}^{6.5 \gevc < p_T < 14.0 \gevc, 2.0<y<4.5}  
	&= \etacPromptAbsoluteXsec.
\end{align*} 

\pt-differential prompt production cross-section, relative and absolute measurements:
\begin{table}[h]
\centering
\small
\begin{tabular}{c|c}
\pt, \gev      &$d\sigma^{prompt}_{\etac}/d\sigma^{prompt}_{\jpsi}$ \\ \hline
6.5 - 8.0     & 1.68 $\pm$ 0.33 $\pm$ 0.06 $\pm$ 0.11 $\pm$ 0.21 \\ 
8.0 - 10.0    & 2.01 $\pm$ 0.28 $\pm$ 0.09 $\pm$ 0.13 $\pm$ 0.25 \\  
10.0 - 12.0   & 2.27 $\pm$ 0.36 $\pm$ 0.13 $\pm$ 0.14 $\pm$ 0.28 \\ 
12.0 - 14.0   & 3.30 $\pm$ 0.62 $\pm$ 0.22 $\pm$ 0.21 $\pm$ 0.41 \\
\end{tabular} 
\caption{The relative \pt-differential \etac prompt production.}
\label{tab:relPromptTable}
\centering
\footnotesize
\begin{tabular}{c|c}
\pt, \gev      & $d\sigma^{prompt}_{\etac}/d\pt$, \nb/\gev  \\ \hline
6.5 - 8.0     & 536.09 $\pm$ 105.04 $\pm$ 19.61 $\pm$ 34.19 $\pm$ 70.67 \\ 
8.0 - 10.0    & 180.92 $\pm$ 24.81 $\pm$ 7.90 $\pm$ 11.35 $\pm$ 24.97  \\  
10.0 - 12.0   & 73.92 $\pm$ 11.57 $\pm$ 4.07 $\pm$ 4.60 $\pm$ 10.32   \\ 
12.0 - 14.0     & 42.12 $\pm$ 7.95 $\pm$ 2.83 $\pm$ 2.62 $\pm$ 6.01   \\ 
\end{tabular} 
\caption{The \pt-differential \etac prompt production.}
\end{table}

\newpage
\textbf{\etac production in \bquark-hadron inclusive decays}

Absolute branching fraction:
\begin{align*}
\BR_{\bToEtacX}/\BR_{\bToJpsiX} &= \etacSecondaryRelativeBR, \\
\BR_{\bToEtacX} &= \etacSecondaryAbsoluteBR. 
\end{align*}

\pt-differential production cross-section in \bquark-hadron inclusive decays, relative and absolute measurements:
\begin{table}[h]
\centering
\small
\begin{tabular}{c|c}
\pt, \gev     & $d\sigma^{\bquark-decays}_{\etac}/d\sigma^{\bquark-decays}_{\jpsi}$ \\ \hline 
6.5 - 8.0     & 0.41 $\pm$ 0.06 $\pm$ 0.01 $\pm$ 0.02 $\pm$ 0.05 \\ 
8.0 - 10.0    & 0.61 $\pm$ 0.05 $\pm$ 0.03 $\pm$ 0.03 $\pm$ 0.08 \\  
10.0 - 12.0   & 0.45 $\pm$ 0.06 $\pm$ 0.02 $\pm$ 0.02 $\pm$ 0.06  \\ 
12.0 - 14.0   & 0.54 $\pm$ 0.07 $\pm$ 0.03 $\pm$ 0.02 $\pm$ 0.07  \\   
\end{tabular} 
\caption{The relative \pt-differential \etac production in inclusive \bquark-decays.}
\footnotesize
\centering
\small
\begin{tabular}{c|c}
\pt, \gev     & $d\sigma^{\bquark-decays}_{\etac}/d\pt$, \nb/\gev     \\ \hline 
6.5 - 8.0     & 27.16$\pm$ 4.23$\pm$ 0.99$\pm$ 1.34$\pm$ 3.74 \\ 
8.0 - 10.0    & 18.82$\pm$ 1.52$\pm$ 0.81$\pm$ 0.91$\pm$ 2.61 \\  
10.0 - 12.0   & 6.56$\pm$ 0.84$\pm$ 0.34$\pm$ 0.32$\pm$ 0.93 \\ 
12.0 - 14.0   & 3.79$\pm$ 0.51$\pm$ 0.23$\pm$ 0.18$\pm$ 0.55 \\  
\end{tabular} 
\caption{The \pt-differential \etac production cross-section in inclusive \bquark-decays}
\label{tab:SecondaryTable}
\end{table} 

\clearpage
\textbf{First or most precise measurements of $\chic_J$ production in \bquark-hadron inclusive decays}

Double ratio measurements:
\begin{align*}
\frac{\BR ( b \to \chiczero X ) \times \BR ( \chiczero \to \phi \phi )}{\BR ( b \to \etac X ) \times \BR ( \etac \to \phi \phi )} 
 &= 0.147 \pm 0.023 \pm 0.011 \ , \\
\frac{\BR ( b \to \chicone X ) \times \BR ( \chicone \to \phi \phi )}{\BR ( b \to \etac X ) \times \BR ( \etac \to \phi \phi )} 
 &= 0.073 \pm 0.016 \pm 0.006 \ , \\
\frac{\BR ( b \to \chictwo X ) \times \BR ( \chictwo \to \phi \phi )}{\BR ( b \to \etac X ) \times \BR ( \etac \to \phi \phi )} 
 &= 0.081 \pm 0.013 \pm 0.005 \ . 
\end{align*}

Single ratio measurements:
\begin{align*}
\frac{\BR ( b \to \chiczero X )}{\BR ( b \to \etac X )} &= 0.615 \pm 0.095 \pm 0.047 \pm 0.149 \ , \\
\frac{\BR ( b \to \chicone X )}{\BR ( b \to \etac X )}  &= 0.562 \pm 0.119 \pm 0.047 \pm 0.131 \ , \\
\frac{\BR ( b \to \chictwo X )}{\BR ( b \to \etac X )}  &= 0.234 \pm 0.038 \pm 0.015 \pm 0.057 \ . 
\end{align*}

Absolute branching fractions:
\begin{align*}
\BR ( b \to \chiczero X ) &= ( 3.02 \pm 0.47 \pm 0.23 \pm 0.94 ) \times 10^{-3} \ , \\
\BR ( b \to \chicone X )  &= ( 2.76 \pm 0.59 \pm 0.23 \pm 0.89 ) \times 10^{-3} \ , \\
\BR ( b \to \chictwo X )  &= ( 1.15 \pm 0.20 \pm 0.07 \pm 0.36 ) \times 10^{-3} \ . 
\end{align*}

\textbf{First measurement of \etactwos production in \bquark-hadron inclusive decays}

Double ratio measurement:
\begin{align*}
\frac{\BR ( b \to \etac (2S) X ) \times \BR ( \etac (2S) \to \phi \phi )}{\BR ( b \to \etac (1S) X ) \times \BR ( \etac (1S) \to \phi \phi )} 
 = 0.040 \pm 0.011 \pm 0.004. \ 
\end{align*}

Product of branching fractions:
\begin{align*}
\BR ( b \to \etac (2S) X ) \times \BR ( \etac (2S) \to \phi \phi )
 &= ( 6.34 \pm 1.81 \pm 0.57 \pm 1.89 ) \times 10^{-7}. \ 
\end{align*}

\newpage
\textbf{Search for $X(3872)$, $X(3915)$, $\chictwo(2P)$ production in \bquark-hadron inclusive decays}

Upper limits on double ratios:
\begin{align*}
\frac{\BR ( \bquark \ra X(3872) X ) \times \BR ( X(3872) \ra \phi \phi )}{\BR ( \bquark \ra \chicone X ) \times \BR ( \chicone \ra \phi \phi )} & < 0.39 (0.34) , \\
\frac{\BR ( \bquark \ra X(3915) X ) \times \BR ( X(3915) \ra \phi \phi )}{\BR ( \bquark \ra \chiczero X ) \times \BR ( \chiczero \ra \phi \phi )} & < 0.14 (0.12) , \\
\frac{\BR ( \bquark \ra \chictwo (3930) X ) \times \BR ( \chictwo (3930) \ra \phi \phi )}{\BR ( \bquark \ra \chictwo X ) \times \BR ( \chictwo \ra \phi \phi )} & < 0.20 (0.16). 
\end{align*}

Upper limits on product of branching fractions:
\begin{align*}
\BR ( \bquark \ra X(3872) X ) \times \BR ( X(3872) \ra \phi \phi ) & < 4.5 (3.9) \times 10^{-7} , \\
\BR ( \bquark \ra X(3915) X ) \times \BR ( X(3915) \ra \phi \phi ) &  < 3.1 (2.7) \times 10^{-7} , \\
\BR ( \bquark \ra \chictwo (3930) X ) \times \BR ( \chictwo (3930) \ra \phi \phi ) &  < 2.8 (2.3)  \times 10^{-7} . 
\end{align*}

\textbf{Measurement of charmonia masses and natural width}

\begin{table}[h]
\centering
\begin{tabular}{l|c|c}
                     & using $\ccbar\to\ppbar$ & using $\ccbar\to\phi\phi$ \\ \hline
$M_{\etac(1S)}$      & $2983.91\pm0.77\pm0.11$& $2982.81 \pm 0.99 \pm 0.45$ \\ 
$M_{\chiczero}$      && $3412.99 \pm 1.91 \pm 0.62$ \\ 
$M_{\chicone}$       && $3508.38 \pm 1.91 \pm 0.66$ \\ 
$M_{\chictwo}$       && $3557.29 \pm 1.71 \pm 0.66$ \\ 
$M_{\etac(2S)}$      && $3636.35 \pm 4.06 \pm 0.69$ \\ 
$\Gamma_{\etac(1S)}$ && $31.35 \pm 3.51 \pm 2.01$ \\ 
\end{tabular}
\caption{Charmonia masses (in \mev) and natural width (in \mev).} 
\end{table}

\textbf{Measurement of the branching fraction $\BR(\etac\to\phi\phi)$}

Ratio of branching fractions:
\begin{equation*}
\frac{\BR ( \etac (1S) \ra \phi \phi )}{\BR ( \etac (1S) \ra \proton \antiproton )} = 1.79 \pm 0.14 \pm 0.09 \pm 0.10_{f_s / f_d} \pm 0.03_{f_{\Lb}} \pm 0.29_{\BR}.
\end{equation*}

\clearpage
\textbf{Measurements of branching fractions for \Bs decays to $\phi\phi$ and $\phi\phi\phi$}

Branching fraction of $\Bs \to \phi \phi$ decay:
\begin{equation*}
\BR ( \Bs \to \phi \phi ) =  ( 2.18 \pm 0.17 \pm 0.11 \pm 0.14_{f_s} \pm 0.65_{\BR}) \times 10^{-5},
\end{equation*}

$\Bs \to \phi \phi \phi$, ratio of branching fractions:
\begin{align*}
\frac{\BR ( \Bs \to \phi \phi \phi )}{\BR ( \Bs \to \phi \phi )} 
 = 0.117 \pm 0.030 \pm 0.015 \ ,
\end{align*}

$\Bs \to \phi \phi \phi$, absolute branching fraction:
\begin{equation*}
\BR ( \Bs \to \phi \phi \phi ) = ( 2.15 \pm 0.54 \pm 0.28 \pm 0.21 ) \times 10^{-6}.
\end{equation*}

\clearpage
\chapter*{Acknowledgements}
\addcontentsline{toc}{chapter}{Acknowledgements}
I want to thank my supervisor Sergey Barsuk, for all his help, support, guidings, useful advices and suggestions during the whole period of my work. Also, I found very important his help with my integration into the life of the LHCb collaboration. 
I appreciate that I had much freedom to do research and to push it forward.

I would also thank Jean-Philippe Lansberg,  Hua-Sheng Shao and Emi Kou for beneficial discussions on the theory of charmonium production.

I would also like to express my great appreciation to my colleagues from LHCb group of LAL - Patrick, Jibo, Jacques, Marie-H\'{e}l\'{e}ne, Fr\'{e}d\'{e}ric, Guy, Michael, Yanxi, Francesco, Laure, Vladik, Vitalii, Fabrice, Renato, Victor, Joao, Carla, Elizabeth and Davide. I was pleased to work with you, and I will be happy to continue collaborating with you at LHCb! 

I would thank students  I had a great pleasure to work with - Valeriia, Taras and Pavlo. I have no doubts that you will become great physicists!

I can't omit my thanks to my school teacher of physics, Yury Pasikhov, and my university teacher, Oleg Bezshyiko. Without your inspiration, I wouldn't start my research in physics.

I thank my friends Denys, Rita, Nefedka, Kolya, Iaroslava, Nastya, Andrii, Igor, Salama, Edoardo, Yacine and Alenka for the time we spent together and for all discussions we had.

Finally, I thank my incredible wife Olga and my family for their strong support.

\clearpage
\chapter*{Synth\`ese en fran\c{c}ais}
\addcontentsline{toc}{chapter}{Synth\`ese en fran\c{c}ais}
\doublespacing
\begin{otherlanguage}{french}
Les études de production de charmonium fournissent les tests rigoureux de modèles théoriques basés sur chromodynamique quantique (QCD) non-relativiste. 
A ce jour les modèles sont confrontés principalement aux mesures expérimentales des etats $J^{PC} = 1^{-\,-}$ de charmonium, comme les mésons $\jpsi$ et $\psitwos$, qui désintègrent à une paire de muons. De plus, les états $\chicone$ et $\chictwo$ sont reconstruits via leurs transitions radiatives en \jpsi, ce qui nécessite cependant une reconstruction de photons des basses énergies.
La reconstruction des désintégrations vers les hadrons permet de réaliser les études de production pour tous les états connus du charmonium. En utilisant les états finaux \ppbar et $\phi\phi$ j'ai étudié la production des états $\etac(1S)$, $\etactwos$ et $\chic_J$ de charmonium avec l'expérience \lhcb.\par

En utilisant les données LHCb prises en 2015 et 2016 à l'énergie du centre de masse $\sqs = 13$~\tev, qui correspondent à une luminosité intégrée de 2.0 \invfb, la section efficace de la production prompte du méson \etac est mesurée pour la première fois avec la désintégration $\decay{\etac}{\ppbar}$. Les taux de production relatifs des mésons \etac et \jpsi dans la région fiduciale de LHCb ($2.0 <y <4.5, 6.5 \gevc < \pt <14.0 \gevc$) sont mesurés à l'aide de techniques \tzfit et \tzcut pour distinguer le charmonium produit dans le vertex de collision proton-proton et le charmonium produit dans les désintégrations des hadrons \bquark. Étant dominée par la statistique, l’incertitude de mesure sera améliorée en utilisant un plus grand échantillon de données. La précision obtenue est meilleure que celle de la mesure effectuée à \sqs = 7 et 8~\tev principalement en raison d'une section efficace de production plus élevée. Puisque la précision de la mesure est déjà meilleure que la prédiction théorique, une nouvelle mesure basée sur un plus grand échantillon de données ne contraindra plus la théorie. Toutefois, l'échantillon complet de données Run II \lhcb potentiellement permettra une extension de la plage \pt étudiée. Les mesures de la section efficace de la production différentielle dans la région du \pt élargie permettront de séparer plus efficacement les contributions des éléments de matrice color-singlet et color-octet.\par

Je mesure le rapport d’embranchement de la désintégration inclusive des hadrons \bquark vers l’état \etac. La précision de la mesure est limitée par l’incertitude systématique, dominée par celle de $\BR(\etac \to \ppbar)$. Par conséquent, les futures améliorations sur la précision de $\BR(\bquark \to \etac X)$ demandent la mesure plus précise de $\BR(\etac \to \ppbar)$, ce qui est attendu dans les usines du charme existantes ou futures.
Un des résultat important de cette thèse porte sur la mesure la plus précise du rapport d’embranchement pour les désintégrations des hadrons \bquark vers le méson \etac, $\BR(\bquark \to \etac X)$.\par

La production d'autres états du charmonium dans les désintégrations inclusives des hadrons \bquark est étudiée avec une luminosité intégrée de $ 3 \, fb ^ {- 1} $, en utilisant les désintégrations de charmonia vers les paires de mésons $\phi$, $\phi \phi$.
L'analyse de la production de charmonium à l'aide de désintégrations de charmonium en $\phi \phi$ nécessite la soustraction de la contribution du fond de kaons non résonants.
Pour cela, j'ai proposé une technique permettant de sélectionner des états finaux avec deux ou trois mésons $\phi$ libres du fond combinatoire de kaons.\par

Avec cette nouvelle technique, la production de tous les états $\chi_c$ dans les désintégrations inclusives des hadrons \bquark est mesurée. Le rapport d’embranchement $\BR(b \to \chiczero X)$ est mesuré pour la première fois.
Le résultat pour la production du méson \chicone est la mesure la plus précise
pour les échantillons des hadrons \bquark avec le mélange de \Bz, \Bp, \Bs et baryons avec le quark \bquark. 
Le rapport d’embranchement de la désintégration des hadrons \bquark en \chictwo est mesurée pour la première fois pour les échantillons des hadrons \bquark avec le mélange de \Bz, \Bp, \Bs et baryons avec le quark \bquark. 
Le résultat est en accord avec la moyenne mondiale du rapport d’embranchement pour les échantillons de mésons \Bz et \Bp uniquement. Les mesures peuvent être encore améliorées en utilisant toutes les données \lhcb acquis en Run II.
La précision des mesures des rapports d’embranchement inclusives est limitée par la connaissance des rapports d’embranchement de désintégrations du charmonium vers $\phi\phi$. Dans cette thèse, il est démontré que la valeur moyenne mondiale actuelle du $\BR(\decay{\etac}{\phi \phi})$ n'est pas fiable et que des nouvelles mesures dans les usines du charme sont nécessaires. Je mesure le rapport d’embranchement de la désintégration $\decay{\etac(1S)}{\phi \phi}$ pour résoudre une tension avec d'autres mesures existantes.\par

La dépendance de la production de charmonia issus des hadrons \bquark en impulsion transverse $\pt$ est étudiée pour les états $\etac(1S)$ et $\chi_c$ pour les désintégrations où les kaons finaux traversent le détecteur \lhcb et pour $\pt> 4 \gevc$.
Une précision d’environ 15\% pour le méson $\etac(1S)$ et entre 20\% et 30\% pour les états $\chi_c$ sont atteints.\par

Les études de l’état  $\etactwos$ dans la thèse ont fourni également des nouvelles mesures. LHCb a mesuré l’état $\etactwos$ dans les désintégrations des hadrons \bquark pour la première fois. Et c’est aussi pour la première fois que LHCb a vu la désintégration $\etactwos \to \phi \phi$. Le rapport d’embranchement $\BR(\decay{\bquark}{\etactwos X} \times \BR(\etactwos \to \phi \phi))$ a été mesuré. Un plus grand échantillon de données améliorera la précision de la mesure. Une mesure de $\BR(\etactwos \to \phi \phi)$ est nécessaire pour l’extraction de $\BR(\bquark \to \etactwos X)$.\par

En outre, une recherche de la production d'autres états de type charmonium dans les désintégrations avec \bquark-hadrons est effectuée. Les états de charmonium avec des nombres quantiques similaires sont utilisés pour normalisation.\par

Les produits du rapport d’embranchement de la production inclusive des états $X(3872)$, $ X(3915)$ et $\chictwo(3930)$ dans les désintégrations des hadron \bquark et le rapport d’embranchement des transitions de ces états vers $\phi\phi$ sont étudiés. Aucun signal est observé et la limite est obtenue pour la production de chaque état étudié.\par

Afin de comparer les résultats obtenus aux calculs théoriques, j'ai proposé d'utiliser un ajustement simultané les mesures de la production des états de charmonium et les éléments de matrice longue distance, les deux pour la production hadronique de charmonium et la production dans les désintégrations inclusives des hadrons \bquark.
Cela permet de restreindre fortement l'espace de phase autorisé pour les éléments de la matrice décrivant la production de charmonium. Cela démontre également une limite d'application de la théorie et appelle à la poursuite du développement des modèles.\par

Une analyse phénoménologique montre que la valeur mesurée de $\BR(\bquark \to \etac X)/\BR(\bquark \to \jpsi X)$ peut être adaptée par les études théorique
\foreignlanguage{british}{\cite{Beneke:1998ks}}. 
Une analyse simultanée de la hadroproduction de l’état \etac utilise la prédiction de Réf.
\foreignlanguage{british}{\cite{Han:2014jya}}; 
la mesure de la section efficace de hadroproduction de l’état \jpsi dans la plage \pt limitée;
la mesure de la production de \etac dans les désintégrations des hadrons \bquark
\foreignlanguage{british}{\cite{Beneke:1998ks}};
et la mesure de la production inclusive de l’état \etac dans les désintégrations des hadrons \bquark
permettent de réduire l'espace de phase des LDME impliqués, ce qui fournit une description consistante de tous les observables. Toutefois, cette dernière est réalisée par l’annulation numérique de contributions importantes en CO. Cela motive des nouvelles prédictions théoriques et éventuellement de nouvelles approches. Une nouvelle description de la hadroproduction de l’état \etac par factorisation \kt doit être validée à l'aide d'autres observables, telles que, par exemple, les sections efficaces de la photoproduction et de la production de \epem, et la polarisation de \jpsi.\par

En utilisant un échantillon de candidats $ \decay{\bquark}{(\decay{\ccbar}{\ppbar}) X} $, la différence de masse entre les états \jpsi et \etac est mesurée.
Le résultat obtenu est en accord avec la valeur moyenne mondiale. Cette mesure représente la détermination de la masse de l’état \etac la plus précise à ce jour. Cette mesure est aussi en accord avec des mesures individuelles les plus précises. Avec l’ensemble de données de Run II et la sélection de candidats $ \decay{\bquark}{(\decay{\ccbar}{\ppbar}) X} $ \lhcb améliorera la précision de mesure de la masse de l’état \etac et fournira la mesure de la largeur naturelle en concurrence avec la moyenne mondiale.\par 

Enfin, les mésons \Bs sont reconstruits via les désintégrations en deux ou trois mésons $\phi$. Cela permet d'effectuer une mesure indépendante de $\BR(\Bs \to \phi \phi)$. En plus, l’évidence de la désintégration $\Bs \to \phi \phi \phi$ est obtenue pour la première fois. Une structure résonance de la désintégration $\Bs \to \phi \phi \phi$ ainsi que la polarisation du méson $ \phi $ sont également étudiés dans la thèse.\par

Le rapport d’embranchement de la désintégration $\Bs \to \phi \phi$ est déterminé en utilisant la mode $\etac\to \phi \phi $ comme une référence. L’évidence de la désintégration $\Bs \to \phi \phi \phi$ est obtenue au niveau d'environ quatre déviations standards avec la mesure d’un rapport d’embranchement. Les études détaillées de la structure résonance de la désintégration, y compris la contribution $\Bs \to (\etac\to \phi \phi) \phi$ deviendra possible avec l'ensemble de données de Run II.\par 

La thèse porte principalement sur les études de la production des états de charmonium, \etac(1S), \etac(2S), \chiczero, \chicone, \chictwo. Les études visent la vérification expérimentale des modèles basée sur la QCD non-relativiste. Production des états de charmonium dans le vertex de collision des faisceaux des proton sur LHC ainsi que la production des états de charmonium issus des hadron \bquark sont étudiés. Les résultats présentés dans la thèse sont les premières mesures ou les mesures les plus précises à ce jour. En plus une nouvelle technique a été développée, qui permet de contraindre les valeurs des éléments de matrice longue distance de QCD non-relativiste. 
Les paramètres des états étudiés, la masse et la largeur naturelle, sont aussi déterminés. 
Finalement le rapport d’embranchement de la désintégration $\B_s \to \phi \phi \phi$ est mesuré pour la première fois. 
\end{otherlanguage}
\begin{singlespace}
\setboolean{inbibliography}{true}
\bibliographystyle{unsrturl}
\bibliography{ms,LHCb-PAPER,LHCb-CONF,LHCb-DP,LHCb-TDR}{}

\begin{thebibliography}{100}

\bibitem{Brambilla:2010cs}
N.~Brambilla et~al.
\newblock {Heavy quarkonium: progress, puzzles, and opportunities}.
\newblock {\em Eur. Phys. J.}, C71:1534, 2011.
\newblock \href {http://arxiv.org/abs/1010.5827} {\path{arXiv:1010.5827}},
  \href {http://dx.doi.org/10.1140/epjc/s10052-010-1534-9}
  {\path{doi:10.1140/epjc/s10052-010-1534-9}}.

\bibitem{GellMann:1964nj}
Murray Gell-Mann.
\newblock {A Schematic Model of Baryons and Mesons}.
\newblock {\em Phys. Lett.}, 8:214--215, 1964.
\newblock \href {http://dx.doi.org/10.1016/S0031-9163(64)92001-3}
  {\path{doi:10.1016/S0031-9163(64)92001-3}}.

\bibitem{Feynman:1969ej}
Richard~P. Feynman.
\newblock {Very high-energy collisions of hadrons}.
\newblock {\em Phys. Rev. Lett.}, 23:1415--1417, 1969.
\newblock [,494(1969)].
\newblock \href {http://dx.doi.org/10.1103/PhysRevLett.23.1415}
  {\path{doi:10.1103/PhysRevLett.23.1415}}.

\bibitem{Bjorken:1969ja}
J.~D. Bjorken and Emmanuel~A. Paschos.
\newblock {Inelastic Electron Proton and gamma Proton Scattering, and the
  Structure of the Nucleon}.
\newblock {\em Phys. Rev.}, 185:1975--1982, 1969.
\newblock \href {http://dx.doi.org/10.1103/PhysRev.185.1975}
  {\path{doi:10.1103/PhysRev.185.1975}}.

\bibitem{Bloom:1969kc}
Elliott~D. Bloom et~al.
\newblock {High-Energy Inelastic $ep$ Scattering at 6-Degrees and 10-Degrees}.
\newblock {\em Phys. Rev. Lett.}, 23:930--934, 1969.
\newblock \href {http://dx.doi.org/10.1103/PhysRevLett.23.930}
  {\path{doi:10.1103/PhysRevLett.23.930}}.

\bibitem{Greenberg:1964pe}
O.~W. Greenberg.
\newblock {Spin and Unitary Spin Independence in a Paraquark Model of Baryons
  and Mesons}.
\newblock {\em Phys. Rev. Lett.}, 13:598--602, 1964.
\newblock \href {http://dx.doi.org/10.1103/PhysRevLett.13.598}
  {\path{doi:10.1103/PhysRevLett.13.598}}.

\bibitem{Aubert:1974js}
J.~J. Aubert et~al.
\newblock {Experimental Observation of a Heavy Particle $J$}.
\newblock {\em Phys. Rev. Lett.}, 33:1404--1406, 1974.
\newblock \href {http://dx.doi.org/10.1103/PhysRevLett.33.1404}
  {\path{doi:10.1103/PhysRevLett.33.1404}}.

\bibitem{Augustin:1974xw}
J.~E. Augustin et~al.
\newblock {Discovery of a Narrow Resonance in $e^+ e^-$ Annihilation}.
\newblock {\em Phys. Rev. Lett.}, 33:1406--1408, 1974.
\newblock [Adv. Exp. Phys.5,141(1976)].
\newblock \href {http://dx.doi.org/10.1103/PhysRevLett.33.1406}
  {\path{doi:10.1103/PhysRevLett.33.1406}}.

\bibitem{Bjorken:1964gz}
J.~D. Bjorken and S.~L. Glashow.
\newblock {Elementary Particles and SU(4)}.
\newblock {\em Phys. Lett.}, 11:255--257, 1964.
\newblock \href {http://dx.doi.org/10.1016/0031-9163(64)90433-0}
  {\path{doi:10.1016/0031-9163(64)90433-0}}.

\bibitem{Glashow:1970gm}
S.~L. Glashow, J.~Iliopoulos, and L.~Maiani.
\newblock {Weak Interactions with Lepton-Hadron Symmetry}.
\newblock {\em Phys. Rev.}, D2:1285--1292, 1970.
\newblock \href {http://dx.doi.org/10.1103/PhysRevD.2.1285}
  {\path{doi:10.1103/PhysRevD.2.1285}}.

\bibitem{Callaway:1982eb}
David J.~E. Callaway and Aneesur Rahman.
\newblock {The Microcanonical Ensemble: A New Formulation of Lattice Gauge
  Theory}.
\newblock {\em Phys. Rev. Lett.}, 49:613, 1982.
\newblock \href {http://dx.doi.org/10.1103/PhysRevLett.49.613}
  {\path{doi:10.1103/PhysRevLett.49.613}}.

\bibitem{Abe:1997jz}
F.~Abe et~al.
\newblock {$\jpsi$ and $\psi(2S)$ production in $p\bar{p}$ collisions at
  $\sqrt{s} = 1.8$ TeV}.
\newblock {\em Phys. Rev. Lett.}, 79:572--577, 1997.
\newblock \href {http://dx.doi.org/10.1103/PhysRevLett.79.572}
  {\path{doi:10.1103/PhysRevLett.79.572}}.

\bibitem{LHCb-PAPER-2014-029}
R.~Aaij et~al.
\newblock {Measurement of the $\etac(1S)$ production cross-section in
  proton-proton collisions via the decay $\etac(1S) \to\proton\antiproton$}.
\newblock {\em Eur. Phys. J.}, C75:311, 2015.
\newblock \href {http://arxiv.org/abs/1409.3612} {\path{arXiv:1409.3612}},
  \href {http://dx.doi.org/10.1140/epjc/s10052-015-3502-x}
  {\path{doi:10.1140/epjc/s10052-015-3502-x}}.

\bibitem{Peter:1997me}
Markus Peter.
\newblock {The Static potential in QCD: A Full two loop calculation}.
\newblock {\em Nucl. Phys.}, B501:471--494, 1997.
\newblock \href {http://arxiv.org/abs/hep-ph/9702245}
  {\path{arXiv:hep-ph/9702245}}, \href
  {http://dx.doi.org/10.1016/S0550-3213(97)00373-8}
  {\path{doi:10.1016/S0550-3213(97)00373-8}}.

\bibitem{Schroder:1998vy}
York Schroder.
\newblock {The Static potential in QCD to two loops}.
\newblock {\em Phys. Lett.}, B447:321--326, 1999.
\newblock \href {http://arxiv.org/abs/hep-ph/9812205}
  {\path{arXiv:hep-ph/9812205}}, \href
  {http://dx.doi.org/10.1016/S0370-2693(99)00010-6}
  {\path{doi:10.1016/S0370-2693(99)00010-6}}.

\bibitem{Eichten:1975ag}
E.~Eichten, K.~Gottfried, T.~Kinoshita, K.~D. Lane, and Tung-Mow Yan.
\newblock {The Interplay of Confinement and Decay in the Spectrum of
  Charmonium}.
\newblock {\em Phys. Rev. Lett.}, 36:500, 1976.
\newblock \href {http://dx.doi.org/10.1103/PhysRevLett.36.500}
  {\path{doi:10.1103/PhysRevLett.36.500}}.

\bibitem{Eichten:1978tg}
E.~Eichten, K.~Gottfried, T.~Kinoshita, K.~D. Lane, and Tung-Mow Yan.
\newblock {Charmonium: The Model}.
\newblock {\em Phys. Rev.}, D17:3090, 1978.
\newblock [Erratum: Phys. Rev.D21,313(1980)].
\newblock \href {http://dx.doi.org/10.1103/PhysRevD.17.3090,
  10.1103/physrevd.21.313.2} {\path{doi:10.1103/PhysRevD.17.3090,
  10.1103/physrevd.21.313.2}}.

\bibitem{Eichten:1979ms}
E.~Eichten, K.~Gottfried, T.~Kinoshita, K.~D. Lane, and Tung-Mow Yan.
\newblock {Charmonium: Comparison with Experiment}.
\newblock {\em Phys. Rev.}, D21:203, 1980.
\newblock \href {http://dx.doi.org/10.1103/PhysRevD.21.203}
  {\path{doi:10.1103/PhysRevD.21.203}}.

\bibitem{Eichten:2004uh}
Estia~J. Eichten, Kenneth Lane, and Chris Quigg.
\newblock {Charmonium levels near threshold and the narrow state $X(3872) \to
  \pi^{+}\pi^{-}\jpsi$}.
\newblock {\em Phys. Rev.}, D69:094019, 2004.
\newblock \href {http://arxiv.org/abs/hep-ph/0401210}
  {\path{arXiv:hep-ph/0401210}}, \href
  {http://dx.doi.org/10.1103/PhysRevD.69.094019}
  {\path{doi:10.1103/PhysRevD.69.094019}}.

\bibitem{Voloshin:2007dx}
M.~B. Voloshin.
\newblock {Charmonium}.
\newblock {\em Prog. Part. Nucl. Phys.}, 61:455--511, 2008.
\newblock \href {http://arxiv.org/abs/0711.4556} {\path{arXiv:0711.4556}},
  \href {http://dx.doi.org/10.1016/j.ppnp.2008.02.001}
  {\path{doi:10.1016/j.ppnp.2008.02.001}}.

\bibitem{Buchmuller:1980su}
W.~Buchmuller and S.~H.~H. Tye.
\newblock {Quarkonia and Quantum Chromodynamics}.
\newblock {\em Phys. Rev.}, D24:132, 1981.
\newblock \href {http://dx.doi.org/10.1103/PhysRevD.24.132}
  {\path{doi:10.1103/PhysRevD.24.132}}.

\bibitem{Barchielli:1986zs}
A.~Barchielli, E.~Montaldi, and G.~M. Prosperi.
\newblock {On a Systematic Derivation of the Quark - Anti-quark Potential}.
\newblock {\em Nucl. Phys.}, B296:625, 1988.
\newblock [Erratum: Nucl. Phys.B303,752(1988)].
\newblock \href {http://dx.doi.org/10.1016/0550-3213(88)90036-3}
  {\path{doi:10.1016/0550-3213(88)90036-3}}.

\bibitem{Brown:1979ya}
Lowell~S. Brown and William~I. Weisberger.
\newblock {Remarks on the Static Potential in Quantum Chromodynamics}.
\newblock {\em Phys. Rev.}, D20:3239, 1979.
\newblock \href {http://dx.doi.org/10.1103/PhysRevD.20.3239}
  {\path{doi:10.1103/PhysRevD.20.3239}}.

\bibitem{Szczepaniak:1996tk}
Adam~P. Szczepaniak and Eric~S. Swanson.
\newblock {On the Dirac structure of confinement}.
\newblock {\em Phys. Rev.}, D55:3987--3993, 1997.
\newblock \href {http://arxiv.org/abs/hep-ph/9611310}
  {\path{arXiv:hep-ph/9611310}}, \href
  {http://dx.doi.org/10.1103/PhysRevD.55.3987}
  {\path{doi:10.1103/PhysRevD.55.3987}}.

\bibitem{Wilson:1974sk}
Kenneth~G. Wilson.
\newblock {Confinement of Quarks}.
\newblock {\em Phys. Rev.}, D10:2445--2459, 1974.
\newblock [,319(1974)].
\newblock \href {http://dx.doi.org/10.1103/PhysRevD.10.2445}
  {\path{doi:10.1103/PhysRevD.10.2445}}.

\bibitem{Gromes:1984ma}
Dieter Gromes.
\newblock {Spin Dependent Potentials in QCD and the Correct Long Range Spin
  Orbit Term}.
\newblock {\em Z. Phys.}, C26:401, 1984.
\newblock \href {http://dx.doi.org/10.1007/BF01452566}
  {\path{doi:10.1007/BF01452566}}.

\bibitem{Bali:2000gf}
Gunnar~S. Bali.
\newblock {QCD forces and heavy quark bound states}.
\newblock {\em Phys. Rept.}, 343:1--136, 2001.
\newblock \href {http://arxiv.org/abs/hep-ph/0001312}
  {\path{arXiv:hep-ph/0001312}}, \href
  {http://dx.doi.org/10.1016/S0370-1573(00)00079-X}
  {\path{doi:10.1016/S0370-1573(00)00079-X}}.

\bibitem{Lucha:1991vn}
W.~Lucha, F.~F. Schoberl, and D.~Gromes.
\newblock {Bound states of quarks}.
\newblock {\em Phys. Rept.}, 200:127--240, 1991.
\newblock \href {http://dx.doi.org/10.1016/0370-1573(91)90001-3}
  {\path{doi:10.1016/0370-1573(91)90001-3}}.

\bibitem{Brambilla:2000gk}
Nora Brambilla, Antonio Pineda, Joan Soto, and Antonio Vairo.
\newblock {The QCD potential at O(1/m)}.
\newblock {\em Phys. Rev.}, D63:014023, 2001.
\newblock \href {http://arxiv.org/abs/hep-ph/0002250}
  {\path{arXiv:hep-ph/0002250}}, \href
  {http://dx.doi.org/10.1103/PhysRevD.63.014023}
  {\path{doi:10.1103/PhysRevD.63.014023}}.

\bibitem{Caswell:1985ui}
W.~E. Caswell and G.~P. Lepage.
\newblock {Effective Lagrangians for Bound State Problems in QED, QCD, and
  Other Field Theories}.
\newblock {\em Phys. Lett.}, 167B:437--442, 1986.
\newblock \href {http://dx.doi.org/10.1016/0370-2693(86)91297-9}
  {\path{doi:10.1016/0370-2693(86)91297-9}}.

\bibitem{Bodwin:1994jh}
Geoffrey~T. Bodwin, Eric Braaten, and G.~Peter Lepage.
\newblock {Rigorous QCD analysis of inclusive annihilation and production of
  heavy quarkonium}.
\newblock {\em Phys. Rev.}, D51:1125--1171, 1995.
\newblock [Erratum: Phys. Rev.D55,5853(1997)].
\newblock \href {http://arxiv.org/abs/hep-ph/9407339}
  {\path{arXiv:hep-ph/9407339}}, \href
  {http://dx.doi.org/10.1103/PhysRevD.55.5853, 10.1103/PhysRevD.51.1125}
  {\path{doi:10.1103/PhysRevD.55.5853, 10.1103/PhysRevD.51.1125}}.

\bibitem{Pineda:1997bj}
A.~Pineda and J.~Soto.
\newblock {Effective field theory for ultrasoft momenta in NRQCD and NRQED}.
\newblock {\em Nucl. Phys. Proc. Suppl.}, 64:428--432, 1998.
\newblock [,428(1997)].
\newblock \href {http://arxiv.org/abs/hep-ph/9707481}
  {\path{arXiv:hep-ph/9707481}}, \href
  {http://dx.doi.org/10.1016/S0920-5632(97)01102-X}
  {\path{doi:10.1016/S0920-5632(97)01102-X}}.

\bibitem{Brambilla:1999xf}
Nora Brambilla, Antonio Pineda, Joan Soto, and Antonio Vairo.
\newblock {Potential NRQCD: An Effective theory for heavy quarkonium}.
\newblock {\em Nucl. Phys.}, B566:275, 2000.
\newblock \href {http://arxiv.org/abs/hep-ph/9907240}
  {\path{arXiv:hep-ph/9907240}}, \href
  {http://dx.doi.org/10.1016/S0550-3213(99)00693-8}
  {\path{doi:10.1016/S0550-3213(99)00693-8}}.

\bibitem{Fleming:2008yn}
Sean Fleming and Thomas Mehen.
\newblock {Hadronic Decays of the X(3872) to $\chi_{cJ}$ in Effective Field
  Theory}.
\newblock {\em Phys. Rev.}, D78:094019, 2008.
\newblock \href {http://arxiv.org/abs/0807.2674} {\path{arXiv:0807.2674}},
  \href {http://dx.doi.org/10.1103/PhysRevD.78.094019}
  {\path{doi:10.1103/PhysRevD.78.094019}}.

\bibitem{Fleming:2007rp}
S.~Fleming, M.~Kusunoki, T.~Mehen, and U.~van Kolck.
\newblock {Pion interactions in the $X(3872)$}.
\newblock {\em Phys. Rev.}, D76:034006, 2007.
\newblock \href {http://arxiv.org/abs/hep-ph/0703168}
  {\path{arXiv:hep-ph/0703168}}, \href
  {http://dx.doi.org/10.1103/PhysRevD.76.034006}
  {\path{doi:10.1103/PhysRevD.76.034006}}.

\bibitem{Braaten:2007dw}
Eric Braaten and Meng Lu.
\newblock {Line shapes of the X(3872)}.
\newblock {\em Phys. Rev.}, D76:094028, 2007.
\newblock \href {http://arxiv.org/abs/0709.2697} {\path{arXiv:0709.2697}},
  \href {http://dx.doi.org/10.1103/PhysRevD.76.094028}
  {\path{doi:10.1103/PhysRevD.76.094028}}.

\bibitem{Collins:1989gx}
John~C. Collins, Davison~E. Soper, and George~F. Sterman.
\newblock {Factorization of Hard Processes in QCD}.
\newblock {\em Adv. Ser. Direct. High Energy Phys.}, 5:1--91, 1989.
\newblock \href {http://arxiv.org/abs/hep-ph/0409313}
  {\path{arXiv:hep-ph/0409313}}, \href
  {http://dx.doi.org/10.1142/9789814503266_0001}
  {\path{doi:10.1142/9789814503266_0001}}.

\bibitem{Brock:1993sz}
Raymond Brock et~al.
\newblock {Handbook of perturbative QCD: Version 1.0}.
\newblock {\em Rev. Mod. Phys.}, 67:157--248, 1995.
\newblock \href {http://dx.doi.org/10.1103/RevModPhys.67.157}
  {\path{doi:10.1103/RevModPhys.67.157}}.

\bibitem{Gribov:1984tu}
L.~V. Gribov, E.~M. Levin, and M.~G. Ryskin.
\newblock {Semihard Processes in QCD}.
\newblock {\em Phys. Rept.}, 100:1--150, 1983.
\newblock \href {http://dx.doi.org/10.1016/0370-1573(83)90022-4}
  {\path{doi:10.1016/0370-1573(83)90022-4}}.

\bibitem{Catani:1990eg}
S.~Catani, M.~Ciafaloni, and F.~Hautmann.
\newblock {High-energy factorization and small x heavy flavor production}.
\newblock {\em Nucl. Phys.}, B366:135--188, 1991.
\newblock \href {http://dx.doi.org/10.1016/0550-3213(91)90055-3}
  {\path{doi:10.1016/0550-3213(91)90055-3}}.

\bibitem{Lipatov:1995pn}
L.~N. Lipatov.
\newblock {Gauge invariant effective action for high-energy processes in QCD}.
\newblock {\em Nucl. Phys.}, B452:369--400, 1995.
\newblock \href {http://arxiv.org/abs/hep-ph/9502308}
  {\path{arXiv:hep-ph/9502308}}, \href
  {http://dx.doi.org/10.1016/0550-3213(95)00390-E}
  {\path{doi:10.1016/0550-3213(95)00390-E}}.

\bibitem{Kuraev:1976ge}
E.~A. Kuraev, L.~N. Lipatov, and Victor~S. Fadin.
\newblock {Multi - Reggeon Processes in the Yang-Mills Theory}.
\newblock {\em Sov. Phys. JETP}, 44:443--450, 1976.
\newblock [Zh. Eksp. Teor. Fiz.71,840(1976)].

\bibitem{Balitsky:1978ic}
I.~I. Balitsky and L.~N. Lipatov.
\newblock {The Pomeranchuk Singularity in Quantum Chromodynamics}.
\newblock {\em Sov. J. Nucl. Phys.}, 28:822--829, 1978.
\newblock [Yad. Fiz.28,1597(1978)].

\bibitem{Collins:1981uw}
John~C. Collins and Davison~E. Soper.
\newblock {Parton Distribution and Decay Functions}.
\newblock {\em Nucl. Phys.}, B194:445--492, 1982.
\newblock \href {http://dx.doi.org/10.1016/0550-3213(82)90021-9}
  {\path{doi:10.1016/0550-3213(82)90021-9}}.

\bibitem{Aybat:2011zv}
S.~Mert Aybat and Ted~C. Rogers.
\newblock {TMD Parton Distribution and Fragmentation Functions with QCD
  Evolution}.
\newblock {\em Phys. Rev.}, D83:114042, 2011.
\newblock \href {http://arxiv.org/abs/1101.5057} {\path{arXiv:1101.5057}},
  \href {http://dx.doi.org/10.1103/PhysRevD.83.114042}
  {\path{doi:10.1103/PhysRevD.83.114042}}.

\bibitem{Ji:2005nu}
Xiang-dong Ji, Jian-Ping Ma, and Feng Yuan.
\newblock {Transverse-momentum-dependent gluon distributions and semi-inclusive
  processes at hadron colliders}.
\newblock {\em JHEP}, 07:020, 2005.
\newblock \href {http://arxiv.org/abs/hep-ph/0503015}
  {\path{arXiv:hep-ph/0503015}}, \href
  {http://dx.doi.org/10.1088/1126-6708/2005/07/020}
  {\path{doi:10.1088/1126-6708/2005/07/020}}.

\bibitem{Fritzsch:1977ay}
Harald Fritzsch.
\newblock {Producing Heavy Quark Flavors in Hadronic Collisions: A Test of
  Quantum Chromodynamics}.
\newblock {\em Phys. Lett.}, 67B:217--221, 1977.
\newblock \href {http://dx.doi.org/10.1016/0370-2693(77)90108-3}
  {\path{doi:10.1016/0370-2693(77)90108-3}}.

\bibitem{Halzen:1977rs}
F.~Halzen.
\newblock {CVC for Gluons and Hadroproduction of Quark Flavors}.
\newblock {\em Phys. Lett.}, 69B:105--108, 1977.
\newblock \href {http://dx.doi.org/10.1016/0370-2693(77)90144-7}
  {\path{doi:10.1016/0370-2693(77)90144-7}}.

\bibitem{Gluck:1977zm}
M.~Gluck, J.~F. Owens, and E.~Reya.
\newblock {Gluon Contribution to Hadronic \jpsi Production}.
\newblock {\em Phys. Rev.}, D17:2324, 1978.
\newblock \href {http://dx.doi.org/10.1103/PhysRevD.17.2324}
  {\path{doi:10.1103/PhysRevD.17.2324}}.

\bibitem{Schuler:1996ku}
Gerhard~A. Schuler and Ramona Vogt.
\newblock {Systematics of quarkonium production}.
\newblock {\em Phys. Lett.}, B387:181--186, 1996.
\newblock \href {http://arxiv.org/abs/hep-ph/9606410}
  {\path{arXiv:hep-ph/9606410}}, \href
  {http://dx.doi.org/10.1016/0370-2693(96)00999-9}
  {\path{doi:10.1016/0370-2693(96)00999-9}}.

\bibitem{Mangano:1992kq}
Michelangelo~L. Mangano, Paolo Nason, and Giovanni Ridolfi.
\newblock {Fixed target hadroproduction of heavy quarks}.
\newblock {\em Nucl. Phys.}, B405:507--535, 1993.
\newblock \href {http://dx.doi.org/10.1016/0550-3213(93)90557-6}
  {\path{doi:10.1016/0550-3213(93)90557-6}}.

\bibitem{Amundson:1996qr}
J.~F. Amundson, Oscar J.~P. Eboli, E.~M. Gregores, and F.~Halzen.
\newblock {Quantitative tests of color evaporation: Charmonium production}.
\newblock {\em Phys. Lett.}, B390:323--328, 1997.
\newblock \href {http://arxiv.org/abs/hep-ph/9605295}
  {\path{arXiv:hep-ph/9605295}}, \href
  {http://dx.doi.org/10.1016/S0370-2693(96)01417-7}
  {\path{doi:10.1016/S0370-2693(96)01417-7}}.

\bibitem{Bodwin:2005hm}
Geoffrey~T. Bodwin, Eric Braaten, and Jungil Lee.
\newblock {Comparison of the color-evaporation model and the NRQCD
  factorization approach in charmonium production}.
\newblock {\em Phys. Rev.}, D72:014004, 2005.
\newblock \href {http://arxiv.org/abs/hep-ph/0504014}
  {\path{arXiv:hep-ph/0504014}}, \href
  {http://dx.doi.org/10.1103/PhysRevD.72.014004}
  {\path{doi:10.1103/PhysRevD.72.014004}}.

\bibitem{Ma:2016exq}
Yan-Qing Ma and Ramona Vogt.
\newblock {Quarkonium Production in an Improved Color Evaporation Model}.
\newblock {\em Phys. Rev.}, D94(11):114029, 2016.
\newblock \href {http://arxiv.org/abs/1609.06042} {\path{arXiv:1609.06042}},
  \href {http://dx.doi.org/10.1103/PhysRevD.94.114029}
  {\path{doi:10.1103/PhysRevD.94.114029}}.

\bibitem{Vogt:2019zmr}
R.~Vogt.
\newblock {Quarkonium Production and Polarization in an Improved Color
  Evaporation Model}.
\newblock {\em Nucl. Phys.}, A982:751--754, 2019.
\newblock \href {http://dx.doi.org/10.1016/j.nuclphysa.2018.08.003}
  {\path{doi:10.1016/j.nuclphysa.2018.08.003}}.

\bibitem{Einhorn:1975ua}
M.~B. Einhorn and S.~D. Ellis.
\newblock {Hadronic Production of the New Resonances: Probing Gluon
  Distributions}.
\newblock {\em Phys. Rev.}, D12:2007, 1975.
\newblock \href {http://dx.doi.org/10.1103/PhysRevD.12.2007}
  {\path{doi:10.1103/PhysRevD.12.2007}}.

\bibitem{Ellis:1976fj}
S.~D. Ellis, Martin~B. Einhorn, and C.~Quigg.
\newblock {Comment on Hadronic Production of Psions}.
\newblock {\em Phys. Rev. Lett.}, 36:1263, 1976.
\newblock \href {http://dx.doi.org/10.1103/PhysRevLett.36.1263}
  {\path{doi:10.1103/PhysRevLett.36.1263}}.

\bibitem{Carlson:1976cd}
C.~E. Carlson and R.~Suaya.
\newblock {Hadronic Production of psi/J Mesons}.
\newblock {\em Phys. Rev.}, D14:3115, 1976.
\newblock \href {http://dx.doi.org/10.1103/PhysRevD.14.3115}
  {\path{doi:10.1103/PhysRevD.14.3115}}.

\bibitem{Chang:1979nn}
Chao-Hsi Chang.
\newblock {Hadronic Production of $\jpsi$ Associated With a Gluon}.
\newblock {\em Nucl. Phys.}, B172:425--434, 1980.
\newblock \href {http://dx.doi.org/10.1016/0550-3213(80)90175-3}
  {\path{doi:10.1016/0550-3213(80)90175-3}}.

\bibitem{Chao:2012upa}
Kuang-Ta Chao and Yan-Qing Ma.
\newblock {Quarkonium production review}.
\newblock {\em PoS}, ConfinementX:003, 2012.
\newblock \href {http://dx.doi.org/10.22323/1.171.0003}
  {\path{doi:10.22323/1.171.0003}}.

\bibitem{Fleming:2000ib}
Sean Fleming, I.~Z. Rothstein, and Adam~K. Leibovich.
\newblock {Power counting and effective field theory for charmonium}.
\newblock {\em Phys. Rev.}, D64:036002, 2001.
\newblock \href {http://arxiv.org/abs/hep-ph/0012062}
  {\path{arXiv:hep-ph/0012062}}, \href
  {http://dx.doi.org/10.1103/PhysRevD.64.036002}
  {\path{doi:10.1103/PhysRevD.64.036002}}.

\bibitem{Andronic:2015wma}
A.~Andronic et~al.
\newblock {Heavy-flavour and quarkonium production in the LHC era: from
  proton–proton to heavy-ion collisions}.
\newblock {\em Eur. Phys. J.}, C76(3):107, 2016.
\newblock \href {http://arxiv.org/abs/1506.03981} {\path{arXiv:1506.03981}},
  \href {http://dx.doi.org/10.1140/epjc/s10052-015-3819-5}
  {\path{doi:10.1140/epjc/s10052-015-3819-5}}.

\bibitem{Brambilla:2014jmp}
N.~Brambilla et~al.
\newblock {QCD and Strongly Coupled Gauge Theories: Challenges and
  Perspectives}.
\newblock {\em Eur. Phys. J.}, C74(10):2981, 2014.
\newblock \href {http://arxiv.org/abs/1404.3723} {\path{arXiv:1404.3723}},
  \href {http://dx.doi.org/10.1140/epjc/s10052-014-2981-5}
  {\path{doi:10.1140/epjc/s10052-014-2981-5}}.

\bibitem{Lansberg:2019adr}
Jean-Philippe Lansberg.
\newblock {New Observables in Inclusive Production of Quarkonia}.
\newblock 2019.
\newblock \href {http://arxiv.org/abs/1903.09185} {\path{arXiv:1903.09185}}.

\bibitem{Collins:1977iv}
John~C. Collins and Davison~E. Soper.
\newblock {Angular Distribution of Dileptons in High-Energy Hadron Collisions}.
\newblock {\em Phys. Rev.}, D16:2219, 1977.
\newblock \href {http://dx.doi.org/10.1103/PhysRevD.16.2219}
  {\path{doi:10.1103/PhysRevD.16.2219}}.

\bibitem{Beneke:1998re}
M.~Beneke, M.~Kramer, and M.~Vanttinen.
\newblock {Inelastic photoproduction of polarized $J / \psi$}.
\newblock {\em Phys. Rev.}, D57:4258--4274, 1998.
\newblock \href {http://arxiv.org/abs/hep-ph/9709376}
  {\path{arXiv:hep-ph/9709376}}, \href
  {http://dx.doi.org/10.1103/PhysRevD.57.4258}
  {\path{doi:10.1103/PhysRevD.57.4258}}.

\bibitem{Abreu:1994rk}
P.~Abreu et~al.
\newblock {$\jpsi$ production in the hadronic decays of the Z}.
\newblock {\em Phys.Lett.}, B341:109--122, 1994.
\newblock \href {http://dx.doi.org/10.1016/0370-2693(94)01385-3}
  {\path{doi:10.1016/0370-2693(94)01385-3}}.

\bibitem{Adriani:1993ta}
O.~Adriani et~al.
\newblock {$\chi_c$ production in hadronic Z decays}.
\newblock {\em Phys.Lett.}, B317:467--473, 1993.
\newblock \href {http://dx.doi.org/10.1016/0370-2693(93)91026-J}
  {\path{doi:10.1016/0370-2693(93)91026-J}}.

\bibitem{Buskulic:1992wp}
D.~Buskulic et~al.
\newblock {Measurements of mean lifetime and branching fractions of b-hadrons
  decaying to $\jpsi$}.
\newblock {\em Phys.Lett.}, B295:396--408, 1992.
\newblock \href {http://dx.doi.org/10.1016/0370-2693(92)91581-S}
  {\path{doi:10.1016/0370-2693(92)91581-S}}.

\bibitem{Alam:1986ic}
M.~S. Alam et~al.
\newblock {A Study of the Decay $B \to \psi$ X}.
\newblock {\em Phys. Rev.}, D34:3279, 1986.
\newblock \href {http://dx.doi.org/10.1103/PhysRevD.34.3279}
  {\path{doi:10.1103/PhysRevD.34.3279}}.

\bibitem{Aubert:2002hc}
Bernard Aubert et~al.
\newblock {Study of inclusive production of charmonium mesons in $B$ decay}.
\newblock {\em Phys.Rev.}, D67:032002, 2003.
\newblock \href {http://arxiv.org/abs/hep-ex/0207097}
  {\path{arXiv:hep-ex/0207097}}, \href
  {http://dx.doi.org/10.1103/PhysRevD.67.032002}
  {\path{doi:10.1103/PhysRevD.67.032002}}.

\bibitem{Anderson:2002md}
S.~Anderson et~al.
\newblock {Measurements of inclusive $B \to \psi$ production}.
\newblock {\em Phys.Rev.Lett.}, 89:282001, 2002.
\newblock \href {http://arxiv.org/abs/hep-ex/0207059}
  {\path{arXiv:hep-ex/0207059}}, \href
  {http://dx.doi.org/10.1103/PhysRevLett.89.282001}
  {\path{doi:10.1103/PhysRevLett.89.282001}}.

\bibitem{PDG2018}
M.~Tanabashi et~al.
\newblock {Review of Particle Physics}.
\newblock {\em Phys. Rev.}, D98(3):030001, 2018.
\newblock \href {http://dx.doi.org/10.1103/PhysRevD.98.030001}
  {\path{doi:10.1103/PhysRevD.98.030001}}.

\bibitem{Chen:2000ri}
S.~Chen et~al.
\newblock {Study of \chicone and \chictwo meson production in \B meson decays}.
\newblock {\em Phys.Rev.}, D63:031102, 2001.
\newblock \href {http://arxiv.org/abs/hep-ex/0009044}
  {\path{arXiv:hep-ex/0009044}}, \href
  {http://dx.doi.org/10.1103/PhysRevD.63.031102}
  {\path{doi:10.1103/PhysRevD.63.031102}}.

\bibitem{Abe:2002wp}
K.~Abe et~al.
\newblock {Observation of $\chictwo$ production in $\B$ meson decay}.
\newblock {\em Phys.Rev.Lett.}, 89:011803, 2002.
\newblock \href {http://arxiv.org/abs/hep-ex/0202028}
  {\path{arXiv:hep-ex/0202028}}, \href
  {http://dx.doi.org/10.1103/PhysRevLett.89.011803}
  {\path{doi:10.1103/PhysRevLett.89.011803}}.

\bibitem{Bhardwaj:2015rju}
V.~Bhardwaj et~al.
\newblock {Inclusive and exclusive measurements of $B$ decays to $\chi_{c1}$
  and $\chi_{c2}$ at Belle}.
\newblock {\em Phys. Rev.}, D93(5):052016, 2016.
\newblock \href {http://arxiv.org/abs/1512.02672} {\path{arXiv:1512.02672}},
  \href {http://dx.doi.org/10.1103/PhysRevD.93.052016}
  {\path{doi:10.1103/PhysRevD.93.052016}}.

\bibitem{PDG2016}
C.~Patrignani et~al.
\newblock {\href{http://pdg.lbl.gov/}{Review of particle physics}}.
\newblock {\em Chin. Phys.}, C40:100001, 2016.
\newblock \href {http://dx.doi.org/10.1088/1674-1137/40/10/100001}
  {\path{doi:10.1088/1674-1137/40/10/100001}}.

\bibitem{Balest:1994jf}
R.~Balest et~al.
\newblock {Inclusive decays of B mesons to charmonium}.
\newblock {\em Phys.Rev.}, D52:2661--2672, 1995.
\newblock \href {http://dx.doi.org/10.1103/PhysRevD.52.2661}
  {\path{doi:10.1103/PhysRevD.52.2661}}.

\bibitem{LHCb-PAPER-2011-045}
R~Aaij et~al.
\newblock {Measurement of $\psitwos$ meson production in $\proton\proton$
  collisions at $\sqrt{s}=7$\tev}.
\newblock {\em Eur. Phys. J.}, C72:2100, 2012.
\newblock \href {http://arxiv.org/abs/1204.1258} {\path{arXiv:1204.1258}},
  \href {http://dx.doi.org/10.1140/epjc/s10052-012-2100-4}
  {\path{doi:10.1140/epjc/s10052-012-2100-4}}.

\bibitem{Chatrchyan:2011kc}
Serguei Chatrchyan et~al.
\newblock {$\jpsi$ and $\psi(2S)$ production in $pp$ collisions at $\sqrt{s}=7$
  TeV}.
\newblock {\em JHEP}, 02:011, 2012.
\newblock \href {http://arxiv.org/abs/1111.1557} {\path{arXiv:1111.1557}},
  \href {http://dx.doi.org/10.1007/JHEP02(2012)011}
  {\path{doi:10.1007/JHEP02(2012)011}}.

\bibitem{Ko:1995iv}
Pyungwon Ko, Jungil Lee, and H.~S. Song.
\newblock {Inclusive S wave charmonium productions in B decays}.
\newblock {\em Phys. Rev.}, D53:1409--1415, 1996.
\newblock \href {http://arxiv.org/abs/hep-ph/9510202}
  {\path{arXiv:hep-ph/9510202}}, \href
  {http://dx.doi.org/10.1103/PhysRevD.53.1409}
  {\path{doi:10.1103/PhysRevD.53.1409}}.

\bibitem{Beneke:1998ks}
M.~Beneke, F.~Maltoni, and I.~Z. Rothstein.
\newblock {QCD analysis of inclusive B decay into charmonium}.
\newblock {\em Phys. Rev.}, D59:054003, 1999.
\newblock \href {http://arxiv.org/abs/hep-ph/9808360}
  {\path{arXiv:hep-ph/9808360}}, \href
  {http://dx.doi.org/10.1103/PhysRevD.59.054003}
  {\path{doi:10.1103/PhysRevD.59.054003}}.

\bibitem{Ma:2000bz}
J.~P. Ma.
\newblock {Effects of the initial hadron in $B \to \jpsi + X$}.
\newblock {\em Phys. Lett.}, B488:55--62, 2000.
\newblock \href {http://arxiv.org/abs/hep-ph/0006060}
  {\path{arXiv:hep-ph/0006060}}, \href
  {http://dx.doi.org/10.1016/S0370-2693(00)00845-5}
  {\path{doi:10.1016/S0370-2693(00)00845-5}}.

\bibitem{Beneke:1999gq}
M.~Beneke, G.~A. Schuler, and S.~Wolf.
\newblock {Quarkonium momentum distributions in photoproduction and B decay}.
\newblock {\em Phys. Rev.}, D62:034004, 2000.
\newblock \href {http://arxiv.org/abs/hep-ph/0001062}
  {\path{arXiv:hep-ph/0001062}}, \href
  {http://dx.doi.org/10.1103/PhysRevD.62.034004}
  {\path{doi:10.1103/PhysRevD.62.034004}}.

\bibitem{Fleming:1996pt}
Sean Fleming, Oscar~F. Hernandez, Ivan Maksymyk, and Helene Nadeau.
\newblock {NRQCD matrix elements in polarization of $J / \psi$ produced from
  b-decay}.
\newblock {\em Phys. Rev.}, D55:4098--4104, 1997.
\newblock \href {http://arxiv.org/abs/hep-ph/9608413}
  {\path{arXiv:hep-ph/9608413}}, \href
  {http://dx.doi.org/10.1103/PhysRevD.55.4098}
  {\path{doi:10.1103/PhysRevD.55.4098}}.

\bibitem{Ko:1999zx}
P.~Ko, J.~Lee, and H.~S. Song.
\newblock {Testing J/psi production mechanisms in $B \to \jpsi + X$}.
\newblock {\em J. Korean Phys. Soc.}, 34:301--305, 1999.

\bibitem{Bodwin:1992qr}
Geoffrey~T. Bodwin, Eric Braaten, Tzu~Chiang Yuan, and G.~Peter Lepage.
\newblock {P-wave charmonium production in B meson decays}.
\newblock {\em Phys. Rev.}, D46:R3703--R3707, 1992.
\newblock \href {http://arxiv.org/abs/hep-ph/9208254}
  {\path{arXiv:hep-ph/9208254}}, \href
  {http://dx.doi.org/10.1103/PhysRevD.46.R3703}
  {\path{doi:10.1103/PhysRevD.46.R3703}}.

\bibitem{Kuhn:1979zb}
Johann~H. Kuhn, S.~Nussinov, and R.~Ruckl.
\newblock {Charmonium Production in B Decays}.
\newblock {\em Z. Phys.}, C5:117, 1980.
\newblock \href {http://dx.doi.org/10.1007/BF01576192}
  {\path{doi:10.1007/BF01576192}}.

\bibitem{Kuhn:1983ar}
Johann~H. Kuhn and R.~Ruckl.
\newblock {Clues on Color Suppression From $B \to \jpsi$ + X}.
\newblock {\em Phys. Lett.}, 135B:477--480, 1984.
\newblock [Erratum: Phys. Lett.B258,499(1991)].
\newblock \href {http://dx.doi.org/10.1016/0370-2693(91)91125-F,
  10.1016/0370-2693(84)90319-8} {\path{doi:10.1016/0370-2693(91)91125-F,
  10.1016/0370-2693(84)90319-8}}.

\bibitem{Abe:1997yz}
F.~Abe et~al.
\newblock {Production of $\jpsi$ mesons from $\chi_c$ meson decays in
  $p\bar{p}$ collisions at $\sqrt{s} = 1.8$ TeV}.
\newblock {\em Phys. Rev. Lett.}, 79:578--583, 1997.
\newblock \href {http://dx.doi.org/10.1103/PhysRevLett.79.578}
  {\path{doi:10.1103/PhysRevLett.79.578}}.

\bibitem{LHCb-PAPER-2012-039}
R.~Aaij et~al.
\newblock {Measurement of $\jpsi$ production in $\proton\proton$ collisions at
  $\sqrt{s}=2.76$\tev}.
\newblock {\em JHEP}, 02:041, 2013.
\newblock \href {http://arxiv.org/abs/1212.1045} {\path{arXiv:1212.1045}},
  \href {http://dx.doi.org/10.1007/JHEP02(2013)041}
  {\path{doi:10.1007/JHEP02(2013)041}}.

\bibitem{LHCb-PAPER-2011-003}
R.~Aaij et~al.
\newblock {Measurement of $\jpsi$ production in $\proton\proton$ collisions at
  $\sqrt{s}=7$\,TeV}.
\newblock {\em Eur. Phys. J.}, C71:1645, 2011.
\newblock \href {http://arxiv.org/abs/1103.0423} {\path{arXiv:1103.0423}},
  \href {http://dx.doi.org/10.1140/epjc/s10052-011-1645-y}
  {\path{doi:10.1140/epjc/s10052-011-1645-y}}.

\bibitem{LHCb-PAPER-2013-016}
R.~Aaij et~al.
\newblock {Production of $\jpsi$ and $\Upsilon$ mesons in $\proton\proton$
  collisions at $\sqrt{s}=8$\tev}.
\newblock {\em JHEP}, 06:064, 2013.
\newblock \href {http://arxiv.org/abs/1304.6977} {\path{arXiv:1304.6977}},
  \href {http://dx.doi.org/10.1007/JHEP06(2013)064}
  {\path{doi:10.1007/JHEP06(2013)064}}.

\bibitem{LHCb-PAPER-2015-037}
R.~Aaij et~al.
\newblock {Measurement of forward $\jpsi$ production cross-sections in
  $\proton\proton$ collisions at $\sqrt{s}=13$\tev}.
\newblock {\em JHEP}, 10:172, 2015.
\newblock \href {http://arxiv.org/abs/1509.00771} {\path{arXiv:1509.00771}},
  \href {http://dx.doi.org/10.1007/JHEP10(2015)172}
  {\path{doi:10.1007/JHEP10(2015)172}}.

\bibitem{Aad:2011sp}
Georges Aad et~al.
\newblock {Measurement of the differential cross-sections of inclusive, prompt
  and non-prompt $\jpsi$ production in proton-proton collisions at $\sqrt{s}=7$
  TeV}.
\newblock {\em Nucl. Phys.}, B850:387--444, 2011.
\newblock \href {http://arxiv.org/abs/1104.3038} {\path{arXiv:1104.3038}},
  \href {http://dx.doi.org/10.1016/j.nuclphysb.2011.05.015}
  {\path{doi:10.1016/j.nuclphysb.2011.05.015}}.

\bibitem{Aad:2015duc}
Georges Aad et~al.
\newblock {Measurement of the differential cross-sections of prompt and
  non-prompt production of $\jpsi$ and $\psi (2\mathrm {S})$ in $pp$ collisions
  at $\sqrt{s} = 7$ and 8 TeV with the ATLAS detector}.
\newblock {\em Eur. Phys. J.}, C76(5):283, 2016.
\newblock \href {http://arxiv.org/abs/1512.03657} {\path{arXiv:1512.03657}},
  \href {http://dx.doi.org/10.1140/epjc/s10052-016-4050-8}
  {\path{doi:10.1140/epjc/s10052-016-4050-8}}.

\bibitem{Sirunyan:2017qdw}
A.~M. Sirunyan et~al.
\newblock {Measurement of quarkonium production cross sections in $pp$
  collisions at $\sqrt{s}=$ 13 TeV}.
\newblock {\em Phys. Lett.}, B780:251--272, 2018.
\newblock \href {http://arxiv.org/abs/1710.11002} {\path{arXiv:1710.11002}},
  \href {http://dx.doi.org/10.1016/j.physletb.2018.02.033}
  {\path{doi:10.1016/j.physletb.2018.02.033}}.

\bibitem{Abelev:2012kr}
B.~Abelev et~al.
\newblock {Inclusive $\jpsi$ production in $pp$ collisions at $\sqrt{s} = 2.76$
  TeV}.
\newblock {\em Phys. Lett.}, B718:295--306, 2012.
\newblock [Erratum: Phys. Lett.B748,472(2015)].
\newblock \href {http://arxiv.org/abs/1203.3641} {\path{arXiv:1203.3641}},
  \href {http://dx.doi.org/10.1016/j.physletb.2012.10.078,
  10.1016/j.physletb.2015.06.058} {\path{doi:10.1016/j.physletb.2012.10.078,
  10.1016/j.physletb.2015.06.058}}.

\bibitem{Aamodt:2011gj}
K.~Aamodt et~al.
\newblock {Rapidity and transverse momentum dependence of inclusive J$/\psi$
  production in $pp$ collisions at $\sqrt{s} = 7$ TeV}.
\newblock {\em Phys. Lett.}, B704:442--455, 2011.
\newblock [Erratum: Phys. Lett.B718,692(2012)].
\newblock \href {http://arxiv.org/abs/1105.0380} {\path{arXiv:1105.0380}},
  \href {http://dx.doi.org/10.1016/j.physletb.2011.09.054,
  10.1016/j.physletb.2012.10.060} {\path{doi:10.1016/j.physletb.2011.09.054,
  10.1016/j.physletb.2012.10.060}}.

\bibitem{LHCb-PAPER-2013-008}
R.~Aaij et~al.
\newblock {Measurement of $\jpsi$ polarization in $\proton\proton$ collisions
  at $\sqrt{s}=7$\tev}.
\newblock {\em Eur. Phys. J.}, C73:2631, 2013.
\newblock \href {http://arxiv.org/abs/1307.6379} {\path{arXiv:1307.6379}},
  \href {http://dx.doi.org/10.1140/epjc/s10052-013-2631-3}
  {\path{doi:10.1140/epjc/s10052-013-2631-3}}.

\bibitem{LHCb-PAPER-2013-067}
R.~Aaij et~al.
\newblock {Measurement of $\psitwos$ polarisation in $\proton\proton$
  collisions at $\sqrt{s}=7$\tev}.
\newblock {\em Eur. Phys. J.}, C74:2872, 2014.
\newblock \href {http://arxiv.org/abs/1403.1339} {\path{arXiv:1403.1339}},
  \href {http://dx.doi.org/10.1140/epjc/s10052-014-2872-9}
  {\path{doi:10.1140/epjc/s10052-014-2872-9}}.

\bibitem{Chatrchyan:2013cla}
Serguei Chatrchyan et~al.
\newblock {Measurement of the prompt $\jpsi$ and $\psi$(2S) polarizations in
  $pp$ collisions at $\sqrt{s}$ = 7 Te }.
\newblock {\em Phys. Lett.}, B727:381--402, 2013.
\newblock \href {http://arxiv.org/abs/1307.6070} {\path{arXiv:1307.6070}},
  \href {http://dx.doi.org/10.1016/j.physletb.2013.10.055}
  {\path{doi:10.1016/j.physletb.2013.10.055}}.

\bibitem{Abelev:2011md}
Betty Abelev et~al.
\newblock {$\jpsi$ polarization in $pp$ collisions at $\sqrt{s}=7$ TeV}.
\newblock {\em Phys. Rev. Lett.}, 108:082001, 2012.
\newblock \href {http://arxiv.org/abs/1111.1630} {\path{arXiv:1111.1630}},
  \href {http://dx.doi.org/10.1103/PhysRevLett.108.082001}
  {\path{doi:10.1103/PhysRevLett.108.082001}}.

\bibitem{Abulencia:2007us}
A.~Abulencia et~al.
\newblock {Polarization of $\jpsi$ and $\psi_{2S}$ mesons produced in $p
  \bar{p}$ collisions at $\sqrt{s}$ = 1.96 TeV}.
\newblock {\em Phys. Rev. Lett.}, 99:132001, 2007.
\newblock \href {http://arxiv.org/abs/0704.0638} {\path{arXiv:0704.0638}},
  \href {http://dx.doi.org/10.1103/PhysRevLett.99.132001}
  {\path{doi:10.1103/PhysRevLett.99.132001}}.

\bibitem{Aaij:2011jh}
R.~Aaij et~al.
\newblock {Measurement of $\jpsi$ production in $pp$ collisions at
  $\sqrt{s}=7~\rm{TeV}$}.
\newblock {\em Eur. Phys. J.}, C71:1645, 2011.
\newblock \href {http://arxiv.org/abs/1103.0423} {\path{arXiv:1103.0423}},
  \href {http://dx.doi.org/10.1140/epjc/s10052-011-1645-y}
  {\path{doi:10.1140/epjc/s10052-011-1645-y}}.

\bibitem{Artoisenet:2010zz}
Pierre Artoisenet.
\newblock {Quarkonium production at the Tevatron and the LHC}.
\newblock {\em PoS}, ICHEP2010:192, 2010.
\newblock \href {http://dx.doi.org/10.22323/1.120.0192}
  {\path{doi:10.22323/1.120.0192}}.

\bibitem{Butenschoen:2010rq}
Mathias Butenschoen and Bernd~A. Kniehl.
\newblock {Reconciling $\jpsi$ production at HERA, RHIC, Tevatron, and LHC with
  NRQCD factorization at next-to-leading order}.
\newblock {\em Phys. Rev. Lett.}, 106:022003, 2011.
\newblock \href {http://arxiv.org/abs/1009.5662} {\path{arXiv:1009.5662}},
  \href {http://dx.doi.org/10.1103/PhysRevLett.106.022003}
  {\path{doi:10.1103/PhysRevLett.106.022003}}.

\bibitem{Lansberg:2008gk}
J.~P. Lansberg.
\newblock {On the mechanisms of heavy-quarkonium hadroproduction}.
\newblock {\em Eur. Phys. J.}, C61:693--703, 2009.
\newblock \href {http://arxiv.org/abs/0811.4005} {\path{arXiv:0811.4005}},
  \href {http://dx.doi.org/10.1140/epjc/s10052-008-0826-9}
  {\path{doi:10.1140/epjc/s10052-008-0826-9}}.

\bibitem{Ma:2010yw}
Yan-Qing Ma, Kai Wang, and Kuang-Ta Chao.
\newblock {$\jpsi (\psi^\prime)$ production at the Tevatron and LHC at ${\cal
  O}(\alpha_s^4v^4)$ in nonrelativistic QCD}.
\newblock {\em Phys. Rev. Lett.}, 106:042002, 2011.
\newblock \href {http://arxiv.org/abs/1009.3655} {\path{arXiv:1009.3655}},
  \href {http://dx.doi.org/10.1103/PhysRevLett.106.042002}
  {\path{doi:10.1103/PhysRevLett.106.042002}}.

\bibitem{Frawley:2008kk}
Anthony~D. Frawley, T.~Ullrich, and R.~Vogt.
\newblock {Heavy flavor in heavy-ion collisions at RHIC and RHIC II}.
\newblock {\em Phys. Rept.}, 462:125--175, 2008.
\newblock \href {http://arxiv.org/abs/0806.1013} {\path{arXiv:0806.1013}},
  \href {http://dx.doi.org/10.1016/j.physrep.2008.04.002}
  {\path{doi:10.1016/j.physrep.2008.04.002}}.

\bibitem{Baranov:2015laa}
S.~P. Baranov, A.~V. Lipatov, and N.~P. Zotov.
\newblock {Prompt charmonia production and polarization at LHC in the NRQCD
  with $k_T$ -factorization. Part I: $\psi (2S)$ meson}.
\newblock {\em Eur. Phys. J.}, C75(9):455, 2015.
\newblock \href {http://arxiv.org/abs/1508.05480} {\path{arXiv:1508.05480}},
  \href {http://dx.doi.org/10.1140/epjc/s10052-015-3689-x}
  {\path{doi:10.1140/epjc/s10052-015-3689-x}}.

\bibitem{Baranov:2016clx}
S.~P. Baranov and A.~V. Lipatov.
\newblock {Prompt charmonia production and polarization at LHC in the NRQCD
  with $k_T$-factorization. Part III: $\jpsi$ meson}.
\newblock {\em Phys. Rev.}, D96(3):034019, 2017.
\newblock \href {http://arxiv.org/abs/1611.10141} {\path{arXiv:1611.10141}},
  \href {http://dx.doi.org/10.1103/PhysRevD.96.034019}
  {\path{doi:10.1103/PhysRevD.96.034019}}.

\bibitem{Affolder:2000nn}
T.~Affolder et~al.
\newblock {Measurement of $\jpsi$ and $\psi(2S)$ polarization in $p\bar{p}$
  collisions at $\sqrt{s} = 1.8$ TeV}.
\newblock {\em Phys. Rev. Lett.}, 85:2886--2891, 2000.
\newblock \href {http://arxiv.org/abs/hep-ex/0004027}
  {\path{arXiv:hep-ex/0004027}}, \href
  {http://dx.doi.org/10.1103/PhysRevLett.85.2886}
  {\path{doi:10.1103/PhysRevLett.85.2886}}.

\bibitem{Chao:2012iv}
Kuang-Ta Chao, Yan-Qing Ma, Hua-Sheng Shao, Kai Wang, and Yu-Jie Zhang.
\newblock {$\jpsi$ Polarization at Hadron Colliders in Nonrelativistic QCD}.
\newblock {\em Phys.Rev.Lett.}, 108:242004, 2012.
\newblock \href {http://arxiv.org/abs/1201.2675} {\path{arXiv:1201.2675}},
  \href {http://dx.doi.org/10.1103/PhysRevLett.108.242004}
  {\path{doi:10.1103/PhysRevLett.108.242004}}.

\bibitem{Aaij:2013nlm}
R~Aaij et~al.
\newblock {Measurement of $\jpsi$ polarization in $pp$ collisions at
  $\sqrt{s}=7$ TeV}.
\newblock {\em Eur. Phys. J.}, C73(11):2631, 2013.
\newblock \href {http://arxiv.org/abs/1307.6379} {\path{arXiv:1307.6379}},
  \href {http://dx.doi.org/10.1140/epjc/s10052-013-2631-3}
  {\path{doi:10.1140/epjc/s10052-013-2631-3}}.

\bibitem{Butenschoen:2011yh}
Mathias Butenschoen and Bernd~A. Kniehl.
\newblock {World data of $\jpsi$ production consolidate NRQCD factorization at
  NLO}.
\newblock {\em Phys. Rev.}, D84:051501, 2011.
\newblock \href {http://arxiv.org/abs/1105.0820} {\path{arXiv:1105.0820}},
  \href {http://dx.doi.org/10.1103/PhysRevD.84.051501}
  {\path{doi:10.1103/PhysRevD.84.051501}}.

\bibitem{Gong:2012ug}
Bin Gong, Lu-Ping Wan, Jian-Xiong Wang, and Hong-Fei Zhang.
\newblock {Polarization for Prompt \jpsi and \psitwos Production at the
  Tevatron and LHC}.
\newblock {\em Phys. Rev. Lett.}, 110(4):042002, 2013.
\newblock \href {http://arxiv.org/abs/1205.6682} {\path{arXiv:1205.6682}},
  \href {http://dx.doi.org/10.1103/PhysRevLett.110.042002}
  {\path{doi:10.1103/PhysRevLett.110.042002}}.

\bibitem{Aaij:2014qea}
Roel Aaij et~al.
\newblock {Measurement of $\psi(2S)$ polarisation in $pp$ collisions at
  $\sqrt{s}=7$ TeV}.
\newblock {\em Eur. Phys. J.}, C74(5):2872, 2014.
\newblock \href {http://arxiv.org/abs/1403.1339} {\path{arXiv:1403.1339}},
  \href {http://dx.doi.org/10.1140/epjc/s10052-014-2872-9}
  {\path{doi:10.1140/epjc/s10052-014-2872-9}}.

\bibitem{Shao:2014yta}
H.~S. Shao, H.~Han, Y.~Q. Ma, C.~Meng, Y.~J. Zhang, and K.~T. Chao.
\newblock {Yields and polarizations of prompt $\jpsi$ and $\psi(2S)$ production
  in hadronic collisions}.
\newblock {\em JHEP}, 05:103, 2015.
\newblock \href {http://arxiv.org/abs/1411.3300} {\path{arXiv:1411.3300}},
  \href {http://dx.doi.org/10.1007/JHEP05(2015)103}
  {\path{doi:10.1007/JHEP05(2015)103}}.

\bibitem{Bodwin:2014gia}
Geoffrey~T. Bodwin, Hee~Sok Chung, U-Rae Kim, and Jungil Lee.
\newblock {Fragmentation contributions to $\jpsi$ production at the Tevatron
  and the LHC}.
\newblock {\em Phys. Rev. Lett.}, 113(2):022001, 2014.
\newblock \href {http://arxiv.org/abs/1403.3612} {\path{arXiv:1403.3612}},
  \href {http://dx.doi.org/10.1103/PhysRevLett.113.022001}
  {\path{doi:10.1103/PhysRevLett.113.022001}}.

\bibitem{Butenschoen:2014dra}
Mathias Butenschoen, Zhi-Guo He, and Bernd~A. Kniehl.
\newblock {$\eta_c$ production at the LHC challenges nonrelativistic-QCD
  factorization}.
\newblock {\em Phys. Rev. Lett.}, 114(9):092004, 2015.
\newblock \href {http://arxiv.org/abs/1411.5287} {\path{arXiv:1411.5287}},
  \href {http://dx.doi.org/10.1103/PhysRevLett.114.092004}
  {\path{doi:10.1103/PhysRevLett.114.092004}}.

\bibitem{Feng:2015cba}
Yu~Feng, Jean-Philippe Lansberg, and Jian-Xiong Wang.
\newblock {Energy dependence of direct-quarkonium production in $pp$ collisions
  from fixed-target to LHC energies: complete one-loop analysis}.
\newblock {\em Eur. Phys. J.}, C75(7):313, 2015.
\newblock \href {http://arxiv.org/abs/1504.00317} {\path{arXiv:1504.00317}},
  \href {http://dx.doi.org/10.1140/epjc/s10052-015-3527-1}
  {\path{doi:10.1140/epjc/s10052-015-3527-1}}.

\bibitem{Sun:2015pia}
Zhan Sun and Hong-Fei Zhang.
\newblock {Reconciling charmonium production and polarization data in the
  midrapidity region at hadron colliders within the nonrelativistic QCD
  framework}.
\newblock {\em Chin. Phys.}, C42(4):043104, 2018.
\newblock \href {http://arxiv.org/abs/1505.02675} {\path{arXiv:1505.02675}},
  \href {http://dx.doi.org/10.1088/1674-1137/42/4/043104}
  {\path{doi:10.1088/1674-1137/42/4/043104}}.

\bibitem{Likhoded:2015qyl}
A.~K. Likhoded, A.~V. Luchinsky, and S.~V. Poslavsky.
\newblock {Production of heavy quarkonia in hadronic experiments}.
\newblock {\em Phys. Atom. Nucl.}, 78(9):1056--1065, 2015.
\newblock [Yad. Fiz.78,no.12,1119(2015)].
\newblock \href {http://dx.doi.org/10.1134/S1063778815090100}
  {\path{doi:10.1134/S1063778815090100}}.

\bibitem{Gao:2016ihc}
Xiangrui Gao, Yu~Jia, LiuJi Li, and Xiaonu Xiong.
\newblock {Relativistic correction to gluon fragmentation function into
  pseudoscalar quarkonium}.
\newblock {\em Chin. Phys.}, C41(2):023103, 2017.
\newblock \href {http://arxiv.org/abs/1606.07455} {\path{arXiv:1606.07455}},
  \href {http://dx.doi.org/10.1088/1674-1137/41/2/023103}
  {\path{doi:10.1088/1674-1137/41/2/023103}}.

\bibitem{Faccioli:2017hym}
Pietro Faccioli, Carlos Lourenço, Mariana Araújo, Valentin Knünz, Ilse
  Krätschmer, and João Seixas.
\newblock {Quarkonium production at the LHC: A data-driven analysis of
  remarkably simple experimental patterns}.
\newblock {\em Phys. Lett.}, B773:476--486, 2017.
\newblock \href {http://arxiv.org/abs/1702.04208} {\path{arXiv:1702.04208}},
  \href {http://dx.doi.org/10.1016/j.physletb.2017.09.006}
  {\path{doi:10.1016/j.physletb.2017.09.006}}.

\bibitem{Butenschoen:2017iks}
Mathias Butenschoen, Zhi-Guo He, and Bernd~A. Kniehl.
\newblock {$\eta_c$ Hadroproduction at Large Hadron Collider Challenges NRQCD
  Factorization}.
\newblock {\em EPJ Web Conf.}, 137:06009, 2017.
\newblock \href {http://dx.doi.org/10.1051/epjconf/201713706009}
  {\path{doi:10.1051/epjconf/201713706009}}.

\bibitem{Han:2014jya}
Hao Han, Yan-Qing Ma, Ce~Meng, Hua-Sheng Shao, and Kuang-Ta Chao.
\newblock {$\eta_c$ production at LHC and indications on the understanding of
  $\jpsi$ production}.
\newblock {\em Phys. Rev. Lett.}, 114(9):092005, 2015.
\newblock \href {http://arxiv.org/abs/1411.7350} {\path{arXiv:1411.7350}},
  \href {http://dx.doi.org/10.1103/PhysRevLett.114.092005}
  {\path{doi:10.1103/PhysRevLett.114.092005}}.

\bibitem{Baranov:2019joi}
S.~P. Baranov and A.~V. Lipatov.
\newblock {Prompt $\eta_c$ meson production at the LHC in the NRQCD with
  $k_T$-factorization}.
\newblock 2019.
\newblock \href {http://arxiv.org/abs/1904.00400} {\path{arXiv:1904.00400}}.

\bibitem{Khachatryan:2015rra}
Vardan Khachatryan et~al.
\newblock {Measurement of \jpsi and \psitwos Prompt Double-Differential Cross
  Sections in pp Collisions at $\sqrt{s}$=7 TeV}.
\newblock {\em Phys. Rev. Lett.}, 114(19):191802, 2015.
\newblock \href {http://arxiv.org/abs/1502.04155} {\path{arXiv:1502.04155}},
  \href {http://dx.doi.org/10.1103/PhysRevLett.114.191802}
  {\path{doi:10.1103/PhysRevLett.114.191802}}.

\bibitem{Aaij:2014bga}
Roel Aaij et~al.
\newblock {Measurement of the $\eta_c (1S)$ production cross-section in
  proton-proton collisions via the decay $\eta_c (1S) \rightarrow p \bar{p}$}.
\newblock {\em Eur. Phys. J.}, C75(7):311, 2015.
\newblock \href {http://arxiv.org/abs/1409.3612} {\path{arXiv:1409.3612}},
  \href {http://dx.doi.org/10.1140/epjc/s10052-015-3502-x}
  {\path{doi:10.1140/epjc/s10052-015-3502-x}}.

\bibitem{Lansberg:2017ozx}
Jean-Philippe Lansberg, Hua-Sheng Shao, and Hong-Fei Zhang.
\newblock {$\eta_c'$ Hadroproduction at Next-to-Leading Order and= its
  Relevance to $\psi^{\prime}$ Production}.
\newblock {\em Phys. Lett.}, B786:342--346, 2018.
\newblock \href {http://arxiv.org/abs/1711.00265} {\path{arXiv:1711.00265}},
  \href {http://dx.doi.org/10.1016/j.physletb.2018.10.009}
  {\path{doi:10.1016/j.physletb.2018.10.009}}.

\bibitem{Bodwin:2015iua}
Geoffrey~T. Bodwin, Kuang-Ta Chao, Hee~Sok Chung, U-Rae Kim, Jungil Lee, and
  Yan-Qing Ma.
\newblock {Fragmentation contributions to hadroproduction of prompt $\jpsi$,
  $\chi_{cJ}$, and $\psi(2S)$ states}.
\newblock {\em Phys. Rev.}, D93(3):034041, 2016.
\newblock \href {http://arxiv.org/abs/1509.07904} {\path{arXiv:1509.07904}},
  \href {http://dx.doi.org/10.1103/PhysRevD.93.034041}
  {\path{doi:10.1103/PhysRevD.93.034041}}.

\bibitem{ATLAS:2014ala}
Georges Aad et~al.
\newblock {Measurement of $\chi_{c1}$ and $\chi_{c2}$ production with
  $\sqrt{s}$ = 7 TeV $pp$ collisions at ATLAS}.
\newblock {\em JHEP}, 07:154, 2014.
\newblock \href {http://arxiv.org/abs/1404.7035} {\path{arXiv:1404.7035}},
  \href {http://dx.doi.org/10.1007/JHEP07(2014)154}
  {\path{doi:10.1007/JHEP07(2014)154}}.

\bibitem{Chatrchyan:2012ub}
Serguei Chatrchyan et~al.
\newblock {Measurement of the relative prompt production rate of $\chi_{c2}$
  and $\chi_{c1}$ in $pp$ collisions at $\sqrt{s}=7$ TeV}.
\newblock {\em Eur. Phys. J.}, C72:2251, 2012.
\newblock \href {http://arxiv.org/abs/1210.0875} {\path{arXiv:1210.0875}},
  \href {http://dx.doi.org/10.1140/epjc/s10052-012-2251-3}
  {\path{doi:10.1140/epjc/s10052-012-2251-3}}.

\bibitem{Aaij:2013dja}
R.~Aaij et~al.
\newblock {Measurement of the relative rate of prompt $\chi_{c0}$, $\chi_{c1}$
  and $\chi_{c2}$ production at $\sqrt{s}=7$TeV}.
\newblock {\em JHEP}, 10:115, 2013.
\newblock \href {http://arxiv.org/abs/1307.4285} {\path{arXiv:1307.4285}},
  \href {http://dx.doi.org/10.1007/JHEP10(2013)115}
  {\path{doi:10.1007/JHEP10(2013)115}}.

\bibitem{Ma:2010vd}
Yan-Qing Ma, Kai Wang, and Kuang-Ta Chao.
\newblock {QCD radiative corrections to $\chi_{cJ}$ production at hadron
  colliders}.
\newblock {\em Phys. Rev.}, D83:111503, 2011.
\newblock \href {http://arxiv.org/abs/1002.3987} {\path{arXiv:1002.3987}},
  \href {http://dx.doi.org/10.1103/PhysRevD.83.111503}
  {\path{doi:10.1103/PhysRevD.83.111503}}.

\bibitem{Baranov:2010zz}
S.~P. Baranov.
\newblock {On the $\sigma(\chi_{c1})/\sigma(\chi_{c2})$ ratio in the
  $k_T$-factorization approach}.
\newblock {\em Phys. Rev.}, D83:034035, 2011.
\newblock \href {http://dx.doi.org/10.1103/PhysRevD.83.034035}
  {\path{doi:10.1103/PhysRevD.83.034035}}.

\bibitem{Baranov:2015yea}
S.~P. Baranov, A.~V. Lipatov, and N.~P. Zotov.
\newblock {Prompt charmonia production and polarization at LHC in the NRQCD
  with $k_T$-factorization. Part II: $\chi_c$ mesons}.
\newblock {\em Phys. Rev.}, D93(9):094012, 2016.
\newblock \href {http://arxiv.org/abs/1510.02411} {\path{arXiv:1510.02411}},
  \href {http://dx.doi.org/10.1103/PhysRevD.93.094012}
  {\path{doi:10.1103/PhysRevD.93.094012}}.

\bibitem{Likhoded:2013aya}
A.K. Likhoded, A.V. Luchinsky, and S.V. Poslavsky.
\newblock {Hadronic production of $\chi_c$-mesons at LHC}.
\newblock 2013.
\newblock \href {http://arxiv.org/abs/1305.2389} {\path{arXiv:1305.2389}}.

\bibitem{Adloff:2002ex}
C.~Adloff et~al.
\newblock {Inelastic photoproduction of $\jpsi$ mesons at HERA}.
\newblock {\em Eur. Phys. J.}, C25:25--39, 2002.
\newblock \href {http://arxiv.org/abs/hep-ex/0205064}
  {\path{arXiv:hep-ex/0205064}}, \href
  {http://dx.doi.org/10.1007/s10052-002-1009-8}
  {\path{doi:10.1007/s10052-002-1009-8}}.

\bibitem{Aaron:2010gz}
F.~D. Aaron et~al.
\newblock {Inelastic Production of $\jpsi$ Mesons in Photoproduction and Deep
  Inelastic Scattering at HERA}.
\newblock {\em Eur. Phys. J.}, C68:401--420, 2010.
\newblock \href {http://arxiv.org/abs/1002.0234} {\path{arXiv:1002.0234}},
  \href {http://dx.doi.org/10.1140/epjc/s10052-010-1376-5}
  {\path{doi:10.1140/epjc/s10052-010-1376-5}}.

\bibitem{Chekanov:2002at}
S.~Chekanov et~al.
\newblock {Measurements of inelastic $J / \psi$ and $\psi^{\prime}$
  photoproduction at HERA}.
\newblock {\em Eur. Phys. J.}, C27:173--188, 2003.
\newblock \href {http://arxiv.org/abs/hep-ex/0211011}
  {\path{arXiv:hep-ex/0211011}}, \href
  {http://dx.doi.org/10.1140/epjc/s2002-01130-2}
  {\path{doi:10.1140/epjc/s2002-01130-2}}.

\bibitem{Chekanov:2009ad}
S.~Chekanov et~al.
\newblock {Measurement of $\jpsi$ helicity distributions in inelastic
  photoproduction at HERA}.
\newblock {\em JHEP}, 12:007, 2009.
\newblock \href {http://arxiv.org/abs/0906.1424} {\path{arXiv:0906.1424}},
  \href {http://dx.doi.org/10.1088/1126-6708/2009/12/007}
  {\path{doi:10.1088/1126-6708/2009/12/007}}.

\bibitem{Abramowicz:2012dh}
H.~Abramowicz et~al.
\newblock {Measurement of inelastic $\jpsi$ and $\psi^\prime$ photoproduction
  at HERA}.
\newblock {\em JHEP}, 02:071, 2013.
\newblock \href {http://arxiv.org/abs/1211.6946} {\path{arXiv:1211.6946}},
  \href {http://dx.doi.org/10.1007/JHEP02(2013)071}
  {\path{doi:10.1007/JHEP02(2013)071}}.

\bibitem{Kramer:1995nb}
Michael Krämer.
\newblock {QCD corrections to inelastic $J / \psi$ photoproduction}.
\newblock {\em Nucl. Phys.}, B459:3--50, 1996.
\newblock \href {http://arxiv.org/abs/hep-ph/9508409}
  {\path{arXiv:hep-ph/9508409}}, \href
  {http://dx.doi.org/10.1016/0550-3213(95)00568-4}
  {\path{doi:10.1016/0550-3213(95)00568-4}}.

\bibitem{Kramer:2001hh}
Michael Krämer.
\newblock {Quarkonium production at high-energy colliders}.
\newblock {\em Prog. Part. Nucl. Phys.}, 47:141--201, 2001.
\newblock \href {http://arxiv.org/abs/hep-ph/0106120}
  {\path{arXiv:hep-ph/0106120}}, \href
  {http://dx.doi.org/10.1016/S0146-6410(01)00154-5}
  {\path{doi:10.1016/S0146-6410(01)00154-5}}.

\bibitem{Butenschoen:2009zy}
Mathias Butenschoen and Bernd~A. Kniehl.
\newblock {Complete next-to-leading-order corrections to $\jpsi$
  photoproduction in nonrelativistic quantum chromodynamics}.
\newblock {\em Phys. Rev. Lett.}, 104:072001, 2010.
\newblock \href {http://arxiv.org/abs/0909.2798} {\path{arXiv:0909.2798}},
  \href {http://dx.doi.org/10.1103/PhysRevLett.104.072001}
  {\path{doi:10.1103/PhysRevLett.104.072001}}.

\bibitem{Artoisenet:2009xh}
P.~Artoisenet, John~M. Campbell, F.~Maltoni, and F.~Tramontano.
\newblock {$\jpsi$ production at HERA}.
\newblock {\em Phys. Rev. Lett.}, 102:142001, 2009.
\newblock \href {http://arxiv.org/abs/0901.4352} {\path{arXiv:0901.4352}},
  \href {http://dx.doi.org/10.1103/PhysRevLett.102.142001}
  {\path{doi:10.1103/PhysRevLett.102.142001}}.

\bibitem{Chang:2009uj}
Chao-Hsi Chang, Rong Li, and Jian-Xiong Wang.
\newblock {$\jpsi$ polarization in photo-production up-to the next-to-leading
  order of QCD}.
\newblock {\em Phys. Rev.}, D80:034020, 2009.
\newblock \href {http://arxiv.org/abs/0901.4749} {\path{arXiv:0901.4749}},
  \href {http://dx.doi.org/10.1103/PhysRevD.80.034020}
  {\path{doi:10.1103/PhysRevD.80.034020}}.

\bibitem{Butenschon:2009zz}
M.~Butenschoen.
\newblock {$\jpsi$ photoproduction at NLO in NRQCD}.
\newblock {\em Nucl. Phys. Proc. Suppl.}, 191:193--202, 2009.
\newblock \href {http://dx.doi.org/10.1016/j.nuclphysbps.2009.03.126}
  {\path{doi:10.1016/j.nuclphysbps.2009.03.126}}.

\bibitem{Butenschon:2010iy}
Mathias Butenschoen and Bernd~A. Kniehl.
\newblock {Direct $\jpsi$ photoproduction at next-to-leading-order in
  nonrelativistic QCD}.
\newblock {\em PoS}, DIS2010:157, 2010.
\newblock \href {http://arxiv.org/abs/1006.1776} {\path{arXiv:1006.1776}},
  \href {http://dx.doi.org/10.22323/1.106.0157}
  {\path{doi:10.22323/1.106.0157}}.

\bibitem{Butenschoen:2012qh}
Mathias Butenschoen and Bernd~A. Kniehl.
\newblock {J/$\psi$ production in NRQCD: A global analysis of yield and
  polarization}.
\newblock {\em Nucl. Phys. Proc. Suppl.}, 222-224:151--161, 2012.
\newblock \href {http://arxiv.org/abs/1201.3862} {\path{arXiv:1201.3862}},
  \href {http://dx.doi.org/10.1016/j.nuclphysbps.2012.03.016}
  {\path{doi:10.1016/j.nuclphysbps.2012.03.016}}.

\bibitem{Baranov:2002cf}
S.~P. Baranov.
\newblock {Highlights from the $k_T$ factorization approach on the quarkonium
  production puzzles}.
\newblock {\em Phys. Rev.}, D66:114003, 2002.
\newblock [,162(2002)].
\newblock \href {http://dx.doi.org/10.1103/PhysRevD.66.114003}
  {\path{doi:10.1103/PhysRevD.66.114003}}.

\bibitem{Kniehl:2006sk}
B.~A. Kniehl, D.~V. Vasin, and V.~A. Saleev.
\newblock {Charmonium production at high energy in the $k_{T}$ -factorization
  approach}.
\newblock {\em Phys. Rev.}, D73:074022, 2006.
\newblock \href {http://arxiv.org/abs/hep-ph/0602179}
  {\path{arXiv:hep-ph/0602179}}, \href
  {http://dx.doi.org/10.1103/PhysRevD.73.074022}
  {\path{doi:10.1103/PhysRevD.73.074022}}.

\bibitem{Martin:2009ii}
A.~D. Martin, M.~G. Ryskin, and G.~Watt.
\newblock {NLO prescription for unintegrated parton distributions}.
\newblock {\em Eur. Phys. J.}, C66:163--172, 2010.
\newblock \href {http://arxiv.org/abs/0909.5529} {\path{arXiv:0909.5529}},
  \href {http://dx.doi.org/10.1140/epjc/s10052-010-1242-5}
  {\path{doi:10.1140/epjc/s10052-010-1242-5}}.

\bibitem{Adare:2009js}
A.~Adare et~al.
\newblock {Transverse momentum dependence of \jpsi polarization at midrapidity
  in $p+p$ collisions at $s^{1/2}$ = 200 GeV}.
\newblock {\em Phys. Rev.}, D82:012001, 2010.
\newblock \href {http://arxiv.org/abs/0912.2082} {\path{arXiv:0912.2082}},
  \href {http://dx.doi.org/10.1103/PhysRevD.82.012001}
  {\path{doi:10.1103/PhysRevD.82.012001}}.

\bibitem{Acosta:2004yw}
D.~Acosta et~al.
\newblock {Measurement of the $\jpsi$ meson and $b-$hadron production cross
  sections in $p\bar{p}$ collisions at $\sqrt{s} = 1960$ GeV}.
\newblock {\em Phys. Rev.}, D71:032001, 2005.
\newblock \href {http://arxiv.org/abs/hep-ex/0412071}
  {\path{arXiv:hep-ex/0412071}}, \href
  {http://dx.doi.org/10.1103/PhysRevD.71.032001}
  {\path{doi:10.1103/PhysRevD.71.032001}}.

\bibitem{Khachatryan:2010yr}
Vardan Khachatryan et~al.
\newblock {Prompt and non-prompt $\jpsi$ production in $pp$ collisions at
  $\sqrt{s}=7$ TeV}.
\newblock {\em Eur. Phys. J.}, C71:1575, 2011.
\newblock \href {http://arxiv.org/abs/1011.4193} {\path{arXiv:1011.4193}},
  \href {http://dx.doi.org/10.1140/epjc/s10052-011-1575-8}
  {\path{doi:10.1140/epjc/s10052-011-1575-8}}.

\bibitem{ATLAS:2010sca}
{A first measurement of the differential cross section for the $\jpsi\to\mu\mu$
  resonance and the non-prompt to prompt $\jpsi$ cross-section ratio with pp
  collisions at $\sqrt{s}$=7 TeV in ATLAS}.
\newblock 2010.

\bibitem{Abelev:2014qha}
Betty~Bezverkhny Abelev et~al.
\newblock {Measurement of quarkonium production at forward rapidity in $pp$
  collisions at $\sqrt{s} = 7$ TeV}.
\newblock {\em Eur. Phys. J.}, C74(8):2974, 2014.
\newblock \href {http://arxiv.org/abs/1403.3648} {\path{arXiv:1403.3648}},
  \href {http://dx.doi.org/10.1140/epjc/s10052-014-2974-4}
  {\path{doi:10.1140/epjc/s10052-014-2974-4}}.

\bibitem{Butenschoen:2011ks}
Mathias Butenschoen and Bernd~A. Kniehl.
\newblock {Probing nonrelativistic QCD factorization in polarized $\jpsi$
  photoproduction at next-to-leading order}.
\newblock {\em Phys. Rev. Lett.}, 107:232001, 2011.
\newblock \href {http://arxiv.org/abs/1109.1476} {\path{arXiv:1109.1476}},
  \href {http://dx.doi.org/10.1103/PhysRevLett.107.232001}
  {\path{doi:10.1103/PhysRevLett.107.232001}}.

\bibitem{Abdallah:2003du}
J.~Abdallah et~al.
\newblock {Study of inclusive $J / \psi$ production in two photon collisions at
  LEP-2 with the DELPHI detector}.
\newblock {\em Phys. Lett.}, B565:76--86, 2003.
\newblock \href {http://arxiv.org/abs/hep-ex/0307049}
  {\path{arXiv:hep-ex/0307049}}, \href
  {http://dx.doi.org/10.1016/S0370-2693(03)00660-9}
  {\path{doi:10.1016/S0370-2693(03)00660-9}}.

\bibitem{Ma:1997bi}
J.~P. Ma, B.~H.~J. McKellar, and C.~B. Paranavitane.
\newblock {$\jpsi$ production at photon - photon colliders as a probe of the
  color octet mechanism}.
\newblock {\em Phys. Rev.}, D57:606--609, 1998.
\newblock \href {http://arxiv.org/abs/hep-ph/9707480}
  {\path{arXiv:hep-ph/9707480}}, \href
  {http://dx.doi.org/10.1103/PhysRevD.57.606}
  {\path{doi:10.1103/PhysRevD.57.606}}.

\bibitem{Japaridze:1998ss}
George Japaridze and Avto Tkabladze.
\newblock {Color octet contribution to $J / \psi$ production at a photon linear
  collider}.
\newblock {\em Phys. Lett.}, B433:139--146, 1998.
\newblock \href {http://arxiv.org/abs/hep-ph/9803447}
  {\path{arXiv:hep-ph/9803447}}, \href
  {http://dx.doi.org/10.1016/S0370-2693(98)00697-2}
  {\path{doi:10.1016/S0370-2693(98)00697-2}}.

\bibitem{Godbole:2001pj}
R.~M. Godbole, D.~Indumathi, and M.~Kramer.
\newblock {$\jpsi$ production through resolved photon processes at $e^{+}
  e^{-}$ colliders}.
\newblock {\em Phys. Rev.}, D65:074003, 2002.
\newblock [,1594(2001)].
\newblock \href {http://arxiv.org/abs/hep-ph/0101333}
  {\path{arXiv:hep-ph/0101333}}, \href
  {http://dx.doi.org/10.1103/PhysRevD.65.074003}
  {\path{doi:10.1103/PhysRevD.65.074003}}.

\bibitem{Klasen:2001mi}
M.~Klasen, Bernd~A. Kniehl, L.~Mihaila, and M.~Steinhauser.
\newblock {$\jpsi$ plus dijet associated production in two photon collisions}.
\newblock {\em Nucl. Phys.}, B609:518--536, 2001.
\newblock \href {http://arxiv.org/abs/hep-ph/0104044}
  {\path{arXiv:hep-ph/0104044}}, \href
  {http://dx.doi.org/10.1016/S0550-3213(01)00318-2}
  {\path{doi:10.1016/S0550-3213(01)00318-2}}.

\bibitem{Klasen:2001cu}
M.~Klasen, Bernd~A. Kniehl, L.~N. Mihaila, and M.~Steinhauser.
\newblock {Evidence for color octet mechanism from CERN LEP-2 $\gamma \gamma
  \to \jpsi$ + $X$ data}.
\newblock {\em Phys. Rev. Lett.}, 89:032001, 2002.
\newblock \href {http://arxiv.org/abs/hep-ph/0112259}
  {\path{arXiv:hep-ph/0112259}}, \href
  {http://dx.doi.org/10.1103/PhysRevLett.89.032001}
  {\path{doi:10.1103/PhysRevLett.89.032001}}.

\bibitem{Chen:2016hju}
Zi-Qiang Chen, Long-Bin Chen, and Cong-Feng Qiao.
\newblock {NLO QCD Corrections for $\jpsi+ c + \bar{c}$ Production in
  Photon-Photon Collision}.
\newblock {\em Phys. Rev.}, D95(3):036001, 2017.
\newblock \href {http://arxiv.org/abs/1608.06231} {\path{arXiv:1608.06231}},
  \href {http://dx.doi.org/10.1103/PhysRevD.95.036001}
  {\path{doi:10.1103/PhysRevD.95.036001}}.

\bibitem{Aubert:2001pd}
Bernard Aubert et~al.
\newblock {Measurement of $\jpsi$ production in continuum $e^+e^-$
  annihilations near $\sqrt{s}=10.6$ GeV}.
\newblock {\em Phys. Rev. Lett.}, 87:162002, 2001.
\newblock \href {http://arxiv.org/abs/hep-ex/0106044}
  {\path{arXiv:hep-ex/0106044}}, \href
  {http://dx.doi.org/10.1103/PhysRevLett.87.162002}
  {\path{doi:10.1103/PhysRevLett.87.162002}}.

\bibitem{Abe:2001za}
Kazuo Abe et~al.
\newblock {Production of prompt charmonia in $\epem$ annihilation at $s^{1/2}$
  is approximately 10.6 GeV}.
\newblock {\em Phys. Rev. Lett.}, 88:052001, 2002.
\newblock \href {http://arxiv.org/abs/hep-ex/0110012}
  {\path{arXiv:hep-ex/0110012}}, \href
  {http://dx.doi.org/10.1103/PhysRevLett.88.052001}
  {\path{doi:10.1103/PhysRevLett.88.052001}}.

\bibitem{Yuan:1996ep}
Feng Yuan, Cong-Feng Qiao, and Kuang-Ta Chao.
\newblock {Prompt $\jpsi$ production at $e^{+} e^{-}$ colliders}.
\newblock {\em Phys. Rev.}, D56:321--328, 1997.
\newblock \href {http://arxiv.org/abs/hep-ph/9703438}
  {\path{arXiv:hep-ph/9703438}}, \href
  {http://dx.doi.org/10.1103/PhysRevD.56.321}
  {\path{doi:10.1103/PhysRevD.56.321}}.

\bibitem{Cho:1996cg}
Peter~L. Cho and Adam~K. Leibovich.
\newblock {Color singlet $\psi_Q$ production at \epem colliders}.
\newblock {\em Phys. Rev.}, D54:6690--6695, 1996.
\newblock \href {http://arxiv.org/abs/hep-ph/9606229}
  {\path{arXiv:hep-ph/9606229}}, \href
  {http://dx.doi.org/10.1103/PhysRevD.54.6690}
  {\path{doi:10.1103/PhysRevD.54.6690}}.

\bibitem{Baek:1998yf}
Seungwon Baek, P.~Ko, Jungil Lee, and H.~S. Song.
\newblock {Polarized $J / \psi$ production at CLEO}.
\newblock {\em J. Korean Phys. Soc.}, 33:97--101, 1998.
\newblock [,225(1998)].
\newblock \href {http://arxiv.org/abs/hep-ph/9804455}
  {\path{arXiv:hep-ph/9804455}}.

\bibitem{Schuler:1998az}
Gerhard~A. Schuler.
\newblock {Testing factorization of charmonium production}.
\newblock {\em Eur.Phys.J.}, C8:273--281, 1999.
\newblock \href {http://arxiv.org/abs/hep-ph/9804349}
  {\path{arXiv:hep-ph/9804349}}, \href
  {http://dx.doi.org/10.1007/s100529900948} {\path{doi:10.1007/s100529900948}}.

\bibitem{Kiselev:1994pu}
V.~V. Kiselev, A.~K. Likhoded, and M.~V. Shevlyagin.
\newblock {Double charmed baryon production at B-factory}.
\newblock {\em Phys. Lett.}, B332:411--414, 1994.
\newblock \href {http://arxiv.org/abs/hep-ph/9408407}
  {\path{arXiv:hep-ph/9408407}}, \href
  {http://dx.doi.org/10.1016/0370-2693(94)91273-4}
  {\path{doi:10.1016/0370-2693(94)91273-4}}.

\bibitem{Liu:2003zr}
Kui-Yong Liu, Zhi-Guo He, and Kuang-Ta Chao.
\newblock {Production of $J / \psi + c \bar{c}$ through two photons in \epem
  annihilation}.
\newblock {\em Phys. Rev.}, D68:031501, 2003.
\newblock \href {http://arxiv.org/abs/hep-ph/0305084}
  {\path{arXiv:hep-ph/0305084}}, \href
  {http://dx.doi.org/10.1103/PhysRevD.68.031501}
  {\path{doi:10.1103/PhysRevD.68.031501}}.

\bibitem{Abe:2002rb}
Kazuo Abe et~al.
\newblock {Observation of double \ccbar production in \epem annihilation at
  $s^{1/2}$ approximately 10.6-GeV}.
\newblock {\em Phys. Rev. Lett.}, 89:142001, 2002.
\newblock \href {http://arxiv.org/abs/hep-ex/0205104}
  {\path{arXiv:hep-ex/0205104}}, \href
  {http://dx.doi.org/10.1103/PhysRevLett.89.142001}
  {\path{doi:10.1103/PhysRevLett.89.142001}}.

\bibitem{Liu:2003jj}
Kui-Yong Liu, Zhi-Guo He, and Kuang-Ta Chao.
\newblock {Inclusive charmonium production via double $c \bar{c}$ in $e^{+}
  e^{-}$ annihilation}.
\newblock {\em Phys. Rev.}, D69:094027, 2004.
\newblock \href {http://arxiv.org/abs/hep-ph/0301218}
  {\path{arXiv:hep-ph/0301218}}, \href
  {http://dx.doi.org/10.1103/PhysRevD.69.094027}
  {\path{doi:10.1103/PhysRevD.69.094027}}.

\bibitem{Pakhlov:2009nj}
P.~Pakhlov et~al.
\newblock {Measurement of the $\epem \to \jpsi \ccbar$ cross section at
  $s^{1/2}$ ~10.6 GeV}.
\newblock {\em Phys. Rev.}, D79:071101, 2009.
\newblock \href {http://arxiv.org/abs/0901.2775} {\path{arXiv:0901.2775}},
  \href {http://dx.doi.org/10.1103/PhysRevD.79.071101}
  {\path{doi:10.1103/PhysRevD.79.071101}}.

\bibitem{Zhang:2006ay}
Yu-Jie Zhang and Kuang-Ta Chao.
\newblock {Double charm production $e^+ e^- \to J / \psi + c \bar{c}$ at
  B-factories with next-to-leading order QCD correction}.
\newblock {\em Phys. Rev. Lett.}, 98:092003, 2007.
\newblock \href {http://arxiv.org/abs/hep-ph/0611086}
  {\path{arXiv:hep-ph/0611086}}, \href
  {http://dx.doi.org/10.1103/PhysRevLett.98.092003}
  {\path{doi:10.1103/PhysRevLett.98.092003}}.

\bibitem{Ma:2008gq}
Yan-Qing Ma, Yu-Jie Zhang, and Kuang-Ta Chao.
\newblock {QCD correction to $e^+e^- \to \jpsi+gg$ at B Factories}.
\newblock {\em Phys. Rev. Lett.}, 102:162002, 2009.
\newblock \href {http://arxiv.org/abs/0812.5106} {\path{arXiv:0812.5106}},
  \href {http://dx.doi.org/10.1103/PhysRevLett.102.162002}
  {\path{doi:10.1103/PhysRevLett.102.162002}}.

\bibitem{Gong:2009kp}
Bin Gong and Jian-Xiong Wang.
\newblock {Next-to-Leading-Order QCD Corrections to $\epem \to \jpsi gg$ at the
  B-Factories}.
\newblock {\em Phys. Rev. Lett.}, 102:162003, 2009.
\newblock \href {http://arxiv.org/abs/0901.0117} {\path{arXiv:0901.0117}},
  \href {http://dx.doi.org/10.1103/PhysRevLett.102.162003}
  {\path{doi:10.1103/PhysRevLett.102.162003}}.

\bibitem{He:2009uf}
Zhi-Guo He, Ying Fan, and Kuang-Ta Chao.
\newblock {Relativistic correction to $e^+ e^- \to \jpsi + gg$ at B-factories
  and constraint on color-octet matrix elements}.
\newblock {\em Phys. Rev.}, D81:054036, 2010.
\newblock \href {http://arxiv.org/abs/0910.3636} {\path{arXiv:0910.3636}},
  \href {http://dx.doi.org/10.1103/PhysRevD.81.054036}
  {\path{doi:10.1103/PhysRevD.81.054036}}.

\bibitem{Jia:2009np}
Yu~Jia.
\newblock {Color-singlet relativistic correction to inclusive \jpsi production
  associated with light hadrons at B factories}.
\newblock {\em Phys. Rev.}, D82:034017, 2010.
\newblock \href {http://arxiv.org/abs/0912.5498} {\path{arXiv:0912.5498}},
  \href {http://dx.doi.org/10.1103/PhysRevD.82.034017}
  {\path{doi:10.1103/PhysRevD.82.034017}}.

\bibitem{Zhang:2009ym}
Yu-Jie Zhang, Yan-Qing Ma, Kai Wang, and Kuang-Ta Chao.
\newblock {QCD radiative correction to color-octet $\jpsi$ inclusive production
  at B Factories}.
\newblock {\em Phys. Rev.}, D81:034015, 2010.
\newblock \href {http://arxiv.org/abs/0911.2166} {\path{arXiv:0911.2166}},
  \href {http://dx.doi.org/10.1103/PhysRevD.81.034015}
  {\path{doi:10.1103/PhysRevD.81.034015}}.

\bibitem{Barsuk:2012ic}
Sergey Barsuk, Jibo He, Emi Kou, and Benoit Viaud.
\newblock {Investigating charmonium production at LHC with the $p \bar{p}$
  final state}.
\newblock {\em Phys.Rev.}, D86:034011, 2012.
\newblock \href {http://arxiv.org/abs/1202.2273} {\path{arXiv:1202.2273}},
  \href {http://dx.doi.org/10.1103/PhysRevD.86.034011}
  {\path{doi:10.1103/PhysRevD.86.034011}}.

\bibitem{PDG2017}
C.~Patrignani et~al.
\newblock {\href{http://pdg.lbl.gov/}{Review of particle physics}}.
\newblock {\em Chin. Phys.}, C40:100001, 2016.
\newblock and {\href{http://pdglive.lbl.gov/}{2017 update}}.
\newblock \href {http://dx.doi.org/10.1088/1674-1137/40/10/100001}
  {\path{doi:10.1088/1674-1137/40/10/100001}}.

\bibitem{Aaij:2018bla}
Roel Aaij et~al.
\newblock {Evidence for an $\eta_c(1S) \pi^-$ resonance in $B^0 \to \eta_c(1S)
  K^+\pi^-$ decays}.
\newblock {\em Submitted to: Eur. Phys. J.}, 2018.
\newblock \href {http://arxiv.org/abs/1809.07416} {\path{arXiv:1809.07416}}.

\bibitem{Aaij:2016kxn}
Roel Aaij et~al.
\newblock {Observation of $\eta_{c}(2S) \to p \bar p$ and search for $X(3872)
  \to p \bar p$ decays}.
\newblock {\em Phys. Lett.}, B769:305--313, 2017.
\newblock \href {http://arxiv.org/abs/1607.06446} {\path{arXiv:1607.06446}},
  \href {http://dx.doi.org/10.1016/j.physletb.2017.03.046}
  {\path{doi:10.1016/j.physletb.2017.03.046}}.

\bibitem{Kato:2017gfv}
Y.~Kato et~al.
\newblock {Measurements of the absolute branching fractions of $B^{+} \to
  X_{c\bar{c}} K^{+}$ and $B^{+} \to \bar{D}^{(\ast) 0} \pi^{+} $ at Belle}.
\newblock {\em Phys. Rev.}, D97(1):012005, 2018.
\newblock \href {http://arxiv.org/abs/1709.06108} {\path{arXiv:1709.06108}},
  \href {http://dx.doi.org/10.1103/PhysRevD.97.012005}
  {\path{doi:10.1103/PhysRevD.97.012005}}.

\bibitem{Aaij:2017tzn}
R.~Aaij et~al.
\newblock {Study of charmonium production in ${b}$-hadron decays and first
  evidence for the decay ${{{B}} ^0_{{s}}} \!\rightarrow \phi \phi \phi $}.
\newblock {\em Eur. Phys. J.}, C77(9):609, 2017.
\newblock \href {http://arxiv.org/abs/1706.07013} {\path{arXiv:1706.07013}},
  \href {http://dx.doi.org/10.1140/epjc/s10052-017-5151-8}
  {\path{doi:10.1140/epjc/s10052-017-5151-8}}.

\bibitem{Ablikim:2011uf}
M.~Ablikim.
\newblock {Observation of $\chi_{cJ}$ decaying into the $p\bar{p}K^{+}K^{-}$
  final state}.
\newblock {\em Phys. Rev.}, D83:112009, 2011.
\newblock \href {http://arxiv.org/abs/1103.2661} {\path{arXiv:1103.2661}},
  \href {http://dx.doi.org/10.1103/PhysRevD.83.112009}
  {\path{doi:10.1103/PhysRevD.83.112009}}.

\bibitem{Jacques}
Jacques Lefrancois.
\newblock {private communication, 2017}.

\bibitem{Ablikim:2013hdv}
M.~Ablikim et~al.
\newblock {Search for $\eta_c(2S) / h_c \to p\overline{p}$ decays and
  measurements of the $\chi_{cJ} \to p\overline{p}$ branching fractions}.
\newblock {\em Phys. Rev.}, D88(11):112001, 2013.
\newblock \href {http://arxiv.org/abs/1310.6099} {\path{arXiv:1310.6099}},
  \href {http://dx.doi.org/10.1103/PhysRevD.88.112001}
  {\path{doi:10.1103/PhysRevD.88.112001}}.

\bibitem{Aaij:2017vck}
Roel Aaij et~al.
\newblock {$\chi_{c1}$ and $\chi_{c2}$ Resonance Parameters with the Decays
  $\chi_{c1,c2}\to \jpsi\mu^+\mu^-$}.
\newblock {\em Phys. Rev. Lett.}, 119(22):221801, 2017.
\newblock \href {http://arxiv.org/abs/1709.04247} {\path{arXiv:1709.04247}},
  \href {http://dx.doi.org/10.1103/PhysRevLett.119.221801}
  {\path{doi:10.1103/PhysRevLett.119.221801}}.

\bibitem{Ablikim:2018ewr}
Medina Ablikim et~al.
\newblock {First observations of $h_c \to$ hadrons}.
\newblock {\em Phys. Rev.}, D99(7):072008, 2019.
\newblock \href {http://arxiv.org/abs/1810.12023} {\path{arXiv:1810.12023}},
  \href {http://dx.doi.org/10.1103/PhysRevD.99.072008}
  {\path{doi:10.1103/PhysRevD.99.072008}}.

\bibitem{Evans:2008zzb}
Lyndon Evans and Philip Bryant.
\newblock {LHC Machine}.
\newblock {\em JINST}, 3:S08001, 2008.
\newblock \href {http://dx.doi.org/10.1088/1748-0221/3/08/S08001}
  {\path{doi:10.1088/1748-0221/3/08/S08001}}.

\bibitem{Christiane:1260465}
Christiane Lefèvre.
\newblock {The CERN accelerator complex. Complexe des accélérateurs du CERN}.
\newblock Dec 2008.
\newblock URL: \url{https://cds.cern.ch/record/1260465}.

\bibitem{Aad:2008zzm}
G.~Aad et~al.
\newblock {The ATLAS Experiment at the CERN Large Hadron Collider}.
\newblock {\em JINST}, 3:S08003, 2008.
\newblock \href {http://dx.doi.org/10.1088/1748-0221/3/08/S08003}
  {\path{doi:10.1088/1748-0221/3/08/S08003}}.

\bibitem{Chatrchyan:2008aa}
S.~Chatrchyan et~al.
\newblock {The CMS Experiment at the CERN LHC}.
\newblock {\em JINST}, 3:S08004, 2008.
\newblock \href {http://dx.doi.org/10.1088/1748-0221/3/08/S08004}
  {\path{doi:10.1088/1748-0221/3/08/S08004}}.

\bibitem{Aamodt:2008zz}
K.~Aamodt et~al.
\newblock {The ALICE experiment at the CERN LHC}.
\newblock {\em JINST}, 3:S08002, 2008.
\newblock \href {http://dx.doi.org/10.1088/1748-0221/3/08/S08002}
  {\path{doi:10.1088/1748-0221/3/08/S08002}}.

\bibitem{Adriani:2008zz}
O.~Adriani et~al.
\newblock {The LHCf detector at the CERN Large Hadron Collider}.
\newblock {\em JINST}, 3:S08006, 2008.
\newblock \href {http://dx.doi.org/10.1088/1748-0221/3/08/S08006}
  {\path{doi:10.1088/1748-0221/3/08/S08006}}.

\bibitem{Anelli:2008zza}
G.~Anelli et~al.
\newblock {The TOTEM experiment at the CERN Large Hadron Collider}.
\newblock {\em JINST}, 3:S08007, 2008.
\newblock \href {http://dx.doi.org/10.1088/1748-0221/3/08/S08007}
  {\path{doi:10.1088/1748-0221/3/08/S08007}}.

\bibitem{Pinfold:2009oia}
James Pinfold et~al.
\newblock {Technical Design Report of the MoEDAL Experiment}.
\newblock 2009.

\bibitem{Alves:2008zz}
A.~Augusto Alves, Jr. et~al.
\newblock {The LHCb Detector at the LHC}.
\newblock {\em JINST}, 3:S08005, 2008.
\newblock \href {http://dx.doi.org/10.1088/1748-0221/3/08/S08005}
  {\path{doi:10.1088/1748-0221/3/08/S08005}}.

\bibitem{Aaij:2014jba}
Roel Aaij et~al.
\newblock {LHCb Detector Performance}.
\newblock {\em Int. J. Mod. Phys.}, A30(07):1530022, 2015.
\newblock \href {http://arxiv.org/abs/1412.6352} {\path{arXiv:1412.6352}},
  \href {http://dx.doi.org/10.1142/S0217751X15300227}
  {\path{doi:10.1142/S0217751X15300227}}.

\bibitem{Abashian:2000cg}
A.~Abashian et~al.
\newblock {The Belle Detector}.
\newblock {\em Nucl. Instrum. Meth.}, A479:117--232, 2002.
\newblock \href {http://dx.doi.org/10.1016/S0168-9002(01)02013-7}
  {\path{doi:10.1016/S0168-9002(01)02013-7}}.

\bibitem{Aubert:2001tu}
Bernard Aubert et~al.
\newblock {The BaBar detector}.
\newblock {\em Nucl. Instrum. Meth.}, A479:1--116, 2002.
\newblock \href {http://arxiv.org/abs/hep-ex/0105044}
  {\path{arXiv:hep-ex/0105044}}, \href
  {http://dx.doi.org/10.1016/S0168-9002(01)02012-5}
  {\path{doi:10.1016/S0168-9002(01)02012-5}}.

\bibitem{Ablikim:2009aa}
M.~Ablikim et~al.
\newblock {Design and Construction of the BESIII Detector}.
\newblock {\em Nucl. Instrum. Meth.}, A614:345--399, 2010.
\newblock \href {http://arxiv.org/abs/0911.4960} {\path{arXiv:0911.4960}},
  \href {http://dx.doi.org/10.1016/j.nima.2009.12.050}
  {\path{doi:10.1016/j.nima.2009.12.050}}.

\bibitem{Patrignani:2004vt}
C.~Patrignani et~al.
\newblock {E835 at FNAL: Charmonium spectroscopy in anti-p p annihilations}.
\newblock {\em Nucl. Phys. Proc. Suppl.}, 142:98--103, 2005.
\newblock \href {http://dx.doi.org/10.1016/j.nuclphysbps.2005.01.017}
  {\path{doi:10.1016/j.nuclphysbps.2005.01.017}}.

\bibitem{Anashin:2013twa}
V.~V. Anashin et~al.
\newblock {The KEDR detector}.
\newblock {\em Phys. Part. Nucl.}, 44:657--702, 2013.
\newblock \href {http://dx.doi.org/10.1134/S1063779613040035}
  {\path{doi:10.1134/S1063779613040035}}.

\bibitem{bbangles}
Christian Els{\"a}sser.
\newblock $\overline{b}b$ production angle plots.
\newblock
  {\url{https://lhcb.web.cern.ch/lhcb/speakersbureau/html/bb\_ProductionAngles.html}},
  {2019}.

\bibitem{Follin:2014nva}
F.~Follin and D.~Jacquet.
\newblock {Implementation and experience with luminosity levelling with offset
  beam}.
\newblock In {\em {Proceedings, ICFA Mini-Workshop on Beam-Beam Effects in
  Hadron Colliders (BB2013): CERN, Geneva, Switzerland, March 18-22 2013}},
  pages 183--187, 2014.
\newblock [,183(2014)].
\newblock \href {http://arxiv.org/abs/1410.3667} {\path{arXiv:1410.3667}},
  \href {http://dx.doi.org/10.5170/CERN-2014-004.183}
  {\path{doi:10.5170/CERN-2014-004.183}}.

\bibitem{LHCb-DP-2014-001}
R.~Aaij et~al.
\newblock {Performance of the LHCb Vertex Locator}.
\newblock {\em JINST}, 9:P09007, 2014.
\newblock \href {http://arxiv.org/abs/1405.7808} {\path{arXiv:1405.7808}},
  \href {http://dx.doi.org/10.1088/1748-0221/9/09/P09007}
  {\path{doi:10.1088/1748-0221/9/09/P09007}}.

\bibitem{LHCb-DP-2013-003}
R.~Arink et~al.
\newblock {Performance of the LHCb Outer Tracker}.
\newblock {\em JINST}, 9:P01002, 2014.
\newblock \href {http://arxiv.org/abs/1311.3893} {\path{arXiv:1311.3893}},
  \href {http://dx.doi.org/10.1088/1748-0221/9/01/P01002}
  {\path{doi:10.1088/1748-0221/9/01/P01002}}.

\bibitem{LHCb-DP-2012-003}
M.~Adinolfi et~al.
\newblock {Performance of the \lhcb RICH detector at the LHC}.
\newblock {\em Eur. Phys. J.}, C73:2431, 2013.
\newblock \href {http://arxiv.org/abs/1211.6759} {\path{arXiv:1211.6759}},
  \href {http://dx.doi.org/10.1140/epjc/s10052-013-2431-9}
  {\path{doi:10.1140/epjc/s10052-013-2431-9}}.

\bibitem{LHCb-DP-2012-002}
A~A Alves~Jr. et~al.
\newblock {Performance of the LHCb muon system}.
\newblock {\em JINST}, 8:P02022, 2013.
\newblock \href {http://arxiv.org/abs/1211.1346} {\path{arXiv:1211.1346}},
  \href {http://dx.doi.org/10.1088/1748-0221/8/02/P02022}
  {\path{doi:10.1088/1748-0221/8/02/P02022}}.

\bibitem{LHCb:2001aa}
P.~R. Barbosa-Marinho et~al.
\newblock {LHCb VELO TDR: Vertex locator. Technical design report}.
\newblock {\em CERN-LHCC-2001-011}, 2001.

\bibitem{veloPlots}
Velo approved conference plots.
\newblock {\url{https://lbtwiki.cern.ch/bin/view/VELO/VELOConferencePlots}},
  {2019}.

\bibitem{Aaij:2013mpa}
R~Aaij et~al.
\newblock {Precision measurement of the $B^{0}_{s}$-$\bar{B}^{0}_{s}$
  oscillation frequency with the decay $B^{0}_{s}\rightarrow
  D^{-}_{s}\pi^{+}$}.
\newblock {\em New J. Phys.}, 15:053021, 2013.
\newblock \href {http://arxiv.org/abs/1304.4741} {\path{arXiv:1304.4741}},
  \href {http://dx.doi.org/10.1088/1367-2630/15/5/053021}
  {\path{doi:10.1088/1367-2630/15/5/053021}}.

\bibitem{LHCB:2000ac}
S.~Amato et~al.
\newblock {LHCb magnet: Technical design report}.
\newblock {\em CERN-LHCC-2000-007}, 2000.

\bibitem{STPlots}
Silicon tracker approved conference plots.
\newblock {\url{https://lhcb.physik.uzh.ch/ST/public/material/}}, {2019}.

\bibitem{Barbosa-Marinho:582793}
P~R Barbosa-Marinho et~al.
\newblock {\em {LHCb inner tracker: Technical Design Report}}.
\newblock Technical Design Report LHCb. CERN, Geneva, 2002.
\newblock revised version number 1 submitted on 2002-11-13 14:14:34.
\newblock URL: \url{http://cds.cern.ch/record/582793}.

\bibitem{Arink:2013twa}
R~Arink et~al.
\newblock {Performance of the LHCb Outer Tracker}.
\newblock {\em JINST}, 9(01):P01002, 2014.
\newblock \href {http://arxiv.org/abs/1311.3893} {\path{arXiv:1311.3893}},
  \href {http://dx.doi.org/10.1088/1748-0221/9/01/P01002}
  {\path{doi:10.1088/1748-0221/9/01/P01002}}.

\bibitem{Amato:494264}
S~Amato et~al.
\newblock {\em {LHCb calorimeters: Technical Design Report}}.
\newblock Technical Design Report LHCb. CERN, Geneva, 2000.
\newblock URL: \url{http://cds.cern.ch/record/494264}.

\bibitem{Adinolfi:2012qfa}
M.~Adinolfi et~al.
\newblock {Performance of the LHCb RICH detector at the LHC}.
\newblock {\em Eur. Phys. J.}, C73:2431, 2013.
\newblock \href {http://arxiv.org/abs/1211.6759} {\path{arXiv:1211.6759}},
  \href {http://dx.doi.org/10.1140/epjc/s10052-013-2431-9}
  {\path{doi:10.1140/epjc/s10052-013-2431-9}}.

\bibitem{Papanestis:2017zcj}
A.~Papanestis and C.~D'Ambrosio.
\newblock {Performance of the LHCb RICH detectors during the LHC Run II}.
\newblock {\em Nucl. Instrum. Meth.}, A876:221--224, 2017.
\newblock \href {http://arxiv.org/abs/1703.08152} {\path{arXiv:1703.08152}},
  \href {http://dx.doi.org/10.1016/j.nima.2017.03.009}
  {\path{doi:10.1016/j.nima.2017.03.009}}.

\bibitem{LHCb:2001ab}
P.~R. Barbosa-Marinho et~al.
\newblock {LHCb muon system technical design report}.
\newblock {\em CERN-LHCC-2001-010}, 2001.

\bibitem{LHCb-DP-2012-004}
R.~Aaij et~al.
\newblock {The \lhcb trigger and its performance in 2011}.
\newblock {\em JINST}, 8:P04022, 2013.
\newblock \href {http://arxiv.org/abs/1211.3055} {\path{arXiv:1211.3055}},
  \href {http://dx.doi.org/10.1088/1748-0221/8/04/P04022}
  {\path{doi:10.1088/1748-0221/8/04/P04022}}.

\bibitem{Gligorov:2012qt}
V.~V. Gligorov and Mike Williams.
\newblock {Efficient, reliable and fast high-level triggering using a bonsai
  boosted decision tree}.
\newblock {\em JINST}, 8:P02013, 2013.
\newblock \href {http://arxiv.org/abs/1210.6861} {\path{arXiv:1210.6861}},
  \href {http://dx.doi.org/10.1088/1748-0221/8/02/P02013}
  {\path{doi:10.1088/1748-0221/8/02/P02013}}.

\bibitem{Sjostrand:2006za}
Torbj\"{o}rn Sj\"{o}strand, Stephen Mrenna, and Peter" Skands.
\newblock {PYTHIA 6.4 physics and manual}.
\newblock {\em JHEP}, 05:026, 2006.
\newblock \href {http://arxiv.org/abs/hep-ph/0603175}
  {\path{arXiv:hep-ph/0603175}}, \href
  {http://dx.doi.org/10.1088/1126-6708/2006/05/026}
  {\path{doi:10.1088/1126-6708/2006/05/026}}.

\bibitem{Sjostrand:2007gs}
Torbj\"{o}rn Sj\"{o}strand, Stephen Mrenna, and Peter" Skands.
\newblock {A brief introduction to PYTHIA 8.1}.
\newblock {\em Comput.Phys.Commun.}, 178:852--867, 2008.
\newblock \href {http://arxiv.org/abs/0710.3820} {\path{arXiv:0710.3820}},
  \href {http://dx.doi.org/10.1016/j.cpc.2008.01.036}
  {\path{doi:10.1016/j.cpc.2008.01.036}}.

\bibitem{LHCb-PROC-2010-056}
I.~Belyaev et~al.
\newblock {Handling of the generation of primary events in Gauss, the LHCb
  simulation framework}.
\newblock {\em {J. Phys. Conf. Ser.}}, 331:032047, 2011.
\newblock \href {http://dx.doi.org/10.1088/1742-6596/331/3/032047}
  {\path{doi:10.1088/1742-6596/331/3/032047}}.

\bibitem{Lange:2001uf}
D.~J. Lange.
\newblock {The EvtGen particle decay simulation package}.
\newblock {\em Nucl. Instrum. Meth.}, A462:152--155, 2001.
\newblock \href {http://dx.doi.org/10.1016/S0168-9002(01)00089-4}
  {\path{doi:10.1016/S0168-9002(01)00089-4}}.

\bibitem{Golonka:2005pn}
Piotr Golonka and Zbigniew Was.
\newblock {PHOTOS Monte Carlo: A precision tool for QED corrections in $Z$ and
  $W$ decays}.
\newblock {\em Eur.Phys.J.}, C45:97--107, 2006.
\newblock \href {http://arxiv.org/abs/hep-ph/0506026}
  {\path{arXiv:hep-ph/0506026}}, \href
  {http://dx.doi.org/10.1140/epjc/s2005-02396-4}
  {\path{doi:10.1140/epjc/s2005-02396-4}}.

\bibitem{Allison:2006ve}
John Allison, K.~Amako, J.~Apostolakis, H.~Araujo, P.A. Dubois, et~al.
\newblock {Geant4 developments and applications}.
\newblock {\em IEEE Trans.Nucl.Sci.}, 53:270, 2006.
\newblock \href {http://dx.doi.org/10.1109/TNS.2006.869826}
  {\path{doi:10.1109/TNS.2006.869826}}.

\bibitem{Agostinelli:2002hh}
S.~Agostinelli et~al.
\newblock {Geant4: A simulation toolkit}.
\newblock {\em Nucl. Instrum. Meth.}, A506:250, 2003.
\newblock \href {http://dx.doi.org/10.1016/S0168-9002(03)01368-8}
  {\path{doi:10.1016/S0168-9002(03)01368-8}}.

\bibitem{LHCb-PROC-2011-006}
M~Clemencic et~al.
\newblock {The \lhcb simulation application, Gauss: Design, evolution and
  experience}.
\newblock {\em {J. Phys. Conf. Ser.}}, 331:032023, 2011.
\newblock \href {http://dx.doi.org/10.1088/1742-6596/331/3/032023}
  {\path{doi:10.1088/1742-6596/331/3/032023}}.

\bibitem{LHCb-PAPER-2016-016}
R.~Aaij et~al.
\newblock {Observation of $\etac(2S)\to \proton\antiproton$ and search for
  $X(3872) \to\proton\antiproton$ decays}.
\newblock {\em Phys. Lett.}, B769:305, 2017.
\newblock \href {http://arxiv.org/abs/1607.06446} {\path{arXiv:1607.06446}},
  \href {http://dx.doi.org/10.1016/j.physletb.2017.03.046}
  {\path{doi:10.1016/j.physletb.2017.03.046}}.

\bibitem{Feng:2019zmn}
Yu~Feng, Jibo He, Jean-Philippe Lansberg, Hua-Sheng Shao, Andrii Usachov, and
  Hong-Fei Zhang.
\newblock {Phenomenological NLO analysis of $\eta_c$ production at the LHC in
  the collider and fixed-target modes}.
\newblock {\em Nucl. Phys.}, B945:114662, 2019.
\newblock \href {http://arxiv.org/abs/1901.09766} {\path{arXiv:1901.09766}},
  \href {http://dx.doi.org/10.1016/j.nuclphysb.2019.114662}
  {\path{doi:10.1016/j.nuclphysb.2019.114662}}.

\bibitem{Zhang:2014ybe}
Hong-Fei Zhang, Zhan Sun, Wen-Long Sang, and Rong Li.
\newblock {Impact of $\eta_c$ hadroproduction data on charmonium production and
  polarization within NRQCD framework}.
\newblock {\em Phys. Rev. Lett.}, 114(9):092006, 2015.
\newblock \href {http://arxiv.org/abs/1412.0508} {\path{arXiv:1412.0508}},
  \href {http://dx.doi.org/10.1103/PhysRevLett.114.092006}
  {\path{doi:10.1103/PhysRevLett.114.092006}}.

\bibitem{LHCb-PAPER-2014-026}
R.~Aaij et~al.
\newblock {Measurement of $\CP$ violation in $\Bs\to\phiz\phiz$ decays}.
\newblock {\em Phys. Rev.}, D90:052011, 2014.
\newblock \href {http://arxiv.org/abs/1407.2222} {\path{arXiv:1407.2222}},
  \href {http://dx.doi.org/10.1103/PhysRevD.90.052011}
  {\path{doi:10.1103/PhysRevD.90.052011}}.

\bibitem{VonHippel:1972fg}
F.~Von~Hippel and C.~Quigg.
\newblock {Centrifugal-barrier effects in resonance partial decay widths,
  shapes, and production amplitudes}.
\newblock {\em Phys. Rev.}, D5:624--638, 1972.
\newblock \href {http://dx.doi.org/10.1103/PhysRevD.5.624}
  {\path{doi:10.1103/PhysRevD.5.624}}.

\bibitem{LHCb-PAPER-2011-002}
R.~Aaij et~al.
\newblock {First observation of $\Bs\to\jpsi f_0(980)$ decays}.
\newblock {\em Phys. Lett.}, B698:115, 2011.
\newblock \href {http://arxiv.org/abs/1102.0206} {\path{arXiv:1102.0206}},
  \href {http://dx.doi.org/10.1016/j.physletb.2011.03.006}
  {\path{doi:10.1016/j.physletb.2011.03.006}}.

\bibitem{Flatte:1976xu}
Stanley~M. Flatte.
\newblock {Coupled - Channel Analysis of the $\pi \eta$ and K anti-K Systems
  Near K anti-K Threshold}.
\newblock {\em Phys. Lett.}, B63:224, 1976.
\newblock \href {http://dx.doi.org/10.1016/0370-2693(76)90654-7}
  {\path{doi:10.1016/0370-2693(76)90654-7}}.

\bibitem{Huang:2003dr}
H.~C. Huang et~al.
\newblock {Evidence for $B \to \phi \phi K$}.
\newblock {\em Phys. Rev. Lett.}, 91:241802, 2003.
\newblock \href {http://arxiv.org/abs/hep-ex/0305068}
  {\path{arXiv:hep-ex/0305068}}, \href
  {http://dx.doi.org/10.1103/PhysRevLett.91.241802}
  {\path{doi:10.1103/PhysRevLett.91.241802}}.

\bibitem{Aubert:2004gc}
Bernard Aubert et~al.
\newblock {Branching fraction measurements of $B \to \eta_c K$ decays}.
\newblock {\em Phys. Rev.}, D70:011101, 2004.
\newblock \href {http://arxiv.org/abs/hep-ex/0403007}
  {\path{arXiv:hep-ex/0403007}}, \href
  {http://dx.doi.org/10.1103/PhysRevD.70.011101}
  {\path{doi:10.1103/PhysRevD.70.011101}}.

\bibitem{ppbar}
Roel Aaij et~al.
\newblock {Measurement of the $\eta_c (1S)$ production cross-section in
  proton-proton collisions via the decay$\eta_c (1S) \rightarrow p \bar{p}$}.
\newblock {\em Eur. Phys. J.}, C75(7):311, 2015.
\newblock \href {http://arxiv.org/abs/1409.3612} {\path{arXiv:1409.3612}},
  \href {http://dx.doi.org/10.1140/epjc/s10052-015-3502-x}
  {\path{doi:10.1140/epjc/s10052-015-3502-x}}.

\bibitem{ShaoPriv}
H.-S. Shao.
\newblock private communication.

\bibitem{CLEO_chic1}
S.~Anderson et~al.
\newblock {Measurements of inclusive $B \to \psi$ production}.
\newblock {\em Phys. Rev. Lett.}, 89:282001, 2002.
\newblock \href {http://arxiv.org/abs/hep-ex/0207059}
  {\path{arXiv:hep-ex/0207059}}, \href
  {http://dx.doi.org/10.1103/PhysRevLett.89.282001}
  {\path{doi:10.1103/PhysRevLett.89.282001}}.

\bibitem{CLEO_chic2}
S.~Chen et~al.
\newblock {Study of $\chi_{c1}$ and $\chi_{c2}$ meson production in $B$ meson
  decays}.
\newblock {\em Phys. Rev.}, D63:031102, 2001.
\newblock \href {http://arxiv.org/abs/hep-ex/0009044}
  {\path{arXiv:hep-ex/0009044}}, \href
  {http://dx.doi.org/10.1103/PhysRevD.63.031102}
  {\path{doi:10.1103/PhysRevD.63.031102}}.

\bibitem{BELLE_chic1}
Kazuo Abe et~al.
\newblock {Observation of $\chi_{c2}$ production in B meson decay}.
\newblock {\em Phys. Rev. Lett.}, 89:011803, 2002.
\newblock \href {http://arxiv.org/abs/hep-ex/0202028}
  {\path{arXiv:hep-ex/0202028}}, \href
  {http://dx.doi.org/10.1103/PhysRevLett.89.011803}
  {\path{doi:10.1103/PhysRevLett.89.011803}}.

\bibitem{BABAR_chic1}
Bernard Aubert et~al.
\newblock {Study of inclusive production of charmonium mesons in $B$ decay}.
\newblock {\em Phys. Rev.}, D67:032002, 2003.
\newblock \href {http://arxiv.org/abs/hep-ex/0207097}
  {\path{arXiv:hep-ex/0207097}}, \href
  {http://dx.doi.org/10.1103/PhysRevD.67.032002}
  {\path{doi:10.1103/PhysRevD.67.032002}}.

\bibitem{L3_chic1}
O.~Adriani et~al.
\newblock {$\chi_c$ production in hadronic Z decays}.
\newblock {\em Phys. Lett.}, B317:467--473, 1993.
\newblock \href {http://dx.doi.org/10.1016/0370-2693(93)91026-J}
  {\path{doi:10.1016/0370-2693(93)91026-J}}.

\bibitem{DELPHI_chic1}
P.~Abreu et~al.
\newblock {$\jpsi$ production in the hadronic decays of the Z}.
\newblock {\em Phys. Lett.}, B341:109--122, 1994.
\newblock \href {http://dx.doi.org/10.1016/0370-2693(94)01385-3}
  {\path{doi:10.1016/0370-2693(94)01385-3}}.

\bibitem{phiphi}
R.~Aaij et~al.
\newblock {Study of charmonium production in ${b}$-hadron decays and first
  evidence for the decay ${{{B}} ^0_{{s}}} \!\rightarrow \phi \phi \phi $}.
\newblock {\em Eur. Phys. J.}, C77(9):609, 2017.
\newblock \href {http://arxiv.org/abs/1706.07013} {\path{arXiv:1706.07013}},
  \href {http://dx.doi.org/10.1140/epjc/s10052-017-5151-8}
  {\path{doi:10.1140/epjc/s10052-017-5151-8}}.

\bibitem{fs}
R.~Aaij et~al.
\newblock {Measurement of $b$-hadron production fractions in $7~\rm{TeV}$ pp
  collisions}.
\newblock {\em Phys. Rev.}, D85:032008, 2012.
\newblock \href {http://arxiv.org/abs/1111.2357} {\path{arXiv:1111.2357}},
  \href {http://dx.doi.org/10.1103/PhysRevD.85.032008}
  {\path{doi:10.1103/PhysRevD.85.032008}}.

\bibitem{fLb}
R.~Aaij et~al.
\newblock {Study of the kinematic dependences of $\Lambda_{b}^{0}$ production
  in pp collisions and a measurement of the $\Lambda_{b}^{0} \to
  \Lambda_{c}^{+}$ $\pi^{-}$ branching fraction}.
\newblock {\em JHEP}, 08:143, 2014.
\newblock \href {http://arxiv.org/abs/1405.6842} {\path{arXiv:1405.6842}},
  \href {http://dx.doi.org/10.1007/JHEP08(2014)143}
  {\path{doi:10.1007/JHEP08(2014)143}}.

\bibitem{Bushm}
Estia~J. Eichten and Chris Quigg.
\newblock {Quarkonium wave functions at the origin}.
\newblock {\em Phys. Rev.}, D52:1726--1728, 1995.
\newblock \href {http://arxiv.org/abs/hep-ph/9503356}
  {\path{arXiv:hep-ph/9503356}}, \href
  {http://dx.doi.org/10.1103/PhysRevD.52.1726}
  {\path{doi:10.1103/PhysRevD.52.1726}}.

\bibitem{ind}
Vineet Kumar and Prashant Shukla.
\newblock {Charmonia production in $p+p$ collisions under NRQCD formalism}.
\newblock {\em J. Phys.}, G44(8):085003, 2017.
\newblock \href {http://arxiv.org/abs/1606.08265} {\path{arXiv:1606.08265}},
  \href {http://dx.doi.org/10.1088/1361-6471/aa7818}
  {\path{doi:10.1088/1361-6471/aa7818}}.

\bibitem{Abulencia:2007bra}
A.~Abulencia et~al.
\newblock {Measurement of $\sigma_{\chi_{c2}}{\cal B}(\chi_{c2} \to \jpsi
  \gamma)/\sigma_{\chi_{c1}} {\cal B}(\chi_{c1} \to \jpsi \gamma)$ in $p
  \bar{p}$ Collisions at $\sqrt{s}$ 1.96 TeV}.
\newblock {\em Phys. Rev. Lett.}, 98:232001, 2007.
\newblock \href {http://arxiv.org/abs/hep-ex/0703028}
  {\path{arXiv:hep-ex/0703028}}, \href
  {http://dx.doi.org/10.1103/PhysRevLett.98.232001}
  {\path{doi:10.1103/PhysRevLett.98.232001}}.

\bibitem{Beneke_B2chicK}
M.~Beneke and L.~Vernazza.
\newblock {$B \to \chi_{cJ} K$ decays revisited}.
\newblock {\em Nucl. Phys.}, B811:155--181, 2009.
\newblock \href {http://arxiv.org/abs/0810.3575} {\path{arXiv:0810.3575}},
  \href {http://dx.doi.org/10.1016/j.nuclphysb.2008.11.025}
  {\path{doi:10.1016/j.nuclphysb.2008.11.025}}.

\bibitem{CDFpsi2s}
R~Aaij et~al.
\newblock {Measurement of $\psi(2S)$ meson production in $pp$ collisions at
  $\sqrt{s}$=7 TeV}.
\newblock {\em Eur. Phys. J.}, C72:2100, 2012.
\newblock \href {http://arxiv.org/abs/1204.1258} {\path{arXiv:1204.1258}},
  \href {http://dx.doi.org/10.1140/epjc/s10052-012-2100-4}
  {\path{doi:10.1140/epjc/s10052-012-2100-4}}.

\bibitem{Butenschoen:2012px}
Mathias Butenschoen and Bernd~A. Kniehl.
\newblock {$\jpsi$ polarization at Tevatron and LHC: Nonrelativistic-QCD
  factorization at the crossroads}.
\newblock {\em Phys.Rev.Lett.}, 108:172002, 2012.
\newblock \href {http://arxiv.org/abs/1201.1872} {\path{arXiv:1201.1872}},
  \href {http://dx.doi.org/10.1103/PhysRevLett.108.172002}
  {\path{doi:10.1103/PhysRevLett.108.172002}}.

\bibitem{Anashin:2015rca}
V.~V. Anashin et~al.
\newblock {Final analysis of KEDR data on $\jpsi$ and $\psi(2S)$ masses}.
\newblock {\em Phys. Lett.}, B749:50--56, 2015.
\newblock \href {http://dx.doi.org/10.1016/j.physletb.2015.07.057}
  {\path{doi:10.1016/j.physletb.2015.07.057}}.

\bibitem{Adams:2005mp}
G.~S. Adams et~al.
\newblock {Measurement of $\Gamma_{ee}(\jpsi)$, $\Gamma_{tot}(\jpsi)$, and
  $\Gamma_{ee}(\psitwos)/Gamma_{ee}(\jpsi)$}f.
\newblock {\em Phys. Rev.}, D73:051103, 2006.
\newblock \href {http://arxiv.org/abs/hep-ex/0512046}
  {\path{arXiv:hep-ex/0512046}}, \href
  {http://dx.doi.org/10.1103/PhysRevD.73.051103}
  {\path{doi:10.1103/PhysRevD.73.051103}}.

\bibitem{Andreotti:2007ur}
M.~Andreotti et~al.
\newblock {Precision Measurements of the Total and Partial Widths of the
  $\psi_{2S}$ Charmonium Meson with A New Complementary-Scan Technique in
  $\bar{p} p$ Annihilations}.
\newblock {\em Phys. Lett.}, B654:74--79, 2007.
\newblock \href {http://arxiv.org/abs/hep-ex/0703012}
  {\path{arXiv:hep-ex/0703012}}, \href
  {http://dx.doi.org/10.1016/j.physletb.2007.08.044}
  {\path{doi:10.1016/j.physletb.2007.08.044}}.

\bibitem{Bai:2002zn}
J.~Z. Bai et~al.
\newblock {A Measurement of \psitwos resonance parameters}.
\newblock {\em Phys. Lett.}, B550:24--32, 2002.
\newblock \href {http://arxiv.org/abs/hep-ph/0209354}
  {\path{arXiv:hep-ph/0209354}}, \href
  {http://dx.doi.org/10.1016/S0370-2693(02)02909-X}
  {\path{doi:10.1016/S0370-2693(02)02909-X}}.

\bibitem{BESIII:2011ab}
M.~Ablikim et~al.
\newblock {Measurements of the mass and width of the $\eta_c$ using
  $\psi^\prime \to \gamma \eta_c$}.
\newblock {\em Phys. Rev. Lett.}, 108:222002, 2012.
\newblock \href {http://arxiv.org/abs/1111.0398} {\path{arXiv:1111.0398}},
  \href {http://dx.doi.org/10.1103/PhysRevLett.108.222002}
  {\path{doi:10.1103/PhysRevLett.108.222002}}.

\bibitem{Ablikim:2005yd}
M.~Ablikim et~al.
\newblock {Precise measurement of spin-averaged $\chi_{cJ}(1P)$ mass using
  photon conversions in $\psi(2S) \to \gamma \chi_{cJ}$}.
\newblock {\em Phys. Rev.}, D71:092002, 2005.
\newblock \href {http://arxiv.org/abs/hep-ex/0502031}
  {\path{arXiv:hep-ex/0502031}}, \href
  {http://dx.doi.org/10.1103/PhysRevD.71.092002}
  {\path{doi:10.1103/PhysRevD.71.092002}}.

\bibitem{Andreotti:2003sk}
M.~Andreotti et~al.
\newblock {Interference study of the $\chi_{c0}(1^{3}P_0)$ in the reaction
  $p\bar{p} \to \pi^0 \pi^0$}.
\newblock {\em Phys. Rev. Lett.}, 91:091801, 2003.
\newblock \href {http://arxiv.org/abs/hep-ex/0308055}
  {\path{arXiv:hep-ex/0308055}}, \href
  {http://dx.doi.org/10.1103/PhysRevLett.91.091801}
  {\path{doi:10.1103/PhysRevLett.91.091801}}.

\bibitem{Andreotti:2005ts}
M.~Andreotti et~al.
\newblock {Measurement of the resonance parameters of the $\chi_1(1^3P_1)$ and
  $\chi_2(1^3P_2)$ states of charmonium formed in $p\bar{p}$ annihilations}.
\newblock {\em Nucl. Phys.}, B717:34--47, 2005.
\newblock \href {http://arxiv.org/abs/hep-ex/0503022}
  {\path{arXiv:hep-ex/0503022}}, \href
  {http://dx.doi.org/10.1016/j.nuclphysb.2005.03.042}
  {\path{doi:10.1016/j.nuclphysb.2005.03.042}}.

\bibitem{Armstrong:1991yk}
T.~A. Armstrong et~al.
\newblock {Study of the $\chi_1$ and $\chi_2$ charmonium states formed in
  anti-p p annihilations}.
\newblock {\em Nucl. Phys.}, B373:35--54, 1992.
\newblock \href {http://dx.doi.org/10.1016/0550-3213(92)90448-K}
  {\path{doi:10.1016/0550-3213(92)90448-K}}.

\bibitem{Ablikim:2012ur}
M.~Ablikim et~al.
\newblock {Study of $\psi(3686)\to\pi^0 h_c, h_c\to\gamma\eta_c$ via $\eta_c$
  exclusive decays}.
\newblock {\em Phys. Rev.}, D86:092009, 2012.
\newblock \href {http://arxiv.org/abs/1209.4963} {\path{arXiv:1209.4963}},
  \href {http://dx.doi.org/10.1103/PhysRevD.86.092009}
  {\path{doi:10.1103/PhysRevD.86.092009}}.

\bibitem{Dobbs:2008ec}
S.~Dobbs et~al.
\newblock {Precision Measurement of the Mass of the $h_c(1P)$ State of
  Charmonium}.
\newblock {\em Phys. Rev. Lett.}, 101:182003, 2008.
\newblock \href {http://arxiv.org/abs/0805.4599} {\path{arXiv:0805.4599}},
  \href {http://dx.doi.org/10.1103/PhysRevLett.101.182003}
  {\path{doi:10.1103/PhysRevLett.101.182003}}.

\bibitem{delAmoSanchez:2011bt}
P.~del Amo~Sanchez et~al.
\newblock {Observation of $\etac (1S)$ and $\etac (2S)$ decays to $\Kp \Km \pip
  \pim \piz$ in two-photon interactions}.
\newblock {\em Phys.Rev.}, D84:012004, 2011.
\newblock \href {http://arxiv.org/abs/1103.3971} {\path{arXiv:1103.3971}},
  \href {http://dx.doi.org/10.1103/PhysRevD.84.012004}
  {\path{doi:10.1103/PhysRevD.84.012004}}.

\bibitem{Ablikim:2013gd}
M.~Ablikim et~al.
\newblock {Evidence for \etactwos in $\psi(3686)→\gamma\KS
  K^{\pm}\pi^{mp}\pip\pim$}.
\newblock {\em Phys. Rev.}, D87(5):052005, 2013.
\newblock \href {http://arxiv.org/abs/1301.1476} {\path{arXiv:1301.1476}},
  \href {http://dx.doi.org/10.1103/PhysRevD.87.052005}
  {\path{doi:10.1103/PhysRevD.87.052005}}.

\bibitem{Godfrey:1985xj}
S.~Godfrey and Nathan Isgur.
\newblock {Mesons in a Relativized Quark Model with Chromodynamics}.
\newblock {\em Phys. Rev.}, D32:189--231, 1985.
\newblock \href {http://dx.doi.org/10.1103/PhysRevD.32.189}
  {\path{doi:10.1103/PhysRevD.32.189}}.

\bibitem{Barnes:2005pb}
T.~Barnes, S.~Godfrey, and E.~S. Swanson.
\newblock {Higher charmonia}.
\newblock {\em Phys. Rev.}, D72:054026, 2005.
\newblock \href {http://arxiv.org/abs/hep-ph/0505002}
  {\path{arXiv:hep-ph/0505002}}, \href
  {http://dx.doi.org/10.1103/PhysRevD.72.054026}
  {\path{doi:10.1103/PhysRevD.72.054026}}.

\bibitem{Olsen:2017bmm}
Stephen~Lars Olsen, Tomasz Skwarnicki, and Daria Zieminska.
\newblock {Nonstandard heavy mesons and baryons: Experimental evidence}.
\newblock {\em Rev. Mod. Phys.}, 90(1):015003, 2018.
\newblock \href {http://arxiv.org/abs/1708.04012} {\path{arXiv:1708.04012}},
  \href {http://dx.doi.org/10.1103/RevModPhys.90.015003}
  {\path{doi:10.1103/RevModPhys.90.015003}}.

\bibitem{Davies:1995db}
C.~T.~H. Davies, K.~Hornbostel, G.~P. Lepage, A.~J. Lidsey, J.~Shigemitsu, and
  J.~H. Sloan.
\newblock {Precision charmonium spectroscopy from lattice QCD}.
\newblock {\em Phys. Rev.}, D52:6519--6529, 1995.
\newblock \href {http://arxiv.org/abs/hep-lat/9506026}
  {\path{arXiv:hep-lat/9506026}}, \href
  {http://dx.doi.org/10.1103/PhysRevD.52.6519}
  {\path{doi:10.1103/PhysRevD.52.6519}}.

\bibitem{Trottier:1996ce}
Howard~D. Trottier.
\newblock {Quarkonium spin structure in lattice NRQCD}.
\newblock {\em Phys. Rev.}, D55:6844--6851, 1997.
\newblock \href {http://arxiv.org/abs/hep-lat/9611026}
  {\path{arXiv:hep-lat/9611026}}, \href
  {http://dx.doi.org/10.1103/PhysRevD.55.6844}
  {\path{doi:10.1103/PhysRevD.55.6844}}.

\bibitem{Okamoto:2001jb}
M.~Okamoto et~al.
\newblock {Charmonium spectrum from quenched anisotropic lattice QCD}.
\newblock {\em Phys. Rev.}, D65:094508, 2002.
\newblock \href {http://arxiv.org/abs/hep-lat/0112020}
  {\path{arXiv:hep-lat/0112020}}, \href
  {http://dx.doi.org/10.1103/PhysRevD.65.094508}
  {\path{doi:10.1103/PhysRevD.65.094508}}.

\bibitem{Peset:2018jkf}
Clara Peset, Antonio Pineda, and Jorge Segovia.
\newblock {P-wave heavy quarkonium spectrum with
  next-to-next-to-next-to-leading logarithmic accuracy}.
\newblock {\em Phys. Rev.}, D98(9):094003, 2018.
\newblock \href {http://arxiv.org/abs/1809.09124} {\path{arXiv:1809.09124}},
  \href {http://dx.doi.org/10.1103/PhysRevD.98.094003}
  {\path{doi:10.1103/PhysRevD.98.094003}}.

\bibitem{Pineda:2011dg}
Antonio Pineda.
\newblock {Review of Heavy Quarkonium at weak coupling}.
\newblock {\em Prog. Part. Nucl. Phys.}, 67:735--785, 2012.
\newblock \href {http://arxiv.org/abs/1111.0165} {\path{arXiv:1111.0165}},
  \href {http://dx.doi.org/10.1016/j.ppnp.2012.01.038}
  {\path{doi:10.1016/j.ppnp.2012.01.038}}.

\bibitem{Brambilla:2010ey}
Nora Brambilla, Pablo Roig, and Antonio Vairo.
\newblock {Precise determination of the $\eta_c$ mass and width in the
  radiative $\jpsi \to \eta_c$ gamma decay}.
\newblock {\em AIP Conf. Proc.}, 1343:418--420, 2011.
\newblock \href {http://arxiv.org/abs/1012.0773} {\path{arXiv:1012.0773}},
  \href {http://dx.doi.org/10.1063/1.3575048} {\path{doi:10.1063/1.3575048}}.

\bibitem{Aaij:2019evc}
Roel Aaij et~al.
\newblock {Near-threshold $D\bar{D}$ spectroscopy and observation of a new
  charmonium state}.
\newblock 2019.
\newblock \href {http://arxiv.org/abs/1903.12240} {\path{arXiv:1903.12240}}.

\bibitem{Suzuki:2005ha}
Mahiko Suzuki.
\newblock {The X(3872) boson: Molecule or charmonium}.
\newblock {\em Phys. Rev.}, D72:114013, 2005.
\newblock \href {http://arxiv.org/abs/hep-ph/0508258}
  {\path{arXiv:hep-ph/0508258}}, \href
  {http://dx.doi.org/10.1103/PhysRevD.72.114013}
  {\path{doi:10.1103/PhysRevD.72.114013}}.

\bibitem{PDG2019}
M.~Tanabashi et~al.
\newblock {Review of Particle Physics}.
\newblock {\em Phys. Rev.}, D98(3):030001, 2018.
\newblock \href {http://dx.doi.org/10.1103/PhysRevD.98.030001}
  {\path{doi:10.1103/PhysRevD.98.030001}}.

\bibitem{Blitz:2015nra}
Samuel~H. Blitz and Richard~F. Lebed.
\newblock {Tetraquark Cusp Effects from Diquark Pair Production}.
\newblock {\em Phys. Rev.}, D91(9):094025, 2015.
\newblock \href {http://arxiv.org/abs/1503.04802} {\path{arXiv:1503.04802}},
  \href {http://dx.doi.org/10.1103/PhysRevD.91.094025}
  {\path{doi:10.1103/PhysRevD.91.094025}}.

\bibitem{Swanson:2015bsa}
E.~S. Swanson.
\newblock {Cusps and Exotic Charmonia}.
\newblock {\em Int. J. Mod. Phys.}, E25(07):1642010, 2016.
\newblock \href {http://arxiv.org/abs/1504.07952} {\path{arXiv:1504.07952}},
  \href {http://dx.doi.org/10.1142/S0218301316420106}
  {\path{doi:10.1142/S0218301316420106}}.

\bibitem{Aaij:2013qja}
R~Aaij et~al.
\newblock {Measurement of the \Lb, \Xibm and \Omegab baryon masses}.
\newblock {\em Phys.Rev.Lett.}, 110:182001, 2013.
\newblock \href {http://arxiv.org/abs/1302.1072} {\path{arXiv:1302.1072}},
  \href {http://dx.doi.org/10.1103/PhysRevLett.110.182001}
  {\path{doi:10.1103/PhysRevLett.110.182001}}.

\bibitem{Lees:2014iua}
J.~P. Lees et~al.
\newblock {Dalitz plot analysis of $\eta_c \to K^+ K^- \eta$ and $\eta_c \to
  K^+ K^- \pi^0$ in two-photon interactions}.
\newblock {\em Phys. Rev.}, D89(11):112004, 2014.
\newblock \href {http://arxiv.org/abs/1403.7051} {\path{arXiv:1403.7051}},
  \href {http://dx.doi.org/10.1103/PhysRevD.89.112004}
  {\path{doi:10.1103/PhysRevD.89.112004}}.

\bibitem{Aaij:2015cxj}
Roel Aaij et~al.
\newblock {Measurement of the $B_s^0 \to \phi \phi$ branching fraction and
  search for the decay $B^0 \to \phi \phi$}.
\newblock {\em JHEP}, 10:053, 2015.
\newblock \href {http://arxiv.org/abs/1508.00788} {\path{arXiv:1508.00788}},
  \href {http://dx.doi.org/10.1007/JHEP10(2015)053}
  {\path{doi:10.1007/JHEP10(2015)053}}.

\bibitem{Bartsch:2008ps}
Matthaus Bartsch, Gerhard Buchalla, and Christina Kraus.
\newblock {$B \to V(L) V(L)$ decays at Next-to-Leading Order in QCD}.
\newblock 2008.
\newblock \href {http://arxiv.org/abs/0810.0249} {\path{arXiv:0810.0249}}.

\bibitem{Beneke:2006hg}
Martin Beneke, Johannes Rohrer, and Deshan Yang.
\newblock {Branching fractions, polarisation and asymmetries of $\B \to V V$
  decays}.
\newblock {\em Nucl.Phys.}, B774:64--101, 2007.
\newblock \href {http://arxiv.org/abs/hep-ph/0612290}
  {\path{arXiv:hep-ph/0612290}}, \href
  {http://dx.doi.org/10.1016/j.nuclphysb.2007.03.020}
  {\path{doi:10.1016/j.nuclphysb.2007.03.020}}.

\bibitem{Cheng:2009mu}
Hai-Yang Cheng and Chun-Khiang Chua.
\newblock {QCD factorization for charmless hadronic \Bs decays revisited}.
\newblock {\em Phys.Rev.}, D80:114026, 2009.
\newblock \href {http://arxiv.org/abs/0910.5237} {\path{arXiv:0910.5237}},
  \href {http://dx.doi.org/10.1103/PhysRevD.80.114026}
  {\path{doi:10.1103/PhysRevD.80.114026}}.

\bibitem{Kagan:2004uw}
Alexander~L. Kagan.
\newblock {Polarization in $\B \to VV$ decays}.
\newblock {\em Phys.Lett.}, B601:151--163, 2004.
\newblock \href {http://arxiv.org/abs/hep-ph/0405134}
  {\path{arXiv:hep-ph/0405134}}, \href
  {http://dx.doi.org/10.1016/j.physletb.2004.09.030}
  {\path{doi:10.1016/j.physletb.2004.09.030}}.

\bibitem{Datta:2007qb}
Alakabha Datta, Andrei~V. Gritsan, David London, Makiko Nagashima, and
  Alejandro Szynkman.
\newblock {Testing explanations of the $\B \to \phi K^*$ polarization puzzle}.
\newblock {\em Phys.Rev.}, D76:034015, 2007.
\newblock \href {http://arxiv.org/abs/0705.3915} {\path{arXiv:0705.3915}},
  \href {http://dx.doi.org/10.1103/PhysRevD.76.034015}
  {\path{doi:10.1103/PhysRevD.76.034015}}.

\bibitem{Chen:2005mka}
Chuan-Hung Chen and Chao-Qiang Geng.
\newblock {Scalar interactions to the polarizations of $\B \to \phi K^*$}.
\newblock {\em Phys.Rev.}, D71:115004, 2005.
\newblock \href {http://arxiv.org/abs/hep-ph/0504145}
  {\path{arXiv:hep-ph/0504145}}, \href
  {http://dx.doi.org/10.1103/PhysRevD.71.115004}
  {\path{doi:10.1103/PhysRevD.71.115004}}.

\bibitem{Huang:2005qb}
Chao-Shang Huang, Pyungwon Ko, Xiao-Hong Wu, and Ya-Dong Yang.
\newblock {MSSM anatomy of the polarization puzzle in $B \to \phi K^{*}$
  decays}.
\newblock {\em Phys.Rev.}, D73:034026, 2006.
\newblock \href {http://arxiv.org/abs/hep-ph/0511129}
  {\path{arXiv:hep-ph/0511129}}, \href
  {http://dx.doi.org/10.1103/PhysRevD.73.034026}
  {\path{doi:10.1103/PhysRevD.73.034026}}.

\bibitem{LHCb-PAPER-2013-007}
R.~Aaij et~al.
\newblock {First measurement of the $\CP$-violating phase in $\Bs\to\phiz\phiz$
  decays}.
\newblock {\em Phys. Rev. Lett.}, 110:241802, 2013.
\newblock \href {http://arxiv.org/abs/1303.7125} {\path{arXiv:1303.7125}},
  \href {http://dx.doi.org/10.1103/PhysRevLett.110.241802}
  {\path{doi:10.1103/PhysRevLett.110.241802}}.

\bibitem{Ali:2007ff}
Ahmed Ali, Gustav Kramer, Ying Li, Cai-Dian Lu, Yue-Long Shen, et~al.
\newblock {Charmless non-leptonic $B_s$ decays to $PP$, $PV$ and $VV$ final
  states in the pQCD approach}.
\newblock {\em Phys.Rev.}, D76:074018, 2007.
\newblock \href {http://arxiv.org/abs/hep-ph/0703162}
  {\path{arXiv:hep-ph/0703162}}, \href
  {http://dx.doi.org/10.1103/PhysRevD.76.074018}
  {\path{doi:10.1103/PhysRevD.76.074018}}.

\bibitem{Acosta:2005eu}
D.~Acosta et~al.
\newblock {First evidence for $\Bs \to \phi \phi$ decay and measurements of
  branching ratio and $A_{\CP}$ for $\Bp \to \phi \Kp$}.
\newblock {\em Phys.Rev.Lett.}, 95:031801, 2005.
\newblock \href {http://arxiv.org/abs/hep-ex/0502044}
  {\path{arXiv:hep-ex/0502044}}, \href
  {http://dx.doi.org/10.1103/PhysRevLett.95.031801}
  {\path{doi:10.1103/PhysRevLett.95.031801}}.

\bibitem{Aaltonen:2011rs}
T.~Aaltonen et~al.
\newblock {Measurement of polarization and search for CP-violation in $B_s^0
  \to \phi\phi$ decays}.
\newblock {\em Phys. Rev. Lett.}, 107:261802, 2011.
\newblock \href {http://arxiv.org/abs/1107.4999} {\path{arXiv:1107.4999}},
  \href {http://dx.doi.org/10.1103/PhysRevLett.107.261802}
  {\path{doi:10.1103/PhysRevLett.107.261802}}.

\bibitem{Abe:1999ze}
Kenji Abe et~al.
\newblock {Search for charmless hadronic decays of \B mesons with the SLD
  detector}.
\newblock {\em Phys.Rev.}, D62:071101, 2000.
\newblock \href {http://arxiv.org/abs/hep-ex/9910050}
  {\path{arXiv:hep-ex/9910050}}, \href
  {http://dx.doi.org/10.1103/PhysRevD.62.071101}
  {\path{doi:10.1103/PhysRevD.62.071101}}.

\bibitem{LHCb-PAPER-2011-018}
R.~Aaij et~al.
\newblock {Measurement of $\bquark$ hadron production fractions in 7\,TeV
  $\proton\proton$ collisions}.
\newblock {\em Phys. Rev.}, D85:032008, 2012.
\newblock \href {http://arxiv.org/abs/1111.2357} {\path{arXiv:1111.2357}},
  \href {http://dx.doi.org/10.1103/PhysRevD.85.032008}
  {\path{doi:10.1103/PhysRevD.85.032008}}.

\bibitem{LHCb-PAPER-2012-037}
R.~Aaij et~al.
\newblock {Measurement of the fragmentation fraction ratio $f_s/f_d$ and its
  dependence on $\B$ meson kinematics}.
\newblock {\em JHEP}, 04:001, 2013.
\newblock \href {http://arxiv.org/abs/1301.5286} {\path{arXiv:1301.5286}},
  \href {http://dx.doi.org/10.1007/JHEP04(2013)001}
  {\path{doi:10.1007/JHEP04(2013)001}}.

\bibitem{LHCb-PAPER-2014-004}
R.~Aaij et~al.
\newblock {Study of the kinematic dependences of $\Lb$ production in
  $\proton\proton$ collisions and a measurement of the $\Lb\to\Lc\pim$
  branching fraction}.
\newblock {\em JHEP}, 08:143, 2014.
\newblock \href {http://arxiv.org/abs/1405.6842} {\path{arXiv:1405.6842}},
  \href {http://dx.doi.org/10.1007/JHEP08(2014)143}
  {\path{doi:10.1007/JHEP08(2014)143}}.

\bibitem{Bisello:1990re}
D.~Bisello et~al.
\newblock {Study of the $\etac$ decays}.
\newblock {\em Nucl.Phys.}, B350:1--24, 1991.
\newblock \href {http://dx.doi.org/10.1016/0550-3213(91)90251-R}
  {\path{doi:10.1016/0550-3213(91)90251-R}}.

\bibitem{Bai:2003tr}
J.Z. Bai et~al.
\newblock {Measurement of branching ratios for $\etac$ hadronic decays}.
\newblock {\em Phys.Lett.}, B578:16--22, 2004.
\newblock \href {http://arxiv.org/abs/hep-ex/0308073}
  {\path{arXiv:hep-ex/0308073}}, \href
  {http://dx.doi.org/10.1016/j.physletb.2003.10.042}
  {\path{doi:10.1016/j.physletb.2003.10.042}}.

\bibitem{Ablikim:2005yi}
M.~Ablikim et~al.
\newblock {Experimental study of $\eta_c$ decays into vector-vector final
  states}.
\newblock {\em Phys.Rev.}, D72:072005, 2005.
\newblock \href {http://arxiv.org/abs/hep-ex/0507100}
  {\path{arXiv:hep-ex/0507100}}, \href
  {http://dx.doi.org/10.1103/PhysRevD.72.072005}
  {\path{doi:10.1103/PhysRevD.72.072005}}.

\bibitem{LHCb-PAPER-2015-028}
R.~Aaij et~al.
\newblock {Measurement of the $\Bs\to\phi\phi$ branching fraction and search
  for the decay $\Bz\to\phi\phi$}.
\newblock {\em JHEP}, 10:053, 2015.
\newblock \href {http://arxiv.org/abs/1508.00788} {\path{arXiv:1508.00788}},
  \href {http://dx.doi.org/10.1007/JHEP10(2015)053}
  {\path{doi:10.1007/JHEP10(2015)053}}.

\bibitem{Aaij:2013qqa}
R~Aaij et~al.
\newblock {Measurement of the fragmentation fraction ratio $f_{s}/f_{d}$ and
  its dependence on $B$ meson kinematics}.
\newblock {\em JHEP}, 04:001, 2013.
\newblock \href {http://arxiv.org/abs/1301.5286} {\path{arXiv:1301.5286}},
  \href {http://dx.doi.org/10.1007/JHEP04(2013)001}
  {\path{doi:10.1007/JHEP04(2013)001}}.

\bibitem{Cowan:2010js}
Glen Cowan, Kyle Cranmer, Eilam Gross, and Ofer Vitells.
\newblock {Asymptotic formulae for likelihood-based tests of new physics}.
\newblock {\em Eur.Phys.J.}, C71:1554, 2011.
\newblock \href {http://arxiv.org/abs/1007.1727} {\path{arXiv:1007.1727}},
  \href {http://dx.doi.org/10.1140/epjc/s10052-011-1554-0,
  10.1140/epjc/s10052-013-2501-z} {\path{doi:10.1140/epjc/s10052-011-1554-0,
  10.1140/epjc/s10052-013-2501-z}}.

\bibitem{Adolph:2008vn}
C.~Adolph et~al.
\newblock {Measurement of the $\eta \to 3 \pi^0$ Dalitz plot distribution with
  the WASA detector at COSY}.
\newblock {\em Phys.Lett.}, B677:24--29, 2009.
\newblock \href {http://arxiv.org/abs/0811.2763} {\path{arXiv:0811.2763}},
  \href {http://dx.doi.org/10.1016/j.physletb.2009.03.063}
  {\path{doi:10.1016/j.physletb.2009.03.063}}.

\end{thebibliography}
\end{singlespace}



\begin{otherlanguage}{french}
\begin{singlespace}

\definecolor{SchoolColor}{rgb}{0.145,0.666,1} 

\usetikzlibrary{calc}
\thispagestyle{empty}

\makeatletter
\newcommand*\mysize{%
  \@setfontsize\mysize{12.5}{12.0}%
}
\makeatother

\newpage
\thispagestyle{empty}
\begin{tikzpicture}[remember picture, overlay]
\node [anchor=north west, shift={(1.2 cm,-0.2cm)}] at (current page.north west) {\protect\includegraphics[width=7.5cm]{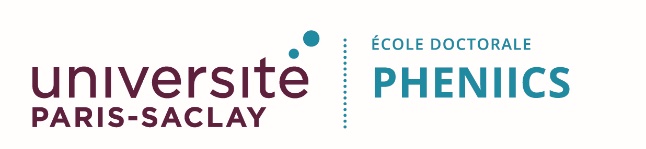}};
 \mysize 
\node [anchor=north, yshift=-2.1 cm, text width=18cm, inner sep=.3cm] (resume) at (current page.north) {
\begin{minipage}{\linewidth}  
\justify{     {\bf Titre:} \'{E}tudes de production des \'{e}tats de charmonium avec leurs d\'{e}sint\'{e}grations vers des hadrons dans l'exp\'{e}rience LHCb\\
{\bf Mots cl\'{e}s:} 
\textit{Analyse de données, Production et désintégrations de charmonium, Désintégrations du méson Bs, Tests QCD, Spectroscopie}\\  
{\bf R\'{e}sum\'{e}:}  
Les \'{e}tudes des propri\'{e}t\'{e}s et du m\'{e}canisme de la production du charmonium ont d\'{e}but\'{e} avec la d\'{e}couverte du m\'{e}son $J/\psi$. Depuis plus de 40 ans, le m\'{e}canisme de production de charmonium n'est toujours pas clair.
Les \'{e}tats de charmonium sont copieusement produits dans les collisionneurs hadroniques, ce qui permet d'\'{e}tudier syst\'{e}matiquement leurs paramètres de r\'{e}sonance, leur production et leurs d\'{e}sint\'{e}grations. En d\'{e}pit des taux de production \'{e}lev\'{e}s, de nombreux \'{e}tats de charmonium sont peu \'{e}tudi\'{e}s suite aux difficult\'{e}s de leur reconstruction avec le bruit de fond important. 
L’exp\'{e}rience LHCb offre une opportunit\'{e} unique d’\'{e}tudier les \'{e}tats S et P de charmonium en utilisant leurs d\'{e}sint\'{e}grations vers des hadrons, et en particulier la production des \'{e}tats $\eta_c$ et $\chi_c$. Selon le formalisme de la QCD non relativiste (NRQCD), les \'{e}l\'{e}ments de matrice d\'{e}crivant l’hadronisation des \'{e}tats S (ou P) du charmonium sont li\'{e}s. Par cons\'{e}quent, les mesures LHCb fournissent les nouveaux tests rigoureux de NRQCD. Dans le cadre de la thèse, la première mesure de la section efficace de production de l’\'{e}tat $\eta_c(1S)$ \`{a} $\sqrt{s}=13\,TeV$ et la mesure la plus pr\'{e}cise de la masse sont effectu\'{e}s, avec la d\'{e}sint\'{e}gration de l'\'{e}tat $\eta_c(1S)$ vers $p\bar{p}$. De plus, la production des \'{e}tats $\chi_c$ et $\eta_c(2S)$ dans les d\'{e}sint\'{e}grations des hadrons $b$ est \'{e}tudi\'{e}e en utilisant leurs d\'{e}sint\'{e}grations vers $\phi\phi$. Les r\'{e}sultats obtenus sont confront\'{e}s aux pr\'{e}dictions de modèles th\'{e}oriques. L’analyse ph\'{e}nom\'{e}nologique originale d\'{e}montre que la description de la production de charmonium dans les collisions hadroniques et de la production dans les d\'{e}sint\'{e}grations inclusives des hadrons $b$ dans la gamme entière des impulsions transverses demeure un d\'{e}fi.
}
\end{minipage}
};

\node [anchor=north, yshift=-0.3 cm, text width=18cm, inner sep=.3cm] (abstract) at (resume.south) { 
\begin{minipage}{\linewidth}
    
\justify{     {\bf Title:} Study of charmonium production using decays to hadronic final states with the LHCb experiment\\
{\bf Key words:} 
\textit{Data analysis, Production and decays of charmonium, Decays of the Bs meson, QCD tests, Spectroscopy} \\
{\bf Abstract:}
Studies of charmonium properties and production mechanism started with the discovery of $J/\psi$ meson. Since more than 40 years the charmonium production mechanism is still not entirely understood.
Following the era of investigations at $e^+e^-$ machines, nowadays, charmonium states are copiously produced at hadron colliders, that allows systematic studies of their resonance parameters, production observables and decays. Despite large production rates, many charmonium states are barely studied due to the complications of their reconstruction against a large background level.
The LHCb experiment provides a unique opportunity to study S-wave and P-wave charmonium states using their decays to hadrons. This allows measuring production observables of $\eta_c$ and $\chi_c$ charmonium states. According to the theoretical formalism of Non-Relativistic QCD (NRQCD), the production observables of the same wave charmonium states are linked. Hence, the LHCb measurements provide a series of stringent tests of NRQCD. In the framework of this thesis, the first measurement of the $\eta_c(1S)$ differential production cross-section at $\sqrt{s}=13\,TeV$ and the most precise to date single mass measurement are performed, where the $\eta_c(1S)$ state is reconstructed via its decay to $p\bar{p}$. In addition, the production of the $\chi_c$ and $\eta_c(2S)$ states in $b$-hadron decays is studied using decays to $\phi\phi$. The obtained results are confronted with existing theory predictions. The original phenomenological analysis concludes that the description of charmonium prompt production and production in inclusive $b$-hadron decays in an entire range of transverse momentum remains a challenge.  
}
\end{minipage}
};
    
\draw[line width=1 pt, violet!80!red] (resume.south west) -- (resume.north west) -- (resume.north east) -- (resume.south east) -- (resume.south west);
\draw[line width=1 pt, violet!80!red] (abstract.south west) -- (abstract.north west) -- (abstract.north east) -- (abstract.south east) -- (abstract.south west);

\node [anchor=south west, violet!80!red, shift={(1.2 cm,0.5cm)}, inner sep=0.2pt] at (current page.south west) {
\begin{minipage}{12cm}
{\bf Universit\'{e} Paris-Saclay} \\
Espace Technologique / Immeuble Discovery \\
Route de l'Orme aux Merisiers RD 128 / 91190 Saint-Aubin, France 
\end{minipage}
};

\node [anchor=south east, violet!80!red!80!black, shift={(-1.5 cm,0.5cm)}, inner sep=0pt] at (current page.south east) {\protect\includegraphics[width=1.6 cm]{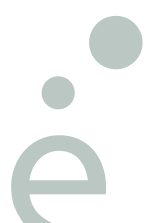}};
\end{tikzpicture}

\end{singlespace}
\end{otherlanguage}
\end{document}